\documentclass[11pt, oneside, a4paper]{book}

\usepackage{color}
\usepackage{graphicx}
\usepackage{epstopdf}
\usepackage{fancyhdr}
\usepackage[square, sort&compress, super, comma, numbers]{natbib}
\usepackage{amsmath}
\usepackage{amssymb}
\usepackage[mathscr]{eucal}
\usepackage{mathrsfs}
\usepackage{bm}
\usepackage[a4paper, textwidth=16cm, textheight=23cm, centering, headsep=.5cm]{geometry} 	
\usepackage{setspace}
\usepackage{sectsty}
\usepackage{rotating}
\usepackage{subfigure}
\usepackage{framed}

\usepackage{commath}

\usepackage{yfonts}

\usepackage{pstool}
\usepackage{grffile} 

\usepackage{array}
\usepackage{booktabs}

\usepackage{pifont}
\usepackage{xcolor,cancel}

\usepackage{dsfont}
\usepackage{relsize}
\usepackage{wrapfig, framed, caption}
\newcommand\hcancel[2][black]{\setbox0=\hbox{$#2$}%
\rlap{\raisebox{.45\ht0}{\textcolor{#1}{\rule{\wd0}{0.5pt}}}}#2}

\allowdisplaybreaks 			

\makeatletter
\def\overbracket{\@ifnextchar [ {\@overbracket} {\@overbracket
[\@bracketheight]}}
\def\@overbracket[#1]{\@ifnextchar [ {\@over@bracket[#1]}
{\@over@bracket[#1][0.3em]}}
\def\@over@bracket[#1][#2]#3{
\mathop {\vbox {\m@th \ialign {##\crcr \noalign {\kern 3\p@
\nointerlineskip }\downbracketfill {#1}{#2}
                              \crcr \noalign {\kern 3\p@ }
                              \crcr  $\hfil \displaystyle {#3}\hfil $%
                              \crcr} }}\limits}
\def\downbracketfill#1#2{$\m@th \setbox \z@ \hbox {$\braceld$}
                  \edef\@bracketheight{\the\ht\z@}\downbracketend{#1}{#2}
                  \leaders \vrule \@height #1 \@depth \z@ \hfill
                  \leaders \vrule \@height #1 \@depth \z@ \hfill
\downbracketend{#1}{#2}$}
\def\downbracketend#1#2{\vrule depth #2 width #1\relax}

\makeatletter
\def\underbracket{%
  \@ifnextchar [ %
    {\@underbracket}%
    {\@underbracket [\@bracketheight]}}
\def\@underbracket[#1]{%
  \@ifnextchar [ %
    {\@under@bracket[#1]}%
    {\@under@bracket[#1][0.4em]}}
\def\@under@bracket[#1][#2]#3{
  \mathop {%
    \vtop {%
      \m@th \ialign {%
	##\crcr $\hfil \displaystyle {#3}\hfil $%
       \crcr \noalign %
       {\kern 3\p@ \nointerlineskip }%
	\upbracketfill {#1}{#2}
       \crcr \noalign %
       {\kern 3\p@ }%
     }%
   }%
  }%
  \limits%
}
\def\upbracketfill#1#2{%
  $\m@th \setbox \z@ \hbox {$\braceld$}
  \edef\@bracketheight{\the\ht\z@}\bracketend{#1}{#2}
  \leaders \vrule \@height #1 \@depth \z@ \hfill 
  \leaders \vrule \@height #1 \@depth \z@ \hfill%
  \bracketend{#1}{#2}$%
}
\def\bracketend#1#2{\vrule height #2 width #1\relax}

\makeatletter
\def\Ddots{\mathinner{\mkern1mu\raise\p@
\vbox{\kern7\p@\hbox{.}}\mkern2mu
\raise4\p@\hbox{.}\mkern2mu\raise7\p@\hbox{.}\mkern1mu}}
\makeatother

\newcommand{\bra}[1]{\langle #1 |}
\newcommand{\ket}[1]{| #1 \rangle}

\newcommand{\expect}[1]{\langle #1 \rangle}

\newcommand{\mi}{\mathrm{i}}

\newcommand{\tr}{\text{Tr}}
\newcommand{\sigmaop}[1]{\hat{\sigma}_{#1}}
\newcommand{\sigmawigop}[1]{\hat{\tilde{\sigma}}_{#1}}
\newcommand{\I}{\text{i}}

\newcommand{\sech}{\text{sech}}
\newcommand{\sinc}{\text{sinc}}

\newcommand{\DeltaS}{\Delta_\text{S}}
\newcommand{\DeltaAS}{\Delta_\text{AS}}
\newcommand{\CS}{C_\text{S}}
\newcommand{\CAS}{C_\text{AS}}

\newfont{\suet}{suet14}

\newcommand{\cs}{Cs} 

\newcommand{\first}{$1^{\text{st}}\,$}

\newcommand{\cmark}{\ding{51}}
\newcommand{\xmark}{\ding{55}}

\newcommand{\scd}[1]{\textit{scd}}
\newcommand{\sd}[1]{\textit{sd}}
\newcommand{\cd}[1]{\textit{cd}}
\newcommand{\diode}[1]{\textit{d}}		

\newcommand{\etamem}{$\eta_\text{mem}$} 	
\newcommand{\etain}{$\eta_\text{in}$} 		
\newcommand{\etaret}{$\eta_\text{ret}$} 		

\newcommand{\gtwo}{$g^{(2)}$}			
\newcommand{\hereff}[1]{$\eta_\text{her}$}	
\newcommand{\Nin}{$N_\text{in}$}			

\newcommand{\tisa}{Ti:Sa}					
\newcommand{\hsp}[1]{HSP}	 				
\newcommand{\hsps}[1]{HSPs}	 				
\newcommand{\coh}[1]{\text{c.s.}}  				
\newcommand{\pockels}[1]{\text{P.C.}}  			
\newcommand{\spcmdt}[1]{$\text{D}_\text{T}$}		
\newcommand{\spcmdh}[1]{$\text{D}_\text{H}$}		
\newcommand{\spcmdv}[1]{$\text{D}_\text{V}$}		
\newcommand{\etapump}{$\eta_\text{pump}$}

\newcommand{\mW}{\, \text{mW}}	
\newcommand{\muW}{\, \mu \text{W}}	

\newcommand{\nm}{\, \text{nm}}
\newcommand{\mum}{\, \mu \text{m}}
\newcommand{\mm}{\, \text{mm}}
\newcommand{\cm}{\, \text{cm}}
\newcommand{\m}{\, \text{m}}

\newcommand{\ps}{\, \text{ps}}
\newcommand{\ns}{\, \text{ns}}
\newcommand{\mus}{\, \mu \text{s}}

\newcommand{\Hz}{\, \text{Hz}}
\newcommand{\kHz}{\, \text{kHz}}	
\newcommand{\MHz}{\, \text{MHz}}	
\newcommand{\GHz}{\, \text{GHz}}	

\newcommand{\pcW}{\, \frac{\%}{\text{W}}}

\newcommand{\degC}{\, ^{\circ} \text{C}}	

\newcommand{\dB}{\, \text{dB}}	

\newcommand{\ppp}{\, \gamma/\text{pulse}} 		

\newcommand{\hpol}{$|H\rangle$}
\newcommand{\vpol}{$|V\rangle$}
\newcommand{\dpol}{$|+\rangle$}
\newcommand{\apol}{$|-\rangle$}
\newcommand{\rpol}{$|R\rangle$}

\chapterfont{\centering}

\fancypagestyle{plain}{
\fancyhead{}
}

\newenvironment{abstract}
{\thispagestyle{plain}
\begin{center}
\vspace*{1.5cm}
\Large{\textbf{Abstract}}
\end{center}
\setstretch{1}
\normalsize
\vspace{0.5cm}}
{}

\begin{document}

\pdfoutput=1


\frontmatter


\begin{titlepage}
\begin{spacing}{1.2}

\begin{center}
\Huge{\textbf{Room temperature caesium quantum memory for quantum information applications}}
\end{center}

\vfill
\begin{center}
\Large{Patrick Michelberger \\ Balliol College, Oxford}
\end{center}

\vfill
\begin{figure}[h]
\begin{center}
\includegraphics[width= 0.35 \textwidth]{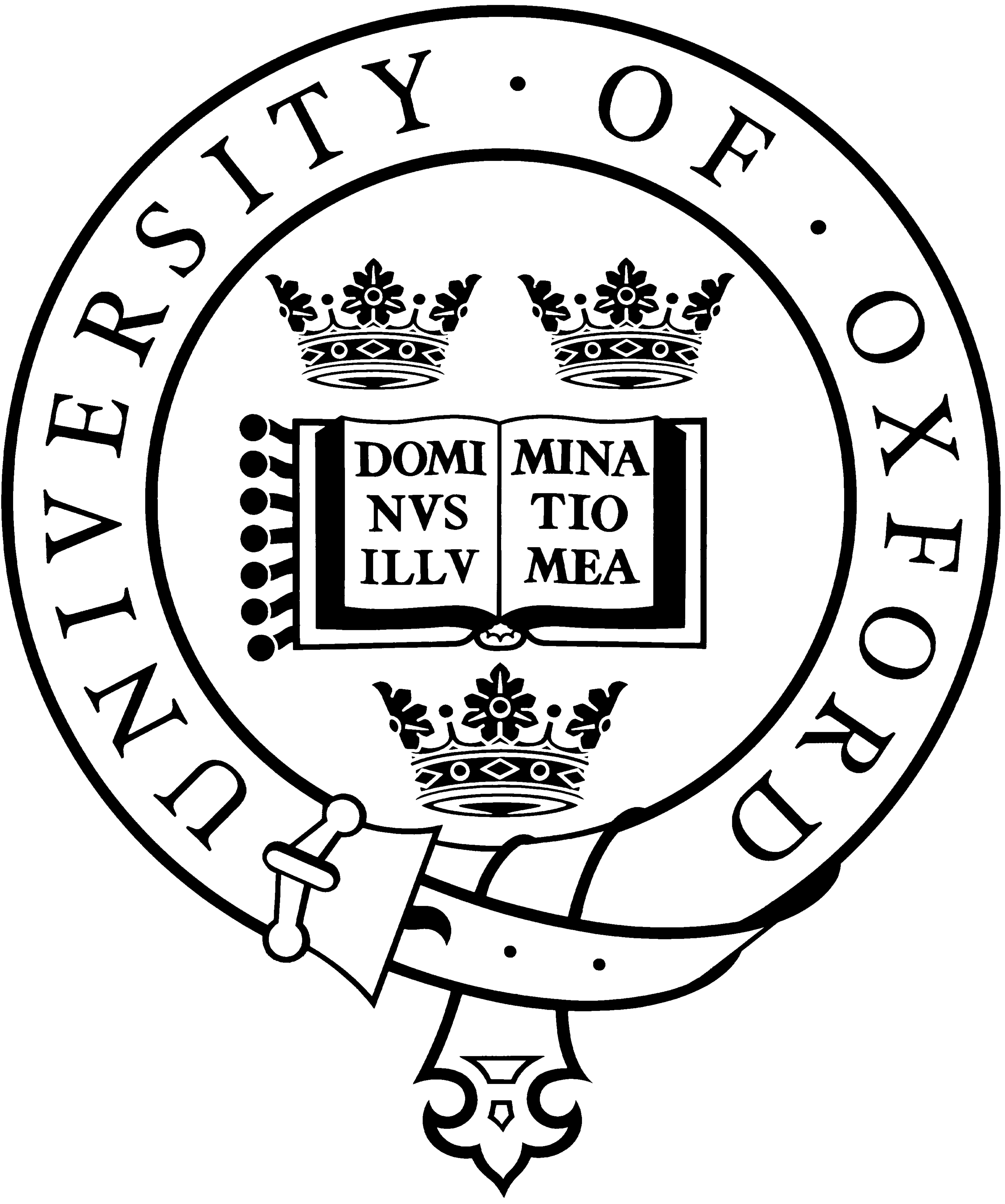}
\label{fig:logo}
\end{center}
\end{figure}

\vfill
\begin{center}
Submitted for the degree of Doctor of Philosophy \\
2015
\end{center}

\vfill
\begin{center}
Supervised by \\
Prof. Ian A. Walmsley
\end{center}

\vfill
\begin{center}
Clarendon Laboratory \\
University of Oxford \\
United Kingdom
\end{center}

\end{spacing}
\end{titlepage}

\begin{spacing}{1.3}
\begin{center}
{\large\textbf{Room temperature caesium quantum memory for quantum information applications}} \\
\vspace{5pt}
Patrick Steffen Michelberger \\
Balliol College, Oxford \\
\vspace{5pt}
Submitted for the degree of Doctor of Philosophy \\
Hilary Term 2015
\end{center}
\vspace{-30pt}

\begin{abstract}

Quantum memories are key components in photonics-based quantum information processing networks. Their  ability to store and retrieve information on demand makes repeat-until-success strategies scalable.
Warm alkali-metal vapours are interesting candidates for the implementation of such memories, thanks to their very long storage times as well as their experimental simplicity and versatility. 
Operation with the Raman memory protocol enables high time-bandwidth products, which denote the number of possible storage trials within the memory lifetime. Since large time-bandwidth products enable multiple synchronisation trials of probabilistically operating quantum gates via memory-based temporal multiplexing, the Raman memory is a promising tool for such tasks.  
Particularly, the broad spectral bandwidth allows for direct and technologically simple interfacing with other photonic primitives, such as heralded single photon sources. 
Here, this kind of light-matter interface is implemented using a warm caesium vapour Raman memory.
Firstly, we study the storage of polarisation-encoded quantum information, a common standard in quantum information processing. High quality polarisation preservation for bright coherent state input signals can be achieved, when operating the Raman memory in a dual-rail configuration inside a polarisation interferometer.
Secondly, heralded single photons are stored in the memory. 
To this end, the memory is operated on-demand by feed-forward of source heralding events, which constitutes a key technological capability for applications in temporal multiplexing. 
Prior to storage, single photons are produced in a waveguide-based spontaneous parametric down conversion source, whose bespoke design spectrally tailors the heralded photons to the memory acceptance bandwidth. 
The faithful retrieval of stored single photons is found to be currently limited by noise in the memory, with a signal-to-noise ratio of $\sim 0.3$ in the memory output. Nevertheless, a clear influence of the quantum nature of an input photon is observed in the retrieved light by measuring the read-out signal's photon statistics via the $g^{(2)}$-autocorrelation function. Here, we find a drop in $g^{(2)}$ by more than three standard deviations, from $g^{(2)} \sim 1.69$ to $g^{(2)}\sim1.59$ upon changing the input signal from coherent states to heralded single photons. 
Finally, the memory noise processes and their scalings with the experimental parameters are examined in detail. Four-wave-mixing noise is determined as the sole important noise source for the Raman memory. These experimental results and their theoretical description point towards practical solutions for noise-free operation.

\end{abstract}
\end{spacing}

\cleardoublepage

\tableofcontents

\mainmatter

\chapter{Introduction}
\label{ch1}

\begin{flushright}
{\tiny \textfrak{Faust}: \,  \textfrak{Da{\ss} ich erkenne, was die Welt; Im Innersten zusammenhŠlt} }
\end{flushright}

\section{The need for quantum memories\label{ch2_intro}}

Quantum information processing\cite{Gisin:2002aa,Bennett:1998,Bennett:2000fk} has experienced immense research focus in the last two decades. 
Experimental implementations of protocols and gates\cite{Nielsen:2004kl} range from trapped ions\cite{Haffner:2008vn} and atoms\cite{Monroe:2002ys} all the way to integrated solid state systems\cite{Wallraff:2004,Clarke:2008zr}.
The optical domain offers a valuable platform for quantum information, where linear optical systems can be used to implement quantum logic gates\cite{Knill:2001nx,Raussendorf:2001}. 
However, the experimental ease of these photonic systems comes at the price of non determininistic operation\cite{OBrien:2003aa, Bouwmeester:1997aa}, i.e. the success probability of such systems is below unity. 
This probabilistic feature restrains the scalability of quantum optical systems\cite{Duan:2001vn}.
On the one hand, the combination of several inefficient logic gates in a quantum information processor lowers the success rate of the overall operation. 
On the other hand, transmission channel loss strongly limits the achievable data rates of quantum communication protocols over long distances\cite{Sanguard2011, Ursin:2007vn}. 
A suggested loss mitigation strategy lies in the insertion of a quantum repeater chain\cite{Duan:2001vn,Sanguard2011,Chen:2008fk} into the communication channel. 
At their heart, quantum repeaters are founded on a series of entanglement swapping operations\cite{Pan:1998}. Since these operations only work probabilistically themselves, low success rates are ultimately also the performance limiting factor for quantum repeater architectures. \\
Quantum memories may facilitate a shift towards deterministic operation in optical QIP.
Quantum memories\cite{Duan:2001vn, Lvovsky:2009ve} are devices capable of storing quantum information carriers on demand. 
A quantum memory placed at the output of a gate would allow for the storage of successful output events for that gate, while waiting for the successful operation of a neighbouring quantum gate\cite{Sanguard2011}. 
Consequently, the development of quantum memories could be central to progress in quantum information science.

\paragraph{Memory media} 
As with other components in quantum information processors, since the initial proposal\cite{Fleischhauer:2000vn,Fleischhauer:2002,Phillips:2001} of quantum memories several experimental systems have been considered as possible realisations. 
Because the general principle of a memory is the transformation of flying quantum bits (qubits) into stationary ones, the most popular storage mechanism is the mapping of photonic qubits onto atomic systems. 
The challenge in the development of such light-matter interfaces lies in the small interaction cross section between an atom and the incident light field. 
Increasing the interaction probability is thus one of the major design criteria for memory media. 
In general, multiplying the number of light-matter interaction events boosts this probability. 
To this end, one can either force the incoming photonic qubit to interact multiple times with the same atom, or increase the number of absorbers by storing light in an atomic ensemble. 

A prime example of memories belonging to the former category is a single atom in a high finesse cavity\cite{Specht:2011}. 
Recently, opto-mechanical systems\cite{Sete:2015} and superconducting circuits\cite{Sirois:2015} have also been considered. 
Memories based on atomic ensembles represent the corner-stone of most systems currently under investigation. 
Here, the main platforms range from atomic vapours, where warm gases\cite{Phillips:2001}, laser-cooled trapped atoms\cite{kuzmich,Choi:2008aa} or BECs\cite{Lettner:2011} are employed, to solid-state systems, such as rare-earth ion doped crystals\cite{Tittel:2010} or diamond\cite{Wrachtrup:2006cr}. 
Recently, hybrid systems using cavity as well as ensemble enhancement have also been developed\cite{Sabooni2013}. 

While each of these systems has its advantages, one particularly interesting class are room-temperature alkali-metal vapours. Their experimental simplicity makes them one of the first media to be considered\cite{Phillips:2001}. The very long storage times \cite{Longdell:2005aa,Dudin:2013aa,Heinze:2013aa} and large time-bandwidth products \cite{Reim:2011ys,Bao:2012aa}, obtainable by optical storage in atoms, 
are ideal prerequisites for their utilisation in temporal synchronisation tasks of probabilistic quantum devices\cite{Nunn:2013ab}.
Moreover, atomic vapours are very versatile and allow for the implementation of a great number of memory protocols.

\paragraph{Memory protocols}
Besides investigating suitable storage media, research efforts over the past decade have also focussed on the actual protocols to be used for most beneficial storage of information in the memory and its subsequent retrieval. 
Here we provide a brief overview of the most relevant protocols that we will be referring to later on; more detailed reviews on quantum memories can be found elsewhere\cite{Nunn:DPhil,Lvovsky:2009ve}. 
Following the initial proposal for a quantum memory\cite{Duan:2001vn}, one of the first storage protocols was based on electromagnetically-induced transparency\cite{Gorshkov:2007,Gorshkov:2007aa,Gorshkov:2007rw,Gorshkov:2008rz,Novikova:2012} (EIT). This protocol enables the storage of an input signal under the simultaneous application of a bright control beam. Both light fields are resonant with the optical transitions in a $\Lambda$-level system\footnote{
	See fig. \ref{fig_ch2_lambda_system} in chapter \ref{ch2} for an exemplary $\Lambda$-level system. 
} of the storage medium. 
Signal storage is obtained by gradually turning off the control field, which initially splits the excited state transition into two separate dressed states. 
Information, stored in a dark-state polariton\cite{Fleischhauer:2002}, can be maintained in the medium for extremely long timescales up to seconds\cite{Heinze:2013aa} and is released by re-application of the control. 
The magnitude of the achievable excited state splitting, determined by the control intensity, defines the usable signal bandwidths. 
These are on the order of MHz, making EIT a rather narrowband protocol for optical input signals with carrier frequencies on the order of hundreds of THz.
Modification of the scheme by moving signal and control fields off-resonance with the excited state in the $\Lambda$-level system circumvents the bandwidth limitation, imposed by the control intensity. This leads to three other types of protocols: 
 
The first two, controlled-reversible inhomogeneous broadening\cite{Nilsson:2005kl,Alexander:2006tg, Staudt:2007hc} (CRIB) and the gradient-echo-memory\cite{Hetet:2008dp,Hetet:2008kx,Hetet:2008jt, Hetet:2008kl} (GEM) are similar to one another. 
In these protocols, an external electric or magnetic field is applied to the storage medium to induce a linear Stark- or Zeeman- shift of the atomic resonances. 
The shift, applied either to the excited state or to the storage state, spectrally broadens the signal absorption line. This increases the memory bandwidth, because the various frequency components of an incoming signal are resonant with a subset of the ensemble, whose resonance has been shifted appropriately by the external field. 
A notable advantage of GEM systems is the achievable memory efficiency, with reported values ranging up to\cite{Hosseini:2011aa} $87\,\%$. 

The third protocol in this category is the Raman memory\cite{Gorshkov:2007,Gorshkov:2007aa,Gorshkov:2007rw,Gorshkov:2008rz, Nunn:2007wj}, which is the subject of this thesis. 
Unlike CRIB and GEM, no external field is necessary. Instead, the control field is produced by an intense short laser pulse. 
Detuned far off-resonance from the excited state of the atomic $\Lambda$-level system (see fig. \ref{fig_ch2_Raman_protocol}), the control pulses induce a virtual excited state, whose bandwidth is now given by the control bandwidth, rather than the control intensity. 
Thanks to the short control pulse duration, storage of broadband signal pulses is possible. 
Depending on the characteristics of the storage medium, control pulse durations of hundreds of pico-second (ps) down to a few tens of femto-seconds (fs) have been used\cite{Reim2010, England:2014}. 
The corresponding bandwidths, ranging from the GHz- to the THz-regime, make it possible to achieve large time-bandwidth products ($B$) on the order of $B \sim 1000$, where $B$ is defined as the product of the spectral acceptance width and the storage time ($\tau_\text{S}$). 
While $\tau_\text{S}$ is still prohibitively short in the fs demonstrations\cite{England:2014}, the reasonable storage times of several micro-seconds ($\mu$s), obtainable with GHz-bandwidth devices, make Raman memories a promising candidate technology for temporal synchronisation tasks\cite{Nunn:2013ab}, where large time-bandwidth products are a key prerequisite. 
Besides showing promising performance numbers, Raman memories are also very versatile in terms of input states, as they can store continuous\cite{Weedbrook:2012} (cv) and discrete (dv) variable quantum states. 
In dv schemes, such as the system studied in this thesis, single photons are the carriers of quantum information and quantum states are expressed in the photon number basis. 
Qubit storage of dv states follows the ideas outlined above: single photons, sent into the storage medium, are absorbed and later re-emitted by application of a pair of control pulses. 
In contrast, in the cv space, information is encoded on the quadrature components of quantum states. 
Instead of absorbing the incoming light, cv Raman memories map the quadrature components of the input signal onto the atomic ensemble via a quantum non-demolition measurement\cite{Kozhekin:2000bs,Hammerer:2010vn,Julsgaard:2004lr}. 

Besides these modifications of EIT, there are two more alternative memory protocols, the revival of silent-echo\cite{Damon:2011} (ROSE) and atomic frequency combs\cite{Afzelius:2009qf,Tittel:2010} (AFCs), both of which rely on photon-echo. 
Several recent demonstrations of quantum operations involving quantum memories featured AFCs\cite{Clausen:2011kx, Rielander:2014aa, Bussieres:2014aa,Sinclair:2014}, which, similarly to Raman memories, can achieve broadband signal storage\cite{Saglamyurek:2011fk}.
For these experiments, the protocol was implemented in rare-earth-ion-doped crystals. 
In these media, interaction with the solid-state crystal field broadens the absorption line of the ions. Selective optical pumping tailors this resonance into a series of absorbing comb lines. 
In turn, the comb teeths' frequency separation ($\Delta \nu_\text{AFC}$) sets the storage time ($\tau_\text{S} = 1/\Delta \nu_\text{AFC}$), after which the ions, initially excited by resonant absorption of the input signal, re-phase to re-emit the signal. 
On-demand AFC operation, using an additional, bright control pulse with arbitrary storage times, has also been 
achieved\cite{Afzelius:2010fk, Timoney:2013, Guendogan:2013, Jobez:2014}.

\section{Structure of this thesis\label{ch1_sec_thesis_structure}}

In this thesis, we continue the investigation of the Raman memory protocol in warm caesium (Cs) vapour, building on an initial set of proof-of-principle experiments. 
In previous experiments, our research group used bright laser pulse signals to demonstrate the storage and retrieval of light\cite{Reim2010}. 
Apart from achieving one of the largest time-bandwidth products of ${B} \sim 10^3$, we also showed first-order interference between light sent into and light recalled from the memory. 
In these experiments we furthermore undertook the first steps towards the quantum regime by studying the storage of laser pulses with intensities at the single photon level\footnote{
	The experiment used an input signal with an average number of 
	$\sim 1.6 \,\frac{\text{photons}}{\text{pulse}}$. 
}. 
The work presented here continues this journey and tries to answer the question, whether our warm caesium vapour Raman system actually has the capability to serve as a quantum memory. 
We demonstrate effects of the quantum features of an input signal on its counterpart retrieved from the memory. 
However, we also find that faithful storage of quantum information carriers in the present system is still hampered by a parasitic noise process, whose suppression we identify as the key remaining challenge for the Raman memory. 
To this end, we characterise this noise, show that it originates from four-wave mixing (FWM), and highlight the extent to which it limits successful memory performance. 

The experiments we discuss here broadly fall into three categories, which form the main parts of this thesis:
\begin{enumerate}
\item \textit{Experiments with bright coherent states}

We study the storage of the input light's polarisation in the memory. 
Encoding information in the polarisation domain is a popular technique, 
making this capability an important property for memory applications in quantum networks. 
Polarisation storage is investigated with bright coherent state laser pulses as input signals. 
It is a logical continuation of our initial proof-of-principle experiment\cite{Reim2010}, and requires only modest setup modifications\footnote{
	It is noteworthy that this strategy has also been followed by other research 
	groups\cite{Kim:2010,Gruendogan:2012,Clausen:2012,Zhang:2011aa}.
}.
We investigate the storage of bright coherent states, using the framework of quantum process tomography. 
We benchmark the memory's performance in terms of purity and fidelity of the storage process. 
Additionally, we evaluate the possibility of storing the polarisation of input signals at the single photon level. 
Anticipating later results, presented in parts 2 and 3, we establish the limits on the memory noise floor that are required to enable good single photon level polarisation storage.

\item \textit{Experiments with single photons}
We realise the storage of single photons. 
To this end, we first introduce the means for their generation, followed by the analysis of their storage in the Raman memory. 
These two projects result in a temporal multiplexer prototype\cite{Nunn:2013ab}, which is the main advance presented in this thesis. \\
To produce single photon input signals, we use heralding of photon-pair emission events from spontaneous parametric down-conversion\cite{Migdall:book} (SPDC), which we implement in a non-linear periodically-poled potassium-titanyl-phosphate (KTP) waveguide.
We identify the requirements the source has to fulfil to produce single photons tailored for optimal interfacing with the memory. 
Subsequently, we describe the source design and its experimental implementation, before we conclude with a characterisation of the source performance. 
We pay particular attention to the spectral characteristics of the photons, since these have to match the acceptance line of the memory's input channel.\\
To interface the source with the memory, we modify the memory apparatus and implement electronic feed-forward of single photon heralding events from the source to the memory. 
In this way, storage and retrieval in the {\cs} vapour is synchronised with single photon production. 
We then proceed to demonstrate single photon storage via mean field measurements. 
To test the preservation of the photons' quantum characteristics, i.e. the memory's capability to faithfully store quantum features, we investigate the statistics of photons retrieved from the memory. 
The statistics are compared with results obtained by storing coherent states at the single photon level, which serve as a benchmark for the memory's performance with classical inputs. 
For both input signals, we find significant memory performance limitations caused by the memory noise floor. 
Importantly, despite the noise, we are able to register the influence of the input signal's quantum nature on the statistics of the retrieved signal.  
Moreover, we present a theoretical model to predict the noise effects on the photon statistics. 
We conclude by using this model to evaluate possible ways for noise floor reduction.

\item \textit{Memory noise floor characterisation}

In the final part of the thesis, we investigate the memory noise floor in detail. 
We begin by dissecting the noise into its constituents, using several experimental parameters. 
This approach allows us to experimentally and theoretically demonstrate that the significant contribution to the noise floor originates from four-wave-mixing. 
We compare the measured noise level against the predictions of our theoretical model. 
Further, we study the system's response to different initial conditions, which allow us to separate the noise into FWM, spontaneous Raman scattering (SRS) and collisional induced fluorescence.
We prove the FWM origin of undesired noise in the Raman memory output by examining its spin-wave dynamics. 
Last, in a series of supplementary experiments, presented in the appendix, we study the influence of the experimentally accessible Raman memory parameters on the scaling between noise level and memory efficiency. 

\end{enumerate}

\chapter{The Raman memory protocol}
\label{ch2}

\begin{flushright}
{\tiny \textfrak{Faust}: \,  \textfrak{Die Botschaft hšr ich wohl, allein mir fehlt der Glaube.} }
\end{flushright}

Before diving into the first part of our work on the Raman memory, we give a brief outline of its operational principles and describe its actual implementation in room-temperature Caesium (Cs) vapour. 
We start by introducing the theoretical basics which lead to the formulation of the Maxwell-Bloch equations. This set of equations describes the dynamics of our system and yields expressions for the memory efficiency, the noise level (see chapter \ref{ch7}), the photon statistics in the memory output (see chapter \ref{ch6}), as well as the initial set of experimental parameters, needed to operate the memory\footnote{
The derivation of these equations is laid out here for the reader to follow it coarsely. 
A detailed presentation is beyond the scope of this thesis and can be found in the DPhil thesis of \textit{Joshua Nunn}\cite{Nunn:DPhil}.
}. 
Thereafter we highlight how the Raman memory operates in caesium ({\cs}) and discuss our experimental implementation.

\section{Theory of the Raman memory protocol\label{ch2_Raman_memory_theory}}

To summarise the Raman memory theory, we first introduce the required level structure for our system. 
We then focus on the light fields involved in the protocol, followed by the description of the atomic system and its interaction with the light pulses. 
We combine these to formulate the Maxwell-Bloch equations and introduce the adiabatic approximation, which will take us into the operational regime of the Raman memory. Finally, we establish the expressions for the memory's storage efficiency and discuss the origins of noise in the Raman system. 

\subsection{Level structure\label{subsec_ch2_level_scheme}}
\begin{wrapfigure}{R}{0.5\textwidth}
\centering
\includegraphics[width=0.5\textwidth]{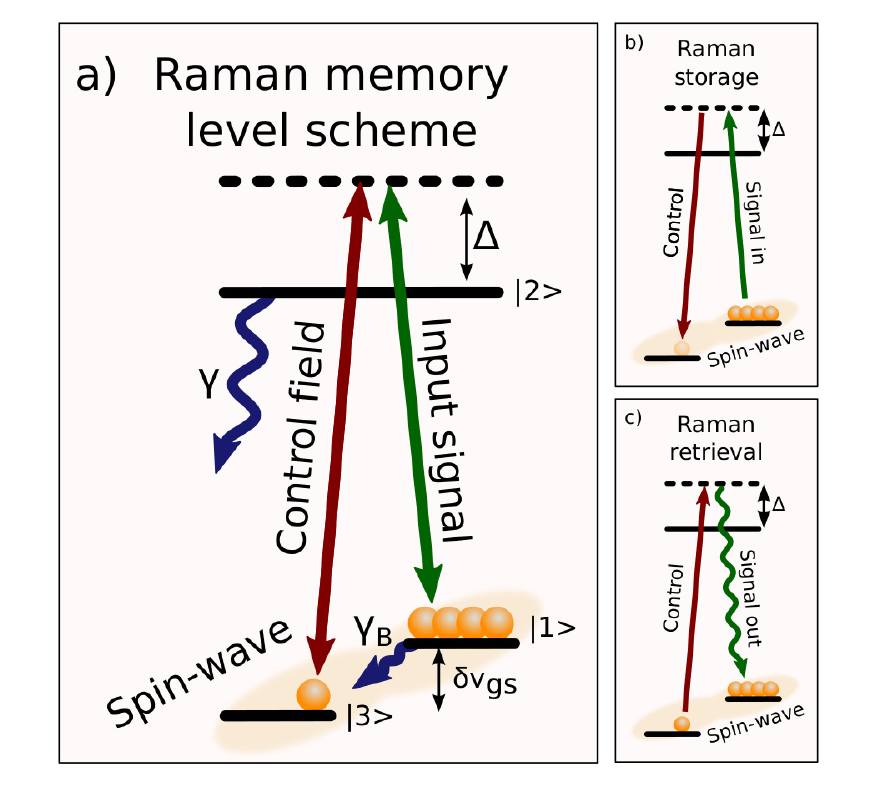}
\caption{Raman memory level schemes. 
\textbf{(a)}: Atomic $\Lambda$-level system. 
\textbf{(b)}: Raman memory read-in sequence. 
\textbf{(c)}: Raman memory retrieval sequence.} 
\label{fig_ch2_lambda_system}
\end{wrapfigure}

The idea behind a quantum memory is to temporarily store quantum information in a well-defined location without risking its loss. 
For our purposes, we restrict ourselves to information encoded on light. 
When considering an atomic vapour system, such as Cs, this translates into two requirements on the atomic levels that are chosen for the memory's empty and charged states.  
These should neither be subject to radiative decay, nor be affected by collisions with other atoms to avoid (de-) excitation channels to other atomic states and therewith information loss. 
Both conditions can be met by selecting atomic hyperfine states with the same parity as the memory's initial and storage states. 
Due to selection rules\cite{Steck:LectureNotes}, these cannot be addressed by an electric dipole transition, so they cannot be coupled efficiently with a single light field\footnote{
	The first allowed transitions is a magnetic dipole transitions. 
}.  
In fact, this absence of state coupling via a single channel is exactly what we desire to avoid radiative decay of the memory's initial and storage state. 
Consequently, to store information in the memory, i.e. flip the atomic state from the initial (empty memory) to the storage state (charged memory) and vice versa, a two-photon process, such as a Raman 
transition\cite{Penzkofer:1979, Raymer:1985bv}, is needed.
One possible atomic energy level structure is the $\Lambda$-system\cite{ShoreBook}, shown in fig. \ref{fig_ch2_lambda_system}. 
The atomic initial state $(\ket{1}$) and the storage state $(\ket{3})$, separated by an energy splitting of 
$\hbar \cdot \delta \nu_\text{gs}$, are both connected to an electronic excited state $(\ket{2})$ via coupling to the optical signal field (frequency $\omega_\text{s} = 2\pi \nu_\text{s}$) and a control field (frequency $\omega_\text{c} = 2\pi \nu_\text{c}$). 
For the Raman memory protocol, both optical fields are in two-photon resonance and detuned by a frequency $\Delta$ from the excited state. 
When applying this configuration in the memory protocol, the detuning can be chosen towards the red\cite{Hetet:2008jt} or to the blue\cite{England:2013,Hosseini:2012} of the excited state $\ket{2}$, which is the configuration depicted in fig. \ref{fig_ch2_lambda_system}. 
While this does neither affect the implementation nor the theoretical description of the system, it can have an effect on the noise background, as we will discuss later. For our experiments, we chose the blue detuned version\cite{Reim2010}.

\subsection{Maxwell equations\label{subsec_ch_Maxwell_equations}}
As we have seen, the Raman memory protocol requires two optical fields: Firstly, the signal field to be stored, which couples to the ${\ket{1} \leftrightarrow \ket{2}}$ transition. Secondly, the control field, which couples the states $\ket{2} \leftrightarrow \ket{3}$ and maps the signal field into the storage state during read-in. For information retrieval, solely the control field is applied. It shuffles the atomic excitation from the storage state $\ket{3}$ back to the initial state $\ket{1}$, under re-emission of the signal field along the $\ket{2} \rightarrow \ket{1}$ leg of the $\Lambda$-system. 

\paragraph{Light fields}
Both optical fields are pulsed. The control field $\vec{E}_c$ is a strong classical laser pulse and thus given by the classical electric field\footnote{
 	c.c. stands for the complex conjugate.
}
\begin{equation}
\vec{E}_c (t,z) = \vec{e}_c E_c(t,z) \exp{\left\{\I \cdot \omega_c \left(t-\frac{z}{c} \right) \right\}} + \text{c.c.},
\label{eq_ch2_Ec}
\end{equation}
with a polarisation $ \vec{e}_c$ and a slowly varying amplitude $E_c(t,z)$.
In the experiment, signal and control will be collinear, and we assume cylindrical symmetry around the propagation direction of the light fields. We furthermore restrict ourselves to modelling only 1-dimensional propagation along this optical axis. 
For these reasons we neglect the transverse dimensions\footnote{
	See \textit{Zeuthen et. al}\cite{Zeuthen:2011} for a 3-dimensional treatment.
}.  
Because we ultimately want to store single photons, we treat the signal field quantum mechanically\cite{Loudon:2004gd} using the field operator\footnote{
	h.c. denotes the hermitian conjugate.
}
\begin{equation}
\hat{\vec{E}}_s(z,\omega_s) = \I \cdot  \vec{e}_s \int g(\omega) \hat{a}(\omega) e^{-\I \omega z} \text{d} \omega + \text{h.c.} = 
\I \cdot \vec{e}_s \cdot g_s \hat{S}(t,z) e^{\I \omega_s \tau } + \text{h.c.}, 
\label{eq_ch2_Es}
\end{equation}
where $\vec{e}_s$ is the signal polarisation, $g(\omega)=\sqrt{\frac{\hbar \omega}{4\pi \epsilon_0 c}}$ is the mode amplitude, $\epsilon_0$ is the permeability of vacuum and $\hat{a}(\omega)$ is the annihilation operator for a signal photon at frequency $\omega$. 
In the second step, we have assumed a constant $g_s = \sqrt{2 \pi}g(\omega_s)$ and defined the broadband annihilation operator 
$\hat{S} = \frac{e^{-\I \omega_s \tau}}{\sqrt{2\pi}} \int  \hat{a}(\omega,t) e^{-\I \omega t} \text{d} \omega$. 
We have also introduced the retarded time $\tau = t-\frac{z}{c}$, which describes a coordinate system moving along with the control and signal pulses\cite{Nunn:DPhil}. 

\paragraph{Maxwell equations}

The propagation of the signal field along the storage medium is described by Maxwell's equations. 
These can be modified to yield the wave equation: 
\begin{equation}
\nabla^2 \vec{E} = \mu_0 \partial_t^2 \vec{D} \Longleftrightarrow  \left[ \nabla^2 - \frac{1}{c^2} \partial_t^2 \right] \vec{E} = \mu_0 \partial_t^2 \vec{P},
\label{eq_ch2_wave_equation}
\end{equation} 
where the dielectric displacement term $\vec{D} = \epsilon_0 \vec{E} + \vec{P}$ contains the macroscopic response of the atomic vapour to the impinging electric field.  
We focus on the interaction of the signal field with the atomic medium here, as it is the signal field, whose change in intensity during the storage and retrieval process we ultimately care about. 
Conversely, we do not care if photons are emitted into the control field channel or absorbed from it. 
We thus insert eq.~\ref{eq_ch2_Es} for the signal's electric field into eq.~\ref{eq_ch2_wave_equation}, where we are faced with single ($\sim \partial_t$) and double time derivatives $(\sim \partial_t^2)$ of its field envelope $\hat{S}(t,z)$. 
In our case, the envelope is slowly varying, so we can neglect terms $\sim \partial_t^2$. This leaves us with the equation
\begin{equation}
\left[ \nabla^2 - \frac{1}{c^2} \partial_t^2\right] \left[ \I \vec{e}_s g_s \hat{S}(t,z) e^{\I \omega_s \tau}\right] = \mu_0 \partial_t^2 \left[ \vec{\mathfrak{P}} e^{-\I \omega_s \tau}  \right] \Rightarrow 
\left[ \partial_z + \frac{1}{c} \partial_t \right] \hat{S} = -\frac{\mu_0 \omega_s^2}{2 g_s k_s} \vec{e}_{s}^{*} \cdot \vec{\mathfrak{P}}_s, 
\label{eq_ch2_wave_equation_signal}
\end{equation}
where we have only used the positive frequency component of eq.~\ref{eq_ch2_Es}, for reasons explained in section \ref{subsec_ch2_maxwell_bloch_eq}. 
Additionally, the polarisation $\vec{P} = \vec{\mathfrak{P}} \cdot e^{-\I \omega_s \tau}$ is decomposed 
into an envelope term $(\vec{\mathfrak{P}})$ and a carrier, oscillating at the signal frequency. 
This macroscopic atomic polarisation is generated by the signal via its coupling to the $\ket{1} \rightarrow \ket{2}$ transition of all atoms located within the beam. 
On the single atom level, we model\cite{Steck:LectureNotes,ShoreBook} this coupling by an induced electric dipole moment 
$\hat{\vec{d}} =  -e \cdot \bra{1} \hat{\vec{r}} \ket{2} \cdot \hat{\tilde{\sigma}}_{1,2} + \text{h.c.}$ (see eq.~\ref{eq_ch2_dipole_moment}), with the electronic charge $e$, the position operator $\hat{\vec{r}}$ and transition projection operator $\hat{\tilde{\sigma}}_{1,2} = \ket{1} \bra{2} e^{\I \omega_{1,2} \tau}$ between atomic states $\left\{\ket{1}, \ket{2}\right\}$, whose phase oscillates at the transition frequency $\omega_{1,2}$ between the two states (see section \ref{subsec_ch2_maxwell_bloch_eq}). 
To connect this microscopic response with the macroscopic polarisation, we consider an atomic density $n$ and sum up the atomic dipole moments 
${\vec{d}_{1,2}=-e\cdot\bra{1}\hat{\vec{r}}\ket{2}}$ of all atoms in a volume element\footnote{
	$\delta V$ is chosen disc-shaped, such that along the optical axis the disc's thickness $\delta z \ll \lambda_s$, 
	and the atoms are subject to an approximately constant signal phase. In the transverse plane, the disc's area 
	$\delta A \gg \lambda_s$, making the atomic separation larger than the induced dipole element and dipole-dipole 
	interactions between atoms negligible. 
} $\delta V$, yielding 
$\vec{P} = \frac{1}{\delta V} \underset{r_i}{\sum} \vec{d}_{1,2} \hat{\tilde{\sigma}}_{1,2}(r_i) + \text{h.c.}$, 
with $r_i$ denoting the position of an atom. 
In fact, this makes the polarisation a quantum mechanical operator, for which a set of creation and annihilation operators can be defined. The latter can be expressed as 
$\hat{P} = \frac{1}{\sqrt{n} \delta V} \underset{r_i}{\sum} \hat{\tilde{\sigma}}_{1,2} e^{\I \Delta \tau}$, 
where $\Delta = \omega_s - \omega_{1,2}$ is the detuning of the signal from the excited state in the $\Lambda$-system and $n$ is the density of atoms in the considered volume element. 
This leaves us with an amplitude of $\hat{\vec{\mathfrak{P}}}_s = \sqrt{n} \cdot \vec{d}_{1,2} \cdot \hat{P}$ for the dielectric response in eq.~\ref{eq_ch2_wave_equation_signal}. 

\subsection{Maxwell-Bloch equations\label{subsec_ch2_maxwell_bloch_eq}}

We now look at the atomic dynamics introduced by the signal and control pulses. 
To this end, we will consider all relevant energy terms and use the resulting Hamiltonian to derive a set of differential equations to  describe the system. 

\paragraph{Hamiltonian} 
To obtain the Maxwell-Bloch equations, we refer to the Heisenberg picture. We consider the projection operators $\hat{\sigma}_{i,j} = \ket{i} \bra{j}$ onto the three atomic states, with $\ket{i}, \ket{j} \in \left\{ \ket{1}, \ket{2}, \ket{3} \right\}$.
The sought after differential equations consequently describe the time evolution $\hat{\sigma}_{i,j}(t)$ of these projectors. They are obtained via the Heisenberg equation 
\begin{equation}
\frac{\partial}{\partial t} \hat{\sigma}_{i,j} = \frac{\I}{\hbar} \left[ \hat{\sigma}_{i,j}, \hat{H}\right], 
\label{eq_ch2_Heisenberg_equation} 
\end{equation}
where $\hat{H} = \hat{H}_\text{atom} + \hat{H}_\text{light} + \hat{H}_\text{int}$ is the system's Hamiltonian. 
It contains three contributions: the energy of the light fields $\hat{H}_\text{light}$, the energy of the atomic system $\hat{H}_\text{atom}$, and the interaction energy between light and atoms $\hat{H}_\text{int}$. \newline
The latter term derives from the dipole moments\cite{Steck:2008qf,Steck:LectureNotes,ShoreBook}  $\hat{\vec{d}} = e \cdot \hat{\vec{r}}$, induced by the light fields in the atomic medium. 
Once again, $e$ is the electronic charge and $\hat{\vec{r}}$ is the position operator for the outmost electron of the atom. 
In this dipole-approximation, we have $\hat{H}_\text{int}~=- \hat{\vec{d}} \cdot \hat{\vec{E}}$, 
with the electric field vector $\hat{\vec{E}}=\hat{\vec{E}}_c + \hat{\vec{E}}_s$ consisting of the contribution from both optical pulses, control and signal. We can express the dipole operator in terms of its projectors onto the three atomic states 
\begin{equation}
\hat{\vec{d}} = \underset{i,j}{\sum} {\vec{d}}_{i,j} \sigmaop{i,j} = {\vec{d}}_{1,2} \sigmaop{1,2} + {\vec{d}}_{2,3} \sigmaop{2,3} + \text{h.c.},
\label{eq_ch2_dipole_moment}
\end{equation}
where the $\vec{d}_{i,j} = \bra{i} \hat{\vec{d}} \ket{j}$ are the matrix elements of transitions from state $j$ to $i$. Only dipole allowed transitions are addressed by the optical fields, so transitions between states $\ket{1}$ and $\ket{3}$ do not feature in $\hat{\vec{d}}$. 

The energy contribution from the atomic system is described by the the population in each state, accessible via the number operator $\hat{N}_j = \ket{j} \bra{j}$ for population the state $\ket{j}$. Expressed in projector terms, this yields 
$$
\hat{H}_\text{atom} = \underset{j}{\sum} \hbar \omega_j \ket{j}\bra{j} = \underset{j}{\sum} \hbar \omega_j \sigmaop{j,j}. 
$$
Similarly, the energy stored in the signal's radiation field is given by 
$\hat{H}_\text{light} = \int \hbar \omega \hat{a}^\dagger(\omega) \hat{a} (\omega) \text{d} \omega$. 
Both, this signal energy and the control energy contribution commute with the $\sigmaop{i,j}$ operators in eq.~\ref{eq_ch2_Heisenberg_equation}, for which reason we can drop $\hat{H}_\text{light}$ from $\hat{H}$ and end up with 
 $$
 {\partial_t \sigmaop{i,j} = \frac{\I}{\hbar} \left[ \sigmaop{i,j}, \hat{H}_\text{atom} + \hat{H}_\text{int}  \right]},
 $$
 yielding the following system of equations for the atomic populations $\sigmaop{i,i}$ and coherences $\sigmaop{i,j}$:
 \begin{align}
\partial_t \sigmaop{1,1} &= -\frac{\I}{\hbar} \vec{E} \cdot \left( \vec{d}_{1,2}\sigmaop{1,2} -\vec{d}_{2,1}^* \sigmaop{2,1}  \right),  \quad
\partial_t \sigmaop{3,3} = \frac{\I}{\hbar} \vec{E} \cdot \left( \vec{d}_{2,3}\sigmaop{2,3} -\vec{d}_{3,2}^* \sigmaop{3,2}  \right), \nonumber \\
\partial_t \sigmaop{1,2} &= \I \omega_{2,1} \sigmaop{1,2}  
- \frac{\I}{\hbar} \vec{E} \cdot \left( \vec{d}_{1,2} \left( \sigmaop{1,1} - \sigmaop{2,2} \right) +\vec{d}_{2,3}^* \sigmaop{1,3}  \right)  
\overset{(**)}{=} \I \omega_{2,1} \sigmaop{1,2}  - \frac{\I}{\hbar} \vec{E} \cdot \vec{d}_{2,3}^* \sigmaop{1,3}, \nonumber \\
\partial_t \sigmaop{1,3} &= \I \omega_{3,1} \sigmaop{1,3}  - \frac{\I}{\hbar} \vec{E} \cdot \left( \vec{d}_{2,3}^* \sigmaop{1,2} - 
\vec{d}^*_{1,2} \sigmaop{2,3}  \right),  \nonumber \\
\partial_t \sigmaop{2,3} &= \I \omega_{3,2} \sigmaop{2,3}  
- \frac{\I}{\hbar} \vec{E} \cdot \left( \vec{d}_{2,3} \left( \sigmaop{2,2} - \sigmaop{3,3} \right) +\vec{d}_{1,2} \sigmaop{1,3}  \right) 
\overset{(**)}{=} \I \omega_{3,2} \sigmaop{2,3} - \frac{\I}{\hbar} \vec{E} \cdot \vec{d}_{1,2} \sigmaop{1,3}, 
 \label{eq_ch2_projectors_1}
 \end{align}
with the transition frequency $\omega_{i,j} = \omega_i - \omega_j$ between states $\ket{i}$ and $\ket{j}$. Furthermore, due to the conservation of the number of electrons, we have $\partial_t \sigmaop{2,2} = - \left( \partial_t \sigmaop{1,1}  + \partial_t \sigmaop{3,3} \right)$.
To simplify and solve these equations, we make several assumptions and approximations that will be discussed in the following. 
One of these, marked by $(**)$ in eqs.~\ref{eq_ch2_projectors_1}, accounts for the initial population distribution in our atomic ensemble. 
For the Raman memory scheme, atoms are always prepared in the initial state $\ket{1}$, which, for red-detuning, corresponds to the lower, and for blue detuning to the higher energetic ground state\cite{Nunn:DPhil} (see fig. \ref{fig_ch2_lambda_system} \textbf{b}). 
In both cases, we start off with expectation values 
$\expect{\sigmaop{1,1}} \rightarrow 1$ and
$\left\{ \expect{\sigmaop{2,2}}, \expect{\sigmaop{3,3}} \right\} \rightarrow 0$ 
for the projectors onto these states. \newline
Because we only store a small number of photons in the Raman memory, we also assume that these numbers do not change significantly. 
Accordingly, we regard the $\left\{ \sigmaop{i,i} \right\}$-projector dynamics as constant. 
This leaves us with the simplified equations for the coherences $\sigmaop{i,j}$, with $i \neq j$ in eqs.~\ref{eq_ch2_projectors_1}. 

\paragraph{Rotating wave approximation, undesired couplings and linear approximation} 
Another frequently utilised simplification is to neglect rapidly oscillating terms. 
To identify these in the eqs.~\ref{eq_ch2_projectors_1}, we transform the projectors into a frame rotating with the respective transition frequencies, introducing the coherences 
${\sigmawigop{i,j} = \sigmaop{i,j} e^{\I \omega_{i,j} \tau}}$. 
On the one hand, these eliminate the leading terms $\sim \I \omega_{i,j} t$ in eqs.~\ref{eq_ch2_projectors_1}. 
On the other hand, they add oscillatory terms to the dipole matrix elements, i.e., the second terms in eqs.~\ref{eq_ch2_projectors_1} now read $\sim \vec{d}_{i,j} e^{-\I \omega_{i,j} \tau}$. 

These dipole terms are driven by the signal and control fields, $\vec{E}_s + \vec{E}_c$, which oscillate with their respective carrier frequencies $\omega_s$ and $\omega_c$. 
When inserting the sum of both electromagnetic fields (eqs.~\ref{eq_ch2_Ec} and \ref{eq_ch2_Es}) into eqs.~\ref{eq_ch2_projectors_1}, their products with the, now rotating, dipole matrix elements introduce terms that oscillate with the differences frequencies $\omega_{i,j}-\omega_c$ and $\omega_{i,j}-\omega_s$, as well as with the sum frequencies $\omega_{i,j}+\omega_c$ and $\omega_{i,j}+\omega_s$.  
We can neglect all terms of the latter category. Due to their rapid oscillations at essentially twice the optical carrier frequency, these average to zero over the time scales of the system's dynamics\footnote{
	The time-scales for the light-matter interaction in the Raman memory are set by the slowly varying envelope of the 
	control, which is on the order of 
	hundreds of pico-seconds (ps). The sum terms however oscillate on time scale of femto-seconds (fs). 
}. 
The left-over terms, oscillating with the difference frequencies, are of three types. 
First, there are terms with frequencies corresponding to the two-photon detuning 
$\Delta$, i.e. $\omega_{2,1} - \omega_s = \omega_{2,3} - \omega_c =  - \Delta$. 
The second group has frequencies $\Delta_1 = \omega_{2,3} - \omega_s$, which represent the coupling of the signal field to state $\ket{3}$. 
Since the signal is weak and the storage state $\ket{3}$ is empty to begin with, and only sparsely populated thereafter, we neglect this process.
Third, there are terms with frequencies 
$\Delta_2 = \omega_{2,1} - \omega_c = - (\Delta + \delta \nu_\text{gs})$. 
These are the coupling of the strong control pulses to the populated initial state $\ket{1}$, leading to spontaneous Raman scattering by the control. 
This is the onset of the FWM noise process, as we will investigate later (see section \ref{ch2_Raman_noise} and chapter \ref{ch7}). 
Because the detuning of this coupling is increased by the hyperfine ground state splitting $\delta \nu_\text{gs}$, the process is suppressed, compared to Raman storage. We thus also ignore it for the moment, but will come back to it later on. 
Nevertheless, we can already conclude that, in order to build a Raman memory with a reasonable signal-to-noise ratio (SNR), a ground state splitting equal or greater than the detuning $\Delta$ is desirable. 
Implementing these steps, the expressions for the coherences in eqs.~\ref{eq_ch2_projectors_1} simplify to\cite{Nunn:DPhil}:  
\begin{align}
\partial_t \sigmawigop{1,2} 	& = 
\frac{g_s}{\hbar} \cdot \vec{d}^{*}_{1,2} \cdot \vec{e}_s  \cdot \hat{S} \cdot e^{-\I \Delta \tau} 
- \frac{\I}{\hbar} \cdot \vec{d}_{2,3} \cdot \vec{e}_c \cdot E_c \cdot e^{-\I \Delta \tau} \cdot \sigmawigop{1,3}, \nonumber \\
\partial_t \sigmawigop{1,3} 	&= 
- \frac{\I}{\hbar} \cdot \vec{d}^{*}_{2,3} \cdot \vec{e}_c^{*} \cdot E^{*}_c \cdot e^{\I \Delta \tau} \cdot \sigmawigop{1,2} 
- \frac{g_s}{\hbar} \cdot \vec{d}^{*}_{1,2}\cdot \vec{e}_s \cdot \hat{S} \cdot e^{-\I \Delta \tau} \cdot \sigmawigop{2,3} , \nonumber \\
\partial_t \sigmawigop{2,3} 	&= 
\frac{g_s}{\hbar} \cdot \vec{d}_{2,3} \cdot \vec{e}^{*}_{s} \cdot \hat{S}^\dagger \cdot e^{\I \Delta \tau} \cdot \sigmawigop{1,3}
\label{eq_ch2_projectors_2}
\end{align}
These equations can be simplified even further, when considering the interaction strengths of the terms involved\cite{Nunn:DPhil}. 
In eqs.~\ref{eq_ch2_projectors_2}, most terms on the right-hand side only involve one operator, i.e. either the signal field $\hat{S}$ or a coherence $\sigmawigop{i,j}$. 
The exceptions are the second term for $\partial_t \sigmawigop{1,3}$ and the third equation for 
$\partial_t \sigmawigop{2,3}$ in eqs.~\ref{eq_ch2_projectors_2}, which contain products $\sim \hat{S} \cdot \sigmawigop{i,j}$ between the signal field and the coherences. 
These represent second order perturbations to the systems dynamics, which we will also ignore in the following. 
 Accordingly, we are left with two equations 
\begin{align}
\partial_t \sigmawigop{1,2} 	& = 
\frac{g_s}{\hbar} \cdot \vec{d}^{*}_{1,2} \cdot \vec{e}_s \cdot \hat{S} \cdot e^{-\I \Delta \tau} 
- \frac{\I}{\hbar} \cdot \vec{d}_{2,3} \cdot \vec{e}_c \cdot E_c \cdot e^{- \I  \Delta \tau} \cdot \sigmawigop{1,3}, \nonumber \\
\partial_t \sigmawigop{1,3} 	&= 
- \frac{\I}{\hbar} \cdot \vec{d}^{*}_{2,3} \cdot \vec{e}_c^{*} \cdot E^{*}_c \cdot e^{\I \Delta \tau} \cdot \sigmawigop{1,2},
 \label{eq_ch2_projectors_3}
\end{align} 
which describe the interaction of the atoms with the signal and control fields. 
Eqs.~\ref{eq_ch2_projectors_3} contain, on the one hand, the excitation of an atomic polarisation $\sim \sigmawigop{1,2}$, generated by the signal field $\hat{S}$, which couples to an atom in the initial state $\ket{1}$ (${\partial_t \sigmawigop{1,2} \sim \hat{S}}$). 
On the other hand, this polarisation $\sigmawigop{1,2}$ also couples to the control field $E_c$, which, in turn, maps it onto an atomic ground state coherence $\sigmawigop{1,3}$. 
These two couplings 
(${\partial_t \sigmawigop{1,2} \sim - E_c  \cdot \sigmawigop{1,3}}$ and 
${\partial_t \sigmawigop{1,3} \sim - E_c \cdot \sigmawigop{1,2}}$) 
correspond to signal read-in. 
Obviously, the time-reversed process can also occur. Here, the control couples to an already excited coherence 
$\sigmawigop{1,3}$ and maps it onto a polarisation coherence $\sigmawigop{1,2}$, involving the excited state $\ket{2}$. 
This then leads to the emission of the signal field $(\sim \hat{S})$ and completes the retrieval of a stored signal. 

\paragraph{Ensemble description}
So far, these equations were developed for a single atom. 
To extend this microscopic description to the atomic ensemble, addressed by the signal and control beams, we apply the same rational used in section \ref{subsec_ch_Maxwell_equations}, when introducing the macroscopic dielectric polarisation $\hat{P}$. 
Once more we consider all atoms in a volume element $\delta V$. 
The coherence terms in eqs.~\ref{eq_ch2_projectors_3}, containing the excited state, i.e. the terms involving operators $\sigmawigop{1,2}$, represent the macroscopic dielectric polarisation. 
For this reason their ensemble operator is also given by $\hat{P}$. 
Consequently, we only have to introduce another such operator for the ground state coherences $\sigmawigop{1,3}$. 
We define a similar ensemble operator 
$\hat{B} = \frac{1}{\sqrt{n} \delta V} \underset{r_i}{\sum} \sigmawigop{1,3}$, 
where $n$ is again the density of atoms. 
Notably, both operators, $\hat{P}$ and $\hat{B}$ inherit\cite{Nunn:DPhil} their boson-like commutation relations from the projectors $\sigmawigop{i,j}$. 
Insertion of these operators into eqs.~\ref{eq_ch2_projectors_3} yields the response of the atomic ensemble to the optical fields, which can now also be combined with eq.~\ref{eq_ch2_wave_equation_signal} for the spatial change in the signal field along the storage medium. 
The resulting three expressions
\begin{align}
\left[ \partial_z + \frac{1}{c} \partial_t \right] \hat{S}	 &= 
-\frac{\mu_0 \omega_s^2}{2 g_s k_s} \sqrt{n} \vec{e}_{s}^{*} \cdot \vec{d}_{1,2} \hat{P} = 
- \kappa^{*} \hat{P}, \nonumber \\
\partial_t \hat{P}			& = 
\I \Delta \hat{P} 
+ \frac{g_s}{\hbar} \left( \vec{d}^{*}_{1,2} \cdot \vec{e}_s \right)  \sqrt{n} \hat{S} e^{-\I \Delta \tau} 
- \frac{\I}{\hbar} \left( \vec{d}_{2,3} \cdot \vec{e}_c \right) E_c e^{- \I  \Delta \tau} \hat{B} 	
= \I \Delta \hat{P} + \kappa \hat{S} - \I \Omega(\tau) \hat{B} , \nonumber \\
\partial_t \hat{B} 					&= 
- \frac{\I}{\hbar} \left( \vec{d}^{*}_{2,3} \cdot \vec{e}_c^{*} \right) E^{*}_c e^{\I \Delta \tau} \hat{P} = 
- \I \Omega^{*}(\tau) \hat{P},
 \label{eq_ch2_maxwell_bloch_1} 
\end{align}
are the three Maxwell-Bloch equations, which describe the propagation of the input signal through the atomic vapour, initially prepared in state $\ket{1}$. 
Here, we have also introduced the control Rabi-frequency $\Omega$, whose temporal shape is that of the control pulse, and the signal field's coupling constant $\kappa$, given by
\begin{equation}
\Omega(\tau) = \frac{\vec{d}_{2,3} \cdot \vec{e}_c}{\hbar} \cdot E_c(\tau) \quad \text{and}
\quad 
\kappa = \frac{\vec{d}^{*}_{1,2} \cdot \vec{e}_s}{\hbar} \cdot g_s \cdot \sqrt{n} = \ \frac{\vec{d}^{*}_{1,2} \cdot \vec{e}_s}{\hbar} \cdot \sqrt{\frac{\hbar \omega_s n }{2 \epsilon_0 c}}.
\label{eq_ch2_Rabi_freq}
\end{equation}
What do these equations mean for the storage process? 
Upon read-in, the information, stored in the ground state coherences $\sigmawigop{1,3}$, is distributed over all atoms in the addressed ensemble.  
We denote such a state, generated by $\hat{B}^\dagger$, as a spin-wave coherence\cite{Fleischhauer:2002, Lukin:2003vn}. 
Storage is thus synonymous with the annihilation of a signal photon, i.e. an optical mode, and the generation of a spin-wave excitation in the atomic ensemble, i.e. a matter mode. Information retrieval is the opposite process. 
Moreover, because it is an excitation delocalised over many atoms, such a spin-wave is also an entangled Dicke state\cite{Dicke:1954fk} between all participating atoms in $\delta V$. %
Due to the robustness of the Dicke state entanglement\cite{Wieczorek:2009kx}, atom loss during information storage does not completely destroy the spin-wave. 
So, decoherence will not lead to total information loss, but rather to a gradual decrease in the memory efficiency.

\paragraph{Decay and decoherence}
Until now, eqs.~\ref{eq_ch2_maxwell_bloch_1} do not yet account for any such decoherence effects. 
However, in reality, the polarisation component $\hat{P}$ is subject to decay. 
It is caused by the electronic excited state $\ket{2}$, whose finite lifetime of $\tau_\text{Cs} \approx 32\ns$ in \cs\cite{Steck:2008qf} leads to spontaneous emission, destroying the $\sigmawigop{1,2}$ coherences. 
The polarisation term only plays a role during the actual read-in and read-out processes. 
Because these happen on the time scales of the signal and control pulse durations, in our case $\sim 300\ps$, spontaneous emission is less of a concern for our Raman memory. 
Nevertheless, we include the excited state decay in eq.~\ref{eq_ch2_maxwell_bloch_1}, by adding\footnote{
	A thorough introduction of these terms, based on Langevin noise operators, is provided in 
	\textit{Joshua Nunn's D.Phil. thesis}\cite{Nunn:DPhil}. Such a treatment is beyond the scope of this introduction. 
} 
an exponential decay term for the projectors, involving the excited state $\ket{2}$, with a rate $\gamma_\text{Cs} = \frac{1}{2 \tau_\text{Cs}}$. 

Apart from the polarisation, the spin-wave can also be subject to decoherence. 
By themselves, the hyperfine ground states $\ket{1}$ and $\ket{3}$  are long lived, so natural decay of population, initially prepared in state $\ket{1}$, into the storage state $\ket{3}$, is irrelevant for us. However, such ground state spin-flips can arise from collisions between two \cs\, atoms\cite{Raymer:1977}, which could occur during the signal's storage time\footnote{
 	We will see in later chapters, that these are on the order of a few micro-seconds. 
}. 
In the experiment, we add buffer gas to the \cs\,. 
A sufficiently high partial buffer gas pressure (see section \ref{ch2_Raman_level_scheme}) leads to preferential collisions between buffer gas and \cs\, atoms, which do not flip the {\cs} ground state spins. 
These extra collisions change the {\cs} velocity and modify the {\cs} transport from ballistic to diffusive, so it takes longer for {\cs} atoms to leave the interaction region with the signal and control beams. 
Because our protocol involves all Zeeman substates of the $\ket{1}$ and $\ket{3}$ hyperfine ground states, the spin-wave can also be subject to magnetic dephasing\cite{Reim:2011ys}. Furthermore, as mentioned above, atoms can be lost as they leave the interaction volume during the storage time. 
We absorb all of these mechanism in another phenomenological decay rate $\gamma_B$ for the spin-wave. 
By adding both decay terms to the Maxwell-Bloch equations, these modify to:
\begin{align}
\partial_z \hat{S}	 &= 
- \kappa^{*} \hat{P}, \nonumber \\
\partial_\tau \hat{P}			& = 
-\gamma \hat{P} + \I \Delta \hat{P} + \kappa \hat{S} - \I \Omega(\tau) \hat{B} = 
- \Gamma \hat{P} +  \kappa \hat{S} - \I \Omega(\tau) \hat{B} , \nonumber \\
\partial_\tau \hat{B} 					&= 
- \gamma_B \hat{B} - \I \Omega^{*}(\tau) \hat{P},
\label{eq_ch2_maxwell_bloch_2} 
\end{align}
where we have now also introduced the complex detuning $\Gamma = \gamma - \I \Delta$ and transformed the variables\footnote{
	The derivatives transform as\cite{Nunn:DPhil}: 
	$$
	\partial_z |_t + \frac{1}{c} \partial_t |_z = \partial_z |_\tau \quad \text{and} \quad \partial_t |_z = \partial_\tau |_z
	$$
} $(z,t)$ into $(z,\tau)$, which are a coordinate system moving along with the signal pulse. 
Eqs.~\ref{eq_ch2_maxwell_bloch_2} illustrate the coupling between both observables $\hat{B}$ and $\hat{S}$, mediated via $\hat{P}$, on the density of atoms $n$. 
To have efficient storage, a dense atomic vapour is required, leading to a high, on-resonance optical depth
$d = \frac{|\kappa|^2 \cdot L }{\gamma} = \frac{|\vec{d}_{1,2} \cdot \vec{e}_s| \omega_s \cdot L \cdot n}{2 \hbar \epsilon_0 \gamma c}$, 
for an ensemble of length $L$. 
This is intuitively clear because higher density $n$ for a fixed length $L$ means more atoms per volume $\delta V$ and thus a higher probability for a signal photon to hit the Raman interaction cross-section\cite{Penzkofer:1979} of a {\cs} atom. 
Moreover, it can be show\cite{Nunn:DPhil} that the optimal efficiency $\eta_\text{opt}$ of the memory is solely determined by the optical depth to $\eta_\text{opt} \approx 1 - \frac{2.9}{d}$.  
It is limited by spontaneous emission ($\gamma$) and increases towards unity as the number of atoms in the ensemble ($n \cdot L$) increases.

\paragraph{Adiabatic limit}
Moving towards the operational regime of the Raman memory, we will now introduce the adiabatic limit. 
In this limit, the polarisation $P$ adiabatically follows the time evolution of the signal and control pulses. 
Changes of $\hat{P}$ over time can be assumed to equal those of the signal ($\hat{S}$) and the control ($\Omega(\tau)$) pulses, such that $\partial_\tau \hat{P} = 0$. 
For this reason, we can eliminate the dynamics of $\hat{P}$ and simplify the system 
to a set of two coupled partial differential equations (PDEs). 
The $2^\text{nd}$ equation in \ref{eq_ch2_maxwell_bloch_2} is solved for $\hat{P}$ and the result is inserted to the other equations, yielding 
\begin{align}
\left[ \partial_z + \frac{d \gamma}{\Gamma} \right] \hat{S}	&=  \I \frac{\Omega \sqrt{d {\gamma}} }{\Gamma} \hat{B},\nonumber \\
\left[ \partial_\tau + \frac{\abs{\Omega}^2}{\Gamma} \right] \hat{B}  &= -\I \frac{\Omega^{*} \sqrt{d {\gamma}}}{\Gamma} \hat{S},  
\label{eq_ch2_maxwell_bloch_3} 
\end{align}
where, for simplicity, we have ignored spin-wave decoherence ($\gamma_B \rightarrow 0$). 
Additionally, we have normalised the spatial propagation $z$ to ${z}/{L} \rightarrow z$, such that $z$ takes values of $z \in \left[0,1 \right]$. To this end, also the spatial derivative 
$\partial_z$ and the spin wave operator $\hat{B}$ are renormalised to 
$\frac{1}{L} \partial_z \rightarrow \partial_z$ and 
$\sqrt{L} \cdot \hat{B} \rightarrow \hat{B}$. 
For this approximation to hold, i.e. to enter the adiabatic regime, 
$1/\abs{\Gamma}$ 
has to be the shortest timescale in the problem. 
Since $\gamma$ is fixed, the detuning $\Delta$ needs to be the dominant frequency and it has to satisfy the following conditions: 
\begin{equation}
\Delta \gg \Delta \nu_c, \quad \Delta \gg \Omega_\text{max} \quad \text{and} \quad \Delta \gg \gamma,\label{eq_ch_adiabatiic_limit}
\end{equation}
Here, $\Delta \nu_c$ is the full width at half maximum (FWHM) spectral bandwidth of the control pulse, $\Omega_\text{max}$ is the peak Rabi frequency and $\gamma$ is the linewidth of the excited state $\ket{2}$. 
For the operational regime of the Raman memory, this is fulfilled, as we will see in section \ref{ch2_Raman_level_scheme}. 

We can achieve some more simplifications, if we firstly measure time in units of the excited state lifetime, re-defining $tau$ as $\tau := \gamma \cdot \tau$, and spatial propagation in terms of the ensemble length, setting $z$ to $z := \frac{z}{L}$, such that $z \in \left[ 0,1\right]$. 
Furthermore, we can view the temporal dynamics in terms of the Rabi frequency, i.e. we switch from the time domain $\tau$ to the energy level of the control pulse that has already interacted with the atomic ensemble at time $\tau$. To this end, we introduce the integrated Rabi frequency 
$$
\omega(\tau) = \frac{1}{W} \overset{\tau}{\underset{-\infty}{\int}} |\Omega(\tau')|^2 \text{d} \tau'\,,
\quad \text{with} \quad
W = \overset{\infty}{\underset{-\infty}{\int}} |\Omega(\tau')|^2 \text{d} \tau' 
=  \frac{1}{2 \epsilon_0 c A_c} \cdot \abs{ \frac{\vec{d}_{2,3} \cdot \vec{e}_c}{\hbar} }^2 \cdot E^p_c.
$$
$W$, which is the fully integrated $\Omega(\tau)$, corresponds to the control pulse energy $E^p_c$ for a beam with diameter $A_c$. With these constants, we can now also define the Raman coupling 
\begin{equation}
C_\text{S} = \frac{\sqrt{W \cdot d \cdot \gamma}}{|\Gamma|} \approx \frac{\sqrt{W \cdot d \cdot \gamma}}{|\Delta|}. 
\label{eq_ch2_Raman_coupling}
\end{equation} 
$C_\text{S}$ represents the mixing angle between the optical mode at the Stokes frequency, i.e. the input signal, and the matter modes\footnote{
	In a semi-classical treatment 
	$C_\text{S} \approx \Omega_\text{eff} \Delta \tau_c$. Here, the effective Rabi-frequency 
	$\Omega_\text{eff} = \frac{\Omega_\text{max} \cdot \kappa \cdot \sqrt{\frac{L}{\Delta \tau_c}}}{\Delta}$ 
	is proportional to the product of the control and signal Rabi-frequencies, and 
	$\Delta \tau_c$ is the control pulse duration with 
	$\Delta \tau_c \sim \frac{1}{\Delta \nu_c}$. 
}. 
As eqs.~\ref{eq_ch2_maxwell_bloch_3} illustrate, in the Raman limit, where $\Gamma \sim \Delta$, the coupling between light ($\hat{S}$) and matter ($\hat{B}$) modes only depends on the Rabi-frequency $\Omega(\tau)$, the optical depth $d$ and the detuning $\Delta$. 
In other words, the interaction strength is completely defined by the Raman coupling $C_\text{S}$.  
In turn, these three variable are the experimental parameters we have at our disposal. 
We can firstly choose the detuning, with the constraint that we need to stay in the adiabatic-regime, sufficiently far off resonance.  
Since $\Omega(\tau)$ and $d$ depend on the control pulse parameters and the atomic vapour density $n$, respectively, we can, on the one hand, also choose the control pulse's intensity ($\sim \Omega_\text{max}$) and duration ($\sim 1/(\Delta \nu_\text{c})$). On the other hand we can set the vapour density $n$ through the ensemble's temperature $T$ (see section \ref{ch2_Raman_level_scheme} below). 
We choose these to optimise the memory performance (see also appendix \ref{appendix_ch7_noise_parameter_dependence}). 

Phenomenologically, the adiabatic approximation causes our $\Lambda$-system to become an effective 2-level system. Incoming photons are absorbed in the ensemble with an effective absorption coefficient $\sim C_\text{S}$. 
The control pulse mediates this absorption, for which reason we can view its effect on the system as creating a virtual absorptive resonance at a frequency detuned by $\Delta$ with respect to $\ket{2}$. 
In contrast to resonant absorption of an atomic excited state, input photons ($\hat{S}$) are instead absorbed into an atomic ground-state coherence ($\hat{B}$). 
Notably, the Raman storage process, described by eqs.~\ref{eq_ch2_maxwell_bloch_3}, is also different to Rabi-oscillations, which one can, for instance, observe in a $\Lambda$-system of a single atom, illuminated by two intense lasers on resonance. 
Contrary to dealing with single atoms and strong lasers, our signal is weak and interacts with an ensemble of atoms, i.e. we have many more atoms than signal photons, so the signal can never saturate the transition. In other words, the first half of a Rabi-cycle, where the full atomic population is shuffled from one state to the other, here from state $\ket{1}$ to $\ket{3}$, is never completed, instead the signal photons are absorbed. 
Even when reading in more intense signals, this still holds. Because we use pulsed fields, 
the coupling between the atomic population to the signal mode is turned off once the signal is absorbed. 
This prevents the second half of a Rabi-cycle. As we will see now, when looking at the solution of eqs.~\ref{eq_ch2_maxwell_bloch_3}, choosing the control pulse shape to maximise read-in efficiency means to effectively tailor it to maximise the first half of a Rabi-cycle\footnote{
	and simultaneously minimise the second half. 
}. 

\paragraph{Propagator based solution}
To find an expression for the memory efficiency, we write down the generic solution to eqs.~\ref{eq_ch2_maxwell_bloch_3}. 
Since these equations are linear impulse-response functions of the ensemble to external fields, 
we can write their solutions in terms of propagators, or Green's functions:
\begin{align}
\hat{S}_\text{out}^t (\omega) 	&= 
\overset{W}{\underset{0}{\int}} \hat{L}(\omega,\omega') \hat{S}_\text{in}^t(\omega') \text{d} \omega' -
\overset{L_z}{\underset{0}{\int}} \hat{K}(\omega,z') \hat{B}^t_\text{in}(z') \text{d} z',\nonumber \\ 
\hat{B}^t_\text{out} 	(z)	&= 
\overset{L_z}{\underset{0}{\int}} \hat{L}(z,z') \hat{B}_\text{in}^t(z') \text{d} z' +
\overset{W}{\underset{0}{\int}} \hat{K}(z,\omega') \hat{S}_\text{in}^t(\omega') \text{d} \omega', 
\label{eq_ch2_maxwell_bloch_4}
\end{align}
Here, $L_z$ is the ensemble length, the superscript $t$ is the time bin, denoting either the read-in ($t_\text{in}$) or retrieval ($t_\text{out}$), time, and the Green's functions $\hat{L}(x,y)$ and $\hat{K}(x,y)$ are the memory kernels. The kernels connect the input (subscript in), i.e. signal and spin-wave prior to the Raman interaction, with the output fields, i.e. signal and spin-wave after the Raman interaction. 
$\hat{K}(x,y)$ represents the storage and retrieval interaction, which maps an incoming optical mode $\hat{S}^{t_\text{in}}_\text{in}$ into the spin-wave $\hat{B}^{t_\text{in}}_\text{out}$, or retrieves an already stored spin-wave $\hat{B}^{t_\text{out}}_\text{in}$ back into the optical mode $\hat{S}^{t_\text{out}}_\text{out}$. 
Conversely, $\hat{L}(x,y)$ describes the transmission of either mode. 
For the optical mode $\hat{S}^{t_\text{in}}_\text{in}$, this is the non-stored fraction of the input signal, transmitted through the memory. 
For the spin-wave $\hat{B}^{t_\text{out}}_\text{in}$ it is the amount of spin-wave not retrieved during read-out. 

Upon retrieval ($t_\text{out}$), we do not send in a signal field and  
$\int \hat{L}(\omega,\omega') \hat{S}_\text{in}^{t_\text{out}}(\omega') \text{d} \omega' \rightarrow 0$. 
Similarly, for a perfect Raman memory, there is no prior excited spin-wave during signal read-in ($t_\text{in}$), so  
$\int \hat{K}(z,z') \hat{B}_\text{in}^{t_\text{in}} (z') \text{d} z' \rightarrow 0$. 
However, in the presence of noise, this does not hold.
As we see in section \ref{ch2_Raman_noise} below, the Raman memory noise processes can actually lead to pre-excited spin-waves. These contribute noise to the transmitted signal $\hat{S}_\text{out}^{t_\text{in}}$ and to the  stored spin wave $\hat{B}_\text{out}^{t_\text{in}}$. Because the latter is the input spin-wave for retrieval, i.e. $\hat{B}_\text{in}^{t_\text{out}}  = \hat{B}_\text{out}^{t_\text{in}}$, any such noise contributions also feed through into the retrieved signal; appendix \ref{app6_coh_model} discusses this in detail. 

The functional form of the kernels can either be found by numerical integration, which is outlined in appendix \ref{app6_coh_model}, or analytically\cite{Nunn:DPhil, Wu:2010} by Fourier transforming the first equation into wavevector space. 
In the Raman limit, one can show\cite{Nunn:DPhil} that the storage kernel has the form: 
\begin{equation}
\hat{K}(x,y) = 
-\I \cdot C_\text{S} \cdot  \exp{ \left\{ -\frac{W (1-y) + d \cdot x}{\Gamma} \right\}} \cdot 
J_0 \left( 2 \I \cdot C_\text{S} \cdot \sqrt{x \left( 1-y \right)} \right).
\label{eq_ch2_Raman_kernel}
\end{equation}
$J_0$ is the $0^\text{th}$ order Bessel function of the $1^\text{st}$ kind. The exponential factor on the right-hand side includes the dynamic Stark shift $\left( \sim e^{-\frac{W (1-y)}{\Gamma}}\right)$ and linear absorption\footnote{
	Note however, that in the Raman limit, the linear absorption term $\sim e^{-\frac{d \cdot z}{\Gamma}}$ 
	becomes just a phase factor $\sim e^{- \I \frac{d \cdot z}{\Delta}}$, as $\Gamma \rightarrow -\I \Delta$.  
} $\left( \sim e^{-\frac{d\cdot x}{\Gamma}}\right)$. 
$\hat{K}(x,y)$ contains the above mentioned control pulse dependence of the interaction via the proportionality of the argument in $J_0$ with $C_\text{S}$. 
As we shall see now, we generally strive to choose a control pulse shape that maximises the overlap between the resulting kernel $\hat{K}(x,y)$ and the signal field $\hat{S}$. 
The Stark shift term $\sim \frac{\abs{\Omega}^2}{\Delta}$ in eqs.~\ref{eq_ch2_maxwell_bloch_3} and \ref{eq_ch2_Raman_kernel} causes an additional detuning of the virtual Raman resonance that follows the control's intensity profile. With signal and control in 2-photon resonance at the beginning of the interaction\footnote{
	i.e. signal and control have the same detuning $\Delta$ with respect to state $\ket{2}$
}, 
the leading edge of the control pulse shifts the Raman line away from resonance with the signal, while it comes back into resonance in the trailing edge. The effects of this Stark shift is small for our free space set-up. It reduces the efficiency slightly, but can greatly be compensated by introducing an additional 1-photon detuning between signal and control\footnote{
	Note that the control pulse frequency is not chirped here. 
	The instantaneous spectral bandwidth of the control does not change throughout the pulse, 
	for which reason the induced virtual resonance linewidth remains constant. 
	The Raman linewidth is large compared to the Stark shift, for which reason the  
	effect of a small shift in the line's central frequency, caused by the Stark shift, on the spectral 
	overlap between signal and storage kernel, is small. 
	This is different to rapid adiabatic passage, where a linear frequency chirp is
	applied across the control pulse, leading to a varying instantaneous frequency bandwidth of the 
	control. In such a situation, the Raman linewidth, and thus the spectral overlap between signal and storage 
	kernel, would vary throughout the interaction time. 
}. 
So we shift the detunings of signal and control from state $\ket{2}$ slightly with respect to one another. 
As a side note, while the Stark shift is unproblematic for our free-space case, it can become sizeable when, e.g., considering an intra-cavity memory\cite{Saunders:2015}. In such a scenario, actual control pulse shaping is necessary for its compensation. 

\subsection{Memory operation in the adiabatic limit\label{subsec_mem_op_adiabatic_lim}}
Assuming noise free operation with 
$\int \langle \left(\hat{B}^{t_\text{in}}_\text{in} \right)^\dagger \hat{B}^{t_\text{in}}_\text{in} \rangle \text{d} z = 0$, we can now easily obtain the read-in efficiencies from eqs.~\ref{eq_ch2_maxwell_bloch_4}, just by evaluating the expectation value for the number of spin-wave excitations ($\hat{N}_B$) that are generated for a set number of input signal photons ($\hat{N}_S$). Writing this in terms of the expectation values for the respective number operators yields 
\begin{equation}
\eta_\text{in}=
\frac{\langle \hat{N}_B \rangle}{\langle \hat{N}_S \rangle} = 
\frac{ \overset{L_z}{\underset{0}{\int}}  \langle \left( \hat{B}_\text{out}^{t_\text{in}}(z) \right)^\dagger \hat{B}_\text{out}^{t_\text{in}} (z)\rangle \text{d} z }
{ \overset{W}{\underset{0}{\int}}  \langle \left( \hat{S}_\text{in}^{t_\text{in}}(\omega) \right)^\dagger \hat{S}_\text{in}^{t_\text{in}}(\omega)  \rangle \text{d} \omega}, \quad \text{with} \quad 
\hat{B}_\text{out}^{t_\text{in}}(z)  = \overset{W}{\underset{0}{\int}} \hat{K}(z,\omega) \hat{S}_\text{in}^{t_\text{in}} (\omega) \text{d} \omega.
\label{eq_ch2_eff_in}
\end{equation} 
So, the read-in efficiency is determined, firstly by the mode overlap between memory kernel and input signal, and, secondly, by the amplitude of $\hat{K}$, set by the Raman coupling $C_\text{S}$.
Assuming perfect mode matching for the moment, the read-in efficiency is maximised along with 
${C_\text{S} \approx \frac{\sqrt{W d \gamma}}{\Delta}}$. 
Because the detuning $\Delta$ is large in the Raman limit, we require a high control pulse energy $W$ and a large optical depth $d$ to achieve any significant interaction strength. 
This means, we need highly energetic control pulses and an optically dense atomic ensemble. 
To achieve the latter, the atoms should, on the one hand, have large dipole moments $\vec{d}_{i,j}$ for the transitions $\left\{ i,j\right\}$ involved in the $\Lambda$-system (see fig. \ref{fig_ch2_lambda_system}), and, on the other hand, allow to obtain sufficient atomic number densities within a reasonable temperature regime\cite{Nunn:DPhil}. 
Atomic \cs\, vapour is a good compromise\footnote{
	Vapour densities achievable in alkali vapour are higher than, for instance, 
	those obtainable with earth alkali systems. 
	This makes alkali atoms a preferred medium when it comes to dense vapours. 
	Amongst the alkalis, caesium's oscillator strengths are higher than those of other 
	stable elements potassium, sodium and rubidium. 
	Furthermore {\cs} has only one stable isotope, simplifying its usage. 
	This makes \cs\, the preferred candidate system
	for the Raman memory. 
} 
to fulfil both of these requirements\cite{Steck:2008qf}. 
Additionally, {\cs} has the largest hyperfine splitting of the stable alkali atoms, allowing for broadband input signals.  
For these reasons it is our atomic medium of choice.  
Obviously, the efficiency cannot be increased to arbitrarily large values. In a previous paper\cite{Nunn:2007wj}, our group has shown that values of $C_\text{S} \sim 2$ are required to converge against the optimal read-in efficiency $\eta^\text{opt}_\text{in} \approx 1-\frac{2.9}{d}$, which is limited by absorption in and spontaneous emission from the atomic ensemble. 
For our system, introduced in section \ref{ch2_Raman_level_scheme}, we achieve $C_\text{S} \approx 0.82$ and $d\approx 1800$, which predicts $\eta_\text{in}^\text{opt} \approx 99.8\,\%$ from the optical depth $d$, but an expected 
$\eta_\text{in} \approx 50\,\%$, 
due to the size of $C_\text{S}$ (see \textit{J. Nunn et. al. 2007}\cite{Nunn:2007wj}). 

So far, we have neglected the mode matching between $\hat{K}$ and $\hat{S}$, which also enters $\eta_\text{in}$. 
For a generic control pulse, we can determine this mode overlap by performing a singular value decomposition (SVD) of the resulting kernel function\footnote{
	Since the kernel in eq.~\ref{eq_ch2_Raman_kernel} for sech-shaped control pulses is hermitian 
	in the Raman limit, the SVD corresponds to the eigenvalue and eigenvector decomposition of 
	$\hat{K}(z,\omega)$.
} $\hat{K}(x,y)$ into a set of light modes $\left\{ \phi_i(1-y) \right\}$ and matter modes $\left\{ \psi_i(x) \right\}$. 
From the conservation of energy\footnote{
	Energy conservation requires a constant 
	total combined number of photons and spin-wave
	excitations. As a result, the kernel functions have to fulfill\cite{Nunn:DPhil} 
	$\hat{L}^\dagger \hat{L} + \hat{K}^\dagger \hat{K} = \mathds{1}$ and 
	$ \hat{L}  \hat{L}^\dagger+ \hat{K} \hat{K}^\dagger = \mathds{1}$. 
} between the signal field and the spin-wave excitations, one can also show, that both kernels $\hat{K}(x,y)$ and $\hat{L}(x,y)$ must have the same sets of eigenmodes\cite{Nunn:DPhil}: 
$$
\hat{K}(x,y) = \underset{i}{\sum} \hat{\psi}_i(x) \lambda_i \hat{\phi}^{*}_i(1-y), \quad \text{and} \quad 
\hat{L}(x,y) = \underset{i}{\sum} \hat{\psi}_i (x) \mu_i \hat{\phi}^{*}_i(1-y), \quad \text{with}\quad \lambda_i^2 + \mu_i^2 = 1.
$$
Moreover, for the Raman kernel in eq.~\ref{eq_ch2_Raman_kernel}, one can show\cite{Nunn:DPhil} that these two sets are identical, due to its persymmetry\footnote{
	A persymmetric matrix is matrix that is symmetric under the reflection of its elements on its anti-diagonal. 
}. 
For a given input signal with amplitude $\hat{S}(t)$, we thus want to choose a control pulse shape $E_c(t)$ 
that leads to the largest overlap 
$ \underset{i}{\sum} \lambda_i {\displaystyle \int} \hat{\phi}^{*}_i(1-\tau) \hat{S}(\tau) \text{d} \tau$. 
Generally, this is obtained when maximising the overlap for the $1^\text{st}$ singular value $\lambda_1$, choosing an appropriately shaped control pulse\cite{Nunn:2007wj, Novikova:2012}. 
When thinking about an atomic 3-level system in terms of Rabi-oscillations, this corresponds to the construction of the control to act as a $\pi$-pulse. In other words, the control pulse duration is tailored such that it is switched off, once the population has been shuffled from the initial state $\ket{1}$ into the storage state $\ket{3}$ during signal read-in and vice-versa for signal retrieval. 

Read-out from the memory can happen in two geometries: Along the same direction as the read-in (forward) or in the opposite direction (backward). 
While backward retrieval can result in higher read-out efficiencies, the forward direction is experimentally simpler to implement, for which reason it our method of choice. 
The performance difference is due to the spatial distribution of the stored spin-wave, whose amplitude is decaying roughly exponentially along the optical axis. In forward retrieval, most of the signal is thus released at the entrance of the atomic ensemble and has to propagate through the entire ensemble thereafter. 
The released signal is subject to higher linear absorption in the storage medium. 
Conversely, the main portion of light retrieved in the backward direction is released just in front of the exit point (see \textit{Nunn et. al. 2008}\cite{Nunn:2008xr} for details). 
To arrive at the read-out efficiency $\eta_\text{out}$, the same arguments as for the derivation of eq.~\ref{eq_ch2_eff_in} apply, only that now we map $\hat{B}_\text{in}^{t_\text{out}}$, which is $\hat{B}_\text{in}^{t_\text{out}}  =  \hat{B}_\text{out}^{t_\text{in}}$, onto $\hat{S}_\text{out}^{t_\text{out}}$ to yield
\begin{equation}
\eta_\text{out}=
\frac{\langle \hat{N}_S \rangle}{\langle \hat{N}_B \rangle} = 
\frac{ \overset{W}{\underset{0}{\int}}  \langle \left( \hat{S}_\text{out}^{t_\text{out}}(\omega) \right)^\dagger \hat{S}_\text{out}^{t_\text{out}}(\omega) \rangle \text{d} \omega }
{ \overset{L_z}{\underset{0}{\int}}  \langle \left( \hat{B}_\text{in}^{t_\text{in}}(z) \right)^\dagger \hat{B}_\text{in}^{t_\text{in}}(z)  \rangle \text{d} z}, 
\quad \text{with} \quad 
\hat{S}_\text{out}^{t_\text{out}}(\omega)  = \overset{L_z}{\underset{0}{\int}} \hat{K}^\text{r}(\omega,z) \hat{B}_\text{in}^{t_\text{out}} (z) \text{d} z.
\label{eq_ch2_eff_out}
\end{equation}
Here, we use the retrieval kernel $K^\text{r}(\omega,z)$, which has the same functional form as eq.~\ref{eq_ch2_Raman_kernel}. 
Consequently, the maximisation of $\eta_\text{out}$ is similar to that of $\eta_\text{in}$. 
We can now define the total memory efficiency $\eta_\text{tot} = \eta_\text{out} \cdot \eta_\text{in}$, which maps 
$\hat{S}_\text{in}$ onto $\hat{S}_\text{out}$ using the product kernel 
${\hat{K}^\text{t}(\omega',\omega) = \overset{L_z}{\underset{0}{\int}} \hat{K}^\text{r}(\omega',z)\hat{K}(z,\omega) \text{d}z}$.

Before looking at the actual experimental implementation of this protocol, we cover one more important aspect and briefly outline the noise processes relevant for our Raman memory. 
These will add another term to eqs.~\ref{eq_ch2_maxwell_bloch_3}, leading to false contributions to the signal transmitted through, and retrieved from the memory.


\section{Memory noise processes\label{ch2_Raman_noise}}

From previous experiments we already knew about the presence of noise in the Raman memory system\cite{Reim:2011ys}. 
One important part of our work will therefore concern its thorough investigation, for which we will determine the noise's consistency, study its parameter dependences and look for ways to suppress it. 
These results form the final part of this thesis (chapter~\ref{ch7} and appendix~\ref{appendix_ch7_noise_parameter_dependence}). 
Here, we introduce the theoretical foundations and the processes behind the relevant noise constituents. 
Our focus lies on the transitions involved in four-wave-mixing, the most significant noise process in our system.  We thus add its description to the system of eqs.~\ref{eq_ch2_maxwell_bloch_3}, with the solutions outlined in appendix \ref{app6_coh_model}. 

\begin{figure}[h!]
\includegraphics[width=\textwidth]{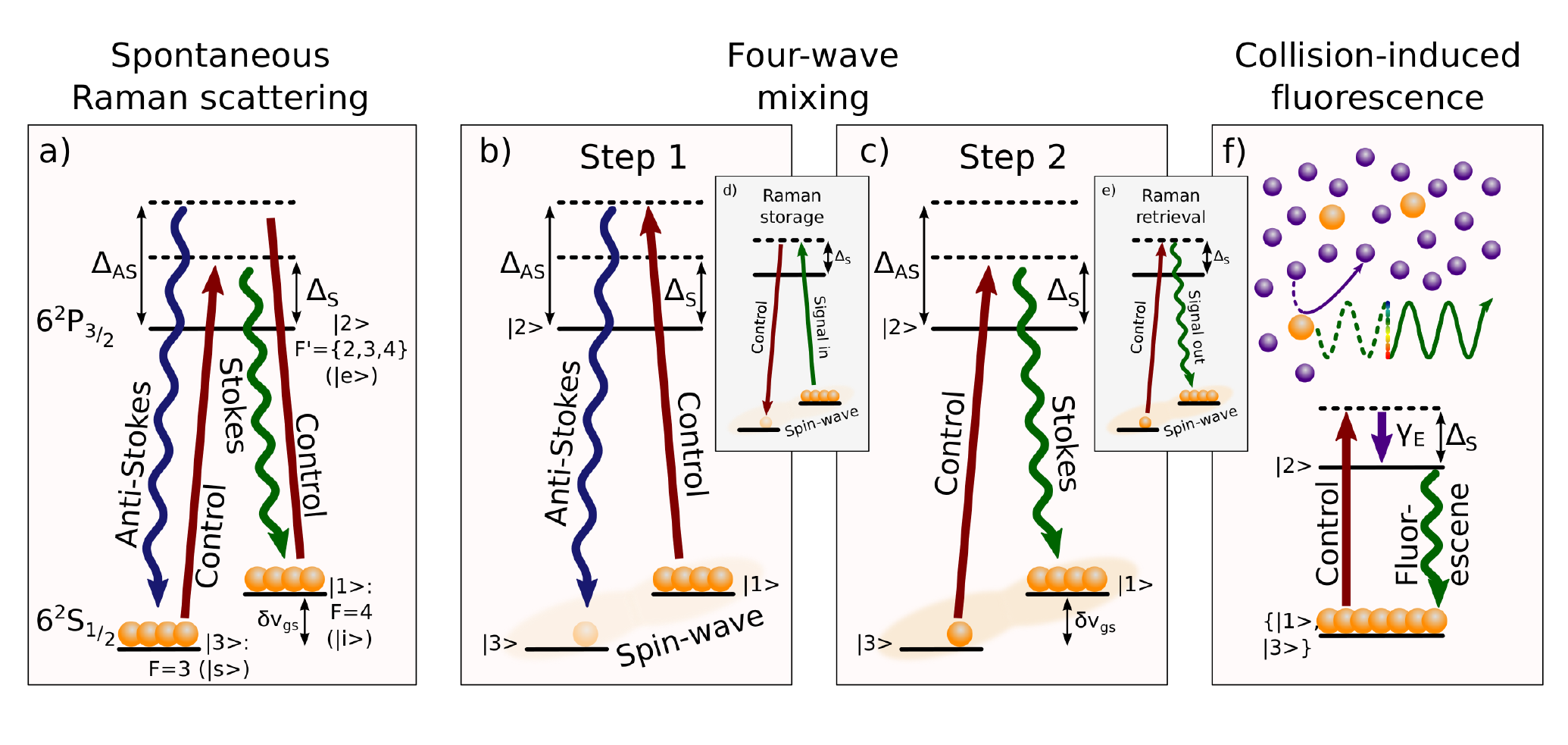}
\caption{Noise processes in the Raman memory. All panels depict the externally applied optical fields by \textit{straight arrows}, and fields generated by nonlinear processes by \textit{wiggly arrows}. 
\textbf{(a)}: SRS for a thermally distributed \cs\, population. 
The control couples to the initial state ($\ket{1} = \ket{\text{i}}$) and the storage state ($\ket{3} = \ket{\text{s}}$). 
To aid the comparison with our implementation of the Raman protocol in atomic {\cs}, shown in fig. \ref{fig_ch2_Raman_protocol}, the relevant {\cs} energy levels are also denoted here. 
\textbf{(b)}: AS noise scattering as the $1^\text{st}$ step in the FWM process. Atomic population is transferred by the control, which couples to the initial state ($\ket{1} = \ket{\text{i}}$) of the spin-polarised ensemble, exciting a spin-wave (\textit{transparent yellow area}). 
\textbf{(c)}: Step 2 of the FWM process; spin-wave retrieval by the control under Stokes noise emission.
\textbf{(d)}: Raman memory signal read-in. 
\textbf{(e)}: Raman memory signal retrieval.
\textbf{(f)}: Fluorescence induced by collisions between \cs\, (\textit{yellow}) and Ne buffer gas atoms (\textit{purple}). 
Top panel illustrates the collision effects on light emitted by \cs\,; bottom panel displays energy-gap bridging by collisions, leading to the excitation of \cs\, atoms. 
}
\label{fig_ch2_FWMlevels}
\end{figure}

In general, noise in the Raman memory falls into two categories: one- and two-photon transition processes. 
The two-photon transition processes are spontaneous Raman scattering (SRS) and four-wave-mixing (FWM). 
Collisional induced fluorescence is the only relevant one-photon transition based noise component. 
Other one-photon processes at the control frequency, such as Rayleigh scattering\cite{Carlsten:1977}, are too weak to contribute significantly\footnote{
	In principle, insufficient control field filtering at the memory output could also add leakage noise into 
	the signal mode; yet, this is eliminated experimentally (see section \ref{ch7_subsec_cold_ensemble}).
}.  
The relevant processes are illustrated in fig. \ref{fig_ch2_FWMlevels}, alongside the Raman memory protocol (fig. \ref{fig_ch2_lambda_system}) added for comparison.

\subsection{Spontaneous Raman scattering\label{ch7_subsec_SRS}}

The simpler two-photon transition process is spontaneous Raman scattering\cite{Penzkofer:1979,Raymer:1985bv,Raymer:1981aa} (SRS), depicted in fig. \ref{fig_ch2_FWMlevels} \textbf{a}. 
It can be generated via the control coupling to population in the initial ($\ket{1}$) or the storage ($\ket{3}$) state, introducing a Raman transition to the other hyperfine ground state. 
Note, this scattering occurs spontaneously and does not require the presence of the input signal. 
Control coupling to state $\ket{3}$ results in the emission of a Stokes (S) photon, whose frequency equals the signal in the Raman memory protocol. This means, it is detuned by $\Delta_\text{S}$ from state $\ket{2}$ and noise is emitted into the frequency mode of the signal. 
Conversely, for the control coupling to the initial state $(\ket{1})$, the detuning increases to $\Delta_\text{AS} = \Delta_\text{S} + \delta \nu_\text{gs}$. 
The emitted anti-Stokes (AS) scattering 
has a central frequency that is 
shifted further off resonance compared to the memory signal frequency. The shift corresponds to the ground state splitting $\delta \nu_\text{gs}$. 
The frequency spectrum of both, S and AS photons, is determined by the control spectrum\cite{Raymer:1979}. 
As shown in fig. \ref{fig_ch2_FWMlevels} \textbf{a}, transitions can occur out of both ground states 
if these are populated. 
In the experiment, SRS is the noise process observed when the ensemble is thermally distributed, i.e., when the
atomic population is not prepared solely in state $\ket{1}$ to start with\footnote{
	In principle, Raman scattering can also be operated in the stimulated regime\cite{Penzkofer:1979}. 
	However, as we will see in section \ref{ch7_ctrl_power}, our operational parameters for the \cs\, 
	vapour and the control field pulse energies are not sufficient to observe stimulated emission 
	in either channel. 
}. 
As we will see now, SRS is also the onset process of FWM noise. However, unlike FWM, SRS does not rely on any spin-wave dynamics. 
We demonstrate this in chapter \ref{ch7}, where we utilise spin-wave related properties to distinguish both noise sources.

\subsection{Four-wave mixing\label{ch7_subsec_FWM}}
Four-wave-mixing\cite{Carman:1966} can be envisaged as a two-step process. 
It starts with an initially spin-polarised \cs\, ensemble, generated by optical pumping. 
Fig. \ref{fig_ch2_FWMlevels} \textbf{b} illustrates this first step. Here, the control couples to the initial state 
$(\ket{1})$, causing AS scattering. Without any prior control field interaction, this initial scattering process is SRS. 
The resulting atomic population transfer to the memory storage state $(\ket{3})$ leads to the generation of a spin-wave coherence between both ground states. 
This is similar to the Raman memory spin-wave creation\cite{Fleischhauer:2000vn}; 
for comparison, the Raman memory read-in step is shown in fig. \ref{fig_ch2_FWMlevels} \textbf{d}.

In the second step, depicted in fig. \ref{fig_ch2_FWMlevels} \textbf{c}, the control retrieves the previously generated FWM spin-wave by coupling to the storage state $(\ket{3})$. 
The corresponding $\Lambda$-level system is thus equal to retrieval from the Raman memory (fig. \ref{fig_ch2_FWMlevels} \textbf{e}). Both processes have the same detuning $\Delta_\text{S}$, so the emitted Stokes (S) noise has the same spectral properties as the signal retrieved from the Raman memory. 
Consequently, the main difference between FWM and the Raman memory protocol is the first step of both processes. 

While in the first step, the control couples to state $\ket{1}$, in the second it couples to state $\ket{3}$. 
Despite having all population in $\ket{1}$ to start with, the FWM AS scattering is weaker than Raman storage. This is the case thanks to the increase detuning of the AS leg when operating the protocol blue-detuned as shown in figs. \ref{fig_ch2_FWMlevels} - \ref{fig_ch2_Raman_protocol}. 
With blue detuning, the AS channel is naturally further away from resonance, so its Raman coupling constant $C_\text{AS} = \sqrt{\frac{W d \gamma }{\Delta_\text{AS}}}$ is lower than $C_\text{S}$ of the Raman storage (see eq.~\ref{eq_ch2_Raman_coupling}), because $\Delta_\text{AS} = \Delta_\text{S} + \delta \nu_\text{gs}$. 
The reduced $\CAS$ results in less FWM spin-wave excitation and lower FWM noise contamination than present when operating the system red-detuned. 
For a red-detuned protocol, the roles of S and AS fields are reversed\footnote{
	For a red-detuned system, which would be operated using $\ket{3}$ as
	the initial state, the role of S and AS channels are flipped, i.e. the memory signal would occupy 
	the AS mode and the $1^\text{st}$ FWM step would be SRS into the S mode.
}. In such a $\Lambda$-system, the noise would dominate as it would be closer to resonance\cite{Hosseini:2012,Philips:2009,Phillips:2011,Camacho:2009ao}. 
Once a spin-wave has been excited by the first FWM step, its retrieval by the control is equivalent to memory read-out. 
As fig. \ref{fig_ch2_FWMlevels} \textbf{c} \& \textbf{e} illustrate, the control-mediated coupling between spin-wave and Stokes mode is the same in both cases. The strengths of both processes are proportional to the Raman coupling constant $\CS = \sqrt{\frac{W d \gamma}{\DeltaS^2}}$.
Accordingly, the ratio between Raman storage and FWM noise, which determines the system's signal-to-noise ratio (SNR), is dominated by the ratio of the Raman coupling constants:  
\begin{equation}
R = \frac{C_\text{S}}{C_\text{AS}} = \frac{\sqrt{\frac{W d \gamma}{\DeltaS^2}}}{\sqrt{\frac{W d \gamma}{\DeltaAS^2}}}
 = \frac{\DeltaAS}{\DeltaS} = 1+\frac{\delta \nu _\text{gs}}{\DeltaS}.
 \label{eq_ch2_Rratio}
\end{equation}
One possibility to increase $R$ is to suppress the anti-Stokes process compared to Raman storage at the Stokes frequency, which is the reason why we operate blue-detuned from the excited state $(\ket{2})$. 
Additionally, choosing a medium with a large hyperfine ground state splitting $\delta \nu_\text{gs}$ optimises $R$. 
\cs\, is thus a good choice, as it has the largest hyperfine ground state splitting amongst all stable alkali metals. 

Since FWM is a third order process, it is also subject to phase-matching restrictions\cite{Kumar1994} similar to those of other non-linear processes such as spontaneous parametric down-conversion \cite{Grice:1997} (SPDC).
Both FWM steps can either occur within the same control pulse or over consecutive control pulses. 
In the former case, S and AS noise are generated within the same control time bin, whereas in the latter scenario some residual FWM spin-wave excitations are stored in the \cs\, ensemble for the time between the control pulses. 
These excitations are retrieved after some storage time $\tau_\text{S}$, just as the signal stored with the Raman memory protocol\cite{Reim2012}. 
Due to this mechanism, FWM can actually build up over a control pulse train (see chapter \ref{ch7}). 
Obviously, FWM noise is generated solely when the control field is present, leading to a temporal pulse shape determined by the control and a confinement to the memory time bins.

\paragraph{Theoretical description}
The phenomenological introduction already exemplifies the similarities between the two FWM steps and Raman storage. 
We will use these to find a simple way of including the two-photon noise processes in our set of eqs.~\ref{eq_ch2_maxwell_bloch_3}. 
Since the $2^\text{nd}$ FWM step of S noise emission through spin-wave retrieval is exactly the same as memory read-out, it is already contained in eqs.~\ref{eq_ch2_maxwell_bloch_2}, so we only have to add the optical AS mode. Generally, this follows the same logic as used for the Stokes mode in sections \ref{subsec_ch_Maxwell_equations} \& \ref{subsec_ch2_maxwell_bloch_eq}. 
Thanks to the symmetry between the detunings, the form of the final set of equations can easily be motivated, without the need of another thorough derivation. With the rotating-wave approximation applied, only terms with frequencies $\omega \sim \delta \nu_\text{gs}$ contribute to the system's dynamics. 
For eqs.~\ref{eq_ch2_maxwell_bloch_3}, these frequencies correspond to the detuning $\DeltaS = \omega_c - \omega_\text{S}\sim \delta \nu_\text{gs}$, where $\omega_\text{S} = \omega_\text{s}$ is the input signal's frequency, which equals the FWM Stokes channel. 
For the AS channel, fig.\ref{fig_ch2_FWMlevels} shows that the detuning 
$\DeltaAS = \omega_\text{AS}-\omega_c = (-1)\cdot  (\omega_c - \omega_\text{AS})$ is just $(-1)$ times the difference between the control and the AS frequency. 
Hence we expect the resulting equations for the coupling between spin-wave $\hat{B}$ and the AS-mode, with annihilation operators $\hat{A}$, to equal that of the S mode $\hat{S}$ in eqs.~\ref{eq_ch2_maxwell_bloch_2}, only with reversed signs and with the complex conjugate for all coupling terms\cite{Michelberger:2014, England:2014}. 
To obtain the full set of equations for a system of S, AS and spin-wave modes, we use both equations for the optical modes and add the coupling terms to the spin-wave $\sim \partial_\tau \hat{B}$. 
The result reads\cite{Michelberger:2014}
\begin{align}
\label{eq_ch2_Maxwell_Bloch_FWM}
\partial_z \hat{S} &= \underbrace{- \frac{d \gamma p_1}{\Gamma_\text{S}}}_{(i) \rightarrow k_\text{S}} \hat{S} - \I \underbrace{\frac{\Omega(\tau)\sqrt{d \gamma}}{\Gamma_\text{S}}}_{(ii)\rightarrow C_\text{S}} \hat{B},\\ 
\partial_z \hat{A}^\dagger &= \underbrace{\frac{d \gamma p_3}{\Gamma_{\text{AS}}^{*}}}_{(i) \rightarrow k_\text{AS}} \hat{A}^\dagger - \I \underbrace{\frac{\Omega^{*}(\tau) \sqrt{d \gamma}}{\Gamma^{*}_\text{AS}}}_{(iii)\rightarrow C_\text{AS}} \hat{B},  \nonumber \\
\partial_\tau \hat{B} =& \underbrace{|\Omega(\tau)|^2 \left(\frac{1}{\Gamma_\text{S}} - \frac{1}{\Gamma_\text{AS}} \right)}_{(iv) \rightarrow \textfrak{S}} \hat{B}
- \I \underbrace{\sqrt{d \gamma} \Omega^{*} \left(\frac{p_1}{\Gamma_\text{S}} + \frac{p_3}{\Gamma_\text{S}^{*}} \right)}_{(ii) \rightarrow C_\text{S}} \hat{S} - 
 \I \underbrace{\sqrt{d \gamma} \Omega(\tau) \left( \frac{p_1}{\Gamma_\text{AS}} + \frac{p_3}{\Gamma_\text{AS}^{*}} \right)}_{(iii) \rightarrow C_\text{AS}} \hat{A}^\dagger, \nonumber
\end{align}
\noindent
where $\Gamma_{ \left\{ \text{S,AS} \right\} } = \gamma - \I \Delta_{ \left\{ \text{S,AS} \right\} }$ denotes the complex detuning for the Stokes (S) or the anti-Stokes (AS) channel. 
The expressions now also depend on the fraction of the total atomic population
 $p_i=\langle  |i \rangle \langle i| \rangle $ initially located in each ground state $\ket{i} \in\left\{ \ket{1},\ket{3}\right\}$, which accounts for the atomic state preparation. 
We will use this parameter in chapters \ref{ch6} \& \ref{ch7} to change the experimental configuration and move from a Raman memory scheme with FWM 
$(p_1 \sim 1$, $p_3 \sim 0)$ to a thermally distributed vapour $(p_1 = 0.5$, $p_3 = 0.5)$, dominated by SRS.  

Eqs.~\ref{eq_ch2_Maxwell_Bloch_FWM} contain four types of effects: 
terms with $(i)$ represent phase-shifts due to dispersion in the \cs\,-vapour.  
The control-induced coupling between spin-wave and S channel is marked by $(ii)$, whereas control coupling to the AS channel is denoted by $(iii)$. Both channels can now contribute a spin-wave excitation. 
Finally, there is also a dynamic Stark shift $(iv)$ for both transitions.
We again consider the adiabatic limit (eq.~\ref{eq_ch_adiabatiic_limit}) and introduce the integrated Rabi-frequencies $\omega$ and $W$. 
Therewith, the dependence on the temporal shape of the control pulses can be removed, by transforming the coordinate system for $\tau$ to the dimensionless time coordinate $\omega$, via $\partial_\tau = |\Omega (\tau)|^2 \cdot \partial_\omega$. 
Additionally, we can use the dimensionless Raman coupling constants 
${ C_{\left\{ \text{S,AS} \right\} }= \sqrt{\frac{W d \gamma}{ \Delta_{\left\{ \text{S,AS} \right\} }^2}} }$ for each optical channel and define the total dynamic Stark shift 
$\textfrak{S} = \frac{W}{\Delta_\text{S}} + \frac{W}{\Delta_\text{AS}}$.
The atomic populations are combined to the population inversion $w = p_1-p_3$. 
For perfect state preparation, with $p_1 =1$, we thus have $w=1$, whereas for thermally distributed populations 
${w\approx 0}$.
Finally, we define the four-wave mixing phase mismatch 
${\textfrak{K} = 2 k_\text{C} - k_\text{S} - k_\text{AS}}$, 
with individual wave-vectors 
${k_\text{C} = \frac{L \cdot \omega_\text{C}}{c} + d \cdot \gamma \cdot \left( \frac{p_1}{\Delta_\text{AS}} + \frac{p_3}{\Delta_\text{S}} \right)}$ 
for the control, 
${k_\text{S} = \frac{L \cdot \omega_\text{S}}{c} + d \cdot \gamma \cdot \left( \frac{p_1}{\Delta_\text{S}} + \frac{p_3}{\Delta_\text{S}-\delta \nu_\text{gs}} \right)}$ 
for light at the Stokes frequency, and  
$k_\text{AS} = \frac{L \cdot \omega_\text{AS}}{c} + d \cdot \gamma \cdot \left( \frac{p_3}{\Delta_\text{AS}} + \frac{p_1}{\Delta_\text{AS}+\delta \nu_\text{gs}} \right)$ 
for light at the anti-Stokes frequency, where $L$ is the ensemble length. These expressions are simply derived by considering the refractive index for each field, due to off-resonant interaction with the atomic transitions. 
Using these definitions, the Maxwell-Bloch equations reduce to\cite{Michelberger:2014}
\begin{equation}
[\partial_z + \mi \textfrak{K} ] \hat{S} =\I \cdot C_\text{S} \cdot \hat{B},  \quad 
\partial_z  \hat{A}^\dagger = -\I \cdot C_\text{AS} \cdot \hat{B}, \quad 
[\partial_\omega + \mi  \textfrak{S}] \hat{B} = \I \cdot w \cdot [C_\text{S} \hat{S} +  C_\text{AS} \hat{A}^\dagger].
\label{eq_ch2_Maxwell_Bloch_FWM_2}  
\end{equation}
\noindent
The solutions to these equations are similar to eqs.~\ref{eq_ch2_maxwell_bloch_4}, whereby an additional kernel is needed for the AS channel. Appendix \ref{app6_coh_model} discussed the solution in more detail.

\subsection{One photon noise processes \label{ch7_subsec_fluorescence}}

One photon noise is fluorescence typically resulting from decay processes. 
We ignore any natural decay of the ground state populations, as transitions between these states are dipole forbidden.
The excited state fluorescence linewidth of an atomic transition however has several components\cite{Demtroeder:ExPhys3}: 
the natural lineshape, set by the excited state lifetime, Doppler-broadening, originating from the velocity distribution of the emitters, and broadening by collisions between atoms\cite{Lewis:1980fk}. Each process contributes a damping rate $\gamma_N$, $\gamma_D$ and $\gamma_C$, respectively, which amounts to a total damping rate of $\gamma = \gamma_N + \gamma_D + \gamma_C$ for the transition. 
Far off-resonance,  
the collisional term is the most significant contribution to the fluorescence signal\footnote{
	Notably, this is only strictly correct for collisional induced fluorescence in an intermediate detuning
	range, which lies inside the impact regime\cite{Carlsten:1977}.
	Yet, it is clear that in the adiabatic limit, with $\Delta \gg \gamma$, the probability for resonant absorption 
	into the excited state, followed by resonance fluorescence decay is small. Similarly, also 
	Doppler absorption is small. Its Gaussian-shaped line quickly falls off far from resonance,
	so the probability to find an atom in an appropriate velocity class becomes negligible. 
}, since its lineshape follows a Lorenzian distribution that is spectrally broader than the natural and Doppler linewidths. 
In a pure \cs\, vapour, it is caused by \cs\,-\cs\, collisions, also called quenching collisions, as they can result in atomic ground state spin-flips.
Such collisions thus redistribute population between the \cs\, ground states, leading to spin-wave decoherence\cite{Liran:1980}. 
Their ability to change the energy states of the {\cs} atoms involved in the collisions has earned them the name inelastic collisions. 
To minimise the number of these events and to simultaneously limit \cs\, diffusion\cite{Sushkov:2008, 
Chrapkiewicz:2014fk} out of the interaction volume between the signal and the control field, a second atomic species can be added as a buffer gas\cite{Novikova:2012}. Usually these are noble gases of approximately similar atomic mass. 
In our system, we choose Neon (Ne) buffer gas, with a partial pressure of $p_\text{Ne} = 20\,\text{Torr}$.  
To operate the memory, we heat the vapour to 
$T_\text{Cs} = 70^\circ\text{C}$. 
Here, the corresponding partial pressure of \cs\, is\cite{Steck:2008qf} $p_\text{Cs} \approx 10^{-4}$ Torr. 
So the predominant collisions are between \cs\, and Ne, which are elastic collisions, preserving the \cs\, ground state spin.
The total damping rate $\gamma_C= \gamma_I + \gamma_E$ is the sum of the rates $\gamma_I$ for the inelastic, and $\gamma_E$ for the elastic collisions\cite{Raymer:1977}.

\paragraph{Collision-induced fluorescence\label{ch7_subsec_fluorescence}}

Fig. \ref{fig_ch2_FWMlevels} \textbf{f} illustrates what happens when such \cs\,-Ne collisions occur during the emission of radiation by the \cs\, atoms\cite{Omont:1972, Krause:1966}: 
The phase of the emitted light is scrambled, leading to a sudden phase jump in the time domain. 
In turn, this gives rise to a wide range of frequencies in the spectral domain, which are responsible for the fat tails of the Lorenzian collisional redistribution line\cite{Raymer:1979}. 
Since collision events are more likely for larger buffer gas pressures, the collision-induced fluorescence noise increases\cite{Rousseau:1975,Raymer:1977} with $p_\text{Ne}$. 
However scattering also occurs when the atoms are subject to a far detuned laser pulse, such as the memory control. 
From the energy level diagram in fig. \ref{fig_ch2_FWMlevels} \textbf{f} we can see how collisions can make up the energy gap between the laser and the excited state, leading to resonant excitation\footnote{
	For the case displayed in fig. \ref{fig_ch2_FWMlevels} \textbf{f}, the Ne atoms would 
	take out energy from the \cs\, - control pulse system to close
	the energy gap. The opposite, Ne atoms contributing energy, can also occur. Elastic collisions are thus
	not necessarily energy balance neutral within the Ne system.
}. The subsequent radiation decay of these excitations is then once more subject to phase scrambling by collisions during the emission. 
Both of these processes are the basis for collision-induced fluorescence\cite{Raymer:1977, Raymer:1983} caused by the control pulses\footnote{
	In principle, at high control energies, there can also be higher order fluorescence 
	excitation paths, which involve a previous Raman transition\cite{Carlsten:1977}. 
	These would lead to a temporal emission with the same temporal shape as the control pulse, which 
	we however do not observe (see section \ref{ch7_sec_fluor_results}). 
}. 
Since the fluorescence emission involves time-scales on the order of the excited state lifetime\cite{Steck:2008qf} $\tau_\text{Cs} = 32\ns$, most of it occurs after the Raman memory interaction, whose duration is proportional to that of the control pulse, here $\tau_c \approx 360\ps$. 
Contrary to narrowband memory protocols\cite{Manz:2007}, whose storage and retrieval also happens on timescales\footnote{
	Note however, that such detrimental effects can be avoided for narrowband protocols by 
	usage of vapour cells without buffer gas. Coating the cell walls with 
	paraffin\cite{Klein:2006,Balabas:2010} and illuminating the entire cell has been found 
	to also enable long coherence times in these systems\cite{Jiang:2009}. 
} similar to the excited state lifetime, collisions actually do not influence the Raman protocol at times when it involves electronic excited states\cite{Krause:1966}.  
Additionally, fluorescence noise can mostly be temporally separated from the memory signal.

Another possibility to reduce this noise is to lower the buffer gas pressure. But, as we still require $p_\text{Ne} \gg p_\text{Cs}$, we can also operate the memory further off-resonance, where fewer Ne atoms with sufficient kinetic energy are available in the Maxwell-Boltzmann velocity distribution of Ne to bridge the energy gap in fig.~\ref{fig_ch2_FWMlevels}~\textbf{f}.
Like SRS, fluorescence noise is emitted isotropically into $4 \pi$ steradian, so single mode fibre (SMF) coupling behind the memory also helps to cut the observed level. 

Similar to the benefit gained in reducing FWM, blue-detuned memory operation is another advantage for minimising contamination by fluorescence. 
Operating in the far off-resonance regime, with $\Delta_\text{S}^2 + \Omega_\text{max}^2 \gg \gamma_N^2$ (see section \ref{subsec_ch2_maxwell_bloch_eq}), an imbalance is expected in the fluorescence emitted towards the red- and the blue-side of the resonance\cite{Carlsten:1977}. 
For our atomic transitions, where the excited state has a larger polarisability than the ground state, scattering reaches out less towards large blue detunings than towards red detunings\footnote{
	The reason for this is an attractive molecular potential for the Cs(P$_\frac{3}{2}$)-Ne(S$_0$) system, 
	which has an energy minimum at some nuclear separation. This gives rise 
	to a Franck-Condon transition line for red detuned light, which does not exist for blue detuning, leading
	to a higher excitation probability and more fluorescence noise when red-detuned. 
	See \textit{Carlsten et. al.}\cite{Carlsten:1977} for details. 
}.

\section{The Raman memory experiment\label{ch2_Raman_level_scheme}}
In this final section, we introduce the implementation of the Raman memory protocol in \cs\, vapour. 
We describe our atomic and laser systems, their operational parameters as well as the resulting constants in our model (eqs.~\ref{eq_ch2_Maxwell_Bloch_FWM_2}). 
We also outline the principle memory configuration, which we will use for our experiments in the remaining chapters.  

\begin{figure}
\centering
\includegraphics[width=0.8\textwidth]{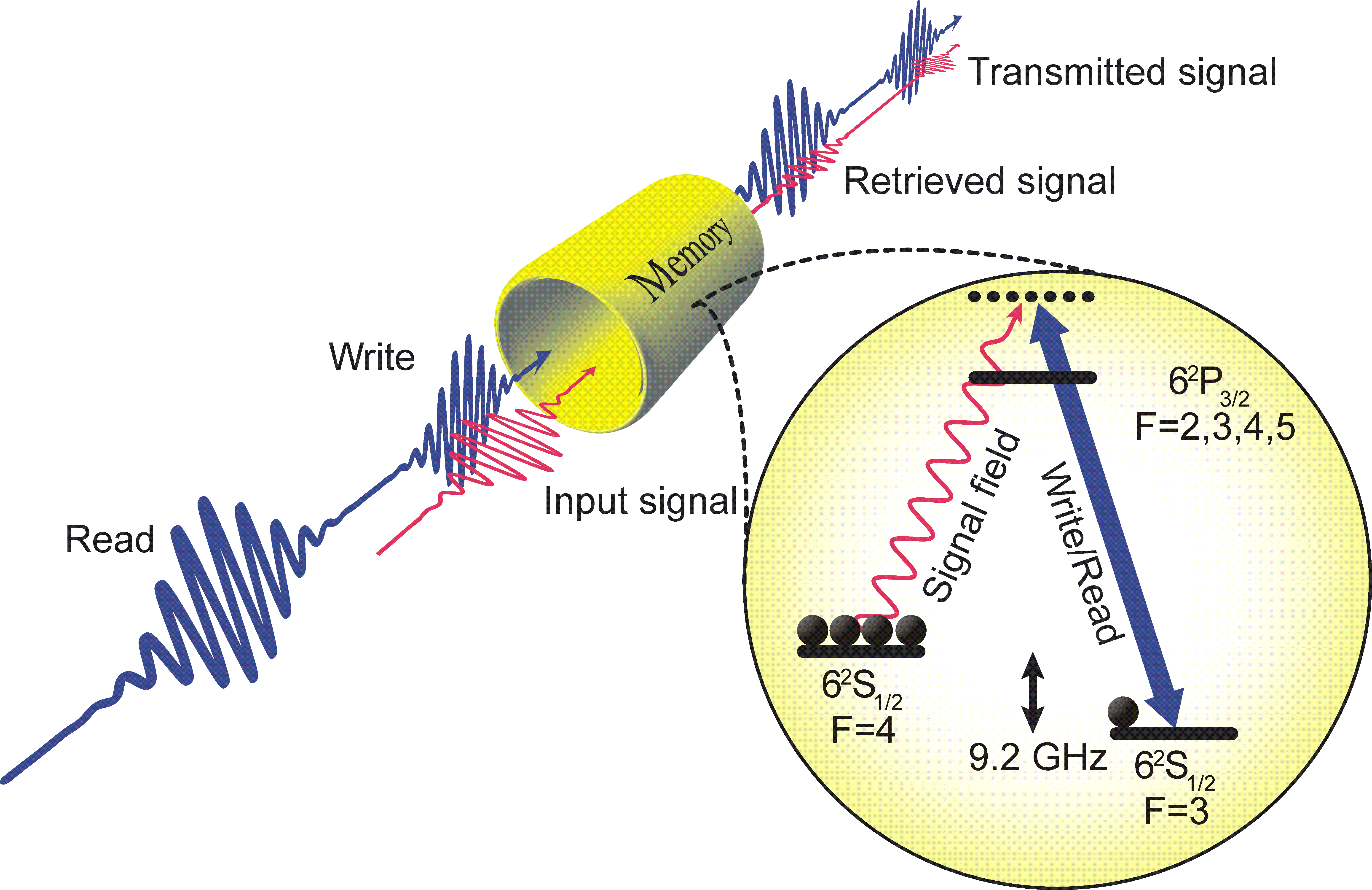}
\caption{Schematic representation of the Raman memory protocol implemented in \cs\, vapour. Shown are the atomic $\Lambda$-level system and the pulse sequence for signal and control. The image is taken from \textit{Reim et. al.}\cite{Reim2010}}
\label{fig_ch2_Raman_protocol}
\end{figure}

\paragraph{Raman memory scheme in caesium}
To implement the Raman scheme, we need to choose an atomic $\Lambda$-level system that resembles fig. \ref{fig_ch2_lambda_system}. In \cs\, atoms, this can be established using the two hyperfine levels F$=4$ and F$=3$ of the $6^2 \text{S}_{\frac{1}{2}}$ ground state as the initial ($\ket{1}$) and storage ($\ket{3}$) state, respectively. 
Both states are separated by the hyperfine ground state splitting of $\delta \nu_\text{gs} \approx 9.2\GHz$. 
We choose to operate at the \cs\, $\text{D}_2$-line, with a wavelength of $\lambda = 852\,\nm$, 
employing the $6^2\text{P}_{\frac{3}{2}}$-manifold as the excited state $\ket{2}$. 
Both optical pulses are blue-detuned from $\ket{2}$ by $\DeltaS \approx 15.2\,\GHz$.
Notably, thanks to the large detuning $\DeltaS$, we effectively do not resolve the excited state manifold, with ${\text{F'} = \left\{2,3,4,5\right\}}$. The optical fields couple to all dipole allowed transitions, which are those involving the $6^2\text{P}_{\frac{3}{2}} \text{F'} = \left\{3,4\right\}$ states. 
Since both states are only separated by\cite{Steck:2008qf} $\delta \nu_\text{es} = 201.2 \MHz$, the effective detuning from each state is similar. 
We can thus approximate the excited state manifold as a single state. 
Our implementation of the Raman protocol is not Zeeman substate selective. 
Fig. \ref{fig_ch2_Raman_protocol} illustrates the relevant \cs\, level structure alongside the employed signal and control pulse sequence.

\paragraph{Storage medium}

The \cs\, vapour is contained in a $7.5\,\text{cm}$ long Pyrex spectroscopy cell, with a $1"$ diameter. It contains pieces of \cs\, alongside the Ne buffer gas at $p_\text{Ne} = 20\,\text{Torr}$ partial pressure. 
\cs\, vapour is created by heating the cell with an electric heater belt, wrapped around the cell body. Temperature increase causes evaporation of \cs\, from the metal pieces, increasing its partial pressure $p_\text{Cs}$ in the cell\footnote{
{	\cs\, melts at $T=25^\circ \text{C} = 298.15\,\text{K}$.
	The dependence between \cs\, vapour pressure $p_\text{Cs}$ and temperature $T$
	\cite{Steck:2008qf}:
	\begin{align}
	\log_{10} p_\text{Cs} &= -219.48200 + \frac{1088.676}{T} - 0.08336185 \cdot T + 94.88752 \cdot 
	\log_{10}(T) \quad \text{(solid phase)} \nonumber \\
	\log_{10} p_\text{Cs} &= 8.22127 + \frac{4006.048}{T} - 0.00060194 \cdot T + 0.19623 \cdot 
	\log_{10}(T) \quad \text{(liquid phase)}, \nonumber
	\end{align}
	with $p_\text{Cs}$ in units of Torr and $T$ in units of Kelvin. 
	}
}. 
Using the ideal gas law $\left(p_\text{Cs} \cdot V = N \cdot k_\text{B} \cdot T\right)$ allows to estimate the number density of \cs\, atoms $n = \frac{N}{V} = \frac{p_\text{Cs}}{k_\text{B} \cdot T}$, and therewith the optical depth $d$. 
In the experiment, the cell is heated to $T=65-70^\circ\text{C}$, which gives an expected on-resonance optical depth of 
$d \sim 1800$ (see section \ref{subsec_ch2_maxwell_bloch_eq}). 
A measurement for the optical depth off-resonance is described in appendix \ref{appA_optical_pumping_efficiency}.
To achieve an approximately constant vapour density in the cell, \cs\, condensation on the cell surfaces needs to be minimised. 
To this end, the \cs\, is thermally insulated, which is discussed in section \ref{ch4_exp_imp}.  
To prepare the initial \cs\, population for the blue-detuned protocol in the higher energetic F$=4$ ground state, 
the initial, thermally distributed \cs\, population is optically pumped\cite{Happer:1972} by a diode laser. 
Details regarding this system are provided in appendix \ref{ch3_diodelaser}, where we also discuss the obtainable state preparation efficiency. 
Since we do not require Zeeman-state polarisation, the pumped ensemble is still distributed over all Zeeman sub-levels of the F$=4$ state.

\paragraph{Signal and control pulses}
The control pulse is derived from a titanium sapphire (\tisa\,) master laser, generating sech-shaped pulses of $P_{\text{\tisa}} \approx 1.2-1.5 \,\text{W}$ average power, with a repetition rate of $80\MHz$ and FWHM pulse duration of $\tau_\text{c} = 300-360\ps$\footnote{
	Due the re-alignment of the laser cavity, the output power and pulse duration changed throughout the experiments.
	For the measurements in chapter \ref{ch4}, it has been $P_{\text{\tisa}} \approx 1.5 \,\text{W}$ and 
	$\tau_\text{c} \approx 300\ps$, while we have $P_{\text{\tisa}} \approx 1.2 \,\text{W}$ and $\tau_\text{c} \approx 360\ps$ 
	for the remaining work in this thesis.
}. 
The latter corresponds to a spectral bandwidth of $\Delta \nu_c \approx 1-1.5\GHz$. 
Details about the laser system, including a characterisation of the pulse duration, is presented in appendix \ref{ch3_tisa}. 
While the actual pulse preparation sequence is discussed in the experimental chapters, we mention here that we end up with a control pulse energy of $E_c^p\approx 10\,\text{nJ}$, going into the storage medium. 
If coherent states are to be prepared as input signals for the Raman memory, these pulses are also derived from the \tisa\, output. 
Otherwise, the signal is generated by an SPDC source, described in chapter \ref{ch5}. 
To initially choose the appropriate \tisa\, pulse duration\footnote{
	In fact, when initially ordering the laser system, we requested a pulse spectrum of $2-3\GHz$ bandwidth.
	Yet, as the system was a prototype and the first of it's kind produced by the manufacturer, 
	the specifications could not be guaranteed and we ended up with a $\sim 1.5\GHz$ wide spectrum.
}, we note, that in the Raman protocol, the control spectral bandwidth must not simultaneously overlap with both ground states.
Consequently, the control pulse bandwidth, which, in turn, determines the storable spectral bandwidth of signal pulses\cite{Nunn:DPhil, Reim2010}, is limited the \cs\, hyperfine ground state splitting\footnote{
	Note here: Whenever we henceforth mention the bandwidth of signal and control pulses,
	we, per default, refer to their spectral bandwidths. Whenever this is not the case,  
	it will be stated explicitly. 
}. 

To store the signal, signal and control pulses are applied simultaneously to drive the Raman transition, shown in fig. \ref{fig_ch2_lambda_system} \textbf{b}. 
In fact, highest efficiency is obtained when the control pulse slightly precedes the signal\cite{Bergman:2001} by approximately $100\ps$.
Signal and control are inserted into the storage medium with orthogonal, linear polarisations. 
The arrangement is needed to prevent destructive interference\cite{Vurgaftman:2013aa}  between the respective transition paths involving the F'$=3$ and the F'$=4$ excited states (see section \ref{ch7_pol_turn_off_principles}). 
As fig. \ref{fig_ch2_Raman_protocol} illustrates, signal and control are also collinear as they propagate through the storage medium. 
Such a spatial arrangement minimises spin-wave dephasing during the storage time $\tau_\text{s}$, as we outline in further detail in appendix \ref{app3_subsec_sig_ctrl_spatial}. 
After the time $\tau_\text{S}$, the signal is retrieved on demand by re-applying the control field, which induces the Raman transition to releases the signal into its original polarisation mode (see fig.~\ref{fig_ch2_lambda_system}~\textbf{c}).

\paragraph{Performance numbers} 

To conclude, we briefly summarise the performance parameters achievable with this system. 
With a control pulse energy of $E_\text{c}^\text{p} \approx 10\,\text{nJ}$ and the \cs\, dipole matrix elements given in \textit{Steck}\cite{Steck:2008qf}, we obtain a peak Rabi-frequency of $\Omega_\text{max} \approx 4.2\,\GHz$. 
Considering sech-shaped pulses with a FWHM duration of $\tau_\text{c} = 360\,\ps$, the integrated Rabi-frequency amounts to $W \approx 1/(0.31) \,\GHz$.  
From the excited state lifetime of\cite{Steck:2008qf} $\tau_\text{Cs} \approx 32\,\ns$, 
we get a decay rate $\gamma = \frac{1}{2\tau_\text{Cs}} \approx 16\,\MHz$. 
In turn, these numbers result in a Raman coupling constant $C_\text{S} \approx 0.82$ for the Stokes signal, which includes the Raman memory process. 
For the anti-Stokes leg of the FWM interaction, the coupling is reduced to $C_\text{AS} \approx 0.51$. 
Utilising our theoretical results for the noise free Raman memory in forward read-out\cite{Nunn:2007wj,Reim2010}, 
the quoted values for $C_\text{S}$ and $E_\text{p}^\text{c}$ would result in an expected total memory efficiency of $\eta_\text{mem} \approx 50\,\%$, which assumes perfect mode matching and no control pulse energy depletion. 

In the experiment, we will be able to achieve $\eta_\text{mem} \approx 30\,\%$, when operating with coherent state input signals, derived from the \tisa\, laser. This ties in with our previous results\cite{Reim:2011ys}. 
For real single photon input signals, increased mode-mismatch reduces this number to $\eta_\text{mem} \approx 21\,\%$ (see chapter \ref{ch6_sec_single_photon_storage}). 
The signal can be stored in the memory with a lifetime of $\tau_\text{s} \approx  1.5\,\mus$, which represents the half-life time. 
Together with the our $\Delta \nu_\text{mem} \approx 1\,\GHz$ spectral bandwidth, the memory operates with a time-bandwith product of $B = \Delta \nu_\text{mem} \cdot \tau_\text{s} \approx 1500$. 
The measurements behind these numbers will feature in the experimental parts of this work, which we move onto now. These contain our actual advances in Raman memory research and introduce the aforementioned benchmarks in greater detail.

\part{Storage of bright laser pulses}


\chapter{Polarisation storage in the Raman memory\label{ch4}}

\begin{flushright}
{\tiny \textfrak{Direktor}: \,  \textfrak{Der Worte sind genug gewechselt; La§t mich auch endlich Taten sehn; Indes ihr Komplimente drechselt; Kann etwas NŸtzliches geschehn..} }
\end{flushright}

We now move on to our first series of experiments, investigating the Raman memory's capability to store information encoded in the polarisation of the optical input signal. 
To this end, a dual-rail configuration of the memory inside a polarisation interferometer is used and the storage process is evaluated with quantum process tomography (QPT). 
We first describe the experimental apparatus and introduce the QPT framework. Thereafter we demonstrate the storage of polarisation encoded, bright coherent states with a process fidelity of up to $\mathcal{F} = 0.93 \pm 0.08$. 
Considering a reduction of the input signal intensity down to the single photon level, 
we show, theoretically, that faithful operation in the quantum regime is currently prevented by the memory's noise floor.

\section{Introduction\label{ch4_intro}\label{ch4_subsec_polstorage}}

Quantum information carriers can be encoded in many different ways\cite{Knill:2001nx, Nielsen:2004kl, Walther:2005uq, OBrien:2007zr, Kok:2007, OBrien:2009fk}. 
One promising approach is the usage of the polarisation of light. Its easy accessibility, manipulation and detection make this degree of freedom a particularly attractive means to implement quantum information processing\cite{Bouwmeester:1997aa, Bouwmeester:1999, Pan:1998,Jennewein:2000, Ursin:2007vn}. 
Since photonics-based schemes operate probabilistically, they rely upon repeat-until-success strategies\cite{Lim:2005,Lim:2006}, whose scalability requires the presence of quantum memories within any such processor\cite{Raussendorf:2001}. 
For these reasons, the capability to faithfully store and retrieve polarisation encoded quantum bits (qubits\cite{Schumacher:1995}) is a key property for quantum memories.
During recent years, storage of polarisation information has been achieved with several memory systems, including AFC- echos in rare-earth ion doped crystals\cite{Clausen:2012,Zhou:2012,Gruendogan:2012}, single atoms in a cavity\cite{Specht:2011}, cold atomic ensembles\cite{Chen:2008fk,Ding:2014}, and warm vapour memories\cite{Kim:2010,England2012}. 
For implementation in the Raman memory, we choose a dual-rail memory architecture, where the memory is placed inside a polarisation interferometer. 
Of course, this is not the only possibility to achieve polarisation storage in this memory type. 
Yet, it is the conceptually and experimentally simplest extension of our system, building on our initial proof-of-principle experiments\cite{Reim2010, Reim:2011ys}.

\begin{figure}[h!]
\centering
\includegraphics[width=\textwidth]{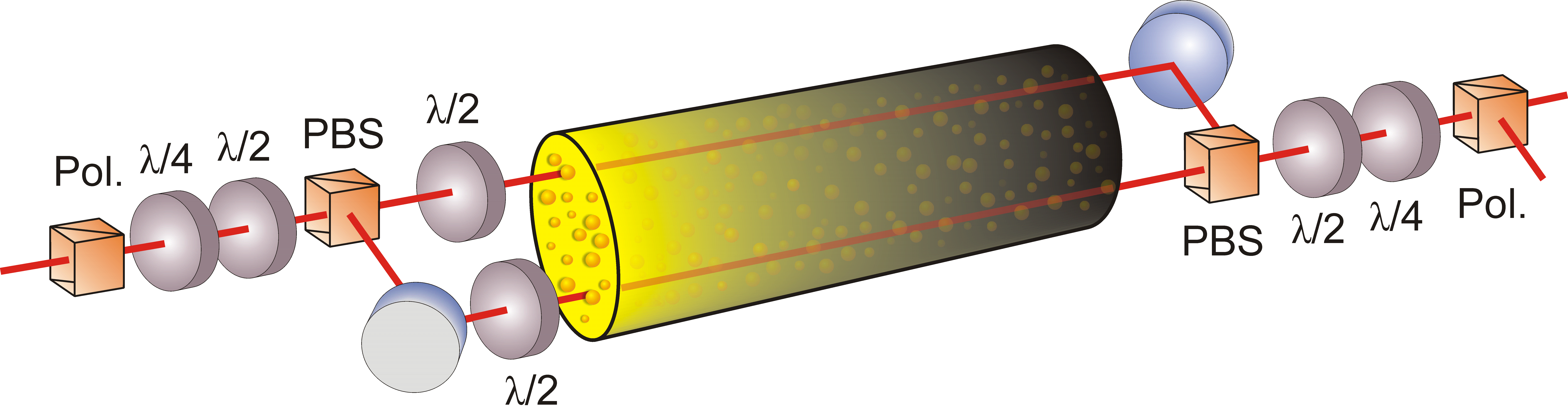}
\caption{Schematic for storage of polarisation qubits in a dual-rail Raman memory using a polarisation interferometer.
Input signals are decomposed into horizontally (\hpol) and vertically (\vpol) polarised components, which are stored in separate modes of the active medium positioned inside the polarisation interferometer.
Recombination of the recalled signals at the interferometer output returns the original signal polarisation state.}
\label{fig_ch4_intro_fig}
\end{figure}

\paragraph{Polarisation storage in the dual-rail Raman memory}
Polarisation information, 
$|\phi\rangle=\frac{1}{\sqrt{2}} \left( |0\rangle + e^{i \theta} |1\rangle \right)$, is encoded in a superposition of the two orthogonal polarisation basis states, $\{|0\rangle, |1\rangle\}$, with a relative 
phase $\theta$. Storage of such information in a memory requires the capability to simultaneously read -in and retrieve both polarisations with equal efficiency, whilst preserving their phase relationship during the storage time. 
For this reason, a suitable memory needs to be multimode in the polarisation domain. 
The Raman protocol, as we currently use it\cite{Nunn:2007wj, Reim2010, Reim:2011ys}, operates single mode, because signal and control have orthogonal linear polarisations.  
One way to store multiple polarisations is thus the use of two separate spatial modes inside the memory's active medium, with orthogonal linear control field polarisations. 
Splitting up the incoming signal state into its horizontally and vertically polarised component enables storage by sending each into the mode defined by the respective, orthogonally polarised control. 
This results in a dual-rail architecture, shown schematically in fig.~\ref{fig_ch4_intro_fig}. 
Notably, this procedure of creating multiple memory rails inside the storage medium is in fact equivalent to spatial mode multiplexing\cite{Higginbottom:2012}, which, in turn, is similar\cite{Nunn:2008oq} to spectral\cite{Sinclair:2014} or temporal\cite{Afzelius:2009qf} multiplexing of the memory.
Thanks to this projection onto the \{\hpol\,,\vpol \}-basis, any arbitrary polarisation state can be decomposed into the two memory modes. 
To faithfully preserve the information, also the phase between both modes must be preserved. 
For this reason, the memory is positioned inside a polarisation interferometer. At the interferometer output port both memory modes are combined to reproduce the original state $\ket{\phi}$ sent into the system.
Furthermore the storage process itself must also be phase stable. 
Accordingly, the phase relation between the spin-waves, excited in both memories, must be preserved during the storage, irrespective of any decoherence mechanisms acting on either memory mode. 

\section{Experimental set-up\label{ch4_subsec_exp_implementation}}

\begin{figure*}[h!]
\centering
\includegraphics[width=15cm]{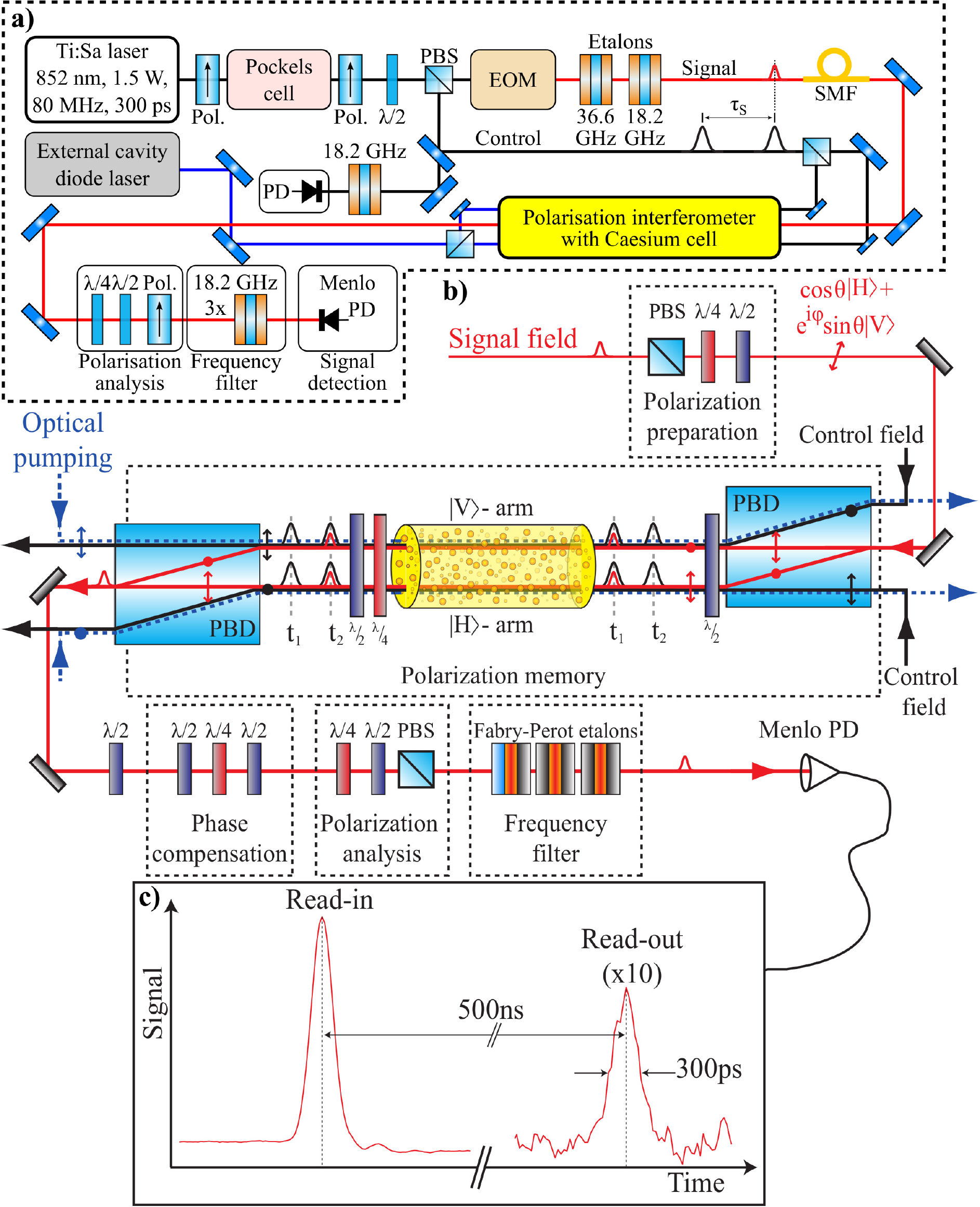}
\caption{Experimental set-up for storing polarisation information in the Raman memory; see text for details. 
\textbf{(a)}: overview of the apparatus, including the preparation stage for the memory pulse sequence. 
\textbf{(b)}: dual-rail memory inside a polarisation interferometer together with the optics required for signal preparation and analysis. 
\textbf{(c)} signal pulse sequence observed on the Menlo PD, showing the transmitted fraction of the input signal pulse, that is not stored, and first retrieved pulse from the memory for a storage time of $\tau_S \approx 500 \ns$.}
\label{fig_ch4_setup_fig}
\end{figure*}

To implement polarisation storage, we modify our initial, proof-of-principle experiment for the Raman memory protocol\cite{Reim2010}. 
When the presented measurements were recorded, the system was still positioned on a small table $(\sim 4 \m^2)$ in a lab corner. 
During this time it became evident, that neither the available workspace nor the lab environment were sufficient to host any further developments of the Raman memory (see appendix \ref{app_ch4_probs_setup}). 
The experimental lay-out discussed here is thus only representative for this chapter. 
Nevertheless, the preparation methods for signal and control pulses, 
as well as the optical pumping also apply to the experiments in chapters~\ref{ch6}~\&~\ref{ch7}.
Fig.~\ref{fig_ch4_setup_fig} \textbf{a} illustrates the setup, where the output of our {\tisa} master laser, here emitting a train of $300\ps$ pulses at $852\nm$ with $80\MHz$ repetition rate and $1.5\,\text{W}$ average output power, is used to generate the memory pulse sequence. A more detailed description of the laser system can also be found in appendix \ref{ch3_tisa}.

\paragraph{Memory pulse sequence}
From the {\tisa} output, pulses are selected with a Pockels cell (\pockels\,) in single pass. The \pockels\, is positioned between two crossed Glan-Laser polarisers and picks two sets of pulses by polarisation rotation\footnote{
The different arrangement with respect to the setup used in chapters \ref{ch5}-\ref{ch7} is a legacy of previous experiments, see appendix \ref{app_ch4_PC} for details.
}. 
The two \pockels\, pulse picking windows can be delayed with respect to one another by an arbitrary time $\tau_S$. Within each window a maximum of 9 pulses can be selected. The first window is used to generate the pulse sequence for memory read-in, so only one pulse is picked. 
In turn, {\tisa} pulses, selected by the second window, become the read-out control pulses. Consequently, $\tau_S$ represents the memory storage time. 
Since the memory efficiency is below unity, multiple pulses in the read-out window can be used to completely deplete the stored spin-wave. In this way, all information is retrieved from the memory in a train of read-out signal pulses\cite{Reim2012}. 
For the shortest possible storage time of $\tau_\text{S} = 12.5\ns$, two consecutive {\tisa} pulses are required for read-in and read-out. Here the second pulse picking window is not used and the first one is opened fully to select 9 pulses. As beforehand, the first pulse from this window defines the read-in time bin, while the remaining 8 pulses are the read-out time bins. 
The \pockels\, is triggered by an $80\MHz$ clock rate signal from an internal photodiode inside the {\tisa} laser cavity. Its electronic driver module divides the {\tisa} clock down to a repetition rate of $f_\text{rep} =  667 \Hz$, with which both pulse windows are generated. 
Accordingly, $f_\text{rep}$ represents the number of conducted memory experiments per second.

Additionally, this \pockels\, trigger signal is also available to gate external devices, such as a scope, used for data acquisition, and an electro-optic modulator (EOM), which prepares the signal pulses for the memory by frequency modulation. 
To obtain signal and control inputs for the memory, the picked {\tisa} pulses are split on a PBS into a strong control arm and a weak signal arm. For the experiments presented here, the splitting ratio was on average $1:50$. 
Notably, by selecting a fraction of the control intensity, we prepare coherent state input signals for the memory. 
In the current experiment, these are bright pulses with an optical power in the $\text{nW}$-regime. Contrary to the work in the remaining chapters of this thesis, the input signals are thus neither true single photons, nor at the single photon level. 

Since signal and control need to correspond to  the longer and shorter wavelength of the Raman interaction, respectively (see fig.\ref{fig_ch2_Raman_protocol} in section \ref{ch2_Raman_level_scheme}), the {\tisa} frequency is set to equal the desired control frequency. 
This corresponds to the resonance frequency of the {${6^2 \text{S}_\frac{1}{2} \text{F}=3 \rightarrow 6^2 \text{P}_\frac{3}{2}}$} transition plus the Raman detuning $\Delta$. 
For two-photon resonance with the control, the signal needs to be down-shifted in frequency by the ground state hyperfine splitting of $\delta \nu_\text{gs}=9.2\GHz$. 
Supplying the EOM with a $9.2\GHz$ radio-frequency (rf) signal results in the generation of sidebands at frequencies $\nu_\text{sig}\pm \delta \nu_\text{gs}$ with respect to the input signal at $\nu_\text{sig}$. From these the desired red sideband is selected by a pair of Fabry-Perot etalons. The first etalon, with a free spectral range (FSR) of $18.2\GHz$, is resonant with both sidebands, attenuating the fundamental at the control frequency.
The second etalon, with $\text{FSR}=38.86\GHz$, resonant with the red-sideband, filters out the undesired blue-sideband\footnote{
	In later experiments (chapters \ref{ch6} \& \ref{ch7}), only the $38.86 \GHz$ etalon will be used to filter the modulated signal beam.
}. 
For the memory pulse sequence (see fig.~\ref{fig_ch2_Raman_protocol}), the signal field must be present only in the read-in time bin. Consequently, the $9.2\GHz$ rf-signal, driving the EOM modulation, is turned off by a fast rf-switch after the first {\tisa} pulse. 
The switching is gated by the \pockels\, trigger signal, that is fed into a digital delay generator (DDG). 
The DDG produces appropriately delayed TTL pulses that are supplied to flip the rf-switch (see also section \ref{ch6_sec:setup} for a detailed explanation of the electronic gating circuitry). 
During the polarisation storage experiment, the rf-switch was triggered active high\footnote{
	Besides an input signal, which the rf-switch can transmit to either of its two output ports, it also receives a $0\,\text{V}$ or $+5\,\text{V}$ dc bias voltage,
	determining to which output port the input signal is routed. Active high means, that the rf-signal is routed towards the EOM when $+5\,\text{V}$ bias are
	applied. 
	For active low, the output ports are flipped and rf-modulation is supplied to the EOM when there is no bias voltage. 
	Importantly, the switching 	from low to high has a sharp rising edge. The reverse process however tails off exponentially, following a capacitor discharge curve. This results in the routing residual rf-modulation signal to the EOM. 
	Thus, active low switching, as used in chapters \ref{ch6} \& \ref{ch7}, results in better extinction of unwanted modulation of any subsequent pulses in the $80\MHz$ {\tisa} pulse train.
}. 
While this is unproblematic for large storage times, the $\sim 5\ns$ fall-time of the rf-switch causes residual modulation of the next pulse at $\tau_S = 12.5\ns$, when picking a single pulse as the input for the memory. 
So there is a small leakage of the input signal in the first memory read-out time bin, which results in a small amount of mixing between storage and read-out. The electronic and optical pulse timing diagrams are discussed in detail later in section \ref{ch6_sec:setup} (see fig.~\ref{fig_6_setup}), where they are contrasted with the modifications for the storage of actual single photons. 

Behind both etalons, whose transmissions are $T_{18\GHz} = 33\,\%$ and $T_{38\GHz} =55 \,\%$ on average, the signal pulses are coupled into a short single-mode fibre (SMF), with an average coupling efficiency of $\eta_\text{SMF}^\text{sig.} = 28 \,\%$, to clean up the spatial mode.
In contrast, the control pulses propagate in free space along a delay line, which ensures temporal overlap between signal and control inside the {\cs} cell. 
Since polarisation storage requires two copies of the control for each memory mode, shown in fig.~\ref{fig_ch4_setup_fig} \textbf{b}, the control is split up on a PBS first prior to entering the polarisation interferometer. 

\paragraph{Frequency stabilisation and Raman detuning}
Contrary to later experiments, the frequency and beam pointing of the {\tisa} laser were not yet actively stabilised here. 
With both freely floating, the laboratory environment, particularly its large intraday temperature gradients, resulted in significant drifts (see appendix \ref{app_ch4_probs_setup}).
To achieve at least partial stability of {\tisa} frequency, a weak control beam pick-off is sent through an additional etalon with $\text{FSR} = 18.2 \GHz$, aligned in resonance with the control frequency, set to the Raman detuning $\Delta$ (see fig.~\ref{fig_ch4_setup_fig} \textbf{a}). 
Its transmission is monitored on a photodiode, whereby the {\tisa} frequency is reset manually. 
However, the etalons drift themselves over time. 
One cause for such drift is again the change in environment temperature.
For this reason, the detuning varied during the measurement time. 
The average detuning over all measurements was $\Delta \approx 18 \GHz$. 
However, it fluctuated within an interval of $\Delta \in \left[ 17\GHz, 20\GHz \right]$ over the coarse of the experiments.

\paragraph{Polarisation interferometer}
The dual-rail memory architecture of fig.~\ref{fig_ch4_intro_fig} is implemented using two polarising beam displacers (PBD) to create a passively stabilised interferometer\cite{Kim:2010,OBrien:2003fk}. The walk-off between light polarised along the orthogonally oriented slow and fast axes in these crystals, here $8\mm$ walk-off for $2\cm$ long calcite crystals, gives rise to high quality polarisation separation $\mathcal{O}(40\,\text{dB}$). 
A combination of two consecutive PBDs, with a $\lambda/2$- polarisation rotation in between, enables accurate separation of the signal into two modes as well as their subsequent recombination. Fig.~\ref{fig_ch4_setup_fig} \textbf{b} shows this schematically. 
Polarisation information is encoded onto the signal in front of the interferometer input with a polarising beam-splitter followed by a $\lambda/4$- and $\lambda/2$- waveplate (see appendix \ref{ch4_subsec_qubits}). 
The need to have reasonably high control pulse intensities\cite{Reim2010} 
does not permit illumination of the entire vapour cell. 
Thus the control beam path also follows the dual-rail configuration. 
Both control arms are equipped with a bespoke set of focussing optics and delay stages for optimising spatial and temporal mode-matching to the signal field inside the {\cs} cell. 
Since the control is orthogonally polarised to the signal to achieve Raman storage, it experiences walk-off in the PBDs whenever its corresponding signal mode is transmitted undeflected, and vice versa. 
Polarisation flipping by the $\lambda/2$- plate, positioned inside the interferometer, results in two control modes at the interferometer output, which are spatially separated from the signal mode. 
To prepare the {\cs} ensemble in the $6^2 \text{S}_{\frac{1}{2}}$ F$=4$ initial state the output of a frequency stabilised diode laser (see appendix \ref{ch3_diodelaser}) is used for optical pumping. 
It is supplied to the setup with a single-mode fibre (SMF) and sent into both interferometer arms in counter propagating geometry, occupying the same spatial and polarisation modes as the control field.
To avoid spin-wave depletion by optical pumping, the diode laser is turned off during signal storage by an acousto-optic modulator (AOM), which is also triggered by the DDG. 
The turn-off procedure is explained in more detail in section \ref{ch6_setup_opticalpumping} later on.

\begin{figure}[h!]
\includegraphics[width=\textwidth]{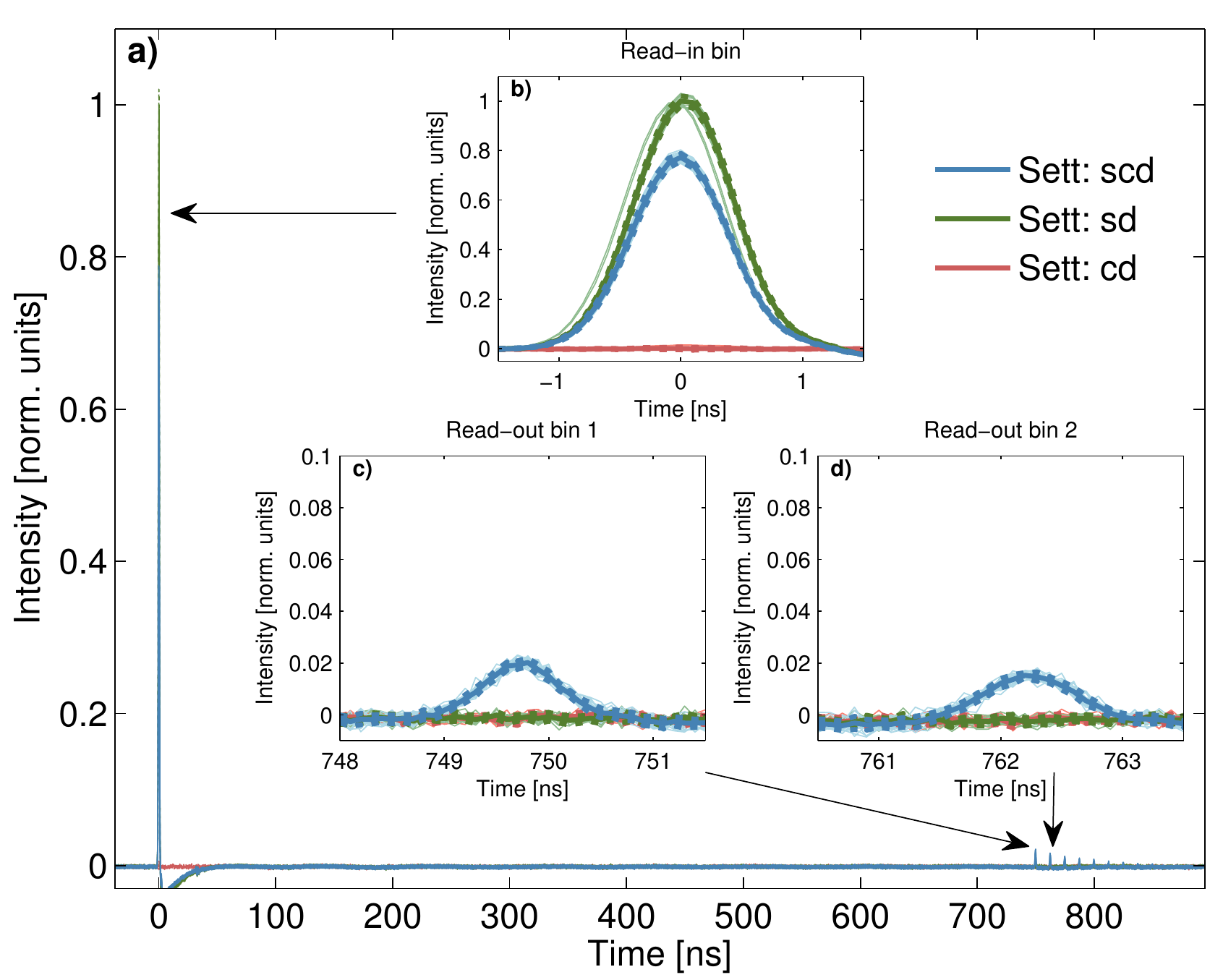}
\caption{\textbf{(a)}: Pulse train sequence observed on the Menlo PD for a \dpol-polarised input signal analysed in the \rpol-basis, with a memory storage time of $\tau_S = 750\ns$. Panels \textbf{(b)} - \textbf{(d)} show a detailed view for the read-in and the first two retrieval time bins. \textit{Blue lines} represent the setting \textit{scd}, where signal and control pulses are sent into the memory simultaneously. \textit{Green lines} are traces for only applying the signal without control, whereas \textit{red lines} are control only. For each category \textit{lightly coloured lines} 
are the individual data traces. The \textit{solid dark coloured lines} are their average pulse envelopes, whereas the \textit{dotted lines} are error bounds on the pulse envelope taken as the standard deviation of the individual traces.
}
\label{fig_ch4_trace_PD_750ns}
\end{figure}

\paragraph{Signal detection}
The signal output from the polarisation interferometer is sent into a phase compensation system, consisting of a sequence of $\lambda/4$-, $\lambda/2$-, $\lambda/4$- 
waveplates for Berry phase\cite{Bhandari:1988} compensation (see section \ref{ch4_exp_imp}). Beforehand an additional $\lambda/2$-plate cancels the polarisation flip introduced by the $\lambda/2$-plate inside the interferometer. Thereafter the polarisation information of the output signal is analysed, using a $\lambda/4$- and a $\lambda/2$- plate, together with a Glan-Laser polariser (see appendix \ref{ch4_subsec_qubits}). 
For detection, the signal is also frequency filtered by three $\text{FSR}=18.2\GHz$ FP etalons. This removes any residual control leakage. Spectral filtering provides $\mathcal{O}(50 \, \text{dB})$ control extinction with an on-resonance transmission of $T^\text{sig.}_\text{filt.}  \approx 37\,\%$ for the filter sequence. 
After filtering the signal is again SMF-coupled with $\eta^\text{sig.filt.}_\text{SMF} \approx 70 \,\%$ efficiency\footnote{
	Notably, this additional SMF-coupling prior to detection is not a fundamental necessity for the measurements 
	with bright input signals, presented here. 
	It is included in the system, because the detection system had to be spatially
	separated from the rest of the experiment, as it did not fit onto the optical table.  
}, and detected on a fast, linear, amplified photodiode (\textit{Menlo Systems PD}).
Fig.~\ref{fig_ch4_setup_fig} \textbf{c} exemplifies the signal traces from the PD for a storage time of $\tau_S\approx 500\ns$, which are observed on a fast oscilloscope (\textit{LeCroy}, $8\GHz$ sample rate). The scope is gated by the \pockels\, trigger output. 
The schematic in fig.~\ref{fig_ch4_setup_fig} \textbf{c} only shows one read-out pulse. 
In fact, the output signal pulse train contains multiple retrieval pulses, with the first three containing sufficient intensity for analysis. 
Fig.~\ref{fig_ch4_trace_PD_750ns} presents this pulse train for an actual dataset.
For each of the polarisation measurements we record 50 independent scope traces, containing the full pulse sequence, i.e. read-in and all read-out time bins\footnote{
	Note, the full set of 50 independent traces are only recorded for the memory pulse sequence, 
	consisting of signal and control pulses applied 
	to the memory simultaneously. For measuring the input signal, i.e. signal pulses sent into the memory without control,
	40 independent traces are recorded. For control leakage measurements, where only the control pulses are sent into the memory 
	without any input signal, 20 independent traces are taken.
}. 
Each of these traces is already an average over 1000 single shot oscilloscope traces.

\paragraph{Data analysis}
Analysis of polarisation storage uses the areas of each pulse in the recorded pulse train. They are determined from the above mentioned, independent scope traces recorded for a specific input polarisation $x$ and analysis polarisation basis $y$, whose total number is denoted by $N^{(x,y)}$. 
First the mean pulse shape is calculated by taking the average over all pulses within one of the time bins ($t$). 
Additionally, also error bounds on the areas are obtained by first taking the standard deviation of all pulse envelopes and subsequent point-wise addition to and subtraction from the mean pulse shape.
Fig.~\ref{fig_ch4_trace_PD_750ns} exemplifies this on the basis of the memory pulse sequence for a storage time $\tau_S = 750 \ns$, measured for diagonally 
polarised input light (\dpol), analysed in the right-circular basis (\rpol); i.e. the polarisation analysis set-up in fig.~\ref{fig_ch4_trace_PD_750ns} is set to transmit \rpol-polarised light (see appendix \ref{ch4_subsec_qubits}).
Each of the sub-panels (\textbf{b} - \textbf{d}) of fig.~\ref{fig_ch4_trace_PD_750ns} shows a pulse time bin, containing all individual traces observed on the Menlo PD as well as the mean pulse shape and its error boundaries.
In a second step, the average pulses are integrated. 
The resulting areas $A^{x,y}_{k,t}$ are the desired results to be used in further analysis.
The errors on these areas are obtained from integration of the upper ($+$) and the lower ($-$) error boundary pulse envelopes. 
The mean difference between both areas ($A^{x,y}_{k,t,\pm}$) yields the error $\Delta A^{x.y}_{k,t}=\frac{1}{2} \left( A^{x,y}_{k,t,+} - A^{x,y}_{k,t,-} \right)$ on the pulse areas, which will be used as the standard deviation for the distribution of pulse areas later on (see appendix \ref{ch4_subsec_QPT}). \newline
For analysing the polarisation storage, we record three different experimental configurations of input signals going into the {\cs} cell, denoted as measurement settings $k$. 
Firstly, it is required to know how well the interferometer itself performs without any signal storage and retrieval. For this reason, the signal (\textit{s}) is measured without the control field (\textit{c}), but with active atomic state preparation (\textit{d}) from the diode laser\footnote{
	This minimises residual linear absorption of the signal field in the {\cs}.
} ($k =sd$). 
Secondly, the effects of storage and retrieval are investigated by sending in the full memory pulse sequence, consisting of input signal, control and diode laser ($k=scd$).
Notably, due to signal pulse leakage in the first read-out bin ($t=\text{out1}$) for $\tau_S=12.5\ns$ storage times, we also need to integrate the leakage pulse for $k=sd$. 
The resulting area $A^{(x,y)}_{sd,\text{out1}}$ is then subtracted from the area of the memory readout pulse $A^{(x,y)}_{scd,\text{out1}}$, obtained in the same time bin, which approximately\footnote{
	Signal leakage in the first read-out time bin does not only contribute additional 
	signal intensity to the retrieved signal fraction. 
	The Raman interaction also causes
	memory read-in of this signal, which results in both, storage and retrieval. In other words, leakage is 
	not linearly separable from the retrieval. 
	This is analogous to the later discussion about signal and noise combination in section 
	\ref{ch6_subsec_g2models} and appendix \ref{app6_coh_model}.
} cancels leakage contributions to the retrieved signal. 
Generally, the comparison between data obtained for settings \textit{sd} and \textit{scd} allows determination of the memory efficiency 
(section \ref{ch4_subsec_exp_conduct}) and evaluation of polarisation storage in the memory.
Finally, we also certify the absence of control pulse leakage and noise in any of the time bins by only sending the control field into the {\cs} cell, after atomic state preparation by the diode ($k=cd$). 
As fig.~\ref{fig_ch4_trace_PD_750ns} exemplifies, none of recorded traces shows any significant contribution from the \cd\,-setting. 
These measurement are nevertheless conducted at the end of data recording for each polarisation combination $(x,y)$, since they certify the stability of the interferometer over the measurement time. 
With the interferometer aligned for maximum control extinction in the signal mode, 
phase drifts between both interferometer arms would result in undesired polarisation rotations, giving rise to control leakage into the signal mode. The absence of any leakage is an experimental check for system stability and the usefulness of the data. 
We will see in the following section \ref{ch4_exp_imp}, why this is needed in the first place. 

Before we continue our discussion with the actual experiment, we note that the remaining parts of this chapter assume knowledge about the basic means of preparing and analysing polarisation information, as well as quantum state and process tomography. 
Due to length restrictions on this document, the introduction of these concepts, alongside the notation for the required Pauli spin-matrices $\sigma_{\left\{ X, Y, Z \right\}}$, the process matrix $\chi$, its purity $\mathcal{P}$ and its fidelity $\mathcal{F}$, are presented in appendix \ref{ch4_qpt_intro}.

\section{Performing the experiment\label{ch4_exp_imp}}

\subsection{Polarisation interferometer stability\label{ch4_subsec_stability}}
\begin{figure}
\includegraphics[width=\textwidth]{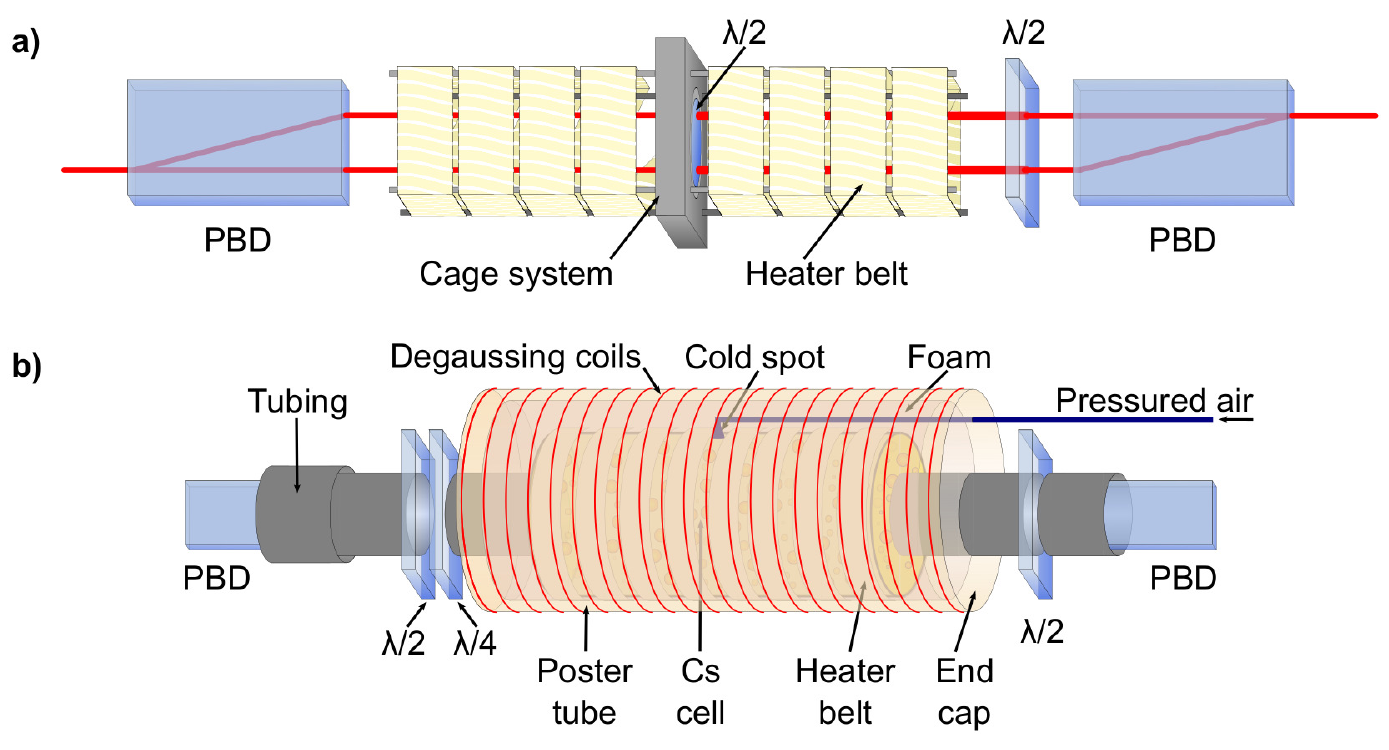}
\caption{
\textbf{(a)}: Test setup for assessing the interferometer instability. 
A cage system is positioned inside the polarisation interferometer, which has a waveplate at its centre and a heater tape wrapped around it.
Heating the system introduces air currents similar to the ones observed with the {\cs} cell inside the interferometer.
\textbf{(b)}: Interferometer with shielding using pipe enclosures to prevent air currents. The {\cs} cell is also thermally insulated by several layers of foam material and a tube with end caps. Pipe tubing reaches into the thermal insulation layers and connects to the {\cs} cell windows enabling optical access. Solidification of {\cs} on the cell windows is prevented by introducing an artificial cold spot along the cell body.
}
\label{fig_ch4_int_instab_setup}
\end{figure}

One of the trickiest and most troublesome elements in the implementation of the experiment is the phase-stability of the polarisation interferometer.
By themselves, passively stable interferometers, using PBDs, have shown great performance in the past\cite{Kim:2010, Fiorentino:2008, Broom:2010}. 
They are experimentally simple and low cost. 
Placing the warm {\cs} vapour cell inside such an interferometer however introduces instability from convection currents and turbulences of hot air. 
The resulting variations in the air's refractive index, due to temperature and density differences, lead to undesired path length fluctuations between the interferometer arms.
Such path differences are synonymous to a phase retardation, as it would be introduced by a waveplate, rotating the output polarisation state at random. 
Unlike drifts, which are directional and occur on minute time scales, rotations from air currents are faster\footnote{Time scales $< 1\,\text{s}$.} 
and undirectional, making their compensation with polarisation optics difficult.
To identify that the instability arises from convection currents, excited by the temperature difference between the $68-70^\circ \text{C}$ warm surfaces of the {\cs} cell and the room-temperature environment, we use the test system shown in fig.~\ref{fig_ch4_int_instab_setup} \textbf{a}. 
Here a heater belt, usually wrapped around the {\cs} cell, is covering an optical cage system which contains a waveplate to resemble one optical interface of the {\cs} cell.

The idea behind this test set-up is to investigate the influence of a temperature gradient in the air between the PBDs and the centre of the interferometer. Fitting a waveplate at the centre of the test set-up introduces a single surface, which cuts off potential laminar air flow. It accordingly simulates one of the {\cs} cell windows and tests, whether these introduce convection or turbulent air flow around their surfaces. Of course, the same effect could also be achieved by, e.g., inserting an empty spectroscopy cell into the 
experiment instead of the {\cs} cell. Yet, in the absence of such a cell, the system depicted in fig.~\ref{fig_ch4_int_instab_setup} \textbf{a} was the cheapest and quickest testing possibility. 

\begin{figure}
\includegraphics[width=\textwidth]{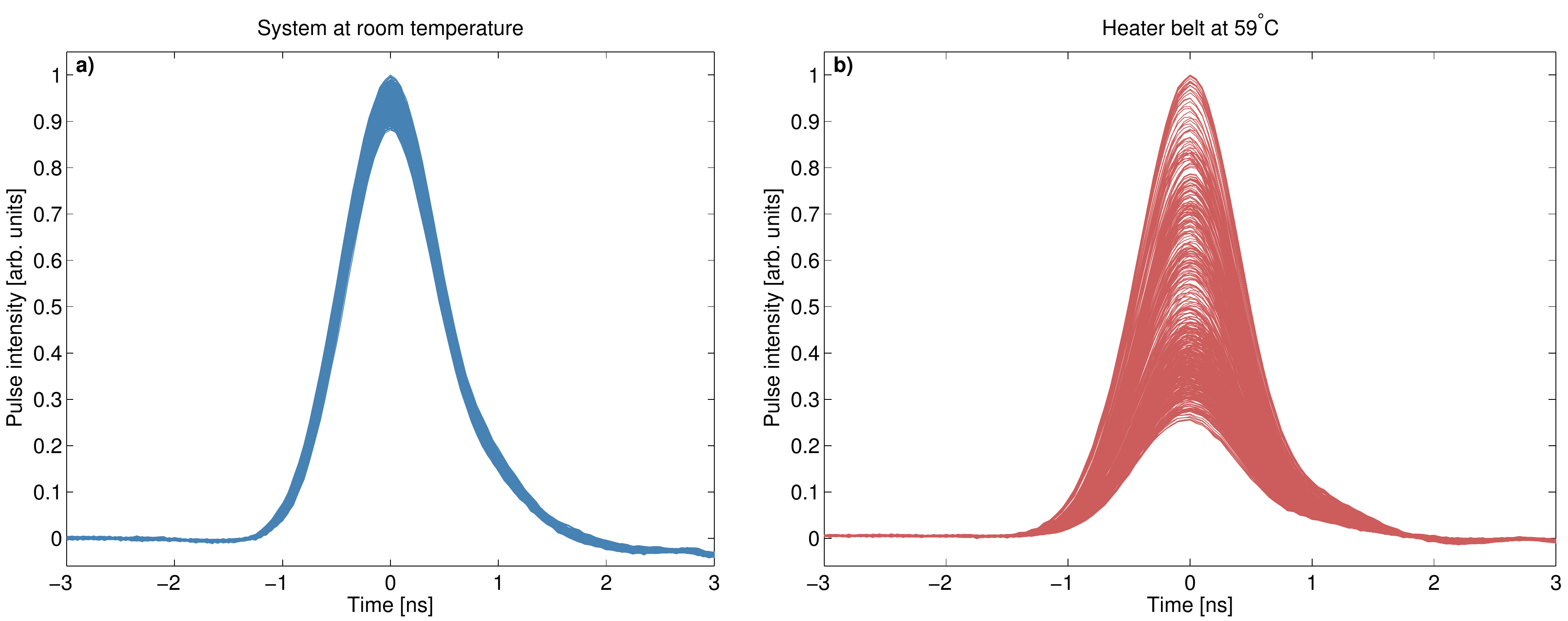}
\caption{Pulse intensity amplitudes observed during testing the interferometer stability. Both panels display the signal observed by the Menlo PD upon sending \dpol-polarised signal pulses, analysed in \rpol-polarisation basis, through the interferometer. 
\textbf{(a)}: pulses transmitted through the empty interferometer, which is reasonably phase stable by itself. 
\textbf{(b)}: pulse intensity variation from phase instability introduced by placing the test set-up, with a heater belt at $59^\circ \text{C}$, inside the interferometer.}
\label{fig_ch4_fluctuations}
\end{figure}

The interferometer stability is assessed by observing signal pulses transmitted through the interferometer on the Menlo PD. This is done firstly for an empty interferometer and secondly, when the test system is inserted between the PBDs.
In both cases, empty interferometer or test set-up in between the PBDs, \dpol-polarised input signal pulses are analysed in the \rpol-polarisation basis,  
which is a combination most sensitive to phase changes between the interferometer arms. 
Fig.~\ref{fig_ch4_fluctuations} \textbf{a} and \textbf{b} shows the recorded traces for both scenarios.
Clearly the interferometer is stable when empty. Notably, for the data in fig.~\ref{fig_ch4_fluctuations} \textbf{a}, air currents have been induced artificially in the area between the PBDs by waving a piece of card over the empty interferometer. 
Because all air is at the temperature of the environment, the phase remains reasonably stable.
Upon introducing the heated test set-up, the respective phase between both interferometer arms immediately becomes unstable. 
As a result the detected pulse amplitudes fluctuate (fig.~\ref{fig_ch4_fluctuations} \textbf{b}). 
Similar intensity fluctuations are observed when removing the waveplate from the cage system, i.e., having solely the heater tape wrapped around the cage rods. 

As a solution to this problem, we introduce shielding pipes around the {\cs} cell and the interferometer arms, including parts of the PBDs. 
These allow us to keep a passively stabilised interferometer. 
Fig.~\ref{fig_ch4_int_instab_setup} \textbf{b} illustrates the resulting setup around the {\cs} cell, which breaks up the convection currents. 
The air inside the pipes is stuck and heated by the adjacent {\cs} cell over time. 
The remaining temperature gradients are rotationally symmetric  around the optical axis and thus affect both interferometer arms similarly, reducing fluctuations in the path length difference. Importantly, this works better the shorter the interferometer arms are. For this reason, the arms lengths are $L_\text{int} \approx 40\cm$ (including PBDs), which is the minimum length required for all optical elements to fit and to have workable conditions.

\paragraph{{\cs} cell insulation}
To implement the thermal shielding, first the {\cs} vapour cell has to be thermally insulated.
Such thermal shielding has proven crucial not only for the polarisation interferometer stability, but also for the long term stability of the memory efficiency. 
As described in appendix \ref{ch7_Coldspot}, cold spots along the cell walls have the ability to significantly reduce the efficiency. 
They lower the vapour density and lead to crystallisation of the atoms at the cold spot locations. 
Such {\cs} deposition on the optical windows is particularly troublesome, because the solidified {\cs} scatters the incoming signal and control light and thus prevents memory experiments. 
Cold spots along the optical beam path must therefore be avoided, which we guarantee by packaging the {\cs} cell in the following manner (see fig.~\ref{fig_ch4_int_instab_setup} \textbf{b}): 
With the heater tape wound around it, it is covered by two layers of foam material and inserted into a poster tube of $\sim 20\cm$ length. 
The {\cs} cell is located at the centre and plastic pipe tubing connects its optical facets to either end of the poster tube. The ends of the poster tube are sealed with plastic caps. The plastic pipes around the optical beam path stick out of these sealing caps by $\sim 5\cm$ on either end.
Improvised magnetic shielding is added by a three-layer wrapping of $\mu$-metal around the poster tube. A magnetic degaussing coil is added on top of the $\mu$-metal shield.
The unfortunate down-side of this arrangement is its inability to shield residual magnetic fields from the electric heater tape. These can lead to a magnetic dephasing of the spin-wave, which we estimate to be one of the limiting factors for the memory lifetime\footnote{This is the 1/e-lifetime.} of $\tau_S \approx 1.3\mus$, as we will see in section \ref{ch4_results} below. 
The optical windows of the cell are not in direct contact with the heater tape. 
Despite the thermal insulation, they would still be the coldest spots of the cell, if no artificial cold spot was introduced somewhere along the cell body. To this end, a tube is inserted into the thermal shield, which allows to blow compressed air onto the cell's sealing nozzle, located at the midpoint of the cell body.
For the experiments in chapters \ref{ch6} \& \ref{ch7}, the cell has additionally also been insulated by layers of aluminium foil. A first layer is introduced around the heater belt, a second one around the outer layer of foam material and on the poster tube end caps.

\paragraph{Polarisation interferometer shielding}
To further minimise air currents, plastic pipe tubing, connected to the vapour cell's optical axis tubing, is extended to cover the entire interferometer, reaching also half way over the PBDs. 
The ends around the PBDs are sealed by stuffing lens tissue into the residual gaps between the cylindrical pipe tubing and the rectangular PBDs. 
The tubes also completely cover either side of all waveplates inserted into the interferometer (see figs. \ref{fig_ch4_setup_fig} and \ref{fig_ch4_int_instab_setup} \textbf{b}). Since access to the interferometer arms is required before every measurement for alignment purposes, the tubing, embracing the interferometer arms, is separated into sections of increasing diameter, which can slide on top of one another.
This remedy makes data collection possible and enables, after an initial thermalisation period, to record data with a stability exemplified by the individual scope traces in fig.~\ref{fig_ch4_trace_PD_750ns} \textbf{b} - \textbf{d}.
Comparison of the pulse intensity fluctuations with fig.~\ref{fig_ch4_fluctuations} \textbf{b} immediately reveals the extent of the improvement.

However, the stability is not perfect; the phase still drifts on a minute timescale.
Apart from residual air currents in the interferometer arms, these phase drifts could also result from turbulences and refractive index fluctuations of the buffer gas in the {\cs} cell. 
Such effects have been ignored in our analysis, for the simple reason, that shielding the air currents enabled us to obtain stability on a time scale long enough to go ahead with our measurements and get results of reasonable quality. 
Due to time constraints for finishing this experiment, we decided to use the system with the presented improvements and cancel any further phase drifts manually, as explained in the following.

\paragraph{Additional phase compensation}
The residual phase drifts are slow enough to be compensated by polarisation optics, for which reason an additional 
$\lambda/2$- and $\lambda/4$- plate are inserted into the interferometer. 
These are reset manually after completing the measurement for one combination of input signal polarisation $x$ and analysis polarisation basis $y$. For each such combination, the measurement settings are always recorded in the sequence \textit{sd}, \textit{scd}, \textit{cd}. 

In practice, the requirement for realignment means that the interferometer itself does not operate in the \{\hpol, \vpol \}-basis anymore. Instead it rotates the input states to some arbitrary location on the Bloch sphere, which is reset to the \{\hpol, \vpol\}-basis by the $\lambda/2$- \& $\lambda/4$- plate. 
Fig.~\ref{fig_ch4_qpt_theory} \textbf{c} in appendix \ref{ch4_subsec_qubits} exemplifies such a rotation\footnote{ 
	Fig.~\ref{fig_ch4_qpt_theory} \textbf{c} of appendix \ref{ch4_subsec_qubits} assumes the rotation 
	as a map from $\{\ket{H}, \ket{V}\}$ to $\{\ket{R}, \ket{L}\}$ (path $1 \rightarrow 2$), followed by subsequent resetting 
	with a $\lambda/4$- plate (path $2 \rightarrow 3$) and a $\lambda/2$- plate (path $3 \rightarrow 4$).
	Over this procedure, the beam picks up a geometrical phase $\Omega_{1-4}$, the Berry phase\cite{Bhandari:1988}, 
	which corresponds to the solid angle of the enclosed path traversed by the initial 
	polarisation state on the Bloch sphere.
}. 
This operation leads to the pick-up of a Berry phase\cite{Bhandari:1988}. 
To allow for faithful detection of the output, the Berry phase also needs to be compensated. 
It can be cancelled by performing the exact opposite sequence of rotations 
experienced by the signal inside the interferometer\footnote{
	i.e.  reversing the path on the Bloch sphere 
}. Generally, such phase-compensation is done using a sequence of three waveplates\cite{Langford:PhD} in the order $\lambda/4$, $\lambda/2$,  $\lambda/4$, which allows arbitrary polarisation rotations on the Bloch sphere (see appendix \ref{ch4_subsec_qubits} for an introduction of the Bloch sphere). 
The three waveplates are added in between the interferometer output and the polarisation analysis (see fig.~\ref{fig_ch4_setup_fig}). 
Likewise to their intra-interferometer counterparts, manual resetting is necessary between measuring consecutive polarisation settings $\left\{x,y\right\}$. 
While the waveplates inside the interferometer are aligned for optimal control extinction in the signal output port, the compensation waveplates are set subsequently by sending in a \dpol\,-pol. signal. 
With the polarisation analysis set to \apol, the waveplates are aligned for minimum transmission. This cancels the Berry phase and allows to faithfully analyse all input states in the basis state set $\text{\suet{B}} \in \left\{ \ket{H}, \ket{V}, \ket{+},\ket{-},\ket{R},\ket{L}\right\}$, required for quantum process tomography (QPT, see appendix~\ref{ch4_qpt_intro}). 
Because the control field is split from the signal by the PBDs beforehand, the Berry phase compensation setup does not affect the measured signal in any way other than cancelling the geometric phase.

\subsection{Measurement procedure\label{ch4_subsec_exp_conduct}}

\paragraph{Memory efficiency}
To measure the memory efficiency, the differences between the observed pulses in the read-in and read-out time bins are recorded upon blocking and unblocking the control field\cite{Reim2010, Reim:2011ys, England2012}. As shown in fig.~\ref{fig_ch4_trace_PD_750ns}, when applying setting \textit{sd}, storage and retrieval are absent.
Here, the Menlo PD records the input signal pulse, with an integrated pulse area $A_\text{in}$ and an output voltage signal amplitude $I_\text{in}$, which is proportional to the pulse peak intensity. 
For setting \textit{scd}, the reduction of the transmitted pulse in the input time bin, with area $A_\text{trans}$ and amplitude $I_\text{trans}$, yields the read-in efficiency \etain\,. 
$A_\text{trans}$ and $I_\text{trans}$ denote the reduction in pulse area and pulse intensity with respect to $A_\text{in}$ and $I_\text{in}$.
Similarly, using the areas $A_{\text{out},i}$ and amplitudes $I_{\text{out},i}$ of the retrieval pulses in read-out bins $i$ allow to calculate the total memory efficiency \etamem\,. 
Both efficiencies are defined as:
\begin{align}
\text{Efficiency in:} \quad & \eta_\text{in} = \frac{A_\text{trans}}{A_\text{in}} \approx \frac{I_\text{trans}}{I_\text{in}}, & \Delta \eta_\text{in} = \sqrt{\frac{\Delta A_\text{trans}^2}{A_\text{in}^2} + \frac{A_\text{trans}^2 \cdot \Delta A_\text{in}^2}{A_\text{in}^4}} \nonumber \\
\text{Efficiency out, bin \textit{i} :} \quad & \eta_{\text{mem},i} = \frac{A_{\text{out},i}}{A_\text{in}} \approx \frac{I_{\text{out},i}}{I_\text{in}}, & \Delta \eta_{\text{mem},i} = \sqrt{\frac{\Delta A_{\text{out},i}^2}{A_\text{in}^2} + \frac{A_{\text{out},i}^2 \cdot \Delta A_\text{in}^2}{A_\text{in}^4}}
\label{eq_ch4_mem_eff}
\end{align}
Similar results are obtained when using the pulse areas or their voltage amplitudes. The former yield the efficiencies stated in table \ref{tab_ch_6}, which lists the observed values for all investigated storage times $\tau_S$. \newline 
Notably, eqs. \ref{eq_ch4_mem_eff} implicitly assume the absence of any noise, whose contribution would additionally require usage of results for setting \textit{cd}.
We will see in chapters \ref{ch6} \& \ref{ch7}, that the Raman memory has a significant noise floor at the single photon level. 
Thus, at the single photon level, recording of all three settings \{\textit{scd}, \textit{sd}, \textit{cd}\} is required. 
For measurements with bright coherent states on linear photodiodes, the noise is however far too weak to be detected (see fig.~\ref{fig_ch4_trace_PD_750ns}) and can be neglected. 

\begin{table}[h!]
\centering
\begin{center}
\begin{tabular}{l | c | c | c | c | c}
\toprule
$\tau_s$		& $\eta_\text{in} [\%]$	& $\eta_{\text{mem},1} [\%]$	& $\eta_{\text{mem},2} [\%]$	& $\eta_{\text{mem},3} [\%]$ & $\eta_{\text{mem},\text{tot}} [\%]$\\
\midrule
$12.5\ns$		& $26.4 \pm 0.1$	& $5.4 \pm 0.2$		& $4.46 \pm 0.02$		& $1.19 \pm 0.02$	& $11.05 \pm 0.2$		\\
$312\ns$		& $22 \pm 0.5$		& $4 \pm 0.3$			& $3.1 \pm 0.3$		& $2.4 \pm 0.2$ 	& $9.5 \pm 0.5$		\\
$500\ns$		& $14.7 \pm 1.1$	& $2.3 \pm 0.3$		& $2 \pm 0.3$			& $1.7 \pm 0.3$	& $6 \pm 0.5$			\\
$750\ns$		& $20.6 \pm 0.4$	& $2.4 \pm 0.2$		& $2.1 \pm 0.1$		& $1.7 \pm 0.2$	& $6.5 \pm 0.3$		\\
$987\ns$		& $20.4 \pm 0.1$	& $3.4 \pm 0.04$		& $2.9 \pm 0.1$		& $2.33 \pm 0.04$	& $8.63 \pm 0.1$		\\
$1512\ns$		& $18.4 \pm 0.9$	& $1.4 \pm 0.2$		& $1.1 \pm 0.2$		& $0.9 \pm 0.2$	& $3.4 \pm 0.3$		\\
\bottomrule
\end{tabular}
\caption{Memory efficiencies obtained during the polarisation storage experiments. The efficiencies are the averages over all input polarisations $x$ for the same analysis polarisation $y=x$, i.e. at maximum transmission through the analysis polariser. Note that these numbers are not used for determining the memory lifetime data, shown in fig.~\ref{fig_ch4_fidelity_lifetime}, which was recorded in a separate measurement.}
\label{tab_ch_6}
\end{center}
\end{table}

\paragraph{Balancing both memory rails}
Faithful polarisation storage in the dual-rail memory also requires equal memory efficiencies in both arms. 
Imbalances artificially rotate the polarisation, since the respective weights between the contributions $\alpha \ket{H}$ and $\beta \ket{V}$ 
in the decomposition of an 
input state $\ket{\phi} = \alpha |H\rangle + \beta |V\rangle$ are changed. 
Efficiency balancing between both arms is firstly achieved by matching the spatial mode profiles and the locations of the beam waists, which are positioned at the centre of the {\cs} cell. 
Signal and control are focussed to waist sizes $w_{0,H}^\text{sig.}= 290 \mum \times 260 \mum$, $w_{0,H}^\text{ctrl.}= 500 \mum \times 690 \mum$ and $w_{0,V}^\text{sig.}= 300 \mum \times 270 \mum$, {${w_{0,V}^\text{ctrl.}= 560 \mum \times 700 \mum}$} for the \hpol\, and \vpol\, arms, respectively. 
Similarly, the optical pumping beam is focussed to beam waists of $w^\text{diode}_{0,H}= 480 \mum \times 430\mum$ and $w^\text{diode}_{0,V}= 390\mum \times 330\mum$ in each arm, whose locations are coincident with those of signal and control.
Temporal mode matching between signal and control is optimised with translation stages positioned in both control arms prior to control insertion into the first PBD.
When actually performing measurements, the power splitting of the control into both arms is adjusted to balance the memory efficiencies, yielding, on average, a splitting ratio of $1.1:1$ between the \hpol\, and \vpol\, arm. 
Similarly, also the available diode laser power of $3.5 \mW$ is split with a ratio of $1.22:1$ between the \hpol- and \vpol-arm to obtain balanced efficiencies. 

The splitting in the available control pulse energy reduces the memory efficiency\cite{Nunn:2007wj}. 
The lower control pulse energy leads to a reduction from an efficiency of $\eta_\text{mem} \approx 30\,\%$, 
observable in a single mode memory, where the full control power is inserted into this single mode (see chapters~\ref{ch6}~\&~\ref{ch7}), 
down to $\eta_\text{mem} \approx 5\,\%$ for the dual rail configuration, when the control power is split between both modes. 
These efficiencies represent the numbers for the first read-out time bin at $\tau_S = 12.5\ns$ storage time.

\paragraph{Input signal intensity level}
Since the experiment uses bright coherent state input signals, the results presented here do not unambiguously prove the operation of our system as a quantum memory (see section \ref{ch4_subsec_outlook} below). 
Nevertheless, experiments with bright coherent states are a good first benchmark to test the capabilities of our system, since the counting statistics of coherent states at the single photon level, when passing through a linear optical system, follows the classical behaviour\cite{Loudon:2004gd,Novikova:2007kc,Kim:2010}. 
Ignoring any contributions from noise, one would thus expect to obtain similar result for the polarisation storage of signals at the single photon level. 
An extension of the experiment down to the single photon level has been omitted, as it is quite challenging for two technical reasons.
The first is specific to the current experimental layout: at the single photon level, small polarisation rotations start to matter, as these increase control field leakage into the signal output mode. The leakage is registered by an avalanche photodiode (APD), but it is not observable on the linear Menlo PD.
Such rotations are introduced by the hot {\cs} cell and they are different between both interferometer arms\footnote{
	In previous work\cite{Reim:2011ys, England2012}, the origin of these rotations was attributed 
	to birefringence in the cell windows. 
	However, it could also result from a Faraday rotation in the pumped {\cs} at higher densities. 
}. Rebalancing would require separate sets of continuously accessible $\lambda/2$ and $\lambda/4$ waveplates in each interferometer arm, which are tricky to combine with interferometer shielding.
Secondly, single photon level measurements necessitate significantly longer integration times per measurement setting ($\Delta t_\text{meas} \sim 10\min$) than the times employed here ($\Delta t_\text{meas} \lesssim1\min$); see also section \ref{ch6_sec_single_photon_storage}.
Consequently, the interferometer phase would need to be stable for $\gtrsim 30\min$ until all three settings are recorded, which is not achievable with the current system. Either active stabilisation or further reduction of heat transport, e.g. by placing the interferometer in a vacuum environment, could be means to achieve these longer stability times\footnote{
	Again, this neglects any effects from the buffer gas, which would have to be studied and eventually
	compensated for as well.
}. 
All of these points are technical challenges, which can be overcome. So, in principle, the experiment can be conducted at the single photon level. 
For the current proof-of-principle study, we however only investigated bright coherent states. 
Besides these technical issues, there is an actual, crucial challenge when operating the system at the single photon level. This is the memory noise floor, as we shall find out in section \ref{ch4_subsec_outlook}.

\section{Results\label{ch4_results}}
Because our Raman memory operates with orthogonally polarised signal and control (see chapter \ref{ch2}), we would expect the stored signal's polarisation to be dependent on the polarisation of the control. 
With equal control pulse timings in both interferometer arms and equal memory efficiencies, the read-out signal's polarisation should hence be unaffected by the storage. 
Accordingly, the process matrix $\chi_{i,j}$, with $i,j \in \left\{X,Y,Z,\mathds{1} \right\}$, where 
$X$, $Y$, $Z$ represent the 3 Pauli spin-matrices for a qubit\cite{Nielsen:2004kl} and $\mathds{1}$ is the identity (see appendix \ref{ch4_subsec_qubits}), should solely contain one non-zero value at $\chi_{\mathds{1},\mathds{1}}$, the identity operation.  
Deviations therefrom, i.e. polarisation rotations of the output, can have different causes. The obvious possibility is interferometer instability. Furthermore, noise emitted by the memory can add to the signal and, if polarised differently, lead to an effectively rotated signal upon detection; yet noise contributions are negligible for bright coherent states. 
Since storage and retrieval are a coherent process, the phase of the output is sensitive to the phase of the spin-wave\cite{Jenkins:2006}. Different phase evolutions of the two spin-waves in each interferometer arm can lead to a phase difference and thus to a rotation in the output signal. A further possibility for polarisation rotations are Faraday rotations\cite{Mohapatra:2008cl} in the {\cs} vapour. 

We can investigate the influence of any such effects by performing full process tomography, using the basis states $\text{\suet{B}} = \left\{\ket{H},\ket{V},\ket{+},\ket{-},\ket{R},\ket{L} \right\}$ for the input and analysis polarisations, as outlined in appendix \ref{ch4_subsec_QPT}. 
We reconstruct the process matrix $\chi$ for the input signal (setting \textit{sd}) and the signals transmitted through and retrieved from the memory (setting \textit{scd}). This is conducted for a set of storage times $\tau_S = \left\{12.5\,\text{ns},312\,\text{ns}, 500\,\text{ns}, 750\,\text{ns}, 987\,\text{ns}, 1512\,\text{ns} \right\}$, until decoherence decreases the memory output to levels too low for determining the read-out signal fraction. 
Using the reconstructed process matrices $\chi$, we also determine the process purity $\mathcal{P}$ and fidelity $\mathcal{F}$ (see appendix \ref{ch4_subsec_QPT}). 

\subsection{Process matrix}
\begin{figure}[h!]
\includegraphics[width=\textwidth]{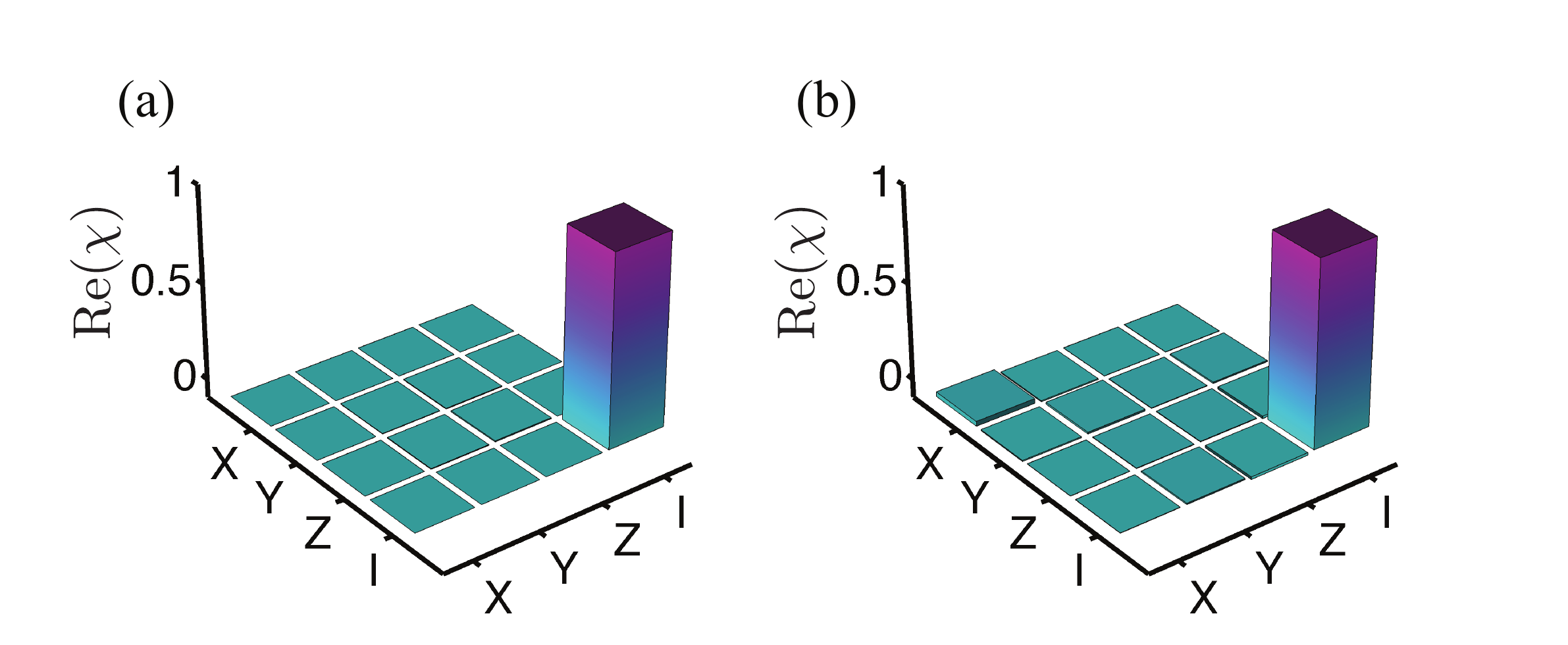}
\caption{Real part of the process matrices $\chi_{i,j}$ for $\tau_S = 750\ns$ storage time. 
\textbf{(a)}: Process matrix $\chi_\text{off}=\chi_{i,j}^{sd}$ for setting \textit{sd}, corresponding to the action of the polarisation interferometer on the input signal. 
\textbf{(b)}: Process matrix $\chi_\text{on}^{\text{out},1} = \chi_{i,j}^{scd,\text{out},1}$ for the signal in the first read-out time bin, i.e. retrieved after storage for $\tau_S$.
The elements $\chi_{i,j}$, with $i,j \in \left\{X,Y,Z,\mathds{1} \right\}$ denote a polarisation rotation, induced by application of the respective Pauli spin-matrices $\sigma_{i}$ and $\sigma_j$ to the input state, with density matrix $\rho_\text{in}$ (see eqs. \ref{eq_ch4_unitary_rot_1} and \ref{eq_ch4_unitary_rot_2} in appendix \ref{ch4_qpt_intro}). So, each column $\chi_{i,j}$ yields an output state $\rho_\text{out} = \sigma_i \rho_\text{in} \sigma_j^\dagger$. 
}
\label{fig_ch4_process_matrix}
\end{figure}

When the memory is off (setting $\textit{sd}$), we detect the input signal transmitted through the interferometer.  
Its process matrix $\chi_\text{off} = \chi^{sd}$ contains solely the effects of the polarisation interferometer and thus allows to benchmark its performance.
With good alignment, the interferometer should leave the signal state unchanged, corresponding to the identity operation\footnote{
	$\chi_\text{off} = \tilde{\mathds{1}} = \left( \begin{matrix} 1 & 0&0&0 \\ 0&0&0&0 \\ 0&0&0&0 \\ 0&0&0&0 \end{matrix} \right)$
} $\chi_\text{off}$. 
With control pulses present, $\chi^{scd,\text{in}}$ for the read-in time bin shows the influence of the Raman interaction on the non-stored signal.
Any deviation of $\chi^{scd,\text{in}}$ from $\chi^{sd}$ arises from imbalances in the read-in efficiencies {\etain} between both arms.
In turn, the process matrices for the retrieved signal $\chi^{t}_\text{on}= \chi^{scd,\text{out},t}_{i,j}$ in time bin $t$ show how well polarisation is preserved. 
Since additional components besides $\chi^{t}_\text{on} = \mathds{1}$ alter and possibly reduce the retrieved information content, they negatively influence the memory's performance. 
Fig.~\ref{fig_ch4_process_matrix} illustrates the obtained results, exemplified by $\chi_\text{off}$ and $\chi_\text{on}^\text{out,1}$ for $\tau_S = 750\ns$ storage time. 
Every element in $\chi$, besides the unity operation $\chi_{\mathds{1},\mathds{1}}$ on the diagonal, corresponds to a change in the polarisation of the input state. For instance, the other diagonal elements $\left\{ \chi_{XX}, \chi_{YY}, \chi_{ZZ} \right\}$ indicate the probability for a polarisation rotation expressed by the respective Pauli-matrix (see eqs. \ref{eq_ch4_unitary_rot_1} and \ref{eq_ch4_unitary_rot_2}). 
So, if the matrix was given solely by the element $\chi = \chi_{XX}$, the process would generate a $\ket{+}$-polarised output whenever a $\ket{H}$-polarised input state is inserted. 

As we would thus expect for good polarisation maintenance, the only considerable contribution to the matrix $\chi^\text{off}$, displayed in fig.~\ref{fig_ch4_process_matrix} \textbf{a}, comes from the identity operation. 
This means, there are no significant influences from the interferometer.
In the first retrieval bin, fig.~\ref{fig_ch4_process_matrix} \textbf{b}, $\chi^{\text{out},1}_\text{on}$ is also dominated by the identity operation. We can clearly see that the $\chi_{\mathds{1}\mathds{1}}$ column is, by far, the highest element of $\chi$. 
Additionally, small diagonal and off diagonal elements appear, most noticeable on the diagonal element $\chi_{XX}$. 
It represents the aforementioned polarisation rotation, which only occurs with low probability, represented by its relative column height. 
Yet, these additional elements do not necessarily result in mixed output states, because off-diagonal coherence elements also appear. For instance, uncompensated unitary rotations can also add such terms.
To investigate their influence, we will next look at the purity of the process matrices.

\subsection{Process purity\label{ch4_subsec_purity}}
The process purity, as defined in appendix \ref{ch4_subsec_QPT}, is shown in fig.~\ref{fig_ch4_purity_fidelity} for both settings \textit{sd} and \textit{scd}.
Within the input time bin (fig.~\ref{fig_ch4_purity_fidelity} \textbf{a}), the input signal transmitted through the interferometer has an average purity of 
$\mathcal{P}_\text{off} = \tr{\left(\chi_\text{off}^2  \right)} = 0.989$.
The high number illustrates that, despite the experimental challenges imposed by thermal instability (see section \ref{ch4_subsec_stability}), the interferometer can nevertheless be aligned quite well and can be kept stable throughout a measurement cycle.
These purity levels are closely matched by the non-stored signal, transmitted through the memory, which on average has $\mathcal{P}_\text{on}^\text{in} = \tr{\left( \left(\chi^\text{in}_\text{on}\right)^2  \right)} = 0.988$. 
So the memory read-in efficiencies between the interferometer arms also remain reasonably stable throughout the measurement. 
Since the location of both sets of data around the mean correlate, 
changes in performance are partially due to interferometer alignment.
In the read-out time bins $1$-$3$, the process purity decreases to averages $\mathcal{P}_\text{on}^\text{out,1} = 0.829$, $\mathcal{P}_\text{on}^\text{out,2} = 0.846$, and $\mathcal{P}_\text{on}^\text{out,3} = 0.811$ for the respective time bins. 
Hence the additional elements in the process matrices $\chi$ give rise to some mixture in the memory output. 
As a result, the memory scrambles the polarisation information at times. 
The purities are time independent, i.e. there is no downward trend for longer storage times. 
So spin-wave decoherence does not increase polarisation state mixture in the output.
Moreover, the purities have nearly equal averages for all read-out bins. Their spreads around their means are also identical and correlated for the different time bins.
Partial spin-wave retrieval hence does not affect the remaining fraction of the spin-wave and the decrease in purity most likely arises within the Raman transition process.
Despite the corresponding reductions in the Bloch vector length (see appendix \ref{ch4_subsec_qubits}), memory storage does not scramble a pure input state. 

\begin{figure}[h!]
\includegraphics[width=\textwidth]{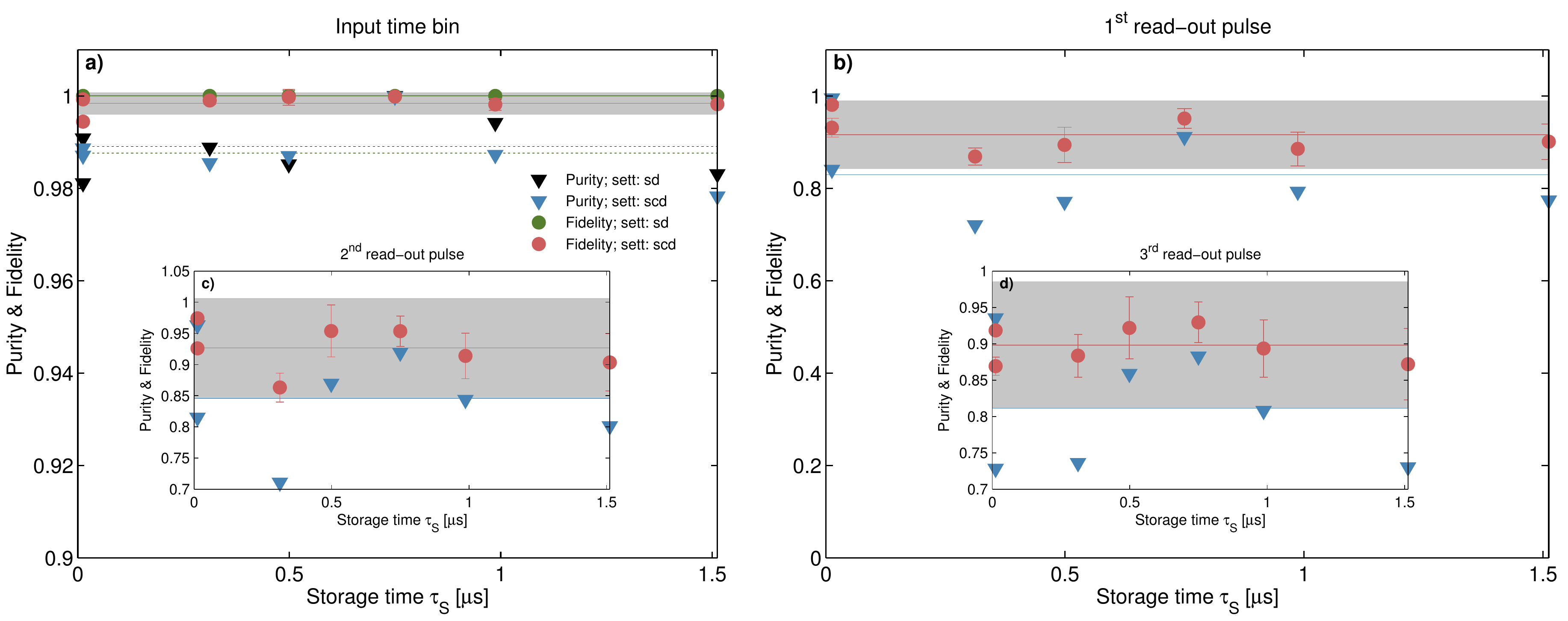}
\caption{Process purity and process fidelity for 
\textbf{(a)}: Input time bin showing the signal transmitted through the memory, i.e. the fraction of the input signal that is not stored.  
\textbf{(b)}: First read-out time bin.  
\textbf{(c)}: Second read-out time bin.  
\textbf{(d)}: Third read-out time bin. The values obtained from the reconstructed process matrices of the individual measurements are shown by \textit{black triangles} for the purity and by \textit{green circles} for the fidelity of the input state transmitted through the interferometer (setting \textit{sd}). The purity and fidelity data points observed in each time bin with active memory interaction (setting \textit{scd}) are drawn with \textit{blue triangles} and \textit{red circles}, respectively. \textit{Horizontal lines} show the average values of the data with equal colour coding, where the \textit{grey areas} denote the error regions of the fidelities obtained for active memory (setting \textit{scd}) by Monte-Carlo simulation (see appendix \ref{ch4_subsec_QPT}).}
\label{fig_ch4_purity_fidelity}
\end{figure}

\subsection{Process fidelity\label{ch4_subsec_fidelity}}
The next question to ask is, whether storage and retrieval change the direction of the Bloch vector (see appendix \ref{ch4_subsec_qubits}).
This is answered by evaluating how closely the process $\chi_\text{on}$ resembles the best possible process\footnote{
	This simplified picture of fidelity only holds true for pure states, see appendix \ref{ch4_subsec_qubits}.
} 
$\chi_\text{off}$, observed for input signal transmission through the interferometer. 
We use the process fidelity $\mathcal{F}(\chi | \chi_\text{off})$ as defined in appendix \ref{ch4_subsec_state_tomo}.
To aid comparison, the results are shown in fig.~\ref{fig_ch4_purity_fidelity} alongside the process purity.
By definition, $\mathcal{F}=1$ for the input signal (\textit{sd}) with $\chi_\text{off}$, since it is the benchmark. 
As the average fidelity $\mathcal{F}^\text{in} = 0.998 \pm 0.002$ demonstrates, 
the transmitted fraction of the signal in the input time bin is nearly unmodified.
Thanks to good initial balancing of the read-in efficiencies between both interferometer arms and their stability throughout the measurement, there are no undesired state rotations from preferred Raman absorption in one of the arms.
The fidelities in the output time bins are reduced to, on average, $\mathcal{F}^\text{out,1}=0.92 \pm 0.07$,  
$\mathcal{F}^\text{out,2}=0.93 \pm 0.08 $ and $\mathcal{F}^\text{out,3}= 0.90\pm 0.09$, respectively. 
As the data points in fig.~\ref{fig_ch4_purity_fidelity} show, purity and fidelity correlate in their variations around their means. 
Mixed states, containing rotated states from operation elements $\chi_{i,j} \neq \chi_{\mathds{1},\mathds{1}}$, do not only reduce the purity, they can also deflect the direction of the Bloch vector.

\subsection{Purity and fidelity during the storage time\label{ch4_ch4_subsec_purity_fidelity_vs_storage_time}}
\begin{figure}[h!]
\includegraphics[width=\textwidth]{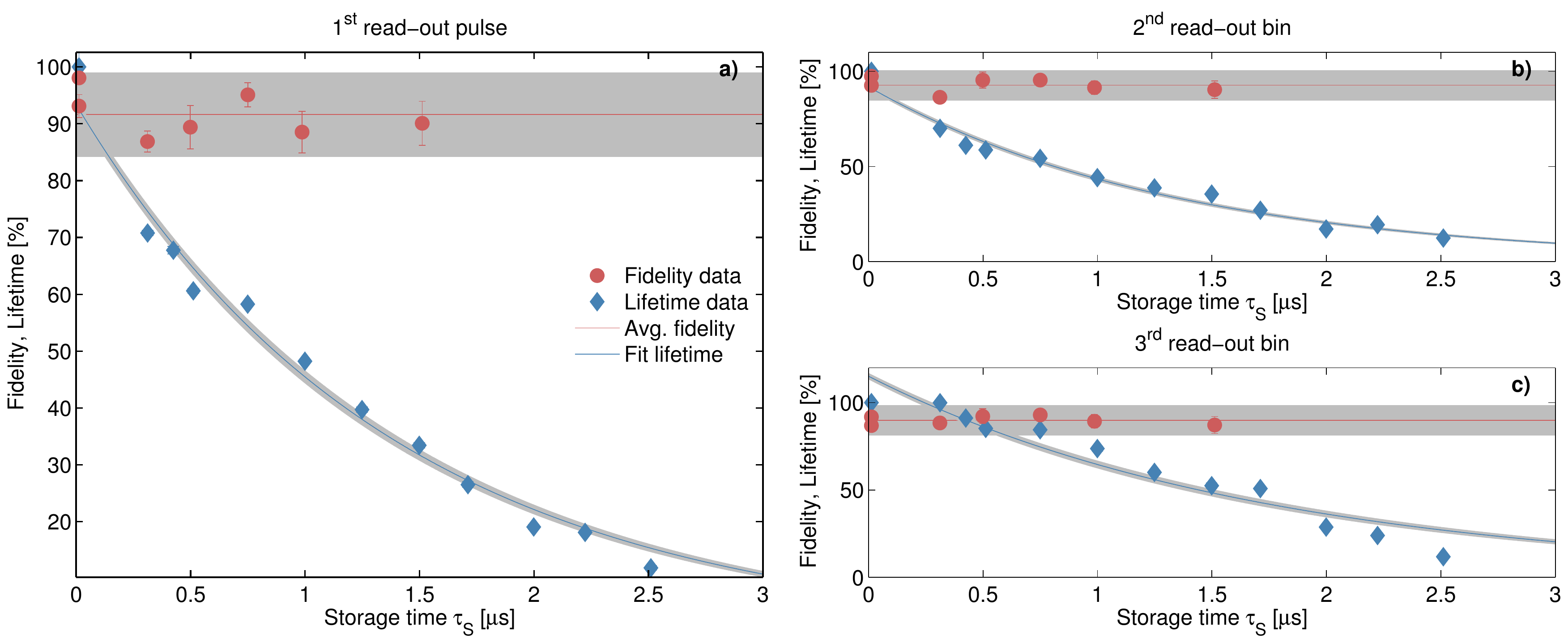}
\caption{\textbf{(a)} - \textbf{(c)}: Process fidelity $\mathcal{F}^{\text{out},i}$ and memory lifetime for read-out time bins 1 - 3. Like fig.~\ref{fig_ch4_purity_fidelity}, \textit{red circles} represent the fidelity data and the \textit{red lines} their averages, whose errors are the \textit{grey shaded area}. \textit{Blue points} are the normalised efficiencies $\frac{\eta_{\text{mem},i}(\tau_s)}{ \eta_{\text{mem},i}(12.5\ns)}$, with an exponential fit as the \textit{blue line}; fit errors are also shaded in \textit{grey}.
}
\label{fig_ch4_fidelity_lifetime}
\end{figure}

As for the purity, the fidelity is approximately independent of the storage time $\tau_s$. 
Fig.~\ref{fig_ch4_fidelity_lifetime}~\textbf{a}~-~\textbf{c} explicitly shows 
that spin-wave decoherence in the dual-rail memory does not visibly affect the polarisation storage quality by comparing the process fidelity with the memory lifetime over storage time $\tau_S$. 
This finding is different to polarisation storage demonstrations in some other systems\cite{Specht:2011,Kim:2010}, where reductions however appear after much longer times $\tau_S$ than currently available with our system.
We determine the lifetime from the pulse areas of $\ket{H}$-pol. input states, analysed also in $\ket{H}$-pol. basis. It was measured before the QPT datasets and after initial degaussing of the memory cell. The normalised reduction in pulse areas is fitted by an exponential decay\footnote{
	For pure magnetic dephasing, caused by a well defined, static B-field, one would expect a Gaussian decay 
	$\sim \exp{\left\{- \frac{t^2}{\tau_\text{men}^2}\right\}}$, which previous
	results suggested to be the dominant decoherence mechanism in our system\cite{Reim:2011ys}. 
	For the present dataset, as well as later measurements (see appendix \ref{ch7_lifetime}), an exponential
	decay better describes the data. Exponential efficiency decay is, e.g., expected for atomic diffusion 
	out of the laser beam.
} 
$\sim \alpha \exp{\left\{-\frac{t}{\tau_\text{mem}}\right\}}$ with memory lifetime $\tau_\text{mem}$. 
Fitting with a variable amplitude factor $\alpha$ yields lifetimes $\tau_\text{men}^\text{out,1} =1.39 \pm 0.01 \, \mu\text{s}$, $\tau_\text{men}^\text{out,2} =1.34 \pm 0.01 \, \mu\text{s}$ and {${\tau_\text{men}^\text{out,3} = 1.73 \pm 0.03 \, \mu\text{s}}$}, obtained from the three output pulses. 
Due to the small pulse areas for the third retrieval pulse, the data quality is reduced compared to the other time bins, which are thus more representative. 
The measured storage times tie in with the previous performance of our system, which, at the time these experiments were conducted, constituted one of the largest time-bandwidth products in the memory space\cite{Reim2010,Reim:2011ys}. 
Taken by themselves, these storage times are however quite short, compared to the minute-long spin-wave lifetimes achievable in atomic vapours\cite{Heinze:2013aa}. Possible improvements\cite{Novikova:2012} are magnetic shielding of the cell, and paraffin-coated cell walls. Furthermore, the use of a top-hat spatial mode intensity profile for the control pulses and a cell of smaller diameter, such that the control illuminates the full cell diameter, can also increase the storage time, as atoms cannot diffuse out of the laser beam anymore. 

While decoherence affects the storage efficiency by reducing the spin-wave amplitude, it does not lead to any reduction in the retrieved quality of polarisation information. 
This means, the global phase relationship between both memory rails is constant throughout the storage time, as any change would lead to a polarisation rotation in the read-out.
Assuming magnetic dephasing is the dominant loss mechanism, as previous results suggest\cite{Reim:2011ys}, 
spin-wave dephasing would not influence the polarisation state if both rails are subject to the same static magnetic field and both memory modes have an equal population distribution across the $6^2 \text{S}_{1/2} \text{F}=3$ Zeeman sub-levels. 
The dephasing rate would be similar between both memories, for which reason the phase change of their spin-waves is correlated.
To our current knowledge, the residual magnetic fields mostly originate from the heater belt, which is wrapped around the Cs cell. 
As such, the H-field is cylinder symmetrically aligned along the optical axis. The memory rails are positioned with approximately equal distances to cylinder axis of the cell. 
Hence to good approximation, both arms experience the same H-field.
In the case of decoherence through atom loss from diffusion out of the vapour volume, covered by the control pulses, no influence on polarisation storage is expected either. Lost atoms, to first approximation\cite{Novikova:2012}, do not contribute to the read-out signal as long as they do not re-enter the cell volume covered by the control. So the remaining atoms still have the same global phase relationship.

\begin{figure}[h!]
\centering
\includegraphics[width=0.94\textwidth]{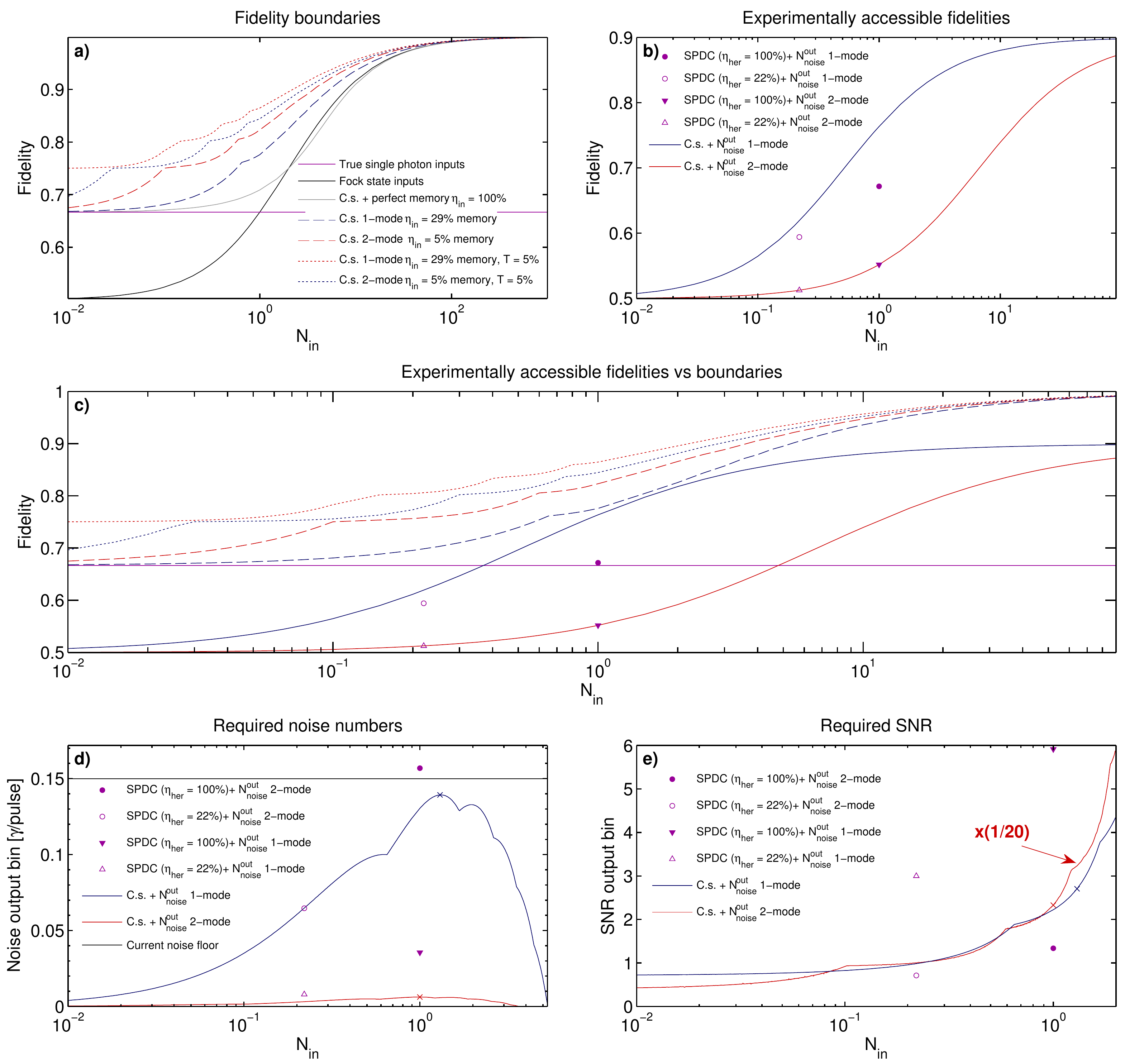}
\caption{
\textbf{(a)}: Boundary fidelities $\mathcal{F}_B$ for Fock (\textit{solid black line}) and \coh\, (\textit{solid grey line}) vs. input photons $N_\text{in}$. The \textit{solid pink line} is the $\mathcal{F}_B=2/3$ bound for single photon states ($N_\text{in} = 1\ppp$). 
\textit{Dashed lines} show the increased $\mathcal{F}_B$ for \coh\,, stored in a single mode memory with $\eta_\text{mem} = 29\,\%$ (\textit{blue}) and a dual-rail memory with $\eta_\text{mem} = 5\,\%$ (\textit{red}). 
Equally coloured \textit{dotted lines} denote $\mathcal{F}_B$ when also considering losses in the output ($T=5\,\%$). 
\textbf{(b)}: Fidelities expected with a constant memory noise floor 
($N_\text{noise}^\text{out}=0.15\ppp$). 
\textit{Blue} and \textit{red solid lines} denote \coh\, signals for the single-mode and the dual-rail memory, respectively. 
Similarly, \textit{pink circles} and \textit{triangles} are the same memory configurations for single photon inputs; 
\textit{filled} and \textit{empty markers} indicate perfect ($\eta_\text{her} = 100\,\%$) and the current ($\eta_\text{her} = 22\,\%$) heralding efficiency. 
\textbf{(c)}: Comparison between the boundary and the expected experimental fidelities at the single photon level (combination of \textbf{(a)} \& \textbf{(b)} with same colour coding). 
\textbf{(d)}: Required memory noise floor to obtain expected fidelities $\mathcal{F}$ equal to the boundary fidelities $\mathcal{F}_B$. 
\textit{Black line} is the current memory noise floor $N_\text{noise}^\text{out} =0.15\ppp$; other colour coding for the different signal types as in \textbf{(b)}. The $\times$ marks the experimentally optimal \coh\, input photon number. 
\textbf{(e)}: Signal-to-noise ratio (SNR) required to obtain the noise levels shown in \textbf{(d)}. 
Colour coding is equal to \textbf{(b)} and \textbf{(d)}. The SNR for \coh\, and a dual-rail memory is divided by a factor of $20$ for better visibility.}
\label{fig_ch4_fidelity_bounds}
\end{figure}

\section{Outlook for memory operation in the quantum regime\label{ch4_subsec_outlook}}
Despite small reductions in the retrieval time bin, the observed purity and fidelity values look promising for the Raman memory's applicability in a polarisation based quantum network at first glance\cite{England2012}. 
However, during our measurements, the system was supplied with bright coherent states. 
So, the question, whether the memory can actually be operated in the quantum regime, still remains. 
While we have neither attempted polarisation storage with input signals at the single photon level, nor with real single photon input states, we can nevertheless assess the memory operation theoretically by comparing 
our fidelity results against boundary fidelities $(\mathcal{F}_B)$. 
The boundary fidelity is the minimum fidelity states,
retrieved from a quantum memory, must possess to exclude the possibility, that these states were produced by a classical memory. 

To do this, we will first briefly outline the relevant boundary levels, then extrapolate the expected performance of our system, to obtain an estimate of the fidelities that can be expected in the single photon regime. We will then compare the expected performance with the boundary fidelities and, in a final step, determine the memory noise characteristics, required to enable operation in the quantum regime.

\paragraph{Boundary fidelities}
$\mathcal{F}_B$ correspond to the fidelities a state, created on-demand by an eavesdropper after interception and measurement of an input quantum state, can possibly possess\cite{Curty:2005,Felix:2001}. 
In this particular example of an intercept-and-resend attack, the eavesdropper could just completely by-pass the memory and supply the copied states directly into the memory's output mode.
Due to the no-cloning theorem\cite{Nielsen:2004kl}, such copied states cannot have all characteristics of the original quantum state; one example for this is the state's fidelity. 
In order to distinguish memory storage from such an attack, the memory needs to preserve the incoming quantum state's characteristics better than one can reproduce by classical state copying. 
In other words, if an eavesdropper could prepare a state with a fidelity $\mathcal{F}_B$ after measurement of the input signal, states, retrieved from the memory, have to possess fidelities $\mathcal{F} > \mathcal{F}_B$. 
If this condition is fulfilled, the memory is said to operate in the quantum regime.  
The exact boundary values depend on the actual experimental configuration and, in case the input signal differs from true single photons, on the number of photons, $N_\text{in}$, contained in the input state. Appendix \ref{app_ch4_fidelity_bounds} outlines the calculation of $\mathcal{F}_B$ in detail. Here we only summarise the results, which are shown in fig.~\ref{fig_ch4_fidelity_bounds} \textbf{a} for coherent state inputs (c.s.) and heralded single photons, obtained with an SPDC source (SPDC). 

For Fock state input signals, containing $N_\text{in}$ photons, the fidelity bound is $\mathcal{F}_B =  \frac{N_\text{in}+1}{N_\text{in}+2}$ (\textit{black line} in fig.~\ref{fig_ch4_fidelity_bounds} \textbf{a}). 
This gives the widely known value $\mathcal{F}_B = 2/3$ for a true single photon input \cite{Massar:1995} with $N_\text{in} = 1$ (\textit{pink line} in fig.~\ref{fig_ch4_fidelity_bounds} \textbf{a}). 
The value is essentially set by the number of projectors $\sigma_{\left\{X,Y,Z\right\}}$ that can be measured simultaneously. 
The fidelity bound increases with $N_\text{in}$, as measurements in different bases can be made by splitting up the photons\cite{Dusek:2000}.
Importantly, this bound changes if the input state is not a Fock state\cite{Specht:2011}. 
For other signals, such as coherent states (\coh\,), the memory's storage efficiency also leads to an additional modification of the bound. 
Both factors increase $\mathcal{F}_B$ with a functional form originally presented by \textit{Specht et. al}\cite{Specht:2011} and \textit{G\"undogan et. al.}\cite{Gruendogan:2012}, which is discussed in appendix \ref{app_ch4_fidelity_bounds}.
With perfect memory efficiency, i.e. $\eta_\text{mem} = 100\,\%$, one obtains a fidelity bound shown by the \textit{grey line} in fig.~\ref{fig_ch4_fidelity_bounds} \textbf{a}. 
When inserting all control pulse energy into a single memory mode, we achieve $\eta_\text{mem}= 29\,\%$ efficiency, depicted by the \textit{blue dashed line} in fig.~\ref{fig_ch4_fidelity_bounds} \textbf{a}. 
Due to the control power splitting in the dual-rail configuration, the efficiency reduces to $\eta_\text{mem} = 5\,\%$, denoted by the \textit{red dashed line} in fig.~\ref{fig_ch4_fidelity_bounds} \textbf{a}. 

Apart from inefficient storage, any transmission losses behind the memory as well as inefficient photon detection increase $\mathcal{F}_B$ even further. 
To obtain an estimate for such losses in an actual single photon experiment, we take the transmission of $T \approx 10\,\%$ for our heralded single photon storage set-up, presented in chapter \ref{ch6} (see fig.~\ref{fig_6_setup}), and assume a single photon detection efficiency of $\eta_\text{det} \approx 50\,\%$, which is the expected value for our single photon counting modules (see section \ref{sec_ch4_photon_detection}). 
In total, this yields an effective transmission of $T = 5\,\%$, which equals the probability of registering a photon once it has been released from the memory. 
The resulting boundary fidelities are depicted by the \textit{blue} and \textit{red dotted lines} in fig.~\ref{fig_ch4_fidelity_bounds} \textbf{a} for $\eta_\text{mem} = 29\,\%$ and $\eta_\text{mem} = 5\,\%$, respectively. 

We clearly see that the requirements on $\mathcal{F}$ become quite stringent, the less efficient the system becomes or the more photons $N_\text{in}$ are sent into the memory. 
Note in particular that all boundary lines for $\mathcal{F}_B$ converge against $1$ for $N_\text{in} \rightarrow 100\, \ppp$. 
Our bright \coh\, results would have to show unrealistically good fidelities to overcome this bound. 
Any experiment with bright coherent state input signals will thus struggle to demonstrate memory operation in the quantum regime. 
Nevertheless, it is worthwhile to examine, what fidelities can be expected from our memory, if 
we reduced the signal strength down to the single photon level. 

\paragraph{Expected fidelities at single photon level}
To estimate $\mathcal{F}$, we use the argument\cite{England2012} that the operation of the dual-rail memory and the interferometer are those of linear optical systems. 
For such systems, the counting statistics of single photon transmission follows the classical behaviour\cite{Loudon:2004gd}. An example are interferometers, showing interference fringes in their outputs for light at the single photon level\cite{Taylor:1909} and single particle Fock states\cite{Joensson:1961} alike. 
Replacing the input signal with a heralded single photon, for which $\mathcal{F}_B= 2/3$, should thus yield the same process matrix $\chi$. 
Consequently, we estimate the retrieved state's fidelity at the single photon level by the process fidelity\footnote{
	Importantly, the fidelity bound analysis works on quantum state fidelities, not process fidelities. 
	Here we use the process fidelity as the average output fidelity obtained for an input state with 
	$\mathcal{F}_\text{state} = 1$. Naively this is motivated from the fact that process 
	tomography is obtained from state tomography over all states of an orthonormal basis. 
	Hence the process should represent the average operation on any input state vector. 
	It can be shown, that this averaging also applies to fidelities\cite{Nielsen:2002}.
} 
of $\mathcal{F} \approx 0.9$, measured for bright coherent state inputs.

Yet, tacitly this assumes the absence of noise, which is not the case for our system. 
As we will see in chapters \ref{ch6} \& \ref{ch7}, likewise to other memory systems\cite{Manz:2007,Kupchak:PhD,Timoney:2013,Hosseini:2012,Novikova:2012}, there is a non-negligible noise floor at the single photon level\cite{Reim:2011ys,Michelberger:2014}. 
It needs to be considered when estimating the fidelity of the output signal. Since noise has the fidelity $\mathcal{F}_\text{n} = 1/2$ of a completely mixed state, it reduces the signal's fidelity with respect to the bright coherent state values, for which the noise contribution is negligible.  
The methodology of adding the noise depends on whether a single memory\cite{Bowdrey:2002} or the dual-rail configuration is considered\cite{Reim:2011ys}. 
It is laid out in appendix \ref{app_ch4_fidelity_bounds} and used to predict the fidelities we can expect to find, when actually performing the experiment with weak coherent states (c.s.), or with heralded single photons (SPDC). 
In the calculation, we use the experimental parameters of chapter \ref{ch6}, with $N_\text{noise}^\text{out} = 0.15\,\frac{\text{photons}}{\text{pulse}} (\ppp)$ for the noise floor in the memory read-out bin (eq. \ref{ch6_eq_NoiseFloor}), 
and $N_\text{in} = 0.22\,\ppp$ for the input number of single photons, which corresponds to a heralding efficiency 
of $\eta_\text{her} = 22\,\%$ (see section \ref{ch5_hereff}). 
Additionally, we also consider an SPDC source with perfect heralding efficiency $\eta_\text{her} = 100\,\%$. 
While this is experimentally unrealistic, it is the optimal performance of a single photon source and thus acts as a benchmark for the minimal requirements on the memory noise parameters. 
Likewise to the calculation for $\mathcal{F}_B$, we further consider memory efficiencies of $\eta_\text{mem}^\text{coh} = 29\,\%$ and $\eta_\text{mem}^\text{SPDC} = 21\,\%$ for a single mode memory, when coherent states and heralded single photons are inserted as input signals, respectively.  
For the dual-rail configuration, we assume an efficiency of $\eta_\text{mem}=5 \,\%$ for both input signal types. 

The \textit{solid blue} and \textit{red lines} in fig.~\ref{fig_ch4_fidelity_bounds} \textbf{b} show the resulting fidelities that can be expected for \coh\,, stored in a single mode ($\eta_\text{mem}^\text{coh} = 29\,\%$) and a dual-rail ($\eta_\text{mem}^\text{coh} = 5\,\%$) memory. 
Additionally, \textit{pink markers} in fig.~\ref{fig_ch4_fidelity_bounds} \textbf{b} illustrate the expected values for \hsp\, inputs. 
Here, \textit{filled symbols} represent $\eta_\text{her} = 100\,\%$, which corresponds to a single photon Fock state of 
${N_\text{in} =1\ppp}$. Open symbols denote our experimental value of $\eta_\text{her} = 22\,\%$. 
The two different memory configurations are represented by \textit{triangles} for the dual-rail, and \textit{circles} for the single mode memory. 

In both cases, the single mode configuration has a higher expected fidelity, because of the higher memory efficiency. It increases the contribution of the signal with respect to the fixed amount of noise in the read-out time bin. So the signal-to-noise ratio (SNR) is greater and the noise fidelity $\mathcal{F}_\text{n}$ affects the states less. 
Similarly, for a fixed $\eta_\text{mem}$ value, a higher number of input photons $N_\text{in}$ also increases the signal fraction in the read-out, leading to a better SNR and a higher expected fidelity. 

To assess, whether these expected fidelities would allow for operation in the quantum regime, we plot them against the fidelity boundaries in fig.~\ref{fig_ch4_fidelity_bounds} \textbf{c}. 
Quantum operation is possible, if the expected values for $\mathcal{F}$ lie above $\mathcal{F}_B$. 
This means, for \coh\,, the \textit{solid red} and \textit{blue lines} have to be compared to the \textit{dashed} and \textit{dotted lines} of equal colour coding. Similarly, for SPDC photons, the \textit{pink symbols} are compared to the \textit{solid pink line}. 
For both, single mode and dual-rail memory, the current noise floor is too high to allow demonstration of quantum regime storage with coherent states. 
Similarly, no quantum storage could be shown using single photons from our SPDC source, described in chapter \ref{ch5}, with $\eta_\text{her} = 22\,\%$ heralding efficiency. 
Yet, when assuming a source with perfect heralding efficiency ($\eta_\text{her} = 1$) and 
storage in a single mode memory, the expected fidelity would lie above the classicality bound. 
Importantly, this is not the case for other benchmark metrics, e.g. the photon statistics. 
As we will see in section \ref{ch6_subsec_g2models}, no non-classical statistics can be expected at the current noise level even for perfect single photon preparation. 

From fig.~\ref{fig_ch4_fidelity_bounds} \textbf{c} we can therefore draw two conclusions: First, even if we conducted the experiments at the single photon level, we would not be able to claim quantum storage of a polarisation qubit.
Second, to demonstrate quantum storage of a polarisation qubit using any realistically available input signal in the dual-rail configuration, the memory noise floor has to be reduced. 
For this reason, our finial step will now investigate how much noise one can actually tolerate before the fidelities can be reproduced by a classical memory. 

\paragraph{Required noise floor for operation in the quantum regime}
We answer this by looking for the noise floor $N_\text{noise}^\text{out}$, which is required to obtain a fidelity $\mathcal{F}$ that exactly matches the boundary $\mathcal{F}_B$ (see appendix \ref{app_ch4_noisy_fidelity}). 
Fig.~\ref{fig_ch4_fidelity_bounds} \textbf{d} shows the required numbers for the signals in fig.~\ref{fig_ch4_fidelity_bounds} \textbf{b} \& \textbf{c}. 
Note, boundary fidelities that take into account the transmission behind the memory have been neglected. Their noise requirements become unrealistically high. 
So only the \textit{dashed lines} in fig.~\ref{fig_ch4_fidelity_bounds} \textbf{a} \& \textbf{c} are considered as boundary fidelities for \coh\,

As we know already from the fidelity predictions in fig.~\ref{fig_ch4_fidelity_bounds} \textbf{c}, only single photons, prepared with $\eta_\text{her} = 100\,\%$, can accommodate a noise floor above the current level (\textit{solid black line} in fig.~\ref{fig_ch4_fidelity_bounds}~\textbf{d}). 
Since $\mathcal{F}_B$ of \coh\, is dependent on $N_\text{in}$, as is the expected fidelity $\mathcal{F}$ via the SNR, there is an optimum signal strength $N_\text{in}$ to probe the fidelity with. 
As the functional form of the required noise level shows, this optimum photon number is $N_\text{in}= 1.3 \,\ppp$ and $N_\text{in} =1 \,\ppp $ for a single mode memory ($\eta_\text{mem} = 29\,\%$) and a dual-rail memory ($\eta_\text{mem} =5\,\%$), respectively, assuming a constant noise level of $N_\text{noise}^\text{out} = 0.15\ppp$ in the read-out bin for both cases. 
Here, the required noise is maximal, i.e., these points require the least reduction in noise\footnote{
	Note, the required noise reduction increases towards higher input photon numbers as $\mathcal{F}_B$ 
	increases as well. This is because more photons in the input signal simplify state reproduction for an 
	eavesdropper. 
}. 
With a noise reduction of only $\approx 9\,\%$, the necessary improvements are small when using \coh\, input signals in a single mode memory. 
Conversely, for the dual-rail configuration, noise would have to suppressed by a factor of $\sim 24$ for \coh\, and $\sim 19$ for single photons, prepared with $\eta_\text{her} = 22\,\%$. 
These latter numbers are similar to those needed to observe non-classical statistics in the memory read-out (see chapter \ref{ch6}). 
Hence, storage of a polarisation qubit in the quantum regime is not necessarily easier to achieve than the preservation of other quantum properties, such as the photon number statistics. 

Likewise to $\eta_\text{mem}$, the noise level is dependent on the Raman coupling (see chapter \ref{ch2}). 
To obtain an apparatus-independent benchmark, the requirements on the noise floor can be expressed by the SNR, whose definition reads 
SNR$=\frac{\eta_\text{mem} \cdot N_\text{in}}{N_\text{noise}}$ (see section \ref{ch6_subsec_SNR}). 
Therewith the minimal SNR, required for reaching the $\mathcal{F}_B$ boundary, is displayed in fig.~\ref{fig_ch4_fidelity_bounds} \textbf{e}. 
To measure a non-classical fidelity with a single mode memory, the SNR for \coh\, at $N_\text{in}=1.3\ppp$ would have to be SNR$=2.71$. For a dual-rail configuration an SNR=$46.6$ is needed. 
For storing single photons, prepared with $\eta_\text{her} = 0.22\,\%$ efficiency in a single mode memory, 
the SNR would have to be 0.71:1. In the dual-rail configuration it would have to be $3:1$. 
Given our current value for the SNR for single photons in a single mode memory of $0.3:1$ (eq. \ref{ch6_eq_NoiseFloor}), faithful storage of polarisation encoded quantum information would require an order of magnitude improvement in the SNR. 
Again, this roughly coincides with the necessary noise level reductions for conservation of the input photon statistics, discussed in section \ref{ch6_conclusion} (see fig.~\ref{fig_ch6_Rscan}). 
Such improvements are quite challenging, but not totally unrealistic. 
For instance, noise suppression using  an intra-cavity memory, an idea mentioned in section \ref{ch6_conclusion}, could already be sufficient.

\section{Conclusion\label{sec_ch4_conclusion}}

In this chapter, we have presented the storage of polarisation encoded information in the Raman quantum memory. The conducted experiments have employed bright coherent state input signals to evaluate the polarisation storage characteristics via quantum process tomography. 
The key achievements have been:

\begin{itemize}
\item Demonstration of one possibility to implement polarisation storage, using a dual-rail memory architecture. The simple design of this system allowed to add the polarisation storage capability to our pre-existing, single mode Raman memory\cite{Reim2010,Reim:2011ys} without significant experimental effort. 
The only challenge was thermal insulation of the interferometer to prevent air currents, excited by the heating of the {\cs} cell.

\item Storage of polarisation encoded information with a process purity of up to $\mathcal{P} \approx 83\,\% $ and a process fidelity of up $\mathcal{F} \approx 93\,\%$. These values were similar in consecutive read-out pulses. Furthermore, both parameters were constant throughout the memory lifetime, showing that signal loss due to decoherence did not affect the quality of the retrieved polarisation state. 

\item Investigation of the possibility to operate polarisation storage in the quantum regime. Here, the current noise floor of the memory has been proven too high for us to show quantum operation by just reducing the number of input photons down to the single photon level. 
To enable faithful quantum operation, the memory noise floor needs to be suppressed. 

\end{itemize}

\part{Storage of single photons}

\chapter{Heralded single photon source\label{ch5}}

\begin{flushright}
{\tiny \textfrak{Mephistopheles}: \,  \textfrak{Grau, teurer Freund, ist alle Theorie, Und grŸn des Lebens goldner Baum.} }
\end{flushright}

We now move to the storage of actual single photons in the Raman memory, for which we first need to assemble a source in order to produce the single photons input signals. 
For this purpose, we utilise heralding of the photon pair emission by spontaneous parametric down-conversion (SPDC). 
SPDC is an experimentally simple technology that enables convenient tailoring of the prepared single photons to the memory.
To present the implementation of this source, we start with a short state-of-the-art overview and introduce the  important metrics for seamless interfacing with the Raman memory. 
Thereafter, we go into the details of how to appropriately engineer the SPDC photons. 
Having determined the design parameters, we lay-out the experimental apparatus and the methodology for measuring single photons. We conclude with a characterisation of the source performance. 
We focus particularly on the spectrum, the heralding efficiency and the photon statistics of the produced single photons, which are the key performance determinants.

\section{Design criteria for a source - memory interface\label{ch5_sec_intro}}

\subsection{Challenges in interfacing a source with a memory\label{ch5_sec_intro}}

Single photon sources are one of the furthest developed building blocks for quantum networks\cite{Lounis:2005, Eisaman2011,Takeuchi:2014}. 
One of the crucial design concerns for sources is their interfacing with other components in such a network\cite{Brecht:2011}. 
While for passive quantum gates\cite{Knill:2001nx,OBrien:2003fk}, based on beam splitters and interferometers, the single photon mode structure is not a major experimental obstacle\cite{Moseley:2008,Brecht:2011}, it is important when single photons are to be used with active light matter interfaces, such as quantum memories\cite{Scholz:2009}. One of the major challenges for interfacing sources with memories is spectral bandwidth matching between the produced photons and the signal storage capabilities of the memory. \\
Amongst many promising source platforms, the nonlinear optical process of spontaneous parametric down conversion\cite{Hong:1986} (SPDC) has become a workhorse technology for single photon production.
Employing SPDC is an attractive possibility, but in the optical domain this can be tricky, because atom-light interfaces are usually narrowband, as they involve atomic resonances. 
Contrary, SPDC-based photon sources are broadband, since they are often operated in the parametric regime with pulsed pump lasers, whose large electric fields are required to obtain sufficient photon productions rates. 
Additionally, the wavelength regimes for optimal operation are different. 
While SPDC-based heralded single photon sources show best performances in the telecom range\cite{Eckstein:2011}, quantum memories operate mostly in the near-IR. 
In principle, this mismatch could be overcome with nonlinear frequency conversions techniques\cite{Fernandez-Gonzalvo:2013}, but only at the expense of increasing the system's signal-to-noise ratio. 
Narrowband single photon production with SPDC-based sources is possible using, for instance, intra-cavity designs\cite{Fekete:2013, Scholz:2009}. 
Because these add experimental complication, many experiments employ atom-based light-matter interfaces, either in the form of warm vapours\cite{Eisaman:2005}, laser cooled atoms or single atoms in cavities\cite{Muecke:2013,Specht:2011}. Here the atoms act as the storage medium as well as the generation medium for single photons\cite{Chaneliere:2005jh,Choi:2008aa, Sangouard:2008xr,Chen:2008fk,Zhang:2011aa}. 
By construction, these systems offer the benefit of built-in matching between a photon's spectral lineshape and the memory absorption line. 
Yet, similar to narrowband sources, the down-side of these systems is that they commonly rely on laser cooling, making them technologically complex and resource-demanding. \\
Looking at the parameters of suggested quantum memories, systems with spectral acceptance bandwidths large enough to enable direct interfacing with travelling-wave, pulsed SPDC sources have been demonstrated in solid state. The first example is a Raman memory in diamond\cite{England:2014, Lee:2012}, whose achievable storage times are, at present, too short to be useful in quantum networks. 
Second, rare-earth ion doped crystals, operated by the AFC protocol, have also shown GHz spectral acceptance bandwidths, allowing for direct single photon storage\cite{Saglamyurek:2014aa}. 
However, currently, on-demand storage and retrieval has not yet been achieved in the specific crystal type used in these experiments\footnote{
	Broadband storage has been shown in Tm-doped LiNbO$_3$ crystals, whereas on-demand memory 
	operation based on AFC was conducted in Pr- and Eu-doped YSiO$_5$ 
	crystals\cite{Timoney:2012, Jobez:2015,Guendogan:2015}. 
}. 
This is an important capability for any time-domain memory synchronisation tasks\cite{Nunn:2013ab} (see also section \ref{ch6_sec_intro}). Without it, the memory resembles a delay line of fixed length and could be replaced by an optical fibre. 

Warm vapour based Raman memories offer a compromise between the challenges on either side, thanks to the large time-bandwidth product of $B = \tau_\text{S} \cdot \Delta_\text{mem}$, which, in our system, reaches values on the order of\cite{Reim:2011ys} $B \gtrsim 1000$. 
On the one hand, the storage times in warm vapour cells, which can, in principle, be as long as minutes\cite{Heinze:2013aa, Balabas:2010}, are sufficient for these devices to have a realistic chance to be used for temporal synchronisation tasks\cite{Nunn:2013ab,Sanguard2011}.
On the other hand, the memory's acceptance bandwidth $\Delta_\text{mem} \sim 1 \GHz$ is broad enough to enable interfacing with travelling wave SPDC sources.  
The SPDC output photons still require spectral filtering to match the memory. 
But, because this can happen after the SPDC process\cite{Migdall:book}, a cavity is neither required for spectral narrowing, nor for boosting the emission rates. 
Building a light-matter interface with the Raman memory thus allows one to benefit from the advanced development of SPDC-based heralded single photons while maintaining technological simplicity.

\subsection{Requirements on the produced heralded single photons\label{ch5_subsec_design_criteria}}
To achieve good performance of the interfaced SPDC source - Raman memory system, we need to consider three aspects that dictate the source design. Here, we explain their origins qualitatively, before we discuss the actual source design in the next section: 

\paragraph{Photon number purity} 
Spontaneous parametric down conversion (SPDC) is a non-linear optical effect, where one pump photon is converted into a pair of lower frequency photons, termed signal and idler\cite{Mandel}. 
SPDC can thus be envisaged as the inverse process of second harmonic generation\cite{Yariv:hc,Trebino:book} (SHG). 
Since signal and idler are strictly produced in pairs\cite{Weinberg:1970}, detection of either photon heralds the presence of the other\cite{Hong:1986}. 
Therewith a single photon is produced\cite{Rarity:1987}. 
Importantly, the resulting state must, at most, contain one photon. 
This is characterised by the state's photon statistics\cite{Loudon:1993}, which shows anti-bunching, a property that can be evaluated with the second order autocorrelation function $g^{(2)}(\tau)$, with $g^{(2)}(\tau = 0)  = 0$ for a perfect single photon state\cite{Loudon:1993,Bocquillon:2009}. 
The increase of the pair production probability with SPDC pump power\cite{Eisenberg:2004} leads to the simultaneous generation of multiple photon pairs\cite{Wieczorek:2009kx}. 
Yet, multiple pair emission is undesirable for heralded single photon production, as it adds photons to the output state and elevates $g^{(2)}(0)$, which eventually results in the generation of a thermal state\cite{Eisenberg:2004,Loudon:1993,Yurke:1987}. 
Heralded single photons must thus be produced with good photon number purity, observable in $g^{(2)}(0) \rightarrow 0$. 

\paragraph{Spectral mode matching}
Apart from the resulting upper bound on the pump power, there is the necessity to generate sufficiently many single photons. 
The exact emission rates are determined by the source brightness, which depends on the spectral and spatial emission profile of the parametric fluorescence\cite{Kurtsiefer:2001,Kurtsiefer:2008}. 
With the pump pulse energy split between signal and idler, their respective frequency distributions result from phase-matching constraints that follow from the conservation of energy and momentum\cite{Grice:1997,Migdall:book}.
In the absence of any spectral filtering, travelling-wave SPDC sources commonly have phase-matching bandwidths in the high GHz to THz frequency range. 
To match the signal photon's bandwidth with the memory's acceptance bandwidth, frequency filtering on the idler photon can be used\cite{Moseley:2008,Branczyk:2010} (see appendix \ref{app_ch5_multimode_spdc_heralding}). 
As a result the number of heralding events reduces significantly, for which reason the unfiltered source needs a high brightness to begin with. 
Another source design criterion is thus set by the ability to filter the signal photon to the correct spectrum, accepted by the memory, while simultaneously achieving reasonable production rates.

\paragraph{Heralding efficiency}
The method of heralding defines the third design parameter. 
Since the presence of a photon in the signal mode can only be known with certainty upon detection of its idler counterpart, it is critical to actually deliver the signal photon to any device further downstream once its presence has been heralded. 
This efficiency of delivery is the heralding efficiency \hereff\., defined as 
${\eta_\text{her} = \alpha \cdot \frac{p_{\text{s}|\text{i}} \cdot p_\text{i}}{p_\text{i}} = \alpha \cdot p_{\text{s}|\text{i}}}$. 
Here, $p_{\text{s}|\text{i}}$ corresponds to the conditional probability for signal detection, given detection of an idler photon, and $p_\text{i}$ is the observation probability for an idler photon. 
Note, \hereff\, is independent of the idler photon detection rate. 
Experimentally, the signal is SMF-coupled, so \hereff\. is the probability for a signal photon to exit the SMF after idler photon detection. 
The constant $\alpha$ accounts for the below-unity optical transmission of the signal photon's beam path\footnote{
	Note, this beam path starts at the location in the nonlinear material, where the SPDC photon pair 
	is generated and includes SMF-coupling and propagation in SMF as well as free space.
} (see also eq. \ref{eq_ch5_her_eff}). 
This sets tight tolerances on the optical losses in the signal path and on the signal's spatial mode quality, as it needs to be collected into SMF with high efficiency. 
Additionally, false heralding signals need to be avoided, since these artificially reduce \hereff\,. 
In the idler mode, low single photon fluorescence noise and detection with low dark counts are thus desirable. Overall we aim to achieve an $\eta_\text{her}$ as high as possible.

\section{Source design and expected performance\label{ch4_source_design_theory}}

From these requirements, we can devise the SPDC source set-up. 
In the following, we first describe our choice for the nonlinear medium. Thereafter we explain how to match the central frequency of single photons with the input signal channel of the memory. 
To this end, we introduce a description of the SPDC in the spectral and temporal domain. This does not only allow us to find the required filtering for the idler photons, but also helps in the source performance characterisation later on.

\subsection{Choice of the nonlinear medium\label{ch5_subsec_intro_waveguide}}

The frequency conversion relies on the second-order nonlinear susceptibility tensor, 
found in non centro-symmetric salt crystals\cite{Boyd:2003nx}.
SPDC has been realised with a variety of these media\cite{HandbookNonlinearCrystals} either in form of bulk crystals or waveguide chips. 
The advantage of waveguide systems is their well defined spatial mode structure\cite{Moseley:2009,Karpinski:2012} and high brightness\cite{Fiorentino:2007} at modest pumping powers\footnote{
	This argument depends on how well the pump can be coupled into the waveguide, which 
	is often neglected as it is wavelength and waveguide design dependent. 
}.
These points make the waveguide our system of choice; on the one hand, we obtain relatively easy spatial collection of the SPDC photons, and, on the other hand, the use of modest pump powers allows us to derive the SPDC pump and the memory control pulses from the same laser source, which simplifies experimental complexity and cuts cost{\footnote{
	{
	High heralding efficiency with good mode matching to SMF and 
	high brightness can also be achieved 
	in bulk crystals\cite{Ramelow:2013} with appropriate pump focussing 
	parameters\cite{Fedrizzi:2007}. 
	Waveguides however make the handling of long crystals easier. These are needed for long interaction 
	times and narrow SPDC phase-matching bandwidths, which are desirable for our relatively narrow-band 
	memory acceptance bandwidth of $\Delta_\text{mem}\sim 1\GHz$.
	}
}}. 

\begin{figure}[h!]
\centering
\includegraphics[width=\textwidth]{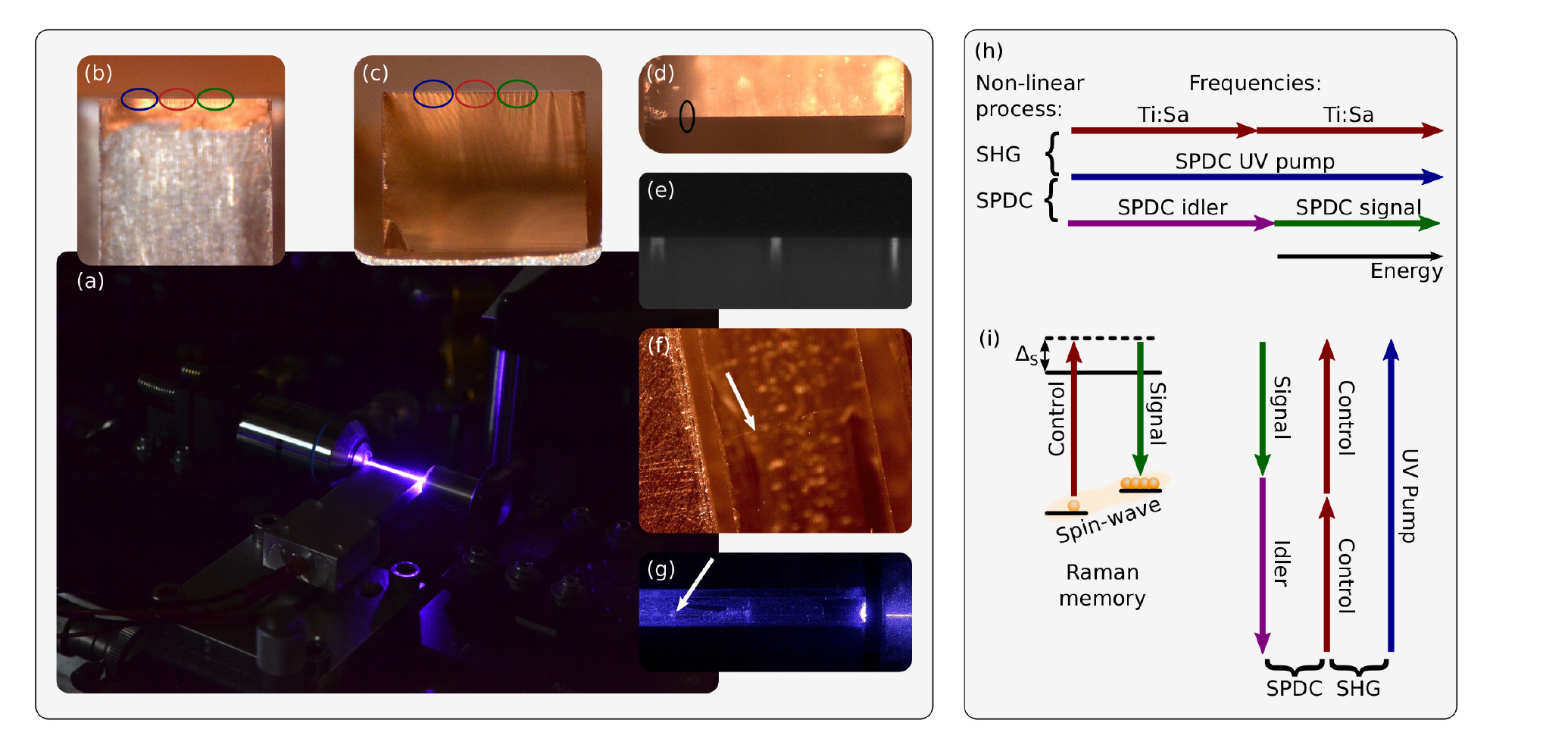}
\caption{\textbf{(a)}: Setup photograph showing the waveguide mount with coupling optics (\textit{photo taken by Annemarie Holleczek}).
\textbf{(b)} - \textbf{(d)}: Waveguide chip top- and edge-view with marked channels. \textit{Blue}, \textit{red} and 
\textit{green circles} mark the three channel families, 
each containing $6$ guides.
\textbf{(e)}: Waveguide end facet showing a set of channels with $4\mum$, $3\mum$ and $2\mum$ widths (left to right). The central guide ($3.2$) is used in our experiment. 
\textbf{(f)}: Top view; the \textit{arrow} marks a scratch running over the chip's surface (see section \ref{ch5_sec_setup} and appendix \ref{app_ch5_insufficiencies}).
\textbf{(g)}: Photograph of light transmitted through the waveguide. The \textit{arrow} marks the position of a scratch, where some of the transmitted light is scattered. 
\textbf{(h)}: Energy conservation between the {\tisa} master laser (memory control and fundamental for SPDC pump via SHG) and the signal and idler photons in the SPDC pair.
\textbf{(i)}: Selection of the SPDC central frequencies, showing the preparation of SPDC photons at the Raman memory's signal frequency by appropriate selection of the idler frequency. 
}
\label{fig_ch5_intro}
\end{figure}

For the nonlinear material, potassium titanyl phosphate\cite{Bierlein:1989} (KTP) is chosen, thanks to its large nonlinear coefficients\cite{Vanherzeele:1992}. 
We implement type-II SPDC\cite{Kwiat:1995}, using the coefficient\cite{Boyd:2003nx} ${d_{24} \approx 3.92}$,  
which results in orthogonally polarised signal and idler photons\footnote{
	The larger coefficients\cite{Bierlein:1989} of $d_{33} \approx 18.5$ and 
	$d_{32} \approx 4.7$ lead to type-I SPDC.
	To generate the UV pump for the SPDC, we use type-I SHG in ppKTP under 
	utilisation of $d_{33}$ (see \ref{ch3_SHG}).
}.
With regard to the large electric fields, resulting from light confinement in the waveguide channels, an additional advantage of KTP is its high damage threshold. 
KTP is produced either by hydrothermal or flux growth\cite{Satyanarayan:1999}. We use flux grown KTP, which suffers less from the dominant damage mechanism of grey tracking\cite{Boulanger:1994}. 
The waveguide chip is produced by ion exchange: 
channels are written onto the top surface of a KTP substrate by replacing titanium (Ti) with rubidium (Rb) ions\cite{Bierlein:1987},  
raising the refractive index from $n_\text{KPT} = 1.83$ to $n_\text{wg} \approx 1.84$. 
The index contrast on the channel boundaries causes guiding of light in the channels. 
These have a tapered, error-function-shaped depth profile\cite{Roelofs:1994} (see fig. \ref{fig_ch5_theory_modes}), resulting from the Rb diffusion into the KTP substrate. 
Fig. \ref{fig_ch5_intro} shows the waveguide channels on top of our $L_\text{KTP} = 2\cm$ long chip, as well as the end facets with the observed channel profiles. 

The entire chip hosts three families of waveguides. Each family contains $6$ guides, spaced by $35\mum$, whose channel widths $w_\text{wg} = \left\{ 2,3,4,2,3,4 \right\} \mum$ vary in sawtooth order.  
As we will discuss shortly, SPDC is pumped with light at $\lambda_\text{p} =  426\nm$ wavelength to produce signal and idler photons at $\lambda_\text{SPDC} = 852\nm$ wavelength. 
To achieve phase-matching for SPDC at these wavelengths, the channels are periodically-poled\cite{Eger:1997} with $\Lambda_\text{KTP} \approx 10.4 \mum$ periodicity. 
Exact phase-matching is obtained by temperature tuning (see appendix \ref{ch5_subsec_SPDC_temp_tuning}). 
All waveguides can support multiple transverse modes\footnote{
	At present, commercially available ppKTP waveguides support multiple transverse modes in 
	the near-IR or the UV regime\cite{Moseley:2009,Christ:2009}. 
	Such waveguides are, to our knowledge, exclusively produced by AdvR. 
	At the time of ordering the waveguide chips, AdvR was producing first test samples for
	single mode operation around $852\nm$, whose performance parameters however could
	not be guaranteed.	
} at both wavelengths\cite{Moseley:2009,Karpinski:2009}. 
We can obtain single mode operation for the SPDC pump, as well as for the SPDC signal and idler photons by appropriately choosing the coupling conditions for the pump (see section \ref{ch5_sec_modes} and appendix \ref{app_ch5_spdc_modes}). 
The optical facets of our waveguide chip are not AR-coated for the two wavelengths $\lambda_\text{SPDC}$ and $\lambda_\text{p}$, leading to a residual Fresnel reflection upon SPDC pair extraction from the guide\footnote{ 
	This feature has been omitted when the chip was originally ordered from the 
	manufacturer \textit{AdvR}. When funds became available in 2013, I have ordered a 
	$2^\text{nd}$, coated waveguide, which however did not arrive in time to be used for 
	my work. Unfortunately, reliable single mode waveguides at $852\nm$ were still not available 
	commercially at the time, for which reason this $2^\text{nd}$ chip has the same 
	characteristics as the one used in this work.
}.

\subsection{Selection of the SPDC pair's central frequencies\label{ch5_subsec_intro_principle}}

To produce heralded single photons (\hsps\,) for storage in the Raman memory, their frequency distribution has to match the signal leg in the memory's $\Lambda$-system (fig. \ref{fig_ch2_Raman_protocol}). 
This entails matching of the \hsps\,' central frequency and their spectral bandwidth\footnote{
	We assume the frequency distribution is entirely described by its first (central frequency) and second 
	moment (bandwidth), i.e., we ignore higher order spectral phases, such as chirp, 
	which leads to shape distortions of the frequency distribution, e.g. kurtosis. 
	Note also, whenever we refer to the width of the heralded single photon spectrum, 
	we mean its spectral bandwidth.  
}. 
With the pump photon's energy split between signal and idler photons in SPDC, the respective distributions of the signal and idler frequencies $\nu_\text{s}$ and $\nu_\text{i}$ result from phase-matching constraints. 
Both distributions build a joint probability space, defined by the joint-spectral-amplitude\cite{Grice:1997} 
$f(\nu_\text{s},\nu_\text{i})$ (see section \ref{ch5_sec_hsp_bandwidth}), whereby signal and idler frequencies must always add up to the pump frequency $\nu_\text{p} = \nu_\text{s} + \nu_\text{i}$ (fig. \ref{fig_ch5_intro} \textbf{h}).
In turn, this allows to select a specific subset of the signal photon's frequency distribution by appropriate filtering of the idler photon\cite{Branczyk:2010}. 
Idler detection after filtering projects the signal frequency into the desired marginal spectrum, which is the core idea behind the spectral engineering we employ here. 

We first deal with the simpler task of choosing the correct central frequencies. 
For the SPDC photons to match the {\cs} D$_2$-line, the SPDC photons should be emitted at 
$\lambda_\text{SPDC} = 852\,\nm$ central wavelength. 
Since SPDC splits the energy of a pump photon between the two SPDC photons, we operate SPDC in degeneracy and require a pump at $\lambda_\text{p} = 426\,\nm$ wavelength.
Experimentally, we will use our {\tisa} laser system at $\lambda_\text{Ti:Sa} =  852\nm$ wavelength\cite{Reim2010} to produce the pump via SHG (see section \ref{ch5_sec_setup} and appendix \ref{ch3}). 
Because the SPDC photons result from of a sequence of frequency up- and down-conversion of the master laser's output, their central frequencies $\nu_\text{s}$ and idler $\nu_\text{i}$ equal the {\tisa} central frequency $\nu_\text{Ti:Sa}$. With $\lambda_\text{Ti:Sa} = 852\,\nm$, we thus automatically match the SPDC pair's central wavelength with the {\cs} D$_2$-line. 

Since we will use the same laser system also for control pulse generation (see chapter \ref{ch6}), it is actually blue-detuned by $\Delta = 15.2\,\GHz$ (see fig. \ref{fig_ch2_FWMlevels}). 
In turn, this also results in the same detuning for the central frequencies of both SPDC photons. 
Consequently, we need some frequency fine tuning for the \hsps\, to match the signal frequency in the memory's $\Lambda$-system. 
As fig. \ref{fig_ch5_intro} \textbf{i} shows, we use the comparatively broad phase-matching bandwidth of the SPDC process, calculated in section \ref{ch5_sec_hsp_bandwidth} below. It allows to select slightly non-degenerate signal and idler frequencies without affecting the \hsp\, production rates.  
The signal's frequency must be lowered by the {\cs} hyperfine ground state splitting 
$\delta \nu_\text{gs} = 9.2\GHz$ with respect to the control frequency, which equals the {\tisa} output at 
$\nu_\text{Ti:Sa}$. 
For this reason, the central frequency of the SPDC signal's marginal spectrum ($\nu_{\text{s},0}$) must have a detuning of $\Delta_\text{s} = \Delta - \delta \nu_\text{gs} = 6\GHz$ from the 
${6^2 \text{S}_\frac{1}{2}, \text{F} = 3 \rightarrow 6^2\text{P}_\frac{3}{2}}$ transition. 
Fig. \ref{fig_ch5_intro} \textbf{i} illustrates how we can filter the idler photon to a detuning, shifted in the opposite direction, to meet this requirement: the idler's centre frequency $\nu_{\text{i},0}$ is filtered to a detuning of $\nu_{\text{i},0} = \Delta + \delta \nu_\text{gs} = 24.4 \GHz$. 

Besides projecting the \hsps\, onto the correct central frequency, their marginal spectrum must also match the memory acceptance bandwidth $\Delta_\text{mem} \sim 1\GHz$. We facilitate this by appropriate choice of the idler filter's spectral bandwidth. 
Bandwidth selection is more subtle and requires knowledge of the actual joint-spectral amplitude $f(\nu_\text{s},\nu_\text{i})$. 
We introduce $f(\nu_\text{s},\nu_\text{i})$ via the frequency domain description of the SPDC process, but beforehand we require to look at the expected waveguide spatial mode structure to obtain the mode refractive index parameters going into the phase-matching function.

\subsection{Expected spatial mode structure in the waveguide\label{ch5_subsec_exp_modes}}

\paragraph{Importance of the spatial modes}
The spatial mode structure is one of the assets, but also one of the challenges in operating SPDC in a  waveguide. 
For nonlinear frequency conversion in the waveguide, the spatial mode of the frequency-converted light is dependent on the spatial mode of the pump beam\cite{Christ:2009,Moseley:2009,Karpinski:2009}. 
In case of SPDC, pump, signal and idler form mode triplets\cite{Christ:2009}, which can be controlled 
by the coupling conditions of the pump into the waveguide\cite{Bierlein:1987,Roelofs:1994,Moseley:2009,Karpinski:2009}. 
The frequency conversion efficiency depends on the respective mode triplets, whereby the highest efficiencies are obtained when all three fields occupy the fundamental mode\cite{Karpinski:2012}. 
Since the fundamental mode also has the largest overlap with the mode of a SMF, we clearly desire to couple the pump beams into it. 
Using a microscope objective as the input coupler, good control over the waveguide mode structure is possible\cite{Karpinski:2009}. 

\begin{figure}[h!]
\centering
\includegraphics[width=\textwidth]{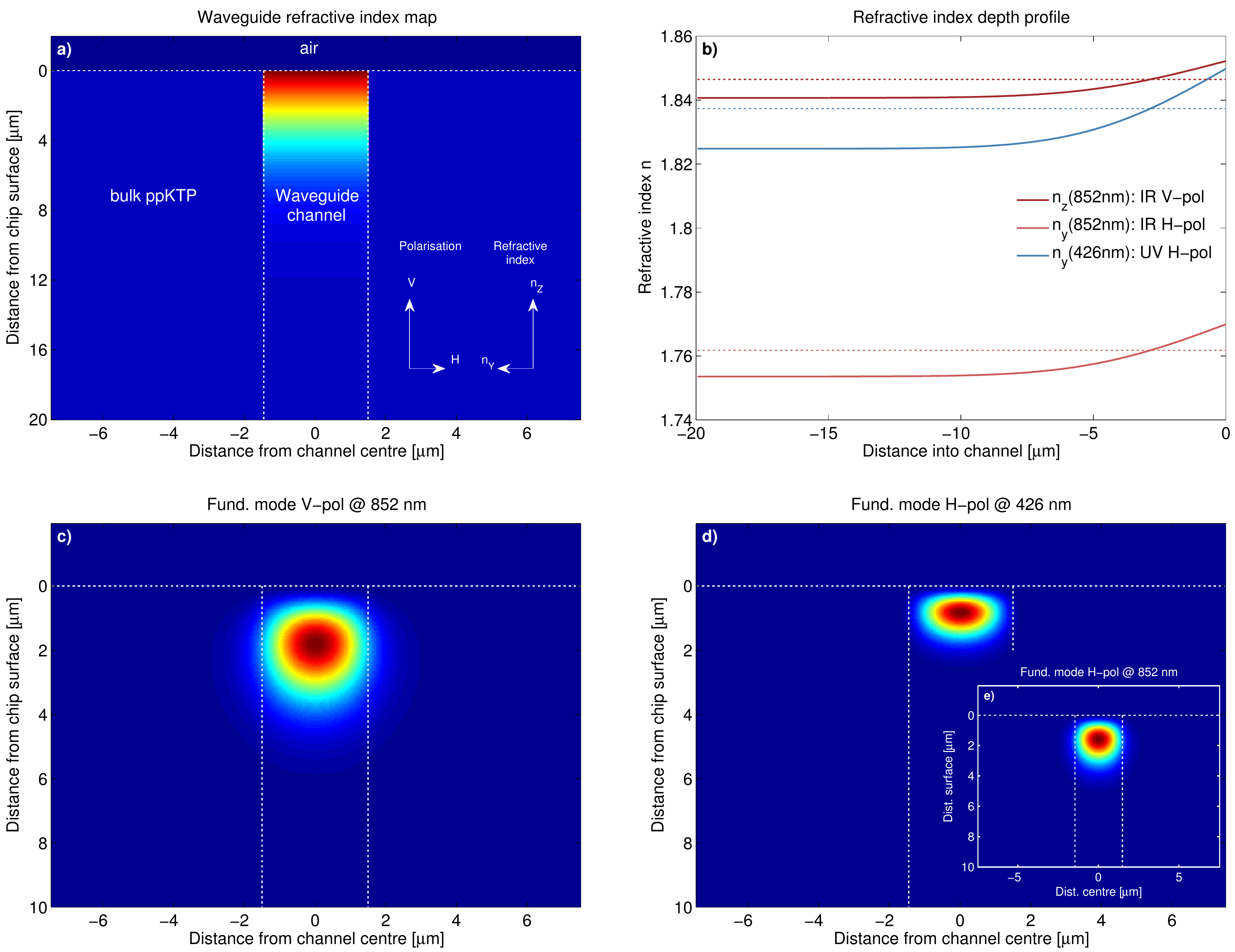}
\caption{Expected waveguide modes by finite-element simulation, \textit{white dotted lines} indicate the waveguide channel boundaries. 
\textbf{(a)}: Transverse refractive index map of the waveguide channel and its surroundings. Increasingly \textit{red colouring} indicates an increase in the refractive index. The coordinate systems illustrate the orientation of the index ellipsoid with respect to the optical polarisations.  
\textbf{(b)}: Refractive index depth profile going into the waveguide channel for all $3$ optical fields involved in SPDC. The index decrease follows an error-function, with \textit{dotted lines} indicating $n^\text{guide}(\tilde{y}) =    n^\text{bulk} + \frac{\Delta n^\text{surf}}{2}$, where the guide's index reduces to half the value at the channel surface. 
\textbf{(c)} \& \textbf{(e)}: Fundamental modes for guided light at $852\nm$ wavelength, showing V- and H- polarisation, respectively. For the former, the light polarisation is perpendicular to the waveguide surface. 
H- and V-polarised modes corresponds to the SPDC signal and idler photons, respectively. 
\textbf{(d)}: Fundamental mode for guided light at $426\nm$ wavelength.
}
\label{fig_ch5_theory_modes}
\end{figure}

\paragraph{Mode simulation}
To determine the required coupling optics and to predict the mode index $n^\text{mode}_j$, for each mode $j$, 
we calculate the fundamental set of modes to be expected for the waveguide channel we use in the actual experiment. It has a width of $3\,\mum$ and a depth of $6\,\mum$ (see section \ref{ch5_sec_setup}).  
The method is based on the work of \textit{Fallahkhair et. al}\cite{Fallahkhair:2008}. 
For the simulation an open source \textit{MatLab} code was used\footnote{
	The \textit{MatLab} library package \textit{waveguidemodesolver} is freely available on \textit{MatLab} 
	File Exchange. Our calculations use the semi-vectorial mode solver supplied with the package.
}. 
The employed algorithm determines the electric field distribution and the mode indices $n^\text{mode}_j$ in a customisable dielectric structure that is supplied as a 2-D map of the transverse dielectric tensor $\epsilon(x,y) = n(x,y)^2$. 
The tensor describes the refractive index profile ($n(x,y)$) of a transverse cut through the waveguide's $x$-$y$-plane. 
The map used here is shown in fig. \ref{fig_ch5_theory_modes} \textbf{a}\footnote{
	The \textit{MatLab} code for producing this map was supplied by Michal Karpinksi.
	It generates the same waveguide structure that was used by 
	\textit{Karpinski et. al.}\cite{Karpinski:2009,Karpinski:2012} 
	for analysing the frequency conversion efficiencies of different spatial modes in a ppKTP waveguide.
}. 
For the simulation, it is important to consider the correct geometry of the refractive index ellipsoid with respect to the waveguide structure. 
In real space, the optical axis is considered as the $z$-direction and the waveguide chip face corresponds to the $x$-$y$-plane, whereby the $x$-direction runs parallel the chip surface (fig. \ref{fig_ch5_theory_modes} \textbf{a}). 
Horizontally (H) and vertically (V) polarised light have electric field vectors pointing in the $x$- and $y$-direction, respectively. 
The refractive index ellipsoid\footnote{
	KTP is a negative, bi-axial crystal.
} is oriented such, that its axes $\left\{ n_x, n_y, n_z \right\}$ lie along the spatial dimensions $\left\{z,x,y \right\}$ (in the given order). Consequently, H-polarised light experiences the index $n_y$, whereas V-polarised light is subject to $n_z$ (fig. \ref{fig_ch5_theory_modes} \textbf{a}). 
The wavelength dependence of the refractive index is modelled using the Sellmeier coefficients\footnote{
	With the wavelength $\lambda$ in units of $\mu$m, these are:
	\begin{align}
	\label{eq_ch5_sellmeier_1}
	n_x^2 &= 3.29100 + \frac{0.04140}{\lambda^2 - 0.03978} + \frac{9.35522}{\lambda^2 - 31.45571}	\\ 
	\label{eq_ch5_sellmeier_2}	
	n_y^2 &= 3.45018 + \frac{0.04341}{\lambda^2 - 0.04597} + \frac{16.98825}{\lambda^2 - 39.43799}	\\ 
	\label{eq_ch5_sellmeier_3}
	n_z^2 &= 4.59423 + \frac{0.06206}{\lambda^2 - 0.04763} + \frac{110.80672}{\lambda^2 - 86.12171} 	
	\end{align}
} 
for bulk KTP ($n^\text{bulk}$), quoted by \textit{Kato et. al.}\cite{Kato:2002}. 
For the UV and IR pulses, these are 
$n^\text{bulk}_{y}(426\nm)= 1.8248$, $n^\text{bulk}_{y}(852\nm) = 1.7536$, and $n^\text{bulk}_{z}(852\nm) = 1.8407$.

The refractive index profile, reaching into the waveguide channel ($y$-dimension), is described by the error-function\cite{Roelofs:1994} ($\text{erfc}(y)$), with 
$n^\text{guide}(y) = n^\text{bulk} + \Delta n^\text{surf} \cdot \text{erfc}(y/d)$,
where the depth parameter $d=6\mum$ has been set to the effective channel depth and $\Delta n^\text{surf}$ is the refractive index contrast between waveguide channel and cladding at the chip's surface. 
The Rb ion exchange raises the refractive index of the channels by 
$\Delta n^\text{surf}_{y}(426\nm) = 0.0281$, 
$\Delta n^\text{surf}_{y}(852\nm) = 0.0250$ and 
$\Delta n^\text{surf}_{z}(852\nm) = 0.0191$ with respect to the bulk ppKTP cladding. 
The channel is surrounded by the ppKTP cladding at the bulk index level on 3 sides and by air ($n^\text{air} = 1$) above the surface. 
Figure \ref{fig_ch5_theory_modes} \textbf{a} shows the vertically decreasing index contrast of the resulting refractive index map. 
The depth dependence of the channel index $n^\text{guide}(y)$ 
is displayed in fig. \ref{fig_ch5_theory_modes} \textbf{b} for the three fields involved. 
Dotted lines indicate the refractive index and the depth $\tilde{y} = 2.9\mum$ at which the index contrast halves.
The structure is capable to support $3$ guided modes at $852\,\nm$ and $2$ modes at $426\,\nm$. 
The transverse intensity distributions $I_j(x,y)$ for the fundamental modes are shown in fig. \ref{fig_ch5_theory_modes} \textbf{c} - \textbf{e} for all three fields involved in SPDC. 
While the mode shape is approximately spherical in the IR, it becomes highly elliptical in the UV with the mode eccentricity\footnote{
	The eccentricity is defined in terms of the FWHM of the mode's intensity distribution. Defining 
	$a = \text{FWHM}_\text{max}/2$  
	and 
	$b = \text{FWHM}_\text{min}/2$ 
	as the larger and smaller mode radii in the horizontal and 
	vertical direction, respectively, the eccentricity is defined by:
	$\epsilon = \sqrt{\frac{a^2 - b^2}{a^2}}$.
} increasing to $\epsilon_{\text{UV,H}} = 0.79$, compared to $\epsilon_{\text{IR,V}} = 0$ and 
$\epsilon_{\text{IR,H}} = 0.3$ in the IR. 
Table \ref{tab_ch5_mode_sizes_exp} states the FWHM diameters of these modes.  
Supplying the $426\,\nm$ SPDC pump by SMF (see fig. \ref{fig_ch5_setup} \textbf{a}) gives it an approximately circular, Gaussian mode shape at the waveguide input. 
Since no beam shaping is applied, significant spatial mode mismatch can be expected. 
Mismatch does not only degrade the coupling efficiency, but also makes it harder to excite solely the fundamental mode with the UV pump. 
The mode indices $n^\text{mode}_j$ are also quoted in table \ref{tab_ch5_mode_sizes_exp}.

\begin{table}[h!]
\centering
\begin{tabular}{c|c|c|c|c}
\toprule
$\lambda$ [nm] & Polarisation 	& $n^\text{mode}$ &	\multicolumn{2}{c}{FWHM of simulates modes}\\
		&  	& 	& FWHM hor. [$\mum$] & FWHM ver. [$\mum$]\\
\midrule
$852$ & V-pol 		& $1.8439$ & $2.4$		&	$2.4$	\\
$852$ & H-pol 		& $1.7594$ & $2.2$ 		&	$2.1$	\\
$426$ & H-pol		& $1.8421$ & $1.7$		&	$1.05$	\\
\bottomrule
\end{tabular}
\caption{Expected modes sizes and mode indices $n^\text{mode}_j$ obtained by finite-element simulation. 
Note that the modes at $852\,\nm$ apply to the generated SPDC photons as well as $852\,\nm$ radiation, 
coupled into the waveguide to generate SHG (see appendix \ref{app_ch5_SHGwaveguide}).}
\label{tab_ch5_mode_sizes_exp}
\end{table}

\subsection{SPDC in the spectral domain\label{ch5_sec_hsp_bandwidth}}	

Moving forward towards finding the filter bandwidth, required to match the \hsps\, to the Raman memory's spectral acceptance bandwidth, we now introduce the necessary expressions for the SPDC spectral output. 
For the Raman memory, the spectral acceptance bandwidth effectively corresponds to the control pulse spectral bandwidth. 
So, coarsely speaking, the SPDC signal photon spectrum, which is initially broadband, due to the large phase-matching bandwidth of the SPDC process\cite{Grice:1997}, needs to be reduced to $\Delta \nu_\text{s} \sim 1\GHz$. 
Similar to selecting the correct central frequencies, the required spectral engineering\cite{Brecht:2011} is implemented by filtering the idler photon, relying on the spectral correlations of the SPDC photon pair. 

\paragraph{Joint spectral amplitude}
The correlation between the spectra of SPDC signal and idler photons can be best understood in terms of the joint spectral amplitude (JSA) of the combined SPDC state\cite{Grice:1997}. 
In general, the JSA is the probability amplitude $f(\nu_\text{s},\nu_\text{i})$ for the observation of a given signal ($\nu_\text{s} = 2\pi\omega_\text{s}$) and idler ($\nu_\text{i}=2\pi\omega_\text{i}$) frequency combination in the SPDC pair. 
Its exact functional dependence can be derived from the Heisenberg equations of motion\cite{Grice:1997}, where the Hamiltonian describes the energy density of the electromagnetic field in the nonlinear medium. 
Since this derivation can be found in most doctoral theses involving SPDC, its formal derivation is skipped here\footnote{ 
	For SPDC in ppKTP waveguide, the theses of \textit{Andreas Christ}\cite{Christ:PhD} and 
	\textit{Alfred U'Ren}\cite{URen:PhD}, as well as \textit{Alan Migdall}'s book 
	\textit{Single photon generation and detection}\cite{Migdall:book} provide good derivations.
}. 
Instead, we straight away use one of the main result: the JSA can be written as the product\cite{Grice:1997,URen:PhD,Moseley:PhD}
\begin{equation}
f(\nu_\text{s},\nu_\text{i}) = \alpha(\nu_\text{p}) \cdot \Phi(\nu_\text{s},\nu_\text{i}) = \alpha(\nu_\text{s} + \nu_\text{i}) \cdot \sinc\left( \frac{\Delta k(\nu_\text{s},\nu_\text{i})}{2} L_\text{KTP} \right)
\label{eq_ch5_jsa}
\end{equation}
between the normalised spectral electric field envelope of the SPDC pump 
$\alpha(\nu_\text{p})$, and the phase-matching function 
$\Phi(\nu_\text{s},\nu_\text{i}) = \sinc\left( \frac{\Delta k(\nu_\text{s},\nu_\text{i})}{2} L_\text{KTP} \right)$ 
for the ppKTP waveguide of length ${L_\text{KTP} = 20\mm}$. 
For $\alpha(\nu_\text{p})$, energy conservation during SPDC as been used to decompose the pump frequency $\nu_\text{p} =\nu_\text{s} + \nu_\text{i}$ into the sum of signal and idler frequencies. 
Therewith, the generated SPDC pair is denoted by the quantum state\cite{URen:PhD,Moseley:PhD} 
\begin{equation}
\ket{\psi}_\text{SPDC} = \ket{0} + B \underset{\nu_\text{s}}{\int} \underset{\nu_\text{i}}{\int} \text{d} \tilde{\nu}_\text{s} \text{d} \tilde{\nu}_\text{i} 
f(\tilde{\nu}_\text{s},\tilde{\nu}_\text{i}) \hat{a}^\dagger_\text{s}  \hat{a}^\dagger_\text{i} \ket{0} + \mathcal{O}(B^2) + \dots,
\label{eq_ch5_spdc_state}
\end{equation}
whose creation operators $\hat{a}^\dagger_\text{s}$ and $\hat{a}^\dagger_\text{i}$ generate one signal and idler photon from the vacuum, respectively. The constant $B \sim d_{24} L_\text{KTP} \tilde{E}_\text{P,0}$ contains the proportionality to the effective nonlinearity $d_{24}$ of the medium and the dependence on the UV pump power $P_\text{UV}$ via the peak electric field amplitude of the pump spectrum 
$\tilde{E}_\text{P}(\nu_\text{p}) \sim \sqrt{P_\text{UV}}$. 
The higher order terms $\mathcal{O}(B^2) + \dots$ correspond to the simultaneous emission of two and more SPDC pairs\cite{Krischek:2010ys, Eisenberg:2004}. 
The number of observable signal and idler photons $\langle \psi_\text{SPDC} | \hat{n}_\text{s/i} | \psi_\text{SPDC} \rangle$, with photon number operator 
${\hat{n}_\text{s/i}=\hat{a}_\text{s/i}^\dagger \hat{a}_\text{s/i}}$, 
is proportional to the absolute square of the terms in eq. \ref{eq_ch5_jsa}. 
The experimentally observable signal and idler frequency distribution is hence set by the joint spectral intensity (JSI), defined as $i(\nu_s,\nu_i) = |f(\nu_s,\nu_i)|^2$. 

\paragraph{Phase-matching function}
While energy conservation during the frequency conversion process is accounted for by the pump envelope function $\alpha(\nu_\text{p})$, the conservation of momentum enters eq. \ref{eq_ch5_jsa} through the phase matching function $\Phi(\nu_\text{s},\nu_\text{i}) = \sinc\left( \frac{\Delta k}{2} L_\text{KTP} \right)$, whose sinc-function dependence originates from the finite interaction volume between the three fields\footnote{
	For the ppKTP waveguide, this corresponds to the integration of the wavevector difference between the 
	three fields along the chip length $L_\text{KTP}$:
	$
	{\overset{L_\text{KTP}/2} {\underset{-L_\text{KTP}/2}{\int}}
	\exp{\left\{-i \left( k_i(\omega_i)+k_s(\omega_s) - k_p(\omega_p) \right) z \right\}} \text{d} z }
	$	
}.
Here, $\Delta k$ represents the phase mismatch
\begin{equation}
\Delta k = \vec{k}_\text{p} - \vec{k}_\text{s} - \vec{k}_\text{i} - \vec{q} = \frac{2 \pi \cdot n_{y}(\lambda_p)}{\lambda_p} -  \frac{2 \pi \cdot n_{y}(\lambda_s)}{\lambda_s} - \frac{2 \pi \cdot n_{z}(\lambda_i)}{\lambda_i}  - \frac{2 \pi}{\Lambda}, 
\label{eq_ch5_phase_mismatch}
\end{equation} 
with wavevectors $k_j= \frac{n_j(\lambda_j) \omega_j}{c}$, wavelengths $\lambda_j=\frac{2\pi c}{\omega_j}$ and refractive indices $n_j$ for the three fields $j \in \left\{ \text{p}, \text{s}, \text{i}\right\}$. 
The constant $\vec{q}$ is the lowest order Fourier component of the refractive index grating, generated by the periodic poling of the KTP chip with a poling period of $\Lambda = 10.4\mum$. 
Since all beams are collinear in the waveguide, critical phase-matching\cite{Boyd:2003nx} by angle tuning of the three optical fields with respect to the nonlinear crystal's axes is not possible. The phase mismatch can only be compensated by the poling period, the appropriate polarisation choice of the optical fields and the crystal temperature, which can change $n_j(\lambda_j)$ by small amounts\cite{Kato:2002}. 
Here, the refractive indices $n_j(\lambda_j) = n_j(\lambda_j)^\text{bulk} + n_j^\text{mode}$ contain the bulk KTP crystal values, given by the Sellmeier equations\cite{Kato:2002} (see eqs. \ref{eq_ch5_sellmeier_1} - \ref{eq_ch5_sellmeier_3}), and the offset $n_j^\text{mode}$, determined by the spatial mode (see table \ref{tab_ch5_mode_sizes_exp}). 
We only consider SPDC in the fundamental waveguide mode triplet\cite{Christ:2009}, with H-polarised pump and SPDC signal photons and V-polarised SPDC idler photons, as shown in fig. \ref{fig_ch5_theory_modes}. 
While $n^\text{mode}_j$ is, in general, wavelength dependent, it is pretty much constant for our narrow spectral regions. Thermal expansion of the waveguide is also negligible.

\begin{figure}[h!]
\centering
\includegraphics[width=0.95\textwidth]{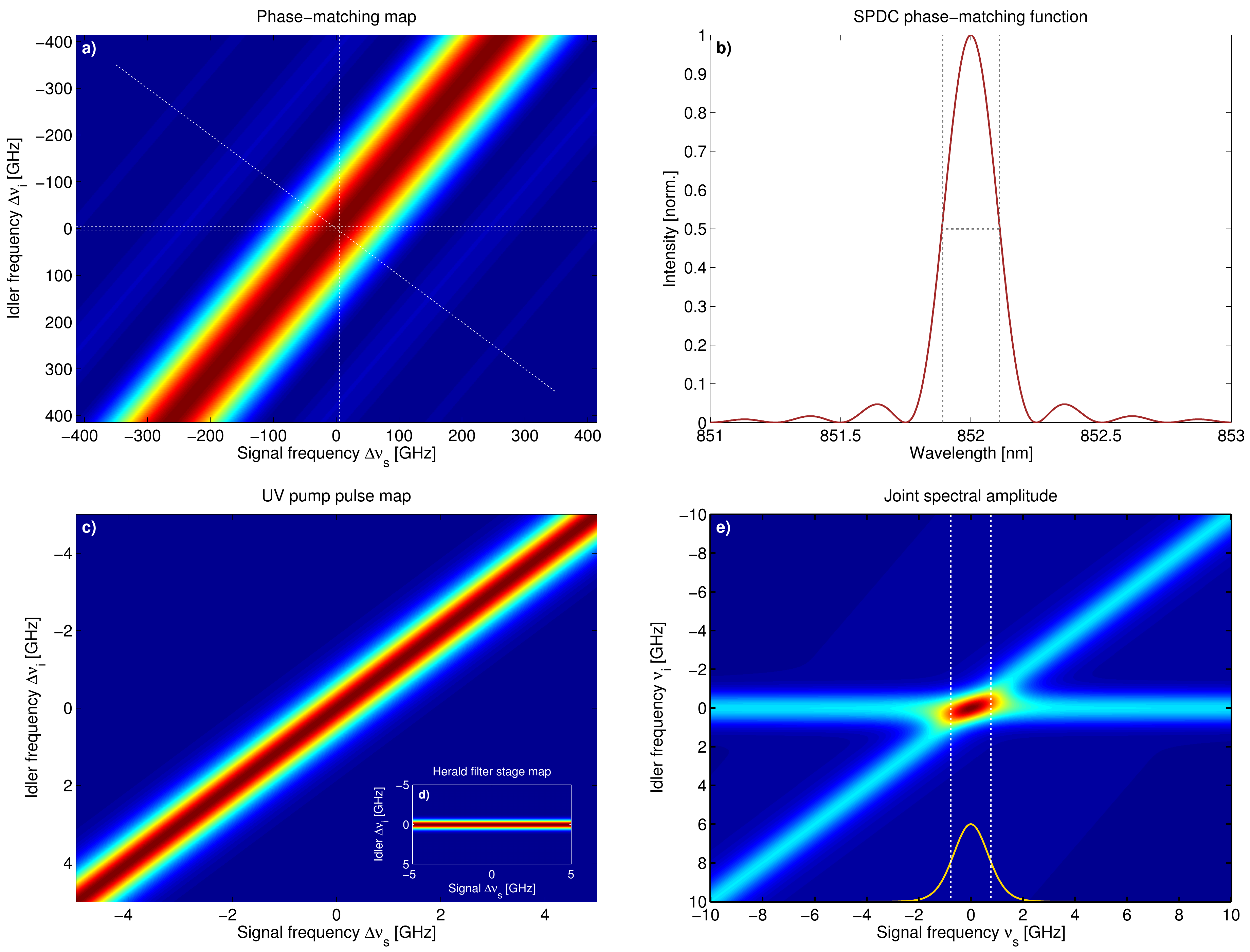}
\caption{
\textbf{(a)}: Phase matching map $\abs{\Phi(\nu_\text{s},\nu_\text{i})}^2$. 
\textit{Vertical} and \textit{horizontal lines} mark a frequency range of $\Delta \nu = \pm 5\GHz$, the \textit{diagonal line} is the cutting plane for determining the phase-matching bandwidth. 
\textbf{(b)}: Phase matching function in the cutting plane of \textbf{(a)}. \textit{Vertical} and \textit{horizontal lines} mark the FWHM bandwidth. 
\textbf{(c)}: Map of the pump envelope function $\abs{\alpha(\nu_\text{s},\nu_\text{i})}^2$. 
\textbf{(d)}: Map of the herald filter stage intensity transmission function $T_\text{filt}^\text{idl}(\nu_\text{s},\nu_\text{i})$ in $\nu_\text{s}$-$\nu_\text{i}$-space. 
\textbf{(e)}: JSI $i(\nu_\text{s},\nu_\text{i})$ of the SPDC pair after idler filtering. The pump envelope 
$\abs{\alpha(\nu_\text{s},\nu_\text{i})}^2$ and the idler filter 
$T_\text{filt}^\text{idl}(\nu_\text{s},\nu_\text{i})$ 
are shown in \textit{light blue colour}, with the phase-matching function $|\Phi(\nu_\text{i},\nu_\text{s})|^2$ as 
\textit{faint broadband background}. 
The resulting marginal signal spectrum, obtained for the SPDC signal photons when detecting the idler photons,  
is indicated by the \textit{yellow line} along the $\nu_\text{s}$-axis. 
}
\label{fig_ch5_spdc_spectra}
\end{figure}

With these parameters, the phase matching function $\Phi(\nu_\text{s},\nu_\text{i})$ can be calculated. Fig. \ref{fig_ch5_spdc_spectra} \textbf{a} shows the phase-matching map 
$\abs{\Phi(\nu_\text{s},\nu_\text{i})}^2$ in the signal and idler frequency space
for our source parameters. 
Here, the dimensions represent $\Delta \nu_\text{s} = \nu_\text{s} - \nu_{\text{s},0}$ and $\Delta \nu_\text{i} = \nu_\text{i} - \nu_{\text{i},0}$, whereby $\nu_{\text{s},0}$ and $\nu_{\text{i},0}$ are the marginal spectra's central frequencies\footnote{
	Since the mode indices used in this calculation are only an approximation by simulation, 
	the periodic poling period has been slightly adjusted in the calculation 
	to achieve perfect phase matching at $852\nm$.
}.
The phase-matching function is quite broad, in fact much broader than the frequency ranges involved in the Raman memory protocol. 
To aid comparison, a frequency interval of $\Delta \nu = 5\GHz$ is marked around the central frequencies $\nu_{\text{s/i},0}$ (\textit{dotted horizontal and vertical lines} in fig. \ref{fig_ch5_spdc_spectra} \textbf{a}). 
Its width can be quantified by looking at an intersection of the map along its diagonal 
(\textit{dotted diagonal white line} in fig. \ref{fig_ch5_spdc_spectra} \textbf{a}) to obtain the phase-matching bandwidth function for the SPDC process. 
This function is obtained from $i(\nu_\text{s},\nu_\text{i})$ by simultaneously varying the frequency/wavelength of signal and idler photons by the same amount, while keeping the UV pump wavelength constant at $426\nm$. 
When observing $i(\nu_\text{s},\nu_\text{i})$ along this diagonal axis, the characteristic sinc$^2$-dependence is observed\cite{Trebino:book}. 
It is shown in fig. \ref{fig_ch5_spdc_spectra} \textbf{b} in wavelength terms 
$\left( \lambda = \lambda_\text{s} = \lambda_\text{i}\right)$. 
Its FHWM of $\Delta \lambda_\text{PM} \approx  0.39 \nm $ 
(\textit{indicated by dotted lines} in fig. \ref{fig_ch5_spdc_spectra} \textbf{b}) 
is commonly referred to as the phase-matching bandwidth. 
In frequency units, the FWHM phase-matching bandwidth is ${\Delta \nu_\text{PM} \approx 161 \GHz}$.

\paragraph{Pump envelope}
Clearly, the phase-matching function is a lot broader than any of the optical fields involved. 
So the shape of the JSA $f(\nu_\text{s},\nu_\text{i})$ will effectively be dominated by the pump bandwidth $\alpha(\nu_\text{p})$, which is a lot narrower than $\Phi(\nu_\text{s},\nu_\text{i})$. 
The exact form of $\alpha(\nu_\text{p})$ depends on the underlying pulse model. 
As discussed in appendix \ref{ch3_tisa}, here we assume sech-shaped pulses\footnote{
	Sech-shaped pulses are chosen, as they correspond to the expected output pulse profile of the {\tisa} laser. 
	A single frequency doubling step of these, needed for creating the SPDC pump 
	(see section \ref{ch5_sec_setup} below), does not yet change
	them into a Gaussian profile, since too few convolutions of the pulses with themselves occur for the 
	central limit theorem to hold. However, for aid of comparison, we will also consider Gaussian pulses 
	when characterising the \hsp\, spectrum. 
	These results are stated in appendix \ref{app_ch5}.
}. 
To obtain $\alpha(\nu_\text{p})$, we can employ the normalised spectral intensity 
$S(\nu_\text{p}) = \sech^2{(\pi^2\Delta t \left( \nu_\text{p}-\nu_{\text{p},0} \right))}$ 
(see eqs. \ref{eq_ch3_pulse_spectra}). For the width parameter $\Delta t$, we use the result of a $g^{(1)}$-interferogram measurement, outlined in appendix \ref{ch3_SHG}, which yield $\Delta t_\text{s}^\text{UV} \approx 150 \ps$ for sech pulses. This corresponds to a spectral bandwidth\footnote{
	Similarly, we get $\Delta t_\text{g}^\text{UV} \approx 205 \ps$ and 
	$\Delta \nu_\text{g}^\text{UV} \approx  1.29 \GHz$ for Gauss pulses.
} of $\Delta \nu_\text{s}^\text{UV} \approx 1.19 \GHz$.
The resulting pump envelope map $\abs{\alpha(\nu_\text{s}+\nu_\text{i})}^2$ is displayed in fig. \ref{fig_ch5_spdc_spectra} \textbf{c}.  
Because it is much narrower than $\abs{\Phi(\nu_\text{s},\nu_\text{i})}^2$, shown in fig. \ref{fig_ch5_spdc_spectra} \textbf{a}, $\alpha(\nu_\text{p})$ limits the spectral bandwidth of the SPDC pair\footnote{
	Note the different axes scalings for the plots in 
	fig. \ref{fig_ch5_spdc_spectra} \textbf{a} and \textbf{c}, respectively.
}, for which reason we can consider the phase-matching part as effectively constant. 
In turn, the join spectral amplitude $f(\nu_\text{s},\nu_\text{i})$ is effectively a stripe under a $45^\circ$ angle in $\nu_\text{s}$-$\nu_\text{i}$-space, with its maximum located at ${\nu_{\text{p},0} = 2 \cdot \nu_\text{Ti:Sa}}$.  
Accordingly, projection of the SPDC signal frequency onto the Raman memory's input channel at $\nu_\text{s} = \nu_\text{Ti:Sa} - 9.2\GHz$  is possible by idler filtering to $\nu_\text{i} = \nu_\text{Ti:Sa} + 9.2\GHz$, as we have illustrated in fig. \ref{fig_ch5_intro} \textbf{i} earlier. Although being slightly off-resonance, we still obtain sufficient count rates, as we will see in section \ref{ch5_subsec_rates}.

\paragraph{Spectral projection by heralding}

Despite the narrowband pump bandwidth, the broad phase matching function still allows SPDC emission into many signal and idler frequencies. 
The SPDC emission is thus spectrally multi-mode. 
Its JSI, displayed in fig. \ref{fig_ch5_spdc_spectra} \textbf{c}, is not separable\cite{URen:2006,Mosley:2008hs}, and signal and idler frequencies are spectrally correlated. 
However, the spectrum of the heralded SPDC signal photons can be modified by filtering of the idler photons. 
The theoretical description\cite{Branczyk:2010} of this spectral shaping by filtering requires an extension of the single mode SPDC state of eq. \ref{eq_ch5_spdc_state} to the multimode case\cite{Law:2000}, which is described in appendix \ref{app_ch5_multimode_spdc_heralding}. 
Phenomenologically, idler filtering corresponds to the multiplication of 
$\alpha(\nu_\text{s},\nu_\text{i})$ 
with a map $t(\nu_\text{s},\nu_\text{i})$, representing the filter transmission function for the light's electric field in the joint $\nu_\text{s}$-$\nu_\text{i}$-space. 
We will see in section \ref{ch5_subsec_filter} that in our case, where multiple filters are used, this function is a Gaussian distribution, given by 
$T_\text{filt}^\text{idl}(\nu) \sim \exp{\left( -\frac{(\nu-\nu_0)^2}{\sigma^2} \right)}$, with a FWHM bandwidth $\Delta \nu_\text{filt}=2 \sqrt{\ln{(2)}} \cdot \sigma$. 
Its map in the $\nu_\text{s}$-$\nu_\text{i}$-space, shown in fig. \ref{fig_ch5_spdc_spectra} \textbf{d}, already uses our experimental value of $\Delta \nu_\text{filt}^\text{idl} = 0.94\GHz$. It contains the dependence $T_\text{filt}^\text{idl}(\nu_\text{i})$ along the vertical axis ($\nu_\text{i}$), but it is constant along the horizontal axis ($\nu_\text{s}$), since it operates only on the idler photons. 
Accordingly, the filter stage map does not affect the JSA in the $\nu_\text{s}$-dimension and forms a horizontal stripe. 
The JSI of the filtered SPDC pair is obtained by the product 
$i(\nu_\text{s},\nu_\text{i}) = \abs{  t(\nu_\text{s},\nu_\text{i}) \cdot f(\nu_\text{s},\nu_\text{i})}^2$, 
illustrated in fig. \ref{fig_ch5_spdc_spectra} \textbf{e}, which has the geometry of a tilted ellipse. 

The marginal spectrum of the \hsp\,, i.e., the SPDC signal spectrum after idler detection, is obtained by marginalisation over the idler frequencies (see appendix \ref{app_ch5_multimode_spdc_heralding}). 
With regard to the JSI map in fig. \ref{fig_ch5_spdc_spectra} \textbf{e}, marginalisation corresponds to the summation over all $\nu_\text{i}$-values for each $\nu_\text{s}$-coordinate point in the JSA $f(\nu_\text{s}, \nu_\text{i})$, followed by taking the absolute square. This accounts for any spectral phase factors in $f(\nu_\text{s}, \nu_\text{i})$ and results in a 1-D projection of the JSI onto the $\nu_\text{s}$-axis, which is also displayed in fig. \ref{fig_ch5_spdc_spectra} \textbf{e} (\textit{yellow line}). 
Normalisation yields the expected \hsp\, spectrum $S_\text{s}^\text{JSA}(\nu)$.

Importantly, the SPDC state is not necessarily spectrally pure. 
We can see this from the shape of the filtered JSA, whose symmetry axes would have to be parallel to the 
$\nu_\text{s}$- and $\nu_\text{i}$-axes for it to be separable\cite{Mosley:2008hs}.  
This means, the \hsp\, state will be a superposition of several Schmidt modes. 
To determine its storage efficiency in the memory, we have to use its decomposition into these Schmidt modes \cite{Eberly:2005oq, Grice:1997} 
\begin{equation}
\tilde{f}(\nu_\text{i},\nu_\text{s}) = \underset{k}{\sum} {\lambda}_k \cdot \zeta_k(\nu_\text{i}) \cdot \xi_k(\nu_\text{s}),  
\label{eq_ch5_JSI_schmidt_decomposition}
\end{equation}
with idler modes $\zeta_k$ and signal modes $\xi_k$ and Schmidt coefficient ${\lambda_k}$.  
As discussed in appendix \ref{app_ch5_multimode_spdc_heralding}, these filtered Schmidt modes are a linear superposition of the original, unfiltered modes \cite{Branczyk:2010}. 
The degree of separability is denoted by the state's purity\cite{Mosley:2008hs} 
$\mathcal{P} = \underset{k}{\sum} \lambda_k^{-2}$. Completely separable states have a purity of $\mathcal{P} =1$, which corresponds to a product state 
$\tilde{f}(\nu_\text{i},\nu_\text{s}) = \zeta_1(\nu_\text{i}) \xi_1(\nu_\text{s})$, 
expressed by a single Schmidt mode pair. Accordingly, decreasing purity corresponds to an increasing number of Schmidt modes. 
In the following, we only consider single photons states, whose vacuum component has been removed by heralding. Additionally, we also neglect higher order emissions. 
The reduced signal density matrix of such a state\cite{Branczyk:2010} 
$\rho_\text{s} = \underset{k}{\sum} \lambda_k \ket{\xi_k,1_\text{s}} \bra{\xi_k,1_\text{s}}$, after idler detection,  
features each signal mode $\ket{\xi_k,1_\text{s}}$ with probability $\lambda_k$, whereby 
$\underset{k}{\sum} \lambda_k =1$.
With this structure, we can now discuss how to optimise the filtering for matching the \hsps\, to the Raman memory. 

\paragraph{Source design parameter optimisation}

\begin{figure*}
\includegraphics[width=\textwidth]{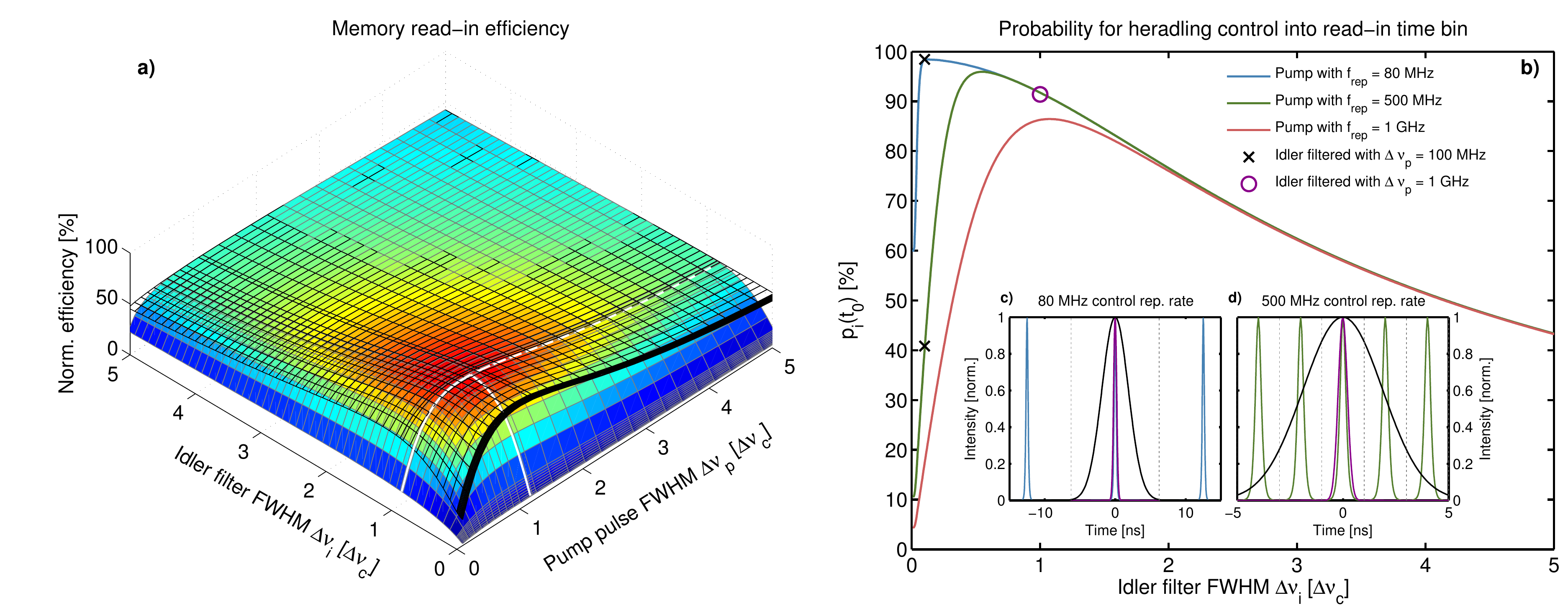}
\caption{ 
\textbf{(a)}: Normalised read-in efficiency as a function of SPDC pump bandwidth $\Delta \nu_\text{p}$ and idler filter bandwidth $\Delta \nu_\text{i}$. The \textit{black grid} shows 
$\tilde{\eta}$ (eqs. \ref{eq_eta_in}), while the \textit{coloured surface} displays 
$\tilde{\eta} \cdot p_\text{i}(t_0)$ (eq. \ref{eq_2}). 
The \textit{white lines} mark our experimental values $\Delta \nu^\text{exp}_\text{p}$ and $\Delta \nu^\text{exp}_\text{i}$. 
\textbf{(b)}: Probability $p_\text{i}(t_0)$ for control field preparation in the correct read-in time bin $t_0$. \textit{Cross} and \textit{circular markers} indicate $\Delta \nu_\text{i} = 100 \,\text{MHz}$ and $\Delta \nu_\text{i} = 1\,\text{GHz}$, respectively. For both filters, the intensities of the corresponding filtered modes $|\zeta_1(\nu_\text{i})|^2$ are shown in inset \textbf{(c)}, for a repetition rate of $f_\text{rep}^\text{exp} = 80\,\text{MHz}$ (used in the experiment), and inset \textbf{(d)} for $f_\text{rep} = 500\,\text{MHz}$, respectively. The \textit{grey} pulse is the filtered idler photon, the \textit{red line} is the spectrum of the heralded single photon, and the control pulse trains are \textit{blue} for $80\,\text{MHz}$ and \textit{green} for $500\,\text{MHz}$, respectively.  \textit{Vertical lines} mark the time bin size of each control pulse. 
}
\label{fig_1}
\end{figure*}

Good matching is achieved, when the \hsp \,'s memory read-in efficiency \etain\, is maximised. 
As we have seen in eq. \ref{eq_ch2_eff_in} of section \ref{subsec_mem_op_adiabatic_lim}, this is the case, when the overlap between the memory kernel $\hat{K}(z,\nu)$ and the input signal's electric field envelope $S_\text{in}(\nu) = E_\text{s}(\nu)$ are maximised. 
For our operational parameters (see section \ref{ch2_Raman_level_scheme}), the kernel is essentially determined by the control field, so we have $\hat{K}(z,\nu)=E_\text{c}(\nu)$. 
This yields a spin-wave excitation of $B(z) = \underset{\nu}{\int} K(z,\tilde{\nu}) E_{\text{s}} \left(\tilde{\nu}\right) \text{d} \tilde{\nu}$ and a memory read-in efficiency of 
${
\eta_\text{in} = \frac{\langle \hat{N}_B \rangle}{\langle \hat{N}_S \rangle}= \frac{\underset{z}{\int} |B(\tilde{z})|^2 d \tilde{z}}{\underset{\nu}{\int} |S(\tilde{\nu})|^2  \text{d}\tilde{\nu}},
}$
with $\langle N_B \rangle$ and $\langle N_S \rangle$ as the number of spin-wave excitations and the number of photons sent into the memory, respectively. 
Storage operates on each mode $k$ of $\rho_\text{s}$ separately,
whereby the mode-overlap between the signal modes and the memory kernel is weighted by $\lambda_k$:
\begin{equation}
\eta_\text{in} = \eta_0 \cdot \sum_{k} \lambda_k \lvert  \int_{\omega} K(\tilde{\omega}) \cdot \xi_k(\tilde{\omega}) \text{d} \tilde{\omega} \rvert^2= \eta_0 \sum_k \lambda_k A_k = \eta_0 \cdot \tilde{\eta}. 
\label{eq_eta_in}
\end{equation}
$\eta_0$ is a constant denoting the read-in efficiency for a perfectly mode-matched system, which is 
defined by the parameters of the storage medium and the control pulse energy. 
The operation of our memory effectively corresponds to a mode filter for the input signal\footnote{
	Accordingly, the memory could also be applied as a mode filter when interfering \hsps\, produced by 
	different SPDC source\cite{Mosley:2008hs}. 
}; 
it effectively resembles a quantum pulse-gate \cite{Brecht:2014}. 
Eq. \ref{eq_eta_in} illustrates how we can change $\eta_\text{in}$ by idler filtering through the associated variables $\lambda_k$ and $\xi_k$.
Experimentally, we actually have two design parameters, the idler filter bandwidth $\Delta \nu_\text{i}$ and the pump spectrum, with a FWHM bandwidth of $\Delta \nu_\text{p}$. 
The normalised memory efficiency depends on both parameters, i.e. we have $\tilde{\eta}(\Delta \nu_\text{i}, \Delta \nu_\text{p})$. 
We now optimise these parameters and assume 
the memory kernel $K(\nu)$ equals the control pulse spectrum. 
The control corresponds to sech-shaped {\tisa} output pulses with amplitude $E_\text{c} (\nu)$, as described in appendix \ref{ch3_tisa_pulse_duration}. 
We use the phase-matching function, shown in fig. \ref{fig_ch5_spdc_spectra} \textbf{a}, and apply eq. \ref{eq_ch5_JSI_schmidt_decomposition} for the JSA. 
The optimisation trade off lies between the signal and control overlap $A_k$ and the Schmidt coefficient $\lambda_k$. Ideally, we would desire a \hsp\, spectrum equal to $E_\text{c} (\nu)$ and a single Schmidt mode, i.e. $\lambda_1 = 1$ and $\lambda_{k>1} = 0$. 
A multimode structure reduces $\tilde{\eta}$, because the $\lambda_k$ decrease for the lower order Schmidt modes, when adding more modes (i.e. the number of $\lambda_{k>1} \neq 0$ increases). However, these lower order modes have the largest overlaps $A_k$. 
The black grid in fig. \ref{fig_1} \textbf{a} illustrates the effects of both design parameters, $\Delta \nu_\text{i}$ and $\Delta \nu_\text{p}$, on $\tilde{\eta}(\Delta \nu_\text{i}, \Delta \nu_\text{p})$.

In terms of the SPDC pump, best efficiency is obtained for $\Delta \nu_\text{p}^\text{opt} \approx 1\,\text{GHz}$, where its spectral bandwidth resembles the bandwidth of the memory control pulses. 
A more narrowband pump makes the filtered JSA (shown in fig. \ref{fig_ch5_spdc_spectra} \textbf{e}) stripier, since it narrows $\alpha(\nu_\text{s},\nu_\text{i})$. 
This reduces the SPDC state purity \cite{Mosley:2008hs} $\mathcal{P}$, leading to an increased number of modes $| \left\{ k \right\} |$ and lower $\lambda_k$ for the modes with large $A_k$. 
Conversely, broader pump spectra also reduce $\tilde{\eta}$ as they lower the spectral overlap $A_k$ with the memory control.
Conveniently, this similarity between pump and memory control bandwidths allows to generate both pulses 
with the same master laser, which is a significant experimental simplification\footnote{
	It prevents the otherwise necessary synchronisation between different pulsed laser systems.  
}. 

The effects of idler filtering on $\tilde{\eta}(\Delta \nu_\text{i}, \Delta \nu_\text{p})$ demonstrate the dominant influence of the $\lambda_k$ in eq.~\ref{eq_eta_in}, and therewith the SPDC state purity $\mathcal{P}$ on $\tilde{\eta}$. 
Since we have $A_k \le 1$, $\forall k$, and $A_{k+1} \le A_k$ for increasing mode numbers\footnote{
	The mode index $k$ denotes the number of modes in the electric field distribution of $\xi_k$, i.e. the 
	fundamental mode $\xi_1$ is uni-modal, $\xi_2$ is bi-modal, with $1$ node of $0$ amplitude, and so on. 
	More such nodes lower the mode overlap with the uni-modal distribution $E_\text{c}$ of the control, 
	thus reducing $A_k$.
}, the largest value for $\tilde{\eta}$ is obtained when $\lambda_1 = 1$ and $\lambda_{k>1} =0$, irrespective of a lower spectral mode overlap $A_k$. 
This is obtained for $\Delta \nu_\text{i} \rightarrow 0$, i.e. for infinitely narrow idler filtering. 
Here the JSA of the SPDC pair becomes a horizontal stripe in $\nu_\text{s}$-$\nu_\text{i}$-space, 
making $f(\nu_\text{s},\nu_\text{i})$ separable\cite{Branczyk:2010}, for which reason the filtered state becomes pure with $\mathcal{P} =1$. 
Even for such an infinitely narrowband idler, $A_k > 0$ because the minimal HSP bandwidth is limited by the pump function $\alpha(\nu_\text{p})$ to\footnote{
	This follows from the geometry of the JSA, where the pump function 
	$\alpha(\nu_\text{p})$ is oriented under a 
	$45^\circ$ angle to the $\nu_\text{s}$-$\nu_\text{i}$ coordinate axes, 
	as shown in fig. \ref{fig_ch5_spdc_spectra}. See also appendix \ref{app_ch5_SPDCspec}.  
} $\Delta \nu_\text{HSP} \approx \sqrt{2} \Delta \nu_\text{p}$.
Obviously, infinitely narrow idler filtering is non-sensible, as it would diminish the \hsp\, production rates to  experimentally insufficient levels. 
However, tighter idler filtering also temporally broadens the idler photon, which effectively bounds $\Delta \nu_\text{i}$ as we will see now. 

\paragraph{Heralding into the correct time bin}
For an interfaced source-memory system an idler detection event does not only herald the presence of a signal photon, but also has to trigger the signal photon's on-demand storage in the memory. 
In practice, the source pump and the memory control will both be based on a pulsed laser system with a fixed repetition rate $f_\text{rep}$. Accordingly, the generated \hsps\, and the memory control pulses fall into well defined time bins $\Delta T$ of duration $\Delta T = \frac{1}{f_\text{rep}}$.
So, feed-forward of idler detection events is used to prepare a control pulse for the time bin, in which a \hsp\, is inserted into the memory. 
In other words, to select the correct control pulse from a continuous pulse train, the filtered idler photon and the control pulse, that is to be selected, must fall into the same pulse train time bin. 
However, if the idler pulse duration is too long, its intensity distribution overlaps with adjacent time bins, leading to a finite probability of preparing control pulses in an earlier or later time bin than that occupied by the HSP. 
Figs. \ref{fig_1} \textbf{c} \& \textbf{d} illustrate this for two pulse train repetition rates $f_\text{rep}$ and two idler filter widths.
If the idler was detected in any of these adjacent time bins, the corresponding \hsp\, and the control pulse would not arrive at the memory simultaneously, so storage would not be possible. 

When choosing $\Delta \nu_\text{i}$, we thus need to account for the temporal idler pulse broadening and the resulting probability of triggering non-synchronised control pulses. 
Given the time bin size $\Delta T$ and SPDC pair generation in time bin $t_0$, the 
effective memory read-in efficiency is modified to 
\begin{equation}
\hat{\eta}_\text{in}(t_0)= \eta_0 \cdot \tilde{\eta}(\Delta \nu_\text{p},\Delta \nu_\text{i}) \cdot p_\text{i}(t_0) =  \eta_0 \cdot \hat{\tilde{\eta}}(\Delta \nu_\text{p},\Delta \nu_\text{i},t_0),  
\label{eq_2}
\end{equation}
$p_\text{i}(t_0)$ is the probability for detecting the idler photon 
in the correct time bin $t_0$. 
It is the sum over the probabilities for each idler Schmidt mode to fall into time bin $t_0$: 
$$
p_\text{i}(t_0) = \sum_k \int_{t_0 - \frac{\Delta T_\text{c}}{2}}^{t_0 + \frac{\Delta T_\text{c}}{2}} |\zeta_k(t)|^2 \text{d} t. 
$$  
Fig.  \ref{fig_1} \textbf{c} shows the influence of $\Delta \nu_\text{i}$ on $p_\text{i}(t_0)$ for different control repetition rates, using the optimal SPDC pump bandwidth $\Delta \nu_\text{p} = \Delta \nu_\text{c}$ in calculating 
$f(\nu_\text{s},\nu_\text{i})$. 
Filtering the idler too tightly, broadens the idler modes $\zeta_k$ too much in the temporal domain and reduces $p_\text{i}(t_0)$, which experimentally results in control pulse generation in an incorrect time bin (see fig. \ref{fig_1} \textbf{d}). 
For this reason $p_i(t_0)$ drops off sharply for $\Delta \nu_\text{i} \rightarrow 0$ in fig. \ref{fig_1} \textbf{b}. 
The resulting reduction in $\hat{\tilde{\eta}}$ effectively establishes a lower bound on $\Delta \nu_\text{i}$, which however depends on $f_\text{rep}$. 
For low $f_\text{rep} \sim 80\,\MHz$, $p_\text{i}(t_0)$ does not fall off before $\Delta \nu_\text{i}$ has been narrowed to $\Delta \nu_\text{i} \sim 100\,\MHz$, which is still too narrow to yield viable \hsp\, production rates. 

Yet, in an actual application, such as temporal multiplexing, one would like the source to run at higher rates $f_\text{rep}$. 
Because the maximal storage time $\tau_\text{s}$ of a memory is ultimately limited by the storage medium properties, it is desirable to maximise the number of photon production trials during $\tau_\text{s}$. 
While the exact value of $f_\text{rep}$ obviously depends on the available technology and one's finances, one possibility to obtain a useful upper limit on the repetition rate can be set by requiring $\mathfrak{P}_\text{i}(t_i)\ge 99\,\%$ of the overall intensity for each pulse, centred at time $t_i$, to fall into its time bin $\left[ t_i - \Delta T/2,  t_i + \Delta T/2\right]$, with the bin size $\Delta T = \frac{1}{f^\text{max}_\text{rep}}$. 
Notably, this approximately corresponds to a $3\cdot \sigma$ pulse separation. 
At higher rates, the pulses start to overlap substantially and one effectively approaches the cw regime. 
For our pulse parameters (see appendix \ref{ch3_tisa_pulse_duration}), this requirement suggests a maximal rate of $f_\text{rep}^\text{max} \sim 1\GHz$ (\textit{red line} in fig. \ref{fig_1} \textbf{b}).   
Here, the optimal idler filter bandwidth becomes 
$\Delta \nu_\text{i}\sim 1\,\GHz$, where $p(t_0)$ has its maximum. 
Accordingly, the normalised read-in efficiency $\hat{\tilde{\eta}}$ also has its maximum at $\Delta \nu_\text{i}\sim 1\,\GHz$. Its dependence on both parameters, $\Delta \nu_\text{i}$ and $\Delta \nu_\text{s}$ is depicted by the coloured surface in fig. \ref{fig_1} \textbf{a}.

For the actual implementation, we will thus attempt to match this optimal idler filter bandwidth. As we use a series of low finesse Fabry-Perot etalons for filtering (see section \ref{ch5_subsec_filter} below), whose effective finesse values influence their performance\cite{Lvovsky:2012}, the resulting filter linewidth does not exactly match this number and ends up at $\Delta \nu_\text{i} = 0.94 \,\GHz$. Similarly, we use SHG of our {\tisa} master laser output to generate the SPDC pump (see section \ref{ch5_sec_source_setup}), which causes slightly broader pump spectra than memory control pulses, with $\Delta \nu_\text{p}=1.19 \,\GHz$ (see table \ref{ch3_table_pulse_durations} in appendix \ref{ch3_SHG}). 
The \textit{white lines} in fig. \ref{fig_1} \textbf{a} indicate these experimental values. 
Additionally, both parameters have also already been used in the JSI plot of fig. \ref{fig_ch5_spdc_spectra} \textbf{e}. 
Marginalisation over the idler frequencies, shown by the \textit{yellow line} in fig. \ref{fig_ch5_spdc_spectra} \textbf{e}, thus represents the predicted spectrum of the \hsp\, in our experiment. 
To illustrate their bandwidth, compared to the memory kernel (here the {\tisa} spectrum) as well as the idler filter, all three of these quantities are plotted in fig. \ref{fig_ch5_meas_spdc_spectra} in section \ref{ch5_sec_exp_hsp_bandwidth}, where they are compared to our experimental results.  
With a FWHM bandwidth of $\Delta \nu_\text{HSP}^\text{pred}= 1.54 \,\GHz$, the expected \hsps\, are slightly broader than the memory kernel. As we have seen in the above discusion, the resulting mismatch in spectral overlap is compensated by the purity of the SPDC state for which we predict $\mathcal{P} \approx 77 \,\%$.

\subsection{Temporal SPDC description\label{subsec_ch5_temporal_SPDC_description}}
To complete our discussion of the SPDC theory, we take a brief look at the SPDC process in the time domain, which we require for the source characterisation measurements in section \ref{ch5_sec_photon_prod} later on. 
The SPDC state $\ket{\psi}_\text{SPDC}$ in the time domain can be derived by investigating the Hamiltonian dynamics in the Schr\"odinger picture instead of the Heisenberg picture, which has led to the state in eq. \ref{eq_ch5_spdc_state}. Here, the two-mode squeezing Hamiltonian 
$H_\text{SPDC} = \exp{\left\{ \text{i} \hbar \kappa  \left( a_\text{s}^\dagger a_\text{i}^\dagger - a_\text{i}  a_\text{s}\right) \right\}}$ 
describes the two-mode-squeezing frequency conversion process, where 
${\kappa  = \alpha \cdot E_\text{UV} \sim d_{24} \cdot L_\text{KTP} \cdot E_\text{UV}}$ accounts for the dependence on the nonlinear crystal parameters and the UV pump power. 
It can be shown\footnote{
	A detailed derivation is for instance provided in the PhD thesis of 
	\textit{Witlef Wiezczorek}\cite{Witlef:PhD}.
} that
unitary evolution in time $t$, described by the operator
${U(t) = \exp{\left\{ - \frac{\text{i}}{\hbar} H_\text{SPDC} \cdot t \right\}}}$, 
yields a quantum state for collinear SPDC emission of\cite{Witlef:PhD,Eisenberg:2004}
\begin{equation}
\ket{\Psi_\text{SPDC}}(t)= 
\sqrt{1- \left(\tanh{\left( |\kappa t |\right)}\right)^2 } \overset{\infty}{\underset{n=0}{\sum}} \left( \tanh{\left( |\kappa t | \right)}\right)^n \cdot \ket{n_\text{s}, n_\text{i}} 
\overset{\kappa t \ll 1}{\longrightarrow} \overset{\infty}{\underset{n=0}{\sum}} \left( |\kappa t |\right)^n \cdot \ket{n_\text{s}, n_\text{i}}.
\label{eq_ch5_SPDCstate_temporal}
\end{equation}  
The Taylor series expansion of the $\tanh$-term in the last step holds in the low pumping power regime\cite{Krischek:2010ys}. 
For our purposes, the important realisation to be taken from eq. \ref{eq_ch5_SPDCstate_temporal} concerns the photon number expectation value  
$\langle n_\text{j} \rangle = \,\langle \Psi_\text{SPDC} | \hat{a}_j^\dagger \hat{a}_j | \Psi_\text{SPDC} \rangle$, where
$j $ can either represent the independent detection of signal ($j = \text{s}$) and idler ($j = \text{i}$) photons, termed singles events, or the combined detection of a signal photon and an idler photon ($j = \text{s},\text{i}$), called coincidence events. 
As discussed in section \ref{sec_ch4_photon_detection} below, we detect the SPDC photons on avalanche photodiodes (APDs), which have a sub-unity detection efficiency of $\eta_\text{det}$. 
These are not capable to resolve the photon number. 
So all terms with $n_j \ge 1$ of eq. \ref{eq_ch5_SPDCstate_temporal} contribute and we obtain
\begin{align}
\label{eq_ch5_photon_number}
\langle n_\text{s} \rangle = \langle n_\text{i} \rangle &=  \overset{\infty} {\underset{n=1}{\sum}}\left( (1-(1-\eta_\text{det})^n) \cdot  |\alpha \cdot t \cdot E_\text{UV}| \right)^{2n}  \\
& = 
 \overset{\infty}{\underset{n=1}{\sum}} \left(  (1-(1-\eta_\text{det})^n) \cdot |\alpha| \right)^{2n} \cdot P_\text{UV}^n 
 = \overset{\infty}{\underset{n=1}{\sum}} \gamma^n \cdot P_\text{UV}^n  \nonumber \\
 \langle n_{\text{s},\text{i}} \rangle &=
 \overset{\infty}{\underset{n=1}{\sum}}  \overset{\infty}{\underset{m=1}{\sum}} \left(\left(  (1-(1-\eta_\text{det})^m) \cdot |\alpha| \right)^{2m} \right) \cdot \left(\left(  (1-(1-\eta_\text{det})^n) \cdot |\alpha| \right)^{2n} \right) \cdot \left(\sqrt{P_\text{UV}}\right)^{n+m}  \nonumber
\end{align}
where the $\left( 1-(1-\eta_\text{det})^k \right)$ terms denote the detection probability of $k$ photons\cite{Migdall:book}.
The scaling of the detected photon number with the SPDC pump power $P_\text{UV}$ is thus directly proportional to the number of produced SPDC pairs\cite{Krischek:2010ys}. 
For \hsp\, production the emission of only a single SPDC pair is desired. 
Due to the mechanics of idler detection without photon number resolution, contributions from higher order emissions are indistinguishable upon heralding. 
Multiple pair emissions can thus result in the presence of two or more photons in the conditionally prepared SPDC signal state, spoiling its single photon character. 

In operating the source, we consequently need to keep $P_\text{UV}$ low enough to avoid multi-pair events, while we simultaneously aim for sufficiently high preparation rates to achieve reasonable photon insertion rates into the memory. 
In our source characterisation later on (see section \ref{ch5_sec_photon_prod}) we analyse these photon count rates within our available pumping power regime (see appendix \ref{ch3_SHG}) and investigate the quality of the \hsps\, in terms of the aforementioned spurious contamination from any potential higher order emissions. To this end, we  measure their photon statistics\cite{Loudon:2004gd} via the $g^{(2)}$ autocorrelation function, introduced in section \ref{sec_ch4_photon_detection}. 

	
\section{Source setup\label{ch5_sec_setup}}
\begin{figure}[h!]
\centering
\includegraphics[width=\textwidth]{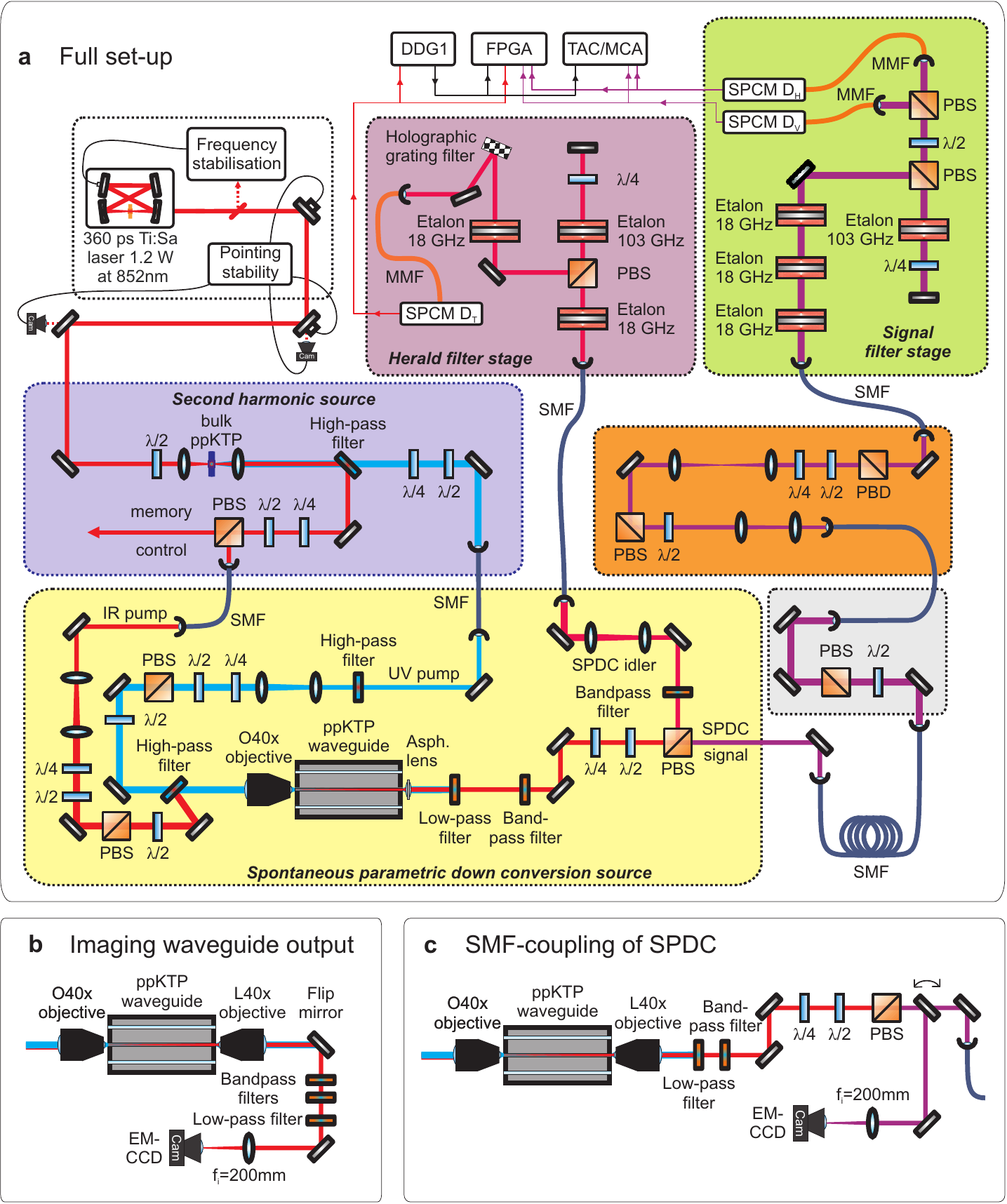}
\caption{Experimental setup of the single photon source. 
\textbf{(a)}: Complete apparatus; coloured panels mark the different segments.
\textbf{(b)}: Modifications for imaging the spatial modes in the waveguide.
\textbf{(c)}: Modifications for optimising SMF-coupling of the SPDC signal and idler modes (see appendix \ref{app_ch5_spdc_modes}).}
\label{fig_ch5_setup}
\end{figure}

Having established the core design parameters for the source, we can consider its experimental implementation. 
We discuss the setup by first outlining the general ideas, and then going into the details of the components, which should allow the reader to reconstruct the system if required. 
Fig. \ref{fig_ch5_setup} \textbf{a} shows the complete experimental set-up for operating the single photon source. 
It contains several separate segments, indicated by coloured panels and discussed individually in the following. These broadly fall into two categories: firstly, the SPDC photon pair generation, centred around the ppKTP waveguide; secondly, the filtering optics required to produce and analyse heralded SPDC signal photons. 
Besides the final experimental lay-out of fig.~\ref{fig_ch5_setup}~\textbf{a}, fig.~\ref{fig_ch5_setup}~\textbf{b}~\&~\textbf{c} display set-up modifications used to characterise critical intermediate steps in building the apparatus.

\subsection{Source set-up principles\label{ch5_sec_source_principles}}

Thanks to the close matching between the spectral bandwidth for the optimal SPDC pump and our {\tisa} master laser, we can generate the required $426\,\nm$ radiation by frequency doubling the {\tisa} pulses. 
This allows for a rather simple source setup, where we just insert the SH source (\textit{blue panel} in fig. \ref{fig_ch5_setup} \textbf{a}) as the first major apparatus component behind the {\tisa} laser system (\textit{white panel}). 
Because we only need UV pump powers in the mW-regime, we can run the SHG at low efficiency and recycle the unconverted $852\,\nm$ radiation, which has pretty much the same pulse energy as the {\tisa} output, to generate the memory control pulses. 
After the SH source, described in detail in appendix \ref{ch3_SHG}, the $426\,\nm$ UV pulses are separated from the $852\,\nm$ light by frequency filtering. 
Using a set of bespoke coupling optics, the UV pulses are inserted into the ppKTP waveguide, pumping the type-II SPDC process (\textit{yellow panel}). 
The generated SPDC emission at $852\,\nm$ wavelength is also separated from the UV pump by frequency filtering, and split into two separate spatial modes according to polarisation. V-polarised SPDC photons become the idler, H-polarised photons the signal. 
Idler photons are frequency filtered immediately and detected (\textit{purple panel}) for spectrally engineering and heralding of the signal photons (see section \ref{ch5_sec_hsp_bandwidth}). 
The signal photons are guided through some free-space optics (\textit{grey} and \textit{orange panels}), which later on will host the Raman memory, before they are also frequency filtered and detected (\textit{green panel}).

Notably, according to the theory arguments we have just presented (see section \ref{ch4_source_design_theory}), we would not require a dedicated signal filter, as filtering and detecting the idler will project the signal photons into the correct spectral modes. 
However, this ignores, for instance, any noise sources in the system, such as single photon fluorescence in the waveguide, caused by the UV pump. 
Moreover, to assure we are measuring the same spectral mode when characterising the source as when running it with the memory (and also as a matter of experimental convenience), we use the same frequency filter stage for analysing the SPDC signal here, as we will apply in chapter~\ref{ch6} to filter the retrieved signal from the Raman memory output.

\subsection{Components in the SPDC source set-up\label{ch5_sec_source_setup}}

\paragraph{Pump beam generation}
The various components, required to generate SPDC photons, feature in the \textit{white}, \textit{purple} and \textit{yellow panels} of fig. \ref{fig_ch5_setup} \textbf{a}. 
Similar to the experiments in chapter \ref{ch4}, we start with our {\tisa} master laser (see appendix \ref{ch3}). 
The first segment in the line of optics thereafter 
is now a frequency and beam pointing stabilisation (\textit{white panel}, see appendix \ref{ch3_tisa}), follow by the preparation of the UV pump pulses for the SPDC process (\textit{violet panel}).
These are produced from the $80\MHz$ {\tisa} pulse train by second-harmonic generation in a bulk ppKTP crystal. The generated SHG at $\lambda_\text{SHG} = \lambda_\text{p} = 426\nm$ wavelength is separated from the fundamental IR radiation at ${\lambda_\text{Ti:Sa} = 852\nm}$ on a dichroic high pass filter (\textit{Semrock BLP01-532R-25}, high-reflective at $\lambda_\text{Ti:Sa} = 852\nm$, and high transmissive at $\lambda_\text{p} = 426\nm$). 
More details regarding the SH-source are provided in the appendix \ref{ch3_SHG}.
Behind the ppKTP crystal, the SH is SMF-coupled and sent into the second segment, which contains the actual waveguide source. 
The non-converted IR light will be processed further to the memory control pulses. 
This procedure is covered in section \ref{ch6_sec:setup}.
Beforehand, a PBS allows to pick-off a small, variable fraction of the IR pulses. 
These $\lambda_\text{Ti:Sa} = 852\nm$ pulses, referred to as IR pump, 
are also SMF-coupled and sent to the waveguide source as an alignment beam (see section \ref{app_ch5_SHGwaveguide}).

\paragraph{Input-coupling optics}
The set of optics required to run the waveguide source is displayed in the \textit{yellow panel} of fig. \ref{fig_ch5_setup} \textbf{a}. 
To achieve good spatial mode quality upon light insertion into the waveguide, high quality fibre couplers\footnote{
	For the UV pump a \textit{Thorlabs NanoMax} stage with a \textit{New Focus 5724 compact aspheric lens} is used. 
	For the IR pump, we employ a \textit{Sch\"after Kirchoff 60FC-F-0-M12-10} fibre coupler.  
} 
collimate the IR and the UV pumps. 
Subsequently, their modes are  expanded using telescopes, chosen empirically to yield the best coupling efficiency into the waveguide. 
Since propagation of UV light in Si-based optical fibre can introduce undesired IR fluorescence noise, an additional high-pass dichroic filter is positioned in the UV pump beam path. 
Additionally, each beam path contains a PBS to define the polarisation coordinate system, whereby horizontal polarisation is parallel to the chip's surface. 
For the desired type-II nonlinear frequency conversion, the UV pump for SPDC needs to be horizontally polarised. Conversely, using the IR pump to generate SH requires diagonal (D)-polarisation, which in turn generates horizontally polarised SH light at $426\,\nm$.
For light insertion into the waveguide chip, both pump fields, IR and UV are overlapped on a high-pass dichroic filter (\textit{Semrock  BLP01-532R-25}).  
All these optical elements introduce loss for the UV pump, reducing its transmission from the SH source to input coupler to $T^\text{prep}_\text{UV} \approx 60 \,\%$. 
Coupling into the waveguide channels is facilitated by a 40-times (40X) microscope objective 
(\textit{Olympus O40X Plan Achromat Objective}) 
with ${f^\text{O40X}_\text{eff} = 4.5\mm}$ effective focal length and $NA^\text{O40X}= 0.65$ numerical aperture. 
The utilised waveguide channel has an approximate numerical aperture of $NA_\text{guide} \approx 0.2$. 
We choose the input beam diameters by empirical optimisation, which also includes optimisation of the telescopes in front of the input coupler. 
The objective has transmissions of 
$T_\text{O40X}(852\nm) = (77\pm2) \,\%$ and 
$T_\text{O40X}(426\nm) = (83\pm 5)\,\%$ for the IR and UV wavelengths, respectively. 

\paragraph{Waveguide mount}
The waveguide chip is positioned in a groove of an aluminium (Al) finger, mounted onto a 3-axis translation stage. The input and output coupling objectives are also positioned on 3-axis stages to allow for maximal translational degrees of freedom (see fig. \ref{fig_ch5_intro} \textbf{a}). The input coupler furthermore sits in a tip and tilt mount to allow angular rotation (see appendix~\ref{app_ch5_insufficiencies} for the reasons why this is necessary). 
The Al finger is heated by a Peltier element. 
Using a temperature sensor (\textit{AD590}), positioned inside the Al mount, the temperature is controlled by a PiD-circuit. 
The maximally achievable temperature is $T_\text{KTP}^\text{max} \approx 55^\circ \text{C}$. 
As mentioned in section \ref{ch4_source_design_theory}, our chip comprises 3 families of waveguide, each with two sets of channels of width $2\,\mum$, $3\,\mum$ and $4\,\mum$. 
We use one of the $3\,\mum$-wide waveguides in the centre of the guide, referred to as channel 3.2\footnote{
	This is the $2^\text{nd}$ channel of the $1^\text{st}$ family, labelled as waveguide group $3$.
}, whose depth\footnote{
	According to AdvR, the depth measurement has been conducted by eye only, for which
	reason the specified value of $6\mum$ is only an approximation. 
} 
is $d\approx 6\mum$. 
While building the experiment, the chip's surface unfortunately got scratched (see fig. \ref{fig_ch5_intro} \textbf{f} \& \textbf{g} and appendix \ref{app_ch5_insufficiencies}). Channel $3.2$ belongs to the waveguide family least affected by these scratches, which motivates its choice.

\paragraph{SPDC output collection optics}
To couple light out of the waveguide, two different lens configurations are used. 
For the actual generation of heralded single photons (\hsps\,), an aspheric lens (\textit{Thorlabs C230TME-B}) with high transmission at $852\nm$ is used to minimise losses for the generated SPDC photon pairs. 
The lens is positioned to collimate $852\nm$ light. 
Radiation at this wavelength is separated from the $426\nm$ SPDC pump with a low-pass filter (\textit{Semrock BLP01-785R-25}) and a $10\nm$ wide bandpass filter centred at $850\nm$ (\textit{Thorlabs FL-850-10}). 

For the characterisation measurements of the spatial modes in the waveguide, presented in 
appendix \ref{app_ch5_spdc_modes}, the aspheric lens is replaced by another 40X microscope objective (\textit{Leyca L-40X Plan Achromatic Objective}), 
with $f^\text{L40X}_\text{eff} = 4.6\mm$ and $NA^\text{L40X} = 0.66$, illustrated in fig. \ref{fig_ch5_setup} \textbf{b} \& \textbf{c}. 
While its reduced transmissions $T(852\nm) =( 92\pm2)\,\%$ and $T(426\nm) =( 77\pm2)\,\%$ are undesirable for \hsp\, production, it allows us to avoid chromatic aberration present in the aspheric lens. 
For this reason we can simultaneously collimate light at $852\,\text{nm}$ and $426\,\nm$ wavelength with reasonable quality, which is helpful when studying the spatial mode structure in the guide. 
We can observe the spatial modes on an \textit{Andor} EM-CCD camera via a flip-mirror.
An $f_i = 200\mm$ focal length lens images the waveguide modes at the output of the guide onto the camera with $M_i = 43.5$ magnification.
Two bandpass filters (\textit{Thorlabs FL-850-10}) and one low-pass filter (\textit{Semrock BLP01-785R-25}) are also inserted in front of the imaging lens.
While not shown in fig. \ref{fig_ch5_setup} \textbf{a}, this mirror is also present in the final set-up. 
It allows us to observe the spatial mode structure whenever necessary during day-to-day operation. 
This is particularly useful when aligning the coupling conditions of the UV pump going into the waveguide. 
These conditions are set for SPDC emission into the fundamental spatial mode (see fig. \ref{fig_ch5_theory_modes} and \ref{fig_ch5_exp_modes}) for the source to achieve the best possible \hsp\, preparation efficiencies. 

\paragraph{Signal and idler separation}
In the waveguide output (right hand side of \textit{yellow panel} in fig. \ref{fig_ch5_setup} \textbf{a}), SPDC signal and idler photons are separated on a PBS. 
Due to the geometry of the waveguide, the SPDC-pair polarisation is commensurate with the pump's polarisation coordinate system (see coordinate system in fig. \ref{fig_ch5_theory_modes} \textbf{a}). 
Any rotations from birefringence in the guide are neutralised by waveplates before the PBS.
The vertically (V) polarised photon forms the idler photon. 
Its detection will generate the heralding events for the horizontally (H) polarised photon, which will be sent into the memory as the input signal. 
An additional bandpass filter in the herald arm rejects undesired single photon fluorescence noise from the waveguide and residual UV leakage. 
Signal and idler photons are both SMF-coupled, using empirically optimised coupling optics (see also appendix \ref{app_ch5_spdc_smf_mode_matching}). 
For this reason, the herald arm contains a telescope with an aspheric SMF-coupling lens (\textit{Thorlabs CME-220-TME-B}), while the signal is directly SMF-coupled (aspheric lens \textit{Thorlabs CME-230-TME-B}).

\paragraph{Signal and idler frequency filter stages}
The idler photon is inserted into a frequency filtering stage (\textit{purple panel}) to spectrally project the SPDC signal photon as described in sections \ref{ch4_source_design_theory} above. 
Thereafter it is detected on a single photon counting module (\spcmdt\,), which is an avalanche photodiode (\textit{Perkin Elmer, SPCM-AQ4C}), operated in Geiger mode. 

The signal is first coupled into an $83\,\text{m}$ long SMF. Thereafter it traverses a short free-space delay line (\textit{grey panel} in fig. \ref{fig_ch5_setup} \textbf{a}) and is coupled into a second, $7.97\,\text{m}$ long SMF, which sends it into another free space set-up (\textit{orange panel}). 
Both free space set-ups are placeholders for optics required to run the system with the Raman memory, as described in chapter \ref{ch6}. 
The former free-space propagation line (\textit{grey panel}) will allow us to switch the input signal type of the memory between \hsps\, and coherent states (\coh\,), which will be inserted into the second port of the PBS in this apparatus segment. 
The latter setup (\textit{orange panel}) represents the optics that will later surround the Raman memory. Eventually, the {\cs} cell will be inserted between the second pair of lenses in this set-up, while the first lens pair shapes the signal's spatial mode appropriately, such that it matches the memory control. 
The $83\,\text{m}$ long SMF introduces a time delay which will be required to prepare the memory control pulses. 
Details about these points are provided in section \ref{ch6_sec:setup}. 
These setup components are already in use here to test the source under the same conditions we will face, when operating it with the Raman memory. 
Behind the memory optics, the SPDC signal is again SMF-coupled and sent into the another frequency filtering stage, which is an updated version of the filter stage we have employed for the experiments in chapter \ref{ch4}. 
After passing through a series of etalons, described below (section \ref{ch5_subsec_filter}), the signal can be split on a PBS and detected by two single photon counting modules  \spcmdh\, and \spcmdv\,, which are again APDs operated in Geiger mode (\textit{Perkin Elmer SPCM-AQRH}). 
For most work in this chapter, besides the photon statistics measurements in section \ref{ch5_subsec_g2}, the signal is only inserted into the H-polarised arm. 
For measuring its photon statistics, it is split 50:50 between both modes and registered on both APDs.

\subsection{Characterisation of the frequency filter stages\label{ch5_subsec_filter}}


Frequency filter stages are required for two reasons: 
First, to correctly project the signal photon's frequency by appropriately filtering the idler photon (see section \ref{ch4_source_design_theory}). 
Second, to analyse the prepared \hsps\, within a narrow frequency range. 
Such frequency selection is necessary to extinguish the single photon fluorescence background in the waveguide, which would otherwise cause false signal counts (see appendix \ref{ch5_subsec_SPDC_temp_tuning}). 
Since this noise is broadband, it would contribute substantially to the counts observed on a single photon counting module without any frequency filtering in the signal arm\footnote{
	Because these noise counts are not necessarily temporally correlated with idler detection events,  
	this is not strictly true, when looking for coincidence detection events between signal and idler photons. 
	However, they can blind the detectors, due to detector dead time. In this regard, note that the
	noise only occurs when UV pump pulses are present, which are 
	time bins of $\sim 1\,\ns$ size. So the noise also concentrates in the same 
	$5\,\ns$ integration time windows
	of our FPGA acquisition system, which we use to look for the signal to be detected 
	(see section \ref{sec_ch4_photon_detection} below). 
}. 
As soon as the single photon source is interfaced with the memory (see chapter \ref{ch6}), frequency filtering of the signal is also essential to reduce memory noise and control field leakage (see chapter \ref{ch7}).  
For these reasons two separate filter stages are used for signal and idler photons. 
The signal filter stage is based on improvements of the earlier version\cite{Reim:2011ys,England2012}, shown in fig. \ref{fig_ch4_setup_fig}, and is customised to also filter the noise floor of the Raman memory, as described in chapters \ref{ch6} \& \ref{ch7}. 
It ensures that we detect the \hsps\, in the same spectral mode we observe later on, when interfacing the source with the memory. 
Consequently, when measuring the source's heralding efficiency $\eta_\text{her}$ in section \ref{ch5_hereff} below, we will look at the same modes as we do when analysing the \hsps\, retrieved from the memory in chapter \ref{ch6}. 
Accordingly, the measured $\eta_\text{her}$ is directly transferrable to a scenario including the memory, without any discrepancies in the spectral overlap between the filters and the incoming \hsps\,. 
The downside of this filtering mechanism is an additionally transmission loss, caused by the numerous elements in the signal filter stage, compared to the bare minimum filtering that would be required to extinguish the single photon fluorescence noise to a practical level. But this is not a major problem, as it can easily be backed out when calculating benchmark metrics such as the heralding efficiency $\eta_\text{her}$ (see eq. \ref{eq_ch5_her_eff}).

\paragraph{Etalon specifications}

Both filter stages contain a series of air-spaced, UV-fused silica Fabry-Perot (FP) etalons. 
These are made of two mirrors, separated by a distance $d$, which defines the free-spectral range FSR$=\frac{c}{2\cdot n_\text{air} \cdot d}$, with the refractive index of air $n_\text{air} =1$. 
The mirror reflectivities $R$ determine the reflectivity finesse $\mathcal{F}_R = \frac{\pi \sqrt{R}}{1-R}$. 
For an ideal etalon, these two parameters set the FWHM filter linewidth $\Delta \nu$ via $\mathcal{F}_R = \frac{\text{FSR}}{\Delta \nu}$. 
Two types of etalons are used: 
The first type has an FSR$=18.4 \GHz$. These were already contained in the previous setup (see fig. \ref{fig_ch4_setup_fig}) and were chosen for suppression of control leakage. Hence, their FSR equals $2\cdot \delta \nu_\text{gs}$, where $\delta \nu_\text{gs}=9.2\GHz$ is the {\cs} ground state detuning. 
The second etalon set has an FSR$=103\GHz$ and is designed mainly for spectral projection of the \hsps\,. Since the $103\GHz$ etalons have also proven useful in memory noise suppression (see chapter \ref{ch7}), they are inserted in both filter stages.
The etalons have the following specifications:
\begin{enumerate}
\item FSR$=18\GHz$: 
$R = (78 \pm 1)\,\%$, 
$d= (8.16 \pm 0.001)\mm$, 
$\mathcal{F}(852\nm) = 12$, 
$15\mm$ aperture 
\item FSR$=103\GHz$: 
$R = (95 \pm 1)\,\%$, 
$d= (1.456 \pm 0.001)\mm$,  
$\mathcal{F}(852\nm) = 68.9$, 
$25\mm$ aperture 
\end{enumerate}
\paragraph{Performance characterisation}
Initially, the signal filter stage was devised for control noise suppression in the memory output, where the main concern was the filter stage's FSR\cite{Reim2010, Reim:2011ys}. 
Yet, for operation with single photons, the actual bandwidths of the filter stages are also relevant, particularly for idler filtering.  
It is thus important to characterise the filter stage performances and take into account the actual experimental inefficiencies of the etalons. Etalon defects reduce the effective finesse and lead to a broader bandwidth\footnote{
	Insufficiencies are taken into account via a defect finesse\cite{Palik:1996} $\mathcal{F}_D$, 
	which is added to the
	reflectivity finesse\cite{Lvovsky:2012}:
	$$
	\frac{1}{\mathcal{F}_\text{eff}^2} = 	\frac{1}{\mathcal{F}_R^2} + \frac{1}{\mathcal{F}_D^2}.
	$$
	Since the etalon FSR remains constant, insufficiencies change the bandwidth to 
	$\Delta \nu = \frac{\text{FSR}}{\mathcal{F}_\text{eff}}$.
} $\Delta \nu$ than expected for the ideal case, when only considering $\mathcal{F}_R$. 

To this end, we firstly examine the expected performance, by, on the one hand, assuming ideal etalons and, on the other hand, taking into account defects. 
The calculation uses a series of FP etalons with transmission lines 
$T_\text{FP}(\nu) = \frac{T_\text{peak} \cdot (1-R_\text{eff})}{1 + R_\text{eff}^2 - 2\cdot R_\text{eff} \cdot \cos{\left(4\pi d \frac{\nu}{c} \right)}}$, whereby $R_\text{eff}$ is chosen to match\cite{McKay:1999} $\mathcal{F}_\text{eff}$. 
For an ideal etalon $R_\text{eff}=R$, i.e., it equals the mirror reflectivity. 
Since defects reduce the etalon transmission, described by 
$T_\text{peak} = \frac{1-R}{1+R} \cdot \frac{1+R_\text{eff}}{1-R_\text{eff}}$, 
we can use the measured, on resonance transmission and the specified reflectivity to extract $R_\text{eff}$ for our etalons and calculate their effective Finesse 
$\mathcal{F}_\text{eff} = \frac{\sqrt{\pi} R_\text{eff}}{1-R_\text{eff}}$.
In terms of the etalon transmission, we assume $T_\text{peak}$ as the average, experimentally achievable transmission for each etalon type 
(${T_\text{peak}^{18\GHz} \approx 70\,\%}$, ${T_\text{peak}^{103\GHz} \approx 76\,\%}$). 
With these, we obtain effective Finesse values of $\mathcal{F}_\text{eff}^{18\GHz}  = 8.7$ and 
${\mathcal{F}_\text{eff}^{103\GHz}  = 46.5}$ for the two classes of etalons. 
In turn, this allows us to estimate the performance for the filter sequences in the ideal and real case. 
We calculate the FWHM spectral bandwidth $\Delta \nu_\text{filt}^\text{real}$ and $\Delta \nu_\text{filt}^\text{ideal}$ for each frequency filter stage, which are stated in table \ref{tab_ch5_filter_fwhm} in appendix \ref{app_ch5_filter_measurement}. 

As a second part of the analysis, we measure the etalon transmission line directly.
Because the measurement of the \hsp\, spectra in section \ref{ch5_sec_hsp_bandwidth} follows the same methodology, we will briefly outline this experiment: 
The linewidth is measured by sweeping the {\tisa} pulses over the filter resonance. 
To this end, the {\tisa} detuning $\Delta$, to the blue of the {\cs} excited state manifold, is scanned, while simultaneously the {\tisa} input power ($P_\text{in}(\Delta)$) and the power transmitted through the filter stage ($P_\text{out}(\Delta)$) are measured. 
The relative power transmission ${\frak{P}_\text{trans}(\Delta) = \frac{P_\text{out}(\Delta)}{P_\text{in}(\Delta)}}$ is given by the convolution 
%
$\frak{P}_\text{trans}(\Delta) = \int_{\tilde{\nu}} T_\text{filt}(\nu) S(\nu - \Delta) \text{d} \nu$
%
between the filter resonance $T_\text{filt}(\nu)$ and the {\tisa} pulse spectrum $S(\nu)$. 
Knowing the {\tisa} pulse duration, this allows to determine the filter resonance line $T(\nu)$. 
However, in doing so, two assumptions must be made:

First, a pulse model has to be assumed for the {\tisa} pulses and the {\tisa} pulse duration must be known. 
As discussed in appendix \ref{ch3_tisa}, a sech pulse profile is a reasonable assumption in our case. 
Its pulse intensity spectrum $S(\nu,\Delta t)$, the FWHM pulse duration $\tau_\text{Ti:Sa}$ and the FWHM spectral bandwidth $\Delta \nu_\text{Ti:Sa}$ are given by\cite{Diplomarbeit}:
\begin{align}
S(\nu, \Delta t) = \sech^2{\left( \pi^2 \Delta t (\nu - \nu_0)\right)} \quad &, \quad
\tau_\text{Ti:Sa} =  2 \cdot \Delta t \cdot \text{arcsech}\left(\frac{1}{\sqrt{2}} \right), \nonumber \\
\Delta \nu_\text{Ti:Sa} &= \frac{2 \cdot \text{arcsech}(1/\sqrt{2})}{\pi^2 \Delta t}.
\label{ch5_eq_sech_pulses}
\end{align}
These depend on the pulse width parameter $\Delta t =183 \,\ps$, which is determined in appendix \ref{ch3_tisa}. 
Since we also require the spectrum of the UV pump pulses, we also assume a sech-profile, which resembles well the expected spectrum after a single SH-conversion of a fundamental sech-shaped {\tisa} spectrum\footnote{
	Notably, SHG corresponds to a convolution of the pump pulse spectrum with itself. 
	Since the sech-distribution is not stable and has finite variance, it convergence against a
	Gaussian pulse for multiple convolutions. 
	A convolution of $3$ sech distributions is already better described by a normal distribution than by 
	a sech distribution.  A convolution between $2$ sech distributions, as in SHG, is on the borderline: a 
	Gauss fit yielding an slightly smaller $R^2 =  0.998$ than a sech fit with $R^2 = 0.999$.	
} (see appendix \ref{ch3_SHG}). 
For comparison, we also consider a Gaussian pulse profile in appendix \ref{app_ch5_SPDCspec}.

Secondly, numerically stable deconvolution\footnote{
	The usage of direct deconvolution algorithms to extract $T(\nu)$ yields non-sensible results 
	and non-smooth filter transmission functions. The instability arises from low intensity values at
	detunings far away from the filter resonance. 
} 
of $\frak{P}_\text{trans}(\Delta)$ effectively requires the assumption of a filter transmission line profile. 
While the transmission line of each FP cavity is described by the Lorenzian function $T_\text{FP}(\nu)$, 
a sequence of $n$ such filters corresponds to the product ${T_\text{filt}^n(\nu) = \left(T_\text{FP}(\nu)\right)^n }$. 
For $n \ge 3$ etalons, $T^n_\text{filt}(\nu)$ is better approximated\footnote{
	Using the parameters for out $18\GHz$ etalons, a fit of $T_\text{FP}(\nu)$ onto
	$T^n_\text{filt}(\nu)$ has $R^2= 0.986$, whereas fitting with $T_\text{FP}^g(\nu)$ 
	yields $R^2= 0.996$
} by a Gaussian distribution, which we therefore use as the assumed filter line. 
The Gaussian lineshape and FWHM linewidth are:
\begin{equation}
T_\text{FP}^g(\nu,\sigma) = T_\text{peak} \cdot \exp{\left( -\frac{(\nu-\nu_0)^2}{\sigma^2} \right)}, \quad 
\Delta \nu_\text{filt}  = 2 \sqrt{\ln{(2)}} \cdot \sigma.
\label{eq_ch5_filter_function}
\end{equation}
Employing both functions, the filter transmission line can be determined by calculating the convolution between $T_\text{FP}^g(\nu)|_\sigma$ and the {\tisa} spectrum $S(\nu)=S(\nu, \Delta t_\text{Ti:Sa})$, as a function of its detuning $\Delta$ from the filter resonance, and fitting the results onto the measured values $\frak{P}_\text{trans}(\Delta)$. 
To this end, the $\frak{P}_\text{trans}(\Delta)$ data is normalised and the optimisation of the fitting routine runs on the filter bandwidth parameter $\sigma$. 
To determine the convolution during fitting, the Fourier transform theorem 
\begin{equation}
{P}_\text{trans}(\Delta) = \int_{\tilde{\nu}} T_\text{filt}(\nu)|_\sigma \cdot S(\nu - \Delta) \text{d} \nu =  
T_\text{filt}(\nu)|_\sigma \ast S(\nu)\,
\Longleftrightarrow\,
\tilde{{P}}_\text{trans}(t) = \left(\tilde{T}_\text{filt}(t)\right)|_\sigma \cdot \tilde{S}(t)
\label{eq_ch5_filter_convolution}
\end{equation}
is used. 
Here, the Fourier transform (FT) $\tilde{{P}}_\text{trans}(t) = \text{FT}\left( {P}_\text{trans}(\Delta) \right)$ of the predicted, convoluted power transmission function ${P}_\text{trans}(\Delta)$ is given by the product 
between the FT of the normalised filter transmission line $\tilde{T}_\text{filt}(t)|_\sigma$, for a fixed bandwidth $\sigma$, and the FT of the pulse spectrum 
$\tilde{S}(t) \sim I(t)$, which corresponds to the normalised pulse intensity envelope. 
Least squares fit optimisation for $\sigma$ yields the most likely filter bandwidth $\Delta \nu_\text{filt}$. 
With this procedure, the experimental values for the filter stage linewidths, stated in table \ref{tab_ch5_filter_fwhm} of appendix \ref{app_ch5_SPDCspec}, are determined\footnote{
	The uncertainties on the bandwidths are obtained from Monte-Carlo simulation as the 
	standard deviation of $\Delta \nu_\text{filt}$, when performing the fitting procedure 
	5000 times under variation of the assumed {\tisa} pulse duration and the 
	datapoints for ${P}_\text{trans}(\Delta)$ within their experimental uncertainty bounds. 
}. 
Notably, measuring the filter lines with the broadband {\tisa} laser introduces measurement uncertainties that could be avoided when performing the same measurement with a narrowband, tuneable laser, such as a diode laser. We have tried to use our stabilised diode laser for this task. However the mode-hop free tuning range was spectrally too narrow to give any useful results. As we had no other, narrowband, tuneable light source at $852\,\nm$ available, we used the {\tisa} laser. 

\begin{figure}[h!]
\centering
\includegraphics[width=\textwidth]{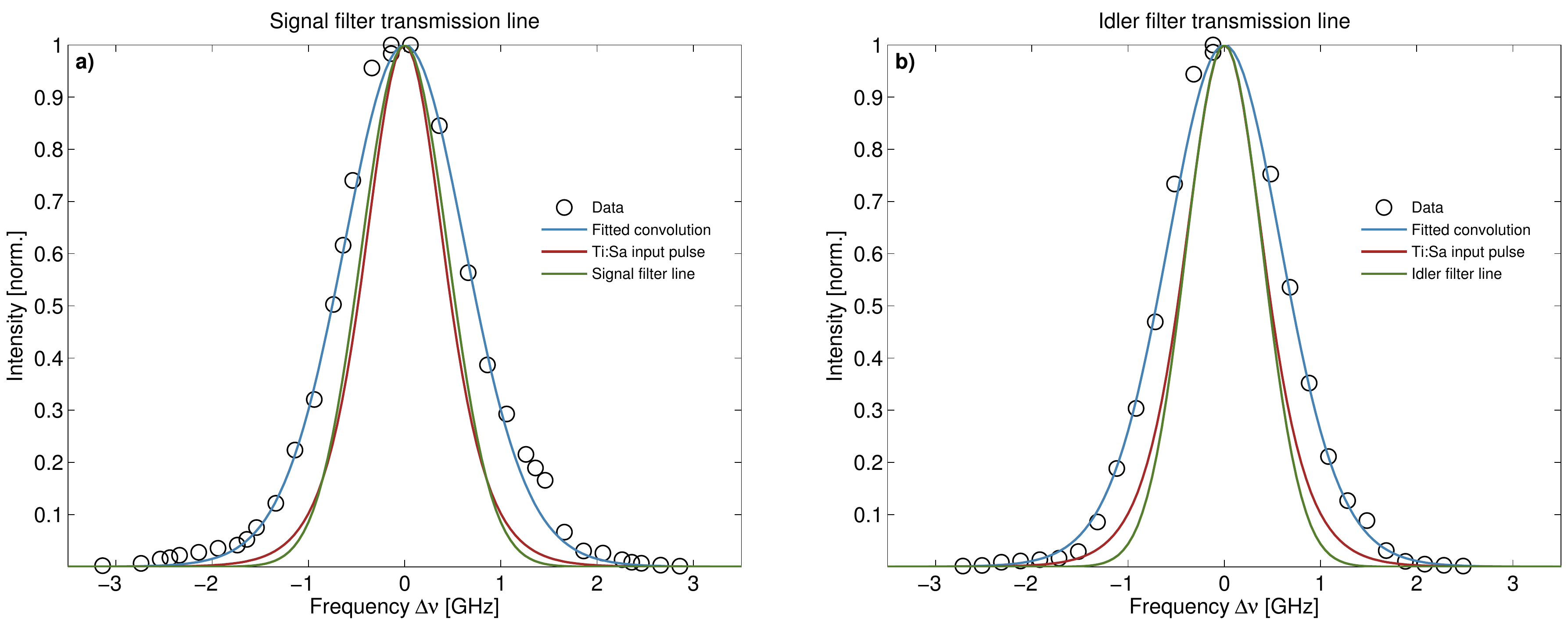}
\caption{Filter stage transmission measurements, for 
\textbf{(a)}: signal filter stage; 
\textbf{(b)}: idler filter stage. 
The displayed datasets assume sech pulses and Gaussian shaped filter lines. \textit{Blue lines} are the fitted convolutions onto the measured data (\textit{black points}). The {\tisa} spectrum is shown by the \textit{solid red lines}, and the resulting filter spectrum by the \textit{solid green lines}. 
}
\label{fig_ch5_filters}
\end{figure}

\paragraph{Signal filter stage}
The signal filter stage consists of three $18\GHz$ etalons and one double-passed $103\GHz$ etalon, as illustrated in the \textit{green panel} of fig. \ref{fig_ch5_setup} \textbf{a}. 
The filter combination is chosen empirically to optimise noise reduction at the memory output. 
The etalon resonance $\nu_0$ is positioned at $\Delta = 15.2\GHz$ detuning with respect to the $6^2 \text{S}_\frac{1}{2} \text{F} = 3 \rightarrow 6^2 \text{P}_\frac{3}{2}$ transition to match the signal frequency in the Raman memory scheme (fig. \ref{fig_ch5_intro} \textbf{i}). 

Fig. \ref{fig_ch5_filters} \textbf{a} shows the measured convolution data and the fit results for the filter bandwidth.   
Using the sech spectrum $S(\nu)|_{\Delta t_s}$ for our $\tau_\text{Ti:Sa} \approx 323 \ps$ long {\tisa} pulses (\textit{red}), which are convolved with the Gaussian filter line $T_\text{filt}(\nu)|_{\sigma_\text{opt}}$ (\textit{green}) yields the predicted power transmission spectrum ${P}_\text{trans}(\Delta \nu)$ (\textit{blue}). 
As can be seen by comparison with the measured, normalised {\tisa} power transmission ${\frak{P}}_\text{trans}(\Delta \nu = \nu_0-\Delta)$ (\textit{black}), the convolution fit agrees well with the data, whereby the optimal fit parameter $\sigma_\text{opt} = 0.64 \GHz$ predicts a FWHM filter bandwidth of\footnote{
	The data is summarised in table \ref{tab_ch5_filter_fwhm} of 
	appendix \ref{app_ch5_filter_measurement} together with the 
	prediction for Gaussian {\tisa} pulses. 
} ${\Delta \nu^\text{sig}_\text{filt} \approx 1.1 \GHz}$. 

Control leakage occurs at $\Delta = 9.2\GHz$ detuning to the blue of the filter resonance, as illustrated by the level diagram in fig. \ref{fig_ch5_intro} \textbf{i}. 
Besides leakage, the filter stage also eliminates fluorescence noise and, importantly, FWM noise in the anti-Stokes channel of our $\Lambda$-system (see fig.~\ref{fig_ch2_FWMlevels}~\textbf{b}), which has a detuning of 
$\Delta = 24.4\,\GHz$ to the blue of the filter's line. 
The expected filter transmission for the control pulses is 
$T_\text{filt}^{\text{sig}}(9.2\,\GHz) \sim 1 \cdot 10^{-9}$, 
whereby the anti-Stokes transmission is 
$T_\text{filt}^{\text{sig}}(24.4\,\GHz) \sim 1 \cdot 10^{-10}$. 
The downside of such heavy filtering is a total, on-resonance stage transmission\footnote{
	For the individual optical components in the stage the following transmissions have been 
	measured on average: 
	$T_{18}^1 \approx 65\,\%$, 
	$T_{18}^2 \approx 77\,\%$, and
	$T_{18}^3 \approx 77\,\%$ for the three $18\GHz$ etalons, 
	$T_{103}^1 \approx 66\,\%$ and 
	$T_{103}^2 \approx 62\,\%$ for the double passed $103\GHz$ etalon, and
	$\eta_\text{MMF} \approx 91\,\%$ for MMF-coupling.
} of $T_\text{filt}^\text{sig.}(\text{15.2 GHz}) \approx 14.5\,\%$ (on average). 
Both numbers yield an extinction factor for control leakage of 
$\epsilon_\text{filt}^{\text{sig}} (9.2\,\GHz)= \frac{T_\text{filt}^{\text{sig}}(9.2\,\GHz)}{T_\text{filt}^{\text{sig}}(\text{0 GHz})} \approx 1.5 \cdot 10^{-8}$.
This value has to be contrasted to the performance of set of ideal etalons with $100\,\%$ on resonance transmission, which would yield  $\epsilon_\text{filt}^{\text{sig}} (9.2\,\GHz)  \approx 2.5\cdot 10^{-10}$. The difference is caused by the below unity transmission and the line broadening, due to 
$\mathcal{F}_\text{eff} < \mathcal{F}_{R}$. 
An improvement would be possible if the double-passed $103\,\GHz$ etalon was replaced by an etalon with FSR$= 48.8\,\GHz$. Its better anti-Stokes channel suppression, resulting from anti-resonance of the AS frequency with the etalon's resonances, could allow operation in single pass, boosting the transmission by approximately a factor of 2. 

\paragraph{Idler filter stage}
The idler filter stage contains two $18\GHz$ etalons and one double passed $103\GHz$ etalon, as shown by the \textit{purple panel} in fig. \ref{fig_ch5_setup} \textbf{a}. 
Additionally, it comprises a $100\GHz$ bandwidth reflective volume holographic grating filter (\textit{Ondax}). 
The latter filters the SPDC idler photon spectrum down to a bandwidth of $100\GHz$, which is narrowed further by the $103\GHz$ etalon to arrive at a single line with a width on the order of $1\GHz$. 
Due to the large FSR of the $103\GHz$ etalon, its transmission function does not tail off rapidly enough when moving off resonance to achieve sufficient suppression of uncorrelated fluorescence noise in the idler mode.
To mitigate this, the $18\GHz$ etalons, which have sharper resonance lines, are inserted. 
As introduced in section \ref{ch5_subsec_intro_principle}, the filter stage resonance $\nu_{\text{i},0}$ is set to $\Delta = +24.4 \GHz$ detuning from the $6^2\text{P}_\frac{3}{2}$ manifold. The total, on-resonance stage transmission\footnote{
	This is measured with bright {\tisa} pulses tuned into resonance with the filters, i.e. 
	$T_\text{filt}^\text{idl} = \frak{P}_\text{trans}(0)$. The measured transmissions of each component 
	are,
	on average, $62\,\%$ and $73\,\%$ for the $18\GHz$ etalons, $90\,\%$ and $77\,\%$ for the 
	double-passed $103\GHz$ etalon, $87\,\%$ for the grating and $97\,\%$ for MMF-coupling.
} amounts, on average, to $T_\text{filt}^\text{idl} \approx 26\,\%$. 

The idler filter's transmission line is shown in fig. \ref{fig_ch5_filters} \textbf{b} (\textit{green line}).
Likewise to the signal filter stage, the fitted convolution $P_\text{filt}(\Delta \nu)$ (\textit{blue}), for $\sigma_\text{opt} = 0.57\GHz$, is in good agreement with the measured {\tisa} pulse transmission $\frak{P}_\text{trans}(\Delta \nu = \nu_0 -\Delta)$ (\textit{black}).
The corresponding FWHM filter linewidth $\Delta \nu^\text{idl}_\text{filt} \approx 0.94\GHz$ closely matches the bandwidth $\Delta \nu_\text{Ti:Sa} \approx 0.98 \GHz $ of the sech-shaped {\tisa} pulses (\textit{red}). 

To test the spectral projection of the SPDC signal photons later on, we also vary the idler filter components to prepare different filtering bandwidth. These different configurations are listed, together with the above results, in table \ref{tab_ch5_filter_fwhm} of appendix \ref{app_ch5_filter_measurement}.

\section{Photon detection - from clicks to performance metrics\label{sec_ch4_photon_detection}}

We continue our set-up discussion with the detection system for single photons. 
Besides the technical components, we introduce the observed signals, the ways they are recorded and how they are transformed into detection probabilities, which are used to determine the heralding efficiency and the photon statistics. 
To this end, we also define the {\gtwo}-autocorrelation measurement, as a metric for the photon statistics. 

\subsection{Single photon detection\label{ch5_subsec_detection_system}}

\paragraph{Single photon counting module (SPCM)}
As mentioned in section \ref{ch5_sec_source_setup}, we use single photon avalanche photodiodes (APD) (\textit{Perkin Elmer}) to detect photons. 
Used in Geiger mode, these devices produce a TTL voltage output when a photon hits the detection area. While the exact physics of this process is described elsewhere\cite{Yariv:hc}, 
we note that detection has a wavelength dependent efficiency; for $852\,\nm$ light, the detection efficiency is 
$\eta_\text{det}\approx 50\,\%$. 
Apart from incoming signals, the detectors also produce dark counts, with a frequency of 
$c_\text{dc} \approx 10-20\, \frac{\text{photons}}{\text{sec}} = 10-20 \,\Hz$. 
Because detection uses a charge avalanche, the detectors have a deadtime of 
$\tau_\text{det} \approx 30\,\ns$ after each photon registration event\footnote{
	This is the time required to firstly remove all electric charges in the semiconductor p-i-n junction, 
	forming the active detection area. Secondly, the p- and n-doped areas need to be 
	re-charged with hole and electrons again to be ready for the next detection event. 
}. 
This time imposes an upper count rate limit of 
$c_\text{det}^\text{max} \sim \frac{1}{\tau_\text{det}} \approx 33 \,\MHz$ 
the detectors can respond to.  
However, our actual detection rates fall well below this limit (see section \ref{ch5_subsec_rates}), so there are no problems with detector saturation. 
Notably, if the signal arm was unfiltered, the waveguide's broadband fluorescence background (see appendix \ref{app_ch5_SHGwaveguide}) would generate enough noise counts for the signal detectors to get blinded\footnote{
	Note that the waveguide source is pumped with $f_\text{rep} = 80\,\MHz$ repetition rate. 
	So, in the absence of loss in the signal arm,  generating a noise photon upon every second pump 
	pulse would be enough to blind the detectors. 
}. 
We can measure the number of output pulses $N_\text{det}$ within a set integration time 
$\Delta t_\text{int}$, yielding the observed count rate $c =\frac{N_\text{det}}{\Delta t_\text{int}}$. 
These counts are registered on a continuous basis during $\Delta t_\text{int}$, i.e., they are not time-gated; we refer to these as singles events. 
Timing information can be obtained, by measuring the output of two or more detectors in coincidence\cite{Kiesel:PhD}. Here, events are only counted when there are simultaneous outputs from all detectors\footnote{
	Simultaneous mean, they are both detected within a given time interval, called the
	coincidence window.
}. 
We term these events double and triple coincidences, depending on whether they include two or all three APDs in our setup (see fig. \ref{fig_ch5_setup}).  
Such counting in possible by feeding the APD TTL outputs into a field-programable gate array (FPGA). 
Detectors \spcmdt\, and \spcmdh\, also feed into a time-to-amplitude converter (TAC), followed by a multi-channel analyser (MCA), allowing to record photon arrive time histograms, which resemble the time series traces in fig.~\ref{fig_ch4_trace_PD_750ns} and are used in chapter~\ref{ch6}. 

\paragraph{TAC/MCA}
The combination of both devices performs histogram binning of signal events with respect to a start trigger pulse. To this end, the time delay between the start trigger and the signal, which acts as a stop trigger, is measured and binned into different channels, according to the time delay between both triggers. 
For each channel, the number of events are counted within a set integration time $\Delta t_\text{meas}^\text{TAC}$, which can take any time greater than 1 sec. 
When used in single photon detection, the start triggers are photon detection events on the idler detector \spcmdt\,, whereas signal detection on \spcmdh\, are the stop triggers.
The resulting recording shows the count histograms of signal photons with respect to a heralding event. Since both are correlated through the simultaneous production of the SPDC photon pair, we obtain a temporal pulse trace, resembling a scope trace of a laser pulse detection signal on a linear photodiode (see fig.~\ref{fig_ch4_trace_PD_750ns}).

On the technical side, the TAC performs arrival time measurements, generating output voltage pulses with amplitudes $V_\text{TAC}$ corresponding to the time differences $\delta t_\text{TAC}$ between the start (\spcmdt\,) and the stop (\spcmdh,) triggers. 
The input signals must be TTL-like voltage pulses with amplitudes $V_\text{TAC} \gtrsim 1.5 \, \text{V}$. 
Such pulses are supplied by the APDs. 
The time ranges $\delta t^\text{max}_\text{TAC}$ over which $\delta t_\text{TAC}$ can vary, are user-definable and range from $5\ns$ up to $3\mus$. 
The TAC output is fed into the MCA unit, which bins it into $n_\text{ch} = 16385$ channels according to the voltage level $V_\text{TAC}$. 
Each bin thus represents a time increment of $\delta t_\text{MCA} = \frac{\delta t_\text{TAC}^\text{max}}{n_\text{ch} }$ with respect to the TAC start trigger. 
Events falling into each time bin are summed over the integration time $\Delta t_\text{meas}^\text{TAC}$.
Notably, the start trigger does not necessarily have to be an APD output. 
In chapter \ref{ch6}, we will also detect coherent states at the single photon level with the TAC/MCA system\cite{Reim:2011ys,Reim:PhD}.   
Here, the start trigger is the {\tisa} reference clock signal (see appendix \ref{ch3_tisa}).

\paragraph{Field programmable gate array}
The FPGA works as a photon counter, which allows registration of single detection events supplied to each of its 8 input channels and also signal coincidences between two or more channels\footnote{
	A more detailed description of FPGA and its coincidence logic can be found in the 
	thesis of \textit{Justin Spring}\cite{Spring:PhD}, who has kindly provided the FPGA 
	program for our experiment. 
}. 
It requires input pulse voltages of $V_\text{FPGA} \gtrsim 1.3 \, \text{V}$ for events to be registered and can thus be directly supplied by the APDs. 
Counting of individual channels yields the aforementioned singles counts, which have no precise timing information within the integration time of $\Delta t_\text{int}^\text{FPGA}$. 
Coincidence counting between the inputs of two or more channels adds this information, allowing to  confine the arrival times into a much smaller time period, set by the coincidence window $\Delta t_\text{coinc}^\text{FPGA}\ge 2.5 \,\ns$. 
For coincidence to be counted, the rising flanks of the input signals have to be registered within the coincidence window, so the pulses need to be temporally synchronised.  
Here, we use $\Delta t_\text{coinc}^\text{FPGA}=5 \,\ns$, unless stated otherwise, which effectively locates events within a single {\tisa} emission time bin. 

The requirement for temporal confinement of the detection events arises from the presence of continuous background noise, e.g. detector dark counts or the optical pumping (see chapter~\ref{ch6}). 
Such noise contributions can overshadow the desired signal counts, because the signal only adds events during a small subsample of $\Delta t_\text{int}^\text{FPGA}$, whereas the noise is on continuously. So, even if the actual frequency of noise photon generation is low, its integrated value over 
$\Delta t_\text{int}^\text{FPGA}$ can heavily exceed the number of signal events. 
Coincidence counting effectively cuts down the integration window size to the much shorter period
$\Delta t_\text{coinc}^\text{FPGA}$. 
Therein, the accumulated counts approximately resemble the actual frequencies of signal photons and background noise, which are dominated by the signal count rates. 

Here, we count the signals from \spcmdh\, and \spcmdv\, in coincidence with events from \spcmdt\,, whereby we record double coincidences $c_\text{H,T}$ and $c_\text{V,T}$ between the APDs \spcmdt\,-\spcmdh\, and \spcmdt\,-\spcmdv\,, as well as triple coincidences $c_\text{H,V,T}$ between all three APDs \spcmdt\,-\spcmdh\,-\spcmdv\,. 
In all cases, heralding events on \spcmdt\, define the $\Delta t_\text{coinc}^\text{FPGA}$-sized time window, within which the signals from \spcmdh\, and \spcmdv\, are observed.  
To this end, events on \spcmdt\, must be delayed appropriately for simultaneous arrival with the other signals, as the longer optical propagation path of the signal photons needs to be cancelled. 
The FPGA sums all counts $N_j$, with $j \in \left\{ T,H,V, (H,T),(V,T),(H,V,T),\right\}$, 
within the integration time of 
$\Delta t_\text{int}^\text{FPGA} \le 10\,\text{s}$, 
yielding the singles, double and triple coincidence count rates 
$c_j=\frac{N_j}{\Delta t_\text{int}^\text{FPGA} }$. 
To span larger measurement times, resembling $\Delta t_\text{meas}^\text{TAC}$, the FPGA records several runs; each run contains all registered counts $N_j$ within one unit of $\Delta t_\text{int}^\text{FPGA}$.

Notably, every FPGA channel can only be used to observe a single time bin. 
So, to account for the read-in and retrieval time bins of the memory later on, we 
need two FPGA channels for each APD \spcmdh\, and \spcmdv\,. 
Both detector outputs are thus split into two copies, each feeding into a separate FPGA channel. 
The delay between the resulting channel pairs for each of the signal detectors corresponds to the memory storage time $\tau_\text{S}$. 
Moreover, to also enable counting of coherent state signals at the single photon level, we will use an additional trigger signal, which derives from the {\tisa} clock rate. The generation of these triggers, which are a variant of the Pockels cell triggers in chapter \ref{ch4}, is detailed in section \ref{ch6_sec:setup}.

\subsection{Conversion of counts to performance parameters\label{subsec_ch5_fpga_count_rate_definition}}

We will now discuss how the detected FPGA counts $N_j$ are used to obtain the count rates and detection probability numbers we require for benchmarking our system's performance below and in chapters \ref{ch6} \& \ref{ch7}. The processing of the TAC/MCA histogram traces is presented in section \ref{ch6_sec3_subsec1}.  

\paragraph{Count rates}
Unless explicitly stated, we use an FPGA integration time of $\Delta t^\text{FPGA}_\text{int} = 10 \sec$ throughout this work. 
While in this chapter we only measure the properties of the \hsp\, as input signals to the memory, for actual \hsp\,-storage, we will observe count rates for different measurement settings $i$, which are combinations of input signal, control and the optical pump applied to the memory. This is similar to chapter \ref{ch4}, where we have selectively un-/blocked the control to observe signal storage 
(see fig. \ref{fig_ch4_trace_PD_750ns}). 
The counts $N_{i,j}$ for each such setting $i$ and detector combination $j$ will be measured for the read-in and read-out time bins, denoted by $t$. 
While in this chapter we only have one time bin, as the \hsps\, are not stored yet, we nevertheless introduce the notation here already. 
So we have counts $N_{i,j}^{t}(t_m)$ for each FPGA run $t_m$, which are integrated for a time 
$\Delta t^\text{FPGA}_\text{int}$, yielding count rates $c_{i,j}^t(t_m) = \frac{N_{i,j}^{t}(t_m)}{\Delta t^\text{FPGA}_\text{int}}$. 

For benchmark parameters that only require a single measurement setting $i$, we can take the average over all points $t_m$ and obtain the average count rates $\bar{c}^{t}_{i,,j} = \underset{m}{\sum} c^t_{i,j}(t_m)$, with a standard error\footnote{
	The standard error is given as 
	$\Delta c_{i,j}^t = \frac{\text{std}\left(c^t_{i,j} (t_m)\right)}{\sqrt{| \left\{ m \right\}|}}$, 
	where $| \left\{ m \right\}| $ is the number of measurement runs $m$ and std stands for the 
	sample standard deviation. 
} $\Delta \bar{c}^t_{i,j}$. 
Notably, the errors $\Delta \bar{c}^t_{i,j}$ implicitly assumes a normal distribution for the average count rates $\bar{c}^t_{i,j}$. The validity of this assumption is demonstrated in section~\ref{ch6_subsec_meas_method} and appendix~\ref{app6_data_aggregation}.

\paragraph{Scaled count rates}
However, when more than one setting is involved, we need to take into account the variation of the experimental repetition rate.
It is set by the idler photon detection rates $c_{i,T}(t_m)$ on \spcmdt\,, which define the coincidence windows.
These vary over different time intervals $t_m$, because the SPDC pair generation is a probabilistic process. 
Consequently, triggering the experiment by idler detection results in a continuous variation in the number of detectable coincidences.
Since different settings $i$ can only be measured sequentially, these variations need to be taken into account to avoid skewing of the counts $c_{i,j}^t(t_m)$. 
To this end, we introduce the multiplicative factor 
$\textfrak{s}_i =  \frac{c_{i,T}(t_m)}{\bar{c}_{i,T}}$ for each setting $i$, 
which is the normalised variation of the heralding rate. 
These factors are applied to the measured counts, leading to updated numbers
$\tilde{c}^t_{i,j} (t_m)= \textfrak{s}_i  \cdot c^t_{i,j}(t_m) $ 
and 
$\tilde{\bar{c}}^t_{i,j} = \textfrak{s}_i  \cdot \bar{c}^t_{i,j}$  
for each setting $i$, detector combination $j$ and time bin $t$. 
Obviously, using a deterministic trigger of constant frequency, such as a divided-down derivative\footnote{
	The $80\,\MHz$ {\tisa} clock signal needs to be divided-down in frequency 
	to be usable in gating the FPGA coincidence windows. This follows from the 
	upper limit on a signal's repetition rate of $\sim 20\,\MHz$ the FPGA is able to process. 
} of the {\tisa} clock rate, the number of experiments is constant, i.e. $\textfrak{s}_i =1, \, \forall i$.
For \hsp\, storage, we consider it understood from now on, that rescaling by $\textfrak{s}_i$ is applied to all count rates and drop the tilde. 

\paragraph{Detection probabilities} 
When dealing with coincidence counts, knowing the probability to detect a photon on the signal APDs, when performing an experimental trial, will become quite useful later on. 
This gives rise to the detection probabilities, which are the number of signal counts, normalised by the experimental repetition rate $f_\text{rep}$. 
For \hsp\, production, we have $f_\text{rep} = \frac{1}{c_T}$. 
When using the APD \spcmdt\, for heralding, we thus end up with detection probabilities of $p^t_{i,j}(t_m) = \frac{c_{i,j}^{t}(t_m)}{c_{T,j}(t_m)}$. Note here, that the repetition rate $\frac{1}{c_{T,j} (t_m)}$ is of course memory time bin independent. 
For $i \in \left\{ (H,T), (V,T), (H,V,T) \right\}$, the $p^t_{i,j}$ correspond to the conditional probabilities of finding signal photons in the H- or/and V-arm, when having an idler photon in mode $T$. 
Consequently, we can also use the notation $p^t_{k,j}$, with $k \in \left\{ H|T, V|T, (H,V)|T \right\}$, for these probabilities. 

\paragraph{Definition of heralding efficiency}
The above count rate scaling by the idler events is also required for determining the source heralding efficiency \hereff\,. 
It is defined as the probability of obtaining an SPDC signal photon upon a heralding event. 
Besides the \hsp\, spectrum, it is the second crucial source parameter, as it determines the effective number of photons sent into the memory upon every experimental trial. 
By definition we must have $\eta_\text{her}\le 1$. We will see the consequences of this relation in section \ref{ch6_subsec_SNR}.
When considering only the actual detection rates, i.e. \hsp\, preparation trials are only counted as successful when the SPDC signal photon makes it to the detector, the heralding efficiency reads\cite{Moseley:2008} 
$\tilde{\eta}_\text{her} = \frac{\bar{c}_{H,T}+\bar{c}_{V,T}}{\bar{c}^\text{her}_T}$. 
Note, when measuring the heralding efficiency, we only need the setting \textit{s}, which represents the unblocked input signal\footnote{
	This is true only when the {\cs} cell is not inserted into the beam path of the SPDC signal photons.
	If it is included, we also require active optical state preparation to avoid linear absorption by the
	warm vapour. Accordingly the diode laser has to be on as well and \hereff\, is measured by the setting
	\textit{sd}, standing for signal \& diode. 
}, so we can use the average count rates $\bar{c}_{i,j}^t$ for $j = \textit{s}$. 
Moreover, the signal can be split between the detectors \spcmdh\, and \spcmdv\, (see chapter \ref{ch6}), so we add up the coincidences observed on each detector. 
In our case, this number amounts to $\tilde{\eta}_\text{her} \approx 1\,\%$. 

However, when quoting \hereff\,, the actual application of the \hsps\, should to be taken into account, which, for us, is the insertion into the Raman memory.
In this regard, it is sensible to define the \hsp\, preparation efficiency with respect to the probability of sending a \hsp\, into the {\cs} cell, rather than with respect to its detection. 
$\tilde{\eta}_\text{her}$ thus needs to be corrected for the detection efficiency $\eta_\text{det}$, 
as well as the signal's transmission $T_\text{sig}^\text{tot}$ from the {\cs} cell's input facet\footnote{
	With the Raman memory present in the signal's beam path, this transmission needs to be 
	measured without Raman absorptions. 
	For the data presented in this chapter, the {\cs} cell has been removed from the beam path, 
	so $T_\text{sig}^\text{tot}$ is essentially the transmission of the signal filter stage. 
}  to the signal APDs. This yields a heralding efficiency of 
\begin{equation}
\eta_\text{her} = \frac{c_{H,T}}{c_T \cdot T_\text{sig}^\text{tot} \cdot \eta_\text{det}} = 
\frac{p_{H|T}}{T_\text{sig}^\text{tot} \cdot \eta_\text{det}}. 
\label{eq_ch5_her_eff}
\end{equation}

\subsection{Photon statistics via {\gtwo} autocorrelation\label{ch6_subsec_g2intro}}
\begin{wrapfigure}{R}{0.5\textwidth}
\centering
\begin{framed} 
\centering
\includegraphics[width=\textwidth]{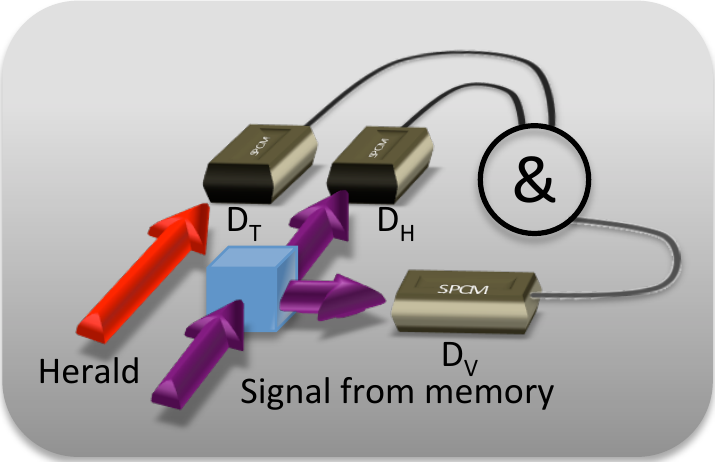}
\caption{Hanbury-Brown-Twiss scheme for heralded $g^{(2)}$-measurements.}
\label{fig_ch6_HBT}
\end{framed}
\end{wrapfigure}
In section \ref{ch5_subsec_design_criteria} we have introduced the necessity for the source to produce \hsps\, with good photon number purity, which is assessed by the photon number statistics. 
Photon number purity preservation is also one of the key memory performance benchmarks, as the quantum characteristics of the input signal have to be maintained for faithful memory operation. 
This is easily understood considering the application of the source-memory system in temporal multiplexing: Clearly, any additional photons coming out of the memory would negatively affect any subsequent quantum gate operations. 
To investigate the signal's photon statistics\cite{Loudon:2004gd}, we measure correlations in a Hanbury-Brown-Twiss-type experiment \citep{Loudon:2004gd,HBT:1956}, illustrated in fig. \ref{fig_ch6_HBT}. 
Here, the signal is split up 50:50 into two spatial modes which are detected in coincidence, while its arrival time in one arm is successively delayed. 
In case of a single photon input, the single particle can only travel along one arm at a time, 
so there are no coincidences for a delay of $\tau=0$, while for $\tau \ne 0$ coincidences can occur. 
Conversely, inputs containing more than one photon can split up into configurations that enable coincidence events at $\tau=0$. 
For an investigation of the single photon character, it is thus sufficient to evaluate $\tau=0$ only. 
The correct time bin $\tau$ is specified by counting triple coincidences between all APDs (\spcmdh\,, \spcmdv\,, \spcmdt\,). 
This measurement is commonly referred to as heralded $g^{(2)}$-autocorrelation, with 
$g^{(2)}(\tau=0) =: g^{(2)}$ defined by
\begin{equation}
g^{(2)}_{j,t}  = 
\frac{\langle \hat{\tilde{a}}^\dagger_{\text{s},H} \hat{\tilde{a}}^\dagger_{\text{s},V} \hat{\tilde{a}}^{}_{\text{s},V} \hat{\tilde{a}}^{}_{\text{s},H}\rangle_{T,j,t}}{\langle \hat{\tilde{a}}^\dagger_{\text{s},V} \hat{\tilde{a}}^{}_{\text{s},V} \rangle_{T,j,t} \cdot \langle \hat{\tilde{a}}^\dagger_{\text{s},H} \hat{\tilde{a}}^{}_{\text{s},H} \rangle_{T,j,t}} 
= \frac{p^t_{((H,V)|T),j}}{p^t_{(H|T),j} \cdot p^t_{(V|T),j}} 
= \frac{c_{(H,V,T),j}^{t} \cdot c_{T,j}}{c^{t}_{(H,T),j} \cdot c_{(V,T),j}^{t}}. 
\label{eq_ch6_g2}
\end{equation}
Here, the SPDC signal annihilation operator $a_{\text{s}}$ from eq. \ref{eq_ch5_spdc_state} is split into the two modes $\tilde{a}_{\text{s},H}$ and $\tilde{a}_{\text{s},V}$, detected on APDs \spcmdh\, and \spcmdv\,, respectively. The subscript $T$ denotes that the signal detection is conditioned on a heralding event\footnote{
	The theory expectation value\cite{Branczyk:2010} needs to be evaluated using the marginalised 
	SPDC state after heralding (eq. \ref{eq_ch5_marg_SPDC_filtered} of 
	appendix \ref{app_ch5_multimode_spdc_heralding}). 
}, the subscripts $\left\{j,t\right\}$ indicate the measurement setting and time bin and 
$p^t_{((H,V)|T),j}$ is the probability of observing triple coincidences. 
The latter is normalised by the product of the double coincidence probabilities 
$p^t_{(H|T),j}$ and $p^t_{(V|T),j}$. 
Similar to \hereff\,, we only observe setting $j = s$ in this chapter, i.e., we only send the input signal onto the APDs \spcmdh\, and \spcmdv\,. 
In chapters \ref{ch6} \& \ref{ch7} we will also study other field combination settings. 
Inserting the memory into the signal arm adds the subtlety of having to count coincidences for the read-in and read-out time bins, which is accounted for in eq. \ref{eq_ch6_g2} via the subscript $t$. 
Note that eq. \ref{eq_ch6_g2} assumes negligible higher order SPDC emissions\cite{Migdall:book}. 
Theoretically \cite{Loudon:2004gd}, one would expect $g^{(2)}= 0$ for true single photons and $g^{(2)} = 1$ for coherent states. Thermal states, such as noise, have $g^{(2)} = 1 + \frac{1}{K}$, where $K$ denotes the number of collected modes; so a single mode thermal state has $g^{(2)} =2$ (see section \ref{ch7_g2noise}). 

Besides the {\gtwo} autocorrelation, there is also the cross-correlation $G^{(2)}$ \citep{Lee:2011, Rielander:2014aa, England:2014}, which uses correlations between the signal and idler modes of an SPDC source\footnote{
	Occasionally, for characterisation of SPDC sources, $G^{(2)}$ is also referred to 
	as $g^{(1,1)}$ to denote that it is a coincidence measurement between 
	the signal and idler arm\cite{Christ:2011}.
}. 
It is based on coincidences between signal and idler detection events, i.e., splitting the signal arm into two modes is not required. 
These coincidences are normalised by the product of the unconditional signal and idler singles counts. $G^{(2)}$ is thus analogous to eq. \ref{eq_ch6_g2} with 
$G^{(2)} = \frac{p_{H|T} + p_{V|T}} {(p_{H} + p_{V}) \cdot p_T}$. 
Since it incorporates count rates of free-running APDs, it is very sensitive to background noise, as mentioned above. 
For this reason, experiments using this metric commonly gate the detection on the preparation of the SPDC pump pulses\cite{Spring:2013ab}, which here would be the {\tisa} clock signal. 
Since the processing bandwidth of our present FPGA system is not high enough to register input signals at $80\,\MHz$, we did not consider a $G^{(2)}$-measurement in this thesis.

\section{Source characterisation measurements\label{sec_ch5_source_characterisation}}

We now move on to the experimental parts of this chapter, where we discuss the source characterisation measurements. 
Before running the system to produce \hsps\,, we have tested and optimised the nonlinear frequency conversion in the reverse direction, generating SH by pumping the waveguide with $852\,\nm$ light from our {\tisa} master laser, and tuning the waveguide's temperature. 
Switching to SPDC thereafter, we have also measured the temperature dependence of the SPDC generation efficiency and observed the full, broadband emission spectrum of the waveguide, including any fluorescence noise. 
These initial measurements are presented in appendix \ref{app_ch5_SHGwaveguide}. In the following we will start our characterisation by looking at the spatial mode structure of the SPDC in the waveguide. Subsequently, we will presented the achievable count rates, the heralding efficiency as well as the {\gtwo} photon statistics measurements on the \hsps\,. 
Last, but not least, we will measure the spectrum of the \hsp\,. 

\begin{figure}[h!]
\centering
\includegraphics[width=\textwidth]{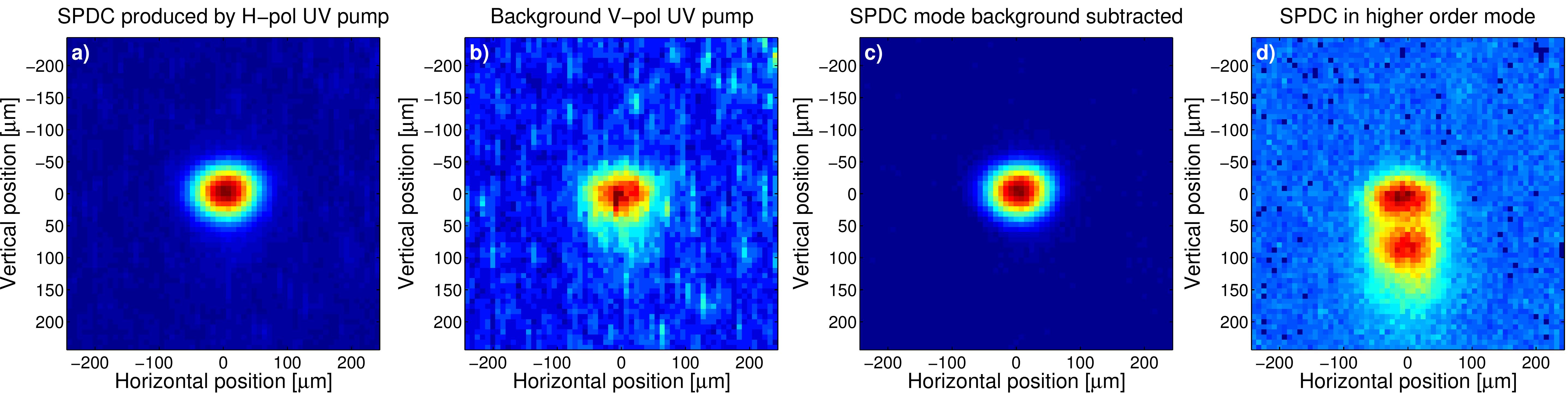}
\caption{SPDC spatial mode structure excited in waveguide 3.2.
\textbf{(a)}: Fundamental mode of the SPDC photons produced by the H-polarised UV pump. 
\textbf{(b)}: Fluorescence mode observed for frustrated SPDC with a V-polarised UV pump.  
\textbf{(c)}: Mode of pure SPDC photons obtained by subtracting the fluorescence background of \textbf{(b)} from the SPDC mode in \textbf{(a)}. 
\textbf{(d)}: First order SPDC mode produced by the H-polarised UV pump. 
The dimensions of the modes are those on the EM-CCD camera, i.e. the modes are magnified by a factor $M_i=43.5$. 
For good visibility of the mode structure, the intensities in all images are rescaled to use the full 8-bit dynamic range of the camera.}
\label{fig_ch5_exp_modes}
\end{figure}

\subsection{SPDC spatial modes in the waveguide\label{ch5_sec_modes}}

To observe the SPDC mode structure in the waveguide, we will replace the regular output coupler lens behind the waveguide by the L40X microscope objective, shown in fig. \ref{fig_ch5_setup} \textbf{b}, and image the output facet of the waveguide with an $f_i=200\mm$ focal length lens on the EM-CCD camera (see section \ref{ch5_sec_source_setup}). The camera thus observes the modes magnified by a factor of $M_i \approx 43.5$ with respect to their actual size inside the waveguide. 
Apart from the SPDC modes themselves, we also analyse the modes of the $852\,\nm$ {\tisa} radiation coupled into the guide 3.2, the SH this IR-light generates, as well as the modes for the $426\,\nm$ UV-pump in the waveguide. These auxiliary results are presented in appendix \ref{app_ch5_spdc_modes}. 
Here, we focus exclusively on the SPDC modes, excited in waveguide channel 3.2. 

By appropriate adjustment of the coupling conditions for the UV-pump into the guide, we can generated SPDC photons in their fundamental spatial mode, as shown in fig. \ref{fig_ch5_exp_modes} \textbf{a}. 
The FWHM mode size is similar to that of the $852\,\nm$ IR radiation from the {\tisa}, coupled into the guide (see table \ref{tab_ch5_mode_sizes} in appendix \ref{tab_ch5_mode_sizes}).
Because the EM-CCD camera is not triggered by a heralding event\footnote{
	This would require an i-CCD camera, which we do not have available.
}, the SPDC mode also contains the single photon fluorescence noise background, which falls into the $852\pm 10\nm$ wavelength range of the bandpass filter, placed in front of the camera (see fig. \ref{fig_ch5_setup} \textbf{b}).
By frustrating the down-conversion, using a V-polarised UV-pump instead of the H-polarisation required to achieve phase-matching, the mode of the noise background is separated from SPDC and observable independently (see fig. \ref{fig_ch5_exp_modes} \textbf{b}), showing a slightly larger mode\footnote{
	While the fluorescence mode is larger than the SPDC mode, 
	the size difference is not sufficient to enable significant spatial filtering of the 
	fluorescence by SMF-coupling behind the waveguide. Fluorescence is thus mainly filtered by the 
	herald and signal filtering stages, whereby 
	uncorrelated noise, falling into the selected frequency bandwidth, is not separated 
	from the actual SPDC photons.
}. 
Due to similar coupling efficiencies for both UV polarisations into the guide, we can directly subtract this fluorescence mode from the SPDC mode for H-polarised UV pump. 
The resulting background-subtracted mode, displayed in fig. \ref{fig_ch5_exp_modes} \textbf{c}, approximates the actual SPDC mode. 
Its FWHM dimensions are smaller than those of the expected mode (see table \ref{tab_ch5_mode_sizes_exp}). 
Nevertheless the overlap of this background subtracted mode with the predicted IR modes in H- and V-polarisation is extremely good, reaching $\mathcal{A}_\text{SPDC} \approx 94\,\%$ (see table \ref{tab_ch5_mode_sizes} in appendix \ref{app_ch5_spdc_modes}).

Because this mode is imaged prior to the polarisation splitting of the SPDC pair (see fig.~\ref{fig_ch5_setup}~\textbf{a}), the mode contains signal and idler photons; it effectively corresponds to the D-polarised IR mode shown in fig.~\ref{fig_app_ch5_exp_modes}~\textbf{a} of appendix~\ref{app_ch5_spdc_modes}.
Good heralding efficiency requires signal and idler SPDC photons to couple individually into SMF with high efficiency. 
To test the mode of each polarisation component, the SPDC mode is also observed after polarisation splitting on the PBS, as shown in fig. \ref{fig_ch5_setup} \textbf{c}, by positioning the EM-CCD camera directly in front of the signal SMF and imaging the incoming collimated mode with the $f_i=200\mm$ lens\footnote{ 
	Notably, here L40X objective at the waveguide output has been replaced 
	with the higher transmissive aspheric lens, which is the component used in the 
	actual experiments. 
}. Mode images are presented in appendix \ref{app_ch5_SPDCmodes_HVpol}, which also contains a description of the SPDC signal and idler mode matching to the SMF modes in each arm.

Direct measurement of the SMF-coupling efficiency $\eta_\text{SMF}$ with SPDC photons is challenging.
For this reason, the H- and V-polarised components of the transmitted IR pump modes (fig.~\ref{fig_app_ch5_exp_modes}~\textbf{b}~\&~\textbf{c} in appendix~\ref{app_ch5_spdc_modes}) are used as a proxy to estimate $\eta_\text{SMF}$. 
Light in both arms can, on average, be coupled with efficiencies of
$\eta_\text{SMF}^\text{signal} = (72 \pm 4) \,\%$ and 
$\eta_\text{SMF}^\text{idler} = (73 \pm 4) \,\%$, 
respectively. 
Notably, the SMF-tips are uncoated, so both modes are subject to $\sim4\,\%$ loss at each SMF-end. 
Additionally, for reasons discussed in section \ref{ch6_setup_electronics}, the SPDC signal photons propagates down an $83 \,\text{m}$ long SMF, which adds $\sim 3.7 \,\%$ propagation loss. 
Without both insufficiencies, the actual SMF-coupling efficiency is estimated to
$\eta_\text{SMF}^\text{signal} =  (81 \pm 4) \,\%$ and 
$\eta_\text{SMF}^\text{idler}  =  (82 \pm 4) \,\%$. 
Due to a lower mode quality of the transmitted IR pump compared to the SPDC emission, these estimates can be considered as  a lower bound. 

We conclude by exemplifying the level of control over the SPDC waveguide modes we can achieve by coupling into the next higher-order mode, TEM$_{(0,1)}$, whose two intensity lobes reach further into the guiding channel. 
Fig. \ref{fig_ch5_exp_modes} \textbf{d} shows this mode\footnote{
	For its excitation the input coupler is displaced vertically downwards 
	from the position used for coupling into the fundamental mode.
}.
Notably, one reason for choosing the $3\mum$ wide guide $3.2$ is the good SPDC mode quality. 
The mode structure of a waveguide with a different channel size (guide $3.1$), which shows a sizeable nonlinear conversion efficiency in the experimentally accessible temperature regime as well (see appendix \ref{ch5_subsec_SPDC_temp_tuning}), is also presented in appendix \ref{app_ch5_guide3_modes}.

\begin{figure}[h!]
\centering
\includegraphics[width=\textwidth]{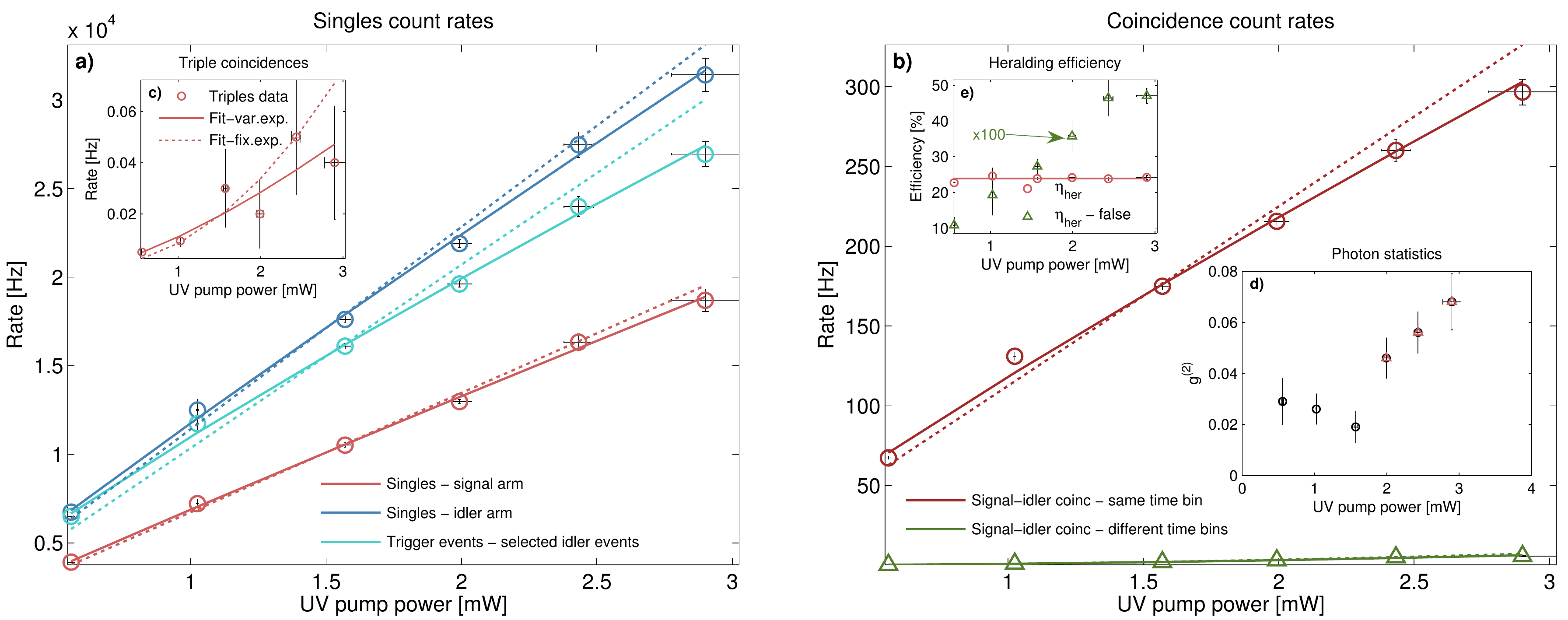}
\caption{SPDC photon production rates: 
\textbf{(a)}: Singles count rates $c_H$ (\textit{red}) and $c_T$ (\textit{blue}) detected in the signal and idler arm, respectively. 
\textit{Mint green} shows the temporally filtered heralding events $c_T^\text{her}$. 
\textit{Solid lines} represent exponential fits for the data's scaling in $P_\text{UV}$, resulting in 
$c_{H} \sim P_\text{UV}^{0.95}$, 
$c_{T} \sim P_\text{UV}^{0.93}$ and 
$c_{T}^\text{her} \sim P_\text{UV}^{0.86}$ 
for the three datasets. \textit{Dotted lines} show the expected scaling 
$c_{H} \sim c_T \sim P_\text{UV}$. 
\textbf{(b)}: Coincidence count rates $c_{H,T}$ between detectors \spcmdt\, and \spcmdh\,. 
\textit{Red} and \textit{green points} are signal and idler photons generated by the same pump pulse $(c_{H,T})$ and subsequent pulses $(c_{H,V}^\text{unc})$, respectively. 
The \textit{solid lines} of similar colour are fits with 
$c_{H,T} \sim P_\text{UV}^{0.89}$ and 
$c_{H,T}^\text{unc} \sim P_\text{UV}^{1.79}$, while the \textit{dotted lines} show 
$c_{H,T} \sim P_\text{UV}$ and 
$c_{H,T}^\text{unc} \sim P_\text{UV}^2$. 
\textbf{(c)}: Triple coincidence rates $c_{H,V,T}$ between APDs \spcmdt\,, \spcmdh\, and \spcmdv\,, with the fitted scaling 
$c_{H,V,T}\sim P_\text{UV}^{1.36}$ (\textit{solid line}) and the expected scaling 
$c_{H,V,T}\sim P_\text{UV}^2$ (\textit{dotted line}). 
\textbf{(d)}: $g^{(2)}$ of the heralded SPDC signal photons.
\textbf{(e)}: Heralding efficiency \hereff\, for SPDC signal photons (\textit{red}), with the \textit{solid line} marking the average of $\eta_\text{her}\approx 24\,\%$. 
False heralding event from uncorrelated SPDC signal photons in the next time bin are shown in \textit{green} (multiplied by 100). 
}
\label{fig_ch5_spdc_count_rates}
\end{figure}

\subsection{Photon count rates\label{ch5_sec_photon_prod}}

In the following we analyse the photon count rates we are able to obtain within our available UV pump power regime (see appendix \ref{ch3_SHG}). 
Here, we will also investigate the achievable heralding efficiencies \hereff\, and determine the photon number purity of the \hsps\, using the {\gtwo} function introduced in eq. \ref{eq_ch6_g2}. 
For measuring the count rates and \hereff\,, all signal photons are sent onto APD \spcmdh\,, only for the {\gtwo} measurement the signal is split 50:50 on a PBS and sent simultaneously to both APDs, \spcmdh\, and \spcmdv\,. 
Importantly, as shown in fig. \ref{fig_ch5_setup} \textbf{a}, the signal is polarised on serval PBSs and a PBD (polarising beam displacer) before it is sent onto the PBS for the intensity splitting to both detectors. 
Its polarisation is thus well defined before it enters the HBT-type setup to measure {\gtwo}, for which reason we can use a PBS for the $50:50$ splitting instead of a non-polarising beam splitter (NPBS). 
This also assures that, when an IR alignment beam is used to adjust the 50:50 intensity splitting, we have the same splitting ratios for single photons. Therefore we effectively resemble the NPBS, which is the default choice for building a HBT-experiment, as shown in fig. \ref{fig_ch6_HBT}.

\paragraph{Singles rates\label{ch5_subsec_rates}}
We start by looking at the SPDC singles count rates $\left\{c_{H}(t_m),c_{V}(t_m)\right\}$ and $c_T(t_m)$ for the SPDC signal and idler photons. 
Note, since we have no storage and retrieval and only send the SPDC signal photons onto our detectors (setting \textit{s}), we drop the time bin and setting indices $t$ and $j$ from the count rate definitions of section \ref{subsec_ch5_fpga_count_rate_definition}.
Fig. \ref{fig_ch5_spdc_count_rates} \textbf{a} shows the mean rates $\bar{c}_i$, 
averaged over all runs $t_m$, for $i \in \left\{T,H,V\right\}$, as a function of the UV pump power $P_\text{UV}$ sent into the waveguide. 
The data is stated in terms of detection frequencies, i.e. counts/second. 
The numbers however derive from a total of $|\left\{ m \right\}| \sim120$ FPGA measurement runs, each with a $\Delta t^\text{FPGA}_\text{int}= 10\sec$ integration time\footnote{
	The data points for $P_\text{UV} = 0.56\mW$ and $P_\text{UV} = 1.02 \mW$ 
	are measured for longer, 
	with a total number of $195$ and $211$ FPGA runs, respectively. 
	Conversely only $70$ runs have
	been recorded for the $P_\text{UV} = 3.01 \mW$ datapoint. 
}. 
Note also that $P_\text{UV}$ is measured directly in front of the waveguide input coupler. The actual UV power 
$P_\text{UV}^\text{coup} = \eta_\text{UV}^{3.2} \cdot P_\text{UV}$, 
available to pump the SPDC, is reduced by the coupling efficiency 
$\eta_\text{UV}^{3.2} \approx 12\,\%$, which accounts for the transmission loss through the O40X microscope objective and the inefficient coupling into the waveguide channel (see appendix  \ref{app_ch5_waveguide_UVmodes}).  

The count rates for both photons, signal and idler (\textit{red} and \textit{blue} in fig. \ref{fig_ch5_spdc_count_rates} \textbf{a}), are the actual detection events on the APD, which show an approximately linear increase with $P_\text{UV}$. This is indicative for the low pumping power regime\cite{Krischek:2010ys}, where at most one SPDC pair is created by each UV pump pulse. 
From eq. \ref{eq_ch5_photon_number}, we thus expect 
$c_j = \left(\gamma^n \cdot P_\text{UV}^n\right)|_{n=1}$. 
The exact dependence can be tested by taking the natural logarithm of both sides to yield
$\ln{\left( c_j\right)} = \alpha + n\cdot \ln{\left( P_\text{UV}\right)}$. 
Fitted the double-logarithmic data by a straight line (\textit{solid lines} in fig. \ref{fig_ch5_spdc_count_rates} \textbf{a}) should thus return $n =1$. 
In practice, we obtain 
$n^\text{s}_H = 0.95 \pm 0.04$ and 
$n^\text{i}_T = 0.93 \pm 0.05$ 
in the signal and the idler arm, respectively, which lie slightly below the expected value of $n = 1$ (\textit{dotted lines} in fig. \ref{fig_ch5_spdc_count_rates} \textbf{a}). 
If higher oder terms contributed to any significant extend, the scaling factor would be modified to super-linear scaling, i.e. $n > 1$, which is not the case. The count rates show that SPDC emission is still well within the  
spontaneous regime\cite{Eisenberg:2004,Christ:2013}. 
The rates are far below $80\MHz$, which is the count rate one would expect if the probability for generating an SPDC pair per UV pulse was approaching $100\,\%$. 
With observed count rates of $\sim 10\,\kHz$, we thus clearly do not generate a photon pair within every UV pump pulse\footnote{
	Since these are detected counts, one can obtain an estimate of the actual production 
	rates by dividing them by the transmission $T_\text{sig}^\text{tot}$ and the
	detection efficiency $\eta_\text{det}$, which amounts to a factor of $\sim 20$.	
}. Because these rates are also far below the maximum rate imposed by detector dead time $c_\text{det}^\text{max}$, detector blinding is not an issue either and $\frac{1}{c_{T}}$ is a good approximate measure for the time $\delta \tau_\text{SPDC}$ between successive SPDC pair emissions, as well as for the experimental repetition rate $f_\text{rep}$. 

Besides SPDC photons the waveguide also emits fluorescence noise (see fig.~\ref{fig_ch5_SHG_vs_temp} in appendix~\ref{app_ch5_guide3_modes}). 
Filtering eliminates all noise outside of the idler's spectral mode. Since the remaining, spectrally indistinguishable fluorescence noise will be mistaken for actual SPDC idler photons by APD \spcmdt\,, its  registration causes false heralding events and will lower \hereff\,. 
Fortunately, the fluorescence noise does not have to occur exactly inside the UV pump time bins that lead to SPDC pair emissions. Some of it is emitted between actual idler generation events. 
To reduce its influence, we can use the limit $\delta \tau_\text{SPDC} \rightarrow c_T$ and apply an external timing filter for the acceptance of heralding events. 
To this end, the idler detection events are fed into a digital delay generator (\textit{Standford Research Systems DG535}, DDG in fig. \ref{fig_ch5_setup}), which delays the APD pulses prior to sending them into the FPGA. Besides adding a delay to an input signal, the DDG can also be issued with a delay to be applied between production of an output pulse and the acceptance of a new input pulse. By setting this second delay to $\mathcal{O}\left(\delta \tau_\text{SPDC}\right)$, heralding events can be forced to have fixed minimum time separations, which reduces the influence of the randomly occurring fluorescence noise on $f_\text{rep}$. 
Fig. \ref{fig_ch5_spdc_count_rates}~\textbf{a} also shows these filtered herald events $c_\text{T}^\text{her}$ (\textit{mint green}), which have a slightly degraded linearity coefficient of 
$n_T^\text{her} = 0.86 \pm 0.06$.
Since SPDC generation is also probabilistic, temporal filtration causes unavoidable loss of some detection events of genuine SPDC idler photons, for which reason we have $c_\text{T}^\text{her} < c_T$. 
For the remainder of this analysis and the work in chapter \ref{ch6}, we will exclusively use $c_\text{T}^\text{her}$ for heralding and triggering purposes.

\paragraph{Coincidence counts}
Using $c_{T}^\text{her}$ for coincidence counting, we obtain the rates 
$c_{H,T}$, 
displayed in fig. \ref{fig_ch5_spdc_count_rates} \textbf{b}. 
They represent the conditional preparation of an SPDC signal photon, contingent on the simultaneous emission of a SPDC idler photon\footnote{
	Obviously, this includes any residual contributions from heralding triggers. 
}. 
As mentioned in section \ref{ch5_subsec_detection_system}, the actual time bin, within which the SPDC signal photon is generated, can be determined by the respective delays between the FPGA channels of signal and herald. 
Besides signal and idler emission within the same UV pump pulse (\textit{red points} in fig. \ref{fig_ch5_spdc_count_rates} \textbf{b}), such events can also occur between different pulses. 
For instance, delaying the FPGA signal channel by an additional $12.5\,\ns$ with respect to the heralding channel looks at SPDC signal photons that are produced by the next UV pump pulse, following the pulse that created the detected idler photon. 
These are accidental coincidences from two uncorrelated photon pair emissions, where one time the signal photon and another time the idler photon got lost. 
Since this requires, at least, the production of two SPDC photon pairs, the count rate $c_{H,T}^\text{unc}$ (\textit{green points} in fig. \ref{fig_ch5_spdc_count_rates} \textbf{b}) of such events is a measure for the amount of higher oder contributions present in the desired correlated coincidence counts $c_{H,T}$.
By calculating the ratio $R_{H,T} = \frac{c^\text{unc}_{H,T}(P_\text{UV})}{c_{H,T}(P_\text{UV})}$, we find the amount of contamination to increase with $P_\text{UV}$ from initially $\sim 0.47\,\%$ to $\sim 2\,\%$ for the highest pump power $P_\text{UV} \approx 3\,\mW$. 

The proportionality of both coincidence types, correlated and accidental events, with $P_\text{UV}$ is again obtained by fitting straight lines onto the double-logarithmic data (\textit{solid lines} in fig.~\ref{fig_ch5_spdc_count_rates}~\textbf{b}), yielding scaling exponents of
$n_{H,T} = 0.89 \pm 0.1$ and
$n_{H,T}^\text{unc} = 1.79 \pm 0.15$. 
As desired for low higher order contamination, the coincidences $c_{H,T}$, for which we expect $n =1$ (\textit{dotted red line}) scale approximately linearly, just as the singles rates. 
In turn, accidentals scale approximately quadratic, as $n=2$ (\textit{dotted green line}) is expected for double pair emission events. 

\paragraph{Triple coincidence counts}
Another test for the presence of more than a single photon in the heralded SPDC state is the detection of triple coincidences. 
For their measurement, the signal is now split $50:50$ between the APDs \spcmdh\, and \spcmdv\, (see fig. \ref{fig_ch5_setup} \textbf{c}), and 
the output pulses of both APDs are counted in coincidence with the herald trigger events $c_\text{T}^\text{her}$. 
The triple coincidence count rates $c_{H,V,T}$ are displayed in fig. \ref{fig_ch5_spdc_count_rates} \textbf{c}. 
The data's scaling with $P_\text{UV}$ of $n_{H,V,T} = 1.36 \pm 0.66$ (\textit{solid line}) is lower than the expected $n=2$ (\textit{dotted line}), which can however result from the larger uncertainties due to low count rates. 
These triple events represent SPDC photons generated within the same UV pump pulse. 
Note that accidental triple coincidences between two different pump time bins are absent, so six photon emission\cite{Wieczorek:2009kx} does effectively not occur.

\subsection{Heralding efficiency\label{ch5_hereff}}

We can now directly use the coincidence count rates $c_{H,T} + c_{V,T}$, together with the heralding events $c_T^\text{her}$, to define the coincidence probabilities 
$p_{H|T}+p_{V|T}$ and therewith the heralding efficiency of 
\begin{equation}
\eta_\text{her} = \frac{p_{H|T} + p_{V|T}}{T_\text{sig}^\text{tot} \cdot \eta_\text{det}} \approx 24\,\%
\label{eq_ch5_her_eff_exp}
\end{equation}
While we send all signal power onto APD \spcmdh\,, for reasons of completeness we also feature the counts on \spcmdv\, in eq. \ref{eq_ch5_her_eff_exp}, despite having $p_{V|T} \rightarrow 0$ at the moment. 
$p_{V|T}$ will be important for the work in chapter \ref{ch6}, where we use a 50:50 intensity splitting between both APDs. 
Fig. \ref{fig_ch5_spdc_count_rates} \textbf{e} displays the measured values for \hereff\, (\textit{red points}), as well as its average over all $P_\text{UV}$ (\textit{red line}, eq. \ref{eq_ch5_her_eff_exp}). 
Notably, this value is quite sensitive to the actual correct spectral projection and relies on good filter stage alignment. 
As shown in fig. \ref{fig_ch5_setup} \textbf{a}, $T_\text{sig}^\text{tot}$ consists of the transmission of the {\cs} cell and the optics behind it ($T_\text{mem} \approx 77\,\%$), the SMF-coupling efficiency to the signal filter stage ($\eta^\text{filt}_\text{SMF} \approx 86\,\%$), and the signal filter stage transmission ($T_\text{filt}^\text{sig} \approx 14.5\,\%$).  
Systematic errors\footnote{
	Note, we do not consider any errors on the detection efficiency $\eta_\text{det}$. 
} cause a variations in the achievable numbers for \hereff\, on the order of 
$\Delta \eta_\text{her} \sim \pm 5\,\%$. 
These are the reason for a smaller value of $\eta_\text{her} \approx 22\,\%$, quoted in chapter \ref{ch6}, which results from a long term average, whereas eq. \ref{eq_ch5_her_eff_exp} represents optimal performance. 

Since frequency filtering cannot eliminate single photon fluorescence in the signal and idler spectral modes, \hereff\, contains a contribution by these photons. 
The amount of which can be estimated by frustrating the down conversion with a V-polarised UV pump, measuring the number of remaining coincidences. 
These amount to $\sim 2\,\%$ of the actual counts obtained when inserting H-polarised UV light. 
In the current form, \hereff\, still contains small amounts of higher order noise, which can be estimated using the accidental coincidences $c_{H,T}^\text{unc}$ (\textit{green points} in fig.~\ref{fig_ch5_spdc_count_rates}~\text{b}). 
Calculating their detection probability $p_{H|T}^\text{unc}$ and inserting the result into eq. \ref{eq_ch5_her_eff_exp}, yields an amount of $\eta_\text{her}^\text{unc} \in \left[ 0.1\,\%, 0.5 \,\% \right]$ for increasing pump powers, which is shown in fig.~\ref{fig_ch5_spdc_count_rates}~\textbf{e} (\textit{green points}). 
Hence the noise contribution lies within the uncertainty range of \hereff\,.

When comparing \hereff\, to other literature values, it is reasonable, but not on the high end, where values up to $80\,\%$ have been reported\cite{Ramelow:2013}.
We expect \hereff\, to be limited by scattering effects inside the waveguide, which can be seen as bright spots (see fig. \ref{fig_ch5_intro} \textbf{g}). 
To estimate the highest \hereff\,-value we can possibly expect, given the measured count rates, we back-out all efficiency factors. 
Besides $\eta_\text{det}$, only the transmission of the \hsps\, between their generation and detection point is relevant for \hereff\,. 
Their path consists, firstly, of their extraction efficiency from the waveguide and, secondly, of any optics behind the waveguide output coupler. 
As illustrated in fig. \ref{fig_ch5_setup} \textbf{a}, 
apart from the above stated $T_\text{sig}^\text{tot}$, this second component also includes the transmission from the waveguide output to the SMF 
($T^\text{SPDC}_\text{sig} \approx 92\,\%$), 
its SMF-coupling efficiency ($\eta_\text{SMF}^\text{sig} \approx 72\,\% $, including the transmission through the fibre),  
as well as the transmission through the second SMF and the optics in between 
($T_\text{SMF} \approx  84\,\%$). 
For extraction from the waveguide, we can also attribute the known Fresnel reflection ($R_\text{KTP,H}(852\nm) \approx 7,5\,\%$, $R_\text{KTP,V}(852\nm) \approx 8,8\,\%$) and the output coupler transmission ($T_\text{C230TME} \approx 95\,\%$). 
Incorporating these additional factors, which amount to a total transmission of $\tilde{T}_\text{tot} \approx 49\,\%$,  into eq. \ref{eq_ch5_her_eff}, yields $\eta^\text{max}_\text{her} \approx 49\,\%$, which is still below the maximal value of $100\,\%$. 

Since the only bit, that is not directly measured in this estimation, is the SPDC photon transmission through the guide, it is therefore likely that this accounts for at least part of the residual discrepancy. 
Under the assumption that all of this residual reduction is due to the scratched waveguide surface (see appendix \ref{app_ch5_insufficiencies}), waveguide replacement by a new chip of the same parameters should cure this problem. 
In the best case scenario, a new chip would allow for $\eta^\text{max}_\text{her} \approx \frac{\eta_\text{her}}{1-\tilde{T}_\text{tot}} \approx 47\,\%$. 
Such a replacement has meanwhile been purchased, however there has not been sufficient time during this project to try out the new device.

\subsection{Photon statistics\label{ch5_subsec_g2}}
With the above count rates, we can also determine the \hsp\, number purity, using the heralded {\gtwo}-autocorrelation function of eq. \ref{eq_ch6_g2}. 
This measurement now uses a 50:50 splitting of the signal intensity onto APDs \spcmdh\, and \spcmdv\,. 
While signal and idler by themselves have thermal statistics\cite{Yurke:1987} with $g^{(2)} = 2$, 
the non-classical correlation between both photons results, optimally, in $g^{(2)} = 0$ for the SPDC signal, when heralding on the idler\cite{Grangier:1986,Loudon:2004gd}. 
This happens because we generate one SPDC pair, whose two photons are split into the signal and the idler arm. In the signal arm, we thus have a single photon, that can only propagate along one of the two paths in the HBT-setup we use to measure {\gtwo} (see fig. \ref{fig_ch6_HBT}). 
Any contamination\cite{Loudon:2004gd} by residual higher order SPDC emissions or other sources of noise, such as single photon fluorescence, will result in $g^{(2)} >0$. 
Since higher order emissions scale at least $\sim P_\text{UV}^2$ or steeper, pump power reduction improves the photon number purity, but comes at the cost of heralding event reduction. 
Fig.~\ref{fig_ch5_spdc_count_rates}~\textbf{e} displays the measured {\gtwo}-values for our UV pump power regime. 
Generally, {\gtwo} is very close to $0$ and increases only for $P_\text{UV} \gtrsim 2 \mW$. 
Regarding the heralding rate, we will see in chapter \ref{ch6} that $c_T^\text{her} \le 10^4$ is more than sufficient for interfacing the source with the Raman memory. 
Hence, we can afford to run the source with $P_\text{UV} \approx 1 \mW$, where {\gtwo} is minimal.
The obtained photon number purity of $\mathcal{O}\left(g^{(2)}\right) \sim10^{-2}$ compares well with the performance of other SPDC-based sources, designed for quantum memories\cite{Scholz:2009}.

\section{Measurement of the heralded single photon spectrum\label{ch5_sec_exp_hsp_bandwidth}}	

The final analysis we present here is the experimental measurement of the \hsp\, spectrum, which allows us to test, whether our expectations from section \ref{ch5_sec_hsp_bandwidth} coincide with the photons we actually prepare. 
Notably, the direct measurement of a spectral bandwidth $\sim 1\,\GHz$ is not simple in the optical domain, particularly when dealing with single photon signals. Regular spectrometer resolutions are not sensitive enough to yield faithful results, because devices usually operate on a nano-meter scale. 
So we have to resort to a different, indirect method, which uses the same principles we have applied in characterising the filter stages (see section \ref{ch5_subsec_filter}). We will now first outline the experimental concepts and then discuss the results.

\subsection{Measurement idea}

To determine the \hsp\, spectrum, we invert the method of our filter stage bandwidth measurements: instead of sweeping a know pulse over an unknown frequency filter, we now do the opposite and scan the unknown \hsp\, spectrum over the known signal filter resonance. 
Once more, the filter function $T_\text{filt}^\text{sig}(\nu)$ and the pulse shape need to be assumed a priori; we use the same functions as in section \ref{ch5_subsec_filter} and utilise the measured FWHM filter bandwidths (see table~\ref{tab_ch5_filter_fwhm} in appendix~\ref{app_ch5_filter_measurement}). 
Probing the \hsp\, spectrum with the signal filter line can be done in two ways: either by detuning the filter resonance frequency, or by changing the central frequency $\nu_{\text{s},0}$ of the \hsps\,. The former is substantial experimental effort and relies on the absence of systematic alignment errors. 
Hence we scan $\nu_{\text{s},0}$. 
To this end, the {\tisa} master laser's output frequency is modified in the same manner as described in section \ref{ch5_subsec_filter}. 
Due to frequency doubling of the {\tisa} pulses, any changes in the {\tisa}  
detuning with respect to the {\cs} $6^2 \text{S}_\frac{1}{2}\text{F}=3 \rightarrow 6^2 \text{P}_\frac{3}{2}$ transition are initially translated into shifts of the centre frequency of the UV pump pulses. 
A change of the {\tisa} frequency $\nu_\text{Ti:Sa}$ by $\Delta \nu$ shifts the UV pump's centre frequency 
$\nu_\text{UV}$ by $2\Delta \nu$, because $\nu_\text{UV} = 2\cdot \nu_\text{Ti:Sa}$ via SHG. 
SPDC of the frequency shifted UV pump will also shift the frequencies of the signal and idler spectra, which have to add to the shifted value of $\nu_\text{UV}$. 
Accordingly, for degenerate SPDC, the centre frequencies of the signal and idler spectra are shifted by the same amounts $\Delta \nu$ as the initial {\tisa} pulses.
Importantly, for this measurement to work, we need constant idler photon production rates, when sweeping the SPDC pump's central frequency. 
In other words, the SPDC phase-matching bandwidth has to be broad enough that shifts $\nu_\text{UV}$ do not change the unfiltered JSA 
$f(\nu_\text{s}, \nu_\text{i}) = \alpha(\nu_\text{s}+ \nu_\text{i}) \cdot \Phi(\nu_\text{s}, \nu_\text{i})$. 
If this was not fulfilled, the recorded data would contain a convolution with the changes in $f(\nu_\text{s},\nu_\text{i})$, which is not straight forward to separate. 
Accordingly, the signal filter resonance would have to be scanned instead. 
However, as fig.~\ref{fig_ch5_spdc_spectra}~\textbf{c} shows, $\Delta \nu_\text{PM}$ is wide enough, such that detuning ranges $|\Delta \nu | \le 5 \GHz$ are phase matched, i.e. 
$\Phi(\nu_\text{s},\nu_\text{i}) \approx 1, \, \forall |\Delta \nu | \le 5 \, \GHz$. 

In the experiment we detect \hsps\, transmitted through the signal filter stage on APD  \spcmdh\, (see fig.~\ref{fig_ch5_setup}~\textbf{a}).
Similar to eq. \ref{eq_ch5_filter_convolution}, the observed photon count rates $c_{H,T}(\Delta \nu)$ represent the convolution between the \hsp\, spectrum $S_\text{HSP}^\text{exp}(\nu)$ and the filter line $T_\text{filt}^\text{sig}(\nu)$, with
\begin{equation}
\tilde{c}_{H,T}(\Delta \nu) = \underset{{\nu}}{\int} T_\text{filt}^\text{sig}(\tilde{\nu}) \cdot  S_\text{HSP}^\text{exp}(\tilde{\nu}-\Delta \nu) \text{d} \tilde{\nu} = 
 T_\text{filt}^\text{sig}(\nu) \ast  S_\text{HSP}^\text{exp}(\nu),
\label{eq_ch5_conv_spdc_spectra}
\end{equation}
where $S_\text{HSP}^\text{exp}(\nu) \sim \langle \hat{n}_{\text{s},H} \rangle$ is given by the photon number expectation value $\hat{n}_{\text{s},H} = \hat{a}^\dagger_{\text{s},H} \hat{a}^{}_{\text{s},H}$ of the \hsps\,\footnote{
	In the language of eq.~\ref{eq_ch5_marg_SPDC_filtered} in 
	appendix~\ref{app_ch5_multimode_spdc_heralding},  
	mathematically this corresponds to the application of 
	a detection operator $\pi_\text{s}$ for the signal mode on the density matrix 
	$\rho_\text{s}^\text{filt}$ (see eq. \ref{eq_ch5_marg_SPDC_filtered}).
} (see eq. \ref{eq_ch6_g2}) and $\tilde{c}_{H,T}(\Delta \nu) =  \frac{{c}_{H,T}(\Delta \nu)}{{c}_{H,T}(0)}$ represent the coincidence rates, normalised to their value for $\Delta \nu = 0$ frequency shift of the {\tisa} laser relative to the filter resonance.

\begin{figure}[h!]
\centering
\includegraphics[width=0.95\textwidth]{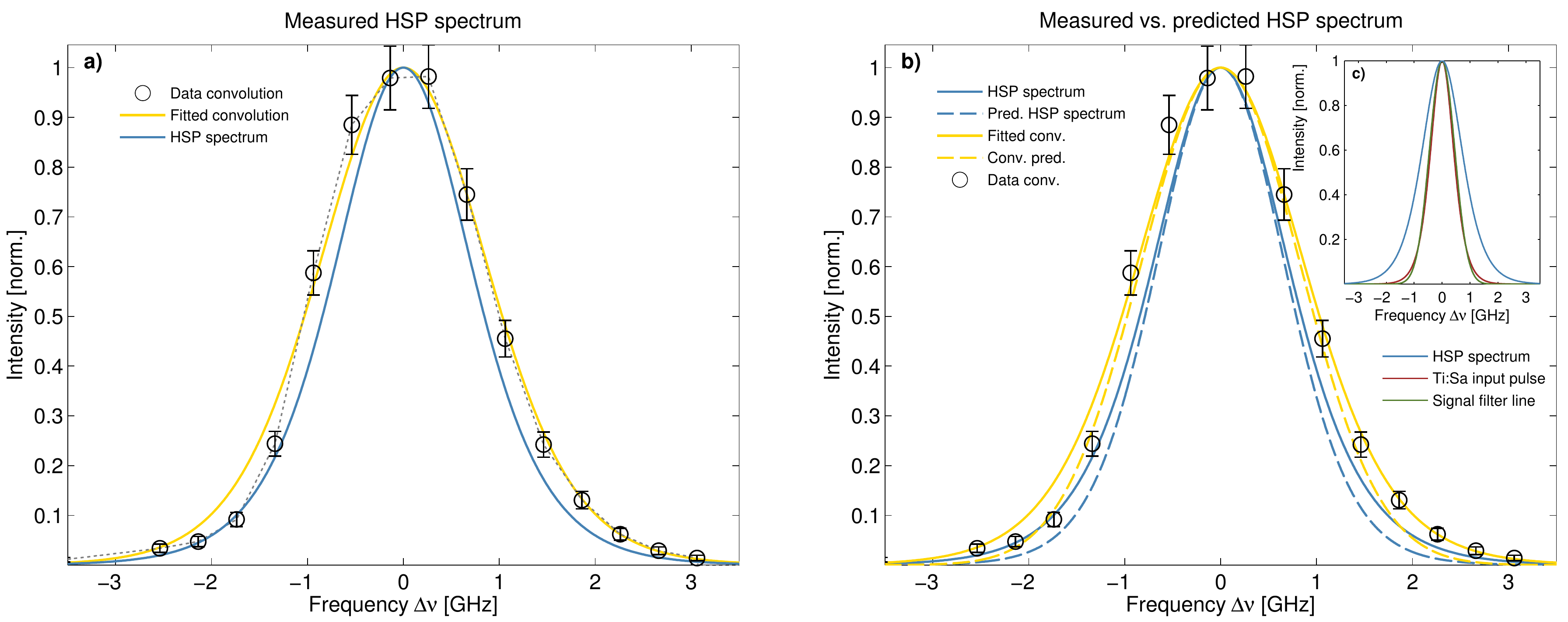}
\caption{
\textbf{(a)}: Measured convolution data for the \hsp\, spectrum, swept over the signal filter line (\textit{black points}), with linear interpolation (\textit{dotted black line}) and fitted convolution function (eq. \ref{eq_ch5_conv_spdc_spectra}, \textit{yellow line}), using the optimal HSP spectrum $S_\text{HSP}(\nu)|_{\Delta t_\text{HSP}^\text{opt}}$ (\textit{blue line}). 
\textbf{(b)}: Comparison of the data (\textit{black points}), the fitted convolution  (\textit{solid yellow line}) and the measured \hsp\, spectrum (\textit{solid blue line}) with the predicted \hsp\, spectrum (\textit{dashed blue line}) and its resulting convolution trace (\textit{dashed yellow line}). 
\textbf{(c)}: Comparison of the measured \hsp\, spectrum (\textit{blue line}) with the {\tisa} spectrum (\textit{red line}) and the signal filter transmission line $T_\text{filt}^\text{sig.}(\nu)$ (\textit{green line}). 
}
\label{fig_ch5_meas_spdc_spectra}
\end{figure}

\subsection{Experimental results\label{ch5_subsec_exp_hsp_spectrum}}

To test our capability of tuning the \hsp\, spectral bandwidth, 
the experiment is conducted with the three different herald filter stage arrangements, quoted in table \ref{tab_ch5_filter_fwhm} of appendix \ref{app_ch5_SPDCspec}. 
Here, we only discuss the results for the configuration applied when interfacing the source with the memory in chapter \ref{ch6}, where the idler filter contains two 18 GHz and two 103 GHz etalons. 
To avoid problems with numerical stability in calculating $S_\text{HSP}^\text{exp}(\nu)$ in eq.~\ref{eq_ch5_conv_spdc_spectra}, we use the same procedure as employed in section \ref{ch5_subsec_filter}: 
We convolve an anticipated pulse spectrum 
$S_\text{HSP}(\nu)|_{\Delta t}$, 
for a given pulse duration parameter $\Delta t$ (see appendix \ref{ch3_tisa_pulse_duration}), under application of the Fourier-transform theorem, with the filter line $T_\text{filt}^\text{sig}(\nu)$ and fit the result onto the normalised, measured count rates 
$\tilde{c}_{H,T}(\Delta \nu)$. 
The fit optimisation runs over the bandwidth parameter $\Delta t$ of the heralded SPDC signal spectrum 
$S_\text{HSP}(\nu)$. 
This effective way of deconvolving eq. \ref{eq_ch5_conv_spdc_spectra} is once more performed for sech-pulses (see eq. \ref{ch5_eq_sech_pulses}) as well as Gaussian pulses; results for the latter reported in appendix \ref{app_ch5_SPDCspec}. 

Fig. \ref{fig_ch5_meas_spdc_spectra} \textbf{a} illustrates the results for sech-shaped \hsp\, wavepackets. 
The best fitting pulse model has an optimal pulse duration parameter of 
$\Delta t_\text{HSP}^\text{opt} \approx 106 \ps$. 
Convolving its pulse spectrum with the filter bandwidth transmission function $(T_\text{filt}^\text{sig}(\nu))$ yields an expected convolution shown by the \textit{yellow line} in fig. \ref{fig_ch5_meas_spdc_spectra} \textbf{a}. 
This trace agrees well with the measured data for $\tilde{c}_{H,V}$ (\textit{black points}). 
A linear interpolation between the measurement points (\textit{dotted black line}) exemplifies the degree of matching between both. 
The actual \hsp\, spectrum 
$S_\text{HSP}(\nu)|_{\Delta t_\text{HSP}^\text{opt}}$ 
is shown by the \textit{blue line}. 

We find a \hsp\, spectral bandwidth of $\Delta \nu_\text{HSP}^\text{exp} \approx 1.69 \pm 0.06 \GHz$. The errors are obtained by Monte-Carlo simulation\footnote{
	The quoted error is the standard deviation of the resulting  
	$\left\{ \Delta \nu_\text{HSP}^\text{exp} \right\}$ sample. We assume normal distributions 
	for the input parameters $\Delta \nu_\text{filt}^\text{sig}$, $\Delta \nu$ and 
	$\tilde{c}_{H,T}(\Delta\nu)$ in eq. \ref{eq_ch5_conv_spdc_spectra}, 
	with the respective distributions' standard deviations given by the experimental 
	uncertainties of the parameters. 
}, varying the signal filter bandwidth and the measured count rates within their uncertainty ranges, as well as assuming a $100\MHz$ error on the detuning $\Delta \nu$.
The result also matches well with the JSA analysis prediction 
${\Delta \nu_\text{HSP}^\text{pred} = 1.54\GHz \sim 0.91 \cdot \Delta \nu_\text{HSP}^\text{exp}}$ (see section \ref{ch5_sec_hsp_bandwidth}). 

To illustrate that the experiment and the prediction return quite similar results, the spectrum 
$S_\text{HSP}(\nu)|_{\Delta t_\text{HSP}^\text{opt}}$ and the predicted spectrum 
$S_\text{HSP}(\nu)|_{\Delta t_\text{HSP}^\text{pred}}$ are plotted in fig. \ref{fig_ch5_spdc_spectra} \textbf{b} by the \textit{solid} and \textit{dashed blue lines}, respectively. 
Moreover, we also calculate the expected convolution trace, using the \hsp\, spectrum from our JSA analysis and inserting it into eq. \ref{eq_ch5_conv_spdc_spectra} to yield the expected convolution trace. 
To compare this expected trace with the aforementioned fitted convolution trace, as well as the direct measurement for $\tilde{c}_{H,T}(\Delta \nu)$, all three convolution datasets are also displayed in fig. \ref{fig_ch5_spdc_spectra} \textbf{b} by the \textit{dashed} and \textit{solid yellow lines} for the expected and fitted convolution, as well as by \textit{black datapoints} for $\tilde{c}_{H,T}(\Delta \nu)$. 
Just as with the \hsp\, spectra $S_\text{HSP}(\nu)$, the convolutions also show good agreement between one another and with the experimental data. 
These results confirm that the \hsp\, spectrum is indeed a bit broader than both, the {\tisa} laser (memory control pulses) and the signal filter line $T_\text{filt}^\text{sig}(\nu)$, which are both shown in 
fig.~\ref{fig_ch5_spdc_spectra}~\textbf{c} (by the \textit{red} and \textit{green lines}, respectively) for comparison with $S_\text{HSP}(\nu)$. 
As we have discussed in section \ref{ch5_sec_hsp_bandwidth}, this spectral mismatch, along with the sub-unity purity of the SPDC state's JSA, will reduce the memory read-in efficiency for \hsps\, compared to the values achievable for perfectly mode matched input signals (see section \ref{subsec_mem_op_adiabatic_lim}). 
A good proxy for the latter type of input signals are coherent states (\coh\,), directly derived from the {\tisa} output pulses by a pick-off, as we have employed them in chapter \ref{ch4}. 
The difference in the read-in efficiencies between both signal types thus reflects the reduction in mode-matching to the memory kernel, given, effectively, by the {\tisa} pulses\cite{Nunn:2007wj}. 
We will come back to this point in section \ref{ch6_subsec:MemEffCalc}, after introducing the interface between source and memory.

\section{Conclusion\label{ch5_sec_Conclusion}}

In this chapter, we have shown the implementation of a technologically simple \hsp\, source, based on SPDC in a ppKTP waveguide. 
The source is made to interface with the Raman quantum memory and produces trigger events in the kHz-regime for the observation of \hsps\, with rates of hundreds of Hz, while maintaining good photon number purity of $g^{(2)} \lesssim 0.02$ for the \hsp\, output. 
The modest SPDC pump powers, required to achieve these rates, furthermore allows to run the SPDC source and the Raman memory in parallel, using the same master laser. This is resource-friendly and limits experimental complexity.
Broadband frequency filtering of the idler arm, implemented with standard, off-the-shelf Fabry-Perot etalons, allows the projection of the central frequency and the spectral bandwidth of the \hsps\, into Raman resonance with the memory control field. 
The employed filter bandwidth of $\sim 1\GHz$ results in \hsps\, with a $\sim 1.69\GHz$ FWHM spectral bandwidth, which are still a bit broader than the $\sim 1\GHz$ memory control. 
With the current waveguide chip, the source heralding efficiency is limited to $\eta_\text{her} \approx 24\,\%$, which, most likely, results from damage-induced scattering. However, a replacement chip should yield better single photon preparation with, theoretically possible, values as high as $\eta_\text{her} \approx 47\,\%$. 

Besides the experimental source characterisation, we have also discussed the trade-off between matching the \hsps\,' marginal spectrum with the Raman memory's signal acceptance line and the spectral purity of the generated SPDC state. This trade-off is inherent when designing an SPDC source for optimal interfacing with the Raman memory. The source design parameters of the SPDC pump bandwidth and the idler filter bandwidth furthermore depend on the desired experimental repetition rates. 
Here, we have picked bandwidth parameters that would, in principle, allow operation with a repetition rate of up to $1\GHz$. 
Ideally, this would require approximately equal spectral bandwidths of the SPDC pump and the memory control. In the current setup, we have $\sim 1.2$ times broader pump pulses. An additional  improvement could thus be gained by also filtering the SPDC pump. 
As another possibility we could change the idler filter bandwidth to narrow the \hsp\, spectrum and to achieve better spectral purity. Despite worse spectral overlap between the \hsps\, and the memory kernel, greater storage efficiencies could be expected, as these are dominated by the spectral purity (see discussion of fig. \ref{fig_1}).  
However, we leave such a modification for future generations to consider and focus in the following on the actual interfacing of this source with our Raman memory.

\chapter{Single photon storage\label{ch6}}
\label{ch6}

\begin{flushright}
{\tiny \textfrak{Faust}: \,  \textfrak{Du kannst! So wolle nur!} }
\end{flushright}

We can now move on to storing the heralded single photons (\hsps\,) in the Raman memory. 
After putting our work in the context of other experiments on memory storage of single photons, we start by discussing the interface between our single photon source and our Raman memory. 
In this regard, we provide a detailed explanation of how to operate the memory, conditioned on the successful production of a single photon, particularly focussing on the required feed-forward logic as the key technological ingredient for a temporal multiplexer. 
Thereafter, we characterise the storage process of heralded single photon input signals, which we benchmark against the storage of coherent state inputs at the single photon level. 
To this end, we first observe \hsp\, storage by a series of mean-field measurements.
Subsequently, we analyse how storage affects the quantum characteristics of the input signal. 
Here, we analyse the photon statistics of the stored signal and compare the results to those obtained for coherent states (\coh\,). 
We see an influence of the \hsps\,' quantum characteristics on the retrieved signal, but find their faithful preservation to be limited by memory noise. 
We show, that noise mitigating is the important next step in Raman memory research, with our results laying out the path towards a solution.

\section{Introduction\label{ch6_sec_intro}}
\noindent
As we have seen in chapter \ref{ch1}, temporal multiplexing\cite{Nunn:2013ab} is a key application area for quantum memories, because memories allow storage of successful quantum gate outputs.  
For many schemes, these outputs are single photons. 
The probabilistic operation of quantum gates requires the memory to store and retrieve the information based on randomly occurring events.
For such applications, a memory consequently needs to meet the following requirements: 
on-demand read-in and read-out must be possible, ideally of broadband single photons to enable high repetition rates (see the discussion in section \ref{ch5_sec_hsp_bandwidth}). 
Furthermore the memory should have a low noise floor, high efficiency and a long lifetime. 
For the sake of technical simplicity, the system ideally also operates at room-temperature. 

The first step towards this desired memory-based synchronisation of single-photon generation is the on-demand storage of single photons.
Recently several groups have investigated the storage of quantum signals with memory light-matter interfaces. 
These works can be categorised into systems using continuous variable quantum state input\cite{Appel:2008aa,Jensen:2010aa}, and systems using single photon input states\cite{Eisaman:2005,Choi:2008aa,Chaneliere:2005jh}. 
For temporal multiplexing, the latter systems are of particular interest; 
for these the current technological status has mainly been established by two types of memories.
The first are solid state systems interfaced with SPDC sources\cite{Rielander:2014aa,Bussieres:2014ab}. These are atomic frequency
comb (AFC) type memories\cite{Afzelius:2009qf} with pre-programmed storage times, i.e. the storage and retrieval is not conditioned on the successful generation of a heralded single photon (\hsp\,).
In their current state, these are not suitable for temporal multiplexing. 
While AFC type memories with arbitrary storage times
exist\cite{Afzelius:2010fk,Timoney:2012,Timoney:2013}, such systems suffered, until very recently\cite{Guendogan:2015}, from a significant noise level. 
Second, single photon storage and retrieval, under preservation of its single-photon character, has been shown in cold atomic ensembles\cite{Eisaman:2005,Novikova:2012}. 
Besides the technological difficulties of cold atom ensembles\cite{Bimbard:2014aa, Bao:2012aa, Choi:2008aa, Zhang:2011aa}, these systems are inherently narrowband. 
It is thus challenging to interface them with pulsed, travelling wave, SPDC-based single photon sources. 
Instead, single photons have to be generated either using the emission of cold atomic ensembles themselves\cite{Eisaman:2005} or technologically complex, intra-cavity SPDC sources\cite{Scholz:2009,Fekete:2013}. 
Moreover, these systems do not allow for high bandwidths and fast operation. 
Their usage in a photonic network would limit the networkÕs repetition rates to MHz speeds, which falls short of the capabilities expected of a photonics based network. 

Our Raman memory is a promising candidate for the temporal multiplexing task, since it automatically ticks the boxes of broad bandwidth and room-temperature operation. 
Both factors enable a technologically simple design, whereby particularly the broad bandwidth in the GHz regime allows for \hsp\, preparation by a travelling wave SPDC source. 
As we have seen in chapter \ref{ch5}, the Raman scheme simplifies building of a matching source. 
Since the Raman memory protocol relies on spin-wave storage, it is also inherently suitable for on-demand operation. In contrast to on-resonant, absorptive systems\cite{Novikova:2012, Tittel:2010}, 
Raman absorption in the, initially transparent, storage medium (\cs\,) must be triggered by the application of a strong control pulse to induce the virtual Raman resonance (see chapter \ref{ch2}). 
Preparation of the control pulses, conditioned on successful SPDC source heralding events, is consequently a convenient way to establish on-demand memory operation.
To this end, the experimental implementation requires an electronic feed-forward logic. 
This logic triggers the memory control pulse generation on the detection event of an SPDC idler photon, which, in our case, can be done by selectively picking control pulses from the {\tisa} pulse train.
SPDC idler detection simultaneously also heralds the presence of a single photon in the SPDC signal mode. 
Appropriate adjustment of the timing delays between the \hsp\, and the picked control pulses thus allows immediate realisation of on-demand single photon storage and retrieval. 

Following the convention used by other quantum memory groups\cite{Sinclair:2014}, 
we define electronic feed-forward as the utilisation of one output signal (\textit{idler detection}) of a first device (\textit{heralded single photon source}), to trigger the operation of a second device (\textit{memory}) on another output (\textit{heralded single photon}) of the first device. The operation of the second device (\textit{control pulse picking}) is not only conditioned on the operation of the first device, but it is also performed before the second output of the first device reaches the second device. Hence the term feed-forward and the aforementioned requirement for appropriately delaying the heralded single photons (\hsps\,).  

We will now see, how we can incorporate the feed-forward into our experiment, when we combine the \hsp\, source of chapter \ref{ch5} with our Raman memory system, which we appropriately modify with respect to the apparatus used in chapter \ref{ch4}.

\section{Experimental setup\label{ch6_sec:setup}}

The key challenge in storing \hsp\, from an SPDC source in a Raman quantum memory is the correct timing of the memory interaction, conditioning it on the source heralding events. 
The probabilistic nature of the SPDC process necessitates communication between the source and the memory, such that the memory control pulse sequence is only prepared when a \hsp\, is inserted into the memory. 
Thus the electronic signals from herald detection need to trigger the entire experimental apparatus downstream of the source; a necessity we meet by using the above mentioned electronic feed-forward logic. 
Moreover, investigation of single photon storage in the memory requires the possibility to tell how well the system performs with \hsp\, input states. 
To this end, our experiment contrasts the storage of \hsps\, with that of coherent state input signals (\coh\,).
While the system must be operated in feed-forward mode for experiments with \hsps\,, operation with coherent state input signals is implemented by triggering the experiment on the internal photodiode of our {\tisa} master laser (see appendix \ref{ch3}), as it has been done in chapter~\ref{ch4}. 
The latter is simpler to implement, since it does not require simultaneous operation of the SPDC source. 

\begin{figure}
\centering
\includegraphics[height=22cm]{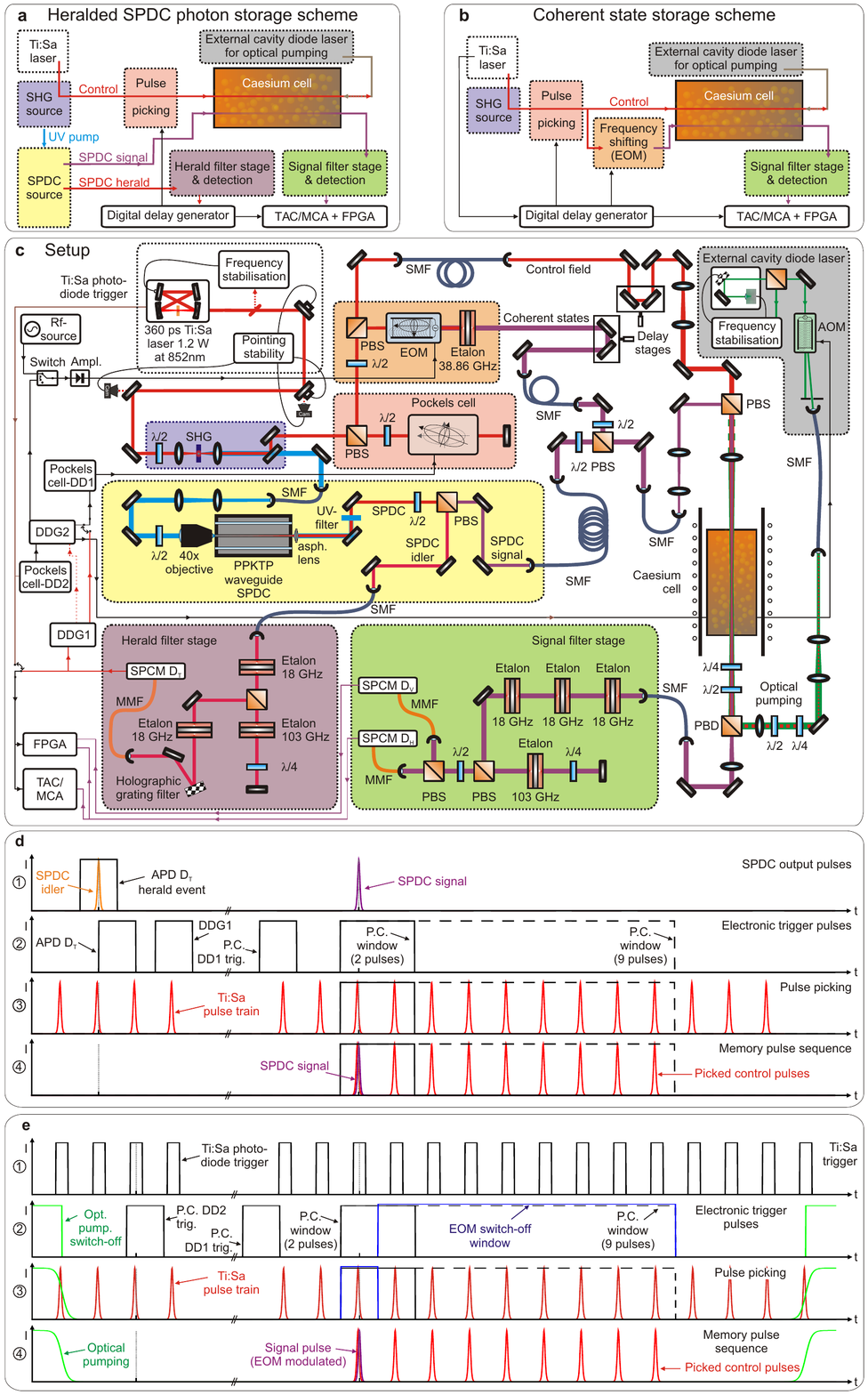}
\caption{Experimental configuration for the storage of heralded single photons and coherent state signals: 
\textbf{a} \& \textbf{b}: Electronic wiring diagrams of the main experimental components for both input signal types. 
\textbf{c}: Complete experimental set-up. \textbf{d} \& \textbf{e}: Pulse timing diagrams for \hsp\, and \coh\, storage, respectively. See main text for more details.}
\label{fig_6_setup}
\end{figure}

\paragraph{Setup design principles}
To implement these two modi operandi, we combine the single photon source with the Raman memory in a way that allows us to easily switch between them. So we end up with two slightly different experimental configurations, which are sketched in fig.~\ref{fig_6_setup}~\textbf{a}~\&~\textbf{b} for \hsps\, and \coh\,, respectively. 
When studying fig. \ref{fig_6_setup} it is key to note that the main differences are the preparation of the input signal and the triggering for the picking of the memory control pulses. 
To have \hsp\, input signals, we obviously need to use our photon source, whereas \coh\, inputs are generated directly from the {\tisa} output. Apart from these differences, both configurations rely on the same Raman memory infrastructure, which consists of the {\cs} cell, the diode laser system, required for optical pumping (see appendix \ref{ch3}), the signal filter stage, as well as the TAC/MCA- and FPGA-based detection systems. 
The latter two are the same systems as used for the \hsp\, source. 
They are described in sections \ref{ch5_subsec_filter} and \ref{sec_ch4_photon_detection}, respectively. 
Note firstly that, similar to chapter \ref{ch4}, we now require the signal filter stage not only to reduce single photon fluorescence from the waveguide, but also to suppress control field leakage and memory noise (see also chapter \ref{ch7} for details). 
Secondly, the diode laser receives different triggering signals, depending on the configuration, so its operation mode is not exactly the same for \hsp\, and \coh\, signal storage (see section \ref{ch6_setup_opticalpumping}). 

The resulting complete experimental set-up, illustrated in fig. \ref{fig_6_setup} \textbf{c}, is thus a combination of the single photon source, shown in fig. \ref{fig_ch5_setup}, and a single mode version\cite{Reim2010,Reim:2011ys} of the Raman memory setup, presented in fig. \ref{fig_ch4_setup_fig}. 
To produce the control pulses, we now use the fraction of the {\tisa} output pulses, that have not been converted to the UV pump for the SPDC source. 
Similar to the previous memory set-up of chapter \ref{ch4}, the control pulses are picked from this non-converted pulse train by our Pockels cell (\pockels\,). 
In turn, \coh\, input signals are once more generated from a pick-off of the picked pulses, which are frequency shifted by $9.2\,\GHz$ using our EOM. 
The main advance from a simple combination of the apparati, introduced in the previous two chapters, is the experimental triggering and routing of the optical and electronic signals. 
In the following, we clarify the resulting added complexity by going through the experimental components step-by-step\footnote{
	Readers only interested in the big picture can skip this technical description, which  
	is designed for readers who intend to rebuild the experiment.
}. 
We first deal with the optical components, then outline the electronic circuit and finally discuss the timings between the various pulses in both configurations, depicted in fig.~\ref{fig_6_setup}~\textbf{d}~\&~\textbf{e}. 

\subsection{Optical components}

\subsubsection{Operation with heralded single photons\label{ch6_setup_optics_spdc}}

For \hsp\, storage fig. \ref{fig_6_setup} \textbf{a} illustrates the general experimental scheme with detailed components shown in fig. \ref{fig_6_setup} \textbf{c}. 
As described in chapter \ref{ch5}, the output of the frequency- and beam-pointing-stabilised {\tisa} oscillator\footnote{
	The {\tisa} laser has $1.2 \, \text{W}$ output power at $852\nm$ wavelength 
	and $f_\text{rep}=80 \MHz$ repetition rate.
} (IR, see appendix \ref{ch3_tisa}) is frequency doubled in the second-harmonic (SH) source (\textit{violet panel}, see appendix \ref{ch3_SHG}). 
A low SH generation (SHG) efficiency (${\eta_\text{SHG} \approx 1.5 \pcW}$) 
is chosen, as only a modest UV power is required for pumping SPDC.
The non-converted, transmitted IR is used as the memory control pulses, for which reason it is separated from the UV on a high-pass filter and sent free-space into the Pockels cell unit (\textit{pink panel}).
The generated UV pump is SMF-coupled and sent into the single photon source (\textit{yellow panel}), which leads to a power of $P_\text{UV}=1\mW$ in front of the input coupling objective. 
When adjusting the objective for single spatial mode SPDC operation, the UV transmittance through waveguide 3.2 is $\eta_\text{UV}^{3.2} \approx 9 \, \%$, so SPDC is effectively pumped by $\tilde{P}_\text{UV}^\text{3.2} \approx 100 \muW$ average power. 
The remainder of the source setup is the same as in chapter \ref{ch5}, i.e. the generated SPDC signal (\textit{purple lines}) and idler (\textit{red lines}) photons are first separated from the pump on a low-pass filter, 
then split into separate spatial modes on a PBS, and finally SMF-coupled with 
$\eta_\text{SMF}^\text{signal}\approx 72\,\%$ and 
$\eta_\text{SMF}^\text{idler}\approx 73\,\%$ efficiency. 
Idler photons are frequency-filtered with the herald filter stage\footnote{
	The idler filter comprises 2 Fabry-Perot (FP) etalons with $\text{FSR}=18.2 \GHz$, 
	1 double-passed etalon with FSR$=103 \GHz$ and a holographic-grating filter (\textit{ONDAX}), 
	followed by a MMF, leading to the APD \spcmdt\,. 
} (\textit{purple panel}, see section \ref{ch5_subsec_filter}) and detected on APD \spcmdt\,. 
Idler detection events on detector \spcmdt\, trigger the memory control field preparation and the signal data acquisition, as explained in section \ref{ch6_setup_electronics} below. 
Memory control pulses are prepared by pulse-picking the unconverted {\tisa} pulses from the SH-source with the \pockels\,, now used in double-pass to obtain a full $\lambda/2$ phase modulation (see appendix \ref{app_ch4_PC}). 
The selected pulses are coupled into a $10\,\m$ long SMF with $\eta^\text{ctrl}_\text{SMF} \approx 53 \, \%$ efficiency and delayed appropriately to obtain temporal overlap with the signal pulses. 
During this procedure the \hsps\, (SPDC signal) are propagating along the $83\m$ and $7.97\m$ long SMFs,  introduced in section \ref{ch5_sec_setup}, before they are combined with the control pulses on a PBS in front of the Cs cell.
Propagation in SMF allows for sufficient time delay to complete the electronic feed forward operation from \spcmdt\, to the \pockels. 
In order to switch easily between \hsp\, and \coh\, signal inputs to the memory, these two SMFs are separated by a PBS, where the \coh\, signal is inserted from the second PBS input port into the spatial mode collected by the second, $7.97\m$ long SMF. 
We chose a free space setup to minimise losses and undesired polarisation rotations for the SPDC signal photons in coupling from SMF to SMF, achieving $\sim 84\,\%$ transmission between the outputs of both SMFs\footnote{This also includes losses from the non IR-coated facets of the $7.97\m$ long SMF.}. 
Notably, this arrangement allows for easy switching between \hsp\, and \coh\, input signals by just blocking one of the two PBS inputs. So, for our experiments, only one input type is applied at any one time, i.e., for \hsp\, storage, the coherent state arm is blocked and vice versa. 
After spatially overlapping with the control, the signal is stored in the Cs cell, which is thermally insulated and surrounded by a magnetic shield (\textit{orange panel}, see section \ref{ch4_exp_imp}). 
The Cs vapour is heated to $70\degC$, with a cell cold spot at $67.5\degC$.
The control beam waist is approximately twice that of the signal, a ratio set to optimise the signal-to-noise ratio (see appendix \ref{ch7_control_focussing}). 
At the memory output, signal and control are separated spatially on a calcite PBD, which provides high-quality polarisation extinction of $\sim 40\dB$ for the undesired mode. The signal is SMF-coupled with $\eta_\text{SMF}^\text{sig.filt.} \approx 88 \,\%$ efficiency and sent into the signal frequency filter stage\footnote{
	The signal filter comprises three $\text{FSR}=18 \GHz$ and the 
	double-passed $\text{FSR}=103 \GHz$ etalon. 
} (\textit{green panel}, see section \ref{ch5_subsec_filter}).
Similar to chapter \ref{ch5}, the transmitted signal is split $50:50$ into two spatial modes to allow the measurement of the {\gtwo} autocorrelation\cite{Loudon:2004gd}. 
Both modes are MMF-coupled with a total transmission of all optical components from memory output to the MMF outputs of $T^\text{sig}_\text{filt}\approx 10\, \%$ for resonant {\tisa} pulses. 
The signal transmitted through the MMFs is observed by the APDs \spcmdh\, and \spcmdv\,, whose outputs feed again into the data acquisition system (see section \ref{sec_ch4_photon_detection}).

\subsubsection{Operation with coherent states}

Besides the triggering mechanism, the key setup difference for producing \coh\, inputs is the signal  generation using a control pulse pick-off, illustrated in fig. \ref{fig_6_setup} \textbf{b}. 
So the optical components are mostly similar to the \hsp\, case.  
For convenience, the $80\MHz$ {\tisa} pulse train is still sent through the SHG source, as the pulse energy loss due to $\eta_\text{SHG}$ is very small. 
From the non-converted IR, the control pulses are again picked by the \pockels\, (\textit{pink panel}), 
which is now triggered by the $80\MHz$ clock rate signal from a photodiode inside the {\tisa} oscillator. 
An internal delay-and-divide unit (\textit{Pockels cell DD2}) divides down the {\tisa} clock signal frequency by a pre-determined factor. This lower frequency signal is used for triggering the \pockels\, pulse picking windows in the same way as in section \ref{ch4_subsec_exp_implementation}.
Behind the \pockels\,, a PBS is used to pick off a small fraction of the selected control pulses, whose intensity is variable. This pick-off is phase-modulated by our EOM, generating $9.2 \GHz$ sidebands around the control field frequency\cite{Reim:PhD}. 
Similar to section \ref{ch4_subsec_exp_implementation}, the red-detuned sideband is selected, using an etalon with $\text{FSR} = 38.86\GHz$. 
As signal pulses are only required to be sent into the memory in the read-in time bin, the rf-modulation signal, driving the EOM, is gated using a fast rf-switch (active-low gating). This ensures that sideband modulation is only applied for the first pulse picked by the \pockels\, (see section \ref{ch6_setup_electronics} below). 
The generated \coh\, input signal is delayed to match the \hsp\, arrival times at the memory input and coupled into two SMFs of $1\,\text{m}$ and $7.97 \, \text{m}$ length, respectively, whereby the latter SMF is shared with the \hsp\, beam path. 
Since both signal types now occupy the same spatio-temporal mode, the \coh\, are also automatically overlapped with the control pulses inside the memory. 
The other optics are the same for both input signal types.

\subsection{Electronics and pulse timing sequence\label{ch6_setup_electronics}}

We now describe the two triggering schemes for operating the experiment with both input signal types, and also introduce the resulting pulse timing sequences (see fig.~\ref{fig_6_setup}~\textbf{d}~\&~\textbf{e}). 
Fig.~\ref{fig_6_setup}~\textbf{c} illustrates the electronics involved in both methods, with electronic units displayed by white boxes and lines with arrows that indicate the path of the electronic signals. 
For storage experiments with \coh\, input signals, the output pulses of the {\tisa} intra-cavity photodiode trigger the experiment (\textit{brown line}), gating the data acquisition and the preparation of memory control pulses. 
For \hsp\, storage, detection events of SPDC idler photons on APD \spcmdt\, form this master trigger (\textit{red lines}) instead, feeding into the same units. 
This enables easy alternation between both configurations. The required swaps in connections of the electronic circuitry are illustrated by the switches in fig. \ref{fig_6_setup} \textbf{c}, which feed into the electronic signal paths common to both triggering methods (\textit{black lines}). 
The remaining detection signals from APDs \spcmdh\, and \spcmdv\, (\textit{purple lines}) both feed into the FPGA. 
Additionally the TAC/MCA unit records signal arrival time histograms of photon detection events on \spcmdh\,\footnote{
	Since we only have one TAC unit available, 
	only the H-pol. arm is used to record the arrival time histograms. 
	Despite only observing $50\,\%$ of the total detection events, due to the $50:50$ 
	splitting of the signal between the H- and V-arm, the information in the MCA 
	traces is independent of this simplification.
}.
We will now give a more detailed description of the pulse timing sequences.

\subsubsection{Triggering using idler detection events (feed-forward operation):} 

\paragraph{Components} 
Output pulses of detector \spcmdt\, feed, in series, into a digital delay generator DDG1 (\textit{Stanford Research Systems DG535}) as well as the FPGA and the TAC. 
The latter two devices use signals from \spcmdt\, directly for gating of detection events registered by \spcmdh\, and \spcmdv\,, respectively.
For the FPGA this means counting of coincidences between events from \spcmdt\,, \spcmdh\, and \spcmdv\,.  To this end, all signals are delayed appropriately to match their arrival times with respect to the FPGA coincidence window (see section \ref{sec_ch4_photon_detection}). 
As explained in section \ref{sec_ch4_photon_detection}, we observe coincidences between the trigger and each signal detector, as well as triple coincidences between all three APDs. 
The TAC uses signals from \spcmdt\, as start triggers, and signals from \spcmdh\, as stop triggers. 
The DDG1 unit triggers the \pockels\, delay-divide unit DD1, which in turn switches the high voltage (HV) supplied to the \pockels\, crystal for pulse picking. 
Notably, for reasons related to optical pumping (see section \ref{ch6_setup_opticalpumping}), the storage time is set to $12.5\ns$. So the DD1 unit only emits one pulse for a single pulse picking window, selecting two consecutive {\tisa} pulses. 

\paragraph{Timing} 
Fig. \ref{fig_6_setup} \textbf{d} illustrates the timing sequence for these operations. 
Panel 1 shows the pulses of an SPDC signal and idler photon pair, as well as the heralding event on \spcmdt\, from idler photon detection. 
The time difference between both pulses is caused by the SMF-propagation delay of $\Delta \tau^\text{SPDC}_\text{sig.} \approx 460 \ns$. 
The gained delay time is needed to operate the electronic switching chain {\spcmdt\, $\rightarrow$ DDG1 $\rightarrow$ DD1}, shown in panel 2 by the output pulses of the three devices and the resulting \pockels\, pulse picking window.  
Since the delay between the pulses of DDG1, DD1 and the onset of the \pockels\, picking window are fixed by the devices' internal response times, the \pockels\, window's starting time is controlled by DDG1 via the time delay set on the respective output channel. 
The width of the \pockels\, pulse picking window determines the number of picked control pulses and is set by the DD1 unit. 
Up to 9 consecutive {\tisa} pulses can be picked (\textit{dashed line} in panel 2 \& 3 of fig. \ref{fig_6_setup} \textbf{d}), whereby the \first pulse defines the memory read-in bin and subsequent pulses are the read-out bins for a multi-pulse addressing of the memory \cite{Reim2012}. 
In this chapter we only consider the first read-out time bin\footnote{
	This limitation is imposed by output voltages of signal detectors 
	\spcmdh\, and \spcmdv\.. Both modules are Perkin Elmer SPCM-AQRH 
	single photon counter modules with an output voltage of 2.4 V, 
	whereas the herald APD is part of a quad module SPCM-AQ4C 
	with 4.5 V output voltage into $50 \, \Omega$. 
	As the signal detectors have to connect into 2 FPGA channels,  $50 \, \Omega$ terminated, 	
	alongside the connection to the TAC, $1 \, \text{M}\Omega$ terminated, for the H-arm, the output 
	voltage of the signal APDs is only sufficient to drive 2 FPGA channels. While it would generally 
	be possible to swap both APDs, technical constraints prevent this at the moment. 
}, and pick two consecutive control pulses only (\textit{solid \pockels\, window line} in panels 2 \& 3).
With appropriate \pockels\, window positioning, the {\tisa} pulse temporally closest to the SPDC signal is selected as the read-in control pulse, which results in the correct pulse sequence for the Raman memory (panel 4). 
Relying on detection events of \spcmdt\, to trigger control pulse creation makes the experiment sensitive to false detection events on \spcmdt\,, e.g. caused by dark counts. 
These events trigger storage experiments without the presence of any input signal and lower the signal-to-noise ration (SNR), as we see in section \ref{ch6_sec_single_photon_storage} below. 
They can be minimised using the temporal filtering introduced in section \ref{ch5_subsec_rates}, where we set the delay of an otherwise unused DDG1 channel to approximately the inverse of the idler photon detection rate to effectively block the output of \pockels\, trigger pulses between SPDC pair emissions\footnote{
	This happens because the DDG1 only accepts trigger pulses once it has gone through its 
	entire sequence of chosen delays. Therewith, undesired detection events that originate from 
	a white noise background in the idler channel and fall 
	into this effective dead time, cannot generate a \pockels\, trigger pulse. 
}. 

Lastly, we briefly mention how this apparatus is useful for temporal multiplexing applications, 
where the read-in and read-out control pulses both have to happen with random timings. 
The answer is that arbitrary timings between both can be obtained using both pulse picking windows of the \pockels\,, instead of only one, where each selects one {\tisa} pulse only. 
Since the DD1 unit receives two separate triggers for each picking window, the first would still be the idler detection signal, triggering \hsp\, storage. 
The second would be the herald of any successful quantum gate operation, triggering the release the single photon on demand.

\subsubsection{Triggering using the {\tisa} clock rate signal:}

For \coh\, storage, the experimental trigger is derived from the $80\MHz$ intra-cavity photodiode signal of the {\tisa} oscillator, fed directly into a second \pockels\, delay-divide-unit DD2. 
The DD2 reduces the repetition rate to $f_\text{rep}=5.72 \kHz$ and triggers the DD1 unit for HV-switching across the \pockels\, crystal. It also triggers a second digital-delay-generator (DDG2), which performs three tasks: 
First, it gates data acquisition on the FPGA and the TAC. 
Second, it gates the switching of the rf-modulation signal of the EOM to determine the generation window of the \coh\, input signal (see also section \ref{ch4_subsec_exp_implementation}). 
Third, it gates the optical pumping switch-off (see section \ref{ch6_setup_opticalpumping}). 
\newline
The resulting timing sequence is displayed in fig. \ref{fig_6_setup} \textbf{e}, with panel 1 showing the {\tisa} photodiode trigger. One of these pulses, marked by the \textit{grey vertical line}, triggers the \pockels\, window generation via {DD2 $\rightarrow$ DD1} 
and the switching of the EOM (panel 2). While the \pockels\, picking window has the same structure as for \hsps\,, the EOM switching window (\textit{blue line, EOM switch-off window}) turns off the EOM-frequency modulation for any control pulse selected by the \pockels\, but the first one (panel 3). 
Since only the first picked pulse is modulated by the EOM, only this pulse experiences a frequency shift into two-photon resonance with the control. 
Combining this EOM-modulated signal pulse and the selected control pulses yields the memory pulse sequence shown in panel 4. 
Here, it is important to have low residual EOM modulation in the subsequent read-out pulse time bins. 
Due to the finite rf switching time of $\approx 5 \ns$, the EOM is switched \textit{active low}, which allows for better extinction (see section \ref{ch4_subsec_exp_implementation}). 
\newline
Notably, for alignment purposes, idler detection events can also be used to prepare \coh\, input signals. Here, the DDG2 is supplied with trigger pulses from DDG1, which, in turn, receives its trigger pulses from detection events on \spcmdt\,. 
Everything else is the same, i.e., DDG2 supplies the gating to the EOM via the rf-switch.

\subsection{Optical pumping\label{ch6_setup_opticalpumping}} 
Another crucial setup part is the optical pumping by the external cavity diode laser 
(ECDL, see appendix \ref{ch3_diodelaser}), 
required to prepare of the Cs atoms in the $6^{2} \text{S}_{\frac{1}{2}}$ F$=4$ ground state.
The diode laser (\textit{grey panel} in fig. \ref{fig_6_setup} \textbf{a} - \textbf{c}) is sent into the Cs cell along the control field beam path, in counter-propagating geometry\footnote{
	A completely collinear arrangement with the control causes 
	back-scattering from the control SMF coupler into the signal detector. 
	To protect the APDs, the diode beam path is angled slightly with respect 
	to the control mode. Its bigger beam diameter ensures that there is no 
	influence on the memory efficiency from the angled geometry.
} 
(\textit{green line}). 
Its spatial mode size is chosen larger than both signal and control 
(see appendix \ref{ch7_control_focussing}). 

\paragraph{Timing}
Since the diode pumps population out of the $6^{2} \text{S}_{\frac{1}{2}}$ F$=3$ state, it also depletes any Raman memory spin-wave excited between the \cs\, hyperfine ground states. Hence for optimal memory operation, the pumping should be switched off during signal storage in the memory. 
The switching task is performed by an AOM in the ECDL set-up, whose modulation signal is turned on and off by an external trigger signal from the DDG2 unit. 
This introduces a significant time delay of $\Delta \tau_\text{diode} \approx 1.3\mus$ between the output of a DDG2 trigger pulse and the diode turn-off at the \cs\, cell\footnote{
	The main contributions come from internal delays in the 
	AOM driver module ($1\mus$) and the acoustic wave decay time ($220 \ns$).
	In the experiment, the diode laser is separated from the memory 
	by about $10\,\text{m}$ distance, due to space constraints. 
	This adds 	approximately $10\,\text{m}$ of BNC and 
	$15\,\text{m}$ of SMF, each adding about 
	$50\ns$ delay to the diode switch-off time. Both 		
	delays are the only ones that could be reduced without exchanging the 
	AOM for a faster device, such as a \pockels\,
}. 
\newline
For experiments using the {\tisa} clock trigger $\Delta \tau_\text{diode}$ is irrelevant. 
Here, the deterministic periodicity of trigger events allows to delay the diode switch-off until the next memory storage experiment is conducted. 
Fig. \ref{fig_6_setup} \textbf{e} displays the resulting diode power entering the memory cell (\textit{green line} in panel 4), where the diode is turned on by an \textit{active high} output of the DDG2, $\sim 1.4\mus$ after the memory control pulse sequence (\textit{green line, opt. pump switch-off} in panel 3). 
The pumping is maintained up to $1.4\mus$ before the read-in time bin, when the DDG2 output is set to low.
The resulting $1\mus$ absence in optical pumping just before the memory pulse sequence does not influence the storage process. 
\newline
When operating the memory in feed-forward mode, the diode laser cannot be turned off during storage at present.
Due to the probabilistic occurrence of herald events and the randomly fluctuating repetition rate of memory experiments, the diode laser would have to be turned off for the exact memory experiment, triggered by a particular idler detection event.
Since $\Delta \tau_\text{diode}$ is larger than the SPDC signal SMF-delay $\Delta \tau^\text{SPDC}_\text{sig.}$, simultaneous switching is currently not possible. 
To have reproducible conditions for every \hsp\, storage attempt, the diode laser is thus left on continuously{\footnote{
	Using a $\ge 260 \m$ long SMF for SPDC signal photon delay could be a potential remedy.
	Assuming lowest loss performance with $3.5 \,\frac{\text{dB}}{\text{km}}$ SMF- attenuation, 
	the transmission loss would amount to $18.9 \,\%$, while it is currently $7.3 \, \%$ for both SMFs. 
	The SPDC source heralding efficiency would accordingly reduce to $\tilde{\eta}_\text{her} = 19.3 \,\%$, 
	which is estimated based on the \textit{Thorlabs SM780} SMF. 
	The numbers are very sensitive to the actual fibre loss. 
	The $5 \,\frac{\text{dB}}{\text{km}}$ for the \textit{Thorlabs SM800} SMF already has a transmission 
	loss of $\approx 18\,\%$, reducing $\eta_\text{her}$ to $15 \, \%$.
}}.
The continuous pumping leads to a reduction in memory efficiency for storage of \hsps\, compared to \coh, as we will see below. 
However, it does not change the signal-to-noise ratio, because it equally reduces the memory noise\footnote{
	Optical pumping depletes the spin-wave during the storage time, which affects the contributions
	to the spin-wave from the input signal and those generated by FWM in the same way. 
}. 
Critically, optical pumping has no effect on the photon statistics of any of the fields involved in the storage process. 

\paragraph{Power}
The pumping power of the diode is set to $P_\text{diode} = 3\mW$ for experiments with \coh\,   
Yet, for \hsp\, storage without diode switch-off, this level is too high, resulting in the termination of memory read-out. For this reason, the pumping power is turned down by polarisation on the PBD at the memory output (fig. \ref{fig_6_setup} \textbf{c}) to a level that maximises the read-out efficiency.

\paragraph{Effects from state preparation} 
Optical pumping in the hot Cs gas also causes a significant amount of background counts scattered into the signal detectors. These counts are minimised by polarisation extinction on the PBD behind the memory, but they still contribute substantially to observed free running detector counts (singles counts). These events have no timing correlation with any of the experimental trigger. Gating the data acquisition on the trigger signal by FPGA coincidence detection between APDs \spcmdt\, and \spcmdh\, or \spcmdv\, thus eliminates this background. 
Accordingly, fluorescence has no influence on the memory efficiency and the {\gtwo} measurements\footnote{
	It would however severely affect a cross-correlation measurement\cite{Spring2013}, 
	where the measurement metric contains the detector singles counts. 
	As mentioned in section \ref{ch6_subsec_g2intro}, a cross-correlation measurement would require 
	the FPGA to count all APD channels in coincidence with the $80\MHz$ {\tisa} clock signal, which 
	exceeds the capabilities of our current FPGA system. 
} 
(see sections \ref{ch6_sec3_subsec1} \& \ref{ch6_subsec_meas_method}).

\section{Observation of single photon storage \label{ch6_sec_single_photon_storage}}

We now turn our attention to operating the interfaced source-memory system, where the observation of single photon storage is our first step. 
To this end we conduct mean-field measurements, i.e., we count photons for an integration time $\Delta t_\text{meas}$, while looking for effects from sequentially blocking and un-blocking the signal input and the memory control pulses. 
Thanks to our two detection systems, TAC/MCA and FPGA, we have two means of recording these measurements. 

As introduced in section \ref{sec_ch4_photon_detection}, we can firstly 
count the FPGA coincidences between the experiment trigger and each of the signal detectors (\spcmdh\, and \spcmdv\,). Their measurements follow the same methodology used in characterising the photon source in chapter \ref{ch5}.
Secondly, we can measure the signal photon arrival time histograms on the TAC/MCA-system, whose start and stop triggers are the experimental trigger (detection events on \spcmdt\, or the DD1 output of the \pockels\,) and the H-pol. signal detector \spcmdh\, (see fig. \ref{fig_6_setup} \textbf{c}), respectively. 
Since the resulting time series traces provide an intuitive picture of what is going on in the system, we will, in the following, study these to demonstrate the storage of \hsps\,. 
As one would expect, the FPGA measurements, presented in appendix \ref{app_ch6_FPGA_data}, yield the same results. 

To benchmark the system, we now also start to compare its performance with \hsp\, inputs to \coh\, input signals. 
To this end, we use \coh\, with input photon numbers of  $N_\text{in}^{\text{c.s.}} \in \left[ 0.23, 2.16 \right] \ppp$. 
Their lower limit is comparable to the \hsp\, input number of $N_\text{in}^{\text{HSP}} = 0.21 \ppp$, which is defined by the average source heralding efficiency throughout our measurements.


\subsection{Measuring single photon storage \label{ch6_sec3_subsec1}}

To visualise the storage and retrieval of single photons we send different combinations between the signal (\textit{s}) and control (\textit{c}) pulses into the atomic ensemble, which is spin-polarised by optical pumping with the diode laser (\textit{d}).
Observing photon storage and determining the memory efficiency requires to selectively block some of these fields, while counting photon detection events at the memory output. 
This is similar to the method used in section \ref{ch4_subsec_exp_implementation} (see fig. \ref{fig_ch4_trace_PD_750ns}), but, at the single photon level, requires to access all contributions to the detected signal:
\begin{enumerate}
\item Memory on (\textit{scd}): signal (\textit{s}), control (\textit{c}) and diode laser (\textit{d}) are sent into the Cs cell 
\item Input signal (\textit{sd}): control is blocked\footnote{
	Optical state preparation by the diode (\textit{d}) is required as an unprepared 
	Cs ensemble shows residual linear absorption from the 
	$6^{2}\text{S}_{\frac{1}{2}} \, \text{F} = 3 \rightarrow 6^{2}\text{P}_{\frac{3}{2}} \, \text{F'} = \{ 2,3,4 \}$-transition, 
	from which the signal is only detuned by $6\GHz$.
} 
\item Noise (\textit{cd}): input signal is blocked\footnote{
	Optical state preparation by the diode (\textit{d}) is required 
	as the noise characteristics depend on the preparation of the 
	atomic state (see chapter \ref{ch7}).
} 
\item Optical pumping background (\textit{d}): signal and control are both blocked 
\end{enumerate}
\noindent
We refer to these combinations as measurement settings and have 
incorporated them already in section \ref{sec_ch4_photon_detection}, when defining count rates and detection probabilities. 
In contrast to bright \coh\, inputs, where the memory efficiency can be determined from settings \textit{scd} and \textit{sd} alone\cite{Reim2010,England2012}, single photon level operation\cite{Reim:2011ys} also requires settings \textit{cd} and \textit{d}. 
Here there are additional contributions to the events detected by the APDs, whose major part comes from the memory noise floor \cite{Reim:2011ys} (setting \textit{cd}\footnote{
	Notably, this assumes the absence of any noise seeding effects from the presence of the signal. 
	We will verify this assumption for low photon numbers in section \ref{ch7_subsec_ASseeding}.
}). 
Its constituents will be discussed in chapter \ref{ch7}. 
Besides the noise, there is also some signal leakage in the read-out bin for \coh\, inputs, originating from residual modulation by the EOM due to its finite switching time. For \hsp\, inputs, the contribution from uncorrelated SPDC signal photons, produced by an SPDC pump pulse in the read-out bin is low (see section \ref{ch5_sec_photon_prod}). 
The presence of the diode laser (\textit{d}) in feed-forward operation for \hsp\, inputs also adds a contribution from scattering, which does not contribute to coincidence counts\footnote{
	This contribution has negligible count rates on the order of $\lesssim 0.1 \, \frac{\text{count}}{\text{sec}}$.
}.

\begin{figure}[h!]
\centering
\includegraphics[width=\textwidth]{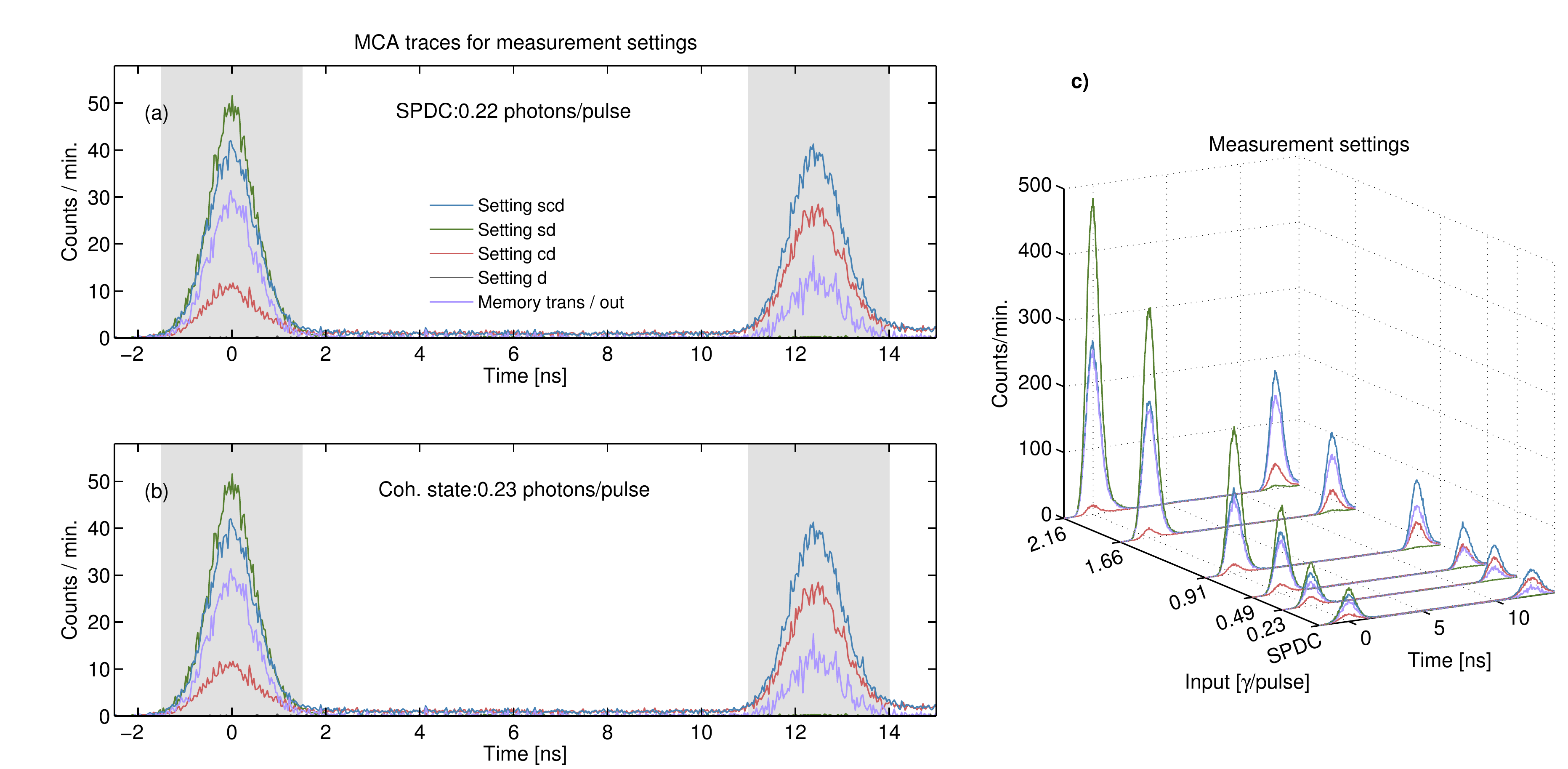}
\caption{
Photon arrival time histograms, showing storage for $\tau_\text{s}=12.5 \, \text{ns}$ of \textbf{(a)}
heralded single photon input and \textbf{(b)} weak coherent states, with input photon numbers of
$N_\text{in} = 0.21\ppp$ and $N_\text{in} = 0.23 \ppp$, respectively, recorded with $10\min$ integration time for each setting. 
\textit{Grey shaded areas} indicate read-in and read-out time bins.  
\textit{Green lines} show setting \textit{sd}, i.e., the input signal field transmitted through the memory for a blocked control (no memory). 
\textit{Red lines} show setting \textit{cd}, corresponding to the noise emitted by the memory. 
Setting \textit{scd}, i.e. active memory interaction with control and input signal applied together, is shown by \textit{blue lines}. 
It includes the signal, transmitted through and retrieved from the memory, and memory noise. 
Subtracting the \textit{red} noise trace from this line yields the non-stored signal, transmitted in the read-in bin, and the retrieved signal (\textit{lilac lines}), whose comparison with the input signal (\textit{sd}, \textit{green line}) results in the the memory efficiency. 
\textbf{(c)}: Traces for the measurement settings \textit{scd}, \textit{sd}, \textit{cd} (same colour coding as in \textbf{a} \& \textbf{b}), showing the effect of an increasing photon number $N_\text{in}$ for the \coh\, input signals. 
For comparison, also the \hsp\, traces are shown. 
The spacing for \coh\, are drawn to scale, but, for better visibility, 
the \hsp\, trace is moved to the foreground, despite having a similar $N_\text{in}$-value as the first \coh\, trace set.
}
\label{fig_6_TACtraces_2D}
\end{figure}

\paragraph{Observation of photon storage} 
We can now directly look at the TAC/MCA count rate histograms for the above settings to detect signal storage in the memory. 
Fig. \ref{fig_6_TACtraces_2D} \textbf{a} \& \textbf{b} show these for \hsp\, and \coh\, input signals with approximately equal input photon numbers ${N_\text{in}=0.21 \ppp}$ and ${N_\text{in}=0.23 \ppp}$, respectively.  
Storage and retrieval is immediately obvious, when comparing the signal input, setting \textit{sd} (\textit{green line}), 
with the memory on setting \textit{scd} (\textit{blue line}): 
Unblocking the control leads to a clear count rate decrease in the input and an increase in the output time bin, which is the signature of photons being stored and retrieved from the memory.
However the residual counts still present upon applying solely the control (\textit{cd}, \textit{red line}) reveal the presence of a non-negligible noise background.
We will demonstrate in chapter \ref{ch7}, that this noise is mainly a result of the four-wave-mixing (FWM) process, introduced in section \ref{ch7_subsec_FWM}. 
This noise is also present in the \textit{scd} traces, thus subtracting the noise (\textit{cd}) from the memory signal (\textit{scd}) yields the transmitted signal in the input and, importantly, the read-out signal in the output time bin (\textit{lilac line}).  
Despite the noise, the recalled signal is still clearly visible for input signal photon numbers of $N_\text{in} = \eta_\text{her} \approx 0.22 \ppp$ obtained from the photon source.
Importantly, this assumes that adding the input signal does not lead to the emission of more FWM noise than observed when just sending in a control pulse. In other words, there is no seeding of the noise by the input signal. We will show in section \ref{ch7_subsec_ASseeding} later on, that for the photon numbers we consider here, this is indeed a good approximation. 

When comparing the duration of the \hsp\, pulses, shown in fig. \ref{fig_6_TACtraces_2D}, with those for \coh\, inputs, we can see that \hsp\, are slightly longer than their \coh\, counterparts. Since the \hsp\, spectral bandwidth is broader than the {\tisa} pulses (see section \ref{ch5_subsec_exp_hsp_spectrum}), one would initially expect the opposite. The reason for the temporal broadening is the triggering of the TAC on the idler detection events for \hsp\, inputs; appendix \ref{app6_TAC_pulse_durations} provides more information about this.

\paragraph{Storage at higher input photon numbers}
Besides applying \coh\, inputs at the same $N_\text{in} \approx \eta_\text{her}$, we have also investigated \coh\, inputs with  ${N_\text{in} = \{0.23, 0.49,  0.91, 1.66, 2.16 \} \ppp}$. 
While stronger inputs are particularly useful to characterise the photon statistics of the stored signal 
(see section \ref{ch6_photon_statistics} below), they can also be employed to demonstrate the independence of the memory efficiency on the input photon number.
The resulting histograms for all input photon numbers are shown in fig. \ref{fig_6_TACtraces_2D} \textbf{c}, 
where they are contrasted with the \hsp\, input. 
Again, we plot the traces for all relevant measurement settings and the non-stored signal portion, transmitted through the memory, as well as the retrieved signal. 
The data illustrates how the amount of read-out signal and therewith the SNR grows for increasing {\Nin}. 
In the limit of bright \coh\, inputs, the noise is negligible, for which reason the memory efficiency can be determined by settings \textit{scd} and \textit{sd} only\cite{Reim2010,England2012}, as it was done in chapter \ref{ch4}.
Comparing the amount of signal stored in the memory, i.e. the difference between the \textit{green} and the \textit{lilac lines} in the read-in bin of fig.~\ref{fig_6_TACtraces_2D}~\textbf{c}, with the retrieved signal (\textit{lilac line} in read-out bin), shows that their ratio is approximately constant. This translates into similar memory efficiencies for all input photon numbers {\Nin} and both signal types.

\subsection{Count rates from TAC/MCA data\label{subsec_ch6_TAC_count_rates}}
When we introduced count rates and detection probabilities in section \ref{sec_ch4_photon_detection}, we focussed on the FPGA recordings. 
We will now briefly outline, how we obtain these numbers from the TAC/MCA histograms of fig. \ref{fig_6_TACtraces_2D}, since we need them to determine the memory efficiency and the signal to noise ratio (SNR). 
Their FPGA counterparts are shown in appendix \ref{app_ch6_FPGA_data}. 
We first determine the area $a^{t}_{i}$ of a pulse in a histogram by integrating, i.e. summing, the counts in each channel of histogram trace that falls into an integration window of size $\Delta t_\text{int}$. 
Here, we choose $\Delta t_\text{int} = 5 \ns$\footnote{
	This is the same time as used by the FPGA, 
	which allows to directly compare the obtained memory efficiencies and SNRs.
	However, while for the FPGA we are limited to $\Delta t_\text{int} \ge 5 \ns$, 
	much smaller integration times can be chosen for the MCA, where $\Delta t_\text{int}$ is limited
	by the duration represented by a single channel. 
} and assign the time bin and the measurement settings with indices  $t$ and $i$, respectively. 
Fig. \ref{fig_6_TACtraces_2D} illustrates $\Delta t_\text{int}$ by the \textit{grey} shaded areas.
Due to Poissonian counting statistics, the errors on the pulse areas are $\delta a^{t}_{i} = \sqrt{a^{t}_{i} }$. 
Importantly, we use different total measurement times  $\Delta t_{\text{meas},i}$ for the different settings. 
As discussed in sections \ref{ch6_subsec:MemEffCalc} \& \ref{ch6_subsec_meas_method}, 
two types of measurements are performed: 
one mainly designed to determine the memory efficiency, and another aimed at determining the photon statistics of the optical fields. 
The data shown in fig.~\ref{fig_6_TACtraces_2D} belongs to the former category. 
Here the settings \textit{scd}, \textit{sd} and \textit{cd} are recorded for $\Delta t_{\text{meas},i} = 10 \min$, 
whereas setting \textit{d} (not shown in fig.~\ref{fig_6_TACtraces_2D}), is only measured for 
$\Delta t_{\text{meas},d} = 5\min$. 
For the second measurement category the integration time is increased to 
$\Delta t_{\text{meas},i} \ge 30 \min$ for \textit{scd} and \textit{cd}, while it is reduced to 
$\Delta t_{\text{meas},sd} = 5\min$ for \textit{sd}. 
To still determine the correct absolute number of counts for calculating the memory efficiency, the integration time difference is accounted for by scaling factors 
$\textfrak{t}_i = \Delta t_{\text{meas},i}^{\text{max}}/ \Delta t_{\text{meas},i}$, 
where 
$\Delta t_{\text{meas},i}^{\text{max}}$ is the longest setting integration time in the respective sequence 
$\{scd, sd,cd,d\}$. 
With these, we define the $\Delta t_{\text{meas},i}$-independent counts $\bar{c}^t_i=  \textfrak{t}_i  \cdot a^t_i $. 
Dividing $\bar{c}^t_i$ by the total measurement time $\Delta t_{\text{meas},i}$ of setting $i$, in turn results in count rates $c_{i}^t =\frac{\bar{c}^t_i}{\Delta t_{\text{meas},i}}$, which are similar to the values defined in section \ref{sec_ch4_photon_detection}.

\subsection{Number of input photons per pulse} 
To determine the number of input photon {\Nin}, sent into the memory for our storage experiments, we can now either use the TAC/MCA data of fig. \ref{fig_6_TACtraces_2D} above, or the FPGA data, presented in appendix \ref{app_ch6_FPGA_data}.
Since the TAC/MCA traces only contain counts from APD \spcmdh\,, whereby the signal is already split $50:50$ between \spcmdh\, and \spcmdv\,, we can either multiply the count rates, extracted from the TAC/MCA histograms, by a factor of 2, or use the sum of the FPGA coincidence counts from both detectors with the experimental trigger. 
In either case, we will also need to know the trigger rate $c_{T,i}$ (see section \ref{sec_ch4_photon_detection}), obtained by the FPGA, and we will use the FPGA coincidences 
$\bar{c}^t_{i} := \bar{c}^t_{i,{H|T}} + \bar{c}^t_{i,{V|T}}$. 
{\Nin} represents the number of photons sent into the memory upon each storage trial, for which reason its definition is essentially the same as that of the heralding efficiency (see eq. \ref{eq_ch5_her_eff}). 
So we require the counts $\bar{c}^\text{in}_{sd}$ for setting \textit{sd} in the read-in time bin ($t=\text{in}$), which are the averages over all FPGA recordings $c^\text{in}_{sd}(t_m)$ within one run (see section \ref{sec_ch4_photon_detection}). 
Moreover, we need the repetition rate of memory experiments $f_\text{rep} = \bar{c}_{sd,{T}}$. 
For \coh\, inputs, the latter is set by the \pockels\, DD1 unit to $f_\text{rep} = 5.722 \kHz$ (see section \ref{ch6_setup_electronics}), while for \hsp\, inputs it is set by the detection rate of idler photons on \spcmdt\, (see section \ref{ch5_sec_photon_prod}). 
Again, these are averages of all recordings $\left\{ c^\text{in}_{i,T}(t_m) \right\}$ within an FPGA run (see appendix \ref{app_ch6_FPGA_data} and  fig. \ref{fig_6_FPGArates_memeff}). 
The other two required parameters are the detection efficiencies of APDs \spcmdh\, and \spcmdv\,, which are both assumed as $\eta_\text{APD,H} \approx \eta_\text{APD,V} \approx 50 \, \%$ at $852\, \nm$, as well as the transmission of the signal field from the memory input facet to the input of the APDs. 
During the alignment of the system, this transmission is measured for each optical element within the transmission line individually, as well as for the total transmission. On average, it amounts to $T_\text{sig} \approx 10 \pm 2\,\%$, where the error is the standard deviation over the  day-to-day variation. 
Moreover, over the course of the measurement time, drifts in the etalons\footnote{
	The misalignment happens over the course of $2-4 \, \text{h}$, 
	although there are particular times during the day, when misalignment 
	happens more quickly. These coincide with times of greater temperature changes, e.g. when people arrive or 
	leave the department or night/ day transitions. 
	Since the etalons are housed in boxes to prevent accidental misalignment, temperature changes and drifts 
	of the mechanical mountings are the only possibilities for their misalignment. 
	Long term deflection measurements of a laser pointer beam on 
	an etalon has not shown any noticeable instability in the mounting.		
} reduce $T_\text{sig}$. 
These drifts are reset periodically by alignment checks of the filter resonance transmission (see section \ref{ch6_subsec:MemEffCalc} below). 
Since the exact transmission is not recorded during each measurement, the transmission determined during the initial filter alignment $T_\text{sig}$ is used. Therewith, we estimate the number of input photons as 
\begin{equation}
N_\text{in} = \frac{\bar{c}^\text{in}_{sd}}{\bar{c}_{sd,T} \cdot T_\text{sig} \cdot \eta_\text{APD, H/V}}.
\label{ch6_eq_Nin}
\end{equation}
\noindent
For measurements that require the combination of data over several days (see section \ref{ch6_subsec_meas_method}), 
the transmission entering eq. \ref{ch6_eq_Nin} is just the average over all $T_\text{sig}$ of the different days. 
Similarly $\bar{c}^\text{in}_{sd}$ and $\bar{c}_{sd,T }$ are obtained by taking the mean over the entire \sd\, data, which is a combination of the coincidence data obtained in each individual measurement. 
The error estimate\footnote{
	Notably, errors on $T_\text{sig}$ and $\eta_\text{APD}$, for APDs \spcmdh\, and \spcmdv\, are neglected, 
	since an error $\Delta \eta_\text{APD}$ is hard to measure 
	in general and the error of $\Delta T_\text{sig}$ is greater intra-day than on a day-to-day basis. 
	Since including a day-to-day-based 
	$\Delta T_\text{sig}$ would still underestimate the error $\Delta N_\text{in}$, we drop it altogether for simplicity.
}
on $N_\text{in}$ follows from Gaussian error propagation on $\bar{c}^\text{in}_{sd}$.

\subsection{Memory efficiency\label{ch6_subsec:MemEffCalc}}

We can now use the FPGA or TAC count rates to define the memory efficiency. 
In the experiment we measure the settings $i$ sequentially, so we use the scaled count rates, defined in section \ref{subsec_ch5_fpga_count_rate_definition}. 
Moreover, we also have to take into account the repetition rate $f_\text{rep} =c_{i,T}$, which can vary between the settings $i$, when idler photon detection events are used as the experiment trigger. 
This is the case, because SPDC is a probabilistic process, whose photon pair generation rate has an inherent uncertainty. More important however are periodic idler count rate drifts, arising from some experimental instability in the photon source (see section \ref{app_ch5_insufficiencies}).  
For this reason, we evaluate the memory efficiency in terms of the detection probabilities 
$p_{i,t} = \frac{\tilde{\bar{c}}_{i}^t}{\tilde{\bar{c}}_{i,T}}$, defined in section \ref{subsec_ch5_fpga_count_rate_definition}, which normalise the coincidence counts to the number of trigger events.

\paragraph{Calculating the memory efficiency}
When discussing the TAC count rate histograms in section \ref{ch6_sec3_subsec1} above, we have essentially already outlined the relevant steps for obtaining the memory efficiency. 
Having one set of measurements for all settings \scd\,, \sd\,, \cd\, and \diode\, available, 
we obtain the amount of read-in signal by subtracting the signal, transmitted through the memory when the Raman interaction is on, i.e. when the control field is applied (\scd\,), from the input signal, obtained when the control is blocked (\sd\,). 
Similarly, the amount of signal read-out of the memory is accessed by subtracting the memory output with active Raman retrieval (\scd\, output bin), from the amount of signal sent into the memory (\sd\,). So far, this is the same as the definition used in chapter \ref{ch4}. However, as fig. \ref{fig_6_TACtraces_2D} shows, at the single photon level, we also have memory noise, which adds undesired counts whenever the control field is on. So we have to subtract the counts for setting \cd\, from those of \scd\,. 
Finally, we also have noise from the diode laser (\diode\,), which has to be subtracted\footnote{
	Note, for the difference $p^t_{scd}- p_{cd}^t$, the diode laser noise drops out automatically, as it contributes
	to the detection probabilities for settings \scd\, and \cd\, by equal amounts. 
}, when measuring the input signal transmitted through the pumped \cs\, ensemble without control (\diode\,). 
So, in terms of the detection probabilities $p_i^t$, the total memory efficiency is
\begin{align}
\label{eq6_memeff}
\eta_{\text{mem}}	&= \left[p_\textit{scd}^{\text{out}}-p_\textit{cd}^{\text{out}}-\left(p_\textit{sd}^{\text{out}}-p_\textit{d}^{\text{out}} \right) \right]/\left(p_\textit{sd}^{\text{in}} - p_\textit{d}^{\text{in}} \right), \\   
\Delta \eta_\text{mem} &=\sqrt{\alpha + \beta},
\label{eq6_err_memeff}
\end{align}
\noindent
with
\begin{align}
\alpha &= \left(\frac{p_{scd}^\text{out} - p_{cd}^\text{out} - p_{sd}^\text{out} + p_{d}^\text{out} }{\left( p_{sd}^\text{in} - p_{d}^\text{in} \right)^2}\right)^2 \cdot 
			\left( \left(\Delta p_{sd}^\text{in} \right)^2 + \left(\Delta p_{d}^\text{in} \right)^2 \right),  \nonumber \\
\beta &= \frac{1}{\left( p_{sd}^\text{in} - p_{d}^\text{in} \right)^2} \cdot 
			\left( \left(\Delta p_{scd}^\text{out}\right)^2 + \left(\Delta p_{cd}^\text{out}\right)^2 + \left(\Delta p_{sd}^\text{out}\right)^2 + \left(\Delta p_{d}^\text{out}					\right)^2 \right). \nonumber
\end{align}
\noindent
The read-in efficiency is
\begin{align}
\label{eq6_memeff_in}
\eta_\text{in} &= \frac{ p^\text{in}_{sd} + p^\text{in}_{cd} - p^\text{in}_{d} - p^\text{in}_{scd} }{p^\text{in}_{sd} - p^\text{in}_{d}}, \\
\Delta \eta_\text{in} &= \sqrt{
\left( \frac{p^\text{in}_{scd}  - p^\text{in}_{cd}}{\left( p^\text{in}_{sd} - p^\text{in}_{d} \right)^2} \right)^2 \cdot 
\left( \left( \Delta p^\text{in}_{sd} \right)^2 +  \left( \Delta p^\text{in}_{d} \right)^2 \right) +
\left( \frac{1}{p^\text{in}_{sd} - p^\text{in}_{d}} \right) \cdot \left( \left( \Delta  p^\text{in}_{scd} \right)^2 +  \left(\Delta  p^\text{in}_{cd} \right)^2 \right)
},
\label{eq6_err_memeff_in}
\end{align}
\noindent
and the retrieval efficiency is  
\begin{align}
\label{eq6_memeff_ret}
\eta_\text{ret} &:=  \frac{\eta_\text{mem}}{\eta_{\text{in}}}, \\
\Delta \eta_\text{ret} &=  \sqrt{ \frac{\Delta \eta_\text{mem}^2}{ \eta_{\text{in}}^2} + \frac{\eta_\text{mem}^2}{ \eta_{\text{in}}^4} \cdot \Delta \eta_{\text{in}}^2}.
\label{eq6_err_memeff_ret}
\end{align}
\noindent
Here, $\Delta \eta$ are the standard errors of the mean on the respective efficiency values, whose functional dependences result directly from Gaussian error propagation. Notably, the retrieval efficiency is not experimentally accessible separately and is hence defined in terms of the measurable value $\eta_\text{mem}$ and $\eta_\text{in}$.

\begin{figure}[h!]
\centering
\includegraphics[width=1\textwidth]{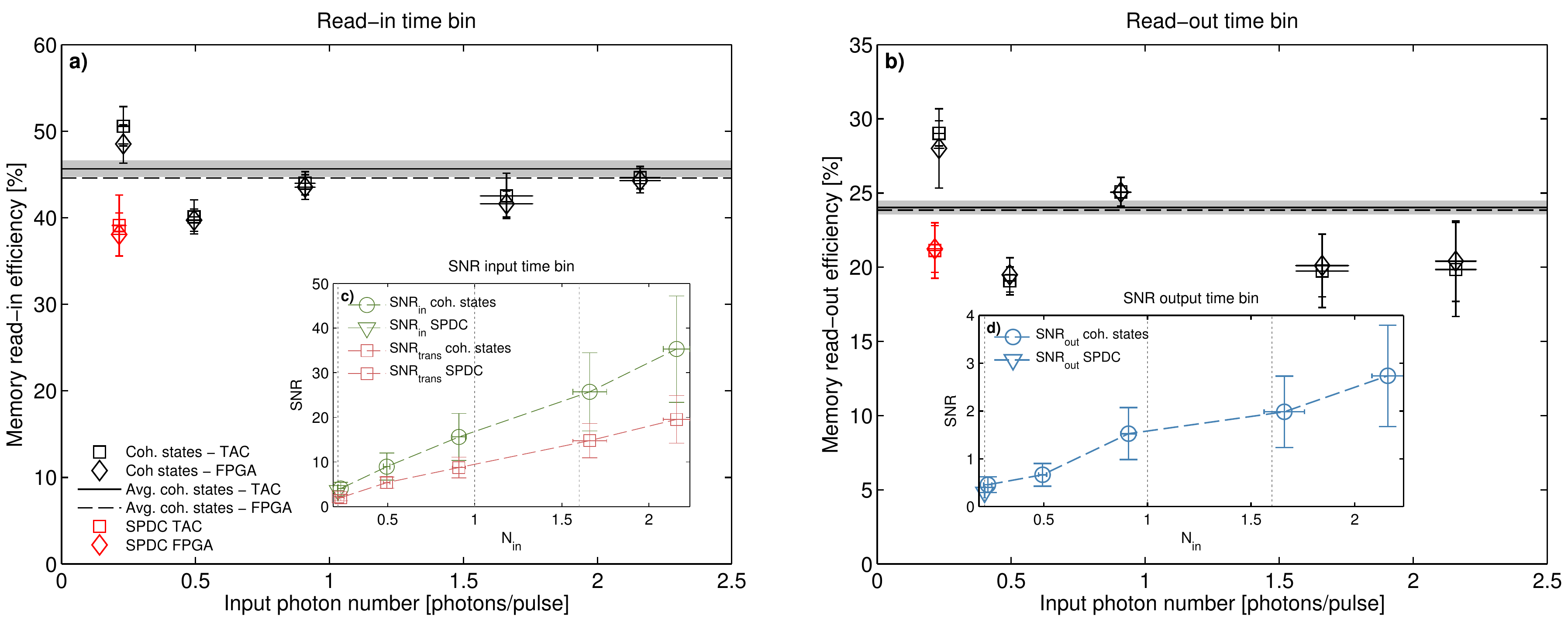}
\caption{
\textbf{(a)} \& \textbf{(b)}: Memory efficiencies for read-in ($\eta_\text{in}$) and read-in followed by retrieval ($\eta_\text{mem})$ for \coh\, input signals for increasing input photon number {\Nin}, marked by \textit{black points}, and \hsp\, inputs, marked by \textit{red points}. \textit{Square} and \textit{diamond markers} represent efficiencies calculated from TAC and FPGA data, respectively.
The \textit{solid} and \textit{dotted black horizontal lines} are the average efficiencies over all \coh\, {\Nin}-values for TAC and FPGA data, respectively. The \textit{grey shaded area} marks the 1-sigma standard error for the average TAC efficiency.  
The insets show the three SNRs: \textbf{(c)} depicts the input time bin with SNR$_\text{in}$ in \textit{green}, and SNR$_\text{trans}$ in \textit{red}. \textbf{(d)} contains SNR$_\text{out}$ for the output time bin in \textit{blue}. In both plots, the \textit{dotted vertical lines} indicate, from left to right, the current heralding efficiency with $N_\text{in}= 0.21 \ppp$, a source with perfect heralding efficiency at $N_\text{in}=1$, and the input photon number used in our first single-photon level experiments\cite{Reim:2011ys} with \coh\, at $N_\text{in}=1.6\ppp$.
}
\label{fig_6_avg_memeff}
\end{figure}

Before we look at any results, it must be stressed that eqs. \ref{eq6_memeff} - \ref{eq6_err_memeff_ret} assume independence of the noise level, measured via \cd\,, from the amount of input signal, sent into the memory. 
In other words, the noise that is present, when measuring setting \scd\,, is the same as when recording \cd\,. For FWM, this is in general not the case and we will study the noise increase, caused by the input signal, in section \ref{ch7_subsec_ASseeding}. 
Yet, we will also prove, that for the signal input photon numbers {\Nin}, considered here, this effect is negligible. We will also see this below, when studying the memory efficiency for increasing values of {\Nin} (see fig. \ref{fig_6_avg_memeff}); so eqs. \ref{eq6_memeff} - \ref{eq6_err_memeff_ret} are applicable.

\paragraph{Measurement sequence and average memory efficiency over the experiment}
The above equations suffice to determine the efficiency of one single measurement run, i.e. one recording sequence of settings \sd\,, \scd\,, \cd\, and \diode\,.
Yet, to investigate the photon statistics later on in section \ref{ch6_photon_statistics},
longer integration for the settings \scd\, and \cd\, are required for each input photon number {\Nin}, which necessitates data accumulation over several days.
Since long measurement times and multiple day data aggregation come inevitably with the consequence of drifts and potential systematic changes in the system alignment, we seek to minimise these by following a set procedure for apparatus alignment and for the collection of data. 
We always measure sequences of separate runs, whereby in each run the set of relevant settings \sd\,, \scd\,, \cd\, is investigated with predefined integration times: 
 \sd\, is recorded for $\Delta t_\text{meas} \approx 5-10\min$ before the settings \scd\, and \cd\, are measured for $\Delta t_\text{meas} \ge 30 \min$ each. 
In addition, at the beginning and the end of each day, runs containing all four settings are recorded, with 
$\Delta t _\text{meas}= 10\min$ for \scd\,, \sd\,, \cd\, and $\Delta t _\text{meas} = 5\min$ for \diode\,. 
Additionally, a \diode\, setting is taken for $\Delta t_\text{meas} = 5\min$ whenever the diode laser had to be relocked to the Cs resonance. 
From that point onwards, this updated dataset for the \diode\, setting is used in the memory efficiency calculation.
Between every measurement run, the system alignment is checked and reset, if required. This includes checking \hereff\,, the signal filter resonance, the overlap between signal and control in the \cs\, cell, and the SMF coupling efficiencies for signal and control\footnote{
	The memory efficiency critically depends on the alignment of the signal-control beam overlap, 
	whereby small drifts in the control field beam pointing lead to an efficiency decrease 
	over long measurement times. The same is true for the SMF-coupling efficiency of the 
	control, which results in a decrease in control pulse energy at the \cs\, cell.
}.
From the resulting sets of data, first the efficiency $\eta_{\text{mem},m}(N_\text{in})$ is evaluated separately for each run $m$. 
Second, the average efficiency $\eta_\text{mem}(N_\text{in})$ is obtained by a weighted average over all  $\eta_{\text{mem},m}(N_\text{in})$, with weighing factors 
\textfrak{m}$_{m,1} = \Delta t_{\text{meas},m}^{scd}/ \sum_{m}{ \Delta t_{\text{meas},m}^{scd}}$.
For simplicity only the measurement durations $\Delta t_\text{meas}^{scd}$ for settings \scd\,, representing active memory interaction, are used.
Last, for \coh\, input signals at $\tau_\text{s} = 12.5\ns$ storage time, the average memory efficiency over all input photon numbers $N_\text{in}$ is calculated again by weighted averaging, using weighing factors 
\textfrak{m}$_{N_\text{in},2} = \Delta t_\text{meas}^{scd}(N_\text{in}) /  \sum_{N_\text{in}}{ \Delta t^{scd}_\text{meas}(N_\text{in})}$ to account for the total measurement time for each photon number {\Nin}. 
The same procedure is applied to calculate the average read-in efficiency {\etain}.

\paragraph{Memory efficiency results}

Averaging the individual efficiencies over all runs $m$ for each photon number {\Nin} yields the efficiency values $\eta_\text{mem}(N_\text{in})$ and $\eta_\text{in}(N_\text{in})$, shown in fig. \ref{fig_6_avg_memeff}. 
The figure contains data for both {\etain} (\textbf{a}) and {\etamem} (\textbf{b}) recorded by FPGA and TAC. 
Notably, we have plotted the results for both recording devices to show, that they agree with one another, as one would expect. 
While detailed results for all {\Nin} are stated in the appendix \ref{app6_memeff}, the efficiencies for \hsp\, and \coh\, input at $N_\text{in} \approx \eta_\text{her}$ are noteworthy enough to be quoted here: 
\begin{framed}
\begin{itemize}
\item \hsp\. inputs at $N_\text{in} = 0.21 \ppp$: $\eta_\text{mem} = 21 \pm 2 \%$, $\eta_\text{in} = 39 \pm 3 \%$ ($\rightarrow$ $\eta_\text{ret} = 54 \pm 7 \%$) 
\item \coh\. inputs at $N_\text{in} = 0.23 \ppp$: $\eta_\text{mem} = 28 \pm 1 \%$, $\eta_\text{in} = 51 \pm 2 \%$ ($\rightarrow$ $\eta_\text{ret} = 57 \pm 3 \%$) 
\end{itemize}
\end{framed}
\noindent
Performing the experiment with the SPDC signal photons results in lower efficiencies than when using \coh\, inputs derived from the {\tisa}\, pulses. 
This drop originates from the already-discussed mismatch between the marginal SPDC signal spectrum and the {\tisa} pulse spectrum (see sections \ref{ch5_sec_hsp_bandwidth} \& \ref{ch5_sec_exp_hsp_bandwidth}), as well as the presence of the optical pumping beam during the storage time in feed-forward operation for \hsps\,\footnote{
	Note, as explained in section \ref{ch6_sec:setup}, we cannot turn off the diode laser for \hsp\, inputs, 
	as we would have to trigger the switch-off on the detection of the same idler photon that heralds
	the signal photon going into the memory. In its current configuration (see appendix \ref{ch3}), 
	the diode laser switch-off takes $\sim 1.5\mus$. In turn, this would require to delay 
	the \hsps\, in optical fibre for at least as long. Since the currently used SMFs are not long enough,
	we run the experiment with the diode laser on. Conversely, for \coh\, inputs this is not a problem,
	thanks to their deterministic repetition rate, allowing to trigger diode switching on one of the {\tisa} pulses
	preceding the pulse that is used for generating the \coh\, input signal. 
}. 
Separating $\eta_\text{mem}$ into read-in ($\eta_\text{in}$) and retrieval efficiency ($\eta_\text{ret}$) allows to estimate the magnitude of both effects. 
Since the presence of the optical pumping does not influence the read-in efficiency, the $10\,\%$ reduction in $\eta_\text{in}$ should predominantly be caused by the \hsp\, spectral mismatch. 
Conversely, since the control works like a spectral filter for the signal\cite{Nunn:DPhil}, one would expect no more influence from this mismatch in $\eta_\text{ret}$. So the $3\, \%$ reduction we see in $\eta_\text{ret}$-value should be purely due to the presence of the optical pump, which is depleting the spin-wave during the storage time. \newline
In an attempt to verify this observation, we have performed an experiment for \coh\, input signals $N_\text{in}=0.23\ppp$, where we have compared storage with the diode on and off. 
While the complete set of results can be found in appendix \ref{app6_coh_pumping_on}, the main findings for the current discussion are the changes in storage efficiency:
\begin{itemize}
\item \coh\, at $N_\text{in}=0.23\ppp$, no pumping: $\eta_\text{mem} =31 \pm 1 \%$,  $\eta_\text{in} =48 \pm 1\%$ ($\rightarrow$ $\eta_\text{ret} = 65 \pm 1\%$) 
\item \coh\, at $N_\text{in}=0.23\ppp$, pumping on: $\eta_\text{mem} =24 \pm 1 \%$,  $\eta_\text{in} =46 \pm 1\%$ ($\rightarrow$ $\eta_\text{ret} = 52 \pm 1\%$) 
\end{itemize}
\noindent
As expected $\eta_\text{in}$ is approximately unaffected, because the input is the same \coh\, signal. 
The magnitude by which optical pumping affects $\eta_\text{ret}$ is significantly larger than the $\sim 3\,\%$ drop we have estimated above, when comparing the retrieval efficiencies for the \coh\, and \hsp\, measurements. The reason for this is unclear at the moment. 
However we can still summarise, that the lower storage efficiencies for \hsps\,, compared to \coh\,, are caused by lower 
$\eta_\text{in}$-values, due to spectral mismatch with the memory control in the read-in bin, and a reduced $\eta_\text{ret}$, due to spin-wave depletion by the optical pumping. 

Fig. \ref{fig_6_avg_memeff} also contains the average efficiencies over all \coh\, input numbers (\textit{solid horizontal line} for TAC, \textit{dotted horizontal line} for FPGA data).
The scattering of the individual efficiencies around the overall average, particularly the lack of an efficiency increase for larger $N_\text{in}$, shows that there are no stimulation effects from the signal input (see also section \ref{ch7_subsec_ASseeding}). 
In fact the efficiencies drop slightly towards larger \footnote{
	This is most likely an effect from worsening conditions in the thermal 
	insulation of the memory cell, whereby the measurements at small 
	$N_\text{in}$ have been performed when the cell system was freshly assembled 
	(see section \ref{ch7_Coldspot}). 
} $N_\text{in}$.
So, the input signal strength neither affects the Raman memory interaction nor the noise generation\footnote{
	In fact, we will see in section \ref{ch7_subsec_ASseeding} that there is a stimulation in the anti-Stokes 
	noise production, which will be accompanied by stimulated Stokes noise creation at the signal 
	frequency. However this effect is small and only present at much larger input photon numbers.
}. 
In other words, increasing $N_\text{in}$ only improves the SNR, yet it does not change the Raman or FWM coupling at the single photon level.
Importantly, we can thus legitimately study the photon statistics of signals retrieved from the memory, as a function of $N_\text{in}$.

\subsection{Noise background and signal to noise ratio (SNR)\label{ch6_subsec_SNR}}
To quantify the memory noise floor the coincidence probabilities $p_{{cd},t}$ in time bin $t$ can be utilised in a similar manner to eq. \ref{ch6_eq_Nin} with: 
 $N^t_\text{noise} = N^t_\text{cd} = p^t_{cd} / \left( \eta_\text{APD} \cdot T_\text{sig.} \right)$.
Again, an APD detection efficiency of 
$\eta_\text{APD} \approx 50 \,\%$ 
is assumed, and $T_\text{sig} \approx 10 \, \%$ represents the transmission of the optics behind the {\cs} cell, including the signal filter stage. 
As before, $T_\text{sig}$ is a weighted average of the individual, daily determined transmissions, for all days when the setting \cd\, has been measured. The weighing factors are the total measurement time for the setting \cd\, on each day.  
Therewith we obtain noise figures of: 
\begin{framed}
\begin{equation}
N^\text{in}_\text{noise} = (6 \pm 2) \cdot 10^{-2} \, \frac{\gamma}{\text{pulse}}, \quad N^\text{out}_\text{noise} = (15 \pm 5) \cdot 10^{-2} \, \frac{\gamma}{\text{pulse}}.
\label{ch6_eq_NoiseFloor}
\end{equation}
\end{framed}
We can now also determine the SNR of the memory as a function of the input signal strength $N_\text{in}$.
To this end, we use the detection probabilities $p^t_{sd}$ and $p^t_{cd}$ and define three SNRs: 
\begin{equation}
\text{SNR}_\text{in} =p_{sd}^\text{in}/p_{cd}^\text{in} \quad , \quad \text{SNR}_\text{trans} =(1-\eta_\text{in}) \cdot p_{sd}^\text{in}/p_{cd}^\text{in} \quad \text{and} \quad \text{SNR}_\text{out} =\eta_\text{mem} \cdot p_{sd}^\text{in}/p_{cd}^\text{out}.
\label{eq_ch6_SNR}
\end{equation}
The first quantity represents the SNR as it would be obtained by an incoherent mixture between the input signal and the noise from the memory, which is the expected SNR if there was no Raman storage but just noise added to the input signal. 
SNR$_\text{trans}$ is the SNR seen in the input time bin, containing the signal transmitted through the memory and the input time bin noise. 
The key metric of the three, SNR$_\text{out}$, is the SNR between the retrieved signal and noise in the output time bin. 
All three ratios are plotted in fig. \ref{fig_6_avg_memeff} \textbf{c} and \textbf{d} for the read-in and the read-out time bin, respectively. 

For \hsps\,, the input with SNR$_\text{in} = 3.5 \pm 1.3$ is quite promising, but unfortunately this number degrades substantially in the output, down to SNR$_\text{out} = 0.3 \pm 0.1$. 
Notably, memory devices should possess an SNR$_\text{out} \gg 1$ in order to be of practical use in quantum networks. 
In our case the higher noise level in the output bin together with the below-unity memory efficiency ($\eta_\text{mem} < 100\,\%$) are responsible for the deterioration. 
As we will see in chapter \ref{ch7}, {\etamem} and $N_\text{noise}$ are linked to one another, whereby the quoted levels already present an experimental optimum for our current Raman memory setup.  
Thus the only actual free parameter\footnote{
	Of course, the choice of the memory medium as well as using a free-space, 
	single-pass gas cell are, so to speak, also free 
	parameters. 
} 
in eqs.~\ref{eq_ch6_SNR} is the number of input photons per pulse $N_\text{in} \sim p_{sd}^\text{in}$, which corresponds to the heralding efficiency \hereff\,. 
This is the reason for the paramount importance of the heralding efficiency in the design considerations for a single photon source, mentioned in section \ref{ch5_subsec_design_criteria}. 

For the \cs\, memory, a perfect heralding efficiency, i.e. $N_\text{in}=1$, would lead to an SNR$_\text{out}\approx 1.45$ (see \textit{vertical lines} in fig. \ref{fig_6_avg_memeff} \textbf{d}). 
Using some of the best numbers for SPDC sources \cite{Ramelow:2013, Eckstein:2011}, for which values of $\eta_\text{her} \approx 80\,\%$ have been reported, an equal mixture between signal and noise would still be possible and one could expect SNR$_\text{out}\approx 1.16$.
Despite lying below unity, our current figures are nevertheless still a substantial improvement compared to the initial measurements for our system. 
The previously quoted value\cite{Reim:2011ys} of SNR$_\text{out} = 1$ for $N_\text{in} \approx 1.6$ has now nearly been doubled, reaching SNR$_\text{out} = 1.98$. 

\noindent
In the end the real question is whether the memory is capable of preserving the quantum nature of non-classical input states. 
While it is already clear from the aforementioned noise level that one can expect a disturbance of the state during storage, we will now investigate how much it is actually impaired, and whether there is still any signature from a non-classical input visible in the memory output.

\section{Photon statistics during storage in the memory \label{ch6_photon_statistics}}

In the following, we apply the Hanbury-Brown-Twiss measurement scheme to determine the {\gtwo} and therewith the photon statistics of the signal stored in the memory. The principle of this measurement and its evaluation in terms of detected coincidence and triple coincidence counts is analogous to the descriptions in sections \ref{ch6_subsec_g2intro} and \ref{ch5_subsec_g2}. 
The difference is that we are now determining {\gtwo} for different field combinations in the memory read-in and read-out time bins, via the measurement settings. 
Before we present the results, we give a brief overview of the quantities we measure and the main experimental parameters. 
To observe modifications in the photon statistics it turns out that we have to determine {\gtwo} with high precision, which results in long measurement times and a multitude of datasets. 
Their combination to a final {\gtwo} number is not straightforward. 
To allow the reader to reconstruct our results from our measurement sequences, we provide an in depth description of the {\gtwo} calculation in appendix \ref{app_ch6_g2_count_aggregation_details}.

\subsection{Photon statistics measurement\label{ch6_subsec_meas_method}}

\paragraph{Measurement procedure}
To investigate the effects of signal storage in the memory, {\gtwo} is measured in the read-in and read-out time bins with active Raman storage.  
Furthermore, knowledge about the statistics of the input signal and the noise by themselves are required.
The measurement settings $i$ allow access to the statistics of all these fields: 
in the input time bin, \textit{sd} gives the statistics of the input signal. 
Ideally it should coincide with the theoretical expectations for the respective signal type, i.e., $g^{(2)} = 0$ for \hsps\, and  $g^{(2)}=1$ for \coh\,  
The noise is analysed in both time bins by measurement of setting \textit{cd}. 
Finally, the memory interaction setting \textit{scd} allows to study the stored and retrieved signal, whereby the noise influence is observable by comparison with the input signal's {\gtwo}. 
To develop an understanding for this influence, we investigate two theoretical models (see section \ref{ch6_subsec_g2models}). 
Since we desire a fair test for both models, we additionally measure an experimental configuration without Raman storage obtained by blocking the optical pumping beam. 
This way, any unaccounted modifications of the input signal statistics by the Raman process, which could potentially influence the model predictions\cite{goldschmidt2013mode}, can be excluded. 
Here, the Cs population is initially in an equal superposition of the states $6^{2}S_\frac{1}{2} \text{F}=3$ and F$=4$.
The population from F$=3$ couples strongly to the control field, which drives spontaneous Raman scattering from the F$=3$ to the F$=4$ ground state under the emission of Stokes photons into the signal frequency mode. 
When measuring setting \textit{c}, the noise process is thus different from the optically pumped configuration one is dealing with when applying setting \textit{cd}. 
This difference is discussed in chapter \ref{ch7_FWM_explanation}. 
The input signal, accessed by setting \textit{s}, should still have a {\gtwo} similar to that for \textit{sd}, since the optical pumping only reduces residual linear absorption of the signal in the Cs, which does not influence the photon statistics \cite{Loudon:2004gd}.
Due to the absence of storage and retrieval, only the input time bin is relevant for analysing the interplay between signal and noise, measured via setting \textit{sc}.

\paragraph{Measurement parameters}

Likewise to the source characterisation measurements in section \ref{ch5_subsec_g2}, we will employ our FPGA to determine the coincidence and triple coincidence detection probabilities, that go into the {\gtwo} metric (see eq. \ref{eq_ch6_g2}). The FPGA again counts coincidences within a coincidence window of $\Delta t_\text{coinc.}^\text{FPGA} = 5 \, \ns$. 
As mentioned in section \ref{subsec_ch6_TAC_count_rates} above, we measure the experimental settings for \sd\,, \scd\, and \cd\, in alteration. This avoids apparatus drifts between the recordings for different settings, and, in turn, allows to determine the memory efficiency alongside the {\gtwo}. For each such sequence, termed measurement run, we record FPGA counts for $\Delta t_{\text{meas},sd} \approx 5-10\,\min$ and 
${\Delta t_{\text{meas},scd} \approx \Delta t_{\text{meas},cd} \gtrsim 30\,\min}$, whereby each datapoint therein contains FPGA counts integrated for $\Delta t^\text{FPGA}_\text{meas} = 10\,\text{min}$. 
To determine {\gtwo} with good precision, i.e. small measurement error, Poissonian count-rate statistics dictates long measurement times. 
Due to a triple count rate frequency in the sub-Hz regime for \hsp\, inputs, we take data over several days. For instance, 
setting \cd\, is measured for a total of $\sim 28$ h; see appendix \ref{app6_coinc_rates} for the total times of all measurements. 
The resulting data has to be concatenated in order to yield the detection probabilities going the {\gtwo}-function of 
eq. \ref{eq_ch6_g2}. 
Correct data aggregation is particularly important, as the SNR already tells us, that we can only expect to see an effect of the non-classicality of the \hsps\, on the {\gtwo} for the retrieved signal, rather than a completely non-classical output. 
So, when comparing \hsps\,, retrieved from the memory, with read-out \coh\, signals at similar input photon number, we must make sure that any differences in {\gtwo} do not arise from a statistical mistreatment. 
The exact procedure of how we obtain the {\gtwo}-values from the raw data is explained in detail in appendix \ref{app_ch6_g2_count_aggregation_details}. 
Here, we omit these intermediate steps and move straight on to the results of our measurements.

\subsection{{\gtwo} results \label{ch6_subsec_g2results}}

\begin{figure}[h!]
\centering
\includegraphics[width=\textwidth]{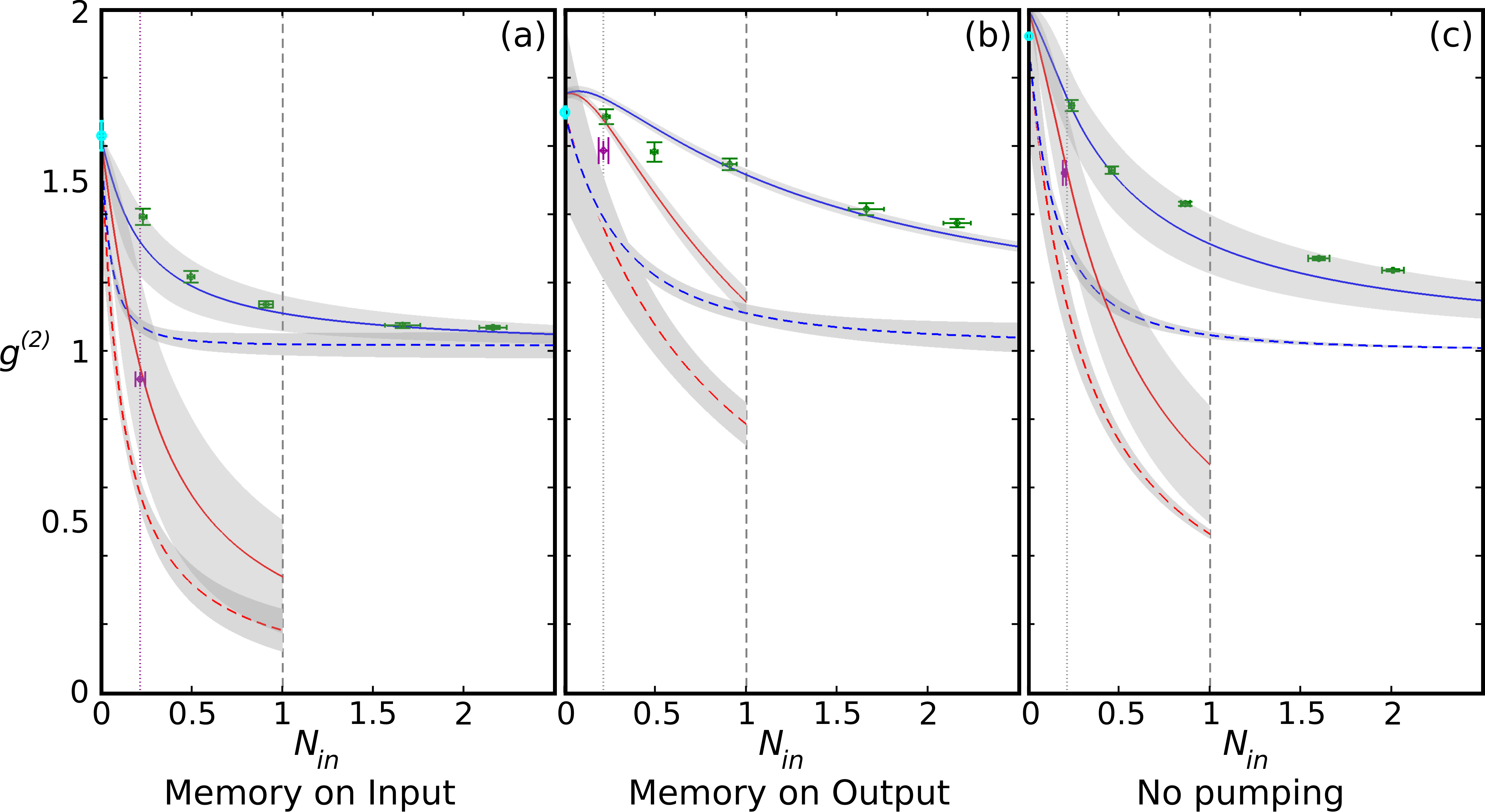}
\caption{{\gtwo} results and predictions of our two theory models, shown for the three experimental configurations. 
\textbf{(a)}: shows the read-in time bin with memory on (\scd\,, in), i.e. with optical pumping active. 
\textbf{(b)}: the same configuration for the read-out time bin (\scd\,, out). 
\textbf{(c)}: the read-in time bin for memory off (\textit{sc}, in), i.e. no optical pumping. The colour coding is as follows:
\textit{Green points} are \coh\, input signal data; 
\textit{magenta points} represent heralded single photon inputs 
($N_\text{in}=\eta_\text{her}$), \textit{cyan points} are the noise (setting \textit{cd} with $N_\text{in} = 0$). 
\textit{Solid blue} and \textit{red lines} show the theoretical predictions obtained by our coherent interaction model for \coh\, and \hsp\, inputs, respectively. \textit{Shaded regions} denote the standard deviation of the model prediction under Monte-Carlo variation of the model parameters. The \textit{dashed lines} illustrate the predictions of the incoherent model, with colours analogous to the \textit{solid lines}.
\textit{Vertical dotted} and \textit{dashed lines} indicate $N_\text{in}=\eta_\text{herald}$ and $N_\text{in}=1$ (perfect heralding efficiency), respectively. 
}
\label{fig_ch6_g2results}
\end{figure}

\paragraph{Observed {\gtwo}-values for stored single photons and coherent states}
The {\gtwo} results for the 3 experimental configurations are shown in fig. \ref{fig_ch6_g2results};  
appendix \ref{app6_data_aggregation} lists the associated set of numbers. 

When optical state preparation is active (fig. \ref{fig_ch6_g2results} \textbf{a} \& \textbf{b}), the input signal statistics, obtained by blocking the control field (\sd\,), results in $g^{(2)}_{sd,\text{coh}}= 1.01 \pm 0.01$ for \coh\, inputs. 
This number is a weighted statistical average over all investigated input photon numbers $N_\text{in}$ 
(see appendix~\ref{app_sub_ch6_g2_count_aggregation_details_data_aggregation}). 
For \hsp\, inputs at $N_\text{in} = \eta_\text{her} = 0.22$, 
$g^{(2)}_{sd,\text{SPDC}} =  0.016 \pm 0.004$ is obtained. 
Both numbers are close to their expectation values of $1$ and $0$, respectively. 
When the input signal is blocked (setting \textit{cd}), we measure 
$g^{(2)}_{{cd},\text{in}}=1.62 \pm 0.04$ and 
$g^{(2)}_{cd,\text{out}}=1.70 \pm 0.02$ for the noise in the input and output time bins, respectively; see section \ref{ch7_g2noise} for a further discussion. 

When signal and control are applied simultaneously (\textit{scd}), the photon statistics of the stored and retrieved signal are modified by the accompanying noise process.
The noise increases the $g^{(2)}$ of both significantly, as fig. \ref{fig_ch6_g2results} illustrates by the \textit{green data points} for \coh\, input and the \textit{magenta point} for \hsp\, inputs.
In the input time bin, transmitted \coh\, signals converge towards the ideal $g^{(2)}=1$ only for large input photon numbers of $N_\mathrm{in} \gtrsim 2.5$, as the amount of signal is increased compared to the fixed amount of noise.
The non-stored fraction of the \hsp\, input shows 
$g^{(2)}_{scd,\text{SPDC},\text{in}}=0.92 \pm 0.02$, just below the classicality boundary. 
Looking at the read-out time bin, we find that coherent states with an input photon number of 
$N_\text{in}=  0.23$ have 
$g^{(2)}_{scd,\text{coh},\text{out}}=1.69 \pm 0.02$. 
Comparison of this value with the noise {\gtwo} (\textit{cyan points} in fig. \ref{fig_ch6_g2results}) reveals no difference; 
\coh\, inputs are thus indistinguishable from the noise. 
Heralded single photons however show 
$g^{(2)}_{scd,\text{SPDC},\text{out}}=1.59 \pm 0.03$, which is a drop in $g^{(2)}$ by more than $3$ standard deviations compared to coherent states and to noise. 
This difference is clearly visible in fig. \ref{fig_ch6_g2results} \textbf{b}. 
The \hsp\, datapoint however lies above the classical boundary, which means that the memory's single photon read-out is currently not suitable for temporal multiplexing applications. 
Unfortunately, this result was not what we had hoped for when we started our series of experiments. 

Yet, the lower $g^{(2)}$, measured for heralded single photons, compared to the observed value for weak coherent states, reveals that there is still an influence of the non-classical SPDC input photon statistics in the memory read-out.
In fact, to obtain a similar {\gtwo} with \coh\, input signals, one would have to double the number of input photons to 
$N_\text{in} \approx 0.49 \ppp$. 
This is a key result as it shows, in principle, the capability of the Raman memory to preserve the photon statistics of the input signal, if the FWM noise floor can be reduced. We will investigate this point further by modelling the measured data and looking at model predictions for a change in noise level in section \ref{ch6_subsec_g2models} below. 

\paragraph{Significance of {\gtwo} difference}
To determine the significance by which we can observe a difference between the {\gtwo}-values for retrieved \hsps\, and \coh\, at $N_\text{in} \approx \eta_\text{her}$, we perform a one-sided, two-sample Welch test\cite{Ruppert:Book} on the $g^{(2)}_{j,\textit{scd},t}$, obtained for the individual measurement runs $j$, for both input signal types (see appendix \ref{app6_stat_tests}). 
Notably, since we are comparing two sets of population data, namely 
$g^{(2)}_{j,\textit{scd},\text{SPDC}}$ and 
$g^{(2)}_{j,\textit{scd},\text{coh}}$, a two-sample location test is required.
A Welch test is chosen since the set of $\left\{g^{(2)}_j\right\}$ for coherent states and heralded single photons have unequal sample sizes, are drawn from different populations and have different variances (see appendix  \ref{app6_g2_measurements}). 
We test the Null hypothesis ($H_0$) that the $g^{(2)}_j$-samples for \coh\, and \hsps\, have the same population mean. 
For the important {\gtwo} difference in the read-out time bin, we obtain a rejection of $H_0$ with a confidence level of 
$\ge 99.7 \, \%$ ($p$-value $=8.7 \cdot10^{-4}$), which corresponds to a significance of $\ge 3$ standard deviations.
Similar results are obtained when replacing the $g^{(2)}_j$ of the coherent state signal with those of the noise.
We also test the \hsp\, $g^{(2)}_j$-values against those obtained for a \coh\, with $N_\text{in} = 0.49\ppp$. 
For this doubling of the \coh\, input photon number, compared to \hsps\,, we do not obtain any violation of $H_0$, neither in a left- nor in a right-handed test. 
To complete the argument we furthermore test the $g^{(2)}_j$ for coherent states at $N_\text{in} = 0.23\ppp$ against those of the noise, again under the $H_0$-hypothesis that the population means are equal. 
In this case, we cannot reject $H_0$ with any reasonable level of confidence ($p$-value $=0.967$). 
Detailed test outcomes, including the input time bin results, can be found in appendix \ref{app6_stat_tests}.
In conclusion, there is a statistically significant difference between the $g^{(2)}_j$-values of retrieved \hsps\, with respect to those of retrieved \coh\, at equal photon number, as well as with respect to the $g^{(2)}_j$-values of the noise. 
On the other hand, no significant difference between the coherent state and noise $g^{(2)}_j$ values is observable.

\paragraph{Effects from optical pumping}
In designing the apparatus, the simplification was made to perform experiments for \hsp\, input signals with active optical pumping by the diode laser. In the analysis of the {\gtwo} results, we have to certify that this difference with respect to \coh\, inputs does not introduce a systematic change, which could modify the photon statistics. 
Since it is experimentally difficult to study these effects on the SPDC signal, we use the \coh\, inputs as a proxy. 
To this end we have studied the photon statistics in our measurement on \coh\, inputs with $N_\text{in}=0.23 \ppp$, where we have swapped between the optical pumping turned on and off during the storage time (see appendix  \ref{app6_coh_pumping_on}). 
While a sizeable reduction in memory efficiency can be seen with optical pumping (see section \ref{ch6_subsec:MemEffCalc}), there are no significant changes in the {\gtwo} for the memory interaction setting \textit{scd}, as we can see from the following set of numbers:
\begin{itemize}
\item continuous optical pumping: $g^{(2)}_{scd,\text{in}}=1.361\pm0.049$, \quad $g^{(2)}_{scd,\text{out}}=1.631 \pm 0.050$.
\item switched optical pumping:  $g^{(2)}_{scd,\text{in}}=1.437 \pm 0.044$, \quad $g^{(2)}_{scd,\text{out}}=1.666 \pm 0.036$.
\end{itemize}
If there was any influence on the output {\gtwo}, then the reduction in $\eta_\text{mem}$, associated with continuous optical pumping, should lead to an increase in $g^{(2)}_{scd,\text{out}}$ towards the noise value $g^{(2)}_{cd,\text{out}}$.
As this is not the case, a systematic error arising from optical pumping can be excluded as an explanation for the {\gtwo}-reduction with SPDC signal photons.

\paragraph{{\gtwo}-results without Cs state preparation}
When Raman storage is absent, the influence of the noise in the input time bin changes, as does the noise {\gtwo}. Fig. \ref{fig_ch6_g2results} \textbf{c} shows the resulting data for all settings. 
Here the noise shows   
$g^{(2)}_{c,\text{in}} = 1.92 \pm 0.01 $ for the input time bin and 
${g^{(2)}_{c,\text{out}} = 1.79 \pm 0.01}$ for the output time bin. 
The different population distribution between the initial atomic states and the associated change in coupling to the control pulses also increases the number of noise photons emitted per control pulse (eq.~\ref{ch6_eq_NoiseFloor}) to 
${N^\text{in}_\text{noise} = 0.29 \pm  0.13 \, \frac{\gamma}{\text{pulse}}}$ 
and 
${N^\text{out}_\text{noise} = 0.29 \pm 0.14 \, \frac{\gamma}{\text{pulse}}}$.
Correspondingly, the SNR-values (eq.~\ref{eq_ch6_SNR}) change, e.g. down to SNR$^{sc}_\text{in} = 0.69 \pm 0.03$ for \hsp\, input signals, which is only $\approx \frac{1}{5}$ of SNR$_\text{in}^{scd}$, obtained during Raman storage\footnote{
	Measuring the setting \textit{sc} is meaningless in the output time bin, 
	given the absence of storage and retrieval, which effectively reduces this setting to \textit{c}. 
}. 
The input photon number for \hsps\, is also slightly reduced to 
${N^{s}_\text{in}= \eta^{s}_\text{her} = 0.2 \pm 0.01\, \frac{\gamma}{\text{pulse}}}$, 
due to residual linear absorption in the \cs\,. 
As anticipated the input signal {\gtwo}-values for setting \textit{s} are basically unaffected by the absence of \cs\, state preparation, showing  
$g^{(2),\text{in}}_{s,\text{coh}} = 1.003 \pm 0.004$ for \coh\, and 
$g^{(2),\text{in}}_{s,\text{SPDC}} = 0.026 \pm 0.006$ for \hsps\,, which agree with the expectations. 
For the combined signal and control input (\textit{sc}), the absence of Raman coupling clearly increases the {\gtwo} to a level which is more similar to the read-out time bin for active Raman interaction than to the corresponding values in the read-in time bin. 
Given the significant deterioration in SNR$^{sc}_\text{in}$, one would already expect such a result.

\subsection{Model for the {\gtwo} results \label{ch6_subsec_g2models}}
The real question now is, whether the observed {\gtwo} behaviour is indeed completely determined by the  unfavourable SNR, or, if there is more to it, such that effects arising from the actual light-matter interaction dynamics play a role in what we see experimentally. To investigate this matter we compare two models.

\paragraph{The incoherent model}
\noindent
This model\cite{goldschmidt2013mode} treats signal and noise as two separate, independent fields with different $g^{(2)}$ values, 
where the combined photon statistics $\left( g^{(2)}_\text{tot} \right)$ is assumed to consist of a mixture  between FWM noise $\left( g^{(2)}_\text{noise} \right)$ and the input signal $\left( g^{(2)}_\text{sig} \right)$.
Both fields are imagined as being combined incoherently into one mode, e.g. by using a beam splitter prior to detection. 
The model is thus based on the incoherent addition of signal and noise. 
The expected value for $g^{(2)}_\text{tot}$ is derived following the argumentation of \textit{Goldschmidt et. al.}\cite{goldschmidt2013mode} to
\begin{equation}
g^{(2)}_\text{tot} = \frac{ \left(N_\text{sig} \right)^2 \cdot g^{(2)}_\text{sig} + 2 N_\text{sig} \cdot N_\text{noise} + \left( N_\text{noise} \right)^2 \cdot g^{(2)}_\text{noise} }{\left(N_\text{sig}+ N_\text{noise} \right)^2} = \frac{\text{SNR}^2 \cdot g^{(2)}_\text{sig} + 2 \cdot \text{SNR} + g^{(2)}_\text{noise}}{\left( 1+\text{SNR} \right)^2},
\label{eq_ch6_goldschmidt}
\end{equation}
which depends on the number of signal photons $\left(N_\text{sig}\right)$ and noise photons $\left(N_\text{noise}\right)$ per pulse contributing to the mixture.  
In the memory-on cases, $N_\text{sig}$ is either the non-stored, transmitted fraction of the signal, 
${N_{\text{sig}}= (1-\eta_{\text{in}} ) \cdot N_\text{in}^{\text{\textit{sd}}}}$, 
or the retrieved fraction of the signal, 
${N_{\text{sig}}= \eta_{\text{mem}} \cdot N_\text{in}^{\text{\textit{sd}}}}$,
with $N_\text{in}^{\text{\textit{sd}}}$ representing the input photon number for the respective signal type; 
$\eta_\text{in}$ is the memory read-in efficiency and $\eta_{\text{mem}}$ is the total memory efficiency (storage and retrieval). 
Eq. \ref{eq_ch6_goldschmidt} mixes the separately measurable input photon statistics with a ratio solely determined by the SNR.
From a faithful prediction of the data in fig. \ref{fig_ch6_g2results} by eq. \ref{eq_ch6_goldschmidt} two conclusions would follow:
Firstly, the SNR is sufficient as a benchmark for the memory performance. 
Secondly, signal and noise can be regarded as unrelated entities, which cannot be separated experimentally only because they occupy the same spatio-temporal mode. 
The predictions from eq. \ref{eq_ch6_goldschmidt} are shown in fig. \ref{fig_ch6_g2results} by the \textit{dashed lines}. 
Clearly the incoherent model significantly underestimates the experimentally measured $g^{(2)}$ data in all three analysed configurations. 

\noindent
Notably, one could argue that the conservation of the photon statistics for the input signal during storage and retrieval in the memory is not necessarily to be taken for granted. 
In this case, it would not be justified to use $g^{(2)}_\text{sig}$ of the input signal in eq. \ref{eq_ch6_goldschmidt} for calculating $g^{(2)}_\text{tot} $ for the configuration with Raman storage. 
Instead, a modified $g^{(2)}_\text{sig}$, which is not directly accessible experimentally, would need to be used to take into account any such modifications. 
This argument however does not apply to the configuration with the optical pumping switched off. 
Since no Raman storage happens, no modification of $g^{(2)}_\text{sig}$ can be expected to occur. Moreover, $N_\text{sig}$ corresponds to the directly measured input photon number $N^{s}_{\text{in}}$, without any additional memory efficiency factors. 
Consequently, the configuration without state preparation represents the ideal testing ground for the incoherent model, where all variables are known from direct measurements. 
So, if the model were applicable, it should at least lead to an agreement here. 
Yet, it also fails totally to describe the data for the absence of optical pumping (see fig. \ref{fig_ch6_g2results} \textbf{c}). \newline
Therefore we can conclude that signal and noise cannot be considered as individual entities in a mixture. 
The SNR also cannot be the only relevant metric for the performance of the memory with respect to noise.

\paragraph{The coherent model} 
Our second model takes into account the full coherent, off-resonant interaction between the incident light fields and the spin-wave excitation in the Cs ground states. 
In this coherent model, signal and noise are generated by the same Hamiltonian, from which the photon statistics in the Stokes output mode are predicted. 
In contrast to the incoherent addition, this model does not contain any free parameters and completely predicts the photon statistics based on the light matter interaction, following from the Maxwell-Bloch equations. 
It is thus based on eqs. \ref{eq_ch2_Maxwell_Bloch_FWM} \& \ref{eq_ch2_Maxwell_Bloch_FWM_2}, introduced in chapter \ref{ch2}. 
Its only inputs are the experimental parameters for the Raman memory\footnote{
	These are the detuning, the control pulse energy, the pulse duration and beam waist, as well as
	the \cs\, cell length and temperature, defining the optical depth and therewith the Raman 
	coupling constants. 
}, stated in section \ref{ch2_Raman_level_scheme}.
The model has been developed by Joshua Nunn. For this reason, only its results are discussed here, and a short description is given in appendix \ref{app6_coh_model}. 
The model's main conclusion is that, while Raman storage and FWM are two different processes, they are intrinsically coupled by sharing the Stokes channel (see eq. \ref{eq_ch2_Maxwell_Bloch_FWM}). 
Signal and noise are thus not a mixture between two different fields. 
Rather there is one memory output state, which, upon read-out from the \cs\,, is pure\footnote{
	Ignoring potential entanglement with residual fractions of the spin-wave.
} and contains contributions at both, Stokes and anti-Stokes frequencies, from both processes. 
We will see this fact also in section~\ref{ch7_subsec_ASseeding}, when studying the amount of anti-Stokes noise 
for varying intensity of memory input signal in the Stokes channel. 
Upon signal detection however, the state is spectrally filtered at the Stokes frequency. 
Accordingly, this traces over the anti-Stokes part and projects the FWM Stokes mode into a thermal state (see also section~\ref{ch7_g2noise}). 
The additional coupling to the stored fraction of the \hsp\, input state still results in a {\gtwo} reduction for the combined Stokes mode, when compared to having no memory input signal, i.e. just FWM noise.
The \textit{solid lines} in fig.~\ref{fig_ch6_g2results} show the resulting model predictions (eq.~\ref{eq_app6_g2_1}), with error bars from Monte-Carlo variation of the model parameters (see appendix~\ref{app6_coh_model_err}). 
Despite a small overestimation of the {\gtwo}-values for low {\Nin}, the predictions show excellent agreement with the experimental data. 
This is particularly noteworthy as this model is an ab-initio calculation, so there is no optimisation with respect to the data.

\begin{figure}[h!]
\centering
\includegraphics[width=\textwidth]{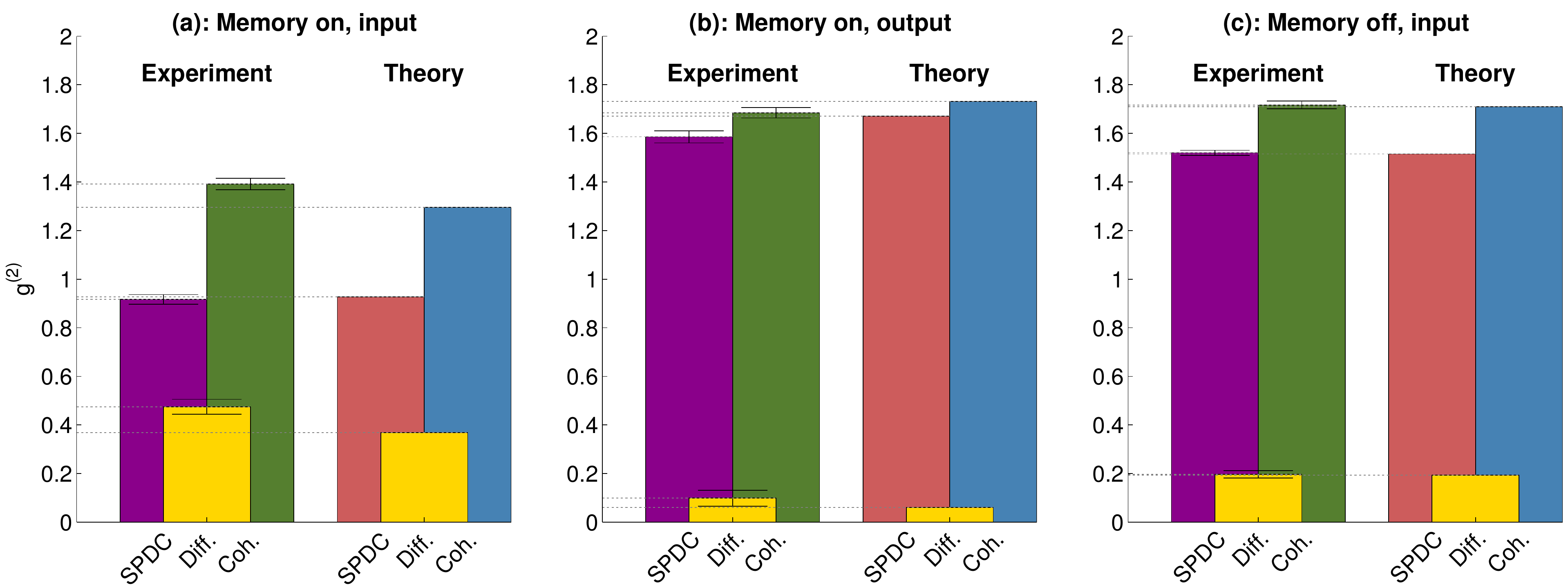}
\caption{Comparison between the coherent model predictions and the measured {\gtwo} data for \hsp\, and \coh\, inputs at $N_\text{in}=0.22\ppp$ and $N_\text{in}=0.23\ppp$, respectively.
The subplots represent the three configurations.
\textbf{(a)}: the input time bin with the memory on; 
\textbf{(b)}: the output time bin with memory on; 
\textbf{(c)}: the input time bin with memory off. 
Colour coding is analogous to fig. \ref{fig_ch6_g2results}, whereby the experimental data 
(\textit{magenta} for \hsps\,, \textit{green} for \coh\,) is shown on the left and the theoretical 
prediction (\textit{red} for \hsps\,, \textit{blue} for \coh\,) on the right. 
The difference between both signal types is displayed by the \textit{yellow bars} for each respective subgroup of bars.
In all three cases, the experimentally observed drop in $g^{(2)}$ between coherent states and heralded single photons is reasonably well replicated by the model.
}
\label{fig_ch6_g2bars}
\end{figure}

Moreover, it is interesting to compare the observed drop in {\gtwo} between \hsp\,  and \coh\, inputs at 
$N_\text{in} = 0.23\ppp$ (see section \ref{ch6_subsec_g2results}) with the {\gtwo} predicted by the model. 
To this end, we can examine the bar graphs in fig. \ref{fig_ch6_g2bars}, which explicitly compare the observed with the predicted {\gtwo}-values alongside the resulting {\gtwo}-differences between \hsps\, and \coh\, 
The measured {\gtwo}-reduction, which we have identified as a signature from the non-classical input statistics, is closely reproduced by the model. 
Consequently, we can conclude that the coherent model offers an accurate and, most likely, realistic description of the Raman memory behaviour under the influence of FWM. 
We can furthermore use the predictions from the model to evaluate possible next steps towards solving the challenge that arises from the Raman memory noise.

\section{Conclusions\label{ch6_conclusion}}

The comparison between the theoretical models and the experimental data reveal that the memory performance cannot be benchmarked solely by the SNR. 
This fact actually applies not only to our Raman memory, but to all systems that suffer from a noise background that arises from a coherent scattering process. Other vapour memory protocols, such as EIT\cite{Gorshkov:2007,Gorshkov:2007aa,Gorshkov:2007rw,Gorshkov:2008rz} and GEM\cite{Hetet:2008dp,Hetet:2008kx,Hetet:2008jt, Hetet:2008kl}, are the obvious candidates for which a thorough investigation of the noise background is also necessary\cite{Phillips:2011,Philips:2009,Hosseini:2012}.
Why is this? Because, as fig. \ref{fig_ch6_g2results} illustrates, the increase in {\gtwo} from the coherent coupling of the noise to the Stokes channel is significantly larger than expected for an incoherent combination. 
Consequently, the real performance is worse than one might expect just from the SNR, and additional information - such as a photon statistics measurement - is mandatory. 
Alternatively, the noise coupling mechanism to the signal requires identification. \newline
The excellent agreement between the coherent model predictions, which assume solely FWM as the noise process in the Raman memory, and the experimental data suggest that FWM is, on the one hand, the only important noise source. 
On the other hand, it is also the main challenge to overcome if the room-temperature vapour Raman memory is ever to be used as a temporal multiplexer. 
This follows directly from the predictions for \hsp\, input signals, shown in fig. \ref{fig_ch6_g2results} (\textit{red line}). 
Even for perfect heralding efficiency of a single photon source, i.e. $\eta_\text{her} = N_\text{in} = 1$, one would not obtain  non-classical photon statistics for the signal retrieved from the memory. 
In other words, improving the SNR from the input signal side will not be successful.
Improving the heralding efficiency of photon sources for the Raman memory is therefore a secondary problem. 
The noise remedy can only come from the memory system, where FWM needs to be tackled. 

\begin{figure}[h!]
\begin{center}
\includegraphics[width=0.6\textwidth]{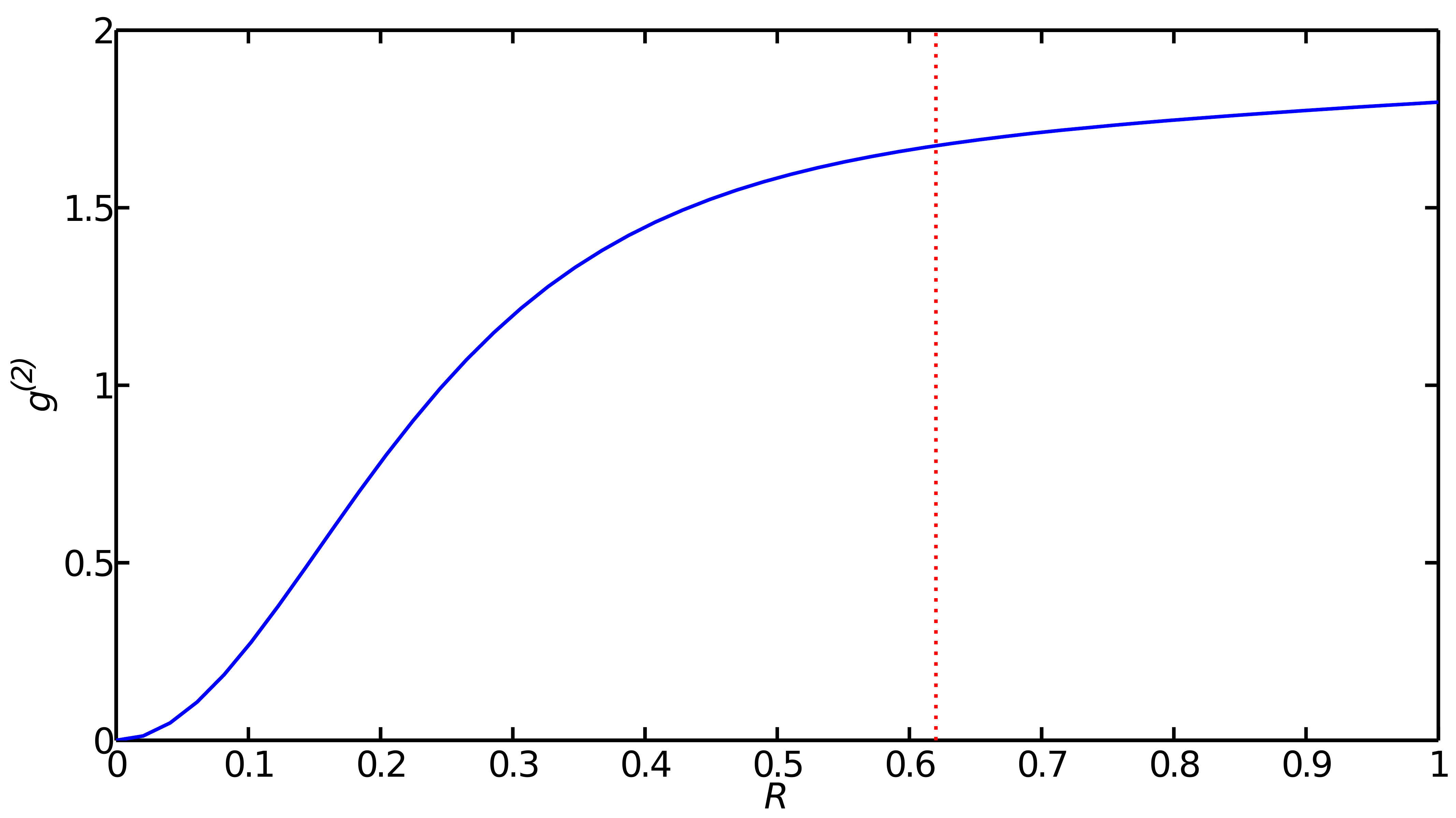}
\caption{Theoretical prediction for the $g^{(2)}$ autocorrelation of the retrieved field after storing a single photon, as the ratio $R = C_\text{AS}/C_\text{S}$ between anti-Stokes and Stokes Raman coupling (eq.~\ref{eq_ch2_Rratio}) is varied. The \textit{vertical dotted line} shows the value $R = 0.625$ that describes our experiments.}
\label{fig_ch6_Rscan}
\end{center}
\end{figure}

Solutions to this challenge are actually possible in case of the Raman memory, because FWM is not intrinsic to the Raman interaction.
Inspection of the couplings between FWM and the Raman process in eqs. \ref{eq_ch2_Maxwell_Bloch_FWM} \& \ref{eq_ch2_Maxwell_Bloch_FWM_2} illustrate this fact.
If the term relating to the emission of an anti-Stokes photon is somehow suppressed, i.e., if the coupling constant $C_\text{AS}$ of the FWM term is reduced such that  $C_\text{AS} \rightarrow 0$, then the system becomes the noise-free ideal Raman memory\cite{Nunn:2007wj} (see eqs. \ref{eq_ch2_maxwell_bloch_3}). 
The coherent model allows to predict how any change in the relative size $R=\frac{C_\text{AS}}{C_\text{S}}$ (eq.~\ref{eq_ch2_Rratio}) between anti-Stokes coupling strength $C_\text{AS}$ and Stokes coupling strength $C_\text{S}$ would influence the resulting photon statistics. Fig.~\ref{fig_ch6_Rscan} shows this for the {\gtwo} of a single photon retrieved from the memory. 
Assuming that FWM is the only important noise source for the Raman memory, we can conclude that the Raman protocol carries the intrinsic possibility to operate in the quantum regime. 

So what can be done about the FWM noise? 
In principle\cite{Walther:2007}, FWM can be suppressed by Zeeman-polarisation of the initial atomic state with an appropriate optical pumping scheme, and the use of circularly polarised signal and control pulses. 
The suppression is based on the fact that population of an extreme Zeeman state, i.e., a state with 
$m_\text{F} = \pm \text{F}$, 
can only couple to one type of circularly polarised light field. 
This is a light field, polarised such that its interaction with the atoms decreases the absolute value of the respective magnetic quantum number of the atoms by its angular momentum.
Excitations from the orthogonally polarised light field are forbidden. Hence, polarising the \cs\, ensemble, such that the control coupling to the initial $\text{F}=4$ state is forbidden, terminates spontaneous anti-Stokes scattering and therewith the FWM noise\cite{Walther:2007}.
The problem with this scheme is the following: for far off-resonance protocols, orthogonally polarised signal and control beams lead to destructive interference between Raman transitions that either involve the 
$6^{2} \text{P}_{\frac{3}{2}} \text{F}'=3$ or $\text{F}'=4$ excited states. 
This has been shown generally for all alkali atom systems\cite{Vurgaftman:2013aa}.
Actually, while it is a nuisance in the quest to erase FWM, we will use it in section \ref{ch7_Fluorescence} to investigate the composition of the noise floor in detail.
Recently, it has been shown\cite{Zhang:2014} that a reduction in FWM can be achieved by only using circular polarisation for the read-out control beam, when using a Zeeman polarised ensemble\cite{Walther:2007}. 
While this terminates the FWM created in the read-out time bin, there is still the contribution left from the excited spin-wave coherence in the read-in time bin. Hence it only reduces the issues arising from FWM noise. 
As FWM is constrained by phase-matching similar to SPDC \cite{Kumar1994}, other alternatives could be to introduce dispersion between the Stokes and anti-Stokes frequencies, or to use a storage medium with larger Stokes shifts \cite{England:2013aa}.
For our system, one approach is to reduce the density of states at the anti-Stokes frequency by placing the memory inside a low-finesse cavity or photonic-bandgap structure \cite{Sprague:2014aa}. As shown in fig. \ref{fig_ch6_Rscan}, FWM suppression by a factor of $\sim 2.5$, achievable with the aforementioned techniques, would preserve the nonclassical signature of retrieved \hsps\,. 

Beside noise reduction, actually viable application of the Raman memory for temporal multiplexing tasks would also require memory efficiencies higher than our values quoted in section \ref{ch6_subsec:MemEffCalc}. 
In this regard, we note that higher memory efficiencies of up to $\eta_\text{mem} \gtrsim 60\,\%$ have meanwhile been observed. 
These higher memory efficiencies were realised in a \cs\, vapour cell with $10\,\text{Torr}$ of nitrogen (N$_2$) buffer gas at $T\approx 77^\circ\text{C}$.  
The work has been done in our group by {Sarah Thomas}, after the experiments in this thesis were completed. 
Additionally, operating the memory system with backward retrieval should also increase the memory efficiency, as it minimises re-absorption\footnote{
	Notably, memory read-in results in an exponentially decreasing spatial distribution of the spin-wave amplitude 
	along the input signal's propagation path in the \cs\, cell. This means, the spin-wave has its highest amplitude 
	at the input and its lowest amplitude at the exit face of the \cs\, cell. Accordingly, during read-out, most of the
	retrieved signal's intensity is located at the former location. 
} of retrieved signal by shortening its propagation through the \cs\, cell\cite{Nunn:DPhil}. 
While the possibility to obtain higher memory efficiencies with other buffer gases, buffer gas pressures or backward retrieval were not explored for the presented work, these possibilities for memory efficiency improvement, together with control over the FWM noise, give hope that it is possible to achieve a highly efficient, low noise atomic vapour Raman memory system for scalable photonics in the near future.

\part{Limitations by noise}

\chapter{Memory noise characterisation\label{ch7}}

\begin{flushright}
{\tiny \textfrak{Mephistopheles}: \,  \textfrak{Ich bin der Geist, der stets verneint! Und das mit Recht; denn alles, was entsteht, Ist wert, da§ es zugrunde geht; Drum besser wŠr's, da§ nichts entstŸnde. So ist denn alles, was ihr SŸnde, Zerstšrung, kurz das Bšse nennt, Mein eigentliches Element.} }
\end{flushright}

Having encountered the influences of noise on the memory performance, we now use this final chapter to examine the noise floor in detail. 
Referring to the noise processes, introduced in section \ref{ch2_Raman_noise}, we determine their relative contributions to the overall noise floor. This will demonstrate the previously claimed dominance of four-wave-mixing (FWM). 
To this end, we examine the noise level in different experimental configurations. 
Firstly, we investigate the noise floor with and without two-photon transitions in the atomic $\Lambda$-system. 
Secondly, we study the noise's scaling for modifications in the atomic density and the populations of the hyperfine levels that  participate in the Raman and FWM interactions. 
These measurements will allow us to experimentally distinguish between noise produced by FWM, spontaneous Raman scattering (SRS) and fluorescence.  
As a final piece of evidence for the FWM origin of our noise floor, we study its spin-wave component via magnetic dephasing measurements, and use noise seeding to demonstrate the coupling between the FWM Stokes (S) and anti-Stokes (AS) channels\footnote{
	These experiments proof the validity of the previous assumption that the Stokes channel 
	noise floor is independent of the input signal intensity for low input photon numbers {\Nin}. 
	As a reminder, we have used this approximation in the definition of the memory efficiencies in 
	eqs. \ref{eq6_err_memeff} - \ref{eq6_err_memeff_ret}.
}.  
We also test the coherent model, introduced in appendix \ref{app_ch6}, further and compare its predictions for the absolute noise levels with the experimental measurements.  
Since our coherent model only assumes FWM noise, we finally also take another look at our {\gtwo} results to account for fluorescence, the second noise source in our system. 
Additionally, in the appendix \ref{app_ch7}, we also present a detailed study of the noise level and Raman memory efficiency dependence on the experimental parameters.
This analysis shows that the parameter set we have chosen for our experiments (see section \ref{ch2_Raman_level_scheme}) approximates well the optimum signal-to-noise performance for our Raman memory setup. 
Before we go into our measurements, we briefly look at known noise sources in various common memory protocols.

\section{Introduction\label{ch7_intro}}

Noise issues relate mostly to on-demand memory operation, facilitated by a strong control field, whose large intensity gives rise to parasitic processes. 
As we have seen, on-demand operation is however a prerequisite for quantum memory application in temporal multiplexing tasks\cite{Nunn:2013ab}, which promises improved scaling of the operational success rates for a network of quantum gates\cite{Duan:2001vn, Sanguard2011}. 
An alternative to this type of system are protocols, which do not incorporate a control field. 
These are spin-echo-based memories with pre-determined storage times\cite{Afzelius:2009qf}. 
Because such devices are not affected by control field initiated noise processes, they offer an 
advantage that has allowed the demonstration of a multitude of quantum effects\cite{Saglamyurek:2011fk, Clausen:2011kx, Rielander:2014aa, Bussieres:2014aa}. 
In light of the challenges imposed by noise processes in on-demand memories, a frequency-domain multiplexing strategy has recently been developed for spin-echo systems\cite{Sinclair:2014}. 
Since this frequency multiplexing does not necessitate on-demand storage, it does not rely on memory protocols incorporating control fields, making it an attractive possibility for quantum repeater tasks\cite{Simon:2007ct}. 
However, other applications of quantum memories, such as the production of high-qubit-number quantum states\cite{Nunn:2013ab, Wieczorek:2009kx}, still require either spatial multiplexing\footnote{
	Spatial multiplexing is a technological path,  
	whose demonstrations so far have not incorporated memories\cite{Migdall:2002, Meany:2014}.
}, or temporal multiplexing\cite{Nunn:2013ab}. 
Amongst the suitable memory systems for temporal multiplexing, the solid-state implementation of the Raman memory in diamond\cite{Lee:2012,England:2013aa} has shown better noise properties than the alkali-vapour-based technology presented here. 
Yet, it comes at the expense of prohibitively short storage times on the order of pico-seconds. 
Similarly, Raman memory in a {\cs} filled hollow-core photonic crystal fibre also shows a lower noise floor\cite{Sprague:2014aa}. 
At present, this technology however suffers from the limited availability of sufficient optical depth\footnote{
	The optical depth in these systems is reduced by the formation of {\cs} molecules on the 
	fibre walls. These have to be blasted off via a light-induced atomic desorption technique\cite{Sprague:2013}
	(LIAD). Each LIAD trial only increases the {\cs} density in the fibre for a few minutes
	before molecule formation starts again and causes a tail-off in the optical depth. 
} to run experiments over time scales longer than a few minutes\cite{Sprague:2013}.
Other incarnations in hot vapours are limited by similar noise sources as the Raman memory:  
In EIT\cite{Phillips:2008uq,Philips:2009,Phillips:2011,Novikova:2012} and GEM memories\cite{Hosseini:2012} FWM noise is present as well. Here, the narrow bandwidths of these protocols\footnote{
	While the limited bandwidth is advantageous for FWM noise reduction, 
	it makes large time-bandwidth products more challenging to achieve
	and complicates interfacing with single photon sources (see chapter \ref{ch5}).
} allow to reduce the influence of FWM noise, when lowering the detuning to the MHz range.
However, in warm vapour EIT memories, the corresponding long signal pulse durations, which are on the order of the excited state lifetime, introduce an additional performance limitation from collisional induced fluorescence\cite{Manz:2007}. 

In terms of possible noise mitigation strategy, one promising and popular approach is to move the experiments to low temperatures. 
Unfortunately, the cryogenic regime, used for rare-earth ion-doped systems, operated with the AFC protocol, still seems insufficient for completely noise free spin-wave storage\cite{Timoney:2013}. However very promising improvements have been made recently by additional control filtering\cite{Guendogan:2015}. 
The best performances were so far obtained with laser-cooled atoms, either in the form of single atoms in a cavity\cite{Specht:2011, Ritter:2012fk}, BECs\cite{Lettner:2011}, or cold atomic clouds in MOTs\cite{Sparkes:2013,Bimbard:2014aa, Choi:2008aa, Zhang:2011aa}. With the latter type, DLCZ-based quantum repeater protocols\cite{Duan:2001vn} have already been demonstrated\cite{Chen:2008fk}. 
Besides their narrow bandwidths, the main challenges imposed by such systems remain their technical complexity and the resources required to operate them. 
Despite significant progress in their miniaturisation\cite{Hansel:2001fk, Brugger:2000} and commercialisation\footnote{
	For instance \textit{Toptica} now sells MOT systems for alkali atoms.
}, cold atoms experiments are still limited in their scalability and integratability into photonic circuits. 
Thanks to their technical simplicity, room-temperature systems, particularly atomic vapours, are still interesting memory candidates. 
To understand their noise properties is thus an important aspect of their characterisation. 

As we can already see with this short discussion, the noise attributes differ depending on the specific operational parameters for the employed protocol. Consequently, noise analysis and mitigation strategies need bespoke tailoring. 
In the following, we conduct this characterisation for our system, with the memory parameter optimisation presented in appendix \ref{appendix_ch7_noise_parameter_dependence}.

\section{Memory noise floor consistency\label{ch7_Fluorescence}} 
Observing the memory noise floor with the spin-polarised ensemble, as we have done in chapter~\ref{ch6}, only yields the combined signal of all noise processes. 
In the first step of the memory noise analysis, we now investigate the contributions of the different noise processes to this overall memory noise floor. 
To this end, we separate the noise floor into its two-photon transition component, which can consist of SRS and FWM, and a fluorescence part. Both parts were introduced in section \ref{ch2_Raman_noise}. 
Apart from noise emitted by the storage medium, the memory noise floor can, of course, also still have a residual contribution from control field leakage\footnote{ 
	For the moment, we attribute any leakage to the fluorescence noise. 
}.
Leakage occurs, if the polarisation and frequency filtering of the control behind the memory is insufficient. 
In sections \ref{ch7_subsec_cold_ensemble} and appendix \ref{ch7_Temp}, we will find 
leakage to contribute only a negligible amount to the overall noise floor, so it can essentially be ignored.
We also note here, that all experiments presented in this chapter were conducted using the experimental setup of 
chapter \ref{ch6}, shown in fig. \ref{fig_6_setup}. The system is operated with coherent state (\coh\,) input signals, 
as described in section \ref{ch6_sec:setup}.

\begin{figure}
\includegraphics[width=\textwidth]{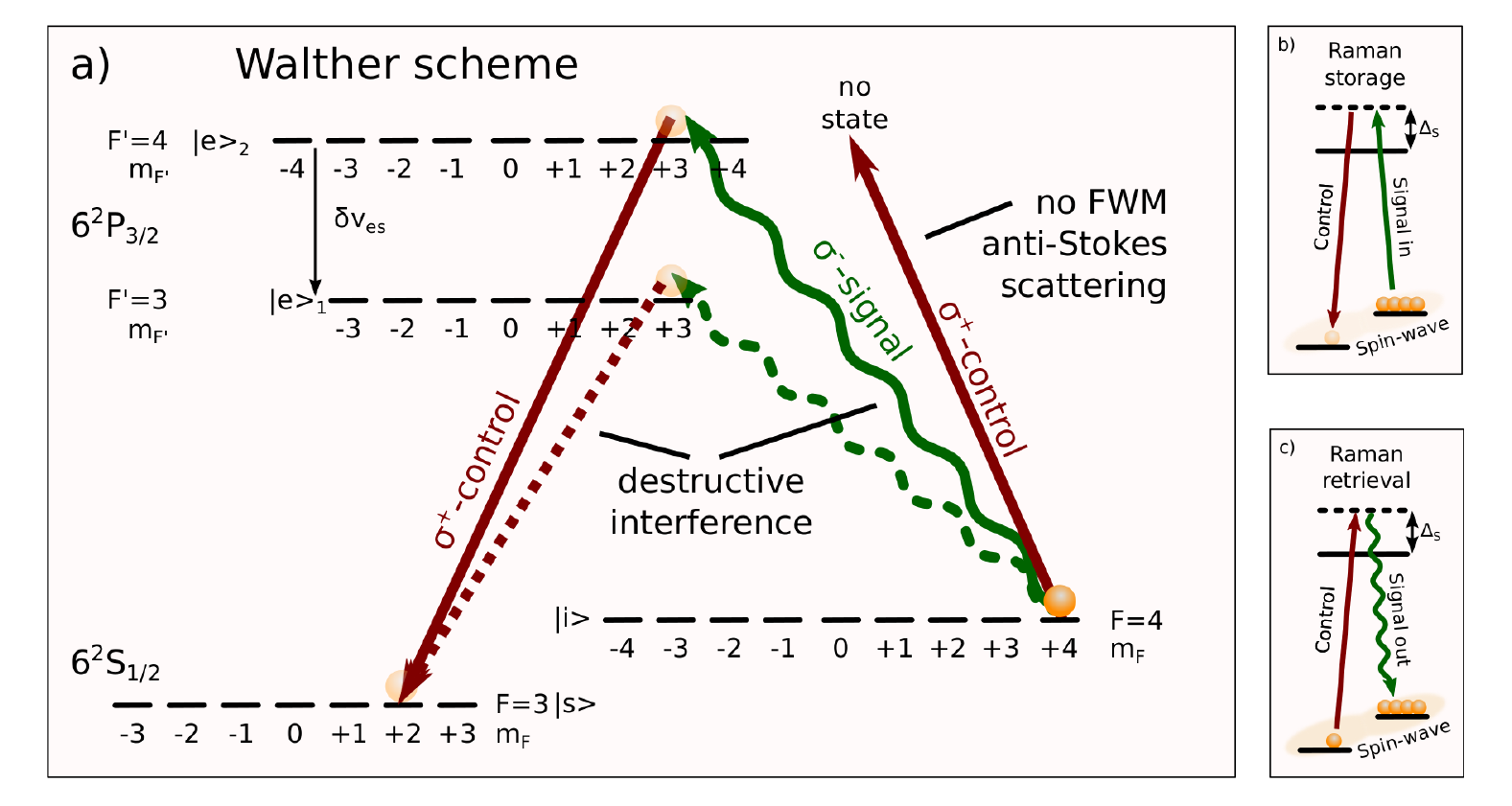}
\caption{\textbf{(a)} Walther scheme. 
\textbf{(b)} \& \textbf{(c)}: Our Raman memory read-in and read-out scheme for comparison. }
\label{fig_ch7_walther_scheme}
\end{figure}

\subsection{Separating fluorescence and two-photon transition-based noise\label{ch7_pol_turn_off_principles}} 

To distinguish fluorescence from the two-photon transition noise processes, we use the trick of changing the polarisation for signal and control. 
We have already mentioned in section \ref{ch6_conclusion}, that the memory operation with orthogonal, circularly polarised signal and control pulses, acting on a Zeeman state polarised atomic ensemble, could in principle be used to turn off the FWM contribution.
This proposal is known as the \textit{Walther}-scheme\cite{Walther:2007}.

\paragraph{Walther scheme idea}
Fig. \ref{fig_ch7_walther_scheme} \textbf{a} illustrates the scheme on the basis of the $\Lambda$-level system in {\cs}, whereby the detuning has been ignored for better readability. 
To allow for comparison with our Raman memory protocol, fig. \ref{fig_ch7_walther_scheme} \textbf{b} \& \textbf{c} show the Raman scheme as we use it. 
The proposal relies on a Zeeman polarised atomic ensemble, with its population prepared in one of the extreme Zeeman levels (here $m_\text{F} = +4$). Instead of linear polarised signal and control fields, which correspond to a superposition of $\sigma^+$- and $\sigma^-$-polarised fields in the atomic quantisation system, the \textit{Walther scheme} only applies circularly polarised signal and control pulses, which are however still orthogonal (here $\sigma^+$-pol. for the control and $\sigma^-$-pol. for the signal). 
The scheme's promised advantage lies in the absence of an appropriate Zeeman-level in the excited state. Such a level would allow SRS by the control, coupling to the population in the initial state $\ket{\text{i}}$. 
Since all other transitions, involving the available Zeeman-levels, are dipole-forbidden, no initial noise scattering can take place. For the FWM process, introduced in section \ref{ch7_subsec_FWM}, this forbidden transition would correspond to AS scattering, the first FWM step. 
While thereby preventing the FWM noise, the two-photon transitions for signal storage and retrieval, are not affected. Signal and control can still couple to the two hyperfine ground states, $6^2 \text{S}_{\frac{1}{2}}$ F$=3$ and F$=4$ via either of the excited states $6^2 \text{P}_{\frac{3}{2}}$ F'$=3$ or F'$=4$. 

\paragraph{Walther scheme in practice}
While it can be of use for near-resonant, narrowband protocols, such as GEM \cite{Hosseini:2011aa} or EIT \cite{Eisaman:2005,Phillips:2011,Kozuma:2009,Lauk:2013,Geng:2014}, the scheme does not work for alkali atoms far off resonance. 
As \textit{Vurgaftman et. al.}\cite{Vurgaftman:2013aa} have shown, in the far off resonance limit, i.e. with a detuning 
$\Delta$ much larger than the energy splitting 
$\delta \nu_\text{es}$ of the excited states (here $\delta \nu_{\text{es}} \approx 201.5\MHz\cite{Steck:2008qf}$), both possible transition paths (\textit{solid} and \textit{dashed lines} in fig. \ref{fig_ch7_walther_scheme} \textbf{a}) interfere destructively. 
For our {\cs} example, the 
$6^2 \text{P}_{\frac{3}{2}}$ F'$=3$ and F'$=4$ states ($\ket{\text{e}}_1$ \& $\ket{\text{e}}_2$) are the only excited states that couple to the initial state
 $6^2 \text{S}_{\frac{1}{2}} \text{F} = 4$ ($\ket{\text{i}}$) 
and the storage state $6^2 \text{S}_{\frac{1}{2}} \text{F} = 3$ ($\ket{\text{s}}$), 
due to selection rules. 
Hence the transition matrix elements 
$\text{M}_{\langle \text{s} | \text{i} \rangle | j}  \sim \frac{X_{\text{s,i} | j}}{\Delta_{j}}$, 
for transitions involving the excited state $\ket{\text{e}_j}$ (with $j \in \left\{1,2 \right\}$) 
only depend on the expectation values of the dipole operators\footnote{
	This is a short-hand notation, which implicitly incorporates the summation over 
	all possible transitions between Zeeman states, scaled by the Clebsh-Gordan coefficients.
} 
${ X_{\text{s,i} | j} = \bra{\text{s}} \vec{d}_{\text{s},\text{e}_j} \cdot \vec{E} \ket{\text{e}_j} \cdot \bra{\text{e}_j} \vec{d}_{\text{e}_j,\text{i}} \cdot \vec{E} \ket{\text{i}} }$. 
In alkali atoms, the dipole operator terms have opposing signs for both transitions, i.e. 
{${X_{\text{s,i}| \text{e}_1} = -X_{\text{s,i}| \text{e}_2}}$}. 
So the resulting destructive interference between the two transition paths terminates all Raman processes. 

\paragraph{Two-photon transition turn-off} 
While the interference makes the \textit{Walther scheme}\cite{Walther:2007} useless for FWM noise reduction in our Raman memory, we can use the polarisation configuration of the optical fields to turn off the two-photon transition processes in the $\Lambda$-system. 
If we use orthogonal, circularly polarised signal and control fields (circ.$\perp$circ.), instead of the usual orthogonal linear polarisations  (lin.$\perp$lin.), we can switch-off all two photon transitions, i.e. the Raman memory interaction, SRS and FWM. 
What we are left with is only the fluorescence noise in the memory time bins. 
Consequently, to determine the amount of fluorescence noise, we operate the system once with orthogonal, linearly polarised signal and control pulses (lin.$\perp$lin.), followed by another run with orthogonal, circularly polarised pulses (circ.$\perp$circ.).
Note, in the latter case, the polarisation of the optical pump, which is counter propagating in the control mode, will also be circularly polarised. 
Due to the absence of repumping\cite{Schmidt:1994}, the atomic population will still be distributed over all Zeeman levels in the initial state $\ket{\text{i}}$, which is different from the configuration studied by \textit{Vurgaftman et. al}\cite{Vurgaftman:2013aa}. 
However, the destructive interference between the transition paths still occurs, as it is independent of the magnetic quantum number\cite{Vurgaftman:2013aa,Nunn:DPhil}. 
 
\begin{figure}
\includegraphics[width=\textwidth]{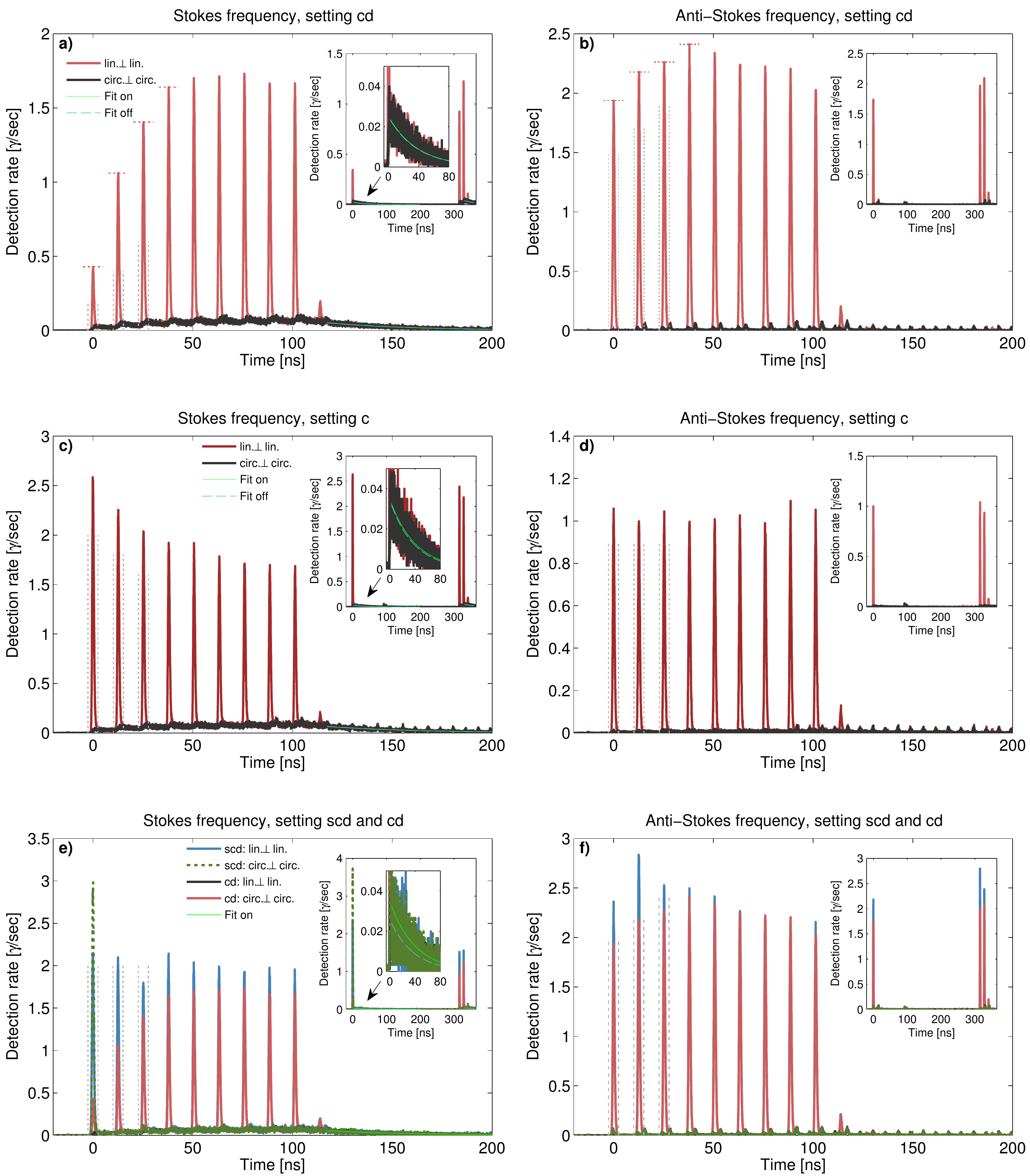}
\caption{TAC count rate histograms for lin.$\perp$lin. and circ.$\perp$circ. signal and control polarisations (see legend in \textbf{(a)}, \textbf{(c)}, \textbf{(e)}), corresponding to active and turned-off two-photon transitions in {{\cs}}. 
Left and right columns contain measurements on the S and AS channel, respectively. 
Main panels show the control pulse sequence for $\tau_\text{S} = 12.5\ns$ storage time; insets their counterpart for $\tau_\text{S} = 312\ns$. 
Secondary insets display exponential fits of the fluorescence decay (\textit{bright green}) in the 
lin.$\perp$lin. (\textit{solid}) and the circ.$\perp$circ. (\textit{dashed}) configuration. 
\textbf{(a)} \& \textbf{(b)}: Noise emitted by the spin-polarised ensemble (setting \cd\,). 
\textbf{(c)} \& \textbf{(d)}: Noise emitted by the thermally distributed ensemble (setting \textit{c}). 
\textbf{(e)} \& \textbf{(f)}: Memory pulse sequence (setting \scd\,, \textit{blue} for lin.$\perp$lin. and \textit{green} for circ.$\perp$circ. polarisation) and noise (setting \cd\,, \textit{red} for lin.$\perp$lin., \textit{black} for circ.$\perp$circ. polarisation). \textit{Dashed, grey vertical lines} mark the 
$\Delta t^\text{TAC}_\text{int} = 5\ns$ pulse integration window. 
\textit{Dashed red vertical lines} in \textbf{(a)} \& \textbf{(b)} illustrate the relative increase in S and AS noise over successive pulses.}
\label{fig_ch7_flourescence}
\end{figure}

\subsection{Determination of the fluorescence noise contribution \label{ch7_sec_fluor_results}}

Using the above mentioned scheme, we now access the amount of noise, resulting from fluorescence and two-photon transitions. To obtain the maximum amount of information, we conduct these measurements, observing both the Stokes (S) and the anti-Stokes (AS) FWM channels. Similar to the work in chapter \ref{ch6}, we also study the ensemble pumped and unpumped, i.e. spin-polarised in $6^2 \text{S}_{\frac{1}{2}}$ F$=4$ and thermally distributed between both ground states. 
These configurations are again denoted by the measurement settings. 

\paragraph{Measurements}
For the measurement, we use the apparatus of chapter \ref{ch6} with coherent state inputs (see fig. \ref{fig_6_setup}). 
Fig. \ref{fig_ch7_flourescence} shows the experimental results for measurements of settings \cd\, (\textbf{a} - \textbf{b}), \textit{c}  (\textbf{c} - \textbf{d}) and \scd\,  (\textbf{e} - \textbf{f}). 
Each measurement is conducted with $\tau_\text{S}=12.5\ns$ storage time (\textit{main panels}), using a train of $9$ successive control pulses\footnote{
	The reduced $10^\text{th}$ pulse in fig. \ref{fig_ch7_flourescence} results from Pockels cell (\pockels\,) 
	leakage, as the subsequent 
	control pulse falls onto the closing edge of the \pockels\, picking window. 
}. 
Additionally, all settings are evaluated for $\tau_\text{S} = 312\ns$ storage time (\textit{insets}), using two successive read-out control pulses\footnote{
	Again these are followed by a reduced pulse, due to Pockels cell leakage.
}. 
For all measurements with active {\cs} state preparation (\textit{panels} \textbf{a}, \textbf{b}, \textbf{e}, \textbf{f}), the optical pumping is turned off $\sim 1.5 \mus$ before the read-in control pulse and is reapplied $\sim 1.4 \mus$ thereafter. 
The experimental repetition rate was set to $f_\text{rep} = 4\kHz$. An integration time of $\Delta t_\text{meas} = 5\min$ ($\tau_\text{S} = 312\ns$) or $\Delta t_\text{meas} =10\min$  ($\tau_\text{S} = 12.5\ns$) per measurement setting was used.
Employing a \coh\, signal with an input photon number of $N_\text{in}= 0.46 \ppp$, the memory efficiencies for ${\tau_\text{S}=12.5\ns}$ storage time are $\eta_\text{in} = 47.3 \pm 0.9 \,\%$ for the read-in and $\eta_\text{mem,1} = 27.7 \pm 0.7\,\%$ for the total efficiency in the first read-out time bin. 
For $\tau_\text{S}=312\ns$ storage time, with ${N_\text{in} = 0.47 \ppp}$, decoherence reduces the total efficiency to ${\eta_\text{mem,1} = 15.1 \pm 0.5\,\%}$, while ${\eta_\text{in} =41.5 \pm 0.8 \,\%}$ remains at a similar level\footnote{
	The residual decrease is due to efficiency drift over the course of the measurement. 
}. 
Each experiment is performed twice, first with the signal filter resonance set to $\Delta=15.2 \GHz$ detuning to analyse the Stokes (S) channel (the Raman memory signal channel), and subsequently to $\Delta=24.4 \GHz$ to analyse the anti-Stokes (AS) channel.

\paragraph{Raman transition turn off} 
Each dataset in fig. \ref{fig_ch7_flourescence} clearly shows the turn-off of all two-photon transition components, i.e. Raman interactions, when switching the input polarisation from lin.$\perp$lin. to circ.$\perp$circ.
For lin.$\perp$lin. polarisation the settings \cd\, (\textit{panels} \textbf{a}, \textbf{b}) and \textit{c} (\textit{panels} \textbf{c}, \textbf{d}) show the noise emitted from the memory. 
The pulse structure completely disappears once circ.$\perp$circ. polarisation is applied. 
Since FWM and SRS always follow the temporal shape of the control pulse\cite{Rousseau:1975}, the absence of pulses in the count rate histograms represents the termination of the two-photon transition processes.
This happens in both channels. 
For linear polarisation, AS noise is emitted simultaneously with S noise; upon switch-off, the pulses in both channels disappear, as expected for FWM.
Circular polarisation also turns off the Raman memory interaction. 
It is illustrated in panel \textbf{e} by the pulse sequences for \scd\, and \cd\,, whose differences yield the memory read-out (see section \ref{ch6_sec3_subsec1}). 
Here, \scd\, pulses are absent in all read-out time bins, so there is no memory. 
Obviously, the increase of the input pulse amplitude for setting \scd\, with circ.$\perp$circ. polarisation
represents the transmitted input signal, which now does not experience Raman absorption. 
Notably, for the lin.$\perp$lin. configuration in the AS channel (panel \textbf{f}), we also observe increased noise emission in setting \scd\, for the first three time bins, i.e. a noise level that is higher than the level in setting \cd\,. This is ``noise-seeding'' by the input signal at the Stokes frequency, which is investigated in section \ref{ch7_subsec_ASseeding} below.

\paragraph{Fluorescence} 
In the Stokes channel (left column in fig. \ref{fig_ch7_flourescence}), the recorded count rates for circ.$\perp$circ. polarisation reduce to the DC-background level, observed between the pulses for lin.$\perp$lin. polarisation.
This background noise is built-up over successive control pulses. 
Within each pulse, noise counts reach their maximum at the pulse end, after which decay sets in until the next control pulse arrives. 
After the last pulse, the background decays exponentially, illustrated by the \textit{insets} for the $\tau_\text{S} =312\ns$ storage time data in fig. \ref{fig_ch7_flourescence}, showing noise decay in the read-in time bin.
This matches the expected count rate time dependence of fluorescence noise\cite{RousseauJCP:1976}, which should show an exponential decay with a 1/e-lifetime of\cite{Steck:2008qf} $\tau_\text{Cs} = 30.5\ns$.

We validate the decay timing by fitting the detected count rates ($c_\text{det}$) 
after the last control pulse in each pulse sequence\footnote{
	For $\tau_\text{S}=312\ns$ storage time, this is done twice: First after the read-in control pulse, 
	and, second, after the $3^\text{rd}$ pulse in the 
	retrieval sequence. In all cases, the free fit parameters are $c_0$ and $\tau_\text{FN}$. 
} with the exponential decay, 
$c_\text{det}(t) = c_0 \cdot \exp{\left( \frac{-t}{\tau_\text{FN}} \right)}$ 
(\textit{light green lines} in fig. \ref{fig_ch7_flourescence}). 
Both polarisation configurations exhibit similar values for
$\tau_\text{FN}$. After the $9^\text{th}$ pulse in the $\tau_\text{S} = 12.5\ns$ control pulse train we get $\tau^{\text{out},12.5}_\text{FN} \approx 44.2 \pm 1.3 \ns$. For the $\tau_\text{S} = 312\ns$ control sequence, fluorescence decays with $\tau^{\text{out},312}_\text{FN} = 39.6 \pm 2.6\ns$ after the read-out pulses, which shortens further to $\tau^{\text{in},312}_\text{FN} = 38.5 \pm 1.1\ns$ for the $\tau_\text{S} = 312\ns$ read-in time bin. 
While these numbers are on the right order of magnitude, they exceed the excited state lifetime $\tau_\text{Cs}$. 

At present, the reason for this discrepancy is unclear. From collisional broadening  one would expect\cite{Rousseau:1975} a reduction with respect to $\tau_\text{Cs}$, i.e. $\tau_\text{FL} < \tau_\text{Cs}$. 
Yet, the numbers for $\tau_\text{FN}$ increase for a longer sequence of control pulses\footnote{
	Notably, this is not an effect from the build-up of noise over the control pulse train. 
	The same decay times $\tau_\text{FN} = \tau_\text{Cs}$ would be expected after both, a single control pulse 
	and a sequence of $9$ consecutive control pulses. 
	This is the case even if the fluorescence generated by each pulse starts to decay directly 
	after the pulse.
	It also does not relate to steady-state atomic diffusion out of the observed interaction 
	region in the {\cs} cell, as this dynamics happens on longer time-scales (see appendix \ref{ch7_lifetime}). 
}.  
One possible reason for this prolongation could be the radiation trapping effects by reabsorption of the fluorescence noise (see section \ref{ch7_subsec_fluorescence}). 
At present, the drivers behind the longer decay times are not completely clear. 

Nevertheless, we can now determine, how much fluorescence actually contributes to the total noise in the S channel.
To this end, each pulse in the \cd\, trace is integrated for both polarisation configurations, using a $\Delta t_\text{int}^\text{TAC} = 5\ns$ integration window, centred on each control pulse. These windows are marked for pulses 1-3 by \textit{grey vertical bars} in fig. \ref{fig_ch7_flourescence}. 
The ratio between areas for circ.$\perp$circ. and lin.$\perp$lin. polarisation gives the fluorescence fraction $R_\text{S,FL}^t$ in each time bin $t$; the remainder originates from two-photon transition processes. 
Table \ref{tab_app_ch7_fluor} in appendix \ref{app_ch7_flour_noise} lists the full set of numbers for all traces in fig. \ref{fig_ch7_flourescence}. 
Most importantly, for $\tau_\text{S} = 12.5\ns$, 
${R}^\text{in}_\text{S,FL} \approx 16 \,\%$ 
and 
${R}^\text{out,1}_\text{S,FL} \approx 14 \,\%$.
This decreases to 
${R}^\text{out,8}_\text{S,FL} \approx 10\,\%$, 
as the FWM contribution increases and eventually saturates. 
In contrast, the anti-Stokes channel (right column in fig.~\ref{fig_ch7_flourescence}) does not contain any significant contribution from fluorescence noise. 
Here, $c_\text{det} \rightarrow 0$ for circ.$\perp$circ. polarisation. 
The AS detuning $\Delta_\text{AS} =24.4\GHz$ is large enough to fall outside the fat tails of the collisional redistribution line\cite{RousseauJCP:1976, Carlsten:1977,Raymer:1977}. 
For $\tau_\text{S}=12.5\ns$ storage, the DC-background only amounts to 
${R}^\text{in}_\text{AS,FL} \approx 0.6 \,\%$ 
and 
${R}^\text{out,1}_\text{AS,FL} \approx 2.3 \,\%$. 
Consequently, the AS channel contains pretty much only FWM noise. 
For the noise in the signal channel of our Raman memory, we have a noise floor consistency of:
\begin{center}
\begin{tabular}{l l r l r}
Read-in bin: 	&  2-photon transitions: 	& 	$84\,\%$, 	& 	 Fluorescence: 	& 	$16\,\%$ \\
Read-out bin: 	&  2-photon transitions: 	& 	$86\,\%$,	& 	 Fluorescence: 	& 	$14\,\%$ 
\end{tabular}
\end{center}

\paragraph{Reducing fluorescence}
Clearly fluorescence noise is not the main component of the memory noise floor, so even if it was eliminated, the memory would still not operate in the quantum regime. 
However, as a side note, we briefly look at possibilities to limit its contribution. 
As the traces in fig. \ref{fig_ch7_flourescence} show, the integration window size 
$\Delta t_\text{int} = \Delta t_\text{int}^\text{TAC} = 5\ns$  
is broader than the actual pulses, causing more fluorescence noise to be picked up than strictly necessary. 
An improvement in the SNR  (eq. \ref{eq_ch6_SNR}) could thus be gained by reducing  $\Delta t_\text{int}$, cutting 
down the fluorescence contribution to $N_\text{noise}$ (eq. \ref{ch6_eq_NoiseFloor}). However too tight an integration window will also lead to signal loss, so there is an optimal value for $\Delta t_\text{int}$. 
Fig. \ref{fig_ch7_SNRintwin} \textbf{a} illustrates the SNR improvement one could achieve for the \coh\, datasets of fig. \ref{fig_ch7_flourescence}, as well as for \hsp\, inputs at $N_\text{in} = 0.22\ppp$, i.e. with $\eta_\text{her} = 22\,\%$ heralding efficiency, when reducing $\Delta t_\text{int}$. 
In both cases, window tightening predominantly reduces the fluorescence contribution until $\Delta t_\text{int} \approx 1\ns$, when further reduction starts to cut into the actual memory pulses. 
At this point, noise reduction is accompanied by signal loss, associated with a reduction in {\etamem} (see fig. \ref{fig_ch7_SNRintwin} \textbf{b}), for which reason the SNR starts to tail off.

For \hsp\, inputs, window reduction to $\Delta t_\text{int} = 1\ns$ would improve the SNR for the $12.5\ns$ memory read-out by a factor of $\sim 1.16$. 
While this would increase the  {\gtwo} separation between \hsps\, and noise, such an improvement is  not sufficient to see non-classical statistics for retrieved single photons (see fig. \ref{fig_ch6_g2results}). 
As we have discussed in section \ref{ch6_conclusion} (see fig. \ref{fig_ch6_Rscan}), to achieve SNR$_\text{out} \sim 1$, we would need an improvement by a factor of $\gtrsim 2.5$ to put us close to the boundary between the classical and the quantum regime. 
Faithful quantum operation would necessitate an order of magnitude improvement in SNR$_\text{out}$, i.e. a factor of $\gtrsim 10$. This is clearly not achievable by reducing $\Delta t_\text{int}$, so this route does not solve the noise problem. 
 
Currently, $\Delta t_\text{int}$ is limited by the FPGA coincidence logic. While it can be modified\cite{Spring:PhD} to achieve $\Delta t_\text{int}=1\ns$, this comes at the expense of losing the ability to modify the delays between the FPGA channels\footnote{
	Due to the additional propagation distance of the signal photons, resulting from the delay required for 
	\pockels\, switching and the memory storage time, the differences in FPGA delays in the experiments 
	with \hsps\, are on the order of $\sim 500\ns$, plus an additional electronic delay in BNC cables of 
	$\sim 200\ns$. 
} by more than $20\ns$, which slightly complicates the synchronisation between channels required for meaningful data acquisition (see section \ref{sec_ch4_photon_detection}). 
Moreover, the signal count rate reduces by a factor of $\sim 1.5$ for \hsps\, and $\sim 1.3$ for \coh\, inputs, which would require longer measurement times.
Since both factors increase the experimental complexity, for no substantial {\gtwo} improvement, 
we did not use a $\Delta t_\text{int} = 1\ns$ coincidence window. 

Another possibility for reducing fluorescence noise is to lower the Ne buffer gas pressure, currently at $p_\text{Ne}= 20$ Torr. 
Far off-resonance, the fluorescence noise intensity $I_\text{FL}$ is linearly proportional\cite{Carlsten:1977} to the collision rate between {\cs} and Ne atoms $\gamma_N \sim n_\text{Ne} \sim p_\text{Ne}$, where $n_\text{Ne}$ is the atomic density of Ne. 
Lower pressures would reduce $\gamma_N$, leading to a lower fluorescence background. It could also offer the possibility to operate the Raman memory closer to resonance (see appendix \ref{ch7_subsec_detuning}). 
Another alternative could be operation at an even larger detuning of $\Delta_\text{S} \sim 24\GHz$, since this regime lies outside the collisional redistribution line (see fig. \ref{fig_ch7_flourescence}). However, this would, \textit{inter alia}, require substantially greater control pulse energies.

\section{Origin of the two-photon transition noise components\label{ch7_FWM_explanation}} 

Knowing the amount of noise that originates from two-photon transitions, we now investigate its actual sources. 
Here, the first step is to understand the noise behaviour, displayed by the count rate histograms in fig. \ref{fig_ch7_flourescence}. 
To this end, we develop a phenomenological explanation which we can compare with the predictions of our coherent  model (see appendix \ref{app_ch6}) later on.

\subsection{Noise in Stokes and anti-Stokes channels for the prepared ensemble\label{ch7_subsec_cd_noise}} 

\begin{figure}[h!]
\includegraphics[width=\textwidth]{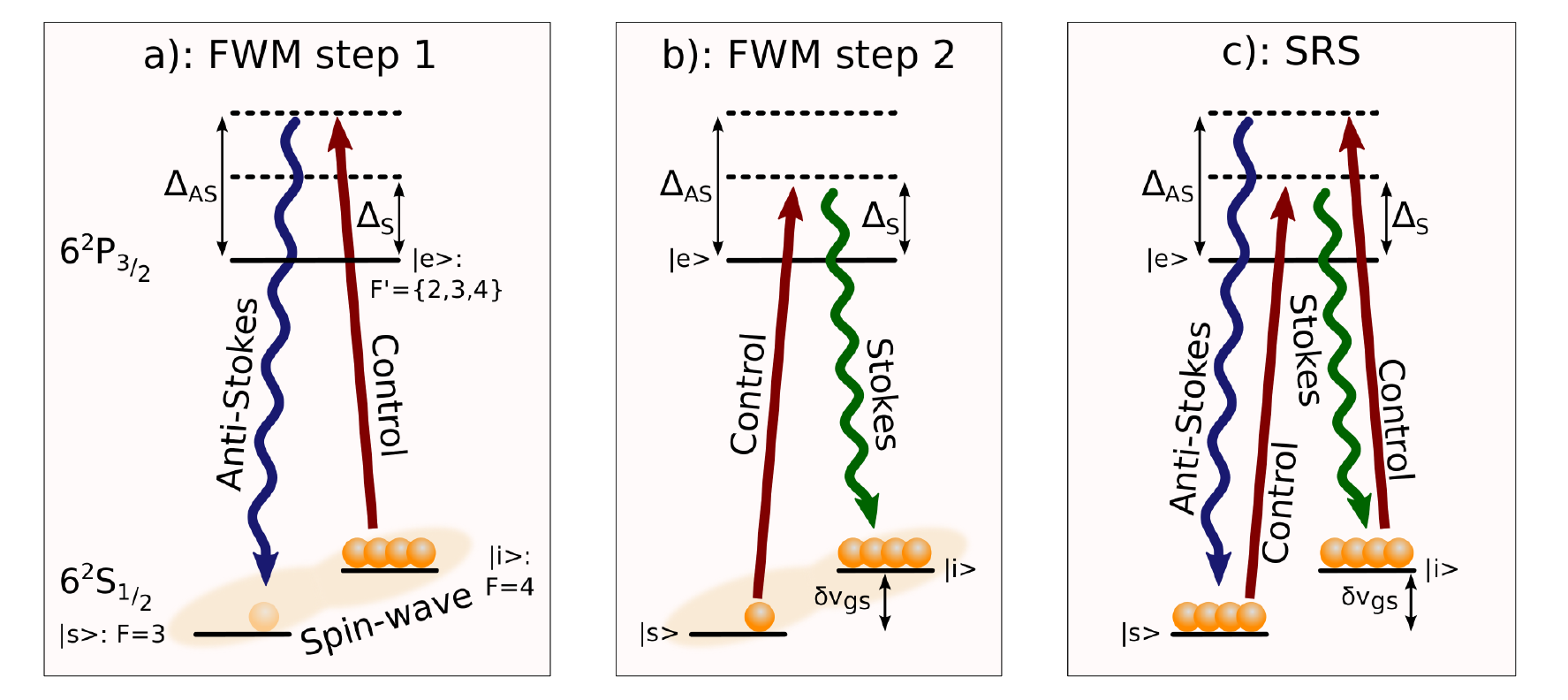}
\caption{Level schemes for FWM and SRS noise processes. 
\textbf{(a)}: Step 1 of the FWM process: AS noise scattering. Atomic population transfer by the control, coupling to the spin-polarised {\cs} ensemble with population initially in $6^2\text{S}_\frac{1}{2} \text{F}=4$, excites a spin-wave (\textit{transparent}). 
\textbf{(b)}: Step 2 of the FWM process: spin-wave retrieval by the control under S noise emission.
\textbf{(c)}: SRS for thermally distributed {\cs} population. 
The control couples to both $6^2\text{S}_\frac{1}{2} \text{F}=\left\{3,4\right\}$ hyperfine ground states. 
}
\label{fig_ch7_FWMlevels_2}
\end{figure}

\paragraph{Noise creation process} 
To determine the origin of the two-photon transition based noise components we investigate the pulses for setting \cd\, in the Stokes (\textit{panel} \textbf{a}) and the anti-Stokes channel (\textit{panel} \textbf{b}) of fig. \ref{fig_ch7_flourescence}.
Noise is emitted into both simultaneously, whereby, for the first pulse, the AS amount is significantly larger than its S counterpart.
Assuming good state preparation (see appendix \ref{ch3_diodelaser}), initially all atoms are in the F$=4$ ground state. 
If we further assume that the noise origin is FWM, emission into the S channel during the first pulse can only occur after a FWM spin-wave is excited by AS scattering. 
Fig. \ref{fig_ch7_FWMlevels_2} \textbf{a} \& \textbf{b} show again these two steps of the FWM process. 
A sub-unity retrieval efficiency for S noise generation, the second FWM step, leaves parts of this FWM spin-wave stored, resulting in lower emission in the S than in the AS channel.
Upon arrival with the next control pulse, the remaining fraction adds to the spin-wave that is freshly excited by this consecutive control pulse, increasing the total spin-wave amplitude the control couples to and therewith also the S emission in this second time bin. 
Hence the S level rises over successive control pulses, saturating after the $4^\text{th}$ pulse.
Due to the coupling between spin-wave and AS channel, the stored spin-wave excitation also results in an increased AS emission.
Noise build-up saturates once spin-wave outflow through retrieval balances new spin-wave generation\footnote{
	Notably, this is essentially a mirrorless Raman laser, with the spin-wave 
	playing the role of the intra-cavity field, and the control pulses 
	providing the gain. The Raman laser reaches steady-state 
	when the gain is balanced by losses through the output-coupler, which 
	here is spin-wave retrieval. 
}. 

\paragraph{Noise scaling between the Stokes and the anti-Stokes channel}
The S and AS couplings to the spin-wave are given by their respective Raman coupling constants $C_\text{S}$ and $C_\text{AS}$ (eq. \ref{eq_ch2_Raman_coupling}), whose ratio $R_\text{S/AS} = \frac{C_\text{S}}{C_\text{AS}} = \frac{\Delta_\text{AS}}{\Delta_\text{S}} $ is determined by the detunings $\Delta_\text{S}$ and $\Delta_\text{AS}$ (see section \ref{subsec_ch2_maxwell_bloch_eq} and appendix \ref{app6_coh_model}).
Since both channels couple to the same spin-wave, the relative increase in noise counts from one pulse to the next in each channel 
should show a ratio of\footnote{ 
	Note, the coupling constants $C_\text{S}$ and $C_\text{AS}$ are defined for the electric field operators $S$ 
	and $A$ for the S and the AS channel, respectively. 
	In the experiment we measure the intensity in each channel, which is proportional to the expectation 
	values of the photon number operators 
	$\langle \hat{n}_\text{S} \rangle= \langle S^\dagger S \rangle \sim C^2_\text{S}$ and 
	$\langle \hat{n}_\text{AS} \rangle = \langle A^\dagger A \rangle \sim C^2_\text{AS}$. So the noise increase 
	ratio is expected to be proportional to $R_\text{S/AS}^2$.
} ${R^2_\text{S/AS} = \left(\frac{\Delta_\text{AS}}{\Delta_\text{S}}\right)^2 = 2.58}$, when comparing relative increases between both channels.
Integrating the TAC count rate histograms within $\Delta t_\text{int}$ for each pulse yields the counts $a^t_i(p)$, for pulse number $t$, channel $i \in \left\{ \text{S},\text{AS} \right\}$ and polarisation configuration $p$. 
The total counts therein, originating from two photon transition noise, are given by 
$\tilde{a}^t_{i} = a^{t}_{i}(\text{lin.} \perp \text{lin.}) -a^{t}_{i}(\text{circ.} \perp \text{circ.})$. 
The relative increases between successive pulses, due to spin-wave coupling, are 
${\Delta a^{t+1,t}_i = \tilde{a}^{(t+1)}_{i} - \tilde{a}^{t}_{i}}$. 
These are illustrated by \textit{dotted vertical lines} at the pulse maxima in fig. \ref{fig_ch7_flourescence} \textbf{a} and \textbf{b}. 
Their ratios yield $\tilde{R}^2_\text{S/AS} (t)= \frac{\Delta a^{t+1,t}_\text{S}}{\Delta a^{t+1,t}_\text{AS}}$, which, in case of FWM, should equal $R^2_\text{S/AS}$.
Averaging $\tilde{R}^2_\text{S/AS}(t)$ over the first 4 pulses, i.e. $t \in [ 1, 3]$, we obtain $\tilde{R}^2_\text{S/AS} = 2.71 \pm 0.14$. This number compares well to the expected $R^2_\text{S/AS}$ and is evidence that confirms the FWM origin of the two-photon transition noise.

 \subsection{Noise in Stokes and anti-Stokes channels for the unpumped ensemble\label{ch7_subsec_c_noise}}

\paragraph{Anti-Stokes scattering} 
By blocking the diode laser, i.e. applying settings \textit{sc} and \textit{c}, the population is equally distributed between the initial ($\ket{\text{i}}$) and the storage state ($\ket{\text{s}}$), as shown in fig. \ref{fig_ch7_FWMlevels_2} \textbf{c}.  
It enables to test the onset of FWM by spontaneous Raman scattering (SRS) into the AS channel from state $\ket{\text{i}}$, which is the first leg of the FWM process. 
To this end, we observe the read-in bin of fig. \ref{fig_ch7_flourescence} \textbf{b} \& \textbf{d}, where effects from previously excited FWM spin-waves are absent. 
Firstly, the Raman coupling constants $C_\text{S}$ and $C_\text{AS}$ for S and AS scattering are proportional to the population of the state they couple to (eq. \ref{eq_ch2_Raman_coupling}), with\cite{Nunn:DPhil} $C_\text{S} \sim \sqrt{N_{\ket{\text{s}}}}$ and $C_\text{AS} \sim \sqrt{N_{\ket{\text{i}}}}$.
Thus, a linear relationship between the count rates of emitted anti-Stokes photons $c_\text{AS} = \frac{d}{dt} \tilde{a}_\text{AS}$ and $N_{\ket{\text{i}}}$ is expected\cite{Penzkofer:1979}. 
Blocking the diode laser approximately halves $N_{\ket{\text{i}}}$, for which reason $c^\text{in}_\text{AS}$ should also half. 
Comparing the histograms in fig. \ref{fig_ch7_flourescence} \textbf{b} \& \textbf{d},
we can indeed observe such a reduction in count rate, where the ratio $R^\text{SRS}_{c/cd,\text{AS}} = \frac{{c}^\text{in}_{c,\text{AS}}}{{c}^\text{in}_{cd,\text{AS}}} \approx 0.53 \pm 0.01$ closely matches this expectation.  

\paragraph{Stokes scattering} 
At the Stokes frequency, the opposite happens. With good optical pumping the only population the control can couple to is the one transferred by preceding anti-Stokes scattering (fig. \ref{fig_ch7_FWMlevels_2} \textbf{a} \& \textbf{b}). 
When thermally distributed (fig. \ref{fig_ch7_FWMlevels_2} \textbf{c}), half of the population will be in state $\ket{\text{s}}$, so the amount of noise at the Stokes frequency $c_\text{S}$ should increase by blocking the diode. 
Like the AS channel, the noise can now be expected to predominantly originate from SRS (see section \ref{ch7_subsec_SRS}). 
Both SRS processes are independent and accordingly the ratio between their respective emissions into the S and AS channel should correspond to the ratio between their Raman coupling constants, which is $R^2_\text{S/AS} = 2.58$. 
From fig. \ref{fig_ch7_flourescence} \textbf{c} \& \textbf{d}, we obtain a ratio for the count rates between the input bins of $\left( R^{\text{SRS}}_{\text{S/AS}}  \right)^2= \frac{c_\text{S}^\text{in}}{c_\text{AS}^\text{in}} = 2.41 \pm 0.02$, which indeed roughly matches $R^2_\text{S/AS}$. 

Unlike the prepared ensemble, for thermally distributed {\cs} the S noise decreases for subsequent time bins, while the AS noise level remains constant. Both thus cannot couple to a common spin-wave. We will experimentally verify this absence of a spin-wave component below, by looking at the magnetic dephasing properties of the noise in section \ref{ch7_subsec_mag_dephasing}. 
While spin-wave absence might be counterintuitive at first, it is actually expected from the system dynamics, as described by our model in appendix \ref{app6_coh_model}. 
Particularly, this decrease is not an effect of population transfer from the F$=3$ to the 
F$=4$ ground state by the control.  
A simple pumping model (see appendix \ref{app_ch7_subsec_flow_model_srs_stokes}), based on steady state population outflow from F$=3$ by SRS into the S channel and population inflow from F$=4$ by SRS into the AS channel, illustrates this.

\begin{figure}[h!]
\centering
\begin{minipage}[l]{.44\textwidth} 
\centering
\includegraphics[width=\textwidth]{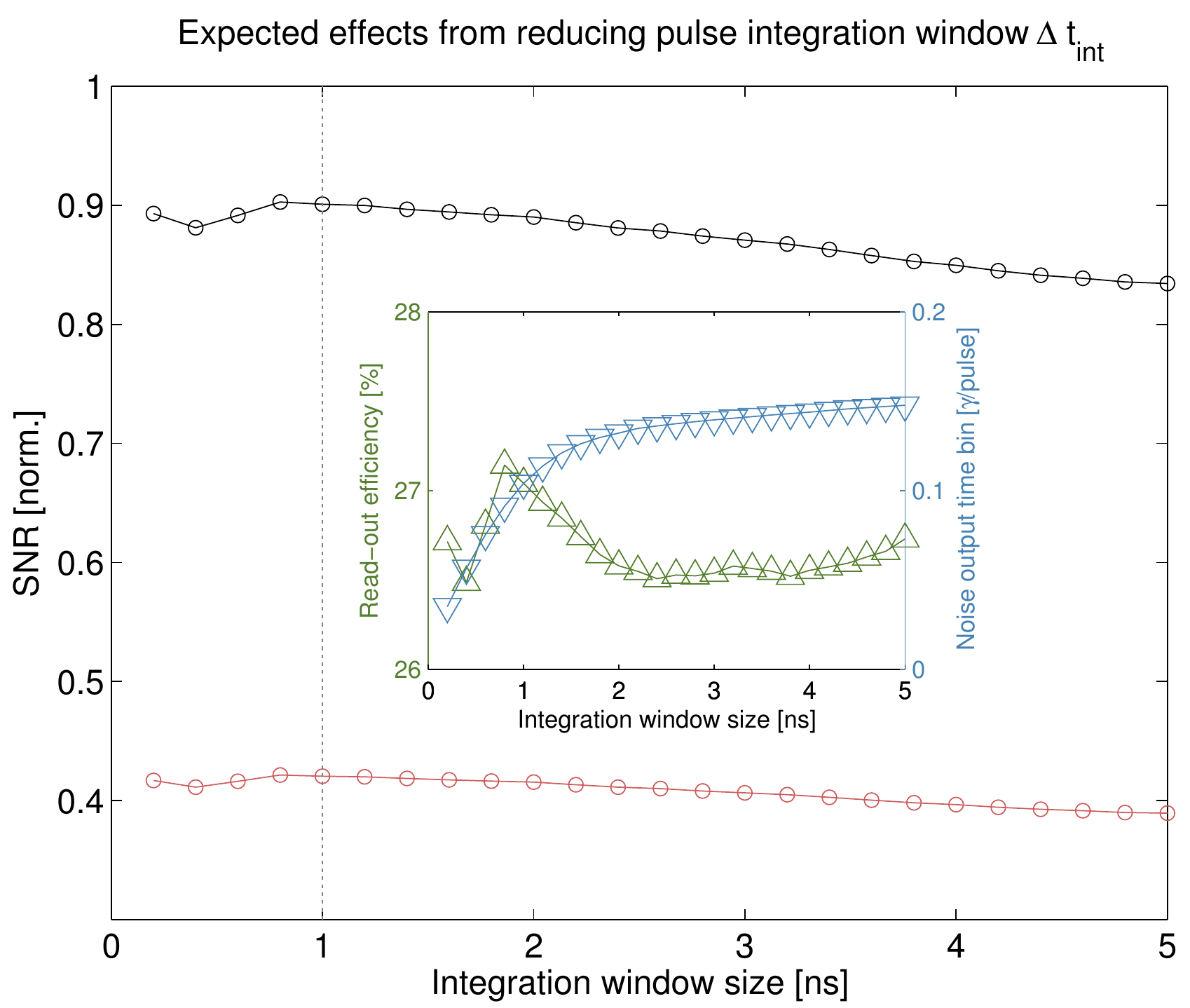}
\end{minipage}
\vspace{0.05\textwidth}
\begin{minipage}[r]{0.48\textwidth} 
\centering
\includegraphics[width=\textwidth]{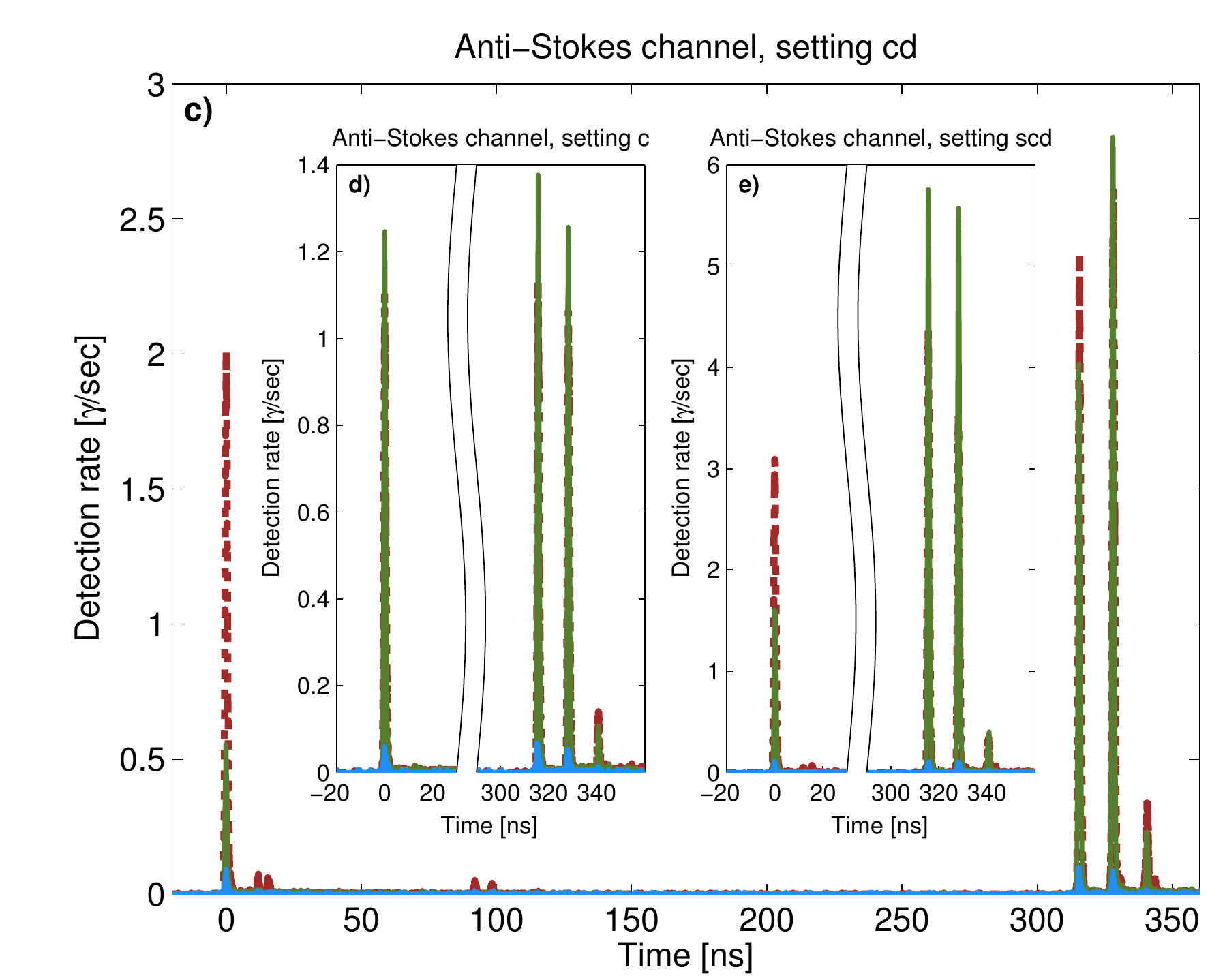}
\end{minipage}
\caption{\textbf{(a)} \& \textbf{(b)}: Improvements from cutting fluorescence noise as a function of integration window size $\Delta t_\text{int}$, centred on the first read-out pulse in the TAC traces. 
\textbf{(a)} displays $\text{SNR}_\text{out}$ of the first retrieval time bin. 
\textit{black line}: \coh\, input signals at $N_\text{in} = 0.47\ppp$; \textit{red line}: \hsp\, inputs at $N_\text{in} = 0.22\ppp$. 
\textbf{(b)} illustrates the memory efficiency $\eta_\text{mem}$ (\textit{green}) for the \coh\, input signal  
and the noise level $N^\text{out}_\text{noise}$ (\textit{blue}). 
\textbf{(c)} - \textbf{(e)}: 
Comparison of the AS noise level for the cold {\cs} cell at $T_\text{Cs} \approx 20^\circ \text{C}$ (\textit{blue line}) with that of the heated cell at $T_\text{Cs} \approx 70^\circ \text{C}$ (\textit{red line}) for $\tau_\text{S} = 312\ns$ storage time. Also shown is the AS noise observed when pumping the {\cs} atoms to the F$=3$ hyperfine ground state 
(\textit{green line}) instead of F$=4$. 
\textbf{(c)}, \textbf{(d)} \& \textbf{(e)} show the spin-polarised ensemble (setting \cd\,), the thermally distributed ensemble (setting \textit{c}) and the memory pulse sequence, consisting of signal, control and diode sent into the spin-polarised ensemble (setting \scd\,), respectively.}
\label{fig_ch7_SNRintwin}
\end{figure}

\subsection{Atomic ensemble at room temperature - control leakage estimation\label{ch7_subsec_cold_ensemble}}

\paragraph{Noise level in the unheated ensemble}
We also investigate the effects of an unheated {\cs} cell on the AS channel, where the vapour temperature is $T_\text{Cs} \approx 20^\circ \text{C}$. 
This allows us to estimate the residual control pulse leakage, which eventually contributes to the fluorescence background\footnote{
	Control leakage is independent of whether lin.$\perp$lin. or circ.$\perp$circ. pol. is applied to
	the {\cs} cell, because it results from the limit on the separability of two orthogonal polarisation states. 
	So it is still contained in the fluorescence fraction we have measured with the two-photon resonance 
	turn-off methodology in section \ref{ch7_pol_turn_off_principles}. 
}.
We observe the AS channel count rate histograms, displayed in fig. \ref{fig_ch7_SNRintwin} \textbf{c} - \textbf{e} for settings $\left\{cd,c,scd \right\}$ (\textit{blue lines}), obtained with \coh\, input signals and $\tau_\text{S} = 312\ns$ storage time. 
These have to be compared to the noise floor for the {\cs} cell at $T_\text{Cs} = 70^\circ \text{C}$, which is also shown (\textit{red lines}).

Despite the lower Raman coupling constant $C_\text{AS}$, caused by atomic density reduction, there is still a small amount of AS noise left in each time bin. 
For setting $i$ and time bin $t$, the integrated pulse areas $a^t_{i,\text{AS}}$ give rise to the fraction 
$R^{t,\text{cold}}_{i,\text{AS}} = \frac{a^t_{i,\text{AS}}(T_\text{Cs} = 20^\circ \text{C})}{a^t_{i,\text{AS}}(T_\text{Cs} = 70^\circ \text{C})}$, which is similar for read-in and read-out time bins. Its mean over both bins amounts to 
$R^{\text{cold}}_{cd,\text{AS}}  = 2.5 \pm 0.5 \%$ and 
$R^{\text{cold}}_{c,\text{AS}}  = 3.4 \pm 0.2 \%$, 
for the pumped and unpumped ensemble, respectively. 
Similar behaviour is seen in the Stokes channel (see section \ref{ch7_Temp}). 
These counts can either result from control leakage or from residual FWM. 

\paragraph{Estimation of control field leakage}
To determine their origin, the AS channel offers an advantage, since here the underlying process, generating AS noise in the $1^\text{st}$ time bin, is predominantly SRS (see fig. \ref{fig_ch7_FWMlevels_2} \textbf{a} \& 
\textbf{c}), which is the case for both pumping configurations (settings \cd\, and \textit{c}). 
We can thus use the ratio 
${R^{\text{cold}}_{c/cd,\text{AS}} = \frac{a^\text{in}_{c,\text{AS}}}{a^\text{in}_{cd,\text{AS}}}}$ 
between both pulses.
In case the noise is due to FWM, this ratio should yield a similar value to the one observed for warm {\cs} vapour. 
For the $\tau_\text{S} = 312\ns$ storage time experiment (insets in fig. \ref{fig_ch7_flourescence}), we have $R^\text{SRS}_{c/cd,\text{AS}} \approx 0.54$ for $70^\circ \text{C}$. 
However, if the counts are control leakage, diode turn off should lead to a count rate reduction due to linear absorption of the control. This reduction should approximately equal the absorption of the signal field\footnote{
	In fact a reduced linear absorption is expected for the control, because its transition with the least possible 
	detuning is $15.2\GHz$ away from the $6^2 \text{P}_\frac{3}{2}$-manifold. In contrast, the
	signal transition nearest to the excited state has only $6\GHz$ detuning 
	(see fig. \ref{fig_ch7_FWMlevels_2}). Consequently, greater linear absorption is 
	expected for signal transmission, rendering the calculated amount of control leakage an upper bound.
}, i.e. 
$R^{\text{cold}}_{c/cd,\text{AS}} \approx R^{\text{cold}}_{s/sd,\text{AS}} = \frac{a^\text{in}_{s,\text{AS}}}{a^\text{in}_{sd,\text{AS}}} =  0.8$. 

Experimentally, we obtain $R^{\text{cold}}_{c/cd,\text{AS}} = 0.62$, which lies in between both expectations. 
The noise is thus likely to result from a linear superposition between both sources, given by:
$$
R^{\text{cold}}_{c/cd,\text{AS}} = (1-\zeta) \cdot R^\text{SRS}_{c/cd,\text{AS}}  + \zeta \cdot R^{\text{cold}}_\text{leak},
$$ 
with $R^{\text{cold}}_\text{leak} = R^{\text{cold}}_{s/sd,\text{AS}}$.
Here, $\zeta = 0.3$ assigns the fraction of control leakage in the signal for the room temperature ensemble\footnote{
	In appendix \ref{ch3_diodelaser} we estimate the amount of linear absorption to $L_\text{abs} = 10\,\%$, whereas
	$R^{\text{cold}}_{s/sd,\text{AS}}$ corresponds to $L_\text{abs} =20\,\%$. 
	This difference can result from insufficient precision. 
	Since the signal transmission is suppressed when the signal filter stage is resonant with the AS frequency, the 
	count rate is low. Moreover, the setting integration times in this measurement has also been short
	 (${\Delta t_\text{meas} =  5\min}$). 
	Using $R^{\text{cold}}_\text{leak} = 0.9$, we however obtain an even
	lower contribution from control leakage of $\zeta = 0.21$, which gives $N_\text{leak} = 1.6\cdot 10^{-3} \ppp$. } (cold), which estimates an absolute value of 
$N_\text{leak} = 2\cdot 10^{-3} \ppp$. 
Leakage remains constant upon heating up the cell, so we estimate  
$\zeta \cdot  R^{t,\text{cold}}_{i,\text{AS}} = 0.75 \,\%$ 
of the AS noise to orginate from control leakage. 
Because the control is always $9.2\GHz$ detuned from the resonance centre of the signal filter stage\footnote{
	It is red detuned when looking at the AS channel and blue detuned for the S channel. 
} and the filter transmission line is symmetric, similar leakage $N_\text{leak}$ can be expected in the S channel. 
Consequently, for $\tau_\text{S} =12.5\ns$ storage, the leakage contribution to $N_\text{noise}^\text{in}$ and  $N_\text{noise}^\text{out}$ (eq. \ref{ch6_eq_NoiseFloor}) are estimated to $3.8\,\%$ and $1.5\,\%$, respectively. 
This is smaller than the measurement uncertainty in eq. \ref{ch6_eq_NoiseFloor}. 

In appendix \ref{ch7_Temp} we present a second method to estimate the control pulse leakage, which predicts an even lower level of approximately half of the number $N_\text{leak}$ estimated here. 
Note that the minimisation of control leakage requires the double-passed FSR$=103\GHz$ etalon in the signal filter stage. 
This element was not included in previous set-up iterations\cite{Reim:2011ys}, which partially explains the improved SNR ratio in this work.

\subsection{Anti-Stokes emission for ensemble preparation in the $6^2 \text{S}_{\frac{1}{2}}$ F$=3$ state\label{ch7_subsec_f3preparation}}

Another testbed for the AS emission is to change the diode laser frequency to the $6^2\text{S}_{\frac{1}{2}} \text{F} = 4 \rightarrow 6^2\text{P}_{3/2}$ resonance, pumping the {\cs} atoms into the F$=3$ hyperfine ground state. 
In this scenario, the roles of the S and the AS channel are reversed in the FWM process, so 
AS emission can only occur after initial spin-wave creation by SRS into the S mode from the F$=3$ level 
(compare to fig. \ref{fig_ch7_FWMlevels_2} \textbf{a} \& \textbf{b}, showing  preparation in F$=4$). 
Simultaneously, the initial SRS into the S mode is stronger than the AS emission for F$=4$ preparation (fig. \ref{fig_ch7_flourescence} \textbf{b}), because it is only detuned by $\Delta = 15.2 \GHz$. 
AS scattering should now show a reduced level, analogue to the one observable for the S channel with F$=4$ preparation (fig. \ref{fig_ch7_flourescence} \textbf{a}). 

Fig. \ref{fig_ch7_SNRintwin} \textbf{c} - \textbf{e} displays the count rate histograms for the F$=3$ preparation (\textit{green lines}), and compares it to the noise level, observed when the ensemble is initially prepared in the F$=4$ ground state (\textit{red line}). 
For the optically pumped ensemble (setting \cd\,), in the read-in time bin, the amount of AS signal is reduced to  
$R_{cd,\text{AS}}^{\text{in}, \text{F}=4/\text{F}=3} = \frac{c_{cd,\text{AS}}^\text{in}(\text{F}=4)}{c_{cd,\text{AS}}^\text{in}(\text{F}=3)}\approx 25\,\%$ 
of the AS noise level observed for ensemble preparation in F$=4$. 
Due to FWM spin-wave excitation, it however increases rapidly over the $2^\text{nd}$ and $3^\text{rd}$ pulse, reaching levels similar to those for F$=4$ preparation. 
Without state preparation, the population is distributed equally between both ground states. So we should not see any effects for the measurements of setting \textit{c}, because no pumping into either ground state (F$=3$ or F$=4$) has happened. This is indeed the case, as a comparison of the similar count rates in figs. \ref{fig_ch7_flourescence} \textbf{b} \& \ref{fig_ch7_SNRintwin} \textbf{d} illustrate\footnote{
	Note that the measurements for F$=3$ and F$=4$ preparation have been recorded on different days.
	The remaining count rate difference is most likely the result of a systematic difference in the day-to-day 
	performance of the system.
}. 

The opposite happens when we look at setting \scd\,, displayed in fig. \ref{fig_ch7_SNRintwin} \textbf{e}, where a \coh\, input signal at the Stokes frequency, containing $N_\text{in} \approx 5 \ppp$, is sent into the {\cs} cell alongside the control.
Here, the AS noise in the read-in time bin is still less than for F$=4$ preparation 
$\left( R_{scd,\text{AS}}^{\text{in},\text{F}=4/\text{F}=3} =0.74 \pm 0.01 \right)$, 
but it surpasses the level for F$=4$ preparation in the read-out time bins 
$\left( R_{scd,\text{AS}}^{\text{out1},\text{F}=4/\text{F}=3} =2.39 \pm 0.02 \right)$. 
Why does this happen? 
When sending in the signal field, Raman scattering into the S mode, the first leg of the FWM process (see fig. \ref{fig_ch7_FWMlevels_2}), becomes stimulated. 
In fact SRS turns into stimulated Raman adiabatic passage\cite{Bergman:2001} (STIRAP), leading to Raman gain in the S channel instead of Raman memory\cite{Reim:PhD}.
The stimulated population transfer also increases the spin-wave. In turn, this enhances the AS emission in fig. \ref{fig_ch7_SNRintwin} \textbf{e}. 
Note that the signal field is only present in the read-in bin, so the count rate boost from stimulated Raman scattering only occurs between the $1^\text{st}$ and the $2^\text{nd}$ control pulse.

\section{Anti-Stokes seeding by the input signal \label{ch7_subsec_ASseeding}}

In fig. \ref{fig_ch7_flourescence} we have already seen that the insertion of a signal field into the S mode can lead to elevated AS noise, when spin-polarising the atomic ensemble. 
For the ensemble prepared in F$=4$, the input signal pulse at the S frequency can seed the $2^\text{nd}$ FWM step, i.e., it can stimulate FWM spin-wave read-out (see fig. \ref{fig_ch7_FWMlevels_2}). 
Thus, sending the input signal into the Raman memory can in fact influence the noise floor of the memory. 
With such FWM noise gain, the noise floor present for setting \scd\, becomes larger than the one measured with setting \cd\,, whose resulting noise numbers, for instance, enter the memory efficiency (eqs. \ref{eq6_memeff} - \ref{eq6_err_memeff_ret}) and the SNR (eq.~\ref{eq_ch6_SNR}) calculation. 
Since S noise is indistinguishable from the signal, any noise gain above the \cd\, background level will be mistaken for signal retrieval from the memory. 
This leads to an overestimation of the memory efficiency {\etamem}, 
since eqs. \ref{eq6_memeff} - \ref{eq6_err_memeff_ret} assume a constant noise background that is independent of the input signal.
Obviously, this is undesirable. We will now investigate the effects the signal input has on the AS channel. 

\paragraph{Noise amplification through seeding} 
The fundamental reason why noise amplification can occur relates to the memory-noise dynamics, which are described by our coherent model (appendix \ref{app6_coh_model}): 
Effectively, the Raman memory corresponds to a beam-splitter interaction\cite{Reim2012,Hosseini:2009lq}, and FWM noise is emitted through a two-mode squeezing interaction on the Stokes and anti-Stokes modes\cite{Loudon:2004gd}.
Without the memory part, the system would be an optical parametric amplifier for FWM noise\cite{Loudon:1993, Loudon:2004gd}. 
Based on these two principles, we derive an upper bound for the amount of noise in the S channel, that can result from FWM gain, in appendix \ref{app_ch7_subsec_ASseeding}. 
An even better estimation can be obtained experimentally, when observing the amount of AS noise as a function of memory input signal \Nin, sent into the S channel in the input time bin. 
Within this time bin, noise gain will lead to an increase in FWM spin-wave retrieval and increased S noise production. 
The results from sections \ref{ch7_subsec_cd_noise} \& \ref{ch7_subsec_theory_prediction} motivate, that this will also lead to an increase in FWM noise in the AS channel, occurring not only in the same, but, due to spin-wave storage, also in the subsequent pulse. 
Since the next time bin is the memory read-out, noise gain herein is even more significant. 
Moreover, we have also seen in section \ref{ch7_subsec_cd_noise} that, in each time bin, FWM S noise production is less than, or at most equal to, the AS level, because S noise can only be generated by retrieving a spin-wave created via AS scattering beforehand. 
Thus, observing the amount of surplus AS noise over the unseeded level (setting \cd\,) directly yields an upper limit for the amount of S noise added by FWM gain.

\paragraph{Measurement procedure for estimating the seeded noise fraction}
Experimentally, the difficulty lies in knowing the exact signal input photon number {\Nin}, when setting the signal filter resonance to the AS frequency (see fig. \ref{fig_6_setup}). 
Since the $\text{FSR} = 103\,\text{GHz}$ etalon in the filter stage blocks any signal transmission, when set to $\Delta_\text{AS} = 24.4\GHz$, direct measurement of $N_\text{in}$ via setting \textit{sd} is not possible.
What is still possible however is signal observation on the Menlo PD, positioned behind the three $\text{FSR}=18.4\GHz$ etalons, but in front of the $\text{FSR}=103\GHz$ etalon. Thanks to the frequency difference between Stokes and anti-Stokes of $2\cdot \delta \nu_\text{gs}=18.4\GHz$ (see fig.~\ref{fig_ch7_FWMlevels_2}), both signals are simultaneously resonant and transmitted with similar efficiency.
With the Menlo PD, the intensity of bright \coh\, input signals can be determined by the pulse amplitude on a scope.
Thereafter, inserting an ND$=7.0$ filter behind the EOM in the signal field preparation path (see fig. \ref{fig_6_setup}) attenuates the signal down to the single photon level and allows to perform the noise seeding measurements using the APD detectors. 
We measure settings $\left\{ scd, cd, sd, c, sc, d \right\}$ (with the signal filter stage resonance set to the AS frequency at $24.4\,\text{GHz}$ detuning). 
Any residual transmission of input photons, observed with settings \sd\, and \textit{s}, is subtracted from the interaction settings \textit{scd} and \textit{sc}, respectively. The optical pumping does not contribute any counts and is henceforth neglected. 

Performing a second set of measurements, now with the signal filter stage resonance set to the S frequency at $\Delta_\text{S} = 15.2\GHz$ detuning, bright signal pulses are prepared such that they show similar pulse amplitudes on the Menlo PD. 
Subsequent insertion of the ND-filter into the signal arm and measurement of setting \sd\, allows to determine the input photon number $N_\text{in}$ at the single photon level (eq. \ref{ch6_eq_Nin}). With these measurements we obtain a calibration between the pulse amplitude voltage on the Menlo PD and the signal input photon number \Nin.
Additionally, the measurements at the S frequency are also used to observe the memory efficiency. 

Notably, in converting pulse amplitude voltages on the Menlo PD to $N_\text{in}$-values for the AS measurement, the slightly different transmission for the signal filter stage of $T_\text{S} = 10.5 \, \%$ and $T_\text{AS} = 9.4\,\%$, for detunings $\Delta_\text{S} = 15.2\GHz$ and $\Delta_\text{AS}=24.4\GHz$, respectively, are taken into account\footnote{
	The transmissions $T$ are the total optical intensity transmissions from the {\cs} cell input to the APD 
	input, under active optical pumping with the diode laser.
}. 
This procedure relies firstly on having the same ND-filter attenuation in both measurements\footnote{
	Experimentally, the ND-filter is a stack of reflective ND-filter plates of smaller attenuation. 
	It has been positioned flush to an alignment mount to prevent changes in the rotational 
	degree of freedom; its height has been fixed as well. 
	The signal filter stage has been realigned before both measurements and its transmission 
	has been checked after each measurement. Notably, equal transmissions have been obtained for 
	the FSR$=18.4\GHz$ etalons, when their resonances have been aligned to the 
	S ($\Delta_\text{S} = 15.2\GHz$) and 
	the AS ($\Delta_\text{AS} = 24.4\GHz$) channel. 
}, and, secondly, on the absence of filter stage drift. 
Both are roughly fulfilled, however the correspondence between the seeding input photon number {\Nin} at the S frequency and the observed AS photon numbers is only approximate. 

\begin{figure}[h!]
\centering
\includegraphics[width=0.9\textwidth]{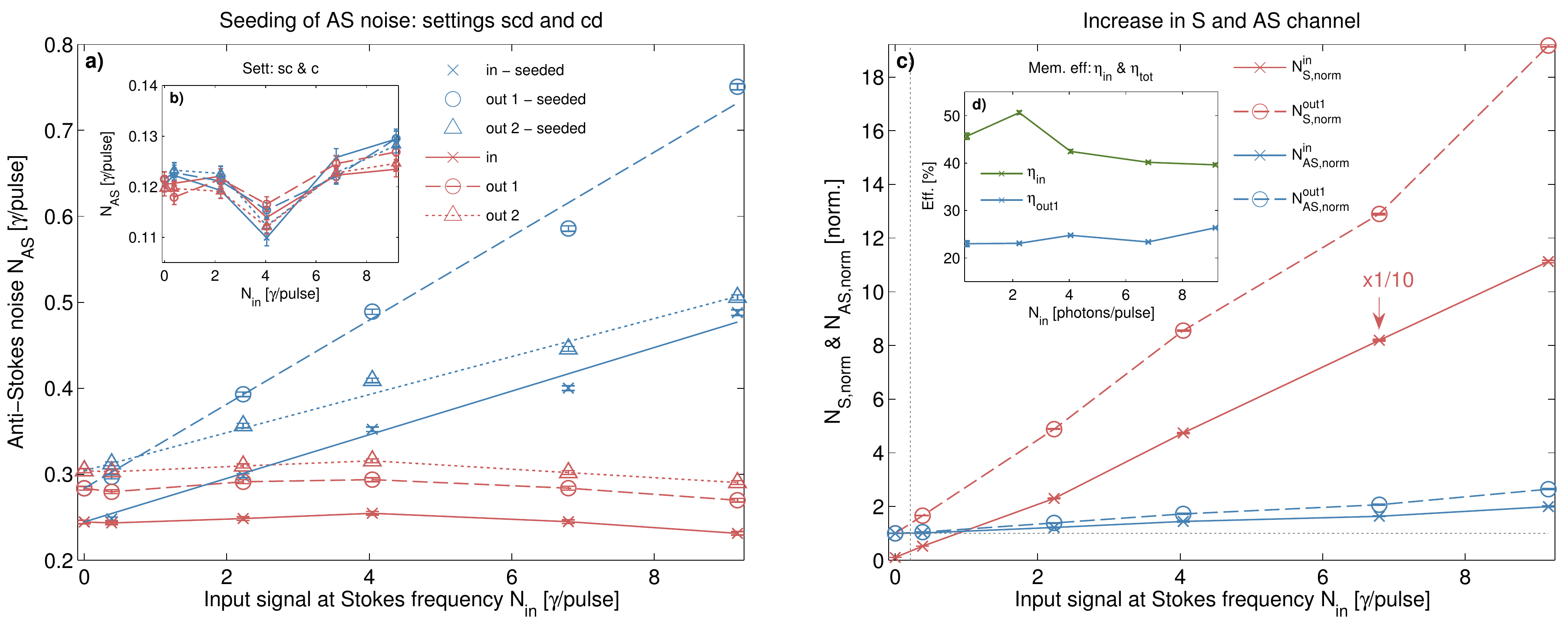}
\caption{AS noise a function of the input photon number $N_\text{in}$ of signal at the S frequency in the read-in time bin. $\times$ and \textit{solid lines} mark the read-in time bin, $\circ$ and \textit{dashed lines} the $1^\text{st}$ read-out bin ($\tau_\text{S} = 312\ns$), and $\bigtriangleup$ and \textit{dotted lines} the $2^\text{ns}$ read-out bin ($\tau_\text{S} = 324.5\ns$). 
\textbf{(a)}: Data with {\cs} ensemble preparation in F$=4$ (diode laser on). 
\textit{Blue lines} show the AS noise when input signal pulses are sent into the {\cs} cell in the S channel (setting \textit{scd}); 
\textit{red lines} show the case without S signal input (setting \textit{cd}), which yields the constant AS noise background. \textit{Blue straight lines} are a fit onto the \textit{scd} data.
\textbf{(b)}: Data with thermally distributed population (diode laser off), with same colour coding as in \textbf{(a)}; setting \textit{sc} in \textit{blue} and \textit{c} in \textit{red}.
\textbf{(c)}: Comparison between the normalised increase of seeded AS noise (setting \text{scd}, \textit{blue}) from panel \textbf{(a)} and the increase of the signal read into (\textit{solid line}) or retrieved from (\textit{dashed line}) the memory at the S frequency (\textit{red}). The \textit{horizontal dashed line} marks the unseeded background noise in the S and AS channels, while the \textit{vertical line} indicates the value for {\Nin} for \hsps\,, as used in chapter~\ref{ch6}.
\textbf{(d)}: Memory read-in $\eta_\text{in}$ and read-out efficiency $\eta_\text{mem} = \eta_\text{out1}$ for $\tau_\text{S} = 312\ns$, during the measurement.
}
\label{fig_ch7_ASseeding}
\end{figure}

The memory is operated with a storage time of $\tau_\text{S} = 312\,\text{ns}$ and $\tau_\text{S} = 324.5\,\text{ns}$  for the $1^\text{st}$ and $2^\text{nd}$ retrieval time bins, respectively, at $f_\text{rep} = 4\kHz$ repetition rate.  
We evaluate the number of detected AS noise photons in all three time bins, read-in and read-out, as a function of the number $N_\text{in}$ of signal photons sent into the memory in the input bin. 
Fig. \ref{fig_ch7_ASseeding} \textbf{a} \& \textbf{b} show the data for settings $\left\{scd, cd\right\}$ and $\left\{sc,c \right\}$, respectively.
In the case of \textit{scd} and \textit{sc} data, any residual signal leakage from \textit{sd} or \textit{s} has been subtracted. 

\paragraph{Results for the amount of seeded noise}

In the spin-polarised ensemble (fig. \ref{fig_ch7_ASseeding} \textbf{a}), we can clearly observe an increase in the AS noise $N_{\text{AS},scd}^t$ over the background level\footnote{
	The variation in the AS background noise level for setting \cd\, is 
	due to apparatus drift over the measurement time, which was $\sim 7\,\text{h}$.
} (setting \cd\,) for all three time bins $t$, when seeding the noise with an input signal pulses (setting \textit{scd}). 
The data shows a linear proportionality $N^t_{\text{AS},scd}(N_\text{in}) \sim \alpha_t \cdot N_\text{in}$ for all three time bins $t$. 
Fitting the data yields the largest gradient $\alpha_t$ for the read-out time bin 1, with $\alpha_{\text{out}1} \approx 1.93 \cdot \alpha_\text{in}$, where $\alpha_\text{in}$ is the gradient in the input bin. 
Noise in $2^\text{nd}$ read-out time bin still increases with a rate $\alpha_{\text{out}2} \approx 0.9 \cdot \alpha_\text{in}$. 
As expected from section \ref{ch7_Fluorescence}, seeded FWM noise, emitted into the S channel, firstly elevates AS emission in the same time bin, but, even more noticeably, also boosts the noise level in the subsequent time bin.
Signal storage leads to greater AS noise in read-out bin 1, and, in turn, also to more S noise during memory retrieval\footnote{
	Note: AS emission is the prerequisite for S noise emission, 
	which are both coupled through their respective coupling constants 
	$C_\text{AS} = \frac{\Delta_\text{S}}{\Delta_\text{AS}} C_\text{S}$ and Greens function 
	$\mathbb{G_\text{A,S}}$ 
	(appendix \ref{app6_coh_model}). 
	An AS noise increase will consequently also elevate the S noise level.
}. 
The calculated values for the SNR and {\etamem} therefore indeed contain a fraction of noise, falsely attributed to the retrieved signal. 
Indicative for this admixture of extra noise are also the behaviours of the memory read-in ({\etain}) and read-out efficiencies ({\etamem}) as a function of {\Nin}, displayed in fig. \ref{fig_ch7_ASseeding}~\textbf{d}.
 While {\etain} is decreasing slightly for larger {\Nin}, {\etamem} increases instead. 
 Both changes are expected by adding noise to the count rates for setting \scd\, (see eqs. \ref{eq6_memeff_in} \& \ref{eq6_memeff_ret}).

In contrast to these findings, no count rate change  between settings \textit{sc} and \textit{c} can be observed for the unpumped {\cs} ensemble with equal ground state populations (fig. \ref{fig_ch7_ASseeding} \textbf{b}).
Such behaviour is to be expected if the process leading to noise in each channel is SRS. 
Since SRS from the F$=3$ and the F$=4$ level are independent processes, stimulating (seeding) one should leave the other unaffected.
Moreover, since neither time bin shows any effect, there is no amplification of subsequent pulses, so there cannot be any spin-wave dynamics.

We can use the seeding measurement to test how severely the additional noise affects the memory read-out signal by 
comparing the count rates for \scd\, between both channels, as shown in fig.~\ref{fig_ch7_ASseeding}~\textbf{c}. 
To determine the contribution size, we assume that every emission event for AS noise will also lead to the emission of a S noise photon. 
So the S noise level is set equal to the AS level, and the FWM spin-wave retrieval efficiency is assumed $100\,\%$. 
Since FWM noise has not yet saturated after $2$ control pulses (see fig. \ref{fig_ch7_flourescence}), this is an overestimation, yielding an upper bound for the amount of seeded noise, added to the read-out signal. 
This amount can be determined by comparing the increase in $N_{\text{AS},scd}^{t}(N_\text{in})$ to the increase in the signal level $N_{\text{S},scd}^{t}(N_\text{in})$, measured at the S frequency. 
The latter corresponds to the count rates for the memory interaction setting \scd\,, when the signal filter stage is resonant with the S channel. 
To make the photon numbers comparable between both channels, they are normalised to 
the background noise level at $N_\text{in}=0$, observed for setting \cd\,: 
$N_{\text{AS,norm}}^{t}(N_\text{in}) =\frac{N_{\text{AS},scd}^{t}(N_\text{in})}{N_{\text{AS},cd}^{t}}$ and
$N_{\text{S,norm}}^{t}(N_\text{in}) =\frac{N_{\text{S},scd}^{t}(N_\text{in})}{N_{\text{S},cd}^{t}}$. 

Fig. \ref{fig_ch7_ASseeding} \textbf{c} shows these relative increases between memory signal and seeded noise. 
For the largest \coh\, input photon number used in chapter \ref{ch6}, $N_\text{in} \approx 2.2 \ppp$, seeding increases the noise in the $1^\text{st}$ read-out bin by a factor of $\sim 1.4$ over the unseeded noise level, while the memory signal is $\sim 4.9$ times larger. 
Therewith $\Delta \eta_\text{mem}^\text{seed} \sim 9\,\%$ of the memory signal can be estimated to arise from additional, 
seeded noise (see appendix \ref{app_ch7_subsec_seedingmemeff}). 
At low signal intensities with $N_\text{in} \approx 0.38 \ppp$, a value close to \hereff\, (\textit{vertical dashed line} in fig. \ref{fig_ch7_ASseeding} \textbf{c}), 
$\Delta \eta_\text{mem}^\text{seed} \sim 7\,\%$ of the observed efficiency are estimated to originate from seeding\footnote{
	The contribution from seeded FWM noise to the retrieved signal is thus 
	smaller than the $15\,\%$ worst case estimation obtained in 
	appendix \ref{app_ch7_subsec_ASseeding} for a pure two-mode squeezing system. 
}.
Importantly, when taking into account the ratio of the Raman coupling constants 
$R^2_\text{S/AS} = \left( \frac{C_\text{S}}{C_\text{AS}} \right)^2 \approx 2.58$ between the S and the AS channel to estimate the contribution of seeded noise, these numbers reduce further to 
$\Delta \eta_\text{mem}^\text{seed} \sim \frac{9\,\%}{2.58} \approx 3.5\,\%$ for 
$N_\text{in} \approx 2.2\ppp$, and 
$\Delta \eta_\text{mem}^\text{seed} \sim \frac{7\,\%}{2.58} \approx 2.7\,\%$ for 
$N_\text{in} \approx 0.38 \ppp$, which are on the order of the measurement uncertainty for $\eta_\text{mem}$. 
In conclusion, we can verify that additional noise from FWM seeding by the memory input signal is negligible at the single photon level.

\section{Magnetic dephasing \label{ch7_subsec_mag_dephasing}}
In our discussion of the FWM noise process so far, we implicitly assumed the excitation of a spin-wave coherence between the two FWM steps, as illustrated in fig. \ref{fig_ch7_FWMlevels_2}. Conversely, for the SRS noise, emitted by the thermally distributed {\cs} ensemble, we relied on the absence of such spin-wave excitations to explain the scaling of the count rates we have seen in fig. \ref{fig_ch7_flourescence}. 
The next point for us to investigate is thus the existence of the FWM spin-wave, when dealing with the spin-polarised atomic ensemble. 
One possibility to do this is the study of the dephasing properties of the noise and the Raman memory upon application of a DC magnetic field ($B$). 
Therewith we show that FWM and the Raman memory couple to the same spin-wave. 
We firstly demonstrate similar scaling for both processes upon modification of the spin-wave amplitude through dephasing and, secondly, observe spin-echoes in the noise and the memory efficiency from magnetic revival. 
While the external $B$-fields can be expected to heavily affect the Raman memory and the FWM noise, they should not influence SRS. Magnetic dephasing measurements are thus another means to distinguish the two-photon transition noise processes. 
Therefore we can also use them to confirm our present categorisation of the noise floor constituents. 
We go through the measurement by first outlining how the $B$-field affects the spin-wave, then we describe the measurement procedure and afterwards discuss the results.

\paragraph{Magnetic field effects on the spin-wave}

Since our memory protocol is not Zeeman-state selective, the spin-wave will include all magnetic sub-levels $m_F$ of the F$=3$ and F$=4$ ground states. 
This can be seen by considering the spin-wave as a Dicke state\cite{Dicke:1954fk, Fleischhauer:2000vn}: its state vector is a coherent sum of all permutations of atoms, excited to the $\text{F}=3$ state, over all remaining atoms in the $\text{F}=4$ state. 
Within the framework of our theory model (appendix \ref{app6_coh_model}), this state is generated by application of the operator 
$\hat{S} = \alpha \cdot \hat{\mathds{1}} + \beta \cdot \hat{ \Sigma}$ to the initial atomic state, which has all atoms in $\text{F}=3$ (see \textit{Reim et. al.}\cite{Reim:2011ys}). 
The constants $\alpha$ and $\beta$ define the spin-wave amplitude, and 
$\hat{\Sigma }=\overset{F_i}{\underset{m_i= -F_i}{ \sum}} \left( \overset{F_f}{\underset{m_f = -F_f}{ \sum}} C(m_i,m_f) \cdot \ket{ F_i,m_i} \bra{F_f,m_f} \right)$ 
is the transition operator\footnote{
	In fact, this is the annihilation operator of the spin-wave Dicke state.
}, 
whose Raman coupling coefficients $C(m_i,m_f)$ between the initial ($\ket{F_i,m_i}$) and final ($\ket{F_f,m_f}$) Zeeman levels depend on the Clebsh-Gordan coefficients.

Application of a $B$-field results in different Lamour precession frequencies around the magnetic field lines, depending on the magnetic quantisation number $m_F$.
It leads to different spin orientations and $m_F$-number dependent phase factors for each spin-wave term. In turn, different phases spoil the constructive interference between the terms in $\Sigma$ upon spin-wave recall, which reduces the memory efficiency. This has been shown previously by a model for bright signal retrieval from the Raman memory\cite{Reim:2011ys, Reim2011_supp}. 
With FWM coupling to the same spin-wave mode, $B$-field application will affect the noise level equally.

\paragraph{Magnetic fields applied to the memory medium}

\begin{figure}[h!]
\includegraphics[width=\textwidth]{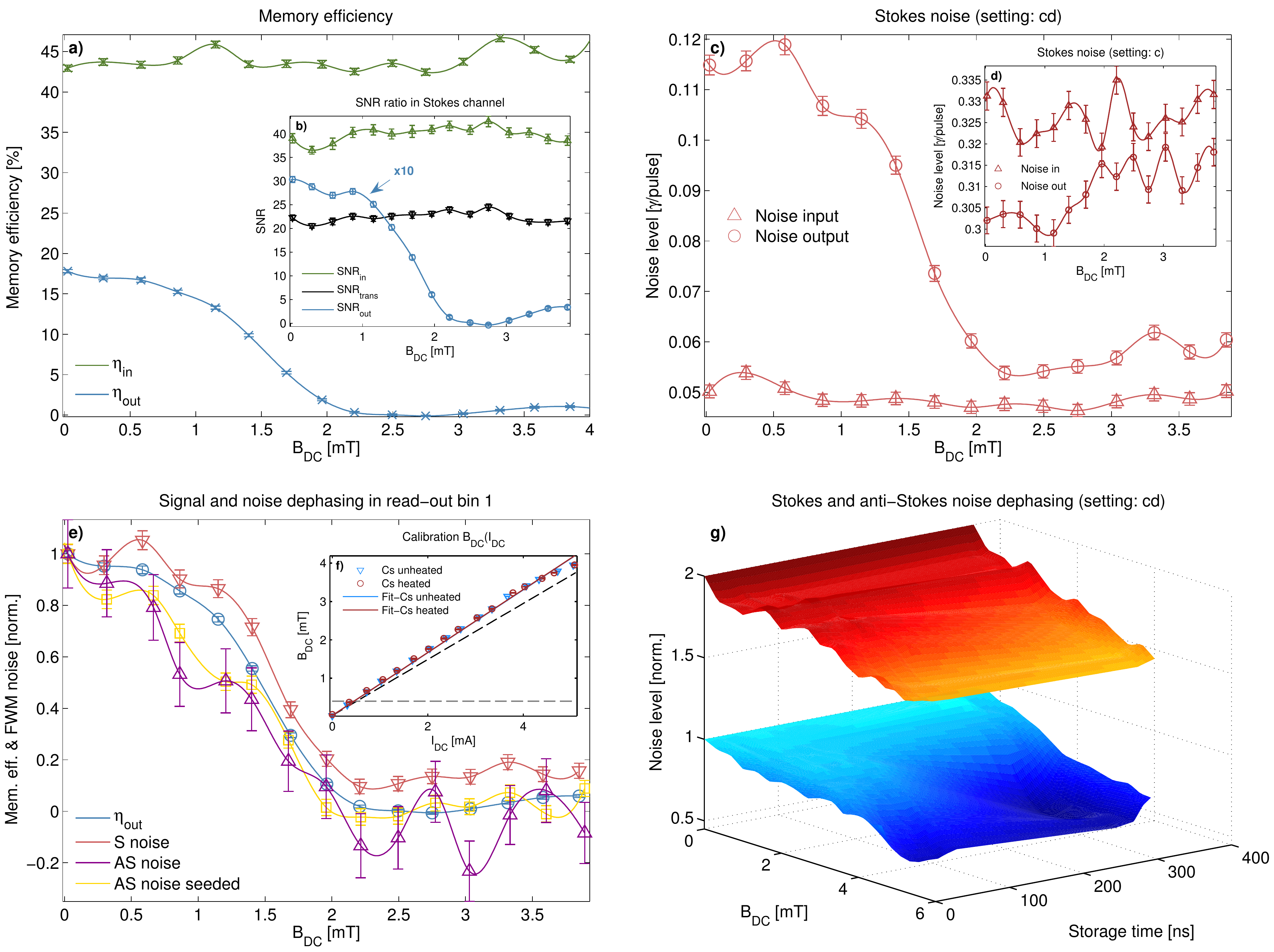}
\caption{Magnetic dephasing of memory efficiency and noise. 
\textbf{(a)}: Memory read-in efficiency {\etain} (\textit{green}) and total read-out efficiency {\etamem} (\textit{blue}).
\textbf{(b)}: Signal-to-noise ratios for the input signal, SNR$_\text{in}$ (\textit{green}), the transmitted signal in the read-in bin, SNR$_\text{trans}$ (\textit{black}), and the signal, retrieved after $\tau_\text{S} = 312\ns$ storage time, SNR$_\text{out}$ (\textit{blue}). 
Note, the SNRs are higher than in section \ref{ch6_subsec_SNR} due to a stronger input signal of $N_\text{in} \sim 2\ppp$. 
\textbf{(c)}: Stokes channel noise level for the pumped ensemble (setting \cd\,) for the input bin ($\bigtriangleup$) and the $\tau_\text{S} = 312\ns$ read-out bin ($\circ$).
\textbf{(d)}: Stokes channel noise level for the unpumped ensemble (setting \textit{c}); markers as in \textbf{(c)}.
\textbf{(e)}: Normalised efficiency {\etamem} (\textit{blue}), Stokes noise $N^\text{out,norm}_{\text{S},cd}$ (\textit{red}), unseeded AS noise $N^\text{out,norm}_{\text{AS},cd}$ (\textit{magenta}) and seeded AS noise $N^\text{out,norm}_{\text{AS},scd}$ (\textit{yellow}) with optical pumping.
\textbf{(g)}: Normalised S and AS noise levels $N^\text{out,norm}_{\text{S},cd}(B,\tau_\text{S})$ (\textit{blue}) and $N^\text{out,norm}_{\text{AS},cd}(B,\tau_\text{S})$ (\textit{red}), respectively, with optical pumping.
\textbf{(f)}: Magnetic field calibration curve for three measurements with the {\cs} cell cold (\textit{blue}) and warm (\textit{red}). The \textit{solid red line} is a linear fit to the data, while the \textit{dashed black line} represents $B(I_\text{DC})$ for an empty solenoid. The \textit{dashed horizontal line} is the remanent B-field.
}
\label{fig_ch7_mag_dephasing}
\end{figure}

Experimental magnetic field generation is simple, thanks to the degaussing coils, wrapped around the {\cs} cell (see fig. \ref{fig_ch4_int_instab_setup}). 
With the AC/DC-power supply, usually employed for degaussing before memory experiments, a DC-current of up to $I_\text{DC,max} = 5.5 \,\text{A}$ can be sent through the coils.
The coils have ${N_\text{coil} = 128}$ turns over a length of ${L_\text{coil} = 22\cm}$, from which a  maximum magnetic flux density of 
$B(I_\text{DC,max}) = \mu_0 \cdot \frac{N_\text{coil}}{L_\text{coil}} \cdot I_\text{DC,max} = 4\,\text{mT}$ 
can be expected. 
Inside the {\cs} cell the $B$-field coincides with the optical propagation axis 
(see figs. \ref{fig_ch4_int_instab_setup} \& \ref{fig_6_setup}). 
To determine the actually applied $B$-field, a Hall probe is placed in front of the {\cs} cell entrance window to record an $I_\text{DC}$ - $B$-calibration curve, shown in fig. \ref{fig_ch7_mag_dephasing} \textbf{f}. 
Since the solenoid is not empty, some ferromagnetic responses is observed and the 
$B$-field values lie above the (empty solenoid) $B(I_\text{DC})$-line. 
The magnetic permeability also leads to a tail-off for large $I_\text{DC} \approx 5\,\text{A}$ and to a remanent field of $B_\text{rem} \approx 0.4 \,\text{mT}$. 
Note, whenever a $B$-field is applied, it lies on the initial magnetisation curve, shown in fig. \ref{fig_ch7_mag_dephasing} \textbf{f}.

\paragraph{Magnetic dephasing measurement procedure}

We record the magnetic dephasing of the memory efficiency and the noise in the S and AS channel, running experiments at $f_\text{rep} = 5.722 \kHz$. 
Besides the unconditional background noise floor, we also investigate the dephasing of AS noise generated by seeding, 
since seeded noise should purely result from spin-wave coupling.
Seeded AS noise is produced by the same procedure as in section \ref{ch7_subsec_ASseeding}, i.e. 
the input signal at the S frequency is sent into the {\cs} cell in the read-in time bin. It boosts the AS noise level in the read-out time bin, whose count rate decrease is observed as a function of the $B$-field. 
Dephasing measurements are conducted with spin-polarised (settings \cd\,) and thermally distributed {\cs} (setting \textit{c}). 
For the measurements on the S channel, the input signal is a \coh\, state with $N_\text{in} =  1.95 \ppp$. 
Since this number is about $10$ times the \hsp\, input, with $N_\text{in} \approx 0.22\ppp$, we expect the SNRs to be order of magnitude higher than the values quoted in section \ref{ch6_subsec_SNR}. 
Using the input photon number calibration method of section \ref{ch7_subsec_ASseeding}, the input signal for the AS seeding measurements is estimated to $N_\text{in} \approx 2 \ppp$.

All measurements store the spin-wave for $\tau_\text{S} = 312\ns$, using the same control pulse sequence\footnote{
	1 pulse for read-in and 2 retrieval pulses; for brevity, only the results 
	for the $1^\text{st}$ read-out pulse are presented.
} as depicted in the insets of fig. \ref{fig_ch7_flourescence}. 
A longer storage time of $\tau_\text{S}=312\ns$ is chosen to minimise effects from fluorescence noise build-up. It also allows for sufficient time for the spin-wave to dephase, despite the modest magnetic field strength reachable with the degaussing solenoid\footnote{
	The additional lifetime dephasing (discussed in detail in appendix \ref{ch7_lifetime}), 
	leads to lower efficiency and noise values than obtained for $\tau_\text{S} = 12.5\ns$, even when $B=0$.
}. 
Having a storage time of $\tau_\text{S}=312\ns$, where the second \pockels\, pulse picking window generates the retrieval control pulses, additionally enables us to study the relationship between magnetic dephasing and storage time. To this end, the first pulse picking window is opened completely to select 9 consecutive control pulses. 
This pulse train generates noise pulses with effective storage times\footnote{
	There are $9$ pulses in the first pulse picking window, generating a pulse train spaced by 
	$12.5\ns$ with storage times from $\tau_\text{S}= \left[ 0\ns, 100\ns \right]$. 
	Additionally the $2^\text{nd}$ \pockels\, window 
	contains pulses at $\tau_\text{S}= \left[ 338\ns, 350\ns\right]$. The shift $312\ns \rightarrow 338\ns$
	is required to avoid interference effects between the voltage pulses of the \pockels\, windows.
} from $\tau_\text{S} = 0\ns$ to $\tau_\text{S} = 350\ns$, whose magnetic dephasing properties can be studied simultaneously.

\paragraph{Results for the spin-wave dependent processes}

The memory efficiency, displayed in fig.~\ref{fig_ch7_mag_dephasing} \textbf{a}, shows constant read-in efficiency {\etain} and decreasing read-out efficiency {\etamem}, as expected for magnetic dephasing. 
Information read-in does not involve interaction with a pre-existing spin-wave and is hence unaffected by the $B$-field. Conversely, the retrieval should have a Gaussian-shaped decay\cite{Reim2011_supp}, as observed in the data\footnote{
	This means: $\eta_\text{mem} \sim \exp{\left\{ -B^2 \right\} }$
}.
For large $B$-fields, around $B_\text{DC} \approx 3.7 \,\text{mT}$, {\etamem} also shows a small amount of magnetic revival\cite{Matsukevich:2006}. 
Similarly, the noise for the spin-polarised atomic ensemble in the S channel (setting \cd\,, fig. \ref{fig_ch7_mag_dephasing} \textbf{c}) is unaffected by the $B$-field in the read-in time bin. 
However, in the read-out bin, it also decays with a Gaussian shape, just like {\etamem}, where it converges against the read-in bin noise level for large $B$-fields.
Both observations are expected for FWM: Noise in the read-in bin does not dephase due to the short pulse durations of $300\ps$. Here, AS spin-wave excitation, followed by retrieval under S noise emission happens too quickly for the $B$-field to cause an effect.
Noise in the subsequent time bin at $\tau_S = 312\ns$ is affected by the magnetic field, since it contains two contributions: 
On the one hand, the amount of noise freshly generated within the read-out bin, and, on the other hand, noise produced by coupling to the spin-wave fraction left over from the read-in time bin. 
It is this latter fraction that can dephase. So the lower bound on the noise level $N^\text{out,1}_{\text{S},cd}(B)$ is set by the noise generated within one control pulse, which in turn equals the read-in bin noise level $N^\text{in}_{\text{S},cd}$. 

Combining memory efficiency and S noise, we calculate the SNR (eq. \ref{eq_ch6_SNR}), plotted in fig. \ref{fig_ch7_mag_dephasing} \textbf{b} for the input, transmission and retrieval from the memory. 
The functional shapes are similar, SNR$_\text{out}$ decays, with some magnetic revival around $B\sim 3.7 \,\text{mT}$, while SNR$_\text{in}$ and SNR$_\text{trans}$ are unaffected by the $B$-field. 
The conservation of the functional form for the dephasing in SNR$_\text{out}$ is indicative for the same scaling of efficiency and noise with the $B$-field.

In order to test this further, and compare it to the results for seeded and unseeded AS noise, data for {\etamem} and all noise datasets for the read-out bin are normalised by their respective values for $B=0$. 
Prior to normalisation, the noise data is background subtracted: 
for unseeded S and AS noise (setting \cd\,), the level in the input bin, 
$N_{{i},cd}^\text{in}$, is subtracted from 
$N_{{i},cd}^\text{out}(B)$, to obtain 
$N^\text{out,norm}_{i,\text{bg}} (B)= \frac{N_{{i},cd}^\text{out}(B) - N_{{i},cd}^\text{in}}{N_{{i},cd}^\text{out}(0)-N_{{i},cd}^\text{in}}$, with $i \in \left\{ \text{S}, \text{AS} \right\}$. 
For the seeded AS signal (setting \scd\,), the unseeded portion in the read-out bin is subtracted instead to give 
{$N^\text{out,norm}_{\text{AS},\text{seed}}(B) = \frac{N_{{\text{AS}},scd}^\text{out}(B) - N_{\text{AS},cd}^\text{out}(B)}{N_{\text{AS},scd}^\text{out}(0)-N_{\text{AS},cd}^\text{out}(0)}$}.
This makes the functional dependence on $B$ comparable across all observables. 
Plotting these normalised data in fig. \ref{fig_ch7_mag_dephasing} \textbf{e} verifies that {\etamem} (\textit{blue}), S noise (\textit{red}), as well as seeded (\textit{yellow}) and unseeded (\textit{magenta}) AS noise all show the same dependence on $B$. 
Magnetic spin-wave dephasing consequently influences the relative levels of each observable equally. 
Such behaviour is expected, if all generating processes, underlying each one of these signals, couple to the same spin-wave. 

\paragraph{Results for the spin-wave independent process} 
The thermally distributed ensemble (setting \textit{c}), whose results are depicted in fig. \ref{fig_ch7_mag_dephasing} \textbf{d} for the S channel, presents an entirely different picture. 
Here, the noise does not decrease in the retrieval bin. 
In fact, there is no obvious deterministic dependence on $B$. Analogue behaviour is observed in the AS channel. 
Such independence is expected for SRS, whose emission intensity only depends on the current, incoherent population in each of the {\cs} ground states at the time of scattering by the control. 
Contrary to FWM, its emission does not depend on the amount of spin-wave coherence previously generated, 
so any dephasing is irrelevant.   
All of these observations for both ensemble preparations tie in nicely with the previous conclusion that FWM is the noise process for spin-polarised {\cs}, whereas SRS occurs without atomic state preparation. 

Lastly, we look at the storage time ($\tau_\text{S}$) dependence of the magnetic noise dephasing for the pumped atomic ensemble (setting \cd\,), for which the above mentioned control pulse train is used. 
Each noise pulse for a particular value of $\tau_\text{S}$ is background subtracted and normalised to yield noise levels $N^{\text{norm}}_{i,\text{bg}}(B,\tau_\text{S})$. 
The noise dependence on both parameters $B$ and $\tau_\text{S}$ is illustrated in fig. \ref{fig_ch7_mag_dephasing} \textbf{g}, whereby the AS noise is offset vertically\footnote{
	For the ease of inspection, fig. \ref{fig_ch7_mag_dephasing} \textbf{g} plots the normalised 
	AS noise by adding $1$, so the graph shows 
	$N^{\text{norm}}_{\text{AS},\text{bg}}(B,\tau_\text{S}) +1 $.
}. 
The 3-dimensional planes show that the magnetic field is actually strong enough to reduce noise pulses within the first \pockels\, pulse picking window as well ($\tau_\text{S} \approx 100\ns$).
Noise on both channels reduces for an increase in either of the variables $B$ and $\tau_\text{S}$. 
AS and S noise approximately show a similar dependence on both, illustrating how decoherence over the memory lifetime affects both channels equally.

\section{Theory model noise level predictions\label{ch7_subsec_theory_prediction}}

\begin{figure}[h!]
\includegraphics[width=\textwidth]{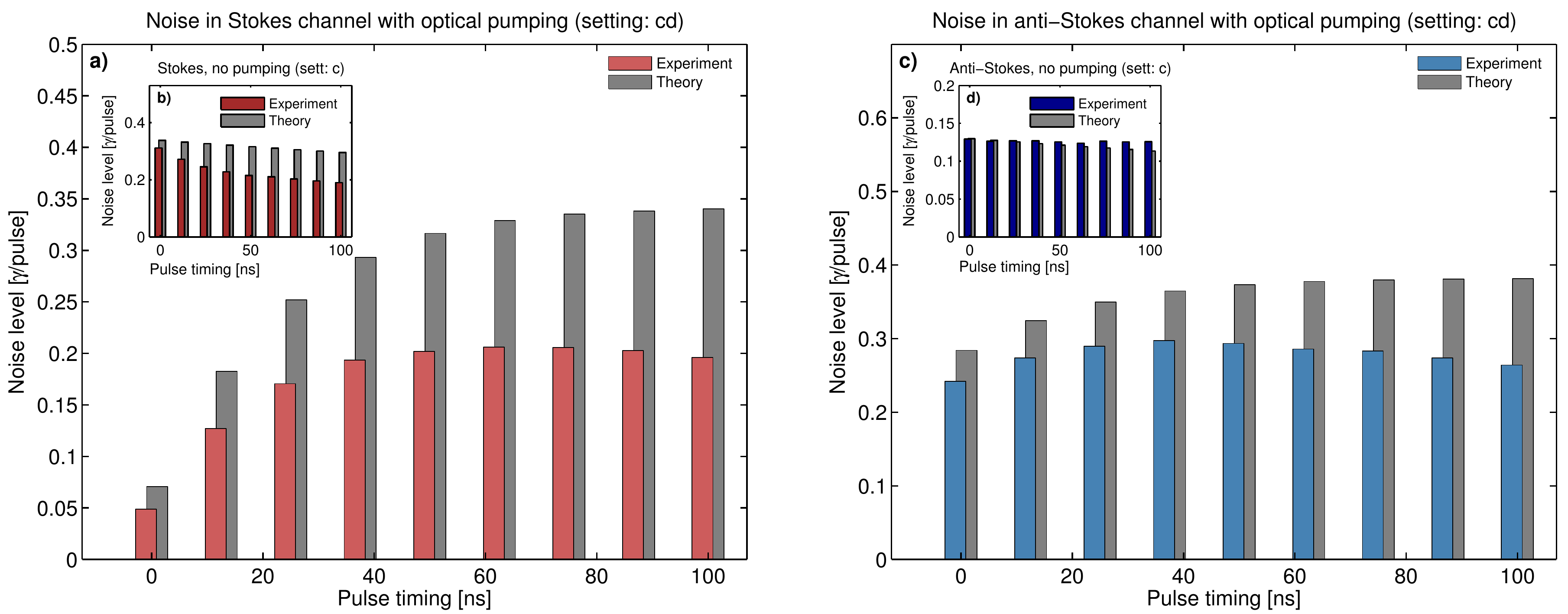}
\caption{Memory noise levels predicted by our theoretical model (\textit{grey bars}) in comparison with the measured values (\textit{coloured bars}). All data are absolute numbers. 
\textbf{(a)} \& \textbf{(b)} show the S channel, \textbf{(c)} \& \textbf{(d)} the AS channel. 
\textbf{(a)} \& \textbf{(c)} contain data for the spin-polarised {\cs} ensemble (setting \textit{cd}), whereas 
\textbf{(b)} \& \textbf{(d)} feature data for the thermally distributed {\cs} population (setting \textit{c}).}
\label{fig_ch7_theory_prediction}
\end{figure}

Having determined the different noise sources and the underlying noise processes experimentally, 
we are now in a position to test these results against the predictions of our theory model, introduced in section \ref{ch2_Raman_noise} \& appendix \ref{app6_coh_model}. 
Our main aim is to interrogate, whether the noise numbers of fig. \ref{fig_ch7_flourescence} tie in with the expectation for the respective noise processes that we have associated with the different experimental settings $i$. 
To this end, we compare the experimental data of fig. \ref{fig_ch7_flourescence} with the theory expectation for the noise photon number, assuming a control pulse train of $9$ consecutive control pulses, each separated by a storage time 
$\tau_\text{S} =12.5\ns$. 
Likewise to the experiment, we switch between FWM and SRS as the dominant noise process using the ground state population inversion $w$, which enters the underlying eqs. \ref{eq_ch2_Maxwell_Bloch_FWM_2}. 
As explained below, to model FWM noise, we set $w=1$. 
This corresponds to {\cs} preparation in $6^2 \text{S}_{\frac{1}{2}}$ F$=4$, while, for SRS, we have $w=0$, which represents the thermally distributed ensemble.
For comparison with the experiment, we also need the measured noise photon numbers $N_i^t$, in terms of photons per pulse, of the two-photon transition components for both setting $i$. 
As usual $t$ counts the control pulse time bins, such that the actual pulse timing $\tau = t \cdot \tau_\text{S}$, shown in fig.~\ref{fig_ch7_flourescence}, corresponds to integer multiples of the storage time $\tau_\text{S}$. 
Using eq. \ref{ch6_eq_Nin}, these numbers are given by 
$N^t_{i} = \frac{\tilde{a}_i^t}{\Delta t_\text{meas} \cdot f_\text{rep} \cdot T_\text{sig} \cdot \eta_\text{det}}$, 
with the integrated pulse areas $\tilde{a}_i^t$, introduced in section \ref{ch7_pol_turn_off_principles}, the total data acquisition time $\Delta t_\text{meas}$, the repetition rate $f_\text{rep}$, the signal filter transmission $T_\text{sig} \approx 9\,\%$ and the APD efficiency $\eta_\text{det}\approx 50\,\%$. 
The resulting experimental noise levels are illustrated by the \textit{coloured bars} in fig. \ref{fig_ch7_theory_prediction}.

\paragraph{Noise photon number prediction}
Using eqs. \ref{eq_app_ch6_Aout1} - \ref{eq_app_ch6_Bout1} of our theory model\footnote{
	The equations read  
	\begin{align}
	\vec{\hat{A}}_\text{out,1} 	&= 	\mathbb{G}_{A,A} \cdot \vec{\hat{A}}_\text{in,1} +\mathbb{G}_{A,B} \cdot 
	\vec{\hat{B}}_\text{in,1}^	\dagger +\mathbb{G}_{A,S} \cdot \vec{\hat{S}}_\text{in,1}^\dagger 
	\nonumber \\
	\vec{\hat{S}}_\text{out,1} 	&= 	\mathbb{G}_{S,A} \cdot \vec{\hat{A}}_\text{in,1}^\dagger +\mathbb{G}_{S,B} 
	\cdot \vec{\hat{B}}_\text{in,1} +\mathbb{G}_{S,S} \cdot \vec{\hat{S}}_\text{in,1}	
	\nonumber \\
	\vec{\hat{B}}_\text{out,1} 	&= 	\mathbb{G}_{B,A} \cdot \vec{\hat{A}}_\text{in,1}^\dagger +\mathbb{G}_{B,B} 
	\cdot \vec{\hat{B}}_\text{in,1} +\mathbb{G}_{B,S} \cdot \vec{\hat{S}}_\text{in,1},
	\end{align}
	see appendix \ref{app6_coh_model} for a definition of all variables. 
	The calculations have been performed by Joshua Nunn. 
	Since they are not part of my personal work, only a brief outline of the methodology is provided
	here.
} in appendix~\ref{app6_coh_model}, we calculate the expected noise level for each of the $9$ control pulses, contingent on the state, the system has been left in, after the previous control pulse. 
Using the notation and results of appendix \ref{app6_coh_model}, noise levels are obtained by the expectation values of the photon number operators 
$\langle N_\text{S}^t \rangle = \langle \hat{S}^\dagger_t \hat{S}_t \rangle$ 
and 
$\langle N_\text{A}^t \rangle = \langle \hat{A}^\dagger_t \hat{A}_t \rangle$ 
for S and AS photons, whereby $\hat{S}$ and $\hat{A}$ denote the field annihilation operators for the respective channel. 
These are evaluated, assuming only the vacuum $\ket{\text{vac}} = \ket{0}$ at the input of each channel, i.e., 
$\hat{S}_\text{in,1}=\hat{A}_\text{in,1} = \ket{\text{vac}}$ in eqs.~\ref{eq_app_ch6_Aout1}~-~\ref{eq_app_ch6_Bout1}.
In the set of matrix equations \ref{eq_app_ch6_Aout1} - \ref{eq_app_ch6_Bout1}, noise is generated in two ways. On the one hand there is the newly produced noise from spontaneous scattering within each control pulse, determined by the Greens functions 
$\mathbb{G}_{A,S}$ (eq. \ref{eq_app_ch6_Aout1}) and $\mathbb{G}_{S,A}$ (eq. \ref{eq_app_ch6_Sout1}). 
On the other hand, there is also the coupling to the spin-wave $\hat{B}_{\text{in},t}$ via Greens functions $\mathbb{G}_{A,B}$  and $\mathbb{G}_{S,B}$. 
This input spin-wave is also set to the vacuum state for the $1^\text{st}$ time bin ($t=1$). 
However, for a spin-polarised ensemble with a population imbalance $w \neq 0$, $B$ changes within the first control pulse. We can also see this from eqs. \ref{eq_ch2_Maxwell_Bloch_FWM_2} in section \ref{ch7_subsec_FWM}, stating 
$\partial_\omega \hat{B} \sim w \hat{A}^\dagger$, 
which corresponds to the generation of a FWM spin-wave by AS noise emission.
For any later time bins $t>1$, the input spin-wave $\hat{B}_{\text{in},t}$ is determined by the output spin-wave $\hat{B}_{\text{out},t-1}$ (eq. \ref{eq_app_ch6_Bout1}) in the previous time bin (neglecting any spin-wave decay during $\tau_\text{S}$). 
So imperfect spin-wave retrieval within the $1^\text{st}$, and any subsequent, time bin leads to an additional contribution to $\langle N_{\left\{\text{S,A}\right\}}^t \rangle$ from the left-over spin-wave $\hat{B}_{\text{in},t} = \hat{B}_{\text{out},t-1}$.
As a result, $\langle N_{\left\{ \text{S,A} \right\} }^t \rangle $ changes dynamically as a function of $t$, leading to the visible noise increase over successive control pulses in fig. \ref{fig_ch7_theory_prediction}. 
The predictions for the noise levels in each of these pulse time bins are obtained by recursive application of eq. \ref{eq_app6_Aout} and its counterpart for the AS channel, using $\hat{B}_{\text{in},t} = \hat{B}_{\text{out},t-1}$ as the only non-constant input parameter. 

\paragraph{Predicted influence of spin-wave on noise floor} 
Similar to the {\gtwo} prediction, the difference between the thermal (setting \textit{c}) and spin-polarised ensemble (setting \cd\,) lies in the population inversion $w$. 
Assuming perfect optical pumping for the calculation, we have set $w_c = 0$ for the former and $w_{cd}=1$ for the latter case to obtain the theory data in fig. \ref{fig_ch7_theory_prediction}. 
Since $\hat{B}_{\text{in},1}  = \ket{\text{vac}}$ for the $1^\text{st}$ control pulse ($t=1$), any non-trivial spin-wave dynamics necessitates $w\neq 0$, i.e. some population imbalance between the {\cs} ground states, as can bee seen\footnote{
	The equation reads: $[\partial_\omega + \mi  \textfrak{S}] \hat{B} = 
	\mi w[C_\text{S} \hat{S} +  C_\text{AS} \hat{A}^\dagger]$
} 
in eqs.~\ref{eq_ch2_Maxwell_Bloch_FWM_2}, where $w$ connects the spin-wave dynamics with the optical input fields in the S and AS channel.
In contrast to the spin-polarised {\cs} ensemble, all spin-wave dynamics are frozen for thermally distributed population, i.e. $\partial_\omega \hat{B} = 0 \rightarrow {\hat{B}(\omega) = \text{const.} = \hat{B}_\text{in,1}}$. 
With $\langle \hat{B}_\text{in,1}\rangle \rightarrow 0$, we expect no spin-wave dynamics for setting \textit{c} and hence no spin-wave related noise floor alteration, for instance caused by dephasing. 
Moreover, eqs. \ref{eq_ch2_Maxwell_Bloch_FWM_2} give 
${\partial_z \hat{S} = \text{const.}}$ and ${\partial_z \hat{A}^\dagger = \text{const.}}$, 
implying that the noise floor $N^t_{i,c}$ should be approximately time bin independent. 
This explains the radically different noise behaviour between the two experimental configurations we have seen so far. 
In terms of the underlying noise process, fig. \ref{fig_ch7_FWMlevels_2} illustrates that without a pre-existing spin-wave, each FWM step just corresponds to SRS into the respective noise channel. FWM in setting \cd\, thus reduces to SRS for setting \textit{c}.

\paragraph{Comparison of the model prediction to the measurement}
The noise prediction is plotted in fig. \ref{fig_ch7_theory_prediction} by \textit{grey bars}. 
Likewise to the $g^{(2)}$ predictions in fig. \ref{fig_ch6_g2results}, it resembles the measurement astonishingly well, not only in describing the functional behaviour over time, but also predicting the exact noise numbers. 
In light of the residual deviations with the experiment, it is important to emphasis again, that this is prediction of absolute numbers, obtained by a fundamental set of equations without any fitting parameters. 
Similarly, the experimental data are also absolute numbers. 
Without taking ratios to eliminate efficiency factors, the numbers for $N^t_{i,k}$ contain, for instance, $\eta_\text{det}$, which is only an estimate based on the manufacturer's specifications. 
Leaving $\eta_\text{det}$ as a free fitting parameter to optimised the measured data with respect to the theory prediction yields an even better resemblance, shown in fig. \ref{fig_app_ch7_theory_prediction_deteff_fit} of appendix \ref{app_ch7_det_eff_fit}. 
The resulting fitted value of $\eta_\text{det} = 37.6\,\%$ is also a reasonable detector efficiency estimate\footnote{
			We have not measured the detection efficiency directly and assume, 
			for historic reasons, the specified value of $\eta_\text{det}\approx 50\,\%$ 
			throughout this work.
}. 
Importantly, these theory predictions also motivate our implicit assumption about the dominance of FWM over SRS in the $1^\text{st}$ control field time bin of the S channel for spin-polarised {\cs}. 
This cannot be proven experimentally, because, to distinguish both, we have to look at spin-wave dephasing. 
As with our magnetic field measurements, such experiments will however only show that any noise in excess of the $N^\text{t=1}_{\text{S},cd}$ level originates from spin-wave coherence and thus must be FWM. 
Yet, the same is not possible for $N^\text{t=1}_{\text{S},cd}$ itself.   
In this regard, the good agreement between theory and experiment for $\tau =0\ns$ can be regarded as evidence for the predominance of FWM S noise in $N^\text{t=1}_{\text{S},cd}$.

\section{Photon statistics of the noise\label{ch7_g2noise}}

\paragraph{Anti-Stokes noise photon statistics} 
To conclude our noise study, we now take another look at the noise's photon statistics. 
Since the FWM interaction can be envisaged as a two-mode squeezing operation, one would expect each output channel to show thermal noise statistics\cite{Yurke:1987}, resulting in $g^{(2)} \rightarrow 2$. 
However, as we have seen in chapter \ref{ch6}, we measure $g^{(2)}_\text{S,in} = 1.63\pm 0.05$ and $g^{(2)}_\text{S,out} = 1.7 \pm 0.02$ for the S channel in the first $2$ time bins of our control pulse train. 
While our theory model predicts a noise {\gtwo} reasonably close to this value, 
it however does not include any incoherent contributions from fluorescence noise, which we have found to contribute by $\sim 15\,\%$ to the total noise floor (see section \ref{ch7_sec_fluor_results}). 
However, we have also seen that noise is effectively absent in the AS channel. 
So, AS noise should be clean FWM noise with a more representative photon statistics  than the values observed in the S channel. 
Hence we measure the {\gtwo} of the AS noise, using the same experimental setup and procedure as described in chapter \ref{ch6}. 
We again probe the spin-polarised (setting \cd\,) and the thermally distributed (setting \textit{c}) {\cs} ensemble, observing the read-in and the $\tau_\text{S} = 12.5\ns$ read-out time bin. 
Both configurations give rise to {\gtwo}-values of 
\begin{align}
g^{(2)}_{\text{AS},cd,\text{in}} = 1.936 \pm 0.028     \quad & \quad 
g^{(2)}_{\text{AS},cd,\text{out}} = 1.943 \pm 0.024 \nonumber \\
g^{(2)}_{\text{AS},c,\text{in}} =  1.932 \pm   0.051   \quad & \quad 
g^{(2)}_{\text{AS},c,\text{out}} = 1.988 \pm 0.051.
\label{eq_ch6_g2_AS_noise}
\end{align}
Clearly, the numbers for the prepared ensemble differ significantly from the noise {\gtwo} data in the S channel. 
Furthermore, unlike the S channel, where settings \cd\, and \textit{c} resulted in different outcomes\footnote{
	$g^{(2)}_{\text{S},c,\text{in}} = 1.92 \pm 0.01$
}, the {\gtwo} in the AS channel is nearly independent of the atomic state preparation. 

The last point can be expected, firstly, because the initial step of FWM is SRS into the AS channel. We have seen (fig. \ref{ch7_Fluorescence}), that SRS, as the first step in the FWM process for setting \cd\,, 
accounts for a significant amount in both time bins, so it is not surprising for the {\gtwo} to be similar to setting \textit{c}. 
Additionally, SRS represent thermal noise and thermal noise should show\cite{Loudon:2004gd} $g^{(2)} = 2$, unless it is multi-modal. 
Generally, if $K$ modes of a multi-modal signal (here the noise) are observed, the {\gtwo} reduces to\cite{Loudon:2004gd,Christ:2011} $g^{(2)} = 1+\frac{1}{K}$. 
Yet, the analysed SRS and FWM noise is not only SMF-coupled at the memory output, but it is also emitted into the control field's spectral-temporal mode, which in turn matches the spectral mode selected by the signal filter stage. 
Hence, both noise types should be single mode ($K=1$) and $g^{(2)}= 2$. 
The measured AS {\gtwo} data agrees with this notion and predicts $K\approx 1.06$ modes for setting \textit{cd} and for the input bin of setting \textit{c}, as well as $K= 1.01$ for the output bin of setting \textit{c}. 
Therefore the AS leg behaves exactly as expected for one mode of a two-mode squeezed state. 
Since there is no immediate reason to suspect the collection of a different number of modes for the S channel, 
the deviations between S and AS {\gtwo} could results from fluorescence noise, which is incoherently added to the FWM noise. 

\paragraph{Photon statistics of the fluorescence noise} 
Fluorescence noise, which is also observed in a single spatial mode, will be temporally multi-mode, because it originates from incoherent {\cs} emissions at different times\footnote{
	This argument is similar to indistinguishability between SPDC photons. In SPDC, spatial walk-off
	between signal and idler photons needs to be compensated to allow for their indistinguishability, 
	when using them for entanglement studies\cite{Kwiat:1995}. In the absence of such compensation,
	the signal and idler wave packets are distinguishable, because they occupy separate temporal modes,
	due to different group velocities. The same happens here, only that the wave-packets are created in 
	different temporal modes to begin with, which the broadband filtering cannot make to overlap. 
}. 
The number of temporal modes, that can, in principle, be analysed, is set by the FPGA coincidence window $\Delta t_\text{coinc}^\text{FPGA} = 5\ns$. 
Since the coincidence window is aligned to the centre of the control pulses in each time bin, 
and fluorescence noise is emitted predominantly after the control (see fig. \ref{fig_ch7_flourescence}), only a window of $\frac{\Delta t_\text{coinc}^\text{FPGA}}{2} = 2.5\ns$ is relevant for resolving the temporal modes of emitted fluorescence. 
This window corresponds to an observable bandwidth of $\Delta \nu_\text{coinc} \approx \frac{2}{\Delta t^\text{FPGA}_\text{coinc}} \approx 400 \MHz$. 
Because any detected light has to pass through the signal filter, its bandwidth of $\Delta \nu_\text{filt} \approx 1\GHz$ defines the resolution of each temporal mode to $\Delta t_\text{filt} \approx \frac{1}{\Delta \nu_\text{filt}} \approx 1\ns$. 
So, in total, $K \approx \frac{\Delta t^\text{FPGA}_\text{coinc}}{2\cdot \Delta t_\text{filt}} \approx 2.5$ modes can be resolved, leading to $g^{(2)}_\text{fluor} \approx 1.16$ for the
fluorescence noise. 

\paragraph{Photon statistics for the two-photon transition noise in the Stokes channel} 
To obtain an estimate for the S noise {\gtwo} without the fluorescence contribution, we use eq. \ref{eq_ch6_goldschmidt}
to express the incoherent addition\cite{goldschmidt2013mode} between fluorescence and FWM according to the fluorescence's fractional contribution $R_\text{S,FL}^t$ (see section \ref{ch7_sec_fluor_results}) to the total noise floor $N_\text{noise}^t$ (eq.~\ref{ch6_eq_NoiseFloor}). 
The measured data $g^{(2)}_{\text{meas},t} = g^{(2)}_{\text{S},t}$ and $g^{(2)}_\text{fluor}$ take the role of $g_\text{tot}^{(2)}$ and $g_\text{noise}^{(2)}$ in eq.~\ref{eq_ch6_goldschmidt}, which we solve for $g_\text{sig}^{(2)}$ to obtain:
\begin{equation}
\tilde{g}^{(2)}_{\text{S},t} = \left(1+r^2 \right) \cdot g^{(2)}_{\text{meas},t} - 2\cdot r - g^{(2)}_{\text{fluor}} \quad \text{with} \quad  r = \frac{(1-R^t_\text{S,FL}) \cdot N_\text{noise}^t}{R^t_\text{S,FL}\cdot N_\text{noise}^t}, 
\label{ch7_g2_noise_exp_golschmidt}
\end{equation}
where $\tilde{g}^{(2)}_{\text{S},t}$ now corresponds to the actual {\gtwo} of pure FWM noise in the S channel.
This yields $\tilde{g}^{(2)}_{\text{S},\text{in}} \approx 1.88$ and $\tilde{g}^{(2)}_{\text{S},\text{out}} \approx 1.93$, in both time bins, which approaches the observed photon statistics in the AS channel and agrees significantly better with two-mode squeezing expectation. 
These values suggest approximately single mode noise, with mode numbers of $K_\text{S,in} \approx 1.13$ and $K_\text{S,out} \approx 1.07$. 
\begin{figure}[h!]
\includegraphics[width=\textwidth]{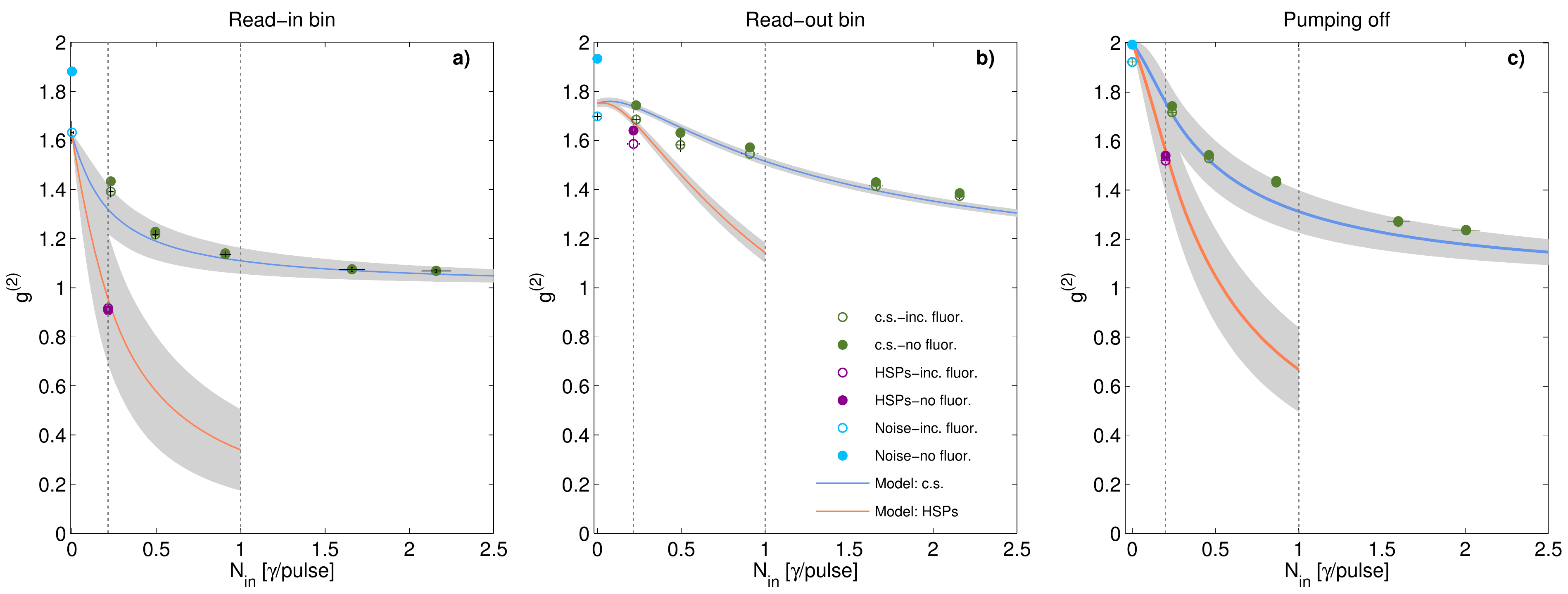}
\caption{Comparison of the directly measured {\gtwo} data (\textit{open markers}) from fig. \ref{fig_ch6_g2results} with the {\gtwo}-values obtained when taking out contributions from fluorescence noise (\textit{filled markers}). 
\textit{Solid lines} are the predictions by our theory model (appendix \ref{app6_coh_model}), which is also fluorescence-free. 
\textbf{(a)} \& \textbf{(b)} show the read-in and read-out time bin for spin-polarised {\cs}, respectively. 
\textbf{(c)} shows the read-in time bin for thermally distributed {\cs} atoms. 
 All other colour coding is equal to fig. \ref{fig_ch6_g2results}.
}
\label{fig_ch7_g2_flour}
\end{figure}

The effect size of the fluorescence noise {\gtwo} on the overall noise {\gtwo} immediately raises the question, by how much it affects the measured values when signal is added into the memory. 
We answer this by re-calculating the photon statistics, shown in fig. \ref{fig_ch6_g2results}, taking into account the fluorescence contribution to the measured unconditional noise floor $N_\text{noise}^t$. 
To this end, we extract the {\gtwo}, expected without the fluorescence noise present, from the measured {\gtwo}-values, following the same procedure as eq. \ref{ch7_g2_noise_exp_golschmidt} for $\tilde{g}^{(2)}_{\text{S},t}$. 
Fig. \ref{fig_ch7_detuning_scan} shows the results for all data contained in fig. \ref{fig_ch6_g2results}. 
The extracted, fluorescence-noise-free {\gtwo} is indicated by \textit{filled markers}. 
Comparison with the directly measured {\gtwo}-values (data from fig. \ref{fig_ch6_g2results}), depicted by \textit{open markers}, shows, that the differences are smaller when signal photons are inserted. 
Here, the {\gtwo} data without fluorescence noise still agrees well with the predictions by our theory model, with a noteworthy improvement in the memory read-out time bin. 
The reason for the difference between the theory model prediction and the fluorescence-free data for pure FWM (\textit{cyan points}) is however not clear yet. 
Importantly, the drop between \coh\, and \hsp\, {\gtwo} in the memory read-out does not change with respect to the observations discussed in chapter \ref{ch6}.

\section{Conclusion}

To summarise, in this chapter, we have studied the composition of the memory noise floor, investigated the behaviour of the noise constituents under different experimental conditions, and compared the measured noise levels with the expectations from our theory model. 

As the first main take away message, we were able to categorise the Raman memory noise floor, identifying its two main contributors. For the spin-polarised ensemble, used for the Raman memory protocol, FWM noise contributes the lions share of $\sim 85\%$ to the noise in the Stokes channel, which is the relevant mode, occupied by the memory signal, with fluorescence accounting for the remaining $\sim 15\,\%$. 
The noise increases over successive control pulses, due to the build-up of a FWM-induced spin-wave coherence in the ensemble, which is not retrieved completely within a single control pulse. 
We furthermore demonstrated that the FWM process can, in principle, be stimulated (i.e. seeded) by an input signal, supplied in its Stokes (or anti-Stokes) channel. However, for Stokes channel input signals at the single photon level, the resulting additional noise was found to be negligible compared to the overall memory noise floor.  

Secondly, we identified SRS as the dominant noise process for a thermally distributed ensemble; fluorescence still contributes the remainder\footnote{
	SRS contributed $\approx 97\,\%$ and $\approx 91\,\%$ of the total noise floor in the read-in time bin and 
	the $12.5\ns$ read-out time bin of the Stokes channel, respectively. 
}. 
Due to the absence of any spin-wave dynamics, this noise process results in an approximately constant noise level over a train of control pulses. 

With further {\gtwo}-measurements we determined the number of modes, each noise component is occupying. FWM and SRS were found to be pretty much mono-modal, occupying $\sim 1.1$ and $\sim 1.01$ modes, respectively, whereas $\sim 2.5$ modes contribute to the observed fluorescence noise. 

Lastly, we calculated the noise level we would expect to see at the memory output, using our theory model. The resulting numbers are in good agreement with the measurements, particularly considering that the prediction is an ab-initio calculation without any fitting parameters. 
This adds further confidence into the appropriateness of our model. Its predictions do not only approximate well the {\gtwo}-data, which are relative observables, but it also yields reasonable values for the absolute numbers of noise photons per control pulse. 

The final, remaining question is, whether we can do any better than the noise floor of ${\sim0.05 \ppp}$ and 
${\sim0.15\ppp}$ we observe at present in the read-in and the $12.5\ns$ read-out time bins of the memory's signal (Stokes) channel? 
To answer this question, we have conducted a detailed study of the experimentally accessible parameters, which are the control pulse energy $E_\text{c}^\text{p}$, the detuning $\Delta_\text{S}$, the memory storage time $\tau_\text{S}$, the {\cs} ensemble temperature, as well as the control beam focussing.    
To this end, we have compared the scalings of memory efficiency, noise as well as resulting SNR for all of these experimental parameters. 
While these measurements do not add any new insights to the problem in terms of the underlying physics, they will nevertheless be of use to the technically oriented reader. 
Due to page limitation, this material is presented in appendix \ref{appendix_ch7_noise_parameter_dependence}. 
The measurements reveal that it is impractical to reduce the noise floor any further. 
Obviously, with our initial aim to build a memory, capable of temporal multiplexing, still prevailing, this result is not very encouraging for our present system. 
In turn, it means, we will have to modify our memory system somehow to achieve this overarching goal. 
After a short summary of the total material, we have presented in this thesis, we will use the following conclusion chapter to give a thought to such possible improvements.


\chapter{Conclusion}
\label{ch8}

\begin{flushright}
{\tiny \textfrak{Faust}: \,  \textfrak{Das war also des Pudels Kern!} }
\end{flushright}

\section{Summary of the work presented in this thesis}

In this thesis, we have presented several steps towards the application of quantum memories 
in photonic quantum information processing networks. 
While quantum memories can perform various tasks in such networks\cite{Sanguard2011}, one key network element is a joint single photon source - memory building block, capable to perform temporal multiplexing tasks\cite{Nunn:2013ab}.
When using single photons, produced by probabilistic sources, as quantum information carriers temporal multiplexers can be employed to synchronise successful single photon generation events. 
Synchronisation allows, for instance, the production of high photon numbers resource states, whose creation would otherwise suffer from prohibitively low probabilities\cite{Wieczorek:2009kx}. 
Raman quantum memories are an interesting technology for such temporal multiplexing devices, thanks to their large time-bandwidth products\cite{Reim2010}. 
The resulting capability to store spectrally broadband, i.e. temporally short, photons for long periods of time enables many synchronisation trials across several sources, or over multiple emission rounds of the same source, while storing the successfully generated outputs. 
To implement a Raman memory, warm atomic vapours are a promising platform that may facilitate the transfer from the optical bench to real-world applications. 

Our work evaluated the potential of the Raman memory protocol, implemented in $70^\circ\text{C}$-warm atomic {\cs} vapour in this regard. 
The main advance was the development of an interfaced quantum memory - single photon source system, which represents a temporal multiplexer prototype.  
To this end, we carried out four series of experiments, each focussing on a particular aspect that is relevant for the viability of our Raman memory in 
quantum information processing applications.

\paragraph{Polarisation storage:}
In the first series of experiments, we researched the possibility of storing polarisation encoded information in the Raman memory. 
Choosing a dual-rail memory architecture inside a polarisation interferometer, 
we analysed the storage of polarisation information encoded on coherent state input signals, using quantum process tomography. 
For the polarisation storage process, we observed fidelities and purities of
 $\mathcal{F} \approx 92\,\%$ and 
 $\mathcal{P} \approx 83\,\%$, respectively. These were also found to be independent of the storage time 
 $\tau_\text{S}$, whose half-life duration is $\tau_\text{S} \approx 1.5\,\mus$. 
Faithful operation with real single photon input signals can however not be expected in the presence of the observed memory noise floor. 

\paragraph{Single photon source:}
Thereafter, we shifted our focus to the construction of an interfaced photon source - quantum memory system. 
Our first step was the design of an experimentally simple heralded single photon source, based on the workhorse technology of spontaneous parametric down-conversion.
The nonlinear medium was a periodically-poled KTP waveguide, operated in travelling wave configuration. The waveguide's well defined spatial mode structure permitted high quality collection of the down-converted photon pairs, leading to a single photon preparation efficiency of $\eta_\text{her} \approx 24\,\%$ upon heralding\footnote{
	Note: the heralding efficiency backs out the detection efficiency for the heralded single photons, 
	which we assume as $50\,\%$.
}. 
Spectrally engineering the signal photons, sent into the memory, by filtering the heralding idler photon, tailored their FWHM spectral bandwidth of $\Delta \nu_\text{HSP} \approx 1.69 \,\GHz$ to our broadband memory acceptance bandwidth of $\Delta \nu_\text{mem} \approx 1\,\GHz$.

\paragraph{Single photon storage in the Raman memory:}
Subsequently, we interfaced the single photon source with our Raman memory, operating the system in feed-forward mode. 
This means, probabilistically occurring heralding events did not only announce the presence of a single photon going into the memory, but also triggered the storage and retrieval processes. 
Such on-demand storage and retrieval with random triggers is the key operating mode for using the system  as a temporal multiplexer. 
Single photons were stored in the Raman memory with a total memory efficiency of 
$\eta_\text{mem} \approx 21\,\%$, which, due to residual mode-mismatch, was slightly below the storage efficiency for coherent states of $\eta_\text{mem} \approx 30\,\%$. 
To test the preservation of the single photon character during storage, we evaluated the photon statistics of the signal, contrasting single photon input signals with coherent state inputs.
While we were able to detect an influence of the input's quantum character in the retrieved signal, the noise background of the memory pushed the photon statistics for all input signal types into the classical regime, with $g^{(2)}\approx 1.59$ for heralded single photons and $g^{(2)} \approx 1.69$ for coherent states. 

\paragraph{Memory noise analysis:}
Finally, we characterised the processes underlying the memory noise floor. 
In a series of experiments, we separated the possible noise processes in the memory and identified FWM as the dominant noise source. 
To this end, we studied the noise's spin-wave dynamics and its dependence on the experimental parameter set, whose optimisation was also investigated. 
We attributed $\sim 86\,\%$ of the total memory noise floor to FWM, with the remainder generated by collisional-induced fluorescence. 
Using the predictions of a simple model, which we developed and that matched well with our experimental observations, we concluded that FWM noise is the only substantial challenge left in the development of Raman quantum memories for quantum information processing applications.

\paragraph{Significance of our findings for the Raman memory research community:}

On the one hand, the presented material is a prototype for the assembly of an interfaced single photon source - quantum memory system. 
On the other hand, it provides a thorough performance analysis of the Raman memory, which serves as a benchmark for future improvements. 
Most importantly however, our results leave us with a holistic understanding of the Raman memory and its performance limitations due to FWM. 
These findings do not only enable us to identify possible noise mitigation strategies but also allow us to quantify the improvements in memory performance we can expect for a given degree of noise floor reduction. 
Our results thus form a cornerstone along the development path of temporal multiplexers based on the Raman memory protocol.

\section{The future of the Raman memory}
Clearly any improvement of the Raman memory needs to overcome the challenge imposed by 
FWM in the memory read-out time bins. This noise needs to be reduced with respect to the retrieved signal. 
To conclude, we will take a brief look at the next steps that could be undertaken to reduce this FWM noise contamination. 

\paragraph{What improvements are required?}
Prior to our experiments, we expected operation in the quantum regime would be possible utilising sources with high heralding efficiencies, highly efficient detectors and some improvements in the memory efficiency, while operating the Raman memory with the noise floor we have achieved here (appendix \ref{app_ch7}). 
Particularly for sources and detectors, technological progress has pushed source heralding efficiencies up to $\sim 80\,\%$ in the near-IR\cite{Ramelow:2013}, while highly efficient detectors, such as transition edge sensors with up to ${\eta_\text{det}\sim 80\,\%}$ detection efficiency\cite{Lita:08}, have also become available. Moreover, our group has recently found that different buffer gases and buffer gas pressures can increase the total memory efficiency up to $\eta_\text{mem} \approx 60\,\%$. 
However, our photon statistics measurements (section \ref{ch6_photon_statistics}) in combination with our theoretical description (appendix \ref{app6_coh_model}) proves that these engineering solutions are insufficient. 
In order to tackle FWM in the Raman memory protocol, we must actively suppress the underlying FWM noise process. 
Because FWM noise generation and Raman storage are intrinsically linked via one of the two $\Lambda$-level systems the Raman memory control pulses can couple to, any fruitful noise mitigation strategy needs to aim at diminishing the Raman coupling in the $\Lambda$-system that is not shared with the Raman memory protocol. In our case this is the anti-Stokes system\footnote{
	As a reminder for the reader,  
	the two $\Lambda$-level systems are 
	the Stokes leg, which is the $\Lambda$-system shared with the Raman memory protocol and 
	the anti-Stokes leg, 
	that is solely relevant for the FWM process (see fig. \ref{fig_ch2_FWMlevels}). 
	The Raman coupling constants $C_\text{S}$ and $C_\text{AS}$ for the 
	respective transitions are linked by the ratio of their excited state 
	detunings $\Delta_\text{S}$ and $\Delta_\text{AS}$ to
	$C_\text{S} = \frac{\Delta_\text{AS}}{\Delta_\text{S}} C_\text{AS}$. 
}. 
To answer the question, how this can be done, we consider three different potential solutions. During the writing process of this thesis, the Raman memory community has already begun to research these.

\paragraph{Storage media with large Stokes shifts:}
One possibility is a change in storage medium, choosing a material with a larger ground state splitting. This results in a detuning increase for the undesired $\Lambda$-system, leading to a lower transition probability and thus suppressing the initial FWM step. 
Such systems can, for instance, be found in solid state materials. Particularly diamond has a level structure with a ground state splitting at an optical frequency ({${\delta \nu_\text{gs}\sim40 \,\text{THz}}$}). 
While in {\cs} vapour the undesired anti-Stokes Raman coupling amounts to $\sim 62\,\%$ of the memory's Stokes Raman coupling constant, it can be suppressed by orders of magnitude in diamond. 
This yields a much improved signal to noise ratio and allows signals with non-classical statistics to be retrieved from the Raman memory\cite{England:2014}. 
However, this particular choice comes at the cost of prohibitively short storage times, which currently prevent actual system applications, apart from proof-of-principle experiments. 

\paragraph{Ladder-type energy state systems:}
Another solution could be a change in the level structure, choosing a ladder system, rather than a $\Lambda$-system. 
Here, the ground state would represent the initial state, while the intermediate state replaces the excited state in the $\Lambda$-system and the storage state is the highest energetic state. The latter state should not decay radiatively. If the transitions to either side of the intermediate state have a frequency difference that is by itself an optical frequency, any undesired noise can easily be spectrally filtered. 
While {\cs} does not have the required level structure, Rb vapour could be a promising candidate. 
For this reason, our group started to investigate this system as a contingency to the {\cs} memory. 

\paragraph{Intra-cavity Cs Raman memory:}
Given our $\Lambda$-system in {\cs} vapour, the most promising noise mitigation approach is to place the memory inside an optical cavity. When resonant with the Stokes frequency but anti-resonant with the anti-Stokes channel in our protocol, the cavity will reduce the density of states at the anti-Stokes frequency and therewith prevent anti-Stokes noise emission into the cavity mode. 
Conversely, thanks to resonance with the cavity, the Stokes frequency will not be suppressed.
Accordingly, the memory input signal can be coupled into the cavity for storage in the {\cs} ensemble. 
The suppression of the anti-Stokes noise emission terminates the onset of the FWM noise process and, in turn, also reduces the amount of Stokes noise emitted into the Raman memory's signal mode. 

To enable operation in the quantum regime with $g^{(2)} \lesssim 1$, our analysis shows that we require a noise suppression factor of $x \approx 2.5$ (fig. \ref{fig_ch6_Rscan} in section \ref{ch6_conclusion}). For faithful operation with a single photon $g^{(2)} \lesssim 0.1$ in the memory retrieval time bin we need $x \approx 10$. 
For an intra-cavity memory this suppression factor is determined by the round-trip losses of signal ($\mu_\text{s}$) and anti-Stokes noise ($\mu_\text{a}$), according to\cite{Saunders:2015} $x=\frac{1-\mu_\text{s}}{1+\mu_\text{a}}$. 
Apart from absorption losses in the {\cs} cell, these factors are determined by the cavity's mirror reflectivities and its length. In other words, they are set by the cavity finesse $\mathcal{F}=\frac{\text{FSR}}{\Delta \nu}$, where $\Delta \nu$ is the cavity's spectral acceptance bandwidth at the Stokes frequency. In the absence of memory bandwidth restrictions, arising from a finite control pulse spectral bandwidth, this acceptance bandwidth now determines the spectral bandwidth of the storable signal. 
Notably, for broadband signal storage, low-finesse cavities are required. This is different from cavity QED experiments, which employ high-quality resonators\cite{Specht:2011}, since, for our kind of application, we do not seek operation in the strong coupling regime. Instead, we only aim to suppress spontaneous anti-Stokes scattering into the cavity mode.

Recently, our group started to investigate this route and constructed a cavity that is double resonant for the Stokes signal and the memory control, but anti-resonant for the anti-Stokes noise. 
A first, promising set of results\cite{Saunders:2015} demonstrates the capability to suppress FWM by a factor of $x\approx 0.24$, which corresponds to an improvement in the SNR by a factor of $\sim 2.9$.
This has been achieved with a cavity of length $L_\text{cav} \approx 40.8\,\text{mm}$, corresponding to an FSR$\approx 7.36\,\GHz$, and acceptance bandwidth $\Delta \nu_\text{cav} \approx 0.95\,\GHz$. 
Accordingly, the cavity finesse was $\mathcal{F} \approx 7$. 

Similar to the other solutions, this system is however also not totally caveat-free. 
The above performance numbers already demonstrate the two main trade-offs: 
Firstly, there is the experimental complexity and the requirement to build cavities of short lengths, which nevertheless have to house a macroscopic {\cs} ensemble. 
For a first order cavity with its resonance at the Stokes frequency and the immediately adjacent anti-resonance at the anti-Stokes frequency, the cavity length would have to be as short as $L_\text{cav} = \frac{c}{\delta \nu_\text{gs}} \approx 3\,\text{mm}$, requiring a monolithic construction. 
Secondly, with the achieved noise suppression, we can expect a retrieved single photon $g^{(2)}$ just at the borderline between the classical and the quantum regime. 
To enable faithful storage of the quantum characteristics, noise suppression still has to be improved. 
In this regard, there is the trade-off between the necessary high noise extinction ratio and the desire 
for a broad signal bandwidth. Both are inversely related through the cavity finesse. 
Consequently, better noise suppression has to be achieved by sacrificing some of the signal bandwidth. 
The next generation of {\cs}-based Raman memories, our group is currently planning, will thus move to signals of 
$\sim 300\,\MHz$ spectral bandwidth. 
For such a system, the signal will still be broadband enough to allow for single photon production using SPDC sources. 
Moreover, we can realistically expect to still achieve reasonable time-bandwidth products if we can extend the memory lifetime appropriately, which means a reduction of decoherence due to magnetic dephasing and diffusive atom loss\cite{Novikova:2012}. 
The required improvements should be feasible with enhanced magnetic shielding, different beam geometries and changes in {\cs} cell design, such as paraffin coated cell walls and a different buffer gas\cite{Thomas:2017}. 
Extending the lifetime to $\tau_\text{s} \approx 5\,\mus$, in order to maintain the current time-bandwidth product of $B\sim 1000$, is a reasonable expectation and has been exceeded by orders of magnitude in similar atomic vapour systems\cite{Heinze:2013aa}. 
This way our memory will maintain its applicability for temporal multiplexing tasks, 
making warm {\cs} vapour Raman memories still an attractive candidate technology for a faithful quantum memory in optical information processing networks.

\appendix

\chapter{Appendix: Experimental apparatus\label{ch3}}

\section{The Ti:Sa laser system\label{ch3_tisa}}

\subsection{Operating principles and parameters}
The master laser for preparation of the memory pulse sequence is a titanium sapphire ({\tisa}) oscillator (\textit{Spectra Physics Tsunami}), which is passively mode-locked, generating pulses with an $80\MHz$ repetition rate.
An intra-cavity acousto-optic modulator (AOM) facilitates the mode-lock, introducing side-bands at $80\MHz$ that coincide with the laser cavity resonances.
Its external controls enable tuning of the sideband phase. 
Phase modification enables fine tuning of the output pulse duration on a time scale of several tens of pico-seconds. 
By itself, such a system would fall into the category of standard pico-second lasers, producing pulse with durations on the order of $10\ps$.
To confine the spectral bandwidth to approximately $1\GHz$, as required for the Raman memory, a Gires-Tournois interferometer (GTI) is inserted into the cavity. 
This thermally controlled Fabry-Perot etalon has a larger free-spectral range (FSR) than the {\tisa} cavity. Thus, some of the cavity resonances fall in between the GTI resonances, which terminates their gain and narrows the output spectral bandwidth through elimination of the respective {\tisa} frequency comb modes. 
Therewith, pulse durations of $\sim 300 \ps$ are possible. 

The central wavelength $\lambda$ of the output spectrum is tuneable on two different scales: 
Firstly, coarse adjustment on the geometric orientation of the GTI moves the spectrum on the order of nano-meters. 
Secondly, fine tuning on the order of GHz is possible by thermal control of the GTI.
The fine tuning allows continuous scanning of the central wavelength $\lambda$ over the entire FSR of the GTI, which corresponds to $\approx 50\GHz$, without the laser dropping out of mode-lock. 
This feature is key for the alignment of the memory experiments. The easy frequency tuning from the 
{\cs} resonances all the way to the idler frequency of our SPDC photons (detuning: $\Delta = 24.4\GHz$, see chapter \ref{ch5}) allows, on the one hand, to distinguish the {\cs} resonances and therefore to determine the exact frequency of the {\tisa}. 
On the other hand, it enables convenient alignment of the extensive frequency filtering stages used in the experiments. 
The Fabry-Perot etalons, contained in these stage, are aligned on the transmission signal of the {\tisa} pulses, which are picked with a Pockels cell (\pockels\,) to yield the memory control. Notably, this tuning capability is not a manufacturer specification, but rather a lucky coincidence.

The oscillator is pumped by a $5.2\,\text{W}$ diode laser at $532\nm$. It produces pulses at $852\nm$ central wavelength of up to $1.5\,\text{W}$ average output power. The output mode of the {\tisa} has $M^2 = 1.1$.
Over the course of the projects presented in this thesis, the {\tisa} laser has been moved between labs, which made repositioning of the intra-cavity AOM and realignment of the cavity mode necessary. 
Unfortunately, these modifications changed the output power down to $1.2\,\text{W}$. The decrease in power coincides with a broadening of the pulses from originally $300\ps$ to $360\ps$. 
Notably, the former number has been determined before my arrival in 2005. In between, the laser cavity mode has been re-optimised many times, so the pulse duration might have been different from the quoted $300\ps$ for the experiments discussed in chapter \ref{ch4}.

\subsection{Stabilisation}
Changes in the laboratory temperature and thermal heating during operation result in drifts of the {\tisa} centre frequency.
For the long measurement durations required for our experiments, such drifts are a severe challenge. 
Initially, for the work described in chapter \ref{ch4}, manual stabilisation onto the resonance of one of the FP frequency filtering etalons has been used to at least have a frequency reference and enable semi-stable conditions. 
However, such manual resetting is not sufficient, since the FP etalon resonances themselves drift over time. 
After moving to a different laboratory, we have implemented an active scheme based on the {\cs} D$_2$-line, onto which also the diode laser is stabilised. 
Here, the {\tisa} laser is locked to the frequency of the diode laser. 
Experimentally low intensity pick-offs of both lasers are inserted into a scanning FP etalon (FSR=$10\GHz$), whose transmission is monitored by a fast photodiode. Its output, together with a voltage ramp signal applied to the scanning FP, are digitalised for computer processing.
The resulting scanning FP spectrum trace defines the \tisa's frequency within a multiple of $10\GHz$ intervals relative to the diode laser\footnote{
 	In other words, $\nu_\text{Ti:Sa} = n*\text{FSR} + (\nu_\text{Ti:Sa} - \nu_\text{d})$, 
	whereby the integer $n$ is unknown. Here $\nu_\text{d}$ and 
	$\nu_\text{Ti:Sa}$ are the centre frequencies of the spectral modes of the diode and the {\tisa} laser, respectively, 
}. 
Its absolute value is determined by simultaneously also sending the {\tisa} through an auxiliary {\cs} vapour cell.
After initial calibration of the absolute frequency with respect to one of the two D$_2$-transitions\footnote{
	$6^2\text{S}_\frac{1}{2} \, \text{F} = 3 \rightarrow 6^2\text{P}_\frac{1}{2} \, \text{F} = \left\{ 3,4 \right\}$, and
	$6^2\text{S}_\frac{1}{2} \, \text{F} = 4 \rightarrow 6^2\text{P}_\frac{3}{2} \, \text{F} = \left\{ 2,3,4,5 \right\}$.
} through observation of fluorescence in the cell, a software PID-loop controls the \tisa's frequency fine adjustment to maintain a constant detuning with respect to the diode laser line, using the scanning FP spectrum traces. 
The control software has been written by \textit{Michael Sprague}\cite{Sprague:PhD}.  
Therewith the detuning can be set appropriately as required for the specific experimental needs and a {\tisa} frequency stability of approximately $100-200\MHz$ can be obtained. 

Besides the need to stabilise its frequency, the tunability of the {\tisa} comes at the price of changes in the beam pointing of its output mode. 
These are caused by the GTI: tuning the {\tisa} frequency modifies the cavity mode and therewith also the pointing of the emitted light. 
Given that memory alignment requires changes in the {\tisa} frequency\footnote{
	Filter alignment uses the bright control pulses, whose frequency has to be tuned to $\Delta = 6\GHz$ detuning for the signal filter stage and 
	$\Delta = 24.4 \GHz$ for the herald filter stage, assuming the memory is to be operated at $\Delta = 15.2 \GHz$ detuning. 
}, this spatio-spectral coupling can be particularly troublesome, if not cancelled by a beam pointing stabilisation system in the {\tisa} output.
For experiments in chapter \ref{ch4}, such a stabilisation was not yet available, for which reason experimental realignment was necessary and constituted a challenging, time consuming task. During the memory system's re-built, a commercial beam pointing stabilisation system, based on software from \textit{KLM Labs}, has been installed. As shown in fig. \ref{fig_6_setup}, two CMOS cameras observe the {\tisa} pulse leakage through two dielectric mirrors in the near and far field.
Beam steering is applied through two motorised mirror mounts, positioned in front of two mirrors whose leakage is collected by the cameras. 
The mirror positioning is controlled by a software-based PID feedback loop.

\subsection{Pulse duration measurements\label{ch3_tisa_pulse_duration}}
\begin{figure}
\centering
\begin{minipage}[l]{0.45\textwidth}
\centering
\includegraphics[width=\textwidth]{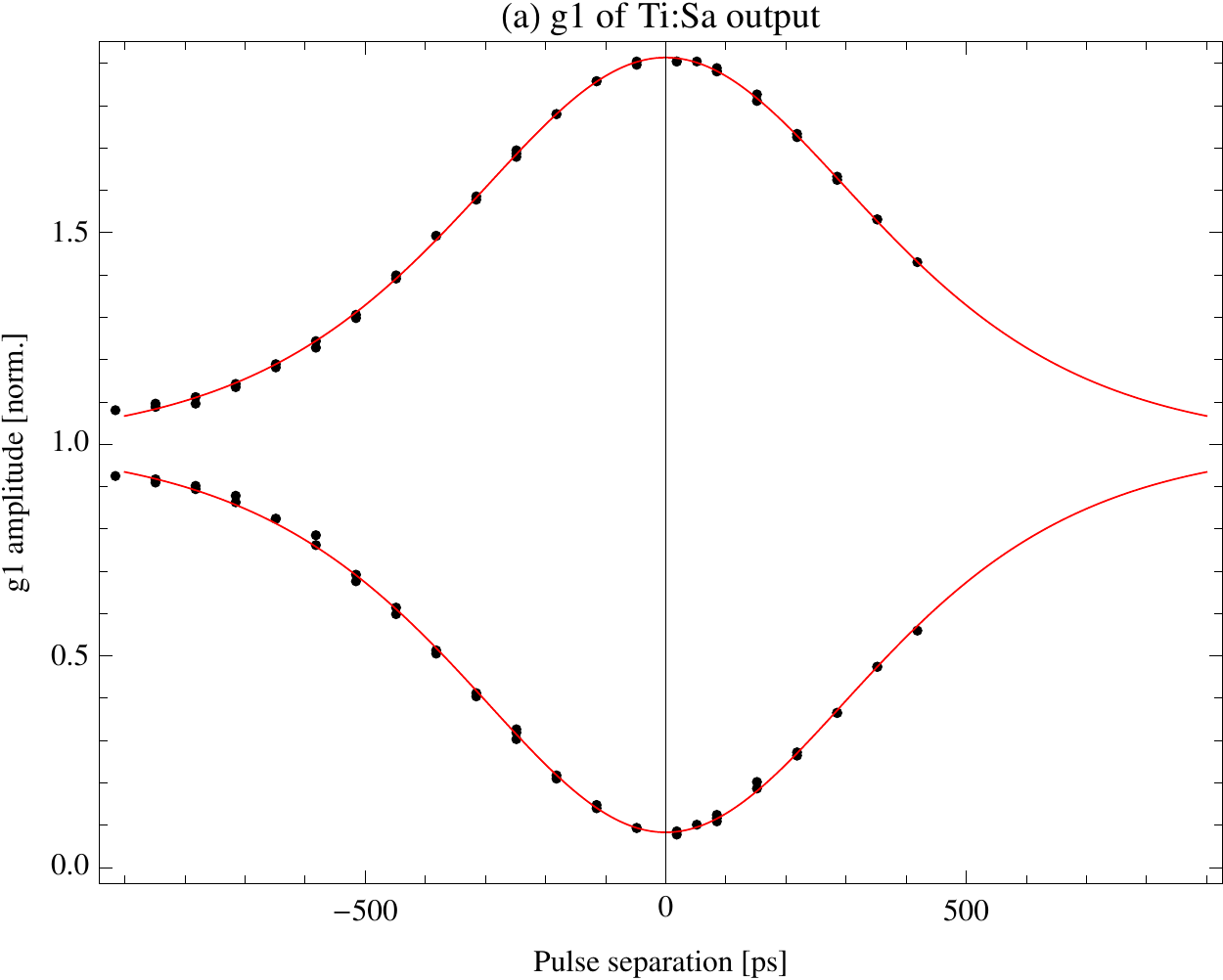}
\end{minipage}
\hspace{0.05\textwidth}
\begin{minipage}[r]{0.45\textwidth} 
\centering
\includegraphics[width=\textwidth]{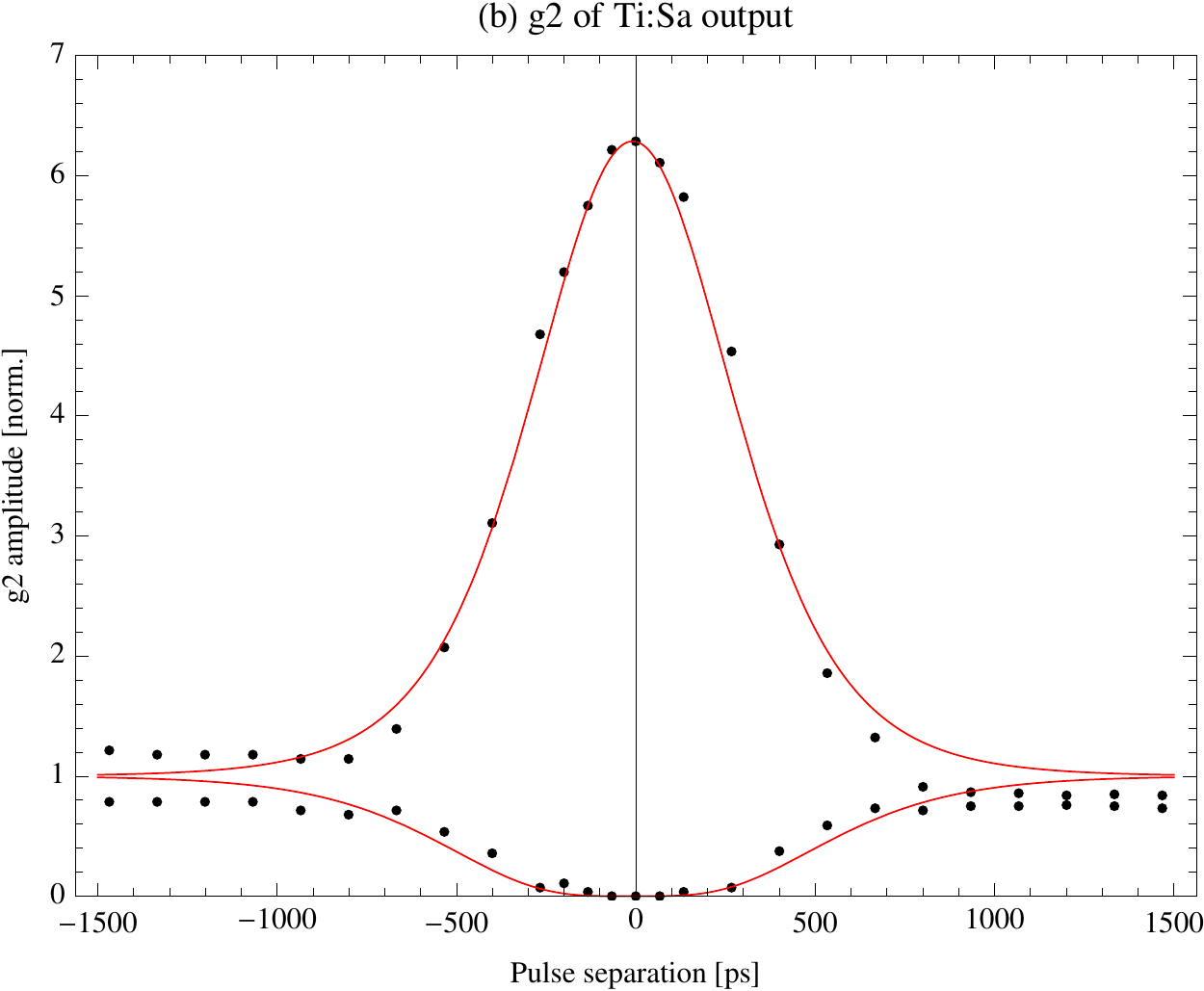}
\end{minipage}
\caption{\textbf{(a)}: $g^{(1)}$ measurement of {\tisa} pulses. 
\textbf{(b)}: $g^{(2)}$ measurement of {\tisa} pulses. \textit{Solid lines} are fits for a sech pulse profile. }
\label{fig_ch3_tisa_autocorr}
\end{figure}

After the lab move, the {\tisa} pulse duration has been measured directly with the method of interferometric autocorrelation\cite{Trebino:book} based on second-harmonic generation (SHG) in a bulk ppKTP crystal (see section \ref{ch3_SHG} below). 
Frequency conversion produces $426\nm$ radiation, which is separated from the $852\nm$ pump pulses by a dichroic filter and detected with a standard slow photodiode\footnote{
	Slow refers to a diode response time which is larger than the {\tisa} pulse duration.
}.
The experiment requires two copies of the {\tisa} pulses, separated by a variable time $\tau$, which are generated by a non-stabilised Michelson interferometer. 
By also detecting the separated fundamental at $852\nm$ with another slow photodiode, one can additionally record the $g^{(1)}$ interferogram. It yields the FT-limited pulse duration, whose information content is equivalent to the spectrum of the {\tisa} pulses. In contrast, the $g^{(2)}$ is sensitive to spectral phase and thus a metric for the real pulse duration. 
Both correlation functions are given by the functions\cite{Diplomarbeit}
\begin{align}
\label{eq_ch3_g1_def}
g^{(1)} &= \frac{\overset{\infty}{\underset{-\infty}{\int}} |E_1(t) + E_2(t-\tau)|^2 \text{d}t }{\overset{\infty}{\underset{-\infty}{\int}} |E_1(t) |^2 \text{d}t + \overset{\infty}{\underset{-\infty}{\int}} |E_2(t-\tau)|^2 \text{d}t } \\
g^{(2)} &= \frac{\overset{\infty}{\underset{-\infty}{\int}} | \left( E_1(t) + E_2(t-\tau) \right)^2|^2 \text{d}t }{\overset{\infty}{\underset{-\infty}{\int}} |E_1(t)^2 |^2 \text{d}t + \overset{\infty}{\underset{-\infty}{\int}} |E_2(t-\tau)^2|^2 \text{d}t },
\label{eq_ch3_g2_def}
\end{align}
whereby $E_{1,2}(t')$ are the electric field amplitudes at time $t'$ of pulses coming from either interferometer arm. 
In the experiment, there is residual spatial mode overlap mismatch, reducing the interference visibility. 
For this reason, a multiplicative factor $a$ is introduced in the above equations, taking into account the reduced interference visibility and degradation in the peak to background ratios\cite{Michelberger:2012}.  
The interferometer does not have to be stabilised. Monitoring the amplitude of the fluctuation in the detected signal as a function of $\tau$, gives both correlation traces, which are shown in fig. \ref{fig_ch3_tisa_autocorr} \textbf{a} and \textbf{b} for the $g^{(1)}$ and $g^{(2)}$ measurements, respectively.
In order to extract the pulse duration, one has to assume a pulse model for the electric fields $E_{1}$ and $E_{2}$. We assume two pulse types, Gaussian pulses and sech-shaped pulses, whose electric fields, centred at time $t_0$, are given by 
\begin{align}
E(t,t_0) = E_0 \cdot \exp{\left( -\frac{t-t_0}{2 \cdot (\Delta t)^2} \right)},\nonumber\\
E(t,t_0)= E_0 \cdot \text{sech}{\left(-\frac{t-t_0}{\Delta t}\right)},
\label{eq_app3_pulse_model}
\end{align}
respectively, with a pulse width parameter $\Delta t$. 
For both types we obtain the envelope functions
\begin{align}
\label{eq_ch3_sech_fit_func}
g^{(1)}_\text{Sech}  =&    1 \pm \frac{2\cdot a}{1 + a^2}\cdot \frac{\frac{\tau - \tau0}{\Delta t} }{\text{sinh} \left( \frac{\tau - \tau0}{\Delta t}\right)},  \\
g^{(2)}_\text{Sech} =&  1+ 18 \cdot \frac{a^2}{1+a^4} \cdot \frac{\frac{\tau-\tau_0}{\Delta t} \cdot \cosh{\left( \frac{\tau - \tau_0}{\Delta t} \right) }- \sinh{\left( \frac{\tau - \tau_0}{\Delta t} \right)}}{\sinh^3{\left( \frac{\tau - \tau_0}{\Delta t} \right)}} \nonumber \\
 \label{eq_ch3_gauss_fit_func}
 & \pm 3 \cdot \frac{a+a^3}{1+a^4}\cdot \frac{\sinh{\left( \frac{2\cdot (\tau - \tau_0)}{\Delta t} \right)} -\frac{2\cdot (\tau - \tau_0)}{\Delta t}}{ \sinh^3\left(\frac{\tau-\tau_0}{\Delta t} \right)}, \nonumber \\
g^{(1)}_\text{Gauss} =& 1 \pm \frac{2a}{1+a^2} \cdot \exp{\left( \frac{-(\tau - \tau_0)^2}{4\cdot \Delta t^2}\right)},  \\
g^{(1)}_\text{Gauss} & =  1+6\cdot \frac{a^2}{1+a^2} \cdot \exp{\left( -\frac{(\tau-\tau_0)^2}{2\cdot \Delta t^2}\right)} 
\pm 4\cdot \frac{a+a^3}{1+a^4} \cdot \exp{ \left( - \frac{3 (\tau-\tau_0)^2}{8 \cdot \Delta t ^2}\right)}, \nonumber
\end{align}
where the $\pm$ signs represent the upper and lower pulse envelopes. 
Fig. \ref{eq_ch3_g2_def} exemplifies their fits onto the respective datasets. Here, the visibility of the interferogram $a$, as well as its width have been fitted via the parameters $a$ and $\Delta t$. Notably, to a account for the reduced interference visibility, the amplitudes of the two interfering pulses have been set to $E_0$ and $a\cdot E_0$ for pulses $E_1(t)$ and $E_2(t)$, respectively\cite{Michelberger:2012}. 

The optimised parameter $\Delta t$ for the $g^{(2)}$-functions determines the full-width-half-maximum (FWHM) pulse duration, which is given as a function of $\Delta t$ by 
$\tau_\text{Gauss} = 2 \cdot \sqrt{\ln{ (2) }} \cdot \Delta t$ and  $ \tau_\text{Sech} = 2 \cdot \text{arcsech} \left( \frac{1}{\sqrt{2}} \right) \cdot \Delta t$, respectively. 
Equally, the optimised parameter $\Delta t$ for fitting the $g^{(1)}$-correlation functions determines the FWHM spectral bandwidth $\Delta \nu$ of the pulses. In turn, $\Delta \nu$ enables to determine the FT limited pulse duration $\tau^\text{FT}$ via the time-bandwidth-products for the respective pulse model. Both relations read: 
\begin{align}
\Delta \nu_\text{sech} = \frac{\sqrt{\ln(2)}}{\pi\cdot \Delta t}, \quad 		& \quad 
\Delta \nu_\text{sech} \cdot \tau_\text{sech}^\text{FT} = \frac{2\cdot \ln{(2)}}{\pi} = 0.441 \\
\Delta \nu_\text{Gauss} = \frac{2\cdot \text{arcsech} \left( \frac{1}{\sqrt{2}}\right)}{\pi^2 \cdot \Delta t}, \quad 	& \quad 
\Delta \nu_\text{Gauss} \cdot \tau_\text{Gauss}^\text{FT} = \frac{4\cdot \text{arcsech}^2\left(\frac{1}{\sqrt{2}} \right)}{\pi^2} = 0.315
\label{ch3_time_bandwidth_products}
\end{align}
The pulse spectra follow from Fourier transform of the underlying pulse modes, which yields\cite{Diplomarbeit}:
\begin{equation}
S_\text{sech}(\nu) = \text{sech}{\left( \pi^2 \Delta t (\nu - \nu_0)\right)}, \quad 
S_\text{Gauss}(\nu)=\exp{ \left( -4 \pi^2 (\Delta t)^2 (\nu -\nu_0)^2 \right)}.
\label{eq_ch3_pulse_spectra}
\end{equation}
The values for pulse duration $\Delta \tau$ and spectral bandwidth $\Delta \nu$, obtained by the data in fig. \ref{fig_ch3_tisa_autocorr} for both models, are stated in table \ref{ch3_table_pulse_durations}.

Gaussian pulses are slightly longer than their sech counterparts. Given the soliton propagation dynamics of pulses inside the laser cavity, the sech-model is the more appropriate one to use for the {\tisa} output. For this reason, we assume a pulse duration of $\tau_\text{Ti:Sa} = 323 \ps$ and a spectral bandwidth of $\Delta \nu_\text{Ti:Sa} = 0.97 \GHz$.
Moreover, the pulses are also longer than their FT-limited duration. This means, the intra-cavity GTI, which is the most dispersive element in the system, chirps the pulses. 

\begin{table}[h!]
\centering
\begin{tabular}{c|c|c|c|c|c}
\toprule
$\lambda$ [nm] & Pulse model & $\Delta t$ [ps] & $\Delta \nu$ [GHz] & $\tau$ [ps] 	& Data for $\tau$ \\
\midrule
$852$		& Sech 		& 	$183$ 	& $0.97$ 			& $323$ 		& $g^{(2)}$\\
$852$		& Gauss 		& 	$249$ 	& $1.06$ 			& $415$  		& $g^{(2)}$\\ 
$426$		& Sech 		& 	$150$	& $1.19$  			& $264$ 		& $g^{(1)}$\\ 
$426$		& Gauss 		& 	$205$ 	& $1.29$ 			& $338$ 		& $g^{(1)}$\\ 
\bottomrule
\end{tabular}
\caption{Pulse durations and spectral bandwidths obtained from $g^{(1)}$- and $g^{(2)}$-interferograms
of the {\tisa} output pulses and the SH, generated thereof, by frequency doubling.}
\label{ch3_table_pulse_durations}
\end{table}

\section{The second harmonic source \label{ch3_SHG}}

\subsection{Experimental lay-out}
To pump the SPDC source, described in chapter \ref{ch5}, frequency-doubled radiation of the {\tisa} pulses at $426\nm$ central wavelength is required.
Since the single photon source has to run in parallel to the Raman memory, sufficient pump power at $426\nm$ has to be available, while also maintaining the highest possible pulse energy for the $852\nm$ pulses which are to be used as the memory control pulses. 
The best compromise between both requirements is to use an inefficient SHG source pumped by the full {\tisa} output power, rather than constructing a highly efficient SHG pumped by a split-off fraction of the {\tisa}. 
For this reason, we place the SHG source as the first significant element behind the stabilised {\tisa} oscillator output (see fig. \ref{fig_6_setup}). 
The unconverted fraction of the {\tisa} pump pulses, separated from the produced second-harmonic (SH) by a high-pass transition edge filter (\textit{Semrock BLP01-532}), is used to prepare the memory control field and the coherent state signal input, as described in chapter \ref{ch6}. 
To achieve sufficient electric field amplitudes for SHG, the {\tisa} pump has to be focussed into the crystal, which introduces a degradation in the spatial mode quality of the unconverted light at $852\nm$ wavelength. 
This ultimately limits the single-mode fibre (SMF) coupling efficiency of the memory control pulses, produced from the transmitted fraction by pulse picking with a Pockels cell, to $\eta_\text{SMF}^\text{cntrl} \approx 53 \,\%$. 
No increase in the coupling efficiency was possible by beam shaping, using for instance a pair of cylindrical lenses or an anamorphic prism pair, placed in the output arm of the Pockels cell.

To produce the SH, we use a bulk periodically-poled KTP (ppKTP) crystal, with dimensions of $2\mm \times 1\mm \times 2\mm$. One of the larger, $2\mm$-long sides is oriented along the optical propagation axis. 
The crystal is optimised for type-I phase matching and is coated for $852\nm$ and $426\nm$ radiation to reduce the, otherwise quite substantial, Fresnel reflection of $R(426\nm) \approx 10 \,\%$ $R(852\nm) \approx 8 \,\%$ per KTP-air interface down to $R \le 1\,\%$ for both wavelengths.
Generated SH is collimated by a $f=50\mm$ focal length lens, positioned in the common beam path of SHG and IR fundamental. 
After separation it is SMF-coupled with $\eta_\text{SMF}^\text{SH} \approx 72\,\%$, to enable a stable beam pointing for the UV pump, when sent into the ppKTP waveguide for SPDC.

\subsection{Characterisation of the second harmonic generation}
The main, tuneable experimental parameters for optimising the SHG process are the focussing into the crystal, the crystal temperature and pump power. 
By using the phase-matching condition of SHG (type I), the pump power is adjustable by polarisation of the fundamental IR input beam. 
Temperature tuning is possible via a Peltier heater element, which is positioned below the crystal mount. 
The mount itself sits on an x-y-z translation stage, which does not only enable transverse crystal positioning but also crystal displacement along the focus of the $f=100 \mm$ lens that is uses to focus the IR pump into the ppKTP crystal. 

\begin{figure}[h!]
\centering
\includegraphics[width=\textwidth]{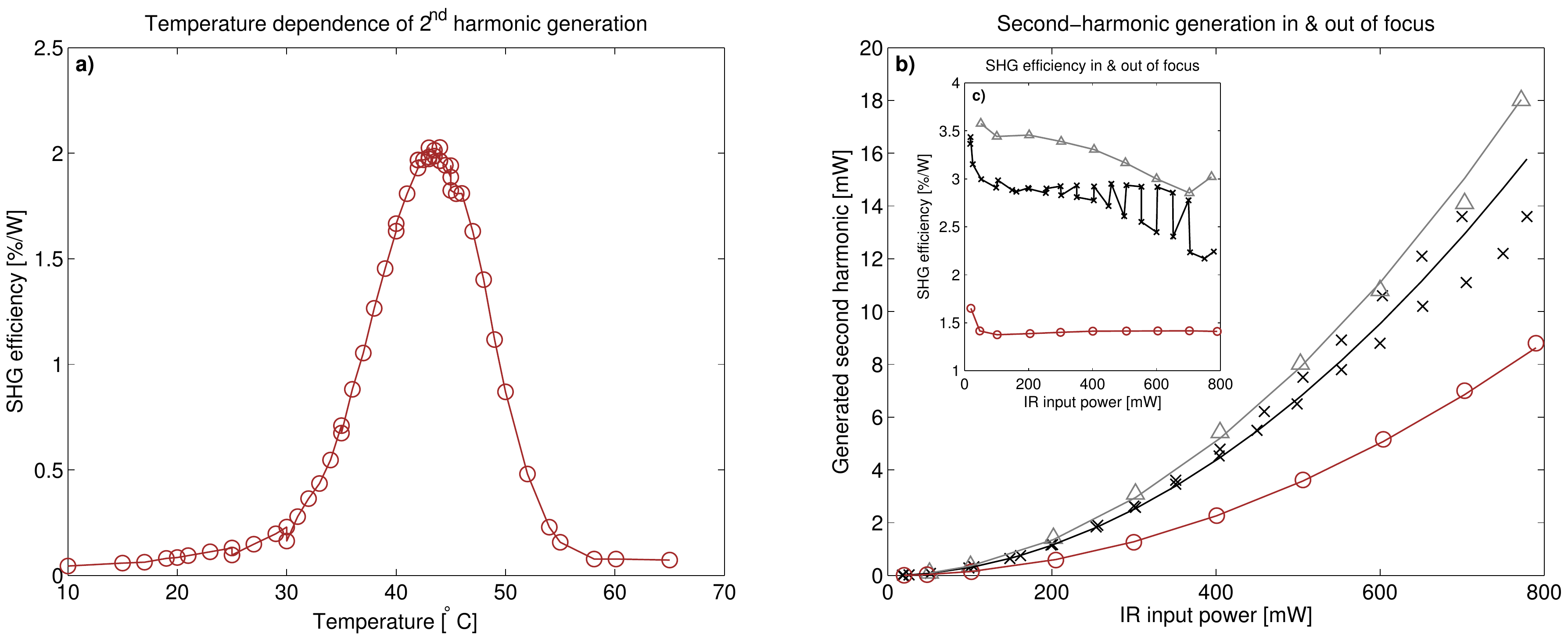}
\caption{\textbf{(a)}: SHG efficiency in our bulk ppKTP crystal as a function of crystal temperature. The \textit{solid line} represents a linear interpolation between the data points. 
\textbf{(b)}: SH, generated in our bulk ppKTP crystal, as a function of {\tisa} power for different longitudinal crystal positions. \textit{Grey} and \textit{black} represent the crystal sitting in the focus of the $f=100\mm$ focal length lens, used to send the collimated {\tisa} pump beam into the crystal. \textit{Red} corresponds to a crystal positioning that is slightly moved out of focus, which is the configuration used for our experiments. 
\textbf{(c)}: SHG efficiency versus {\tisa} input power for the crystal positions shown in panel \textbf{(b)} with equal colour coding.}
\label{fig_ch3_shg}
\end{figure}

The temperature tuning is used to achieve phase matching\cite{Grechin:1999}. 
The corresponding nonlinear conversion efficiency as a function of temperature, expressed by the IR power independent measure of $\eta_\text{SHG} =\frac{P_\text{SH}}{P_\text{IR}^2}$ (see section \ref{ch5_subsec_SHG}), is shown in fig. \ref{fig_ch3_shg} \textbf{a}. 
Optimal SHG is achieved for $T_\text{SHG}^\text{opt} = 44.5^\circ\text{C}$ with a FWHM temperature tuning range of $\Delta T \approx 16\,\text{K}$. 
Stabilising the crystal temperature at $T_\text{SHG}^\text{opt}$, the actual amounts of SH light, produced for the available {\tisa} input powers, are displayed in fig. \ref{fig_ch3_shg} \textbf{b}. The \textit{solid lines} are fits onto the data using the expected proportionality $P_\text{SH} \sim \alpha P_\text{IR}^2$ for the produced SH ($P_\text{SH}$) as a function of the power in the fundamental ($P_\text{IR}$). 
The measurement is conducted thrice: twice with the crystal positioned in the focus of the $f=100\mm$ lens (\textit{black and grey}), and a third time with the crystal slightly moved out of focus (\textit{red}). 
The latter is investigated because the amount of SH power, produced by the crystal, actually exceeds the requirements for the UV pump power of the SPDC process. As discussed in chapters \ref{ch5} and \ref{ch6}, the necessary UV power is limited firstly by the photon number purity of the heralded SPDC signal photons, and secondly by the repetition rate limitations of the Pockels cell. 
In light of these two conditions, production of $P_\text{SH} \approx 2\mW$ is actually sufficient\footnote{
	The difference with respect to the number of $P_\text{SH} \approx 1\mW$, quoted for the SPDC UV pump
	in chapter \ref{ch6} is due to the SMF-coupling efficiency $\eta_\text{SMF}^\text{SH}$ and transmission losses
	of the UV-optics in front of the waveguide input coupler.
}. 
When running the SHG source on a day-to-day basis, the IR pump power is consequently reduced by polarisation, but  the crystal is also moved to the out-of-focus position. This is done for reasons of precaution, since we want to avoid possible damages on the crystal from high power densities. 
While these should neither structurally damage the crystal itself, nor its coating, the imperfect lab conditions can lead to dust particles being fried onto the optical surfaces\footnote{
	The usual avoidance strategy of blowing air onto the crystal introduces beam pointing fluctuations, for 
	which reason we did not adopt it. 
}; a risk we aim to avoid. 
Additionally, at high-power, ppKTP starts to show signs of grey-tracking\cite{Boulanger:1994}, which reduces the SHG efficiency through the formation of colour centres in the crystal. While the effects of most amounts of grey-tracking are temporary and anneal when the crystal is heated, they can result in an undesired tail-off in the SHG efficiency over the course of our measurements. 
We can see these effects when operating the SHG in the focus of the $f=100\mm$ lens and inserting the full IR pump power. 
For illustration purposes, fig.~\ref{fig_ch3_shg}~\textbf{c} shows $\eta_\text{SHG}(P_\text{IR})$ as a function of power in the fundamental ($P_\text{IR}$) for the crystal positions of fig. \ref{fig_ch3_shg} \textbf{b}. 
While the out of focus position is independent of $P_\text{IR}$, as expected for $\eta_\text{SHG}$, the in-focus efficiency shows signs of tail-off at high powers. 
Notably, moving the crystal out of focus does not introduce any noticeable deterioration, neither in the spatial mode profile nor in the SMF-coupling efficiencies. 

\subsection{Pulse duration and spectral bandwidth of the second harmonic}

Similar to the IR pulse, we require to know the spectral bandwidth $\Delta \nu_\text{SH}$, expressed as the FWHM, of the generate SH radiation in order to estimate the spectra of the heralded SPDC signal photons in chapter \ref{ch5}. 
In general, one expects the spectrum of the SH to be proportional to the convolution of the fundamental's electric field envelope with itself. 
The underlying reason is the proportionality of the electric field for the SH, produced by a nonlinear medium with susceptibility $ \chi^{(2)} $, to the square of the IR electric field, $E_\text{SH} \sim \chi^{(2)} \cdot E_\text{IR}^2$, which holds in the time and the frequency domain. 
For a simple Gaussian pulse, with pulse envelope $E_\text{IR}(\nu) = \tilde{E}_0 \exp{\left( -\frac{(\nu-\nu_0)^2}{\sigma^2}\right)}$ and bandwidth (variance) $\sigma^2$, the spectrum of the SH would thus have a variance $\sigma_\text{SH}^2 = 2\cdot \sigma_\text{IR}^2$, yielding a FWHM bandwidth of $\Delta \nu_\text{SH} = 2 \sqrt{\ln(2)} \cdot  \sigma_\text{UV} = \sqrt{2} \cdot 2 \sqrt{\ln(2)} \sigma_\text{IR}^2 = \sqrt{2} \cdot \Delta \nu_\text{IR}$. 

\begin{figure}
\centering
\includegraphics[width=0.6\textwidth]{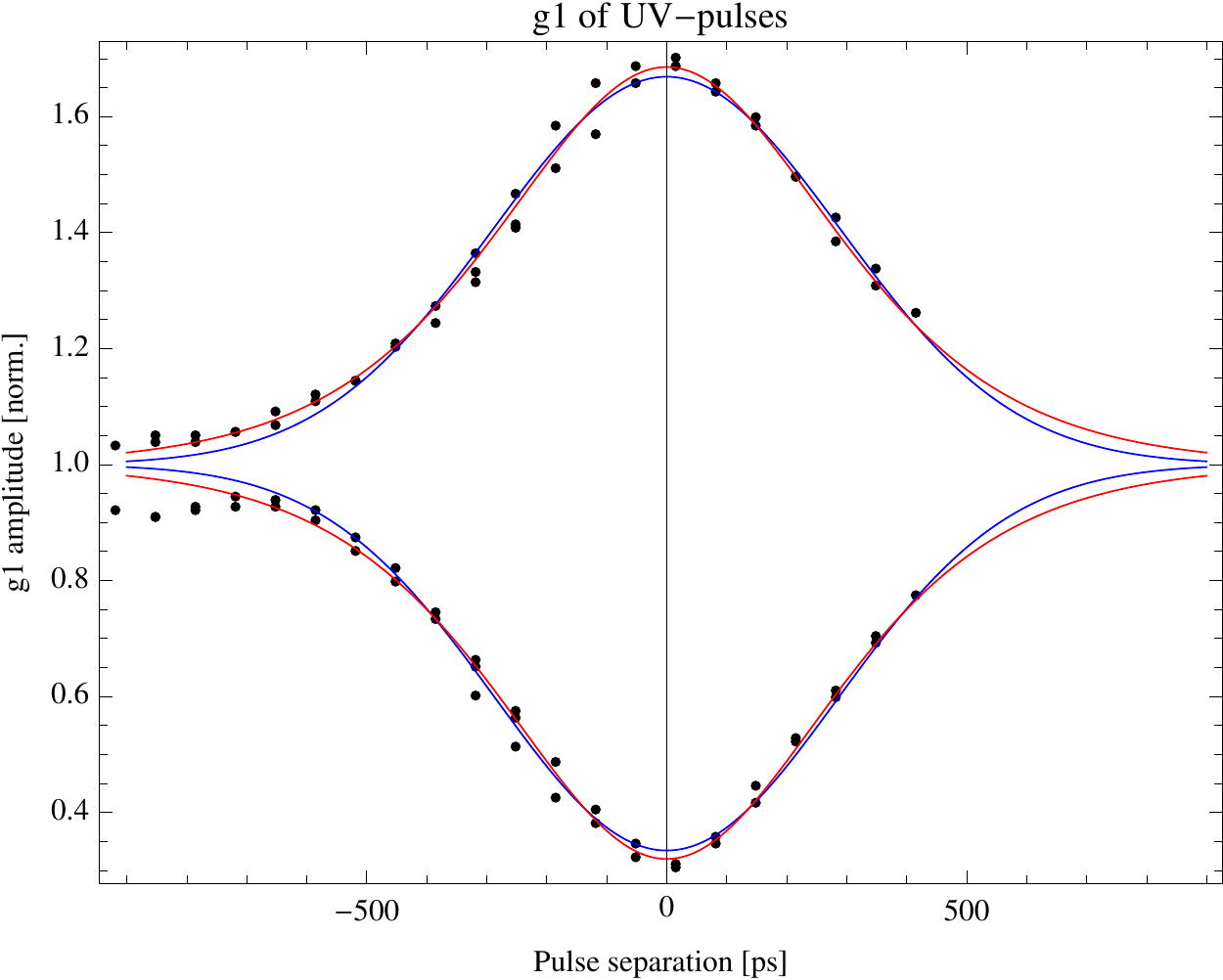}
\caption{$g^{(1)}$ measurement of SH generated by {\tisa} pulses. The \textit{solid red} and \textit{blue lines} represent the fitted $g^{(1)}$ correlation functions for sech- and Gaussian-shaped UV pulses, respectively.}
\label{fig_shg_g1}
\end{figure}

To measure the actual UV bandwidth, we record a $g^{(1)}$-interferogram, using the same technique as presented for the IR in section \ref{ch3_tisa_pulse_duration} above. 
To this end, the SMF-coupled SH is inserted into a Michelson interferometer, whose output is observed by a slow photodiode.  
Data, representing the envelope of the $g^{(1)}$-interferogram, is shown in fig. \ref{fig_shg_g1}. Similar to the {\tisa} pulses, we again assume a sech (red) and a Gauss (blue) pulse model and fit the measured points with the functions stated in eq. \ref{eq_ch3_sech_fit_func}. 

The results for both pulse models are also stated in table \ref{ch3_table_pulse_durations}. 
For sech-shaped pulses the extracted bandwidth is $\Delta \nu_\text{SH} = 1.19\GHz$, which is synonymous for a FT-limited pulse duration of $\tau^\text{FT}_\text{SH} = 264 \ps$.  
So, SHG spectrally broadens the pulses by a factor of $1.22$, which is a bit below the expected factor of $1.41$.

\section{Optical pumping of the caesium vapour \label{ch3_diodelaser}}

\subsection{Frequency stabilisation of the diode laser for optical pumping\label{ch3_subsec_diode_laser}}

\begin{figure}
\includegraphics[width=16cm]{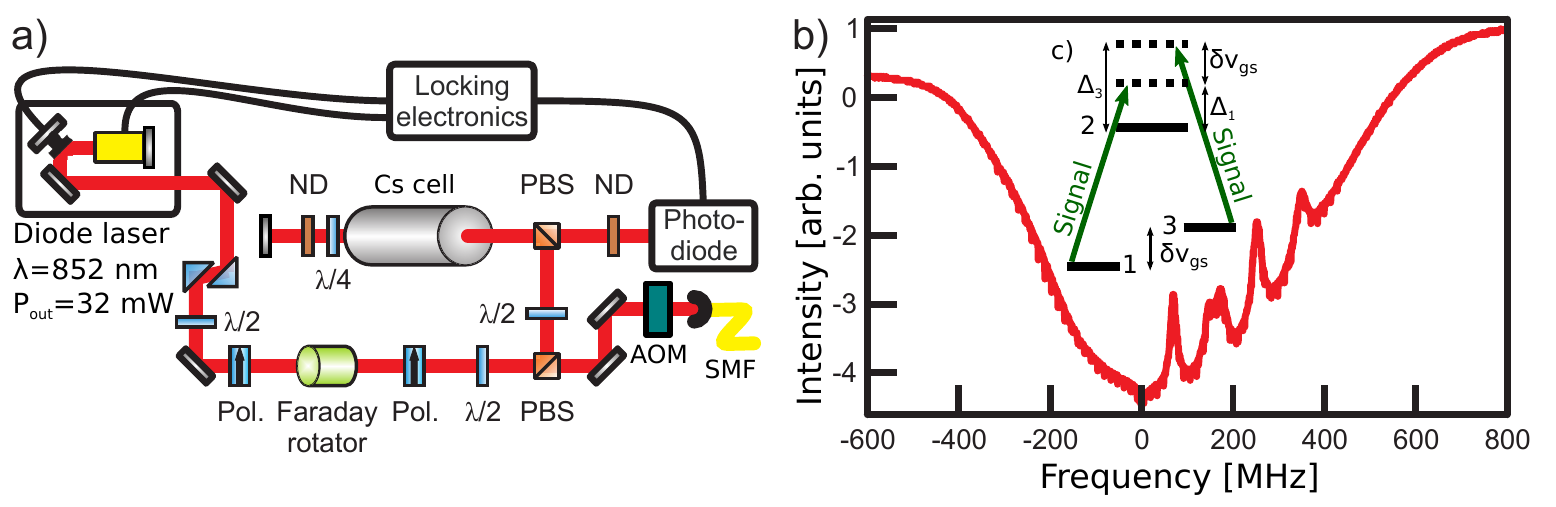}
\caption{\textbf{(a)}: External cavity diode laser and Doppler-free spectroscopy set-up. 
\textbf{(b)}: Doppler-free spectroscopy absorption spectrum of the 
$6^{2} \text{S}_\frac{1}{2} \text{F}=3 \rightarrow 6^{2} \text{P}_\frac{3}{2} \text{F}'=\left\{2,3,4,5\right\}$ Cs-resonances, used for locking the diode laser.
\textbf{(c)} {\cs} atomic $\Lambda$-level system, used for estimating the optical pumping efficiency.} 
\label{fig_ch3_diode}
\end{figure}

To implement optical state preparation in the {\cs} ensemble and to simultaneously have a reliable frequency standard for stabilisation of the {\tisa} master laser's central frequency, we employ a home-built external-cavity diode laser system. 
A schematic representation of the set-up is presented in fig. \ref{fig_ch3_diode} \textbf{a}, which follows the designs of \textit{MacAdam et. al}\cite{MacAdam:1992} and \textit{Hori et. al.}\cite{Hori:1983}. 
The output of its $852\nm$ laser diode is frequency stabilised with respect to the {\cs} D$_2$-line, using the 
$6^{2} \text{S}_\frac{1}{2} \text{F}=3 \rightarrow 6^{2} \text{P}_\frac{3}{2}$ transition to deplete the Raman memory's storage state. 

The stabilisation mechanism is based on Doppler-free saturation spectroscopy\cite{Schmidt:1994}. 
To this end the laser's output frequency is locked onto an error signal, generated by modulations in the driving current and by angular variations of a grating in the external cavity, adjusting the frequency component back-coupled into the active medium. 
The error signal derives from observing the Doppler-free spectrum of the {\cs} atoms, shown in fig. \ref{fig_ch3_diode} \textbf{b}, which is obtained by sweeping the laser frequency over the full
$6^{2} \text{S}_\frac{1}{2}, \text{F}=3 \rightarrow 6^{2} \text{P}_\frac{3}{2}  \text{F}'=\left\{2,3,4,5\right\}$ resonance manifold. 
The detected intensity comprises small peaks within a larger trough, corresponding to resolved hyperfine transitions and cross-over resonances in the Doppler profile\cite{Schmidt:1994}.
To obtain the error signal, a fraction of the diode laser's intensity is split off and transmitted through a Cs gas cell. The initially strong, horizontally polarised beam (pump) is attenuated and back-reflected with the polarisation rotated to vertical (probe). The transmitted probe is detected on a photodiode and the obtained voltage signal is mixed with the dithering signal, which is used to modulate the laser frequency initially, in a lock-box\footnote{
	A lock-box basically consists of phase-sensitive detection and subsequent 
	frequency filtering plus phase shifting, providing a negative feed-back 
	signal of which the high frequency part modulates the diode current 
	and the low frequency part the grating position via a piezo-electric crystal.
},  
generating the error signal. 

For the laser's application in Raman memory state preparation, it is not necessary to lock it to one of the sub-Doppler features, due to Doppler broadening in the $70^{\circ}\text{C}$ warm {\cs} vapour. 
Since the equipment incorporated in this set-up is an old, abandoned piece of kit, left-over from a previous cold atom experiment, locking it to narrow resonances is challenging. 
When stabilising its frequency on the sub-Doppler peaks, the locking signal become very fragile. This severely limits the timescales, over which the laser stays in lock. 
To obtain better and longer term stability, it is therefore stabilised with respect to the bottom of the Doppler absorption profile.  
The flatness of the Doppler profile around its maximum introduces residual frequency uncertainty in stability of the {\tisa} laser, since it follows any drifts in the diode laser frequency (see section \ref{ch3_tisa}). 

The majority of diode laser pump power, which is not used for stabilisation means, is SMF-coupled and sent into the {\cs} memory cell for optical pumping, as illustrated in fig. \ref{fig_6_setup}. 
In front of the SMF, an AOM is placed to turn the pumping beam on and off, as described in chapter \ref{ch6}. 
The maximum amount power, delivered to the memory is $P_\text{diode}^\text{max} \approx 3\mW$.

\subsection{Optical pumping efficiency\label{appA_optical_pumping_efficiency}}
We can now estimate the actual pumping efficiency $\eta_\text{pump}$ the diode laser achieves in the {\cs} ensemble. 
The numbers we calculate here hold for the experimental configuration used for chapter \ref{ch5} - \ref{ch7}. For the work in chapter \ref{ch4}, no explicit measurement has been conducted. However the diode laser power and its focussing into the {\cs} cell were similar, so a similar $\eta_\text{pump}$ can be expected. 

The pumping efficiency is defined as the population difference between the two {\cs} hyperfine ground states $6^2 \text{S}_\frac{1}{2}$ F$=3$ (state 1) and F$=4$ (state 3) of the $\Lambda$-level system. 
The population in each state is denoted as $N_1$ and $N_3$, respectively, whereby the total population $N_\text{tot} = N_1 + N_3$ is the number of {\cs} atoms in the active volume of the Raman memory, i.e. the volume covered by the signal, control and diode beams.
Therewith $\eta_\text{pump} = \frac{N_3 - N_1}{N_\text{tot}} =  \frac{N_3 - N_1}{N_3 + N_1}$. 
In case of thermally distributed population ($N_3 =N_1 = N_\text{tot}/2$), $\eta_\text{pump} = 0$, and $\eta_\text{pump} = +1$ for total {\cs} spin-polarisation in $N_3$. 

One possibility to determine \etapump\, is by 
absorption measurement for the {\tisa} signal pulses. 
Generally, the signal power $P$ is attenuated to $P_\text{out} = P_\text{in} \cdot \exp{(- \tilde{d}_j)}$, whereby $\tilde{d}_j$ is the optical depth for the optical pumping on/off, with $j \in \left\{ \text{on}, \text{off} \right\}$.
It can be obtained from the measured intensities as $\tilde{d}_j = \ln{\left( \frac{P_\text{in}}{P_\text{out}} \right)}$.

During propagation through the cell, the signal couples to both ground states, i.e. $6^2 \text{S}_\frac{1}{2} \text{F}=3 \rightarrow 6^2 \text{P}_\frac{3}{2}$ and
$6^2 \text{S}_\frac{1}{2} \text{F}=4 \rightarrow 6^2 \text{P}_\frac{3}{2}$ transitions can occur with detunings of $\Delta_1$ and $\Delta_3 = \Delta_1 +\delta \nu_\text{gs}$, respectively (see fig. \ref{fig_ch3_diode} \textbf{c}); $\delta \nu_\text{gs} = 9.2 \GHz$ is the ground state hyperfine splitting. 
The total optical depth, including both transitions, is $\tilde{d}_j = d_1 \cdot N_1 \cdot \left(\frac{\gamma}{\Delta_1 + \gamma}\right)^2 + d_3 \cdot N_1 \cdot \left(\frac{\gamma}{\Delta_3 + \gamma}\right)^2$, whereby $\gamma$ is the excited state linewidth of the $6^2 \text{P}_\frac{3}{2}$ state and $d_{i}$ is the coupling constant to state $i \in \left\{1,3\right\}$. 
For warm {\cs}, $\gamma$ is limited by the collisional broadening ($\mathcal{O}(500\MHz)$). 
The constants $d_i$ include the Rabi-frequency $\Omega = \frac{\vec{\mu} \cdot \vec{E}_\text{signal}}{\hbar}$, with oscillator strength $\vec{\mu}$ and a signal electric field amplitude of $\vec{E}_\text{signal}$, as well as the sum over the Clebsh-Gordan coefficients for all allowed transitions between Zeeman-states.

The absorption measurements use the {\tisa} signal on resonance, i.e. $\Delta_1 = 0 \GHz$. Since $\delta \nu_\text{gs} \gg \gamma$, the increased detuning $\Delta_3$ reduces the $2^\text{nd}$ term by a factor of $\sim \frac{1}{81}$ with respect to the $1^\text{st}$ term. To good approximation, the optical depth hence only depends on the population $N_1$ in the F$=3$ ground state: 
$\tilde{d}_j \approx d_1 \cdot N_1 \cdot \left(\frac{\gamma}{\Delta_1 + \gamma}\right)^2 = \alpha \cdot N_1$.

Optical pumping results in most of the atoms being transferred to the F$=4$ level. In the aforementioned approximation, the absorption $\tilde{d}_\text{on}$ of the transmitted {\tisa} signal is only sensitive to $N_1$, so $N_1 \sim \tilde{d}_\text{on}$.
Without the optical pumping, thermal distribution results in equal populations between both hyperfine ground states, such that $N_1 = \frac{N_\text{tot}}{2}$, from which follows $N_\text{tot} \sim 2 \cdot \tilde{d}_\text{off}$, again using $\tilde{d}_\text{off} \sim N_1$. 
Finally, the population in F$=4$ can be written as $N_3 = N_\text{tot} - N_1 \sim 2\cdot \tilde{d}_\text{off} - \tilde{d}_\text{on}$, and the pumping efficiency can be expressed as:

\begin{equation}
\eta_\text{pump} = \frac{N_3-N_1}{N_3 + N_1} = \frac{2\cdot \tilde{d}_\text{off} -2 \cdot \tilde{d}_\text{on}}{2\cdot \tilde{d}_\text{off}} = 1- \frac{\tilde{d}_\text{on}}{\tilde{d}_\text{off}} = 
1-\frac{\ln{\left(P^\text{on}_\text{out}/ P^\text{on}_\text{in} \right)}}{\ln{\left( P^\text{off}_\text{out}/ P^\text{off}_\text{in} \right)}}.
\label{ch3_diode_etapump}
\end{equation}

On resonance ($\Delta_1 = 0 \GHz$), the large optical depth of the warm {\cs} makes it experimentally challenging to measure $P^\text{off}_\text{out}$ precisely. 
In fact, sending a low intensity signal, with an average input power of $P^\text{off}_\text{in} = 330\muW$, through the cell results in complete absorption. 
For this reason, the background reading of the power meter will be taken as $P^\text{off}_\text{out}$. This is on the oder of $\mathcal{O}(100 \,\text{nW})$ but depends on the collected background light. We use a conservative estimate\footnote{
	This is the precision with which the output power of the MMF leading to the APD detectors in the signal
	filter stage can be determined during signal filter stage alignment
	(see fig. \ref{fig_6_setup}), when sending $P^\text{on}_\text{in} = 330\muW$  into the cold {\cs} cell. 
} and assume a background of $P^\text{off}_\text{out} \approx 3\muW$ to obtain $\tilde{d}_\text{on} =4.7$.

Direct measurement of transmitted power with optical pumping ($P^\text{on}_\text{out}$) will contain two reduction effects: on the one hand linear absorption from {\cs}, and on the other hand losses from optic components. The total transmission of the cold cell\footnote{
	Note, the transmission has been improved for the measurement in chapter \ref{ch6} and \ref{ch7} to $T\approx 80\,\%$. 
}  
yields $T_\text{loss}\approx 75\,\%$. 
To show that this number is pretty much only determined by the transmissions of the optical components, we also estimate it from a transmission measurement with pumped atoms at $15\,\text{GHz}$ detuning. 
Here, we can determine the amount of linear absorption separately, by using the single photon level measurements discussed in chapter \ref{ch6}. 
To this end, we calculate the ratio between the count rates $c_{sd}$ and $c_s$ for settings \textit{sd} and \textit{s}, respectively. For both settings the signal is subject to the same set of optics. A difference between the counts can thus only be caused by linear absorption in the {\cs} atoms, which we can determine to 
$L_\text{abs} = 1-\frac{c_s}{c_{sd}} \approx 10\,\%$. 
The total, bright signal power transmission ($T_\text{tot,warm}$) through the warm {\cs} cell at this detuning is again a product between linear absorption and transmission loss in optics. The measured value of $T_\text{tot,warm}  = T_\text{loss} \cdot (1-L_\text{abs}) \approx 68\,\%$ can thus be used in combination with $L_\text{abs} \approx 10\,\%$ to arrive at a transmission loss of the optical components of $T_\text{loss} \approx 75\,\%$, which is the same as the transmission of the cold cell. 
On resonance, $P^\text{on}_\text{in} \approx 210\muW$ are transmitted for $P^\text{on}_\text{in} = 330\muW$ input power. 
Backing out losses from optics $T_\text{loss}$, a transmitted power of $P^\text{on}_\text{in} \approx 271\muW$ is to be expected, which yields a slightly larger amount of linear absorption of $L_\text{abs} \approx 16\,\%$ and an optical depth of $\tilde{d}_\text{off}\approx 0.2$. 

Both optical depth figures lead to an estimated optical pumping efficiency of $\eta_\text{pump} \approx 96 \,\%$.

\section{Spatial alignment of signal and control in the caesium cell\label{app3_subsec_sig_ctrl_spatial}}

Fig. \ref{fig_ch2_Raman_protocol} illustrates the spatial arrangement of signal and control, which are collinear as they propagate through the storage medium. 
Similarly, the diode laser is also collinear with both beams, but propagates through the {\cs} cell in the reverse direction. 
While the latter represents experimental convenience, collinear signal and control fields can in general help to achieve high optical depth and long spin-wave coherence times. 
For our $\sim 1\,\GHz$ pulse bandwidth, the second point is of greater importance. 
Yet, we briefly consider both, as each one relates to the effects, caused by the thermal motion of the {\cs} atoms, 
when there is an angle $\theta$ between signal and control. 
Firstly, such an angle introduces a Doppler shift between the signal and the control frequency. When considering for instance atoms that counter-propagate the control beam path with velocity $V_\text{Cs}^\text{c}$, their velocity component along the signal beam path will be $V_\text{Cs}^\text{s} = V_\text{Cs}^\text{c} \cdot \cos{(\theta)}$. So the signal frequency they experience is shifted by $\Delta \nu \sim \frac{V_\text{Cs}^\text{c} \left(1-\cos{(\theta)} \right)}{\lambda_\text{Cs}}$. 
This effective one-photon detuning reduces the atomic number density, as signal and control do not address exactly the same {\cs} velocity classes. However, the effect is rather small for our parameters. 
For an expected velocity\footnote{
	In these approximative calculations, we assume the absence of Ne buffer gas. 
	The {\cs} atoms are Maxwell-Boltzmann distributed, with an expectation value for the velocity of 
	$V_\text{Cs}=\sqrt{\frac{3 k_\text{B} T}{m_\text{Cs}}}=253 \,\frac{\text{m}}{\text{s}}$, 
	where $k_\text{B}$ is Boltzmann's constant, $T=70^\circ\text{C}$, and $m_\text{Cs}\approx 2.21 \cdot 10^{-25}\,\text{kg}$. 
} 
of $V_\text{Cs}^\text{c} \approx 250 \,\frac{\text{m}}{\text{s}}$, we get $\Delta \nu \rightarrow 300\MHz$, as $\theta \rightarrow 90^\circ$, while for realistic angles\footnote{
	We consider cases, where we do no want to send in the signal through the {\cs} cell's side-walls. So the largest possible 
	angle occurs, when the displacement, orthogonal to the optical axis, between signal and control equals the cell's optical 
	aperture of $D=12.5\,\mm$, as both beams traverse the cell of length $L_\text{Cs} = 7.5\,\cm$. 
	This yields an angle of $\vartheta = \cos^{-1}{ \left( \frac{D}{L_\text{Cs}} \right) } \approx 10^\circ$. Since both beams can enter from
	opposite sides on the entrance aperture, their respective angle is $\theta = 2\vartheta$.
} of $\theta \le 20^\circ$, $\Delta \nu \approx 18\,\MHz$. 
So this shift is negligible compared to spectral bandwidths of signal and control. Note however, an angle would also reduce the effective volume, covered by both beams. Because both beams also have diameters $w \ll 1\,\mum$, the resulting decrease in atoms, illuminated by the two beams, reduces the amount of read-in signal substantially. 

Secondly, even without any effects on the read-in efficiency, an angle would influence the memory lifetime. Spin-wave excitation with angled signal and control leaves the spin-wave with a transverse wave-vector 
${k_B^\perp =\frac{2\pi}{\lambda_B^\perp} \sim k_\text{c} \sin{(\theta)}}$, orthogonal to the optical axis, whereby $k_c \approx \frac{2\pi}{\lambda_\text{Cs}}$. 
Atomic motion along this transverse direction can move atoms, excited into the spin-wave at a peak of $\lambda_\text{B}^\perp$ to a trough in ${\lambda_B^\perp}$. 
Therewith they would have dephased and the information initially stored in the spin-wave would be lost. 
The resulting dephasing time $t_\text{dp} = \frac{\lambda_\text{B}^\perp}{2 V_\text{Cs}}$ would then limit the memory storage time. Again it is proportional to the atomic velocity. 
Assuming the absence of any viscosity increase from the Ne buffer gas, {\cs} atoms would dephase in $\tau_\text{dp}\approx 5\,\ns$ for $\theta \approx 20^\circ$. Even for $\theta \approx 1^\circ$, the expected storage times of $\tau_\text{dp} \approx 100\,\ns$ are a lot shorter than the $\tau_\text{s} \approx 1.5\,\mus$, which we currently achieve in the collinear configuration with moderate experimental effort (see figs. \ref{fig_ch4_fidelity_lifetime} and \ref{fig_ch7_lifetime}). 

For these reasons, we implement a collinear set-up. 
Note however, this also means, that FWM noise will be emitted into the same spatial mode.
While FWM is not phase-matched \cite{Boyer:2013} at our detuning of $\Delta = 15.2\GHz$, there is nevertheless still a finite amplitude for collinear Stokes and anti-Stokes emission. Noise from both channels, Stokes and anti-Stokes, can thus occupy the spatial mode of the memory signal. 
In the experiment, the anti-Stokes noise can be eliminated by frequency filtering\footnote{
	To this end, we will use two Fabry-Perot filter etalons with a free-spectral range of 
	FSR=$103\GHz$ (see fig.~\ref{fig_6_setup})
}, but any Stokes noise will not separable from the signal.

\chapter{Appendix: Polarisation storage in the Raman memory\label{app_ch6}}

\section{Polarisation storage analysis with quantum process tomography\label{ch4_qpt_intro}}

To analyse the polarisation storage quality, we use the framework of quantum process tomography (QPT). 
QPT is a well developed tool for analysing processes operating on polarisation encoded quantum bits. While qubits consist of only a single quantum information carrier, e.g. a single photon, our experiment uses coherent states containing many photons. Yet, the description of their polarisation degrees of freedom is similar, for which reason the framework of polarisation encoded information can be applied to coherent states as well. For this reason, we can use the significant experimental simplification of probing the memory's storage capabilities with coherent states.
In this section, we will briefly review the theoretical properties of polarisation qubits and the concept of QPT, covering the aspects necessary to understand our experiment\footnote{
	The presented discussion is by no means complete and assumes prior basic knowledge of polarisation optics and quantum
	 information processing. More details can be obtained from references \cite{Nielsen:2004kl,Chuang:1997, Langford:PhD}, which 
	 are the basis for the following overview.
}.

\subsection{Encoding, manipulating and analysing polarisation information\label{ch4_subsec_qubits}}

\begin{figure}
\centering
\includegraphics[width=0.9\textwidth]{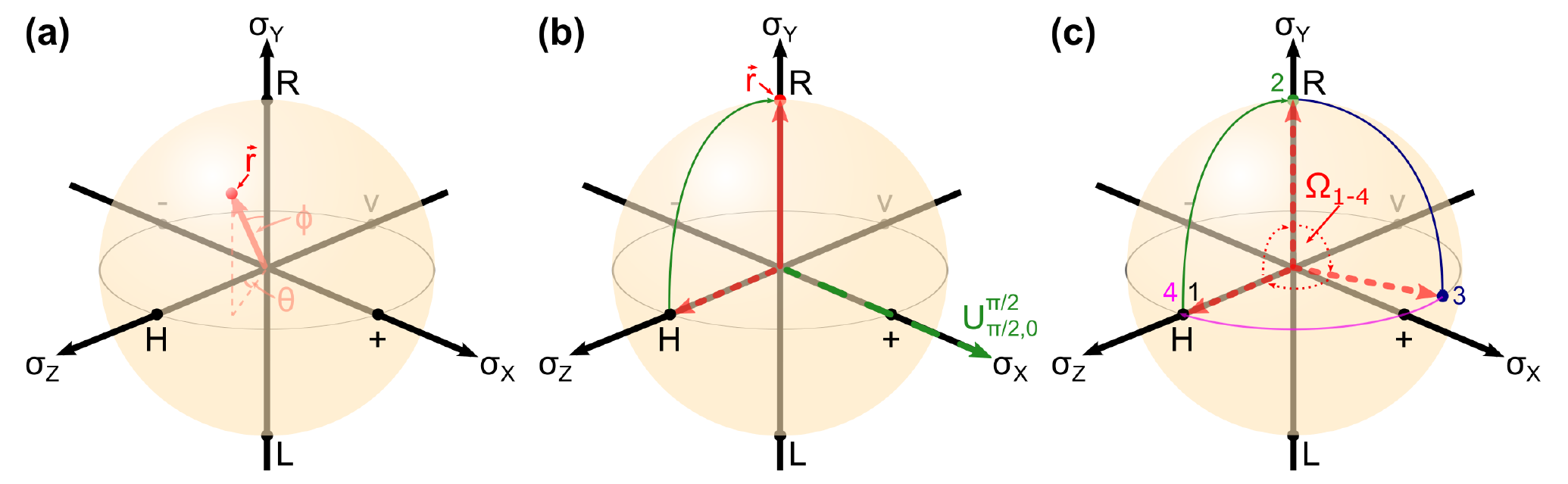}
\caption{\textbf{(a)} Bloch sphere with cartesian coordinates given by the Pauli matrices 
$\left\{ \sigma_X, \sigma_Y, \sigma_Z \right\}$. Each qubit state, represented by a Bloch vector $\vec{r}$, is completely determined by its projections onto the coordinate axis as well as by the spherical coordinate angles $\theta$ and $\phi$. Bloch vectors of pure states point onto the sphere's surface, mixed states are located inside the sphere.
\textbf{(b)} Illustration of a unitary transformation, rotating an \hpol-polarised state to \rpol\, polarisation. The rotation by angle $\omega$ happens in the plane orthogonal to the rotation vector $U^\omega_{\theta,\phi}$ (eq. \ref{eq_ch4_unitary_rot_1}), whose orientation in the Bloch sphere is given by the angles $\theta$ and $\phi$. The displayed example corresponds to a rotation by a $\lambda/4$-plate at a laboratory angle of $\vartheta = 45^\circ$ with respect to the linear input polarisation.
\textbf{(c)}: Bloch sphere representation of the Berry phase picked up by a signal transmitted through the polarisation interferometer. The undesired rotation $1\rightarrow 2$ is compensated by a $\lambda/4$-plate, $2 \rightarrow 3$ and a $\lambda/2$-plate, $3 \rightarrow 4$, whereby the full solid anlge $\Omega_{1-4}$ is the picked-up Berry phase.} 
\label{fig_ch4_qpt_theory}
\end{figure}

\paragraph{Polarisation qubits}
Polarisation information is usually encoded in the computational basis, which uses the horizontal (\hpol) and vertical (\vpol) directions of the light's electric field vector as the basis states for the polarisation vector space. 
Any arbitrary superposition of both basis states represents one possible information state $\ket{\phi} = a_H \ket{H} + a_V \ket{V}$, with $a_H, a_V \in \mathbb{C}$ and $\langle{\phi}\ket{\phi} = |a_H|^2 + |a_V|^2 = 1$, forming a polarisation qubit.
A convenient way to represent these states is the Bloch sphere\cite{Nielsen:2004kl}, shown in fig. \ref{fig_ch4_qpt_theory} \textbf{a}. 
All pure states $\ket{\phi}$ lie on the surface of this sphere. Its equator line contains all linearly polarised states. States positioned in the northern and southern hemispheres are elliptically polarised, with the poles representing right- and left-handed circularly polarised light. 
The location of a qubit on the Bloch sphere can be determined by spherical coordinates with respect to a Cartesian coordinate system, whose origin is located at the centre of the sphere. The state $\ket{\phi}$ is thus given by 
$\ket{\phi}=\cos(\frac{\theta}{2})\ket{H} + \sin{(\frac{\theta}{2})} \exp{(\text{i} \phi)}\ket{V}$.
Here $\theta$ and $\phi$ are the angles with respect the cartesian system's $z$- and $x$-axis, respectively. The intercepts of the coordinate axes with the sphere define the poles of the Bloch sphere. The polar states are the 3 standard polarisation basis states 
$\ket{\phi_{H,V}}~\in~\left\{ \ket{H}, \ket{V} \right\}$, 
$\ket{\phi_{+,-}}~\in~\left\{ \frac{1}{\sqrt{2}} \left(\ket{H} \pm \ket{V} \right) \right\}$ and 
$\ket{\phi_{R,L}}~\in~\left\{ \frac{1}{\sqrt{2}} \left(\ket{H} \pm \text{i} \cdot \ket{V} \right) \right\}$, 
which correspond to linear horizontal or vertical, diagonal or anti-diagonal, and right- or left-circular electric field vectors in the light's polarisation plane.
Defining the coordinate system in this manner is convenient, as the projections of any state $\ket{\phi}$ onto the cartesian axes are determined by its expectation value under application of the one of the Pauli operators 
$\sigma_X~=~\left( \begin{matrix} 0 & 1 \\ 1 & 0 \end{matrix} \right) $, 
$\sigma_Y~=~\left( \begin{matrix} 0 & -i \\ i & 0 \end{matrix} \right)$ and 
$\sigma_Z~=~\left( \begin{matrix} 1 & 0 \\ 0 & -1\end{matrix} \right)$ for 
$x$-, $y$- and $z$-axis, respectively. 
Since the location of $\ket{\phi}$ is completely defined by knowing these projections, measuring a polarisation qubit reduces to determining $\langle \sigma_j \rangle = \bra{\phi} \sigma_j \ket{\phi}$, $\forall j \in \left\{X, Y, Z \right\}$. The resulting values $\langle \sigma_j \rangle \in \left[-1,1\right]$ determine how far along the coordinate axis the state is positioned, e.g. $\bra{\phi} \sigma_Z \ket{\phi} = +1$ corresponds to \hpol-polarised light. 
We will see this mathematically below when discussing density matrices.
These measurements of the Pauli-spin operators are the basis of state tomography\cite{James:2001,Paris:2004kx}.

\paragraph{Polarisation rotations} 
The question is now, how the states $\ket{\phi}$ can be encoded experimentally in the first place and further, how the expectation values for $\sigma_j$ can be measured.
Experimentally, arbitrary qubit states are encoded by starting out in one of the computational basis states, e.g. with horizontally-polarised light, and subsequently rotating the polarisation using waveplates. 
Waveplates\cite{Langford:PhD, Kiesel:PhD} are birefringent uni-axial crystals\cite{Hecht}, whose refractive index ellipsoid is oriented such that the ordinary ($\vec{n}_o$) and extraordinary ($\vec{n}_e$) axes both lie in the transverse plane of the transmitted light. The refractive index contrast between the two orthogonal axes results in a phase shift $\omega$ 
between the components of light polarised along either axis. For $\lambda/2$- and $\lambda/4$-waveplates $\omega = \pi$ and $\omega = \pi/2$, respectively. 
The observed phase shift depends on the rotation angle $\vartheta$ of the waveplate's $\vec{n}_o$-$\vec{n}_e$-coordinate system with respect to the $\left\{ \ket{H}, \ket{V} \right\}$-coordinate system of the light polarisation. The maximum phase shift $\omega$ is experienced if linearly polarised input light is oriented at $\vartheta = 45^\circ$ with respect to the waveplate's $\vec{n}_o$ and $\vec{n}_e$ axes. 
Mathematically, polarisation rotations correspond to unitary transformations $U^\omega_{\theta, \phi}$, whose operators satisfy $\left(U^\omega_{\theta, \phi}\right)^{\dagger} \cdot U^\omega_{\theta, \phi} = \mathds{1}$. 
For single qubits the operators $U^\omega_{\theta, \phi}$ are $2\times2$-matrices. 
One possible basis set for the vector space $\left\{U^\omega_{\theta, \phi} \right\}$ of all such matrices is the Pauli matrices $\left\{ \mathds{1}, \sigma_X, \sigma_Y, \sigma_Z \right\}$. 
Arbitrary polarisation rotations $U^\omega_{\theta, \phi}$ can thus be expressed as a linear combination of Pauli-matrices\cite{Kiesel:PhD}, with
\begin{align}
\label{eq_ch4_unitary_rot_1}
U^\omega_{\theta,\phi} 	&= \text{i} \cos(\omega/2) \mathds{1} + \sin(\omega/2)\sigma_{\theta,\phi} \\
\sigma_{\theta,\phi}		&= \cos(\theta) \sigma_Z + \cos(\phi)\sin(\theta) \sigma_X + \sin(\phi) \sin(\theta) \sigma_Y.
\label{eq_ch4_unitary_rot_2}
\end{align}
In the Bloch sphere picture, the state $\ket{\phi}$ is rotated by an angle $\omega$ around an axis oriented at angles $\theta$ and $\phi$ with respect to the Bloch sphere's Cartesian axes (see fig. \ref{fig_ch4_qpt_theory} \textbf{b}). 
With the above equations, the effects of experimental rotations by waveplates can be predicted. 
In our experiments $\phi=0$, so the rotation axis is located in the equatorial plane\footnote{
	Generally, the waveplate's $\vec{n}_o$-$\vec{n}_e$ coordinate system is expressed 
	as a superposition of the electric fields coordinate system in the $\left\{\ket{H}, 	
	\ket{V} \right\}$-basis. 
	Since the waveplate can only be rotated around the optical propagation axis of the light, 
	i.e. in its transverse plane, 
	the coefficients in the superposition must be real numbers, so $\phi\overset{!}{=}0$.
}. 
Its angle $\theta$ with respect to the $\sigma_Z$-axis is given by the experimental angular settings $\vartheta$ as $\theta = 2\cdot\vartheta$. For example, a $\lambda/4$-plate at $\vartheta=45^\circ$ has its rotation axis along the $\sigma_X$ direction and all states on the Bloch sphere are rotated by $\omega = 90^\circ$. 
In the particular case shown in fig. \ref{fig_ch4_qpt_theory} \textbf{b}, the angle $\omega$ coincides with the angle $\phi$.
Using the following settings, computational basis states can be mapped to all 3 standard bases, i.e. to all poles of the Bloch sphere: 
\begin{itemize}
\item  $\lambda/2$-plate at $\vartheta = 45^\circ$: 		$\ket{H} \leftrightarrow \ket{V}$, 
$U^{\pi}_{\theta=\pi/2,0}= \sigma_Z = \left( \begin{matrix} 1& 0 \\ 0 & -1 \end{matrix}\right)$, 
\item  $\lambda/2$-plate at $\vartheta = 22.5^\circ$: 		$\ket{H} \leftrightarrow \ket{+}$, $\ket{V} \leftrightarrow \ket{-}$, 
$U^{\pi}_{\theta=\pi/4,0}=\frac{1}{\sqrt{2}} \left( \sigma_X + \sigma_Z \right) = \frac{1}{\sqrt{2}} \left( \begin{matrix} 1 & 1 \\ 1 & -1 \end{matrix} \right)$ 
\item  $\lambda/4$-plate at $\vartheta = 45^\circ$: 		$\ket{H} \leftrightarrow \ket{R}$, $\ket{V} \leftrightarrow \ket{L}$,
$U^{\pi/2}_{\theta=\pi/2,0}= \frac{1}{\sqrt{2}} \left(\text{i} \cdot \mathds{1} + \sigma_{X} \right) = \frac{-i}{\sqrt{2}} \left( \begin{matrix} 1 & \text{i} \\ \text{i} & 1 \end{matrix}\right)$.
\end{itemize}
To prepare a polarisation state $\ket{\phi}$, we use an arrangement of optics consisting of a PBS, for initial state preparation in $\ket{H}$, followed by a $\lambda/2$- and a $\lambda/4$-plate as a polarisation preparation set-up. 
Therewith, any arbitrary polarisation state can be generated, particularly the 6 pole states $\text{\suet{B}} = \left\{\ket{H},\ket{V},\ket{+},\ket{-},\ket{R},\ket{L} \right\}$. 
Reversing the optics allows mapping of any state $\ket{\phi}$ onto the computational basis.
In fact, it allows to measure any observable $\langle \sigma_{\theta,\phi}\rangle$ of state $\ket{\phi}$. Detecting the fraction of the signal at each output port of the PBS with respect to the input yields the probability of observing each eigenvalue of the observable $\sigma_{\theta,\phi}$. 
This is expressed by the projector $\mathbb{P}^{\pm}_{\theta,\phi} = \ket{\phi_{\theta,\phi}}  \bra{\phi_{\theta,\phi}} = \frac{1}{2}\left( \mathds{1} \pm \sigma_{\theta,\phi} \right)$, where $\pm$ denotes transmittance of either one of the eigenstates of $\sigma_{\theta,\phi}$ through the PBS.
Notably, $\ket{\phi_{\theta,\phi}}$ is the state on the Bloch sphere that would be produced by going through the analyser backwards\footnote{
	Going through the analyser backwards means, in a Gedankenexperiment, the 
	analyser is used as a polarisation preparation stage, which produces $\ket{\phi_{\theta,\phi}}$ 
	from an \hpol\,-polarised input state. 
	Here the \hpol\,-polarisation would first be transmitted through the PBS and then rotated by the 
	$\lambda/2$- and $\lambda/4$- plate onto $\ket{\phi_{\theta,\phi}}$.
	The waveplate settings applied in this case will correspond to the required angles for mapping
	the incoming state $\ket{\phi_{\theta,\phi}}$ onto $\ket{H}$, when using the system as a polarisation
	analyser in the actual experiment. 
}.
With the above stated waveplate settings, a qubit's projection onto each of the 3 Bloch sphere coordinate axes can be analysed.

\paragraph{Density matrix}
So far qubits have been expressed as pure states $\ket{\phi}$. Since we aim to study their storage in a memory, any shortcomings in the storage process will degrade the polarisation information by introducing statistical mixtures between polarisation states. For this reason, we need to describe $\ket{\phi}$ by its density matrix
$$
\rho = \ket{\phi}\bra{\phi} = |a_H|^2 \ket{H}\bra{H}  +  a_H a_V^* \ket{H}\bra{V} +  a_V a_H^* \ket{V}\bra{H} + |a_V|^2 \ket{V}\bra{V} = \left(
\begin{matrix}
  |a_H|^2 & a_H a_V^* \\
   a_V a_H^* & |a_V|^2
 \end{matrix}
 \right).
$$ 
Its diagonal terms are the state populations, i.e., the probabilities of observing $\phi$ in either one of the basis states. 
Since these must sum to unity the process matrix must have $\tr(\rho)=1$. The population probabilities also cannot be negative, for which reason $\rho$ has to be positive semi-definite, i.e., its eigenvalues must be positive real numbers. 
The off-diagonal elements are called coherences. They describe interference effects between the populated states which can appear when they are superimposed. Reducing the coherences will result in the state becoming increasingly mixed. This is illustrated by the following example, where decoherence reduces a diagonal state to white noise: 
\begin{equation}
\rho_+ =\ket{+}\bra{+} = \frac{1}{1} \left( \begin{matrix} 1 & 1 \\ 1 & 1 \end{matrix} \right) 
\xrightarrow{\text{coh} \rightarrow 0}  
\frac{1}{2} \left(\begin{matrix} 1& 0 \\ 0 & 1 \end{matrix} \right) =
\frac{1}{2} \ket{H} \bra{H} + \frac{1}{2} \ket{V}\bra{V} =  \frac{1}{2} \rho_H + \frac{1}{2} \rho_V.
\label{ch4_eq_rho_decohere}
\end{equation}
A mixed state $\rho_\text{mix}$ is thus separable into a sum of different states $\rho_i$, each observable with probability $p_i$, such that $\rho_i = \underset{i}{\sum} p_i \rho_i$.
Expectation values of operators $\mathbb{O}$ are given by 
$\langle \mathbb{O}_\rho \rangle = \langle \mathbb{O} \rho \mathbb{O}^\dagger \rangle = \tr(\mathbb{O}\rho)$.

\subsection{State tomography\label{ch4_subsec_state_tomo}}
Since the density matrix describes the polarisation qubit it also represents a state on the Bloch sphere. 
Simultaneously, the density matrix is also an operator, whose form for a single qubit is a $2 \times 2$-matrix, irrespective of the chosen basis. 
Consequently, it can be expressed as a linear combination of the Pauli-matrices, just as the unitary transformations $U^\omega_{\theta, \phi}$: 
\begin{equation}
\rho = \frac{1}{2}\mathds{1} + \vec{r} \cdot \vec{\sigma} = \frac{1}{2} \mathds{1} + r_X \sigma_X + r_Y \sigma_Y + r_Z \sigma_Z = \left( \begin{matrix} \frac{1}{2}  + r_Z & r_X - \text{i} \cdot r_Y \\ r_X + \text{i} \cdot r_Y & \frac{1}{2}  - r_Z  \end{matrix} \right)
\label{eq_ch4_rho_bloch}
\end{equation}
Here $\vec{r}$ is the Bloch vector\cite{Nielsen:2004kl}, whose entries are the projections of the density matrix onto the coordinate axes of the Bloch sphere. 
Eq. \ref{eq_ch4_rho_bloch} is the mathematical counterpart to fig. \ref{fig_ch4_qpt_theory}, illustrating that the state of a single qubit is completely defined by knowing $\vec{r}$. 
Measurements of the expectation values for each Pauli-matrix $\langle \sigma_i\rangle = \tr(\sigma_i \rho)$ 
yields the respective component\footnote{
Written as projectors onto the Cartesian coordinate system in the Bloch sphere, these are:
 \begin{align}
 r_X = \langle \sigma_X \rangle = \tr(\sigma_X \rho) &= \bra{+} \rho \ket{+} - \bra{-} \rho \ket{-} \nonumber \\
 r_Y = \langle \sigma_Y \rangle = \tr(\sigma_Y \rho) &= \bra{R} \rho \ket{R} - \bra{L} \rho \ket{L} \nonumber \\ 
 r_Z = \langle \sigma_Z \rangle = \tr(\sigma_Z \rho) &= \bra{H} \rho \ket{H} - \bra{V} \rho \ket{V}  \\
 \end{align}
}
$r_i$. 
Thus, measuring the projectors on all three axes ($\forall i \in \left\{ X,Y,Z \right\}$), on multiple copies of the polarisation qubit, allows direct reconstruction of its density matrix $\rho$.
This is called quantum state tomography (QST). 
Notably, density matrices cannot be determined with a single-shot measurement, since the non-zero commutation relation\cite{Nielsen:2004kl}
$\left[ \sigma_i,\sigma_j \right] = \text{i} \cdot \epsilon_{i,j,k} \cdot \sigma_k$ prevents accurate, simultaneous recording of all terms in eq. \ref{eq_ch4_rho_bloch}.
We will see below, that QST is an integral part of QPT. In fact, QPT characterises a quantum system by performing state tomography on the system's output for a set of input states, which form a complete basis set. 
Since density matrix reconstruction for QST is simpler to understand, we will briefly outline the algorithm we use to obtain the state $\rho$ from the experimental data. 
Extending this concept will then enable to find the process matrix $\chi$ (see appendix \ref{ch4_subsec_QPT}).

\paragraph{From counts to projectors} 

The expectation values $\langle \sigma_i \rangle$ are determined by projective measurements $\mathbb{P}^{\pm}_{\theta,\phi}$ with the polarisation analysis set-up, using waveplate settings $\theta$, $\phi$ to project onto all basis states in $\text{\suet{B}}$. 
The resulting eigenvalue of $\langle \sigma_{\theta,\phi} \rangle$ can either be $\pm1$. 
Experimentally, each analysis setting leads to a count rate $c_{\theta,\phi}$ on a detector at the output of the polarisation analysis set-up. 
When used to observe actual single photon states, one records the number of detection events $c_{\theta,\phi}$ per measurement time $\Delta t_\text{meas}$.
For our bright coherent state signals, the detected pulse intensity on the Menlo PD is the equivalent to this count rate. 
The intensity is proportional to the integrated pulse areas $A_{\theta, \phi} := A^{x,y}_{k,t}$ of the scope traces, where we momentarily drop the indices $k$ and $t$ for the measurement setting and the time bin\footnote{
	Indices $x$ and $y$ corresponds to the settings of the polarisation preparation and analysis stages, 
	respectively, as defined in section \ref{ch4_subsec_exp_implementation}.
}.
Assigning with $+$ and $-$ the results that correspond to either eigenvalue of $\langle \sigma_{\theta,\phi}\rangle$, i.e. the expectation value when applying the respective projector $\mathbb{P}^{\pm}_{\theta,\phi}$ to the output state $\rho$ to measure along an axis at $i=\left\{ X, Y, Z \right\}$ in the Bloch sphere (under the respective andlges $\left\{ \theta, \phi \right\}$, 
we obtain the probability for either eigenvalue\footnote{
	The experimentally determined values for $p$ are frequencies of occurrence for a specific measurement outcome. They are only 
	estimates of the actual probability one would obtain for projection of $\rho$. In the limit of infinite measurement time and under absence 
	of measurement errors, both values would be identical.
}  $p^{\pm}_i = \langle \mathbb{P}^{\pm}_i \rangle = \tr{\left(\mathbb{P}^{\pm}_i \rho\right)} \approx \frac{A^\pm_i}{ A^{+}_i + A^{-}_i} $ and $r_i = \langle \sigma_i \rangle =  (+1) \cdot p^{+}_i + (-1) \cdot p^{-}_i$.

Assuming a constant photon flux, i.e., $A_i^+ + A_i^- = \text{const.} \forall i$, 
$\vec{r}$ can be determined with only the four settings\cite{Nielsen:2004kl, James:2001, Langford:2004} 
$\text{\suet{B}}_1 = \left\{\ket{H},\ket{V},\ket{+},\ket{R} \right\}$, since $A^{-}_{\left\{X,Y\right\}} = \left( A_{Z}^+ + A^-_{Z} \right) - A^+_{\left\{X,Y\right\}}$. However this assumption is not good here, due to experimental instabilities causing count rate drifts during the measurement sequence (see section \ref{ch4_exp_imp}). For this reason, the full, over-complete measurement set\cite{Kiesel:PhD, Wieczorek:2009kx} $\text{\suet{B}}$ is used. With the over-complete set, we can additionally set $\langle \mathds{1}\rangle~=~\underset{i}{\sum} \frac{1}{3}\left(A^{+}_i + A^{-}_i \right)$.

\paragraph{Maximum likelihood reconstruction}
The direct reconstruction for $\rho$ described thus far is still sensitive to experimental uncertainties from fluctuations and drifts in $A^\pm_i$ between measurements on different polarisation settings. This can lead to density matrices $\rho$, which are not necessarily positive and therefore unphysical. 
A remedy is to fit a theory prediction for the projection probabilities 
$\textfrak{p}^\pm_i = \langle \mathbb{P}^\pm_i \rangle = \tr{\left( \mathbb{P}^\pm_i \rho_\text{th} \right)}$, obtained by assuming a density matrix $\rho_\text{th}$, to the actually observed probabilities $p^\pm_i$. 
The optimisation for $\rho_\text{th}$ runs over the set of all physical density matrices. 
This means, we look for the physical density matrix that describes the observed data the closest. 
For this purpose a maximum likelihood algorithm\cite{James:2009} is used, built on the assumption of normally distributed datapoints $A^\pm_i$ with a standard deviation\footnote{
	The Poissonian error assumption arises from the origins of the code used here. 
	Since the algorithm was originally designed for single photon measurements, 
	it assumes Poissonian counting statistics.	
} $\sigma^A_i = \sqrt{A^\pm_i}$. 
The algorithm first calculates a set of count rates $\mathcal{A}^\pm_i = A^\text{tot}_i \cdot \textfrak{p}^\pm_i$, which one would expect to observe with the assumed density matrix $\rho_\text{th}$. 
Here, $A^\text{tot}_i = A^{+}_i + A^{-}_i$ represents the total number of counts for each analysis basis $\sigma_i$. 
These are combined to a likelihood function \cite{Langford:PhD,James:2009} $\mathcal{L}$, describing the probability that the predicted count rates $\mathcal{A}^\pm_i$ correspond to the measured rates ${A}^\pm_i$: 
\begin{equation}
\mathcal{L}(p|\rho_\text{th}) \sim \underset{i,j }{\Pi} \exp{\left( -\frac{ \left( A^j_i - \mathcal{A}^j_i \right)^2 }{2 \left( \sqrt{\mathcal{A}^j_i} \right)^2 }\right)} = \underset{i,j }{\Pi} \exp{\left( -\frac{ A^\text{tot}_i \left( p^j_i - \textfrak{p}^j_i \right)^2 }{2 \textfrak{p}^j_i } \right)},
\label{eq_ch4_penalty_func}
\end{equation}
with $ i \in \left\{  \mathds{1}, X,Y,Z \right\}$, $j \in \left\{ +, - \right\}$. 
The algorithm maximises this likelihood function. 
To start the optimisation process, an initial guess value for $\rho_\text{th}$ is required, for which the density matrix, obtained by direct reconstruction, is used\cite{Gilchrist:2005}. 
In this case its potential unphysicality is irrelevant, because it is only used as a starting point. 
To allow for efficient reconstruction, the applied numerical algorithm\cite{Gilchrist:2005,Langford:PhD} employs convex\footnote{
	Note $\mathcal{L}\sim \exp{(x)}$ is a convex function since $\frac{\text{d}^2}{\text{d}x^2}\mathcal{L} = \text{const.}  >0$. 
} optimisation\cite{boyd:convexopt} with semi-definite programming\cite{Vandenberghe:1996}, where the constraints imposed on to the maximisation function are the three physicality requirements\cite{Nielsen:2004kl} for density matrices $\rho$ (see appendix \ref{ch4_subsec_qubits} above). 
The algorithm has been developed by 
\textit{Nathan Langford}\cite{Nielsen:2004kl, Langford:2004, Barreiro:2005, Langford:2005}, who thankfully made it available for the analysis of our data. 
It has not been modified during the course of this work, for which reason further details are left to the listed references.

\paragraph{Purity} 
White noise states, i.e., absolutely mixed states, such as eq. \ref{ch4_eq_rho_decohere}, have $\vec{r}=0$ and are positioned at the origin of the Bloch spehere\footnote{
	This can readily be seen for the state $\rho_\text{mix} = \frac{1}{2} \left(\rho_H + \rho_V\right)$ in eq. \ref{ch4_eq_rho_decohere}, where $\rho_H = \ket{H}\bra{H}$ and $\rho_V=\ket{V}\bra{V}$. 
	Here, $\langle \sigma_X \rangle =  \langle \sigma_Y \rangle = 0$ because $\rho_\text{mix}$ has no off diagonal elements; $\langle \sigma_Z \rangle = 0$ since 
	both diagonal elements are $1$.
}, while pure states lie on the surface of the Bloch sphere.
Consequently, the distance between the sphere's origin and surface is a measure for the purity of a state. 
Mathematically, the purity is defined as $\mathcal{P} = \tr{(\rho^2)}$, whereby $\mathcal{P}=1$ for pure states and $\mathcal{P}=\frac{1}{d}$ for white noise, where $d$ is the dimensionality of the system ($d=2$ for the example in eq. \ref{ch4_eq_rho_decohere}). 
Accordingly, every physical density matrix has $\tr{(\rho^2)} \in \left[\frac{1}{d},1\right]$.

\paragraph{Fidelity}
Another benchmark for a quantum state is the fidelity, which is a distance measure\cite{nielsen2000qca} for determining how far a quantum state $\ket{\phi}$ is away from a target state $\ket{\psi}$ within the same underlying Hilbert space. 
The definition of fidelity is very simple and instructive in the case of pure states. Here it just corresponds to the projection onto the target state, i.e., $\mathcal{F} = | \langle \phi | \psi \rangle|^2$. 
For mixed states with density matrices $\rho_\phi$ and $\rho_\psi$, the expression becomes more complicated. 
It can be shown \cite{Jozsa:1994, nielsen2000qca} that the fidelity is generally  given by\cite{OBrien:2004} 
$\mathcal{F} = \left( \tr{\left(  \sqrt{ \sqrt{\rho_\phi} \rho_\psi \sqrt{\rho_\phi} } \right)} \right)^2$. 
The fidelity is bounded by $0\le \mathcal{F} \le 1$, with $\mathcal{F}=1$, if both states are identical.\newline
In order to operate in the quantum regime, a memory must preserve the input signal. The fidelity between input and read-out must overcome\cite{Massar:1995} $\mathcal{F}_B=2/3$. Importantly, this boundary is only valid for single qubits in a Fock state, stored in a noise-free memory, with unity efficiency. 
As far as real-world systems are concerned, both of these assumptions do not hold and the boundary fidelity needs to be modified\cite{Gruendogan:2012}. Appendix \ref{app_ch4_fidelity_bounds} discusses the relevant modifications and establishes $\mathcal{F}_B$ for the dual-rail Raman memory under the presence of noise. In section \ref{ch4_subsec_outlook} below, we will utilise these updated bounds to test our memory's performance and to derive the noise level required for faithful operation in the quantum regime.

\subsection{Quantum process tomography (QPT)\label{ch4_subsec_QPT}}

QPT describes the operation of a quantum system, e.g. the memory, as a linear map $\mathcal{E}$ of an input state $\rho_\text{in}$ onto an output state $\rho_\text{out}$, so $\rho_\text{in} \overset{\mathcal{E}}{\rightarrow} \rho_\text{out}$. 
This type of black-box approach is a method to treat open quantum systems, where a principal system, comprising the signal of interest $\rho_\text{sig}$, interacts with an environment $\rho_\text{env}$, whose state changes are ignored. 
Including the environment results in a closed system, within which the process is represented by a unitary transformation\footnote{
	Since no information leaves the combined system, the effects of the process can be considered as state 
	rotations between the principal system and the environment. These are reversible, for which reason the process is unitary.
} 
$U$, operating on the state $\rho = \rho_\text{in} \otimes \rho_\text{env}$ in the combined Hilbert space of both systems. 
Subsequent restriction to observations on the principle system will result in an open quantum system, i.e., all states of the environment are traced out of the combined system's output state after the unitary process dynamics.  
In our case, $\rho_\text{in}$ corresponds to the polarisation state of the input signal, while the memory medium and control represent the environment. 
Using this black-box mapping formalism is an \textit{ex-post} description of the storage process, neglecting time dynamics of the Raman interaction.
Storage and read-out in each time-bin are consequently treated as unrelated events, resulting from independent processes acting on the input signal $\rho_\text{in}$. 
So, effects measured on the transmitted signal cannot be used to infer any properties of the retrieved signal. The same applies between the read-out time bins. 
Consequently, the map $\mathcal{E}$ is determined separately for all time bins $t$.
Assuming an orthonormal basis $\left\{ \ket{e_k} \right\}$ for the environment, which is initially prepared in $\rho_\text{env} = \ket{e_0}\bra{e_0}$,
the process can be written as
\begin{align}
\rho_\text{out} &= \mathcal{E}(\rho_\text{in})= \tr_\text{env} \left( U \left( \rho_\text{in} \otimes \rho_\text{env} \right) U^\dagger \right)= \underset{k}{\sum} \bra{e_k} U \left(\rho_\text{in} \otimes \rho_\text{env} \right) U^\dagger \ket{e_k} \nonumber\\
&= \underset{k}{\sum} \mathbb{E}_k \rho_\text{in} \mathbb{E}_k^\dagger = \underset{k}{\sum} p_k \mathcal{E}_k \left(\rho_\text{in} \right),
\label{eq_ch4_qpt1}
\end{align}
where the resulting operation elements $\mathbb{E}_k$, acting on the principal system, are given by $\mathbb{E}_k = \bra{e_k} U \ket{e_0}$. 
Ignoring the environment leads to a mixed output state $\rho_\text{out}$, even for a pure input state, because each operation $\mathcal{E}_k$ occurs with probability $p_k = \tr_\text{sys}{\left(\mathbb{E}_k \rho_\text{in} \mathbb{E}_k^\dagger \right)}$.
Eq. \ref{eq_ch4_qpt1} is referred to as the operator sum representation or Kraus representation\cite{Kraus:1983}.
If no signal information is lost, the process is trace preserving with $\tr_\text{sys} \left(\rho_\text{out} \right) = 1$, i.e., all probabilities sum to one with $\underset{k}{\sum} p(k)=1$ and the set of operators is complete such that $\underset{k}{\sum} \mathbb{E}_k^\dagger \mathbb{E}_k = \mathds{1}$. 
In our case, the process is not necessarily trace preserving, because individual photons can be lost from the coherent state input signal. This can, e.g., be due to resonant absorption by the \cs\, atoms or just by residual signal leakage into the control spatial modes on the output PBD.
Assigning $N = |\left\{ \mathbb{E}_k \right\}|$, there can generally be up to $1\le N \le d^2$ operation elements in total for an input signal of dimensionality $d$, leading to $N$ Kraus operators\footnote{
	Since the dimensionality $d$ of the principle system's Hilbert space is the 
	same before and after the interaction, there can at most be $d$ 
	linearly independent projectors for each one of the basis states in the principle system's Hilbert 
	space onto another basis state. In tracing 
	out all environment dimensions, each summand in eq.~\ref{eq_ch4_qpt1} corresponds to one such 
	projection, for which reason there are at most $d^2$ summands and operators $\mathbb{E}_k$.
}. For a single qubit, there are thus four linearly independent $\mathbb{E}_k$, each of which is a  $2\times2$ matrix. 

\paragraph{Measurement}
Similar to state tomography, there is unitary freedom\cite{nielsen2000qca} in the choice for a basis representation of the operators $\mathbb{E}_k$. 
So they can be expressed in terms of the Pauli-matrices $\left\{ \sigma_i \right\}$, i.e., $ \mathbb{E}_k = \underset{i}{\sum} \mu_{k,i} \sigma_i$. 
In the resulting operator sum representation 
$$
\mathcal{E}\left( \rho_\text{in} \right)~=~\underset{k,k'}{\sum} \underset{i}{\sum} \mu_{k,i} \mu^{*}_{i,k'} \sigma_k \rho_\text{in} \sigma^{\dagger}_{k'}=\underset{k,k'}{\sum} \chi_{k,k'} \cdot \sigma_k \rho_\text{in} \sigma_{k'}
$$ 
we are now left with the process matrix $\chi$, whose elements represent the probability $p_k$ for a specific Bloch sphere rotation to occur on the input state. Since the rotations $\sigma_k$ are defined, knowledge of $\chi$ is sufficient to characterise the process completely. 
For any input state, we can determine these probabilities by just performing polarisation analysis with the same settings $\text{\suet{B}}$ used in QST.
Moreover, the linearity of $\mathcal{E}$ allows to determine the map for any arbitrary input state by knowing $\mathcal{E}$ for a linearly independent basis set of input states. Since we are investigating single qubits, for which $\rho$ is a $2\times2$-matrix and hence in the span of $\left\{ \sigma_i \right\}$, the polarisation states in $\text{\suet{B}}$ are one such set. 
Therewith, the process map can be established using each pure polar state of the Bloch sphere as a memory input signal. 

In summary, single qubit QPT can be performed by sending in all Bloch sphere pole states and analysing each of the respective outcomes with the basis settings $\text{\suet{B}}$, which also correspond to all Bloch sphere pole sates. 
These measurements suffice to determine $\chi$ and therewith $\mathcal{E}$.
In other words, we perform state tomography on the output states $\rho_\text{out}$ for all input states $\rho_\text{in} \in \text{\suet{B}}$.
This amounts to a total sequence of $36$ measurements, each yielding a pulse area $A^{\pm}_{k,k'}$ for input polarisation state $k \in \text{\suet{B}}$ and analysis polarisation basis $k' \in \text{\suet{B}}$. 

Notably, usage of the integrate pulse areas $A^{\pm}_{k,k'}$ without any normalisation to the {\tisa} repetition rate $f_\text{rep}$, i.e. the number of conducted experiments, effectively makes us conduct measurements in post-selection. Post-selection means that experimental trials are considered only if they have resulted in the registration of an output event; here the detection of a photon on the Menlo PD. 
Experimental trials corresponding to signal photons which were lost, for instance, due to memory loss or the finite transmission of the signal path between memory and detector, are disregarded. 
If our experiment was a photon counting measurement, post-selection would translate into the direct usage of the detected number of photons, without transforming these into a detection rate through normalisation by the number of experimental trials (see chapter~\ref{ch6}). 
For the bright coherent states used here, this is equivalent to usage of the integrated pulse areas $A^{\pm}_{k,k'}$.

In contrast to QST, data evaluation is not concerned about trying to determine the output state $\rho_\text{out}$. The tomography rather aims to reconstruct the elements of $\chi$ which cause the observed events $A^{\pm}_{k,k'}$ under input of a density matrix $\rho_{\text{in},k}$. 
Notably, the process matrix must satisfy similar physicality constrains as state density matrices. It must be Hermitian, positive and, in case $\mathcal{E}$ is trace preserving, it must have unit trace\footnote{
	$\mathcal{E}$ is basis invariant and $\chi$ Hermitian, so there is a basis in which $\chi$ is diagonal. Here each diagonal element, i.e. each eigenvalue, corresponds to the 
	probability of the process returning one of the eigenstates of $\chi$ as the output. For this reason, $\chi$ must be positive. If no signal information is lost, i.e.
	$\mathcal{E}$ is trace preserving, there must be a process output $\rho_\text{out}$ for each trial of sending in an input. So the probabilities for different operations 
	$\mathbb{E}_k$ must sum to one and $\tr{\left( \chi \right)} \overset{!}{=} 1$.  
}
$\tr{\left(\chi \right)} = 1$.
Direct reconstruction from count rate data\cite{nielsen2000qca,Chuang:1997} can thus once again return unphysical process matrices. Similar to QST, the directly reconstructed matrix serves instead as an initial guess for fitting a physical matrix $\chi$ to the measured values $A^\pm_{k,k'}$, using the same maximum likelihood approach as outlined for QST in appendix \ref{ch4_subsec_state_tomo} above. 
Using the same notation, the theoretical expectation values are given by $\mathcal{A}_{k} = A^\text{tot}_k \cdot \chi_{k,k'} \cdot \langle \sigma_k \rho_\text{in} \sigma_{k'} \rangle$, with the total measured pulse area $A^\text{tot}_{k} = A^{+}_k + A^{-}_k$ and $k \in \left\{ X, Y, Z, \mathds{1} \right\}$.
The important difference to QST is, that here the $\chi_{k,k'}$ are the fit variables and the expectation values $\langle \sigma_k \rho_\text{in} \sigma_{k'} \rangle$ are constants, since the states $\rho_\text{in}$ are known with certainty. 
These quantities enter the maximum likelihood algorithm using the cost function $\mathcal{L}$ of eq. \ref{eq_ch4_penalty_func}. Again the problem is solved by convex optimisation with semi-definite programming, where the constraints are the physicality conditions of $\chi$.

\paragraph{Process purity}

Thanks to their similarity to density matrices\cite{nielsen2000qca}, the definitions of purity and fidelity can be applied analogously\cite{Gilchrist:2005} to process matrices $\chi$. 
Here, the process purity $\mathcal{P} = \tr{\left\{ \chi^2 \right\}}$ measures how much mixture the quantum process adds to an input state. A purity of $\mathcal{P}=1$ means, that pure input states $\rho_\text{in} = \ket{\phi} \bra{\phi}$ remain pure at the output. Conversely, for a purity of 
$\mathcal{P}=1/2$  
the output of a process, operating on a single qubit, will be white noise, irrespective whether the input state was pure. 

\paragraph{Process fidelity}
The fidelity\cite{Gilchrist:2005} 
$\mathcal{F}\left( \chi, \tilde{\chi} \right) = \left( \tr{\left( \sqrt{\sqrt{\chi} \tilde{\chi} \sqrt{\chi} } \right)}\right)^2$ 
is a distance measure between two processes. 
Analogously to the state fidelity, it describes to which degree a process $\chi$ resembles a benchmark process $\tilde{\chi}$. 
For the memory, negligible influence from the storage process is desired. 
So optimal performance corresponds to a process matrix that is the identity,
i.e., $\tilde{\chi} = \chi_\text{ideal} = \mathds{1}$. 
Besides the storage itself, the propagation of light through the polarisation interferometer can also result in a modification of the polarisation state. Interferometer transmission may thus also be a non-ideal process. 
Hence, comparing the process matrix $\chi_{scd}$ for setting \textit{scd}, i.e. when the Raman interaction is turned on, 
with $\chi_\text{sd}$, obtained for light transmitted through the interferometer under setting \textit{sd}, is representative for the quality of polarisation storage. 
This gives the best performance one can obtain when turning on the memory. 
We reconstruct both matrices from experimental data and express the memory process fidelity as 
$\mathcal{F} = \left(\tr{\left( \sqrt{\sqrt{\chi_{scd}} \chi_{sd} \sqrt{\chi_{scd}}} \right)}\right)^2$.

\paragraph{Errors}
Propagating the errors in the observed pulse intensities $\Delta A_{k,k'}$ through the tomographic reconstruction is already complicated for QST\cite{James:2001} and becomes prohibitively complex for QPT. Stochastic error estimation using Monte-Carlo simulations of the reconstructed matrices $\chi$ under variation of the experimental input parameters 
$A^\pm_{k,k'}$ on the other hand is reasonably straight forward. Using the latter technique, we run the numerical algorithm 1000 times for each tomography measurement by drawing the input parameters $A^\pm_{k,k'}$ to eq. \ref{eq_ch4_penalty_func} from a normal distribution, whose mean and standard deviation are set to the experimentally measured average pulse area $A_{k,k'}^\pm$ and error $\Delta A_{k,k'}^\pm$, respectively (see \ref{ch4_subsec_exp_implementation}). 
From the resulting set of process matrices, we calculate the distribution of fidelity values. Their respective standard deviation is used as error onto the process fidelities. 
Note that the value for $\mathcal{F}$ is obtained from reconstruction using the mean pulse areas $A_{k,k'}$, which avoids small deviations from the finite number of Monte-Carlo samples.

\section{Shortcomings in experimental setup\label{app_ch4_probs_setup}}
The experimental lay-out for the polarisation storage experiment differs from that used in chapters \ref{ch6} \& \ref{ch7}. As polarisation storage was the first major experiment, conducted just after the initial proof-of-principle demonstration of our {\cs}-based Raman memory, the experimental apparatus naturally still suffered from insufficiencies that were resolved in the second generation of Raman memory experiments, described in chapters \ref{ch6} \& \ref{ch7}. 
Most notably, the active beam and frequency stabilisations of the \tisa\, laser were not yet implemented. On the one hand, this meant that the \tisa\, frequency was drifting continuously; an effect which was magnified by the temperature instability of the lab environment. 
On the other hand, also the beam pointing of the \tisa\, output changed, whenever the \tisa\, frequency was modified. While the size of these effects were not completely known yet when performing these experiments, the free-space arrangement for the control meant that the overlap between signal and control drifted.
Furthermore, the \pockels\, system had not been optimised yet, for which reason the spatial mode coming out of the \pockels\, was distorted, due to having too large a beam diameter inside the crystal. 
While the SMF-coupling of the signal cleaned up its mode and stabilised the signal's beam path further downstream, it reduced the mode overlap with the still distorted control. Unfortunately, SMF-coupling the control was not possible due to the associated power loss, which would have severely reduced the obtainable memory efficiencies.
For budget reasons, at the time of conducting the experiment, there was no possibility of getting the required active stabilisation. The results were a measurement time, which was limited by system drifts and the requirement to completely realign the entire experimental apparatus on a daily basis.

\section{Pockels cell alignment\label{app_ch4_PC}}
The \pockels\, setup during the polarisation storage experiment differs from the system employed in chapters \ref{ch6} and \ref{ch7}. 
At the time when performing the former experiment, the \pockels\, had not yet been characterised thoroughly, but was just taken over from the initial proof-of-principle demonstration of the Raman memory\cite{Reim2010, Reim:2011ys}. 
As it turned out later on, there existed unnoticed inefficiencies, which were corrected only when the experiment was rebuilt during a laboratory move.
In this section, we briefly describe the issues that were found with the initial \pockels\, setup. It is intended for the benefit of someone, who plans to rebuild the experimental apparatus and tries to avoid running into similar problems. 

\begin{enumerate}
\item
The \pockels\, was used in single pass, although it is designed only for $\lambda/4$ phase retardation. 
Using the maximum available voltage of $5.2 \, \text{kV}$, applied to the \pockels\, crystal, pulses could only be picked with $< 90\,\%$ efficiency. 
The remedy for this problem was to move to the double pass configuration used in chapters \ref{ch6} and \ref{ch7}. 
There the \pockels\, performs a $\lambda/4$ phase rotation upon every pass, which together with the $\pi$-phase shift, picked up by light reflected off a dielectric surface of higher refractive index, amounts to a $90^\circ$  rotation of linearly polarised input light. 
In this configuration, only the quarter-wave voltage of $\sim 3 \, \text{kV}$ needs to be applied across the \pockels\, crystal.
Notably, when using such a double-pass configuration in the same spatial mode, i.e. reflecting the beam path back into itself with a mirror, all non-picked pulses are sent straight back into the \tisa\, laser cavity. This breaks up the mode locking. Both paths in the \pockels\, crystal must nevertheless be parallel. For this reason the double-pass is implemented using a retro-reflecting prism instead of a mirror, which introduces a constant displacement between the beams going in either direction. Also the beam divergence between both passes should be small, so the prism needs to be placed closely behind the \pockels\, output face.

\item
There was rotational misalignment between the \pockels\, crystal axes and the pair of crossed polarisers surrounding the \pockels\, in single pass configuration. This means, the uniaxial \pockels\, index ellipsoid was not aligned with respect to the coordinate system (\hpol-\vpol\,-system) defined by the two polarisers. 
Without any voltage applied, the ellipsoid needs to lie along the optical axis, i.e. the refractive index corresponds to the ordinary index $n_o$ and is rotationally symmetric in the transverse polarisation plane of the transmitted light. 
Voltage application tilts the ellipsoid with respect to the optical axis, leading to a non-circularly symmetric index profile in the transverse plane. 
As a result, one polarisation component of the transmitted light sees the extraordinary index (direction $\vec{n}_e$) while the orthogonal direction still experiences the ordinary index (direction $\vec{n}_o$).
To achieve pulse picking, the $\vec{n}_e$-$\vec{n}_o$ coordinate system must be rotated by $45^\circ$ with respect to the \hpol-\vpol system, such that a linearly polarised input state sees the maximum index contrast between its two components polarised along $\vec{n}_e$ and $\vec{n}_o$, respectively. 
In practice, rotational misalignment between both coordinate systems cannot be avoided, since the \pockels\, crystal mount offers no rotational control. 
For this reason, $\lambda/2$ waveplates need to be inserted at either end of the \pockels\,, when it is used in single pass. 
The first plate rotates linear input polarisation from the \hpol-\vpol system into the appropriate orientation in the $\vec{n}_e$-$\vec{n}_o$ system, whereby the second plate reverses the rotation.
In double pass, only one ${\lambda}/2$ waveplate is required behind the input polariser (see fig. \ref{fig_6_setup}). 

\item
The original \pockels\, arrangement suffered from too large a beam diameter when going through the crystal. 
It used the \tisa\, output mode (beam waist $w_0 \approx 1 \, \mm$, $M^2 \approx 1.1$) directly, i.e. without any telescopes, in combination with a \pockels\, crystal of a $1/2 \,$'' diameter. 
Having large beam diameters in the \pockels\, crystal bears the problem, that electric field inhomogeneities distort the spatial mode profile of the picked pulses.
This can be avoided the closer the input beam's confocal parameter $b = \frac{2 \pi w_0^2}{\lambda}$, with the light's wavelength $\lambda$, equals the \pockels\, crystal length $L_\text{PC}$. 
Since $L_\text{PC}$ is small, on the order of $2 \, \text{cm}$, exact matching $b=L_\text{PC}$ would require a beam waist $\sim 50 \, \mu\text{m}$, resulting in too much beam divergence in front and behind the \pockels\, crystal. The strong divergence would, for instance, significantly degrade the polarisation extinction ratios obtained by the  polarisers.
Hence a compromise needs to be found between good spatial mode quality and polarisation extinction.
When moving to the double-passed arrangement in chapters \ref{ch6} and \ref{ch7},  we simplified these requirements by using a \pockels\, cell with a larger aperture of 1''.

\item
To provide good polarisation extinction, the experiment contained Glen-Laser polarisers. 
These separate orthogonal polarisations by multi-pass Brewster reflection between glass-air interfaces. 
To this end they contain two prisms with an air gap in between.  
While enabling $40\,\text{dB}$ intensity extinction of the undesired polarisation component, Fresnel reflection limits their transmission to $T_\text{GL} \approx 88 \,\%$, which is an undesirable loss in control pulse energy for the Raman memory. 
Such losses can be avoided when using polarising beam displacers (PBD), which have comparable extinction ratios, instead of the Glen-laser polarisers. Since it is non-trivial to build a double-pass \pockels\, setup with a PBD, a simple PBS is used to separate picked from non-picked pulses. Picked pulses are subsequently polarisation filtered using the PBD positioned in the reflected arm of the PBS. 

\item
The crystal had a wrong anti-reflection (AR) coating for $1064\,\text{nm}$, which must have gone unnoticed in previous experiments. 
When this was realised during a characterisation measurement, following the polarisation storage experiment, 
the \pockels\, crystal was replaced by a version AR-coated for $852 \, \text{nm}$.

\end{enumerate}

\section{Fidelity boundaries\label{app_ch4_fidelity_bounds}}

In the following, we will briefly describe how to calculate the boundary state fidelities $\mathcal{F}_B$ that have to be overcome in order to claim memory operation in the quantum regime. The arguments presented here are a summary of published material\cite{Gruendogan:2012,Specht:2011,Specht:PhD}, applied to the parameters of the Raman memory. 
Additionally, we also discuss the fidelities one has to expect at the memory output under consideration of the memory noise floor, introduced in chapters \ref{ch6} and \ref{ch7}. From this we can estimate the noise reduction required to show quantum storage of polarisation information in the memory. 

\subsection{Boundary fidelities}
The boundary fidelity $\mathcal{F}_B$ is the state fidelity a qubit, retrieved from the quantum memory, must at least posses in order to certify that this qubit could not have been prepared by a classical memory.
The classical memory follows the strategy of performing measurements on the input qubit to firstly determine its state. 
Subsequently, it uses this knowledge to prepare a new qubit, which it releases as the retrieved signal. 
Due to the no-cloning theorem\cite{Nielsen:2004kl}, such a classical measure and prepare strategy\cite{Curty:2005,Felix:2001} is insufficient to reproduce a quantum state, for which reason the state fidelity $\mathcal{F}_\text{cl}$ of the classically prepared state must be $<1$. The largest value for $\mathcal{F}_\text{cl}$ equals $\mathcal{F}_B$, since any state with higher fidelity can only be obtained from a quantum memory.
To demonstrate that a memory truly operates in the quantum regime, the fidelity of the retrieved signal thus has to exceed $\mathcal{F}_B$.

\paragraph{Fock state fidelities}
For a single input qubit, a Fock state with $N_\text{in}=1$, three projection measurements are necessary to completely determine its state, one onto each Bloch sphere axis (see appendix \ref{ch4_subsec_state_tomo}). Yet only one measurement can be made, so only one component of the Bloch vector $\vec{r}$ can be determined with certainty. This component can be prepared with fidelity $\mathcal{F}=1$. The other two components require guessing and will, on average, be prepared with the fidelity of white noise, $\mathcal{F}_\text{ns} = 1/2$. Consequently, the whole cloned qubit will show a fidelity of $\mathcal{F} = 2/3$.
In case Fock states with larger $N_\text{in}$ are sent into the memory, a classical memory could split up the state and perform simultaneous measurements on individual qubits. Thereby it can retrieve more information about the state and prepare a better clone in the output. Hence the boundary fidelity increases to\cite{Massar:1995} $\mathcal{F}_B = \mathcal{F}_\text{Fock} = \frac{N_\text{in} +1}{N_\text{in} + 2}$.

\paragraph{Coherent state fidelities\cite{Gruendogan:2012,Specht:2011,Specht:PhD}}
Contrary to Fock states, coherent states have a distribution $P(\mu,n) = \frac{\exp{\left\{-\mu\right\}} \mu^n}{n!}$ of the probability to observe a photon number $n$ around a mean photon number $\mu$, whose variance is $\mu$. 
If coherent states are used instead of Fock states for testing a memory's performance, a classical memory could obtain additional information about the input state. 
This added information comes from times, when the coherent state input signal is found to contain more photons $n$ than its mean $\mu$, which is interpreted as a Fock state of input photon number $N_\text{in}=n$.
As a result, the classically reproduced fidelity needs to take into account the probabilities of obtaining all different values of $n$ in a coherent state. 
This is done by averaging over all Fock state fidelities $\mathcal{F}_\text{Fock}(n)$, weighing them by the probabilities $P(\mu,n)$ to observe the respective photon number $n$. 
We only look at post-selected fidelities, i.e., fidelities of states for which an actual photon detection event has occurred. The total probability of observing a photon is thus $1-P(\mu,0)$ and we obtain a classical boundary fidelity of 
$$
\mathcal{F}_B = \mathcal{F}_\text{coh} = \underset{n \ge 1}{\sum} \mathcal{F}_\text{Fock} \cdot \frac{P(\mu,n)}{1-P(\mu,0)} = \underset{n \ge 1}{\sum} \frac{n+1}{n+2} \cdot \frac{P(\mu,n)}{1-P(\mu,0)}.
$$
The fidelity bound now depends on the strength of the coherent state $\mu$ and converges towards $\mathcal{F}_B = 1$ for states with $\mathcal{O}(\mu) \sim 100$. Its dependence on $\mu$ is illustrated by the \textit{grey line} in fig.~\ref{fig_ch4_fidelity_bounds}~\textbf{a} of section~\ref{ch4_subsec_outlook}.

\paragraph{Imperfect memory efficiency}

If an input signal with a  photon number distribution of non-zero variance, such as coherent states, is sent into the memory, there is additional leeway for classical state reproduction.
For an observed memory efficiency of $\eta_\text{mem} <1 $, a classical memory, which can be assumed to have an internal efficiency of $\eta_\text{class} = 1$, does not have to release an output signal upon every retrieval trigger. An output only has to be generated on average with a probability of $P_\text{out} = \eta_\text{mem} (1-P(\mu,0))$, for which reason the classical memory can, in principle, wait and observe the remaining proportion of $(1-\eta_\text{mem})(1-P(\mu,0))$ storage trials. 
This means the classical memory has additional trials available. It can use these trials to selectively pick signals which have a high photon number. 
From these, it can gain additional information by following the above described methodology that led to $\mathcal{F}_\text{coh}$.
Assuming coherent state signals of mean photon number $\mu$, a classical output is thus generated from all inputs with a photon number above a threshold $n_\text{min}$, i.e., the memory releases an output with probability $P_\text{out} = \underset{n \ge n_\text{min} +1}{\sum} P(\mu,n)$.
Since the output probability of the classical system must mimic that of a quantum memory operating at $\eta_\text{mem}$, the minimum photon number is bound by the quantum memory's retrieval probability and $n_\text{min}$ is found as
\begin{equation}
n_\text{min} = \text{min.} \left( \{ \tilde{n} \} \right): \, \underset{n \ge \tilde{n}}{\sum} P(\mu,n) \le  \eta_\text{mem} \left( 1-P(\mu,0)\right)
\label{eq_app4_nmin}
\end{equation}
Any memory only retrieves photons when there is an input, i.e. $\left( 1-P(\mu,0) \right) > 0$. So there is always a solution with the minimum 
threshold $ \text{min} \left( n_\text{min} \right) = 0$, in which case no additional information would be gained and the fidelity bound would equal $\mathcal{F}_\text{coh}$. 
For stronger input signals, $n_\text{min}$ increases. However, photon numbers are discrete, so the above equation can me made smaller than $P_\text{out} =\eta_\text{mem} \left( 1-P(\mu,0)\right)$, but not necessarily equal. 
To achieve exact equality in eq. \ref{eq_app4_nmin} for values of $\eta_\text{mem}$ for which no $n_\text{min}$ can be found, the classical memory also has to produce outputs at input photon numbers of $n_\text{min}$ or below some of the time. 
This adds an additional probability\cite{Gruendogan:2012} $\gamma = \alpha P(\mu,n_\text{min})$ to $\underset{n  > n_\text{min}}{\sum} P(\mu,n)$. It is a fraction of the probability $P(\mu,n_\text{min})$ for input photons at $n_\text{min}$, such that both sides of eq. \ref{eq_app4_nmin} are matched up. 
The size of this ''fudge" factor is thus determined by:
\begin{equation}
\gamma + \underset{n  \ge n_\text{min} + 1}{\sum} P(\mu,n) = \eta_\text{min} \left(1-P(\mu,0) \right) \Longleftrightarrow
\gamma = \eta_\text{mem} \left(1-P(\mu,0) \right) - \underset{n  \ge n_\text{min} + 1}{\sum} P(\mu,n) 
\label{eq_app4_gamma}
\end{equation}
As before, with both numbers $n_\text{min}$ and $\gamma$ determined, the classical memory will be able to reproduce input states with a fidelity given by $F_\text{Fock}(n)$ for each respective photon number $n$ in the input coherent state, with appropriate weights for each photon number $n \ge n_\text{min}$. We thus obtain a boundary fidelity of:
\begin{equation}
\mathcal{F}_B = \mathcal{F}_\text{imp} = \frac{
\frac{n_\text{min} +1}{n_\text{min} +2 } \gamma + \underset{n \ge n_\text{min} +1}{\sum} \frac{n+1}{n+2} P(\mu,n)}
{\gamma + \underset{n \ge n_\text{min}+1}{\sum} P(\mu,n)}.
\label{eq_app4_Fimp}
\end{equation}
The functional behaviour of $\mathcal{F}_\text{imp}$ is shown in fig. \ref{fig_ch4_fidelity_bounds} \textbf{a} by \textit{dashed lines} for memory efficiencies $\eta_\text{mem}$  of $\eta^\text{coh}_{1\,\text{mem}}= 29 \,\%$ (\textit{blue}) and $\eta^\text{coh}_{2\,\text{mem}} = 5 \,\%$ (\textit{red}), as experimentally obtained for our single mode and dual-rail Raman memory. 
Similar to $\mathcal{F}_\text{coh}$, the fidelity boundary converges against $\mathcal{F}_B =1$ for large coherent state inputs with $\mathcal{O}(\mu) \sim 100$. 
However, its convergence against $\mathcal{F}_B = 2/3$ is a lot slower, due to the additional information the classical memory can obtain by waiting for the occasional larger input photon numbers. 
Since lower efficiency $\eta_\text{mem}$ allows for longer waiting times, the fidelity bound increases as $\eta_\text{mem}$ decreases. 

In a real-world system, the memory output signal is also transmitted through additional optics prior to any detection (see figs. \ref{fig_ch4_setup_fig} \& \ref{fig_6_setup}), whose finite transmission $T_\text{sig} < 1$ adds photon loss. Additionally, photon detection is also inefficient with efficiencies $\eta_\text{det} <1$. 
In a conservative scenario, these additional components could be attributes of the memory system. Moreover, a classical memory could be assumed, which does not suffer from both loss mechanisms. The classical memory could, for instance, optically detect the input states and generate appropriate electronic output pulses, which feed directly into the electronic detection systems, rather than synthesising output photons for subsequent detection on a photodiode.
Such a classical memory would be able to use any additional reduction in signal count rates to increase waiting times and therewith the probability of receiving input states with higher photon numbers. 
In this case the memory efficiency $\eta_\text{mem}$ in eqs. \ref{eq_app4_nmin} and \ref{eq_app4_gamma} will be reduced to the total efficiency 
${\eta_\text{tot} = \eta_\text{det} \cdot T_\text{sig} \cdot \eta_\text{mem} \ll \eta_\text{mem}}$, 
increasing $\mathcal{F}_\text{imp}$ even further. Fig. \ref{fig_ch4_fidelity_bounds} \textbf{a} displays the updated fidelity curves (\textit{dotted lines}) for $\eta^\text{coh}_\text{1 mem}$ and $\eta^\text{coh}_\text{2 mem}$, using the parameters of $T_\text{sig} = 0.1$ and $\eta_\text{det} = 0.5$ obtained for single photon storage in a single mode memory (see chapter \ref{ch6}).
Already at the sub single photon level, with $\mu \le 1$, these parameters result in a noticeable increase for the boundary $\mathcal{F}_B$ for the output state.

Notably, no advantage can be obtained from waiting for Fock state input signals, since these always have the same photon number $N_\text{in}$ with zero variance. Heralded single photons from SPDC, whose $g^{(2)}$ values are sufficiently small, such that higher order terms can be ignored, fall into this category. Post-selection eliminates any vacuum components from an imperfect heralding efficiency $\left( \eta_\text{her} < 100 \,\% \right)$.
Single photons from our source, prepared with a heralding efficiency of $\eta_\text{her} = 22\,\%$ and $g^{(2)} = 0.016$, can hence be regarded as Fock state inputs with $\mathcal{F}_B = \mathcal{F}_\text{Fock} = 2/3$ boundary fidelity (see chapters \ref{ch5} \& \ref{ch6}).

\subsection{Effect from noise\label{app_ch4_noisy_fidelity}}

As discussed in chapters \ref{ch6} \& \ref{ch7}, the memory has a non-zero noise floor. At the single photon level, where the number of noise photons is not negligible compared to the number of signal photons, the noise admixture into the retrieved signal has to be taken into account. 
We assume the noise to be in a completely mixed state with a density matrix of white noise $\rho_{wn}$, with fidelity $\mathcal{F}_\text{ns}=1/2$. 
It will hence reduce the obtained fidelities at the single photon level. 
Following the argument of section \ref{ch4_subsec_outlook}, the signal fraction at the single photon level is furthermore assumed to possess the same fidelity of $\mathcal{F}_\text{sig} \approx 0.9$ as for bright coherent state pulses, whose fidelities are also assumed to be independent of the input photon number \Nin. 
The mixing mechanism between this signal and the noise depends on the memory configuration in terms of the number of spatial modes in the memory and, therewith, the control field pulse energy.   
For polarisation storage with the dual-memory configuration (see section \ref{ch4_subsec_polstorage}), both memories can contribute to the noise floor. 
If a single mode configuration is sufficient\cite{Lobino:2008, Lobino:2009fk,Lvovsky:2011,Kupchak:2014,Hosseini:2011zr}, e.g. for investigating the signal's Wigner function or its photon number distribution, the signal to noise ratio (SNR) of the memory can be used to obtain $\mathcal{F}$ for the noisy output. 
For both cases, we consider a constant noise floor of $N^\text{out}_\text{noise} = 0.15 \, \ppp$ as obtained during heralded single photon storage in our system (see chapter \ref{ch6}). 

In chapter \ref{ch7} we show that signal and control have the same dependance on the control pulse energy. Hence we can expect the SNR for a single mode memory with $\eta_\text{1 mem}^\text{coh} = 29 \,\%$ efficiency to also approximate that of each memory in the dual-rail configuration.
The efficiency reduction to $\eta_\text{2 mem}^\text{coh} = 5\,\%$, through splitting of the control pulse energy, will also reduce the FWM noise.
To obtain a worst case scenario, we also investigate the dual-rail configuration with $5 \, \%$ efficiency but with unmodified noise floor at $N^\text{out}_\text{noise} = 0.15\,\ppp$. Since mode mismatch reduces the memory efficiency for heralded single photons from SPDC to $\eta_\text{1 mem}^\text{SPDC} = 21\,\%$ 
(see section \ref{ch6_sec_single_photon_storage}), this lower value is used for fidelity calculations. 
\\
For coherent state input, the number of input photons \Nin\, is given by the average photon number $\mu$. Using heralded single photons, \Nin\, equals the heralding efficiency $\eta_\text{her} = 0.22\,\ppp$ (see chapter \ref{ch5}).

\paragraph{Single-mode memory configuration}
If the output signal of only one memory is observed, the signal-to-noise ratio, with SNR$=\frac{N^\text{out}_\text{sig}}{N^\text{out}_\text{noise}}$ (see section \ref{ch6_subsec_SNR}) represents the respective probabilities, with which signal or noise photons are detected in the output state. 
The mixed state fidelity is given by\cite{Bennett:1998} 
\begin{equation}
\mathcal{F}_\text{1 mem} = \frac{N^\text{out}_\text{sig} \mathcal{F}_\text{sig} + N^\text{out}_\text{noise} \mathcal{F}_\text{noise}}{N^\text{out}_\text{sig} + N^\text{out}_\text{noise}} = \frac{\eta_\text{mem} N^\text{in}_\text{sig} \mathcal{F}_\text{sig} + N^\text{out}_\text{noise} \mathcal{F}_\text{noise}}{\eta_\text{mem} N^\text{in}_\text{sig}  + N^\text{out}_\text{noise}}.
\label{app_ch4_fid_1mem}
\end{equation}
%
The \textit{solid blue line} in fig. \ref{fig_ch4_fidelity_bounds} \textbf{b} shows the expected $\mathcal{F}_\text{1 mem}$ for coherent states as a function of $\mu$ with $\eta^\text{coh}_\text{1 mem} = 29\,\%$ efficiency.
Its counterpart for single photons, with $\eta_\text{1 mem}^\text{SPDC}$ and $N_\text{in} = \eta_\text{her}$, is given by the \textit{open pink circle}. 
The best possible result can be achieved assuming a perfect single photon source, which generates a heralded single photon upon every demand trigger, i.e., $\eta_\text{her} =1$. Such an ideal system would give $\mathcal{F}_\text{1 mem}$, drawn by the \textit{filled purple symbols} in fig. \ref{fig_ch4_fidelity_bounds} \textbf{b}.

\paragraph{Dual-rail memory configuration\cite{Reim:2011ys}}

To obtain a signal output with two independent memories, photon storage has to happen in both memories simultaneously. We consider this process to be successful, whenever no noise photon is generated by the memory, corresponding to a probability of $(1-N^\text{out}_\text{noise})^2$. However not every such noise free trial actually leads to a retrieval of a signal photon. These events only occur, when photons are sent into the system and are successfully stored, which happens in $\eta_\text{mem} \cdot N_\text{in}$ of all cases. Hence signal retrieval occurs with a probability $p_\text{sig} =\eta_\text{mem} \cdot N_\text{in}\cdot (1-N^\text{out}_\text{noise})^2$.

To obtain a probability for noise detection, we assume that only states with exactly one photon are detected. While simplifying the expressions for the noise term, experimentally this would require photon number resolving detectors. The noise value is thus a lower bound, which means the obtained fidelity values are the best possible numbers that can be expected.
Noise output occurs, whenever one of the memories does not store the signal photon, but emits a noise photon instead. 
Since this can happen in either of the two memories, the probability for noise detection is 
$p_\text{noise} =  2 N^\text{out}_\text{noise} \left(1-N^\text{out}_\text{noise}  \right) \left( 1-\eta_\text{mem} \right)$. Again averaging over both contributions to the detected signal yields the expected fidelity\cite{Reim:2011ys}:  
\begin{align}
\mathcal{F}_\text{2 mem} =& p_\text{sig} \cdot \mathcal{F}_\text{sig} + p_\text{noise} \cdot \mathcal{F}_\text{noise} \nonumber \\
= & 
\frac{\eta_\text{mem} N_\text{in}\cdot (1-N^\text{out}_\text{noise})^2 \mathcal{F}_\text{sig} + 2 N^\text{out}_\text{noise} \left( 1- N^\text{out}_\text{noise} \right) \left( 1-\eta_\text{mem} \right)\mathcal{F}_\text{noise}}
{\eta_\text{mem} N_\text{in}\cdot (1-N^\text{out}_\text{noise})^2 + 2 N^\text{out}_\text{noise} \left( 1- N^\text{out}_\text{noise} \right) \left( 1-\eta_\text{mem} \right)} 
\label{app_ch4_fid_2mem}
\end{align}
Analogue to the single mode memory, the expected fidelities $\mathcal{F}_\text{2 mem}$ are shown in fig. \ref{fig_ch4_fidelity_bounds} \textbf{b} for coherent states (\textit{red solid line}) and for single photons from our SPDC source (\textit{purple symbols}). For both input signal types, the memory efficiency is reduced to $\eta_\text{mem} = 5\,\%$, while the noise level remains at $N^\text{out}_\text{noise}  = 0.15 \, \ppp$. 

\paragraph{Required noise floor} 
In both memory configurations the current noise level is too high to unambiguously proof memory operation in the quantum regime. 
With eqs. \ref{app_ch4_fid_1mem} and \ref{app_ch4_fid_2mem}, we can however determine the size of the noise level required to obtain an expected fidelity that equals the boundary fidelity $\mathcal{F}_B$ for the respective input signals. 
Solving for $N^\text{out}_\text{noise}$ yields:
\begin{itemize}
\item For the single mode configuration:
\begin{equation}
N^\text{out}_\text{noise} = \frac{\eta_\text{mem} \cdot N_\text{in} \cdot \left(\mathcal{F}_\text{sig} -\mathcal{F}_B \right)}{\mathcal{F}_B - \mathcal{F}_\text{noise}}
\label{eq_app_ch4_Nnoise_2mem}
\end{equation}
\item For the dual-rail configuration:
\begin{equation}
N^\text{out}_\text{noise} = \frac{\eta_\text{mem} \cdot N_\text{in} \cdot \left(\mathcal{F}_B - \mathcal{F}_\text{sig} \right)}
{2 \mathcal{F}_B (\eta_\text{mem} - 1) + 2 \mathcal{F}_\text{noise} (1 - \eta_\text{mem} ) + \eta_\text{mem} N_\text{in} (\mathcal{F}_B - \mathcal{F}_\text{sig})}.
\label{eq_app_ch4_Nnoise_2mem}
\end{equation}
\end{itemize}
Fig. \ref{fig_ch4_fidelity_bounds} \textbf{d} displays these noise levels for coherent states by the \textit{blue line} for the single mode and the \textit{red line} for the dual-rail memory. The levels for heralded single photons are again displayed by \textit{circles} for the former and by \textit{triangles} for the latter memory configuration.
Noise floor levels below these values are mandatory to be able to expect any non-classical fidelities from the memory.

\paragraph{Model validity}
By investigating the preservation of the signal's photon statistics in chapter \ref{ch6}, we find that signal storage and noise generation are not two independent processes in the Raman memory. For this reason, the linear addition of the noise fidelities in eqs. \ref{app_ch4_fid_1mem} and \ref{app_ch4_fid_2mem} will, most likely, not accurately describe the system; in the same way as an independent superposition of the photon statistics of signal and noise fails to describe the photon statistics of the combined output (see section \ref{ch6_subsec_g2models}).
To obtain a thorough prediction, one would have to use the coherent model introduced in sections \ref{ch6_subsec_g2models} and \ref{app6_coh_model} instead. 
Here, the output density matrices $\rho \sim \tr_\text{AS} \left( S^\dagger_\text{out,2} S_\text{out,2} \right)$ for the signal output mode (eq. \ref{eq_app6_Aout}), traced over all states in the anti-Stokes leg of the FWM noise, which are lost through filtration behind the memory, needs to enter the fidelity operator $\mathcal{F}\left(\rho_\text{out},\rho_\text{in}\right)$ (see appendix \ref{ch4_subsec_state_tomo}).
The input state would be given by $\rho \sim \tr_\text{AS} \left( S^\dagger_\text{out,1} S_\text{out,1} \right)$ (eq. \ref{eq_app_ch6_Aout1} in appendix \ref{app6_coh_model}), whereby
$S_\text{out,1}$ and $S_\text{out,2}$ are the annihilation operators for the signal transmitted through the memory in the read-in time bin and the retrieved signal from the memory in the first read-out time bin, respectively.
The independent treatment discussed here is thus only an approximation for the fidelities that can be expected from the memory at single photon level.

\chapter{Appendix: Single photon source in a ppKTP waveguide\label{app_ch5}}

\section{Spectral shaping by idler filtering\label{app_ch5_multimode_spdc_heralding}}

Here a brief description of spectrally multi-mode SPDC emission is provided. 
We first show, how the description of the SPDC emission in section \ref{ch5_sec_hsp_bandwidth}
can be extended to include multiple, correlated spectral modes of the SPDC signal and idler photons. Thereafter, the effects of heralding are introduced and we show, how filtering of the idler photon can be used to modify the state of the \hsp\,. 

\paragraph{SPDC into multiple spectral modes}
Fig. \ref{fig_ch5_spdc_spectra} \textbf{a} - \textbf{c} show, that the SPDC emission can span a frequency bandwidth of several GHz. 
To incorporate these spectral components into the SPDC quantum state $\ket{\psi_\text{SPDC}}$ (eq. \ref{eq_ch5_spdc_state}), it can be rewritten in terms of broadband creation operators\cite{Rhode:2007}. 
To this end, the JSA is first decomposed into a set of orthonormal eigenmodes 
$\left\{h_\text{s}(\nu_\text{s})\right\}$ for the signal, and
$\left\{g_\text{i}(\nu_\text{i})\right\}$ for the idler subsystem. 
This corresponds to a Schmidt decomposition\cite{Eberly:2006}, which is applicable because $f(\nu_\text{s},\nu_\text{i})$ is a continuous function\footnote{
	In terms of the discretised maps shown in fig. \ref{fig_ch5_spdc_spectra}, $f(\nu_\text{s},\nu_\text{i})$ is a 
	rectangular 2-D matrix in the $\nu_\text{s} \otimes \nu_\text{i}$ space, which can be decomposed by 
	singular-value decomposition (SVD). 
} of $\nu_\text{s}$ and $\nu_\text{i}$. 
In the Schmidt-mode basis $\left\{h_\text{s}(\nu_\text{s}), g_\text{i}(\nu_\text{i})\right\}$, the JSA takes the form
${f(\nu_\text{s},\nu_\text{i}) = \sum_j \lambda_j \cdot g_j(\nu_\text{i}) \cdot h_j(\nu_\text{s})}$, 
where the Schmidt coefficients $\lambda_j$ satisfy $\sum_j \lambda_j ^2= 1$ and $j$ runs over all Schmidt-mode combinations between the signal and idler sub-systems. 
The sum expresses the correlation between different eigenmodes of both sub-systems. 
Correlations are absent only for a pure single photons state\cite{Law:2000,URen:2006,Mosley:2008hs}, for which the JSA reduces to a single term $f=\lambda_1 \cdot g_{1}(\nu_\text{i}) \cdot h_{1}(\nu_\text{s})$. 
Here, the joint spectral intensity (JSI), defined as $i(\nu_\text{s}, \nu_\text{i}) = |f(\nu_\text{s},\nu_\text{i})|^2$, has the shape of a circle\cite{Mosley:2008hs} in the $\nu_\text{s}$-$\nu_\text{i}$ space\footnote{
	As the phase-matching and pump maps in fig. \ref{fig_ch5_spdc_spectra} \textbf{a} \& \textbf{c} show, 
	this cannot be expected for the SPDC state of our source. 
}. 
We now introduce the aforementioned broadband mode operators as the set of all frequencies, which are included in each Schmidt-mode, i.e. 
$$
\hat{A}^\dagger_j = \int \text{d} \nu_\text{i} g_{j}(\nu_\text{i}) \hat{a}^\dagger_\text{i} \quad \text{and} \quad
\hat{B}^\dagger_j = \int \text{d} \nu_\text{s} g_{j}x(\nu_\text{s}) \hat{a}^\dagger_\text{s},
$$ 
whereby $\hat{a}^\dagger_\text{i}$ and $\hat{a}^\dagger_\text{s}$ are the creation operators for idler and signal photons (see eq. \ref{eq_ch5_spdc_state}). 
Accordingly, the JSA and the SPDC state reduce to: 
\begin{equation}
f(\nu_\text{s}, \nu_\text{i}) =\sum_{j} \lambda_j \hat{A}^\dagger_j \hat{B}^\dagger_j \ket{0} \Rightarrow 
\ket{\Psi_\text{SPDC}} =  \ket{0} + \sum_{j} \lambda_j \hat{A}^\dagger_j \hat{B}^\dagger_j \ket{0}. 
\label{eq_ch5_broadband_spdc_state}
\end{equation}
\paragraph{Heralding on spectrally multi-mode idler photons}
When heralding the presence of an SPDC signal photon, APD \spcmdt\, detects the idler photon. 
Therewith it projects the SPDC state onto a single photon\footnote{
	In fact, the APD projects onto any photon number $n\ge 1$, because it is not photon number resolving. 
	This is important for the {\gtwo} of the \hsps\,, but not critical for the spectral shaping argument.
} in the idler arm $\ket{1}\bra{1}$.
Since our APDs have a constant detection efficiency $\eta_\text{APD}$ over the spectral region of the SPDC photons,  the projection operator 
${\pi_\text{APD} = \int_{\nu_\text{i}} \text{d} \nu_\text{i} \eta_\text{APD} \ket{1,\nu_\text{i}} \bra{1,\nu_\text{i}}}$ 
considers all idler frequencies $\nu_\text{i}$.
Heralding on the direct output state $\ket{\Psi_\text{SPDC}}$ of the ppKTP waveguide would consequently result in a marginalised signal state of\cite{Migdall:book} 
\begin{align}
\rho_s &= \frac{1}{\mathcal{N}} \tr_i{\left( \pi_\text{APD} \ket{\Psi_\text{SPDC}} \bra{\Psi_\text{SPDC}}\right)}\nonumber\\
&= 
\frac{\eta_\text{APD}}{\mathcal{N}} \int \int \text{d} \nu_\text{s} \text{d} \nu_\text{s}' \left( 
\int \int \text{d} \nu_\text{i} \text{d} \nu_\text{i}' f(\nu_\text{s},\nu_\text{i}) f^{*}(\nu_\text{s}',\nu_\text{i}') \right) \ket{\nu_\text{s}} \bra{\nu_\text{s}'} \nonumber\\ 
&= 
\frac{\eta_\text{APD}}{\mathcal{N}}  \int \int \text{d} \nu_\text{s} \text{d} \nu_\text{s}' i(\nu_\text{s},\nu_\text{s}') \ket{\nu_\text{s}} \bra{\nu_\text{s}'},
\label{eq_ch5_marginal_SPDC}
\end{align}
where normalisation constant $\mathcal{N}$ is given by the probability of detecting an idler photon with our inefficient APD. Eq. \ref{eq_ch5_marginal_SPDC} includes all possible frequencies $\nu_\text{s}$ in the overlap region of the pump map $\alpha(\nu_p)$ and the phase matching map $\Phi(\nu_\text{s},\nu_\text{i})$ in the marginal SPDC signal spectrum $i(\nu_\text{s},\nu_\text{s}')$. Figs. \ref{fig_ch5_spdc_spectra} \textbf{a} \& \textbf{c} show, that this region is lot wider than the {\tisa} pulse bandwidth. 

\paragraph{Narrowband idler filtering}
This frequency range can be reduced by frequency filtering the idler photon\cite{Branczyk:2010}.
For a completely monochromatic filter, transmitting only a single mode $\nu_\text{i}$, herald filtering would reduce the signal state into pure state\cite{Migdall:book,Branczyk:2010}, whose spectrum equals the 1-D cut through the JSI map along the $\nu_\text{s}$-dimension. 
If we consider Gaussian pulses with spectra 
$S(\nu) = \exp{\left\{-\frac{(\nu-\nu_0)}{\sigma_\text{UV}^2}\right\}}$ 
for our system, the SPDC spectral bandwidth variance 
$\sigma_\text{s}^2 = \sqrt{2} \cdot \sigma_\text{UV}^2$ 
would solely be determined by the pump variance $\sigma_\text{UV}^2$, since the pump map has a $45^\circ$ angle to the $\nu_\text{s}$-$\nu_\text{i}$-coordinate axes (see fig. \ref{fig_ch5_spdc_spectra} \textbf{c}). 
Additionally, $\sigma_\text{UV}^2 = 2\cdot \sigma_\text{Ti:Sa}^2$, because SHG of the {\tisa} pulses convolves the {\tisa} spectrum with itself. 
So the FWHM bandwidth 
$\Delta \nu_\text{s} = 2 \sqrt{\ln{(2)}} \cdot \sigma_\text{s}$ 
of the marginal SPDC spectrum would be double the {\tisa} pump 
$\Delta \nu_\text{s} = 2 \Delta \nu_\text{Ti:Sa} \sim 2\GHz$.

\paragraph{Idler filtering with a broadband filter stage}
In practice however, 
$\sigma_\text{UV} \le \sqrt{2} \cdot \sigma_\text{Ti:Sa}$ (see appendix \ref{ch3_SHG}), and idler filtering selects more than one mode. 
To include this effect we follow the argumentation presented by 
\textit{Braniczyk et. al.}\cite{Branczyk:2010} for the heralded SPDC state with idler filtering. 
A broadband filter can be described by an intensity transmission function $T(\nu)$, as discussed in section \ref{ch5_subsec_filter}. 
Placing our broadband filter stage into the idler arm results in the transmission of all idler photons, whose frequencies fall into the filter's transmission line. 
Everything else is reflected with a reflectivity of $R(\nu_\text{i}) = 1-T(\nu_\text{i})$. 
The effect of the filter is thus equal to a beam-splitter interaction, transforming the idler photon's creation operator according to 
$\hat{a}^\dagger(\nu_\text{i}) \rightarrow t(\nu_\text{i}) \hat{c}^\dagger(\nu_\text{i}) + r(\nu_\text{i}) \hat{d}^\dagger(\nu_\text{i})$, 
with the electric field transmittance and reflectance of 
$t(\nu_\text{i}) = \sqrt{T(\nu_\text{i})} \cdot \sqrt{\eta_\text{APD}}$ and 
$r(\nu_\text{i}) = \sqrt{ \left(1-T(\nu_\text{i}) \right) \cdot \eta_\text{APD}}$. 
In this definition, the sub-unity APD detection efficiency has already been included, which is possible, because inefficient detection can also be modelled by a beam-splitter transformation in the photon number basis\cite{Loudon:2004gd}.
Accordingly, filtering selects only a subset of the idler Schmidt modes. This loss of modes is incorporated in the broadband creation operator for idler photons by applying the above beam-splitter rotation and substituting: 
$\hat{A}^\dagger \rightarrow \frak{T}_{g_k(\nu_\text{i})} \hat{C}^\dagger_{t g_k(\nu_\text{i})} + \frak{R}_{g_k(\nu_\text{i})} \hat{D}^\dagger_{t g_k(\nu_\text{i})}$, with
\begin{align}
\label{eq_ch5_rot_idl_trans_operator}
\hat{C}^\dagger_{t g_k(\nu_\text{i})} &= \frac{1}{\frak{T}_{g_{k}(\nu_\text{i})}} \int \text{d} \nu_\text{i} t(\nu_\text{i}) \cdot g_k(\nu_\text{i}) \hat{a}^\dagger_i 
\quad
\frak{T}_{g_k(\nu_\text{i})} = \sqrt{\int \text{d} \nu_\text{i} | t(\nu_\text{i}) \cdot g_k(\nu_\text{i})|^2} \quad \text{and}
\\
\hat{D}^\dagger_{t g_k(\nu_\text{i})} &= \frac{1}{\frak{R}_{g_{k}(\nu_\text{i})}} \int \text{d} \nu_\text{i} r(\nu_\text{i}) \cdot g_k(\nu_\text{i}) \hat{a}^\dagger_i   
\quad 
\frak{R}_{g_k(\nu_\text{i})} = \sqrt{\int \text{d} \nu_\text{i} | r(\nu_\text{i}) \cdot g_k(\nu_\text{i})|^2}
\label{eq_ch5_rot_idl_refl_operator}
\end{align}
Here, the spectral modification is incorporated into the rotated creation operators 
$\hat{C}^\dagger_{t g_k(\nu_\text{i})}$ and 
$\hat{D}^\dagger_{t g_k(\nu_\text{i})}$, whereby normalisation constant $\frak{T}_{g_k(\nu_\text{i})}$ and $\frak{R}_{g_k(\nu_\text{i})}$ represent the transmission and reflection of the Schmidt mode $g_k(\nu_\text{i})$. 

Notably, the multiplication of the Schmidt modes with the filter transmittance in eq. \ref{eq_ch5_rot_idl_trans_operator} corresponds to the multiplication of the JSA with the herald filter map in fig. \ref{fig_ch5_spdc_spectra} \textbf{d}. 
Because the filter function $t(\nu)$ is not necessarily a Schmidt mode itself, it is not diagonal in the Schmidt mode basis $\left\{g_k(\nu_\text{i}) \right\}$. 
In other words, filtering introduces a mixture between different idler modes, which makes the functions 
$\frak{t}_k(\nu_\text{i}) = t(\nu_\text{i}) \cdot g_k(\nu_\text{i})$ and 
$\frak{r}_k(\nu_\text{i}) = r(\nu_\text{i}) \cdot g_k(\nu_\text{i})$ 
non-orthogonal. 
This can be fixed by another application of the Schmidt orthogonalisation to find an orthonormal basis sets 
$\left\{\phi_n\right\}$ for the transmitted spectral modes $\left\{ \frak{t}_k(\nu_\text{i}) \right\}$ 
and 
$\left\{\psi_m\right\}$ for the reflected modes $\left\{ \frak{r}_k(\nu_\text{i}) \right\}$. 
Accordingly, the broadband creation operators in eqs. \ref{eq_ch5_rot_idl_trans_operator} \& \ref{eq_ch5_rot_idl_refl_operator} for transmitted and reflected idler photons can be expressed as a linear combination of these basis functions:  
\begin{align}
\label{eq_ch5_rot_idl_trans_operator_orthogonal}
\frak{T}_{g_k(\nu_\text{i})}  \hat{C}^\dagger_{t g_k(\nu_\text{i})} \ket{0} = \underset{n}{\sum} u_{k,n} \hat{C}^\dagger_{\phi_n} \ket{0}
 \quad &\text{with}\quad
u_{k,n} = \int \text{d} \nu_\text{i} \phi^{*}_n (\nu_\text{i}) g_k(\nu_\text{i}) t(\nu_\text{i}) \\
\frak{R}_{g_k(\nu_\text{i})}  \hat{D}^\dagger_{t g_k(\nu_\text{i})} \ket{0} = \underset{m}{\sum} v_{k,m} \hat{D}^\dagger_{\psi_m} \ket{0}
\quad &\text{with}\quad
v_{k,m} = \int \text{d} \nu_\text{i} \psi^{*}_m (\nu_\text{i}) g_k(\nu_\text{i}) t(\nu_\text{i}). 
\label{eq_ch5_rot_idl_refl_operator_orthogonal}
\end{align}
These result in a beam-splitter transformation of the original broadband creation operators $\hat{A}^\dagger_{g_k}$ given by 
$\hat{A}^\dagger_{g_k} \rightarrow \underset{l}{\sum} u_{k,l}  \hat{C}^\dagger_{\phi_l}  + v_{k,l}  \hat{D}^\dagger_{\psi_l}$.
Finally, the state vector for the SPDC photon pair after idler filtering becomes
\begin{align}
\ket{\Psi_\text{SPDC}^\text{filt}} &= \ket{0} + \sum_{k} \lambda_k  \left( \int \text{d} \nu_\text{i}  t(\nu_\text{i}) \cdot g_k(\nu_\text{i}) +  r(\nu_\text{i}) \cdot g_k(\nu_\text{i}) \right) \hat{a}^\dagger_i  \hat{B}^\dagger_k \ket{0} \nonumber \\
& = \ket{0} + \sum_{k} \lambda_k 
\left(\underset{l}{\sum} u_{k,l}  \hat{C}^\dagger_{\phi_l}  + v_{k,l}  \hat{D}^\dagger_{\psi_l}\right) \hat{B}^\dagger_k \ket{0},  
\label{eq_ch5_SPDCstate_filtered}
\end{align}
where the first line explicitly shows the multiplication between the filter line and the idler frequencies in the JSA. 

Since $\eta_\text{APD}$ has been included in the transformation already, idler detection is now described by the projection operator 
$\tilde{\pi}_\text{APD} = \int_{\nu_\text{i}} \text{d} \nu_\text{i} \ket{1,\nu_\text{i}} \bra{1,\nu_\text{i}}$. 
The quantum state of the heralded SPDC signal photons is again given by applying $\tilde{\pi}$ on $\ket{\Psi_\text{SPDC}^\text{filt}}$ and tracing over all idler modes. Notably, when doing so, all reflected frequencies $\sim r(\nu_\text{i}) \cdot g_k(\nu_\text{i})$ are traced out, since they cannot be detected. This loss is already included in the definition of $r(\nu_\text{i})$. 
It can be shown\cite{Branczyk:2010}, that the density matrix of the marginalised SPDC signal photons is now given by 
\begin{align}
\rho_\text{s}^\text{filt} 
=& \frac{1}{\mathcal{\tilde{N}}} \tr{\left( \tilde{\pi} \ket{\Psi_\text{SPDC}^\text{filt}} \bra{\Psi_\text{SPDC}^\text{filt}}\right)} 
= \frac{1}{\mathcal{\tilde{N}}} \underset{k,k'}{\sum} \lambda_k \lambda^{*}_{k'} \underset{l}{\sum} u_{k,l} u^{*}_{k',l} \cdot \ket{1_i,h_k}\bra{1_i,h_{k'}}  \nonumber \\
=& 
\frac{1}{\mathcal{\tilde{N}}} \underset{k,k'}{\sum} \lambda_k \lambda^{*}_{k'} \cdot \ket{1_i,h_k}\bra{1_i,h_{k'}} \cdot\nonumber\\
& \underbrace{\left( \sqrt{\int \text{d} \nu_\text{i} | t(\nu_\text{i}) \cdot g_k(\nu_\text{i})|^2} \cdot 
 \sqrt{\int \text{d} \nu_\text{i} | r(\nu_\text{i}) \cdot g_k(\nu_\text{i})|^2} \right) \langle 1_i, t \cdot g_{k'} | 1_i, t \cdot g_{k} \rangle}_{S_\text{s}(\nu_\text{s})}, 
 \label{eq_ch5_marg_SPDC_filtered}
 \end{align}
whereby $\ket{1_i}$ and $\ket{1_s}$ are states in the photon number basis, representing a single signal or idler photon, and $\mathcal{\tilde{N}}$ is again a normalisation constant to make $\rho_\text{s}^\text{filt}$ a physical density matrix, which includes the detection probability of an idler photon.
Eq. \ref{eq_ch5_marg_SPDC_filtered} shows, how the selection of idler modes by frequency filtering with a filter function $T(\nu_\text{i})$ can be used to manipulate the marginal spectrum $S_\text{s}(\nu_\text{s})$ of the \hsp\,.

\paragraph{Spectral purity}

The spectral purity of the heralded single photon state is given by
\begin{equation}
\mathcal{P} = \tr{(\rho_\text{s}^\text{filt})^{2}} = \underset{k}{\sum} \lambda_k^{2} = \frac{1}{K}, 
\label{eq_ch5_purity}
\end{equation}
and results in the sum over all Schmidt coefficients $\lambda_k^2$. It is equal to the inverse of the Schmidt number $K$. 
For a separable state, described by a single Schmidt mode, i.e. $\lambda_1 = 1$ and $\lambda_{k>1} = 0$, we get a
Schmidt coefficient $K=1$ and perfect purity $\mathcal{P} = 1$. 
With any higher number of Schmidt modes, the SPDC state is spectrally entangled and the purity falls below $1$. 

Experimentally\cite{Moseley:PhD}, the Schmidt decomposition can be obtained using the JSI map, shown in fig.~\ref{fig_ch5_spdc_spectra}~\textbf{e}, represented by the matrix $\mathbb{F}_{\nu_\text{s}, \nu_\text{i}}$ that describes the discretised JSA $f(\nu_\text{s}, \nu_\text{i})$. 
Employing a singular-value decomposition (SVD), a general JSA matrix can be written as
$\mathbb{F}_{\nu_\text{s}, \nu_\text{i}}= \mathbb{G} \cdot \mathbb{D} \cdot \mathbb{H}$, 
whereby the matrices $\mathbb{G}$ and $\mathbb{H}$ contain the discretised Schmidt modes for idler and signal in their rows and columns, respectively. The diagonal matrix $\mathbb{D}$ carries the Schmidt coefficients 
$\lambda_k$, which need to be normalised such that $\sum_k \lambda_k ^2= 1$. 
The purity of the state $\mathcal{P} = \tr{\left( \underset{k}{\sum} (\lambda_k ^2) \right)}$ is determined by summing the diagonal elements of the square of $\mathbb{D}$.

\section{Nonlinear frequency conversion\label{app_ch5_SHGwaveguide}}

\subsection{Second-harmonic generation\label{ch5_subsec_SHG}}

When setting up an SPDC source, the first step is to operate the system with the inverse process of SHG. 
Unlike spontaneous parametric fluorescence, requiring single photon counting, due to its inherent weakness, 
the bright fields in SHG simplify system characterisation and alignment considerably. 
Apart from day-to-day alignment, one important application for SHG is the identification of the appropriate waveguide channel to be used. 
Due to the waveguide geometry, the wavevector of guided light shows a dependence on the exact spatial mode that is excited in the waveguide\cite{Migdall:book,Christ:2009}, which influences its effective refractive index\cite{Roelofs:1994} and therewith the phase-matching condition for frequency conversion (see section \ref{ch5_subsec_exp_modes} and eq. \ref{eq_ch5_phase_mismatch}). 
With the refractive index geometry fixed, the temperature dependence\cite{Kato:2002} of ppKTP's refractive indices is used to cancel phase-mismatch (temperature critical phase-matching\cite{Grechin:1999}).
Experimentally, the modification range of the chip's temperature is limited, at the low end, by water condensation on the chip and, at the high end, by the power of the employed Peltier heater element (see appendix~\ref{app_ch5_insufficiencies}). 
We thus need to find the correct set of channels that allow for frequency conversion within the temperature range 
$T \in \left[ 8^\circ \text{C}, 55^\circ \text{C} \right]$. 
Apart from good conversion efficiency, the channel also has to enable single spatial mode operation (see section \ref{ch5_sec_modes}). 

\begin{figure}[h!]
\centering
\includegraphics[width=\textwidth]{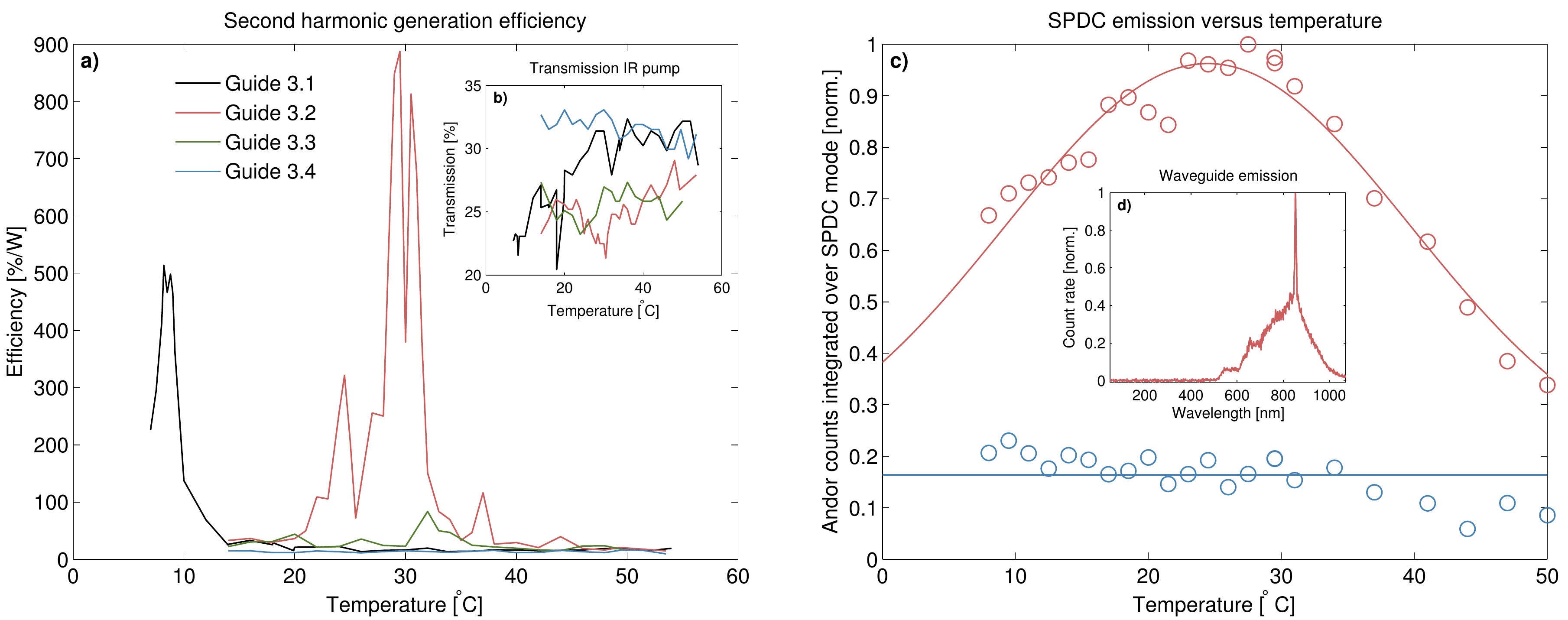}
\caption{\textbf{(a)}: SHG efficiency $\eta_\text{SHG}$ as a function of temperature for the first $4$ waveguides in family $3$. 
\textbf{(b)}: Coupling efficiency for D-polarised IR pump pulses into the waveguide channels shown in panel \textbf{(a)}. 
\textbf{(c)}: Temperature dependence of SPDC and single photon fluorescence noise within a spectral range of 
$(852\pm10)\nm$ generated in waveguide channel $3.2$. \textit{Red points} represent the waveguide emission for H-polarised UV pump, with a Gauss-fit thereon denoted by the \textit{solid red line}. 
\textit{Blue points} are the emission for V-polarised pump pulses, with their mean as the \textit{blue solid line}. 
\textbf{(d)}: Single photon spectrum of the emission from waveguide $3.2$. The spectrum is cut-off by a low-pass filter at $532\nm$.
}
\label{fig_ch5_SHG_vs_temp}
\end{figure}

Unfortunately our present chip suffers from several scratches, running across parts of the waveguide surface (see appendix \ref{app_ch5_insufficiencies}). Fig. \ref{fig_ch5_intro} \textbf{f} shows one of these.
Amongst the three poled waveguide families, we thus choose channel set $3$ (\textit{blue circles} in fig. \ref{fig_ch5_intro} \textbf{b} \& \textbf{d}) , which is the least affected family. 
The temperature response for SHG in the first 4 of channels of family $3$ is shown in fig. \ref{fig_ch5_SHG_vs_temp} \textbf{a}. These guides have widths of $w_{3.1} = 2\mum$,  $w_{3.2} = 3\mum$,  $w_{3.3} = 4\mum$,  $w_{3.4} = 2\mum$, respectively. 

The IR pump power, available inside the waveguide, depends on the waveguide coupling efficiency $\eta_\text{IR}$ (see also section \ref{ch5_subsec_exp_modes}), which is shown in fig. \ref{fig_ch5_SHG_vs_temp} \textbf{b}. 
$\eta_\text{IR}$ is reasonably similar for all channels and also temperature independent, as expected for negligible thermal expansion of the channels. Higher $\eta_\text{IR}$ values for the $2\mum$ wide guides result from simultaneous coupling into multiple spatial modes (see appendix \ref{app_ch5_guide3_modes}). 

To benchmark the conversion efficiency, it is desirable to use a definition for the SHG efficiency $\eta_\text{SHG}$ that is independent of the actual IR power coupled into the guide. 
To this end, we set $\eta_\text{SHG} = \frac{P_\text{SH}^\text{trans}}{(P_\text{IR}^\text{trans})^2}$. 
It is determined by measuring the non-converted IR and the generated $\lambda_\text{SHG} = 426\,\nm$ UV light behind the output coupling lens, for which we use the aspheric lens shown in fig. \ref{fig_ch5_setup} \textbf{a} to limit UV transmission loss. 

The available temperature range is insufficient to phase-match the conversion in guides $3.3$ and $3.4$, so only channels $3.1$ and $3.2$ show sizeable $\eta_\text{SHG}$-values\footnote{
	Note that each waveguide family consists of $6$ channels, whereby the first and the last $3$ 
	have increasing widths of $2\mum$, $3\mum$ and $4\mum$, respectively. The difference 
	between both subsets is their poling period $\Lambda$. For this reason, the $2\mum$-wide
	channel $3.1$ phase-matches at $T=9^\circ \text{C}$, while the same-sized channel $3.4$ 
	remains unresponsive. 
}. 
These have their optimal phase-matching temperatures at $T_\text{opt}^{3.1} = 9^\circ \text{C}$ and $T_\text{opt}^{3.2} = 29.5^\circ \text{C}$, respectively.
Guide $3.2$ displays about twice the nonlinear conversion efficiency of guide $3.1$ and better spatial mode quality, for which reason we choose it for running the SPDC.

\subsection{Spontaneous parametric down-conversion - temperature tuning\label{ch5_subsec_SPDC_temp_tuning}}

Pumping with UV pulses reverses up- to down-conversion. 
In order to phase-match SPDC, the UV-pump pulses must be polarised along the ppKTP crystal's y-axis, which translates into H-polarisation in the laboratory frame and an electric field vector parallel to the waveguide surface (see fig. \ref{fig_ch5_theory_modes} \textbf{a}). 
Besides the creation of SPDC pairs, the UV pump can also cause single photon fluorescence in the near-IR regime. So despite filtering for wavelengths below $532\nm$ at the waveguide output (see fig. \ref{fig_ch5_setup} \textbf{a}), a broadband noise background can be expected around the $852\nm$ SPDC emission \cite{URen:2006,Chen:2009}. 
This single photon fluorescence relates to grey-tracking and originates from the formation of colour-centre in ppKTP\cite{Boulanger:1994}. 
We investigate the entire single photon level emission spectrum of the waveguide by inserting its SPDC output into a single photon spectrometer (\textit{Andor SR163} and \textit{DV420A}). 
The observed count rate histogram, presented in fig. \ref{fig_ch5_SHG_vs_temp} \textbf{d}, displays the SPDC emission peak at $852\nm$, which is surrounded by a broadband background\footnote{
	Note, that SPDC collection has not been optimised for this measurement, so the
	ratios between the background intensity and the SPDC peak are not indicative 
	for the signal-to-noise ratio of the SPDC process.
}, typical for waveguide SPDC sources\cite{Moseley:2009}. 
With this broadband noise background present, the SPDC signal clearly requires filtering. 
Most of the undesired noise is eliminated placing several bandpass filters with $10\,\nm$ FWHM spectral transmission windows into the output path (fig. \ref{fig_ch5_setup} \textbf{a}). 

To analyse the remaining signal we image the output mode of the waveguide on an EM-CCD camera, as shown in fig. \ref{fig_ch5_setup} \textbf{b}. As discussed section \ref{ch5_sec_modes} of the main text, we image the SPDC mode at the waveguide exit face. These are exemplarily shown in figs. \ref{fig_ch5_exp_modes} of the main text and fig. \ref{fig_app_ch5_exp_modes} below. 
Since the camera also records the number of counts each CCD-pixel registers within a set integration time, one can effectively estimate the photon count rate in the output by integrating over the imaged mode. 
By recording these mode images for different chip temperatures, the temperature dependence of the SPDC's phase-matching condition can be measured\footnote{ 
	The same is not possible when coupling the waveguide output into SMF, as illustrated 
	in fig. \ref{fig_ch5_setup} \textbf{a}. Because our waveguide chip is glued into its mount
	(see appendix \ref{app_ch5_insufficiencies}), temperature variation dislocates the waveguide
	position, which results in alignment walk-off for SMF-coupling. 
	Coupling of the pump into the waveguide is thus re-optimised for every datapoint shown in 
	fig. \ref{fig_ch5_SHG_vs_temp} \textbf{c}.
}, 
whereby SPDC pair creation will seize upon moving the temperature far enough away from $T_\text{opt}^{3.2}$. 
Unlike SPDC, single photon fluorescence does not underly any phase-matching restrictions and will consequently show no temperature dependence. 
When inserting H-polarised UV light, SPDC and fluorescence occur simultaneously. 
Rotating the pump polarisation to vertical however spoils the phase-matching and essentially turns off the SPDC process\footnote{
	Other, not phase-matched non-linear processes, such as type-I down-conversion, 
	can potentially also contribute to this background. Such contributions cannot 
	be excluded concisely by subtracting $852\nm$ emission for
	V-polarised pump from that obtained with H-polarised pump.
}; a method we will refer to as frustrating the down-conversion. 
This firstly allows to separate the noise background from the SPDC and also verifies, that the emission peak in the single photon spectrum of fig. \ref{fig_ch5_SHG_vs_temp} \textbf{d} is in fact SPDC. 
The temperature tuning measurements for both pump polarisations are shown in fig. \ref{fig_ch5_SHG_vs_temp} \textbf{c}. 
Clearly, V-polarised UV pulses (\textit{blue}) only result in an approximately constant background. 
Contrary, for the H-polarised pump, a clear temperature dependence is observable, which resembles the central lobe of the expected $\sinc^2$-functional form of critical phase-matching\cite{Grechin:1999} (see eq. \ref{eq_ch5_jsa}). 
Towards high temperatures, the count rate converges against the background level observed for V-polarised UV input. 
We can thus conclude that the surplus over this background is indeed SPDC pair emission. 
Fitting it with a Gaussian distribution\footnote{
	A Gaussian distribution $I = I_0 \cdot \exp{\left( - \frac{T-T_\text{opt}^{3.2}}{\sigma^2}\right)}$ is a 
	good approximation\cite{URen:2006} to the central lope of the $\sinc^2$-function.
} (\textit{solid line}) yields a FWHM temperature bandwidth of $\Delta T_\text{SPDC} = 35.8 \,\text{K}$. 
The broader width compare to $\Delta T_\text{SHG} = 2.5  \,\text{K}$ for SHG 
results from the broader phase-matching bandwidth of the SPDC process\footnote{
	SHG is expected to have a FWHM phase-matching bandwidth of $\Delta \lambda \approx 0.08\nm$, 
	which corresponds to $\Delta \nu \approx 34\GHz$ at $852\nm$.
}(see section \ref{ch5_sec_hsp_bandwidth} of the main text).

\section{Measurements of the heralded single photon spectrum\label{app_ch5_SPDCspec}}

In addition to the characterisation measurement of the signal and idler filter stages in section \ref{ch5_subsec_filter} and the \hsp\, spectrum in the section \ref{ch5_sec_exp_hsp_bandwidth} of the main text, we provide here the summary of the results for different filter stage arrangements and both pulse models, sech and Gauss pulses.

\subsection{Filter measurements\label{app_ch5_filter_measurement}}

In table \ref{tab_ch5_filter_fwhm}, we list the results for the measurements of the signal and idler frequency filter stages, using configurations other than the 4 etalon idler filter and the 5 etalon signal case, discussed in section \ref{ch5_subsec_filter} of the main text. 
We also consider Gaussian shaped pulses here, whose pulse intensity spectrum is given by\cite{Diplomarbeit}:
\begin{equation}
S(\nu, \Delta t_\text{g})=\exp{ \left( -4 \pi^2 (\Delta t_\text{g})^2 (\nu -\nu_0)^2 \right)}, \quad
\tau_\text{Ti:Sa}^g = 2 \sqrt{\ln{(2)}} \Delta t_\text{g}, \quad
\Delta \nu_\text{Ti:Sa}^g = \frac{\sqrt{\ln{(2)}}}{\pi \Delta t_\text{g}},
\label{ch5_eq_gauss_pulses}
\end{equation}
with a pulse duration parameter $\Delta t_g = 414 \,\ps$ (see appendix \ref{ch3_tisa}).

\paragraph{Idler filter}

The idler filtering is also conducted by firstly 
taking out one of the $18\GHz$ etalons.
In a second experiment, the $103\GHz$ etalons is by-passed\footnote{
	This will lead to an effective increase of false heralds, 
	due to the additional resonances of the remaining two $18\GHz$ etalons 
	falling into the $100\GHz$ reflection band of the holographic grating filter. 
	However, this is unproblematic for the SPDC spectral measurements conducted here and  
	in section \ref{ch5_sec_hsp_bandwidth}.
}. 
The values for the fitted width $\Delta \nu_\text{idler}$ are stated in table \ref{tab_ch5_filter_fwhm} for all herald stage configurations and both pulse models.
\begin{table}
\centering
\begin{tabular}{c|c|c|c|c}
\toprule
Filter 	& 	Etalon numbers & 	Filter model 	& 	{\tisa} pulse  & $\Delta \nu_\text{filt}$ [GHz]\\
\midrule
signal	&	$2 \times18 \GHz,2 \times 103\GHz$ & Gauss 			& 	Sech 	&  1.06\\
signal	&	$2 \times18 \GHz,2 \times 103\GHz$ & Gauss 			& 	Gauss 	&  1.10\\
signal	&	$2 \times18 \GHz,2 \times 103\GHz$ & Ideal FP etalons 	& 	- 		& 0.59  \\
signal	&	$2 \times18 \GHz,2 \times 103\GHz$ & Real FP etalons 	& 	- 		&  0.83\\
\hline
idler	&	$2 \times18 \GHz,2 \times 103\GHz$ & Gauss 				& 	Sech 	&  0.94\\
idler	&	$2 \times18 \GHz,2 \times 103\GHz$ & Gauss 				& 	Gauss 	&  0.96\\
idler	&	$2 \times18 \GHz,2 \times 103\GHz$ & Ideal FP etalons 		& 	- 		&0.68  \\
idler	&	$2 \times18 \GHz,2 \times 103\GHz$ & Real FP etalons 		& 	- 		&0.94  \\
\hline
idler	&	$1 \times18 \GHz,2 \times 103\GHz$ & Gauss 				& 	Sech 	& 1.01 \\
idler	&	$1 \times18 \GHz,2 \times 103\GHz$ & Gauss 				& 	Gauss 	& 1.04 \\
\hline
idler	&	$2 \times18 \GHz$ & Gauss 							& 	Sech 	&  1.18\\
idler	&	$2 \times18 \GHz$ & Gauss 							& 	Gauss 	&  1.21\\
\bottomrule
\end{tabular}
\caption{FWHM filter line widths $\Delta \nu_\text{filt}$ for signal and idler, obtained by direct measurement under the assumption of sech and Gaussian {\tisa} pulses. 
Additionally, the expected linewidths for the etalon filter chains, considering perfect and imperfect etalons, are listed.}
\label{tab_ch5_filter_fwhm}
\end{table}

\subsection{Heralded single photon spectra for different idler filters}

In addition to the measurements of the \hsp\, spectrum, discussed in section \ref{ch5_sec_exp_hsp_bandwidth} of the main text, we present here the results for different idler filter stage configurations and also for Gaussian shaped pulses. 
To this end, we conduct the experiment described in section \ref{ch5_sec_exp_hsp_bandwidth} for a total of three idler filter modifications, containing the following etalon sequences (note, the holographic grating filter is always present in the filter stage): 
\begin{enumerate}
\item Two $18\GHz$ etalons, two $103\GHz$ etalons
\item One $18\GHz$ etalon, two $103\GHz$ etalons
\item Two $18\GHz$ etalons
\end{enumerate}	
It is interesting to see, how $\Delta \nu_\text{HSP}$ changes upon modifying the idler frequency filtering. 
As outlined in section \ref{ch5_sec_hsp_bandwidth} of the main text and appendix \ref{app_ch5_multimode_spdc_heralding} above, for an infinitely narrow herald filter and ideal SHG and SPDC processes, involving Gaussian pulses, we would expect\footnote{
	This follows from geometry, considering that the JSA of the unfiltered SPDC output is dominated by the 
	UV pump pulse map $\alpha(\nu_\text{s} + \nu_\text{i})$, which forms a stripe under a $45^\circ$ angle in the
	$\nu_\text{s}$-$\nu_\text{i}$ coordinate system (see fig. \ref{fig_ch5_spdc_spectra} and appendix \ref{app_ch5_multimode_spdc_heralding}). 
}
$\Delta \nu_\text{HSP} = 2\cdot\Delta \nu_\text{Ti:Sa}$. 
Broader idler filters should increase $\Delta \nu_\text{HSP}$, as the selected part of the pump function $|\alpha(\nu_\text{p})|^2$ in the JSI also increases. This enlarges the post-filtering JSI ellipse (see fig. \ref{fig_ch5_spdc_spectra}), stretching its projection onto the $\nu_\text{s}$-axis, which is commensurate with a higher $\Delta \nu_\text{HSP}$-value. 

Experimentally, we can see this effect, when taking either the one of the $18\GHz$ or the double-passed $103\GHz$ etalon out of the idler filter stage. In both cases, $\Delta \nu_\text{filt}^\text{idl}$ increases (see table \ref{tab_ch5_filter_fwhm}). 
Correspondingly, also the \hsp\, bandwidth increases, as expected. The measured values are stated in table \ref{tab_ch5_SPDC_bandwidth}. 
The data reproduces the general trend in the dependence on $\Delta \nu_\text{filt}^\text{idl}$, however the absolute amounts by which $\Delta \nu_\text{HSP}^\text{meas}$ is modified are not exactly linear with broadening $\Delta \nu_\text{filt}^\text{idl}$. 
While increasing $\Delta \nu_\text{filt}^\text{idl}$ from $0.9 \GHz \rightarrow 1 \GHz$ broadens $\Delta \nu_\text{HSP}^\text{meas}$ by $\sim 200\MHz$, further $\Delta \nu_\text{filt}^\text{idl}$ broadening to $1.2 \GHz$ only adds $\sim 200\MHz$ to $\Delta \nu_\text{HSP}^\text{meas}$, although $\sim 400\MHz$ would be expected from the first iteration. 
Note however that our measurement technique has quite a limited precision and is hence error sensitive, because it uses of a broadband signal to measure small bandwidth changes in another broadband signal. 
Moreover, since experimentally determined filter bandwidth are required for the deconvolution of eq. \ref{eq_ch5_conv_spdc_spectra} any errors thereon multiply. 
It is thus not unreasonable to assume that the variation in the heralded single photon bandwidth lies within the precision of the measurement. 
\begin{table}
\centering
\begin{tabular}{c|c|c|c|c}
\toprule
Pulse model & Data source & $\Delta \nu_\text{HSP}$ [GHz] & Idler filter & $\Delta \nu_\text{filt}^\text{idl}$ [GHz] \\
\midrule
Sech 	&  JSA	& $1.54$ 	& 	$2 \times18 \GHz,2 \times 103\GHz$ 	&  $0.94$\\
Sech 	&  Meas.	& $1.69$ 	& 	$2 \times18 \GHz,2 \times 103\GHz$ 	&  $0.94$\\
Sech 	&  Meas.	& $1.88$ 	& 	$1 \times18 \GHz,2 \times 103\GHz$ 	&  $1.01$\\
Sech 	& Meas. 	& $2$ 	& 	$2 \times18 \GHz$ 					&  $1.18$\\
\hline
Gauss 	& JSA 	& 	$1.61$ 	& 	$2 \times18 \GHz,2 \times 103\GHz$ 	&  $0.96$\\
Gauss 	& Meas.	& 	$1.78$ 	& 	$2 \times18 \GHz,2 \times 103\GHz$ 	&  $0.96$\\
Gauss	& Meas.	&	$1.98$ 	& 	$1 \times18 \GHz,2 \times 103\GHz$ 	&  $1.04$\\
Gauss	& Meas. 	& 	$2.13$ 	& 	$2 \times18 \GHz$ 					&  $1.21$\\
\bottomrule
\end{tabular}
\caption{Spectral bandwidth of the heralded SPDC signal photons for different idler filter configurations. JSA refers to the expected FWHM bandwidth by marginalising the JSA (see section \ref{ch5_sec_hsp_bandwidth})}
\label{tab_ch5_SPDC_bandwidth}
\end{table}


\section{Spatial modes in the waveguide\label{app_ch5_spdc_modes}}

With the telescopes and the O40X input coupler objective shown in fig. \ref{fig_ch5_setup} \textbf{a}, we achieve input beam sizes for the IR and the UV pump pulses, that give good control over the waveguide mode structure. 
Here, we investigate the modes structure for the $852\,\nm$ {\tisa} radiation coupled into the employed waveguide channel 3.2, the SH this IR-light generates, as well as the modes for the $426\,\nm$ UV-pump transmitted through the guide. 
We also look at the H- and V-polarised components of the SPDC modes, discussed in section \ref{ch5_sec_modes} of the main text, and their coupling into SMF. 
Furthermore, we present the SPDC mode structure for waveguide 3.1, which has a smaller channel width of $2\,\mum$. 

\begin{figure}[h!]
\centering
\includegraphics[width=\textwidth]{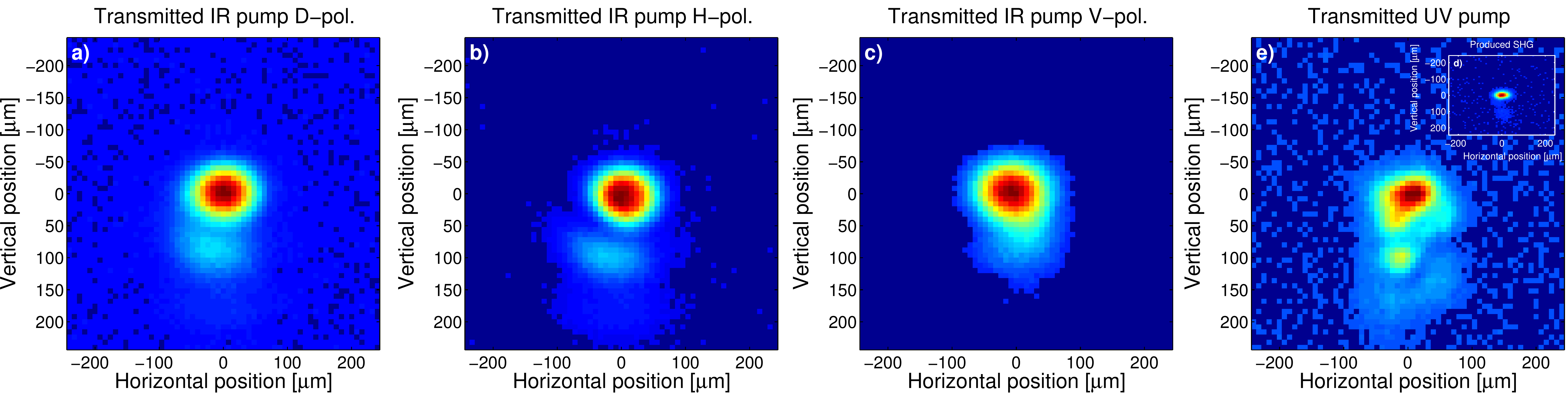}
\caption{Modes excited in the waveguide. 
\textbf{(a)} - \textbf{(c)}: Modes for the IR pump, coupled into the waveguide to produce SH, showing D-, H- and V-polarisation, respectively. 
\textbf{(d)}: SH mode produced by D-polarised IR pump, for which the SHG-process is phase-matched. 
\textbf{(e)}: Mode of the H-polarised UV pump, coupled into the waveguide for SPDC. 
}
\label{fig_app_ch5_exp_modes}
\end{figure}

The actually excited waveguide modes are analysed using the test set-up depicted in fig.~\ref{fig_ch5_setup}~\textbf{b}. 
The waveguide exit face is imaged onto the EM-CCD camera using the L40X microscope objective as output coupler and an $f_i=200\mm$ focal length imaging lens. Observed modes on the camera are consequently magnified by $M_i \approx 43.5$ with respect their actual size inside the waveguide. 
Thanks to the achromatic O40X objective, used as waveguide input coupler, collimated IR and UV pump beams can be coupled into the waveguide simultaneously with only minor longitudinal objective repositioning. This makes it easy to approximate the correct UV coupling conditions for SPDC by running SHG first.
Different modes can be excited by vertical repositioning of the input coupler.
Notably, the objective positioning that yields the highest SPDC emission is commensurate to the best mode. However it does not correspond to the position for greatest pump coupling efficiency. 
Fig. \ref{fig_app_ch5_exp_modes} depicts the modes in guide 3.2, when the input coupler position is optimised for coupling to the fundamental pump mode.

\subsection{Spatial modes for IR radiation at 852 nm wavelength} 
Because type-II phase-matching is used, SHG requires D-polarised IR, such that one 
H- and one V-polarised IR photon can convert to one H-polarised UV photon, yielding the inverse mode triplet of SPDC\cite{Karpinski:2009, Karpinski:2012}.
The mode for D-polarisation is shown in fig.~\ref{fig_app_ch5_exp_modes}~\textbf{a}, while fig.~\ref{fig_app_ch5_exp_modes}~\textbf{b}~\&~\textbf{c} illustrate its decomposition into H- and V-polarised components, respectively. 
Most of the intensity is located inside the lowest order mode, whose slight horizontal eccentricity contrasts the small vertical eccentricity of the predicted mode (see fig. \ref{fig_ch5_theory_modes} \textbf{c} \& \textbf{e}).
A small fraction of the IR is also guided in the first order mode, characterised by a $2^\text{nd}$ intensity lobe further inside the channel. 
Higher order mode components are located on opposite sides of the fundamental mode for H- and V-polarisations. 
Since more light is coupled into the first order mode for the H component (fig. \ref{fig_app_ch5_exp_modes} \textbf{b}), 
the D-polarised mode is slightly asymmetric.

However the intensity of these higher order mode contaminations is small, such that the generated SH (fig. \ref{fig_app_ch5_exp_modes} \textbf{d}) occupies the fundamental mode\footnote{ 
	Due to the quadratic scaling\cite{Boyd:2003nx} of SHG with the IR pump power, any low intensity, 
	higher order IR mode components are suppressed in the SH mode. 
}. 
As expected, the UV mode is highly elliptical. It has approximately only half the size of its IR counterparts, which are smaller than the simulated modes (see table \ref{tab_ch5_mode_sizes_exp}). 

To obtain the modes in fig. \ref{fig_app_ch5_exp_modes} \textbf{a} - \textbf{d} the input coupler position was adjusted for the highest quality SH mode. At this position, the IR coupling efficiency, 
$\eta_\text{IR}^{3.2} = \frac{P_\text{IR}^\text{trans}}{P_\text{IR}^\text{in}}$, 
defined by the input and transmitted IR intensities $P_\text{IR}^\text{in}$ and $P_\text{IR}^\text{trans}$ through the waveguide, yields $\eta_\text{IR}^{3.2}(\text{D}) \approx 27\,\%$ on average.
The SHG generation efficiency for these conditions is 
${\eta_\text{SHG}^\text{3.2}(T_\text{Cs}=29.5^\circ \text{C}) \approx 542 \frac{\%}{\text{W}}}$ on average.

To obtain the FWHM diameters of all modes in fig. \ref{fig_app_ch5_exp_modes}, their intensity distributions are fitted by 2-D Gaussians. 
The resulting FWHM values are stated in table \ref{tab_ch5_mode_sizes}. 
To determine the degree of matching with the simulated modes, the mode overlap between both is calculated. To this end, we determine the electric field distributions $E_i^j(x,y) = \frac{1}{N_j} \sqrt{I_i^j(x,y)}$ for the simulated ($j = \text{theo}$) and the measured ($j = \text{exp}$) mode intensity distributions $I_i^j(x,y)$ of each signal type ($i$) in the table \ref{tab_ch5_mode_sizes}, which are normalised such that $\underset{x,y}{\sum} |E_i^j(x,y)|^2 = 1$. For $j=\text{exp}$ this corresponds to normalisation by the total number of counts $N_j$, registered on the EM-CCD. 
The overlap $\mathcal{A}_i$ is give by 
$\mathcal{A}_i = \underset{x,y}{\sum} \left( E_i^\text{theo}(x,y) \right)^{*} \cdot \left(E_i^\text{exp}(x,y)\right)$. Here $\left\{ x,y \right\}$ run over all points in the image in a $4\time 4 \mum^2$ area around the mode centre. 
The overlaps $\mathcal{A}_i$ for all modes $i$ are also stated in table \ref{tab_ch5_mode_sizes}.
Clearly, the smaller than expected SH mode reduces the mode overlap between the measured and the expected SH mode to $\mathcal{A}_\text{SH,H} \approx 67\,\%$ compared to the transmitted IR with a measurement to expectation overlap of $\mathcal{A}_\text{IR,D} \approx 76\,\%$.

\subsection{Spatial modes for UV pump of SPDC\label{app_ch5_waveguide_UVmodes}} 
To achieve SPDC emission into the fundamental waveguide mode, the UV must be coupled into the mode of the SH (fig. \ref{fig_ch5_theory_modes} \textbf{d}).
To resembles the SH mode, the input coupler 
has to focus the collimated UV input down to a beam waist size $w_0 \in \left[ 0.77\mum, 1.1 \mum \right]$\footnote{
	In table \ref{tab_ch5_mode_sizes} the SH mode is quoted as the 
	FWHM, whereas here the beam waist $w_0 = w(z=0)$ in the 
	focus of the input coupler lens system is used. 
	For a Gaussian spatial mode with intensity distribution\cite{Siegman}
	$I(r,z) = I_0\cdot \frac{w_0^2}{w(z)^2} \cdot \exp{\left( - \frac{2\cdot r^2}{w(z)^2}\right)}$, 
	and $w(z) = w_0 \cdot \sqrt{1+\frac{\lambda \cdot z}{\pi \cdot w_0^2}}$,
	the waist size $w_0$ and the FWHM mode diameter are related by 
	$w_0  = \sqrt{2 \ln{(2)}}\cdot  \text{FWHM}$.
}. 
We measure the size of the UV pump at the waveguide input face by removing the waveguide and placing the output coupler objective in confocal configuration with the O40X input coupler objective. 
The combination of the L40X objective and the $f_i=200\mm$ lens behind the SPDC source images the beam waist, obtained in the focus of the O40X objective, onto the EM-CCD camera (see fig.~\ref{fig_ch5_setup}~\textbf{b}).  
With the empirically optimised UV input beam collimation (fig.~\ref{fig_ch5_setup}~\textbf{a}), the focussed beam has a  FWHM mode size of 
FWHM$_\text{UV} \approx1.5\mum$ in the horizontal and the vertical dimension. 
When reinserting the waveguide and optimising the UV coupling for the best SPDC mode, the transmitted UV intensity shows the mode profile illustrated in fig.~\ref{fig_app_ch5_exp_modes}~\textbf{e}. 

Clearly, the mode quality is worse compared to the SH mode (fig. \ref{fig_app_ch5_exp_modes} \textbf{d}). 
The $2^\text{nd}$ intensity lobe below the fundamental, which represents higher order mode contributions, carries more intensity than for IR input (fig. \ref{fig_app_ch5_exp_modes} \textbf{a}). 
As mentioned above, the circular-symmetric UV input is not mode matched and can couple to higher order modes. 
Additionally, light initially in the fundamental can be scattered into higher order modes at the location of the scratches on the chip surface (see fig. \ref{fig_ch5_setup} \textbf{f}). 

Despite this higher order component, the mode overlap of the transmitted UV pump with the expected UV mode (fig. \ref{fig_ch5_theory_modes} \textbf{d}) is $\mathcal{A}_\text{UV} \approx 61\,\%$, which is not too much below 
$\mathcal{A}_\text{SH,H}$.
With the input coupler set for the best SPDC mode, the coupling efficiency for the H-polarised UV pump 
$\eta_\text{UV}^{3.2} = \frac{P_\text{UV}^\text{trans}}{P_\text{UV}^\text{in}}$ through guide 3.2 is 
$\eta_\text{UV}^{3.2}\approx 12 \,\%$ on average, which is similar for V-polarised UV input. 
As expected $\eta_\text{UV}^{3.2}  < \eta_\text{IR}^{3.2}$. 
Similar to the IR coupling, higher throughput can be achieved upon changing the input coupler position to couple into more modes. 
\begin{table}[h!]
\centering
\begin{tabular}{c|c|c|c|c|c}
\toprule
Signal 	 &	\multicolumn{2}{c|}{FWHM on EM-CCD} & \multicolumn{2}{c|}{FWHM in guide 3.2} & Overlap $\mathcal{A}$\\
		&  	 hor. [$\mum$] & ver. [$\mum$] & hor. [$\mum$] & ver. [$\mum$]  &  [$\%$]  \\
\midrule
D-pol IR pump 		& $92 \pm 7$	&	$87 \pm 12$	&	$2.11 \pm 0.17$ & 	$1.99 \pm 0.28$ & $76.4$\\
H-pol IR pump 		& $84 \pm 8$	&	$91 \pm 17$	&	$1.94 \pm 0.19$ & 	$2.09 \pm 0.38$ & $75.5$ \\
V-pol IR pump 		& $102 \pm 12$ &	$101 \pm 16$	&	$2.35 \pm 0.27$ & 	$2.31 \pm 0.37$ & $77.3$ \\
SH gen. by IR 		& $56 \pm  8$	&	$39 \pm 7$	&	$1.3 \pm 0.19$	& 	$0.91 \pm 0.16$ & $66.4$ \\
SPDC gen. by UV 	& $90 \pm 4$	&	$83 \pm 6$	&	$2.08 \pm 0.09$ & 	$1.9 \pm 0.13$   & $73.2$ \\
Fluor. gen. by UV 	& $112 \pm 15$	 &	$131 \pm 36$	&	$2.58 \pm 0.35$ & 	$3.01 \pm 0.83$ & $64.9$ \\
SPDC backg. subtr. 	& $ 79 \pm 16$ & 	$ 66 \pm 16 $ 	& 	$1.82 \pm 0.37$    & 	$1.52 \pm 0.38$ & $94.2$ \\
\bottomrule
\end{tabular}
\caption{Mode sizes of guided light measured by imaging the waveguide output facet onto the EM-CCD with a $43.5:1$-magnifying telescope.  Also stated are the mode overlaps between the measured with the simulated modes (table \ref{tab_ch5_mode_sizes_exp}).}
\label{tab_ch5_mode_sizes}
\end{table}

\subsection{H- and V-polarised components of SPDC mode\label{app_ch5_SPDCmodes_HVpol}}

\begin{figure}[h!]
\centering
\includegraphics[width=\textwidth]{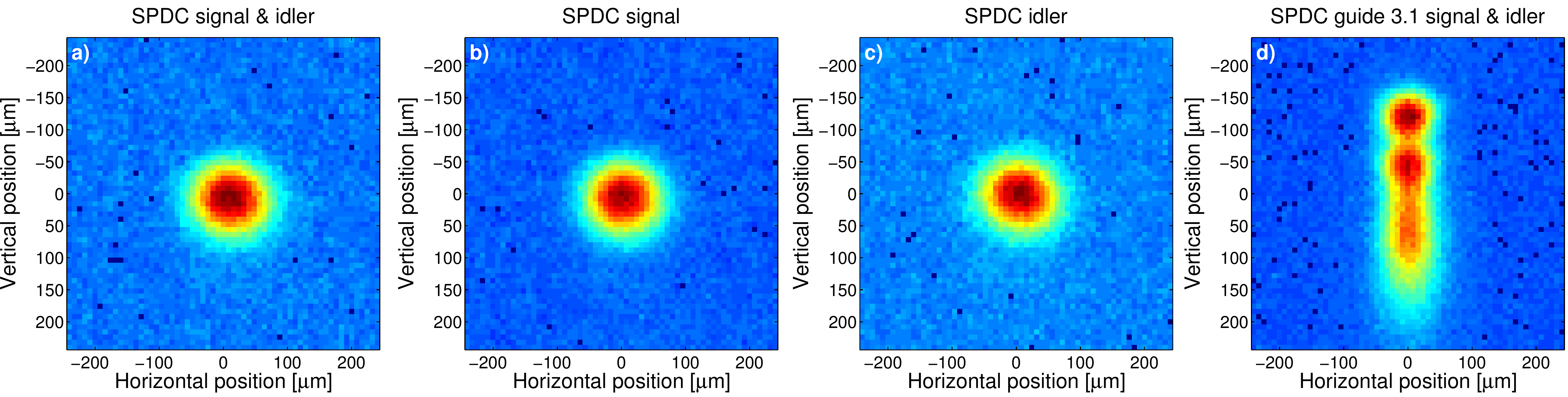}
\caption{Modes for SPDC, observed with the apparatus shown in fig.~\ref{fig_ch5_setup}~\textbf{c} behind PBS, 
\textbf{(a)}: D pol,
\textbf{(b)}: H pol,
\textbf{(c)}: V pol.
\textbf{(d)}: Multiple SPDC modes excited in waveguide 3.1}
\label{fig_app_ch5_more_modes}
\end{figure}

Besides the SPDC mode, shown in fig. \ref{fig_ch5_exp_modes}, which contains signal and idler photons, the individual modes of signal and idler have also been observed individually.
To this end, the configuration in fig. \ref{fig_ch5_setup} \textbf{c} is used, where the collimated SPDC output mode of the waveguide is first polarisation-split on a PBS and then imaged with a $f_i =200\mm$ lens onto the EM-CCD camera, positioned in front of the signal SMF. 
Each polarisation component of the SPDC pair, i.e. signal and idler, is accessible by rotation of a $\lambda/2$-waveplate in front of the PBS.
The spatial modes obtained after polarisation separation, are shown in fig. \ref{fig_app_ch5_more_modes} \textbf{a} - \textbf{c} for D-polarised, H-polarised, and V-polarised transmission through the PBS, respectively. 
D polarisation contains signal and idler photons and is thus directly comparable to fig. \ref{fig_ch5_exp_modes} \textbf{a} (i.e. it is the same mode).
Each polarisation component can furthermore be contrasted with the modes of the transmitted IR pump when running SHG, as shown in fig.~\ref{fig_app_ch5_exp_modes}~\textbf{a}~-~\textbf{c}.

\subsection{SPDC mode matching to single-mode fibre\label{app_ch5_spdc_smf_mode_matching}}

Determining the mode size on the CCD camera allows to estimate the required focussing optics for SMF coupling. 
Using the relation $w_0 = \frac{\lambda f_i}{\pi w_i}$ between the Gaussian beam waists ($w_0$) at the focus of the imaging lens ($f_i$), and the collimated input beam waist ($w_i$), the FWHM mode sizes at the lens input are\footnote{
	These are the geometric averages over the FWHM beam diameters in horizontal and 
	vertical direction, i.e. FWHM$=\sqrt{\text{FWM}_\text{hor} \cdot \text{FWM}_\text{ver}}$.
	For Gaussian beams, the FWM is related to the beam waist $w$ via: FWHM$=\sqrt{2\cdot \ln{(2)}}\cdot w$.
}: 
FWHM$_\text{D} = 781 \mum$ for the combined mode, FWHM$_\text{H} = 789 \mum$ for signal photons and FWHM$_\text{V} = 789 \mum$ for idler photons. 
The optimal focal length $f_\text{SMF}$ of an aspheric lens, used for SMF-coupling, can be determined by matching the input mode convergence upon focussing to the SMF to{\footnote{
	The constant $0.61$ is a mode overlap factor, which is introduced because the 
	optical input mode is expressed in terms of its beam waist diameter $2\cdot w_\text{in}$, 
	which contains the mode up to intensities of $13.53\,\%$ of the on-axis peak intensity,
	whereas the NA in SMF is defined for mode diameters reaching out to intensity values
	of $5\,\%$ of the on-axis peak intensity.
}}
$f_\text{SMF}=0.61\cdot \frac{\sqrt{2} \cdot \text{FWHM}_\text{in}}{\sqrt{\ln{(2)}}\cdot \text{NA}_\text{SMF}}$, with the SMF numerical aperture NA$_\text{SMF}\in [0.1,0.14]$. These beam diameters predict a focal length\footnote{
	Alternatively, the required focal lengths can be estimated by matching 
	the beam waist diameter in the focus of $f_\text{SMF}$ to the mean-field diameter in SMF 
	MFD$_\text{SMF}$, 
	requiring a focal length of $f_\text{SMF} = \frac{\sqrt{2}\pi \cdot 
	\text{FWHM}_\text{in}\cdot \text{MFD}_\text{SMF}}{4\sqrt{\ln{(2)}}\lambda}$. 
	For the SPDC signal and idler modes, this yields
	$f^\text{signal}_\text{SMF} \in [4.6,6.8]\mm$ and
	$f^\text{idler}_\text{SMF} \in [9.3,13]\mm$, which are similar to the values
	obtained by matching the NA.
} of $f^\text{signal}_\text{SMF} \in [5.8,8.2]\mm$ for the SPDC signal photons. 
Since the idler contains a $2:1$ telescope in its actual beam path (see fig. \ref{fig_ch5_setup} \textbf{a}), its input mode is expanded prior to SMF-coupling, so the focal length of the coupling lens needs to be longer, requiring $f^\text{idler}_\text{SMF} \in [11.6,17.1]\mm$.  
Note, the beam expansion is a result of experimental optimisation of the coupling efficiency. 
Testing aspheric lenses with different focal lengths\footnote{
	The empirical tests on the coupling efficiency included aspheric lenses with the following focal lengths
	${f_\text{SMF}  \in \left\{4.5\mm, 6.25\mm, 7.5\mm, 8\mm, 11.2\mm, 15.3\mm, 18.4\mm \right\}}$.
}, optimal coupling has been achieved with $f^\text{signal}_\text{SMF} =  7.5\mm$ and 
$f^\text{idler}_\text{SMF} =  18.4\mm$ for signal and idler, respectively.

\subsection{Spatial modes of waveguide 3.1\label{app_ch5_guide3_modes}}

In addition to the analysis of the spatial modes in waveguide channel 3.2, we also briefly investigate SPDC in waveguide 3.1. Besides guide 3.2, this guide also allows phase-matched frequency conversion in a temperature regime accessible with the current Peltier heater element (see fig.~\ref{fig_ch5_SHG_vs_temp}). 
The SPDC output mode, measured with the EM-CCD camera directly behind the waveguide output coupler (see fig.~\ref{fig_ch5_setup}~\textbf{b}), is shown in fig.~\ref{fig_app_ch5_more_modes}~\textbf{d}.
This corresponds to a higher order mode, which is excited because the mode measurements are performed with the same set of focussing optics used to couple into channel 3.2 (see fig. \ref{fig_ch5_setup} \textbf{a}). 
For this reason, the beam waist sizes of the IR and UV pump beams in the focus of the O40X input coupling objective match the $3\mum$ wide guide 3.2. 
Channel 3.1 has a narrower width of only $2\mum$, so the diameters of the supported modes are smaller than for guide 3.2. Consequently, the input beam waist is too large to excite solely the fundamental mode. 
Likewise to the intensity distribution between the fundamental and higher order SPDC modes of guide 3.2 (see fig. \ref{fig_ch5_exp_modes}), the top lobe's intensity distribution can be used to estimate the shape of the fundamental mode. 
Thanks to its circular symmetry, the horizontal {FWHM$_\text{hor}^{3.1} = 1.66\mum$} should also be a good approximation for its vertical dimension.

\section{Experimental insufficiencies\label{app_ch5_insufficiencies}}
There are a few design flaws in the waveguide mount, which shall briefly be mentioned. 
These are relevant for our experimental results, as they limit performance, and are of concern to someone, who wishes to run the system in the future.
\begin{itemize}
\item
The Al finger, in which the waveguide sits (see fig. \ref{fig_ch5_intro} \textbf{a}) is too long, such that the waveguide is positioned too far away from the mounting base. This makes the waveguide positioning sensitive to vibrations. In turn, these cause misalignment of light coupling into the waveguide as well as in the SMF-coupling of light collected at the waveguide output. 
\item
The Al mount is also too large for the heating capacity of the Peltier heater element used to temperature tune the ppKTP chip. 
The waveguide and the Peltier are unfortunately positioned on opposite sides of the mount, which limits the available temperature range. 
While the guide itself can be heated sustainably to $\tilde{T}_\text{KTP}^\text{max} \approx 150^\circ \text{C}$, the maximally achievable temperature is $T_\text{KTP}^\text{max} \approx 55^\circ \text{C}$. 
This is too low to achieve phase-matching in some channels. 
The guides affected are $X.3$ and $X.4$ in each of the three families $X$ (see fig. \ref{fig_ch5_SHG_vs_temp}), making them useless for frequency conversion. 
Moreover, the waveguide positioning shows an oscillatory positioning misalignment, which is likely to originate from the temperature stabilisation cycle. Over the course of $\sim 20\min$ the waveguide position changes between an initially set optimum and a slightly shifted position, which causes an oscillatory variation in SPDC pair collection and therewith in the production rates of \hsp\,.
\item
Another set of problems arises due to some initial negligence in handling the waveguide chip, resulting in it falling onto the optical table. 
Aiming to prevent such accidents, the guide has unfortunately been glued into the mount. 
The glue introduced a rotational tilt and fixed the guide with its optical facets angled with respect to the optical beam path defined by straight line propagation through the two coupling lenses. 
This complicates the coupling into the channels, making an additional tip and tilt stage necessary for the input coupler to achieve the excitation of the fundamental waveguide mode. 
Due to glue expansion with temperature changes, the absolute waveguide position is also sensitive to the set temperature $T_\text{KTP}$. 
This does not only result in the aforementioned oscillatory drifts, it also prevents the possibility to optimise any SMF-coupled signal at the waveguide output with respect to $T_\text{KTP}$, since any change therein results in immediate misalignment of the SMF-coupler position with respect to the waveguide.
\item
Lastly, the dropping accident caused scratches on top of the waveguide chip. Four such scratches run transversely over the entire chip surface, illustrated in fig. \ref{fig_ch5_intro} \textbf{f}. 
Each scratch cuts into the top of the waveguide channels, introducing scattering spots for the guided UV and IR light, which in turn result in transmission loss. 
This is doubly troublesome. Not only does it reduce the effectively available pump powers, but, more importantly, photons from the SPDC pairs, generated before the scratches, can be lost. 
Fig.~\ref{fig_ch5_intro}~\textbf{g} shows UV light guided in one of the channels, where bright spots along the propagation path (arrow in fig.~\ref{fig_ch5_intro}~\textbf{g}) coincide with the scratch locations. 
Scattering losses of SPDC signal photons, whose idler partner photons are still transmitted, reduce the heralding efficiency. 
We currently assume, that these losses are limiting the achievable heralding efficiency. However, they are unfortunately challenging to quantify, and a precise investigation of transmission losses has not been undertaken. 
Nevertheless, the existence of an intra-waveguide loss mechanism is already clear from the manufacturer specification, which quotes a conservative bound on the IR transmission of $T_\text{IR}^{5.1} \approx  T_\text{IR}^{5.2} \approx 40 \,\%$ in the fundamental mode (for guides in family 5). The same transmission can be expected for the respective channels in the other two guide families.  
In our experiments, the transmission is however limited to $T^{3.1}_\text{IR}\approx 22-32\,\%$ and $T^{3.1}_\text{IR}\approx 32\,\%$ (see fig. \ref{fig_ch5_SHG_vs_temp}), which clearly falls short.
\end{itemize}
Unfortunately, fixing the waveguide onto the mount does neither allow to solve the problems with the Al mount nor to replace the waveguide. For this reason, the experiments in this thesis had to deal the situation as it was. 
As mentioned in section \ref{ch5_subsec_intro_waveguide}, meanwhile a $2^\text{nd}$ waveguide has been purchased. On this occasion, also a new mount has been manufactured\footnote{
	This piece of work has been carried out by Krzysztof Kaczmarek with input on the 
	design from my end. 
}, whose Al extension arm, reaching from 3-axis stage into the optical beam path, forms the heat sink. The waveguide sits in a smaller cooper piece, on top 
the Peltier element, whereby both are located at the far end of the Al piece, i.e. at the position of the optical beam path. The waveguide is helt in place, by generating an under-pressure environment inside a channel within the cooper piece. The channel sits underneath the waveguide groove with small holes connecting it to the surface. These allow to fix the waveguide position in the groove by sucking the waveguide chip onto the groove's bottom. 
Both, the replacement waveguide and the mount, were available too late in the day to be used for the experiments presented in this thesis.

\chapter{Appendix: Single photon storage\label{app_ch6}}

\section{Coherent model for {\gtwo}\label{app6_coh_model}}

\subsection{Derivation of the {\gtwo} prediction from the Maxwell-Bloch equations}

Here we provide a brief overview of the coherent interaction model. The model has been developed by Joshua Nunn, and is based on his previous work on the theory for the noise free Raman memory\cite{Nunn:2007wj}, which has been extended\cite{Reim:2011ys,England:2013} to include the effects of FWM noise. 
Since I have not contributed personally to the development of the model, I will only present a summary of its main results, such that the reader can understand where the prediction of the {\gtwo} values in section \ref{ch6_subsec_g2results} come from. 

The description will start from the Maxwell-Bloch equations, including FWM noise, which we have already introduced in section \ref{ch7_subsec_FWM} of the main text. Using the notation of chapter \ref{ch2}, the final, most simplified version of these equations read: 
\begin{align}
[\partial_z + \mi \textfrak{K} ] \hat{S} &=\I \cdot C_\text{S} \cdot \hat{B} \nonumber \\ 
\partial_z \hat{A}^\dagger &= -\I \cdot C_\text{AS} \cdot \hat{B}, \nonumber \\ 
[\partial_\omega + \mi  \textfrak{S}] \hat{B} &= \I \cdot w \cdot [C_\text{S} \hat{S} +  C_\text{AS} \hat{A}^\dagger].
\label{eq_ch6_Maxwell}  
\end{align}
Here, we use the annihilation operators for photons in the Stokes channel ($\hat{S}$), the anti-Stokes channel ($\hat{A}$) and the spin-wave coherence ($\hat{B}$). 
The constants in eqs. \ref{eq_ch6_Maxwell} represent the dynamic Stark shift 
$\textfrak{S} = \frac{1}{\alpha \Delta_S} + \frac{1}{\alpha \Delta_{AS}}$, 
the Raman coupling constants for the anti-Stokes (AS) and the Stokes (S) channels, 
$C_\text{AS} = \sqrt{\frac{d\gamma}{\alpha \Delta_{AS}^2} }$ and 
$C_\text{S}=  \sqrt{\frac{d\gamma}{\alpha \Delta_{S}^2}} =  R \cdot C_\text{AS}$,  
as well the population inversion $w=p_1 - p_3$. The constant $\alpha = \frac{1}{W} = 0.31\,\ns$ is the inverse of the integrated Rabi-frequency, introduced in chapter \ref{ch2}. 
In the expression for $w$, the expectation values 
$p_i=\langle \left(  |i \rangle \langle i| \right) \rangle$ 
denote the total atomic population in the higher (F$=4$) and the lower (F$=3$) lying {\cs} hyperfine ground state (see fig. \ref{fig_ch2_Raman_protocol}).
With these parameter, the experimental configuration (see section \ref{ch6_subsec_meas_method}) can be accounted for. 
Accordingly, this is where the atomic state preparation enters the model: 
When the diode laser is sent into the {\cs} cell and the ensemble is optically pumped, we have $w=1$. For absent optical pumping, we have thermally distributed {\cs} atoms with roughly equal populations in both ground states, and $w=0$. 

\noindent
The Maxwell-Bloch equations in eqs. \ref{eq_ch6_Maxwell} can either be solved analytically\cite{Wu:2010} or numerically. 
Here, the numeric solution is implemented, whose version for the noise free case is presented in detail in Joshua Nunn's D.Phil. thesis\cite{Nunn:DPhil}. 
It uses a second order Runge-Kutta solver, where the PDE is solved on a grid in $\tau$-$z$ space. See chapter \ref{ch2} for an explanation about the time and space dimensions of eqs. \ref{eq_ch6_Maxwell}. 
Since the PDEs are linear, irrespective of the solution method, the solutions can be written in terms of Greens functions.
Knowing the Greens functions, the output fields after the interaction can be obtained from integration of the product between the respective Greens function and an input field.
This means the Greens functions $G_{i,j}$ link the input field $j$ to the output field $i$, with both $i,j \in \{A,B,S \}$.
For example, the spin-wave excitation created by Raman storage is given by the Greens function $G_{B,S}$ and the input signal field $\hat{S}_0(\tau)$: $\hat{B}(z) = \int G_{B,S}(z,\tau') \cdot \hat{S}_0(\tau') d\tau'$.
Note that this notion of Greens functions is equivalent to the memory kernel functions $K$ and $L$, introduced in section \ref{subsec_ch2_maxwell_bloch_eq}, whereby $G_{B,S}$ corresponds\footnote{
	For the remainder of this section, we drop the operator hats for the Greens functions $G_{i,j}$ as we 
	consider it understood, that these represent operators acting on the optical fields and the spin-wave. 
} to $\hat{K}(z,\omega')$ in chapter \ref{ch2}.  

With the aid of the $G_{i,j}$ we can immediately see how the term $C_\text{AS} \hat{A}^\dagger$ in eqs. \ref{eq_ch6_Maxwell} adds noise to the signal mode $S$. Having only vacuum present at the memory input, the spin-wave term is
$\hat{B}(z) \sim \int G_{B,A}(z,\tau) \hat{A}_0^\dagger (\tau) d\tau$, which leads to a Stokes output mode of 
$\hat{S}(\tau) \sim \int G_{S,B} \hat{B}(z) dz \sim \int G_{S,B} \left(G_{B,S} \hat{S}_0(\tau)+ G_{B,A} \hat{A}_0^\dagger (\tau)\right) d\tau \sim \int G_{S,B} G_{B,A} \hat{A}_0^\dagger (\tau) d\tau$. 
With 
$M_{S,A }= G_{S,B} G_{B,A}$ 
the expected number of photons in the Stokes mode turns out as
\begin{align}
N_\text{out} 	&= \left\langle \int \hat{A}_\text{out}^\dagger (\tau) \hat{A}_\text{out}(\tau) d\tau \right\rangle \nonumber \\
			&= \int d\tau \int d\tau' \int d\tilde{\tau} M^{\dagger}_{S,A}(\tau, \tau') M_{S,A}(\tau, \tilde{\tau}) \cdot \underbrace{ \langle \hat{A}_0 (\tau') \hat{A}_0^\dagger(\tilde{\tau}) \rangle}_{=\delta(\tau' - \tilde{\tau})} \nonumber \\
			& = \underbrace{\int d\tau \int d\tau' |M_{S,A} (\tau,\tau')|^2}_{\neq 0}. 
\label{eq_app6_Nnoise}
\end{align}
\noindent 
So even without any signal input, the presence of the anti-Stokes creation operator in eqs.~\ref{eq_ch6_Maxwell} results in noise added to the system. 
Importantly, the noise comes from spontaneous anti-Stokes scattering induced by the control, whose coupling strength to the $6^2S_{\frac{1}{2}} F=4$ ground state enters through $\sim C_\text{AS}$. If this can be prevented, eqs.~\ref{eq_ch6_Maxwell} reduce to the noise free case\cite{Nunn:2007wj}, given by eqs.~\ref{eq_ch2_maxwell_bloch_3}, where no more noise photons are emitted into the Stokes mode because $G_{B,A} \rightarrow 0$. 

The Greens functions are impulse response functions. For this reason, they can be obtained by probing the system with a set of successively displaced $\delta$-functions as inputs\footnote{
	Or indeed any other complete set of orthogonal input functions.
}, i.e. solving eqs.~\ref{eq_ch6_Maxwell} with 
$\mathcal{\hat{O}}= \delta(\tau-\tau_0), \forall \tau_0$, whereby $\mathcal{\hat{O}} \in \left\{ \hat{A}, \hat{B}, \hat{S} \right\}$.
In the numerical solution, the Greens functions $G_{i,j}$ are matrices with rows and columns running over the grids of the respective variables for $i$ and $j$, i.e. time 
$\tau$ (or $\omega$, as introduced in chapter \ref{ch2}) and space $z$. 
The above equation thus has the following vector form:
$$
\vec{B} = \mathbb{G}_{A,B} \cdot  \vec{A}_0 \quad \longleftrightarrow \quad \left( \begin{array}{c} B \\ \downarrow \\ z \end{array}\right) = \left( \begin{array}{ccc}  & \rightarrow & \tau \\ \downarrow & \mathbb{G}_{B,A}  & \\ z & & \end{array} \right) 
\cdot \left( \begin{array}{c}  A_{0} \\ \downarrow \\ \tau \end{array} \right)
$$
\noindent
With this method, the numerical solution can be computed by solving eqs. \ref{eq_ch6_Maxwell} using input vectors $\vec{\mathcal{O}}_0$ with moving ones along the vector elements as the initial condition:
$$
\left\{ \vec{A}_0 =  \left( \begin{array}{c} 1 \\ 0 \\ \vdots \end{array} \right), \vec{A}_0 =  \left( \begin{array}{c} 0 \\ 1 \\ \vdots \end{array} \right),\dots \right\} 
$$
Once the complete set of Greens functions $\mathbb{G}_{i,j}$ is obtained, the Maxwell-Bloch equations reduce to a set of coupled equations, whose coefficients are matrices\footnote{
	In case of an analytical solution, each matrix-vector product corresponds to an 
	integral over the appropriate variable $\tau$ or $z$ of the field, which is multiplied by the Greens function.
}. 
For the input time bin, these are
\begin{align}
\vec{A}_\text{out,1} 	&= 	\mathbb{G}_{A,A} \cdot \vec{A}_\text{in,1} +\mathbb{G}_{A,B} \cdot \vec{B}_\text{in,1}^\dagger +\mathbb{G}_{A,S} \cdot \vec{S}_\text{in,1}^\dagger 
\label{eq_app_ch6_Aout1}
\\
\vec{S}_\text{out,1} 	&= 	\mathbb{G}_{S,A} \cdot \vec{A}_\text{in,1}^\dagger +\mathbb{G}_{S,B} \cdot \vec{B}_\text{in,1} +\mathbb{G}_{S,S} \cdot \vec{S}_\text{in,1}
\label{eq_app_ch6_Sout1}
\\
\label{eq_app_ch6_Bout1}
\vec{B}_\text{out,1} 	&= 	\mathbb{G}_{B,A} \cdot \vec{A}_\text{in,1}^\dagger +\mathbb{G}_{B,B} \cdot \vec{B}_\text{in,1} +\mathbb{G}_{B,S} \cdot \vec{S}_\text{in,1},
\end{align}
\noindent
where the number $1$ signifies the input time bin, given that the index ``out" here refers to the field after the Raman interaction.
The expressions for the output time bin are more complicated. 
Here, the input parameter $\vec{B}_\text{in}$ is given by the spin-wave excited during read-in, i.e., $B_\text{in,2}$ has to be substituted with the output $B_\text{out,1}$ from the above set of equations. 
The retrieved signal from the memory is thus given by:
\begin{equation}
\vec{S}_\text{out,2} =\underbrace{\mathbb{G}_{S,A} \cdot \vec{A}_\text{in,2}^\dagger}_{*} + \mathbb{G}_{S,S} \cdot \vec{S}_\text{in,2} +
\underbrace{ \mathbb{G}_{S,B}  \mathbb{G}_{B,A} \cdot \vec{A}_\text{in,1}^\dagger}_{**} + \underbrace{\mathbb{G}_{S,B}  \mathbb{G}_{B,S} \cdot \vec{S}_\text{in,1}}_{***} + \mathbb{G}_{S,B}  \mathbb{G}_{B,B} \cdot \underbrace{\vec{B}_\text{in,1}}_{****}
\label{eq_app6_Aout}
\end{equation}
Notably, this expression contains two noise terms: while (*) is the noise generated by the control during retrieval, (**) represents residual noise from the read-in process, where a FWM spin-wave has been excited in the {\cs}, and is now read-out simultaneously with the signal (***). 
In terms of the possible noise reduction strategies, mentioned in section \ref{ch6_conclusion}, eq. \ref{eq_app6_Aout} illustrates how the scheme\cite{Zhang:2014} of using a circularly polarised read-out control pulse, which is interacting with an initially Zeeman polarised atomic ensemble, leads to reduction of the term (*), but would not influence (**). 
The term denoted by (****) contains the information about the measurement configuration, i.e., whether the atomic ensemble has been prepared by optical pumping. 
Since $B \sim |3\rangle \langle 1 |$, the expectation value $\langle B_\text{in,1}^\dagger  B_\text{in,1} \rangle \sim n_\text{thermal} \sim w$ is proportional to the fraction of the population initially located in the upper hyperfine ground state $|1\rangle$ (see fig. \ref{fig_ch2_lambda_system}). 

With the expressions for $S_\text{out,k}$, $k \in \{1,2 \}$, the expected {\gtwo} can be determined as 
\begin{equation}
g^{(2)}_\text{out,k} = 
\frac{\int d \tau \int d \tau' \langle S_{\text{out},k}^\dagger (\tau') S_{\text{out},k}^\dagger (\tau) S_{\text{out},k}(\tau) S_{\text{out},k}(\tau') \rangle}{ \int d \tau' \langle S_{\text{out},k}^\dagger (\tau') S_{\text{out},k}(\tau')\rangle \cdot \int d \tau \langle S_{\text{out},k}^\dagger (\tau) S_{\text{out},k} (\tau) \rangle}, 
\label{eq_app6_g2_1}
\end{equation}
\noindent
where the vector signs have been dropped and the integrals over $\tau$ and $\tau'$ take into account the integration time of our slow detectors\footnote{
	Slow here means a long integration time with respect to the pulse duration of the signal.
}.
Substitution of the Stokes signal $S_{\text{out},k}$ into eq. \ref{eq_app6_g2_1} results in a rather complicated, lengthy expression. However some of these terms simplify or cancel entirely, when considering the actual states the operators are applied to.
Firstly, using the fact that all involved operators satisfy bosonic commutation relations, i.e. 
${\left[\mathbb{{O}}(x), \mathbb{{O}}^\dagger(x') \right] = \delta (x-x')}$, 
with $\mathbb{{O}} \in \{S, A \}$, and 
$\left[\mathbb{B}(x), \mathbb{B}^\dagger(x') \right] = w \delta (x-x')$, with 
$x \in \{\tau,z \}$, all terms in eq.~\ref{eq_app6_g2_1} can be transferred into normally ordered products. 
Secondly, in the output time bin, there is no signal field applied to the memory, i.e. $S_{\text{in},2} = S_\text{vac}$ is operating on the vacuum state. 
So terms with normally ordered products, involving the expression 
$\sim \langle S_{\text{vac}}^\dagger S_{\text{vac}} \rangle$, drop out.
For the initial spin-wave in the read-in time bin, we have already seen that the expectation value of the normally ordered operators $\langle B_{\text{in},1}^\dagger B_{\text{in},1} \rangle$ will be given by the ground state population inversion $w$.
There are also no anti-Stokes photons inserted into the memory within either time bin. 
Hence $A^\dagger_{\text{in},k}=A^\dagger_{\text{vac}}$ also operates on the vacuum state. While application of the bosonic commutator eventually results in normally ordered products $\sim \langle A^\dagger_{\text{vac}} A_{\text{vac}} \rangle\rightarrow 0$, some terms with $\delta$-functions remain to produce the noise contributions from FWM, as we have seen above. 
Integrating out all $\delta$-functions, resulting from the application of the bosonic commutators, will leave products between the Greens function matrices $\mathbb{G}_{i,j}$ remaining. \newline
We can furthermore also define the input signal state $|\Psi \rangle$ at the Stokes frequency. In numerical form it describes the temporal shape of the pulse, which is to be multiplied by the Greens function matrices $\mathbb{G}_{i,S}$, and the quantum state of the input light\cite{Loudon:2004gd}: $S(|\Psi \rangle )= \vec{\psi}(\tau) |\psi \rangle$. For \hsp\, input this would be a single photon Fock state, 
$S(|\psi\rangle) \sim |1\rangle$, 
whereas for \coh\, inputs, with $|\psi\rangle = |\alpha\rangle$, we have $\alpha(\tau) = \psi (\tau)$.
To simplify the resulting expressions from eq. \ref{eq_app6_g2_1} further, the following set of projection operators are introduced for the read-in time bin:
\begin{equation}
\Pi_1 = \vec{\phi} \times \vec{\phi}^{*},\, \text{with }\, \vec{\phi} = \mathbb{G}_{S,S} \vec{\psi}\, ,  
\quad \Pi_2 =  \mathbb{G}_{A,S} \cdot \mathbb{G}_{A,S}^\dagger
\, , \quad \Pi_3 =  \mathbb{G}_{B,S} \cdot \mathbb{G}_{B,S}^\dagger \quad \text{and} \quad \mathbb{P} = \Sigma_j \Pi_j.
\label{eq_app6_help_ops_1}
\end{equation}
The definition of $|\Psi \rangle$ also allows to determine the number of Stokes photons obtained after the Raman interaction, given by $n_\text{out} =\langle \Psi | \Psi  \rangle$. This is the number of non-stored, transmitted signal photons plus the noise in the read-in time bin. Besides $w$, which takes account of the experimental configuration (see section \ref{ch6_subsec_meas_method}), and the Raman coupling strengths $\{ C_\text{S}, C_\text{AS}'\} $, $n_\text{out}$ is the only experimental parameter going into the coherent model. Notably, to predict the noise {\gtwo}, $n_\text{out}$ can just be set to 0 in the resulting expression for the {\gtwo}, which reads
\begin{equation}
g^{(2)} = 1 + \frac{\text{tr} \{\mathbb{P}^2 \} - \left(2-g^{(2)}_\text{in} \right) \cdot n_\text{out}^2}{\tr{\left\{\mathbb{P}^2 \right\}}}.
\label{eq_app6_g2_input_cohmod}
\end{equation}
For the output bin, the expressions are more complicated, requiring a modification of the projection operators in eq. \ref{eq_app6_help_ops_1}. With the updated set of operators
\begin{align}
\Pi_1 &= | \Phi \rangle \langle \Phi |\, , \quad \text{with}\,\, |\Phi\rangle = \mathbb{G}_{S,B} \mathbb{G}_{B,S} |\Psi\rangle\, \nonumber \\
\Pi_2 &=  \mathbb{M}_{2}  \cdot \mathbb{M}_{2}^\dagger \, , \quad \text{with}\,\, \mathbb{M}_2 = \mathbb{G}_{A,B} \mathbb{G}_{B,S}, \nonumber \\
\Pi_3 &=  \mathbb{M}_{3} \cdot \mathbb{M}_{3}^\dagger, \quad \text{with}\,\, \mathbb{M}_3 = \mathbb{G}_{B,S} \mathbb{G}_{B,S}, \nonumber \\
\Pi_4 &= \mathbb{M}_4 \cdot \mathbb{M}_4 ^\dagger,  \quad \text{with}\,\, \mathbb{M}_4 = \mathbb{G}_{A,S}, \nonumber \\
\mathbb{P} &= \Sigma_j \Pi_j,
\label{eq_app6_help_ops_2}
\end{align}
the {\gtwo} in the output time bin can also be expressed by eq. \ref{eq_app6_g2_input_cohmod}.
For both time bins, eq. \ref{eq_app6_g2_input_cohmod} contains $g^{(2)}_\text{in}$, which represents the {\gtwo} of the input signal. For the model predictions, we assume the theoretical values $g^{(2)}_{\text{in},\text{SPDC}} = 0$ and 
$g^{(2)}_{\text{in},\text{coh}} =1$ for \hsp\, and \coh\, input signals, respectively.

\subsection{Monte-Carlo error estimations\label{app6_coh_model_err}}
\label{monte}
We used a Monte-Carlo approach to generate the shaded error regions around the theoretical predictions plotted in fig. \ref{fig_ch6_g2results} of the main text. To this end we computed the theory predictions 1000 times, with each input parameter drawn from a Gaussian distribution with a standard deviation set to match its experimental uncertainty. For each value of $N_\mathrm{in}$, the standard deviations of the predictions were then used to set the vertical width of the error regions in the plots. The errors on the various input parameters were estimated as follows:
\begin{center}
\begin{tabular}{lc|lc}
\toprule
\multicolumn{1}{c}{Variable} & Error & \multicolumn{1}{c}{Variable} & Error \\
\midrule
Optical depth $d$ & $100$ 				& Natural linewidth $\gamma$ & $0.5 \MHz$ \\
Detuning $\Delta_\text{S}$ & $200 \MHz$ 		& {\tisa} pulse duration $\Delta t_\text{\tisa}$ & $20 \ps$ \\
Peak Rabi frequency $\Omega_\text{max}$ &  $100\MHz$ & & \\
\bottomrule
\end{tabular}
\end{center}

\section{Single photon storage in the memory}

\subsection{Observation of single photon storage with the FPGA data\label{app_ch6_FPGA_data}}

To allow comparison of the memory efficiency results from the TAC/MCA count rate histograms, discussed in section \ref{ch6_sec_single_photon_storage} of the main text, with the outcomes of the count rate measurements by the FPGA, we briefly summarise the FPGA measurement here. 
It follows the same logic as the count rate measurements in section \ref{ch5_sec_photon_prod}, using the notation for count rates, introduced in section \ref{sec_ch4_photon_detection}. 

\begin{figure}
\centering
\begin{minipage}[b]{0.44\textwidth}
\centering
\includegraphics[width=\textwidth]{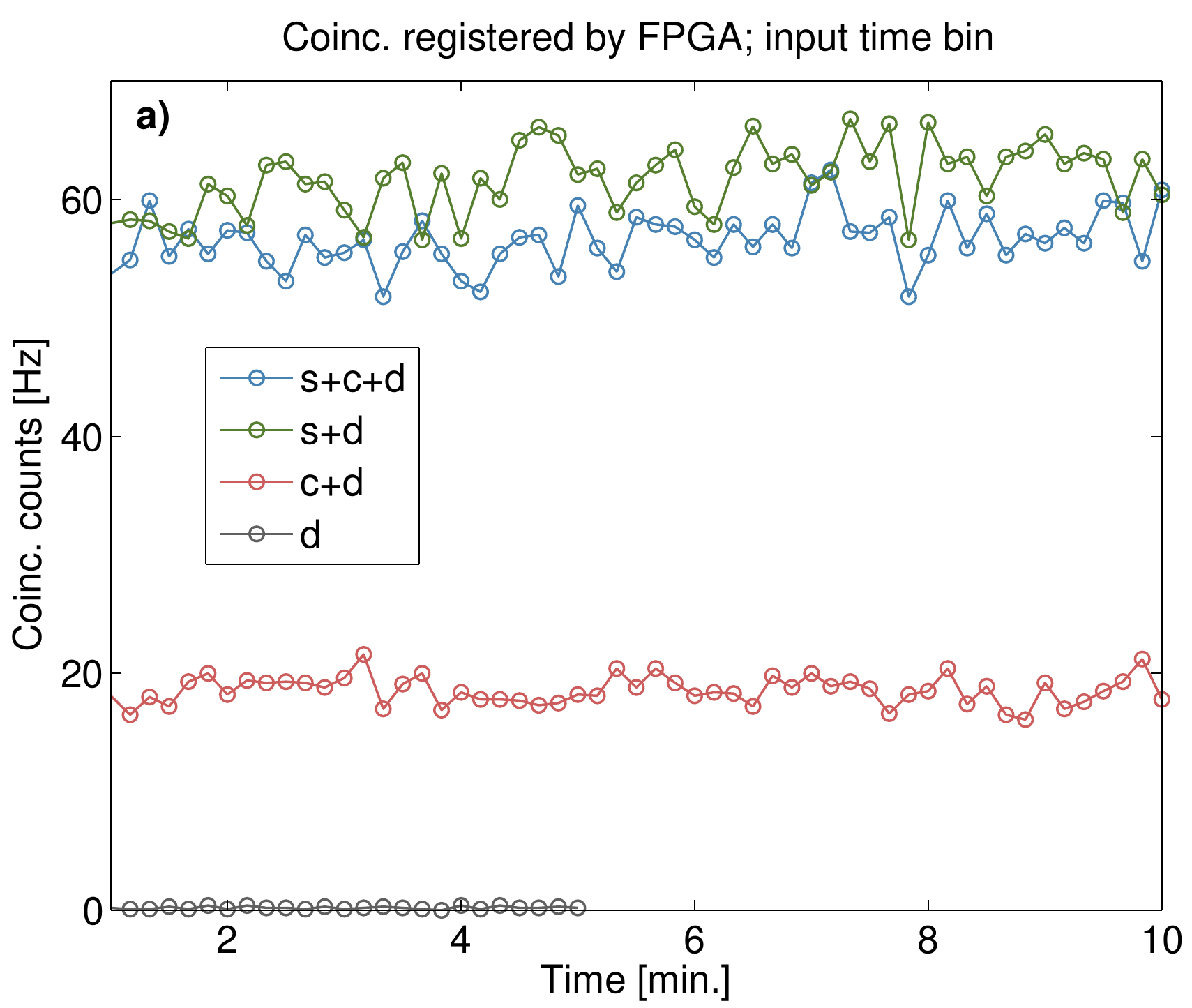}
\end{minipage}
\hspace{0.05\textwidth}
\begin{minipage}[b]{0.44\textwidth} 
\centering
\includegraphics[width=\textwidth]{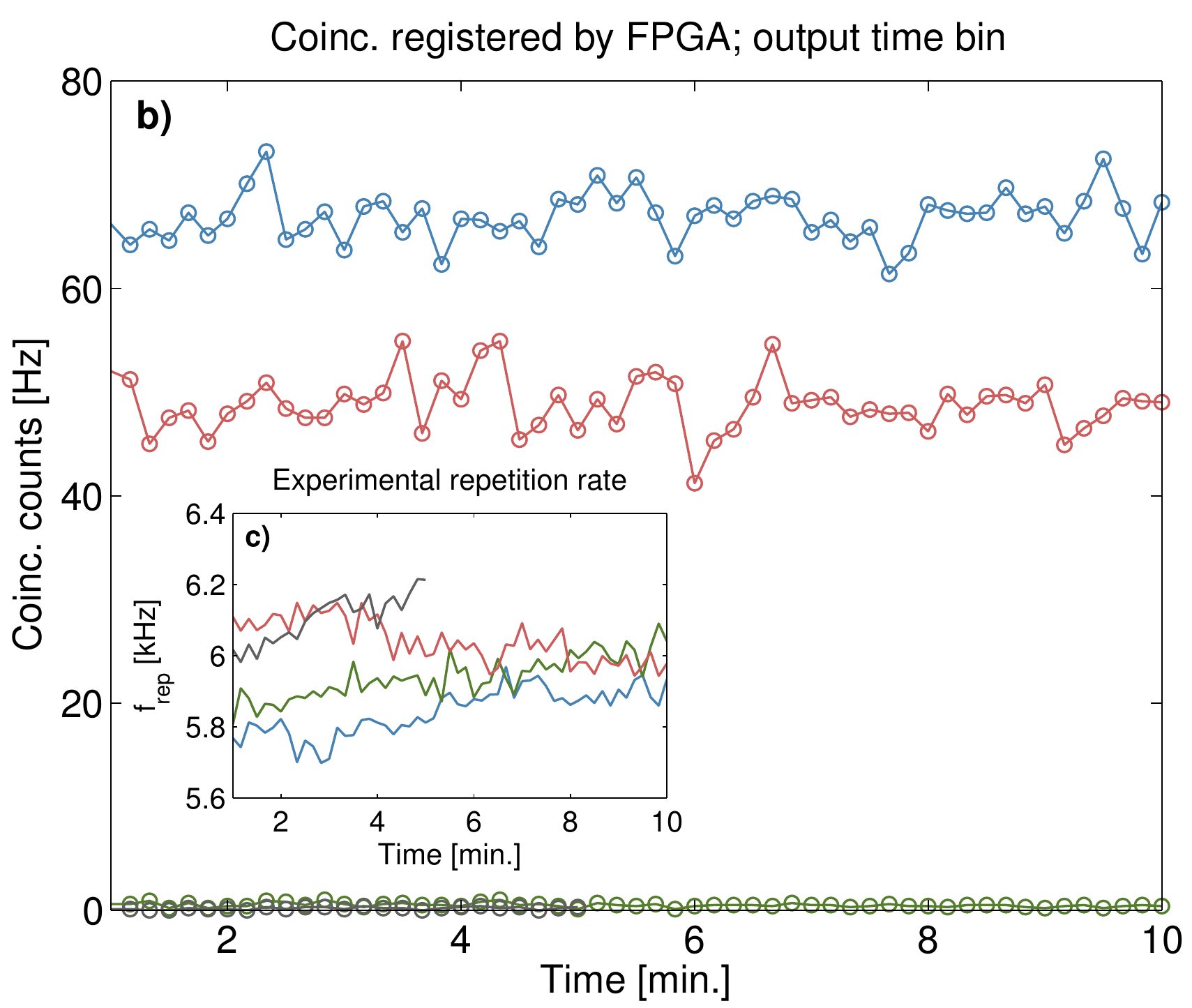}
\end{minipage}
\caption{FGPA coincidence count rates $c^t_i(t_m)$ for the both time bins ($t$) and all four measurement settings ($i$), recorded for \hsp\, input signals. 
The data shows the sum of the coincidence counts between the herald APD \spcmdt\, and the two signal APDs \spcmdh\, and \spcmdv\,, respectively. 
\textbf{(a)} represents the input and 
\textbf{(b)} the output time bin. 
The inset 
\textbf{(c)} displays the experimental trigger rates $c_{i,T}$ for each setting, which corresponds to the singles counts registered by \spcmdt\,, with the same colour coding.
The colour coding for the measurement settings shows setting \textit{sd} in \textit{green}, \textit{scd} in \textit{blue}, \textit{cd} in \textit{red} and 
\textit{d} in \textit{grey}. 
Each setting has been measured for $\Delta t_\text{meas} = 10\min$ except \textit{d}, for which $\Delta t_\text{meas} = 5\min$.
}
\label{fig_6_FPGArates_memeff}
\end{figure}

The FPGA accumulates the number of APD photon detection events registered within an integration time window of 
$\Delta t^\text{FPGA}_\text{int} = 10 \sec$. 
We display the FPGA count rate data as a time series, plotting detected events over measurement time, such that each point represents the counted events in a time bin of size $\Delta t^\text{FPGA}_\text{int}$. 
To observe storage and determine the memory efficiency, the relevant channels to observe are coincidence events between the trigger pulses on \spcmdt\, and signal detection events on APDs \spcmdh\, and \spcmdv\,. 
The resulting count rate time series are illustrated in fig. \ref{fig_6_FPGArates_memeff} for all four settings. 
Each point $c^t_{i}(t_m)$ in fig. \ref{fig_6_FPGArates_memeff} represents the accumulated coincidences, registered for setting $i$ in the input or output time bin ($t$) at time $t_m$, which represents an interval of size $\Delta t^\text{FPGA}_\text{int}$ in the total measurement time $\Delta t_\text{meas}$ (see section \ref{sec_ch4_photon_detection}).
Note, fig. \ref{fig_6_FPGArates_memeff} represents the counts registered within one measurement run $j$. All count rates  
$c_i^t(t_m) = c_{i,(H,T)}^t(t_m)+c_{i,(V,T)}^t(t_m)$ are the sum of the \spcmdt\,--\spcmdh\, and \spcmdt\,--\spcmdv\, coincidences.
Storage and retrieval is observed by the vertical displacement of count rate lines: 
Read-in leads to a count rate drop, for which reason the 
\scd\, line lies below the \sd\, line in fig \ref{fig_6_FPGArates_memeff} \textbf{a}. 
Conversely, read-out is visible by the \scd\, line, lying above the sum\footnote{
	Setting \textit{sd} only has a small number of counts in the output time bin, which come from residual EOM modulation in case of \coh\, inputs or uncorrelated SPDC pair generation for \hsp\, inputs. Counts from \textit{d} are negligible, otherwise these would have to be subtracted from \cd \, + \sd\,.
} 
of the \cd\, and the \sd\, lines. 
For calculating the efficiency, the average over all recorded points $m$ of the 
$\left\{c^t_i(t_m)\right\}$ is taken, yielding $\bar{c}^t_i$ with the standard error $\Delta \bar{c}^t_i$. 
In contrast to the TAC-data, there is no need for count integration here, since all $c^t_i(t_m)$ have been integrated for $\Delta t^\text{FPGA}_\text{int} = 10\sec$ already, irrespective how long each setting has actually been measured for, i.e. how many samples $|\left\{m \right\}|$ were recorded. 
Longer measurement durations $\Delta t_\text{meas}$ only contribute to a reduction in the standard error $\Delta \bar{c}^t_i$, but should not affect $\bar{c}^t_i$. Thus no scaling factors $\textfrak{t}_i$ are required.
Notably, the errors $\Delta \bar{c}^t_i$ implicitly assume a normal distribution for the average count rates $\bar{c}^t_i$. The validity of this assumption is demonstrated in appendix \ref{app6_g2_measurements}.

\subsection{Pulse durations in TAC/MCA arrival time histograms \label{app6_TAC_pulse_durations}}

\paragraph{Reasons for different pulse durations in the histograms}
Besides revealing the memory interaction, the histogram traces in fig. \ref{fig_6_TACtraces_2D} of the main text also show a slight broadening in the pulse durations for \hsp\, input with respect to \coh\, inputs. 
The difference originates from two sources: 
Firstly from the spectral mismatch between the \hsp\, and the {\tisa} pulses (see section \ref{ch5_subsec_exp_hsp_spectrum}).
Since the SPDC signal spectrum is broader than the {\tisa} spectrum, the \hsp\, pulses in the read-in bin should be shorter. Yet, this is not the case, since, secondly, the timing jitter of the TAC start trigger broadens the histogram pulses.
For \coh\, inputs the TAC start trigger has a deterministic repetition rate as it is derived from the \tisa-oscillator clock signal. 
In feed-forward operation, the TAC trigger is subject to the timing jitter involved in SPDC idler detection.
This jitter results from the pulse duration of the idler photon and the inherent timing jitter\cite{Stipcevic:10} of APD \spcmdt\,. 
The pulse durations in the \hsp\, histograms are thus convolutions between this start trigger timing jitter and the signal photon arrival times, which are, in turn, proportional to the duration of the measured signal pulses.
Thus pulses in fig. \ref{fig_6_TACtraces_2D} \textbf{a} are longer than in \textbf{b}, although the marginal SPDC signal spectrum is broader than the {\tisa} spectrum. \newline
Moreover, the different settings can also be used to estimate the bandwidth difference between \hsp\, and the {\tisa} pulses. To this end the histograms in fig. \ref{fig_6_TACtraces_2D} have been fitted with Gaussian pulse shapes, with detailed results stated below. 
The fit shows that settings \textit{sd} and \textit{cd} for \coh\, inputs have nearly equal widths, i.e., the FWM noise has the same temporal profile as the control pulses. 
The noise pulses can thus be used as a proxy to determine the broadening from feed-forward operation, which widens the pulses by a factor of\footnote{
	For an idler photon with a pulse duration equal to that of the {\tisa} laser pulses, and noise pulses with pulse
	durations also equal to {\tisa} pulse duration (see chapter \ref{ch7}), we would expect a factor of $\sqrt{2}$.
} 
$\textfrak{f}_\text{trig} \approx 1.37$.
Yet, the signal input pulse (setting \textit{sd}) for \hsp\, has a width which is only broadened by $\textfrak{f}_\text{sd} \approx 1.11$. Hence we can conclude that the SPDC signal photons only have $0.82$ times the duration of the {\tisa} pulses, which means their spectrum, observed after transmission through the signal filter stage, is broadened by a factor 
$\textfrak{f}_\text{SPDC} = 1.23$ with respect to that of the control pulses. 
This coincides well with the experimental value of $\textfrak{f}_\text{exp} = 1.25$ for the measured spectral bandwidth of the \hsps\, in section \ref{ch5_sec_exp_hsp_bandwidth}.
Furthermore, the pulse duration of the read-in pulse for setting \textit{scd}, which is $1.13$ times longer for feed-forward operation, equals the expectation from a superposition of the transmitted, not-stored signal and the noise pulses, mixed by the signal-to-noise ratio ($\text{SNR}_\text{trans}$ in section \ref{ch6_subsec_SNR} of the main text). On the other hand, the retrieved signal has the same duration as the control pulses\cite{Novikova:2007kc}. 

\paragraph{Pulse broadening by feed-forward operation}
To determine the durations of TAC histogram pulses, shown in fig. \ref{fig_6_TACtraces_2D} for \hsp\, and \coh\, inputs with ${N_\text{in} = 0.21 \ppp}$ and $N_\text{in} = 0.23 \ppp$, respectively, we fit these with a Gaussian intensity profile 
$I(t) = I_0 \cdot \exp{\left\{ -\frac{\left(t-t_0 \right)^2}{\sigma^2} \right\}}$. 
The aim is to determine the pulse duration $\sigma$. 
Since TAC measurements are performed constantly alongside the recording of FPGA data, we calculate the average duration $\sigma^t_i$ for each measurement setting $i$ and time bin $t$, over all recorded TAC traces for the particular input signal type.
These yield $\sigma_i^t$ as stated below, where \hsp\, inputs are listed in the \textit{feed-forward mode} column and \coh\, signals in the \textit{{\tisa} clock trigger} column:
\begin{center}
\footnotesize
\begin{tabular}{@{} c | c |  c | c @{}} 
\toprule
Setting			& Time bin	&Feed-forward (\hsp\,): $\sigma_{i,\text{SPDC}}^t$ [ps]	& {\tisa} clock trigger (\coh\,): $\sigma_{i,\text{coh}}^t$ [ps]		\\ 
\midrule
$scd$ 			& in 			& $719 \pm 6$		& $642 \pm  5$			\\ 
$sd$ 			& in 			& $683 \pm 3$		& $613 \pm 3$			\\ 
$cd$ 			& in 			& $834 \pm 4$		& $611 \pm 2$			\\ 
\midrule
$scd$ 			& out 		& $822 \pm 7$		& $608 \pm 3$			\\ 
$cd$ 			& out 		& $838 \pm 5$		& $614 \pm 2$			\\ 
\bottomrule
\end{tabular}
\end{center}

From these values, the following conclusions can be drawn: 
\begin{itemize}
\item $\sigma_{cd,\text{coh}}^\text{in} \approx \sigma_{sd,\text{coh}}^\text{in}$: The pulse shapes of the FWM noise produced by the control and those of the {\tisa} pulses, which produce the control and the \coh\, input signal, are similar. 
 $\sigma_{cd}$ is thus representative of the {\tisa} pulse duration and can be used to determine the broadening factors arising from feed-forward triggering.
\item The pulse broadening from the additional timing jitter in the TAC start trigger for feed-forward operation can be determined by comparing the pulse durations of the noise pulses between the two triggering configurations. A broadening factor of 
$\textfrak{f}_\text{trig} = \frac{\sigma_{cd,\text{SPDC}}^{in} + \sigma_{cd,\text{SPDC}}^{out}}{\sigma_{cd,\text{coh}}^{in} + \sigma_{cd,\text{coh}}^{out}} \approx 1.37$ 
is obtained when going from the {\tisa} clock rate to triggering off SPDC idler detection.
\item The SPDC input photons only show a broadening factor of 
$\textfrak{f}_\text{sd} = \frac{\sigma_{sd,\text{SPDC}}^\text{in} }{\sigma_{sd,\text{coh}}^\text{in}} = 1.11$.  
The reduction is due to the spectral mode mismatch between heralded SPDC signal photons and the {\tisa} pulses. 
From this one can determine the mismatch factor, i.e. the factor by which the pulses of SPDC photons are shorter, as:
$\textfrak{f}_\text{mis} = \frac{\sigma_{sd, \text{SPDC}}^\text{in}}{\textfrak{f}_\text{trig} \cdot \sigma_{sd,\text{coh}}^\text{in}} = 0.82$. 
This means, the spectrum of the SPDC signal photons is broadened with respect to the {\tisa} spectrum by 
$\textfrak{f}_\text{SPDC} = \frac{1}{\textfrak{f}_\text{mis} } = 1.25$. \newline
To compare this broadening with the measured \hsp\, spectral bandwidth of section \ref{ch5_sec_exp_hsp_bandwidth}, one needs to consider the spectrum of the \hsps\, and the control pulses after transmission through the signal filter. Similar to chapter \ref{ch5}, we assume sech-shaped spectra $S_\text{HSP}(\nu)$ and $S_\text{ctrl}(\nu)$ for the \hsps\, and the control, as well as a Gaussian filter function $T(\nu)$, with the parameters stated in section \ref{ch5_subsec_filter}, respectively. The transmitted signal 
${\tilde{S}_\text{HSP}(\nu) = S_\text{HSP}(\nu) \cdot T(\nu)}$ 
and control intensity profile 
${\tilde{S}_\text{ctrl}(\nu) = S_\text{ctrl}(\nu) \cdot T(\nu)}$ spectrum are fitted by the normalised Gaussian intensity profile
$I(t)$ from above to yield width parameters $\sigma_\text{HSP}$ and $\sigma_\text{ctrl}$. Their ratio 
$\textfrak{f}_\text{mis}^\text{exp} = \frac{\sigma_\text{HSP}}{\sigma_\text{ctrl}}$ 
corresponds to the spectral mismatch factor 
$\textfrak{f}_\text{SPDC}$, obtained from the count rate histograms. With 
$\textfrak{f}_\text{mis}^\text{exp} = 1.25$, the broadening observed in the input time bin of fig. \ref{fig_6_TACtraces_2D} \textbf{a} agrees well with the expectation from the measured \hsp\, pulse spectra.
\item The broadening of the not-stored signal pulses, transmitted through the memory in the read-in bin, is $\textfrak{f}_{scd} = \frac{\sigma_{scd,\text{SPDC}}^\text{in}}{\sigma_{scd,\text{coh}}^\text{in}} = 1.12$, and lies in between the duration of the input signal and the noise. 
We can model, whether this is the expected pulse duration for a combination of the input signal pulse (with $\sigma_{sd,\text{SPDC}}^\text{in}$) and the noise pulse (with $\sigma_{cd,\text{SPDC}}^\text{in}$), assuming a mixing ratio set by the signal to noise ratio in the read-in bin SNR$_\text{trans} = 2.184$ (see section \ref{ch6_subsec_SNR} of the main text).
The expected pulse has an intensity profile given by
\begin{equation}
I_{scd}(t) = \text{SNR}_\text{trans} \cdot \exp{ \left\{ - \frac{t^2}{\left( \sigma_{sd,\text{SPDC}}^\text{in} \right)^2} \right\} } + \exp{ \left\{ - \frac{t^2}{\left( \sigma_{cd,\text{SPDC}}^\text{in}\right)^2} \right\} }.
\label{eq_app6_pulse_duration}
\end{equation}
Fitting $I_{scd}(t)$ with the initial pulse model $I(t)=I_0 \cdot \exp{\left\{ - \frac{t^2}{\sigma_\text{exp}^2}\right\}}$, whereby $I_0 = 1+ \text{SNR}_\text{trans}$, gives $\sigma_\text{exp} = 758 \ps$ and a ration with respect to the pulses observed for {\tisa} triggering of 
$\textfrak{f}_{scd,\text{exp}} = \frac{\sigma_\text{exp}}{\sigma_{scd,\text{coh}}^\text{in}} = 1.13$. 
This expected broadening of $\textfrak{f}_{scd,\text{exp}}$ is nearly identical with the observed value for $\textfrak{f}_{scd}$. The observed difference between $\textfrak{f}_{scd}$ and $\textfrak{f}_\text{trig}$ is thus a result from supplying an input signal with the shorter pulse duration of the SPDC signal photons. 
\item 
The same procedure can be applied to the output time bin for setting \textit{scd}. Here the interesting question is, whether the control bandwidth limits the amount of signal bandwidth, that was stored in the memory. One would expect this limitation\cite{Novikova:2007kc,Nunn:2007wj}, as the bandwidth of the stored signal is limited by the bandwidth of the induced virtual Raman resonance, which, in turn, is determined by the spectral width of the control. 
When using eq. \ref{eq_app6_pulse_duration} with the parameters $\sigma_{cd,\text{SPDC}}^\text{out}$ and SNR$_\text{out}$ for the output bin, one should find $\sigma_\text{exp}^\text{out} < \sigma_{scd,\text{SPDC}}^\text{out}$. This means the output pulses from a combination of input pulse and noise should be shorter in duration than the observed pulse width, whose signal fraction has been spectrally narrowed by storage. 
Yet, the low value of SNR$_\text{out} = 0.296$ strongly limits the contribution from the input signal duration, yielding an expected pulse duration of 
$\sigma_\text{exp}^\text{out} = 881 \ps > \sigma_{cd,\text{SPDC}}^\text{out}$ and 
$\textfrak{f}_{\text{exp},scd}^\text{out} =1.39 \approx \textfrak{f}_\text{trig}$. 
So, with these measurements, we cannot unambiguously determine the spectral projection onto the control field spectrum during Raman storage and retrieval (see chapter \ref{ch2}). 
\end{itemize}

\subsection{Coherent state storage with and without optical pumping \label{app6_coh_pumping_on}}

In the \hsp\, storage experiment, feed-forward operation has been implemented without a trigger for switching-off the diode laser, which results in continued optical pumping during storage. By contrast, when operating the experiment with the {\tisa} clock-rate trigger, the diode laser has been turned off during the storage of \coh\, input signals. 
To ensure that this difference does not alter the observed photon statistics, a test measurement was conducted on \coh\, input signals at $N_\text{in} = 0.23 \ppp\,$, where the photon statistics and the memory efficiency have been evaluated for both optical pumping configurations, i.e., one measurement was taken with the diode pumping on and another one with it off during the storage time.
Both configurations were measured in series within the same day to minimise effects from systematic setup changes. Measurements with active diode laser switching were performed first. 
Since there is still a residual degradation in the memory performance present over time, despite continuous alignment checks of the system, there is a systematic reduction in the memory efficiency values for the measurements without diode switching. 
We have quantified this effect by checking the memory efficiency with bright coherent state inputs on the Menlo Systems photodiode (see fig. \ref{fig_ch4_setup_fig}) for both configurations at the end of the measurement. Observation on the Menlo Systems photodiode is preferred, as bright \coh\, measurements\cite{Reim2010,Reim:2011ys, England2012} do not require long integration times and can be conducted in quick succession ($\sim 30 \sec$), for which reason there are no system drifts during data acquisition.
To allow comparison between the Menlo PD data and the efficiencies obtained by photon counting, efficiency measurements using the Menlo PD have also been conducted after measuring each setting sequence for both optical pumping experiments\footnote
{
	In fact, this has been the standard procedure for all measurements. 
	This means, every time the system alignment has been checked 
	for a measurement going into the {\gtwo} analysis, the bright 
	coherent state efficiency has been determined on the Menlo photodiode. 
}. 
All efficiencies are quoted in the table below, where the first part contains the averaged efficiencies from the FPGA, TAC and Menlo photodiode. The second part shows the control measurements with bright coherent states at the end of the data acquisition. 
\begin{center}
\footnotesize
\begin{tabular}{c|c|c|c|c|c|c}
\toprule
\multicolumn{7}{c}{\textbf{Memory efficiency for optical pumping on/off}} \\
\toprule
Detector  	&   \multicolumn{2}{c|}{$\eta_\text{in}$  [\%]}	&  \multicolumn{2}{c|}{$\eta_\text{ret}$  [\%]}	& \multicolumn{2}{c}{$\eta_\text{mem}$  [\%]} \\ \hline
		& Pump on		& Pump switched		& Pump on			& Pump switched		& Pump on		& Pump switched	\\
\midrule
FPGA	& $47 \pm 0.31$	& $48.82 \pm 0.24$		& $48.36 \pm 0.64$ 		& $63.84 \pm 3.85$		& $22.73 \pm 0.26$		& $31.16 \pm 0.27$ \\
TAC		& $46.19 \pm 0.31$	& $48.39 \pm 0.26$		& $52.09 \pm 0.78$ 		& $64.72 \pm 0.72$		& $24.06 \pm 0.32$		& $31.32 \pm 0.31$ \\	
Menlo PD	& $47.29$			& $50.19$				& $53.99$ 			& $59.67$				& $25.53$				& $29.95$ \\
\bottomrule
\multicolumn{7}{c}{Efficiencies with pumping on/off on Menlo PD at the end of the measurement} \\
\midrule
Menlo PD & $50.21$			& $53.35$ 			& $79.06$				& $81.96$				& $23.61$				& $25.73$ \\
\bottomrule
\end{tabular}
\end{center}

As expected, the results show, that there is little effect from optical pumping on the read-in efficiency \etain\,, while a significant effect $\sim 8-9\,\%$ can be seen on the total efficiency \etamem\,. 
However, comparison with the final cross check on the Menlo PD shows, that there is a drift reducing the memory efficiency during the measurement time. This leads to a reduction of approximately $4\, \%$ in total efficiency over the entire measurement duration and of $\ge 2\,\%$ over the course of measuring the configuration with the optical pump on. 
Taking this into account, the observed difference between both configurations ends up in the region of $\sim 4-7\,\%$. 
This is roughly what is observed when comparing the difference in retrieval efficiency between \hsp\, and \coh\, storage 
(see section \ref{ch6_subsec:MemEffCalc}). \newline 
Despite the difference in memory efficiency, the change in the optical pumping configuration does not have any effect on the {\gtwo}-values, observed for the \textit{scd} setting. 
If there were any influence, one would expect an increase in the {\gtwo} of the retrieved signal, since the lower memory efficiency reduces the amount of signal in the output. 
This would mean a larger fraction of the output would be contributed by noise, for which reason it should show a value closer to the {\gtwo} observed for the \textit{cd} setting.
The table below contains the {\gtwo} results for both optical pumping configurations. It also contains the {\gtwo}-values of the noise (setting \textit{cd}) for comparison.
\begin{center}
\footnotesize
\begin{tabular}{c|c|c|c}
\toprule
Config. 			& Sett. 		& $g^{(2)}_\text{in}$ 		& $g^{(2)}_\text{out}$ \\
\toprule
Pump on			& \textit{scd}	& $1.36 \pm 0.05$	 	& $1.63 \pm 0.05$ \\
Pump off			& \textit{scd}	& $1.44 \pm 0.04$	 	& $1.67 \pm 0.04$ \\
\midrule
Pump on			& \textit{cd}	& $1.69 \pm 0.13$	 	& $1.65 \pm 0.050$ \\
Pump off			& \textit{cd}	& $1.52 \pm 0.17$	 	& $1.74 \pm 0.07$ \\
Overall avg.$^{(*)}$	& \textit{cd}	& $1.62 \pm 0.04$	 	& $1.71 \pm 0.02$ \\
\bottomrule
\end{tabular}
\end{center}
\noindent
Here the overall average ($^{(*)}$) represents the {\gtwo}-results for the noise over all contributing measurement runs, i.e. not just during this comparison measurement (see section \ref{ch6_subsec_g2results} of the main text and appendix \ref{app6_g2results} below). 
All quoted numbers for {\gtwo} are obtained from summation over all counts (see appendix \ref{app6_g2_measurements} below).

\subsection{Memory efficiency results\label{app6_memeff}}

Here we quote the average memory efficiency values for each input photon number \Nin\,, shown in fig. \ref{fig_6_avg_memeff} of the main text. Following the explanation in section \ref{ch6_subsec:MemEffCalc}, each efficiency is obtained by calculating the weighted average over the memory efficiencies observed in each measurement run. The weighing factors are the fractional measurement times for setting \textit{scd} with respect to the total measurement time for all runs contributing to the measurement of a particular input photon number \Nin\,.
\begin{center}
\footnotesize
\begin{tabular}{c|c|c|c|c|c}
\hline
\toprule
Type & $N_\text{in}$ [$\frac{\gamma}{\text{pulse}}$]  &   \multicolumn{2}{c|}{$\eta_\text{in}$  [\%]}	& \multicolumn{2}{c}{$\eta_\text{mem}$  [\%]} \\ \hline
	&		&	TAC					& 	FPGA				&	TAC		& 	FPGA		\\
\midrule
\hsp\, & 0.22	& 	$39.10 \pm 3.53$		& $38.08 \pm 2.5$ 	& $21.13 \pm 1.87$ 	& $ 21.23 \pm 1.57$  \\
\midrule
\coh\, & tot. avg.	&	$50.57 \pm 2.27$		& $48.52 \pm 2.21$	& $29.02 \pm 0.85$	& $28.01 \pm 2.67$ \\
\midrule
\coh\, & 0.23	&	$50.57 \pm 2.27$		& $48.52 \pm 2.21$	& $29.02 \pm 0.85$	& $28.01 \pm 2.67$ \\
\coh\, & 0.49	&	$40.11 \pm 1.96$		& $39.72 \pm 1.28$	& $19.08 \pm 0.94$	& $19.49 \pm 1.16$ \\
\coh\, & 0.91 	&	$43.98 \pm 1.33$		& $43.54 \pm 1.42$	& $25.08 \pm 0.96$	& $25.05 \pm 1.01$ \\
\coh\, & 1.66 	&	$42.53 \pm 2.62$		& $41.60 \pm 1.51$	& $19.74 \pm 2.46$	& $20.11 \pm 2.11$ \\
\coh\, & 2.16	&	$44.63 \pm 1.31$		& $44.3 \pm 1.42$	& $19.85 \pm 3.16$  	& $20.41 \pm 2.71$ \\
\bottomrule
\end{tabular}
\end{center}

\section{{\gtwo} measurements\label{app6_g2_measurements}}

\subsection{Measurement details and count rate aggregation for {\gtwo}-measurement\label{app_ch6_g2_count_aggregation_details}}

\paragraph{Coincidence probabilities\label{app_sub_ch6_g2_count_aggregation_details_coinc_prob}}
As introduced in section \ref{subsec_ch5_fpga_count_rate_definition} of the main text, the coincidence probabilities $p^t_{j,k,i}(t_m)$, with $k \in \{ (H,V)|T, H|T, V|T \}$, are used to determine the {\gtwo} (eq.~ \ref{eq_ch6_g2}).
They are obtained from the coincidence counts $c^t_{j,k,i}(t_m)$ between trigger T and signal arms 
${\text{H}\, \& \, \text{V}}$, H or V, respectively, which are observed for setting $i$ in time bin $t \in \{ \text{in}, \text{out} \}$. 
The probabilities represent the coincidence counts $c^t_{j,k,i}(t_m)$ normalised by the number of trigger events 
$c_{j,T,i}(t_m)=f_\text{rep}$, which are the same for both time bins. 
As described in section \ref{ch6_sec_single_photon_storage} of the main text, we measure sequences of all measurement settings  $i \in \left\{sd, scd, cd, d \right\}$, which we refer to measurement runs. 
We alternate through all settings, recording data for measurement times of 
${\Delta t_{\text{meas},sd} \approx 5-10\,\min}$, ${\Delta t_{\text{meas},scd} \gtrsim 30\,\min}$, ${\Delta t_{\text{meas},cd} \gtrsim 30\,\min}$, ${\Delta t_{\text{meas},d} \approx 5\,\min}$.  
The resulting data for each measurement setting thus forms a time series, whose index $t_m$ denotes the time within each run $j$, when the respective datapoint was recorded. 
Each such point contains the FPGA counts, integrated over $\Delta t^\text{FPGA}_\text{int} = 10\,\sec$. 
All of this is similar to the {\gtwo} analysis in section \ref{ch5_subsec_g2}. 
For all measurements, the FPGA coincidence window time is set to $\Delta t_\text{coinc.}^\text{FPGA} = 5 \, \ns$.

\paragraph{Measurement time\label{app_sub_ch6_g2_count_aggregation_details_meas_time}} 
The \pockels\, sets an upper limit on the repetition rate $f_\text{rep}$ for memory experiments, since its pulse picking windows start to get distorted and lose contrast at too high a repetition rate ($f_\text{rep} \gtrsim 15\kHz$). 
The resulting control leakage outside the pulse picking window is undesirable, as it, e.g., increases the noise floor by building up FWM spin-wave excitations (see section \ref{ch7_FWM_explanation}). 
To avoid this problem, experiments were conducted with ${f_\text{rep} = 5.3 - 7.3  \, \text{kHz}}$ for \hsp\, and ${f_\text{rep} = 5.722 \kHz}$ for \coh\, inputs. Besides setting an upper bound on the brightness requirement of the SPDC source, these rates also limit the observable coincidence rates $c^t_{j,k,i}(t_m)$. 
Particularly the triple coincidences in the read-out time bin show low rates. 
For instance \hsps\, retrieved from the memory yield 
${\bar{c}^\text{out}_{scd,(H,V,T)} = 0.229 \pm 0.004 \, \frac{\text{counts}}{\sec}}$, while noise counts are even lower with 
${\bar{c}^\text{out}_{cd,(H,V,T)} = 0.12 \pm 0.001 \, \frac{\text{counts}}{\sec}}$. 
Tables with the average count rates obtained over the full measurement time $\Delta t_\text{meas}$ for all settings $i$ and values of {\Nin} are stated in the appendix \ref{app6_coinc_rates}. 
As described in section \ref{ch6_subsec_g2results} of the main text, the {\gtwo}-values, particularly for the memory interaction setting \textit{scd}, require high precision to enable observation of effects from the non-classical statistics of the \hsp\, input.
Consequently, small error bounds on the {\gtwo} are necessary, which means small errors on the coincidence rates $c^t_{i,k}$. 
In turn, these translate into long measurement times, which is immediately obvious when applying Poissonian counting statistics to the coincidence rates 
$c^t_{j,i,k} = \sum_{m} c^t_{j,i,k}(t_m)$, whose errors are $\Delta c^t_{j,k} = \sqrt{c^t_{j,k}}$. 
Since the rates $\bar{c}^\text{out}_{j,cd,(H,V,T)}$ are in the sub-Hz regime, the integration times required to achieve sufficient statistical significance on the respective {\gtwo}-values end up on the order of several hours in total (see also appendix \ref{app6_coinc_rates}). 
To facilitate these integration times, the experiment has to run over several days for each {\Nin}, because 
photon source and memory must perform at their optimal parameters \hereff\, and {\etamem} to enable the observations presented below. 
Moreover, systematic changes in the experiment have to be absent between different settings and within the datasets contributing to the measurement for each individual setting. 
These constraints limit the amount of tolerable drift for the apparatus and thereby the available daily measurement time. 
Data was aggregated until sufficient precision in the  {\gtwo} had been reached. 
The stability requirements were one of the most challenging aspects in performing this experiment, with the precise system alignment resulting in a daily adjustment time of $5 - 8\, \text{h}$ prior to any data taking. 
To certify a constant and reproducible performance level, we followed a fixed alignment procedure. 
Furthermore all settings for a given {\Nin} have been measured in alteration, as described in section  \ref{ch6_subsec:MemEffCalc}, to simultaneously also allow for measurements of the memory efficiency. 
The result is a series of measurement runs $m$, with each containing data recorded for all settings $i$. 
Between each run, the spatial overlap between signal, control and optical pumping was optimised for maximum memory efficiency. Additionally, the source heralding efficiency (for \hsp\, inputs), the signal filter stage alignment and the SMF-coupling into the signal filter stage were inspected and re-optimised, if required. This procedure minimised systematic drifts in the experimental apparatus.

\begin{figure}[h!]
\centering
\includegraphics[width=\textwidth]{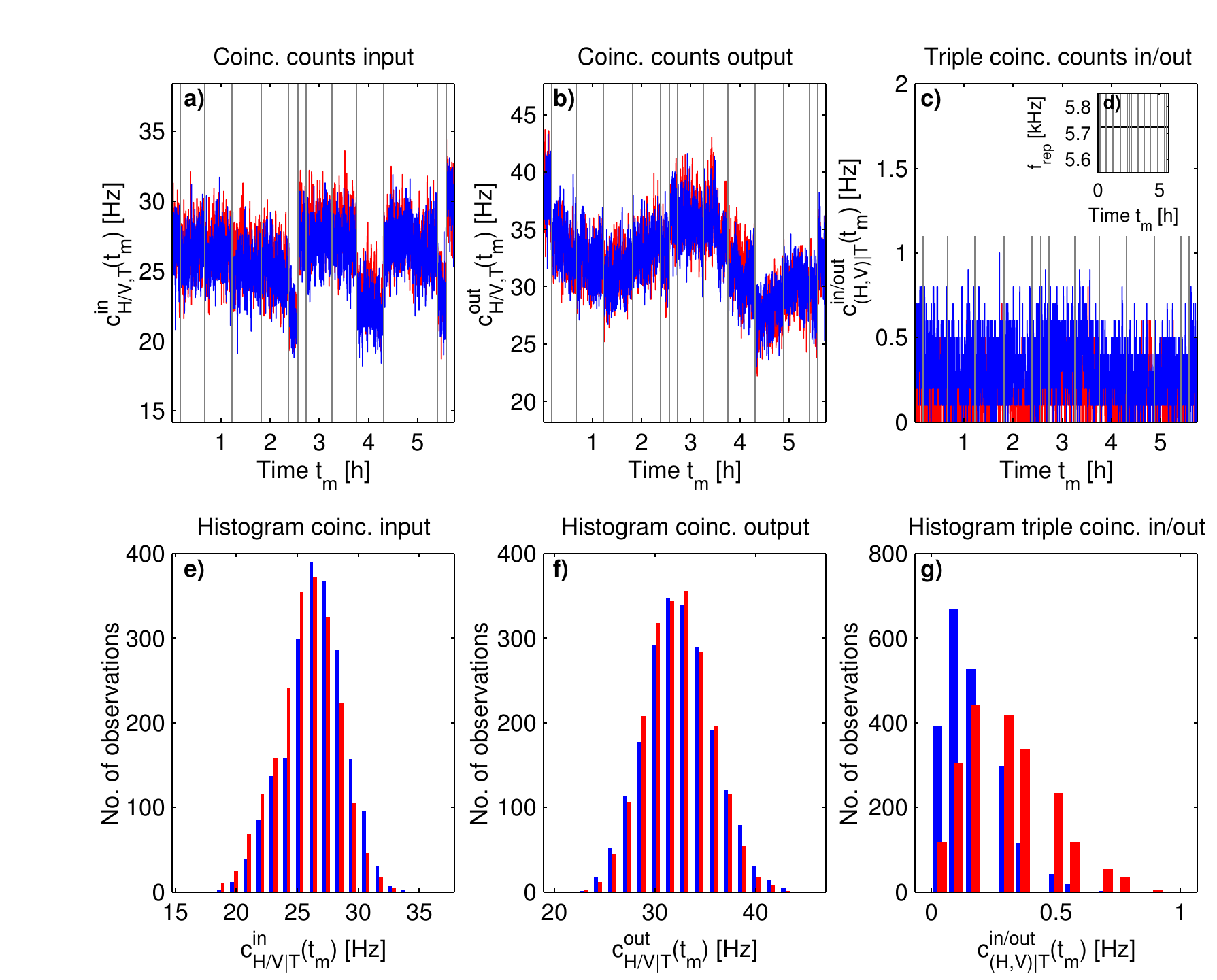}
\caption{Count rates for the measurement setting \scd\, with a \coh\, input at $N_\text{in}=0.23 \, \ppp$. 
\textbf{(a)}: coincidence count rates in the read-in time bin, measured by the FPGA, as a function of time $t_m$. 
\textit{Red} and \textit{blue lines} are coincidences $c_{(H,T)}(t_m)$ and $c_{(V,T)}(t_m)$, respectively. 
\textbf{(b)}: analogue coincidence count rates observed in the read-out time bin, colour coding as in \textbf{(a)}.
\textbf{(c)}: triple coincidence rates $c^\text{in/out}_{(H,V,T)}(t_m)$ as a function of time $t_m$ in both time bins, with the read-in time bin shown in \textit{red} and the read-out time bin in \textit{blue}. 
The inset \textbf{(d)} displays the constant experimental repetition rate $f_\text{rep}=c_T = 5.722\,\kHz$.
\textit{Grey vertical lines} in \textbf{(a)} - \textbf{(c)} mark the boundaries between different measurement runs $j$.
\textbf{(d)} - \textbf{(f)}: histograms of the count rate traces of \textbf{(a)} - \textbf{(c)}, showing the count rate distributions; same colour coding as in \textbf{(a)} - \textbf{(c)}.
}
\label{fig_ch6_coh600_coinc_rates}
\end{figure}

\begin{figure}[h!]
\centering
\includegraphics[width=\textwidth]{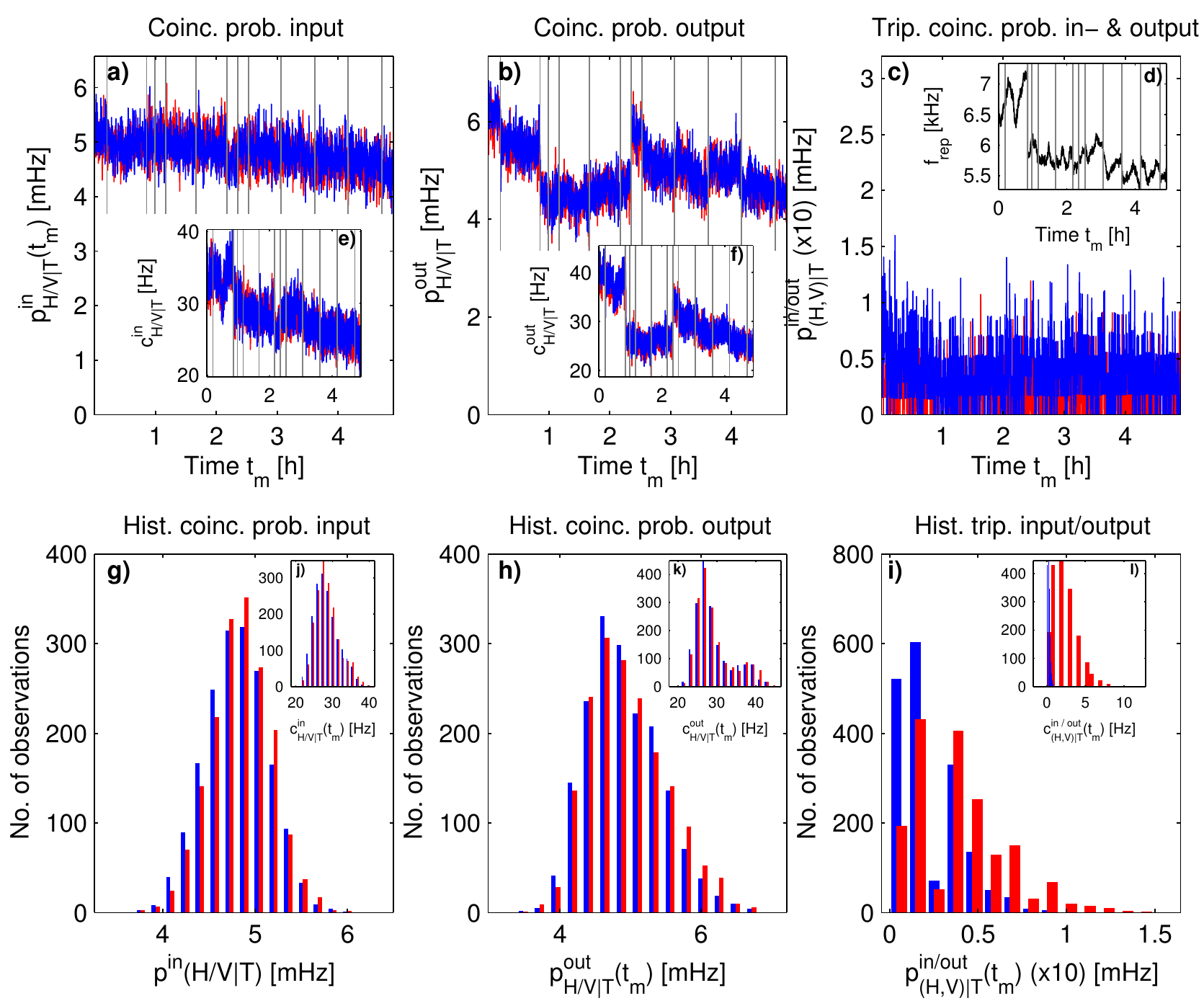}
\caption{Count rate and detection probability data for the setting \textit{scd} with \hsp\, input. 
\textbf{(a)} Detection probabilities 
$p^{\text{in}}_{(H/V|T)}$
for the read-in time bin, derived from the detected coincidence count rates 
$c^{\text{in}}_{(H/V,T)}$, shown in the inset \textbf{(e)}, divided by the experimental repetition rate ${f_\text{rep}=c_T}$, displayed in the inset \textbf{(d)}. 
\textit{Red} and \textit{blue lines} indicate coincidences between arms $\text{H}\, \& \,\text{T}$ and $\text{V} \, \& \, \text{T}$, respectively.
\textbf{(b)}: Analogous probabilities for the read-out time bin, with the corresponding coincidence count rates in inset \textbf{(f)}.
\textbf{(c)}: Triple coincidence probabilities for the read-in (\textit{red}) and the read-out (\textit{blue}) time bin.
\textbf{(g)}: Distribution of the detection probabilities $p^{\text{in}}_{(H/V)|T}$ for the read-in time bin; the inset \textbf{(j)} shows the distribution for the corresponding coincidence count rates $c^{\text{in}}_{(H/V,T)}$.
\textbf{(h)} \& inset \textbf{(k)}: Same datasets for the read-out time bin. 
Colour coding for \textbf{(g)}, \textbf{(h)}, \textbf{(j)}, \textbf{(k)} as in \textbf{(a)}.
\textbf{(l)}: Distribution of the triple coincidence probability $p^{t}_{(H,V)|T}$; inset \textbf{(l)} shows the triple coincidence count rates $c^{t}_{(H,V,T)}$. Colour coding as in \textbf{(c)}.
}
\label{fig_ch6_spdc_coinc_rates}
\end{figure}

\begin{figure}[h!]
\centering
\includegraphics[width=\textwidth]{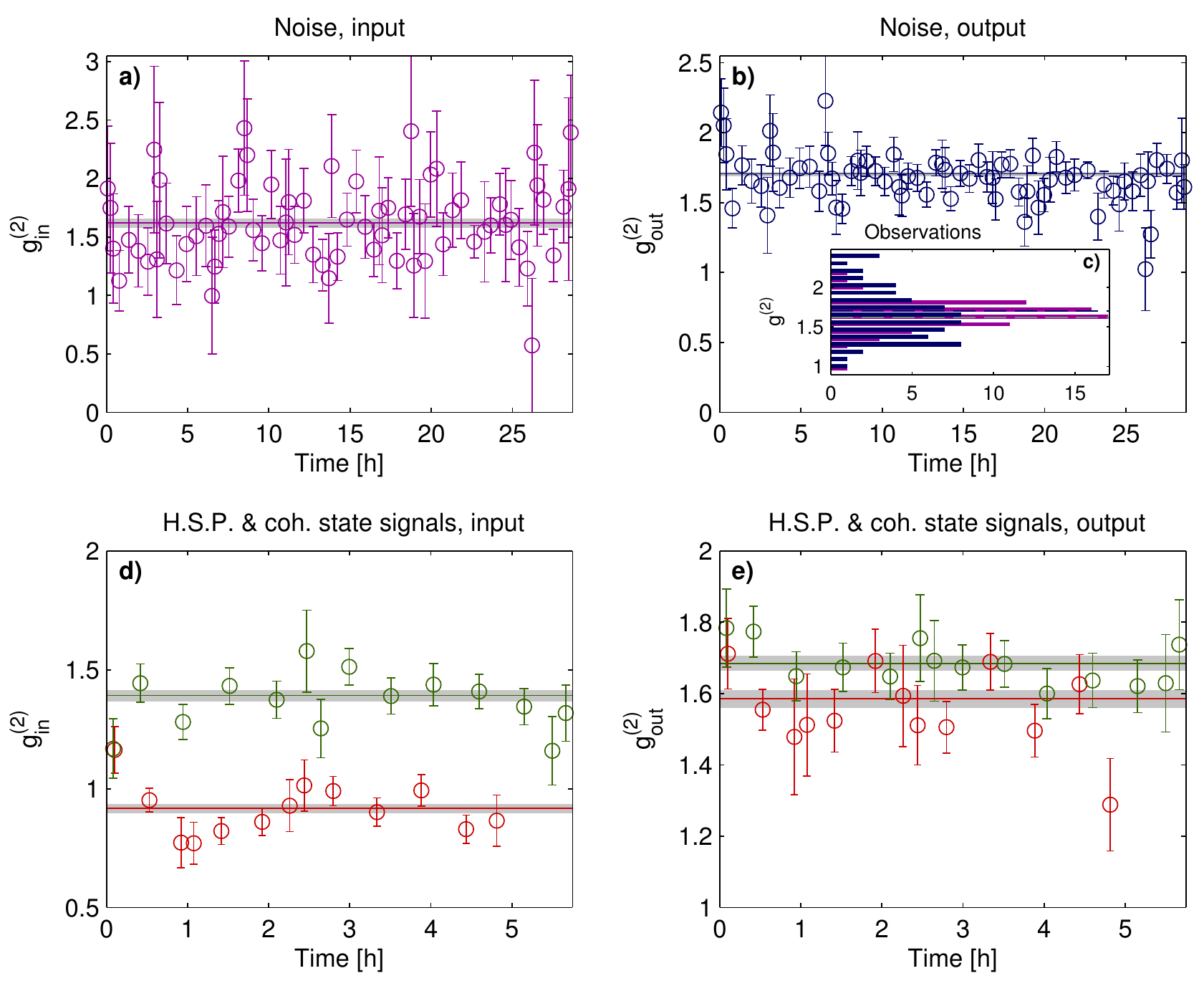}
\caption{Comparison between the {\gtwo}-values obtained from Poissonian summation (eq. \ref{ch6_g2}) and the individual $g^{(2)}_{j,i,t}$ for each measurement run $j$ (eq. \ref{ch6_g2_ind}). The former values are illustrated by constant \textit{vertical lines}, with \textit{grey shaded areas} accounting for the error on the {\gtwo}-value. 
The individual $g^{(2)}_{j,i,t}$ values are displayed as data points. 
\textit{Pink} and \textit{blue points} represent the noise in the input and output time bins, respectively. 
\textit{Red points} represent \hsp\, data and \textit{green points} \coh\, data.
\textbf{(a)} and \textbf{(b)} show the noise {\gtwo}-values in the input and output time bins, 
whose $g^{(2)}_{j,cd,t}$ distributions are depicted in the inset \textbf{(c)}. 
The histogram in \textbf{(c)} also contains the {\gtwo}-values from Poissonian summation as \textit{dotted lines}, where the \textit{grey} and \textit{blue lines} represent its value for the input and output time bin, respectively. 
\textbf{(d)} and \textbf{(e)} contrast the data for \hsp\, and \coh\, inputs, when transmitted through and retrieved from the memory, respectively (setting \scd\,). 
Note here the drop in {\gtwo} between \coh\, and \hsps\,. It illustrates the effect from the quantum nature of the \hsp\, input, when storing the signal in the memory. 
}
\label{fig_ch6_g2ind_g2avg}
\end{figure}

\paragraph{Data aggregation\label{app_sub_ch6_g2_count_aggregation_details_data_aggregation}}
The resulting fragmentation of data into sub-sets for individual runs necessitates its posterior combination. 
Since the effects in the {\gtwo} we aim to measure are small (see section \ref{ch6_subsec_g2results}), 
the statistics applied to data processing needs to be sound. 
For this reason, a detailed description is given based on the three datasets: \coh\, with $N_\text{in} = 0.23 \,\ppp$ 
(setting \textit{scd}, fig. \ref{fig_ch6_coh600_coinc_rates}), 
\hsp\, (setting \textit{scd}, fig. \ref{fig_ch6_spdc_coinc_rates}) 
and noise (setting \textit{cd}, fig. \ref{fig_ch6_g2ind_g2avg}). 
The data processing is best understood when examining the \coh\, case, since here the experimental repetition rate is constant at  
$f_\text{rep}=5.722 \kHz$ (fig. \ref{fig_ch6_coh600_coinc_rates} \textbf{d}). 
Starting from the count rates $c^t_{j,k,i}(t_m)$, detected in each 
$\Delta t^\text{FPGA}_\text{int} = 10\,\sec$ FPGA integration window time bin $t_m$ 
(fig. \ref{fig_ch6_coh600_coinc_rates} \textbf{a} - \textbf{c}), firstly the total number of counts per measurement run $j$, 
$c^t_{j,k,i} = \sum_{m} c^t_{j,k,i}(t_m)$, is computed. 
As a reminder, $t \in \{\text{in}, \text{out}\}$ is the time bin, $j$ is the measurement run number, $k \in \{(H,T), (V,T), (H,V,T) \}$ is the coincidence type and $i \in \{scd, sd, cd, d\}$ is the measurement setting. 
The summation relies on the applicability of Poissonian count rate statistics. 
So the inherent Poissonian count rate fluctuations are the main source of variation in the count rates within each run $j$. 
In particular no other systematic effects, such as changes in experimental parameters or drifts, are dominating the variation in count rate. 
For the data in fig. \ref{fig_ch6_coh600_coinc_rates} this is a good assumption. 
Moreover, thanks to the alignment and continuous re-optimisation of the experiment, this argument also holds true reasonably well throughout the entire measurement, i.e., the rates $c^t_{j,k}$ are approximately dominated by Poissonian fluctuations over all measurement runs $j$.
In fact, due to the constant number of experiments $c_{T}$ (fig. \ref{fig_ch6_coh600_coinc_rates} \textbf{d}), the count rates can be used to tell how well the experimental conditions can be reproduced on a daily basis. 
For the example measurement in fig. \ref{fig_ch6_coh600_coinc_rates}, the rates in the input time bin (\textbf{a}) are approximately constant, besides 3 runs showing an offset that is nevertheless within the Poissonian fluctuation of the dataset. In the output bin (\textbf{b}) there are small count rate oscillations. 
However, the count rate distributions (histograms in fig. \ref{fig_ch6_coh600_coinc_rates} \textbf{e} - \textbf{g}) are clearly uni-modal, i.e., there is only one mean around which events are distributed. 
This is a different formulation of the aforementioned requirement about the dominance of Poissonian fluctuations. 
Due to the high number of coincidence counts, the distributions are normal, while the triple coincidences still follow a  Poisson-distribution.
Thanks to the uni-modal distributions, 
we can calculate the overall number of counts $c^t_{k,i} = \Sigma_{j=1}^{N_\text{r}} c^t_{j,k,i}$, 
where $N_\text{r}$ denotes the total number of measurement runs. 
The same is done for the trigger pulses $c_{T,i} = \sum_{j=1}^{N_\text{r}} c_{j,T,i} =  \sum_{j=1}^{N_\text{r}}  \sum_{m} c_{j,T,i} (t_m)$. 
Errors are simply given by the Poissonian errors $\Delta c^{t}_{j,k,i} = \sqrt{c^{t}_{j,k,i}}$ for the individual runs and $\Delta c^{t}_{k,i} = \sqrt{c^{t}_{k,i}}$ overall. \newline
Notably, in applying the summation over $N_\text{r}$, the dataset is confined to a single input photon number, as different values for {\Nin} naturally result in jumps in the count rates, and consequently multi-modal count rate distributions.
For this reason, summation over $N_\text{r}$ can be applied to all measurements with the exception of the \coh\, input photon statistics (setting $i=$\textit{sd}). 
Here, data is aggregated over all {\Nin}, since the input statistics of \coh\, is independent of their mean photon number.\newline
With the summed-up counts, the coincidence probabilities $p^{t}_{k,i} = \frac{c^{t}_{k,i}}{c_{T,i}}$ are obtained, with errors $\Delta p^{t}_{k,i} = \frac{\sqrt{c^{t}_{k,i}}}{c_{T,i}}$.
Importantly, these do not contain any errors from $c_{T,i}$, since $c_{T,i}$ represents the total number of experiments conducted, which is a constant\footnote{
	For \hsp\, inputs, $c_{T,i}$ also contains contributions from dark counts of \spcmdt\,. 
	These must not be subtracted, because the false \pockels\, triggers from dark counts 
	still represent conducted experimental trials.
}. 
Finally, using eq. \ref{eq_ch6_g2}, the probabilities are used to get the {\gtwo} for setting $i$ and time bin $t$ at the input photon number $N_\text{in}$, according to 
\begin{equation}
g^{(2)}_{i,t} (N_\text{in})= \frac{p^{t}_{((H,V)|T),i}}{p^{t}_{(H|T),i}\cdot p^{t}_{(V|T),i}}. 
\label{ch6_g2}
\end{equation}
\noindent
Its measurement error follows from Gaussian error propagation and is given by
\begin{equation}
\Delta g^{(2)}_{i,t} (N_\text{in}) = \sqrt{
	\frac{\left( \Delta p^{t}_{(H,V)|T,i} \right)^2}{\left( p^{t}_{H|T,i} \cdot p^{t}_{V|T,i} \right)^2} + 
    \frac{\left( p^{t}_{(H,V)|T,i} \cdot \Delta p^{t}_{H|T,i} \right)^2}{\left( \left(p^{t}_{H|T,i} \right)^2 \cdot p^{t}_{V|T,i} \right)^2}+ 
    \frac{\left( p^{t}_{(H,V)|T,i} \cdot \Delta p^{t}_{V|T,i} \right)^2}{\left(  p^{t}_{H|T,i} \cdot \left(p^{t}_{V|T,i} \right)^2 \right)^2}}.
\label{ch6_g2err}
\end{equation}

In contrast to \coh\, input signals, the measurements for \hsps\, have a varying repetition rate $c_T$. 
The rate is subject to the following sources of fluctuation: 
\begin{enumerate}
\item Poissonian count rate fluctuations in idler detection. 
\item Variations in the coupling efficiency of the SPDC source's UV pumping beam into the ppKTP waveguide. 
\item Systematic drifts due to changes in the SPDC generation efficiency, relating to phase-matching by temperature tuning the ppKTP waveguide.
\end{enumerate} 
The corresponding variations in the number in experimental trials translate into variations of the detected coincidence rates $c^t_{j,k,i}(t_m)$, shown for setting \textit{scd} in fig. \ref{fig_ch6_spdc_coinc_rates} \textbf{e} - \textbf{f}. 
As a result these rates contain jumps coinciding with the step changes in the repetition rate $f_\text{rep}$, displayed in fig. \ref{fig_ch6_spdc_coinc_rates} \textbf{d}. 
Correspondingly, their count rate distributions 
(fig. \ref{fig_ch6_spdc_coinc_rates} \textbf{j} \& \textbf{k}) 
are no longer uni-modal but contain two regimes associated with the two different repetition rate regimes 
$f_\text{rep} \approx 6.8 \,\kHz$ and 
$f_\text{rep} \approx 5.8 \,\kHz$.
To process the data one has to look at the rates without the effects from changes in $c_T$. 
For this reason, the coincidence probabilities per registered FPGA datapoint 
$p^t_{j,k,i}(t_m) = \frac{c^t_{j,k,i}(t_m)}{c_{j,T,i}(t_m)}$ are required, which are displayed in fig. \ref{fig_ch6_spdc_coinc_rates} \textbf{a} - \textbf{c}.
These rates are approximately constant and, once more, dominated by Poissonian fluctuations. 
The distribution of the probabilities is thus again reasonably uni-modal 
(fig. \ref{fig_ch6_spdc_coinc_rates} \textbf{g} - \textbf{i}). 
This makes the probabilities comparable to the \coh\, data in fig. \ref{fig_ch6_coh600_coinc_rates}, whose coincidence probabilities are just a rescaled version of their count rates.
It is thus again justified to sum first over all data points $t_m$ per run $j$ and second over all $N_\text{r}$ runs, with the {\gtwo} obtained by using eqs. \ref{ch6_g2} \& \ref{ch6_g2err}.
The numerical outcomes for all {\gtwo} measurements are stated in appendix \ref{app6_g2results}. 

Notably, the results obtained by this summation of counts are the same as one would obtain if the mean of all detection probabilities over the entire dataset $N_r$ was computed, i.e., the variables 
$\bar{p}^t_{j,k,i} = \frac{\underset{j,m}{\Sigma} p^t_{j,k,i}(t_m)}{\underset{j,m}{\Sigma} 1}$ were used in eq. \ref{ch6_g2} instead. 
Errors are comparable to the results from count summation, yet they slightly differ as the standard error\footnote{
	The calculation of the standard error implicitly uses the applicability of the 
	central limit theorem to the dataset. This will be discussed in the text below and 
	is explicitly shown in the appendix \ref{app6_data_aggregation} for the noise data 
	(setting \textit{cd}), where the convergence of the triple coincidence detection 
	probability $p^t_{j,(H,V)|T,i}$ to a normal distribution is illustrated in fig. \ref{fig_ch6_distr_g2ind_noise}.	
} derives from the variance of the counts around their mean. 
These numbers are also presented in the appendix \ref{app6_data_aggregation}.

There is another way to process this data, which has less stringent requirements on experimental reproducibility. 
This method uses the somewhat more standard procedure of taking the mean over the {\gtwo}-values for the individual runs $j$. Notably, these are also needed for the hypothesis tests performed to determine the statistical significance of the {\gtwo} results in appendix \ref{app6_stat_tests}. 
We still utilise the dominance of Poissonian fluctuations within each measurement run and sum the count rates over all recorded points $t_m$ for each run $j$. The resulting $c^t_{j,k,i}$ are used to compute individual {\gtwo}-values per run $j$ via
\begin{equation}
g^{(2)}_{j,i,t} (N_\text{in}) = \frac{p^{t}_{j,((H,V)|T),i}}{p^{t}_{j,(H|T),i}\cdot p^{t}_{j,(V|T),i}} = \frac{c^{t}_{j,(H,V,T),i} \cdot c_{j,T,i}}{c^{t}_{j,(H,T),i}\cdot c^{t}_{j,(V,T),i}}.
\label{ch6_g2_ind}
\end{equation}
Errors are calculated by applying eq. \ref{ch6_g2err} to the individual count rate errors $\Delta c^{t}_{j,k,i}$. 
To get the overall {\gtwo} for the setting $i$, the mean of the individual $g^{(2)}_{j,i,t}$ is determined, $\bar{g}^{(2)}_{i,t} = \frac{\sum_{j=1}^{N_\text{r}} g^{(2)}_{j,i,t}}{N_\text{r}}$. 
As the averaging works on the {\gtwo}-values of individual runs, it is not sensitive to parameter changes between runs, as long as the investigated signal has the same statistical properties. For this reason, it is the method utilised to determine the {\gtwo} of the \coh\, input (setting $i=sd$), where the combined data for all {\Nin} are used. 
The overall error is obtained as the standard error on the mean as $\Delta \bar{g}^{(2)}_ {i,t} = \frac{\sigma_{i,t}}{\sqrt{N_\text{r}}}$, with the sample standard deviation (std) $\sigma_{i,t} = \sqrt{\frac{\sum_{j=1}^{N_\text{r}}\left( g^{(2)}_{j,i,t}-\bar{g}^{(2)}_{i,t} \right)^2}{N_\text{r}-1}}$. 
The standard error takes into account the spread of the different $g^{(2)}_{j,i,t}$-values around their mean $\bar{g}^{(2)}_{i,t}$. 
However it does not account for the individual errors on the $g^{(2)}_{j,i,t}$. 
This means it weighs every datapoint equally, irrespective of the underlying integration time and associated precision of the count rates going into eq. \ref{ch6_g2_ind}. 
To address their varying significance, we can apply weighing factors to each $g^{(2)}_{j,i,t}$, which can be based on the Poissonian error or the measurement time of the respective run (see appendix \ref{app6_data_aggregation}). 

Irrespective of weighing, the calculation rests on the important condition that the central limit theorem (CLT) is applicable\footnote{
	This is also a requirement for the applicability of the statistics tests, 
	used in appendix \ref{app6_stat_tests}.
} to the $g^{(2)}_{j,i,t}$.  
This is not immediately clear, as not all distributions of all variables $p_{j,k,i}$, entering eq. \ref{ch6_g2_ind}, are normal. 
As fig.~\ref{fig_ch6_spdc_coinc_rates} shows, the normality is obvious for the coincidence terms, whose count rates are high enough to have their Poissonian distributions converge to Gaussians. Yet, the triple coincidences are still Poissonian distributed. For them, the normality comes from the CLT\footnote{
	The central limit theorem states that the sum $S_m =\sum_m x_m$ 
	of random variables $x_m$, each with the same distribution, 
	converges to a normal distribution in the limit of large $m$. 
	For the theorem to hold, the $x_m$ must have a finite mean and variance, 
	which is fulfilled for Poissonian and normal distributions. 
	Convergence can usually already be observed 
	after summation of $\sim 5$ variables\cite{Steck:LectureNotes}. 
	In our case, each measurement run $j$ contains at least $10$ data points. 
	The major part of the runs was integrated for $\gtrsim 30\, \min$ and 
	thus has at least $180$ data points, i.e. $| \{ m \}| \ge 180$, which is more than sufficient. 
}: 
Since we use the sum of the count rates $c^{t}_{j,k,i}(t_m)$, over all $t_m$, each factor going into eq. \ref{ch6_g2_ind} is in fact a sum of random variables, for which reason these sums $c^{t}_{j,k,i}$ are normally distributed. 
Fig. \ref{fig_ch6_distr_g2ind_noise} in appendix \ref{app6_data_aggregation} illustrates this explicitly for the 
$c_{j,(H,V,T),cd}$ triple coincidences of the noise measurements. 
However, since the denominator of the $g^{(2)}_{j,i,t}$ in eq. \ref{ch6_g2_ind} contains a product of two normal distributions it is described by a modified Bessel function of $2^\text{nd}$ kind\cite{Steck:LectureNotes}, while the numerator is Poissonian distributed. 
Irrespective of what the exact distributional form of the resulting $g^{(2)}_{j,i,t}$ is, 
it must not have fat tails\footnote{
	The Cauchy or Lorentzian distribution has the functional form $f(x) \sim \frac{1}{x^2}$, 
	which leads to a diverging variance $\sigma^2$, since 
	$\sigma^2 \sim \langle x^2 \rangle - \langle x \rangle^2= \int x^2 f(x) dx + \left(\int x f(x) dx \right)^2$, 
	giving the distribution fat tails.  
	The central limit theorem however only holds for random variables with finite mean and variance.
}.
We certify this by running a Shapiro-Wilk test\cite{Ruppert:Book} on the $g^{(2)}_{j,i,t}$ for all settings $i$ (appendix \ref{app6_stat_tests}). 
With the Null hypothesis $H_0$ that the $g^{(2)}_{j,i,t}$ are normally distributed, the test cannot reject $H_0$ with 
$\gtrsim 95 \, \%$ confidence for all $i$ in both time bins $t$. While this does not prove normality of the $g^{(2)}_{j,i,t}$, it excludes fat tails. 
Hence, the CLT applies and the means $\bar{g}^{(2)}_{i,t}$ are normally distributed. 

An illustration of the $g^{(2)}_{j,i,t}$ data is provided in fig. \ref{fig_ch6_g2ind_g2avg}, where \textbf{a} - \textbf{b} contain plots of the $g^{(2)}_{j,cd,t}$ of the noise for each run $j$ as a function of measurement time. Subplot \textbf{c} shows their resulting distributions as histograms. It is difficult to guess the distributional form graphically. 
Yet, one can nicely see that {\gtwo}-values, obtained from eq. \ref{ch6_g2} through count summation over the entire measurement and plotted by \textit{vertical lines}, coincide well with the centres of the histograms. 
Equally, also the $g^{(2)}_{j,scd,t}$ data for \hsp\, and \coh\, inputs, which are displayed in fig. \ref{fig_ch6_g2ind_g2avg} \textbf{c} \& \textbf{d} for the read-in and read-out time bins, respectively, distribute nicely around the $g^{(2)}_{scd,t}$ values.
Hence, the $g^{(2)}_{i,t}$ are good expressions for the {\gtwo} of the whole sample, as are the means $\bar{g}^{(2)}_{i,t}$. 
Both methods are applicable and yield similar numerical {\gtwo} values (see appendix \ref{app6_data_aggregation}). 
In a final step, we demonstrate that this graphical impression is actually correct, by running Student T-tests (appendix \ref{app6_stat_tests}) on the data samples $\{g^{(2)}_{j,i,t} \}$ of individual {\gtwo}-values. 
These are tests against the $g^{(2)}_{i,t}$ data, obtained by count summation from using eq.~\ref{ch6_g2}, 
which are treated as the assumed means of the populations. 
To this end, a double-sided, one-sample T-test\cite{Ruppert:Book} is applied, where the Null hypothesis $H_0$ is that the $g^{(2)}_{i,t}$ are the mean of the individual $\left\{ g^{(2)}_{j,i,t} \right\}$. 
The T-test cannot reject $H_0$ with at the least $95 \, \%$ confidence for all settings and time bins (see appendix \ref{app6_stat_tests}).
It is hence fair to use $g^{(2)}_{i,t} (N_\text{in})$, obtained by eq. \ref{ch6_g2} through summation of all counts under Poissonian statistics, to benchmark the photon statistics of the signal and noise fields.  
 

\begin{figure}[h!]
\centering
\includegraphics[width=12cm]{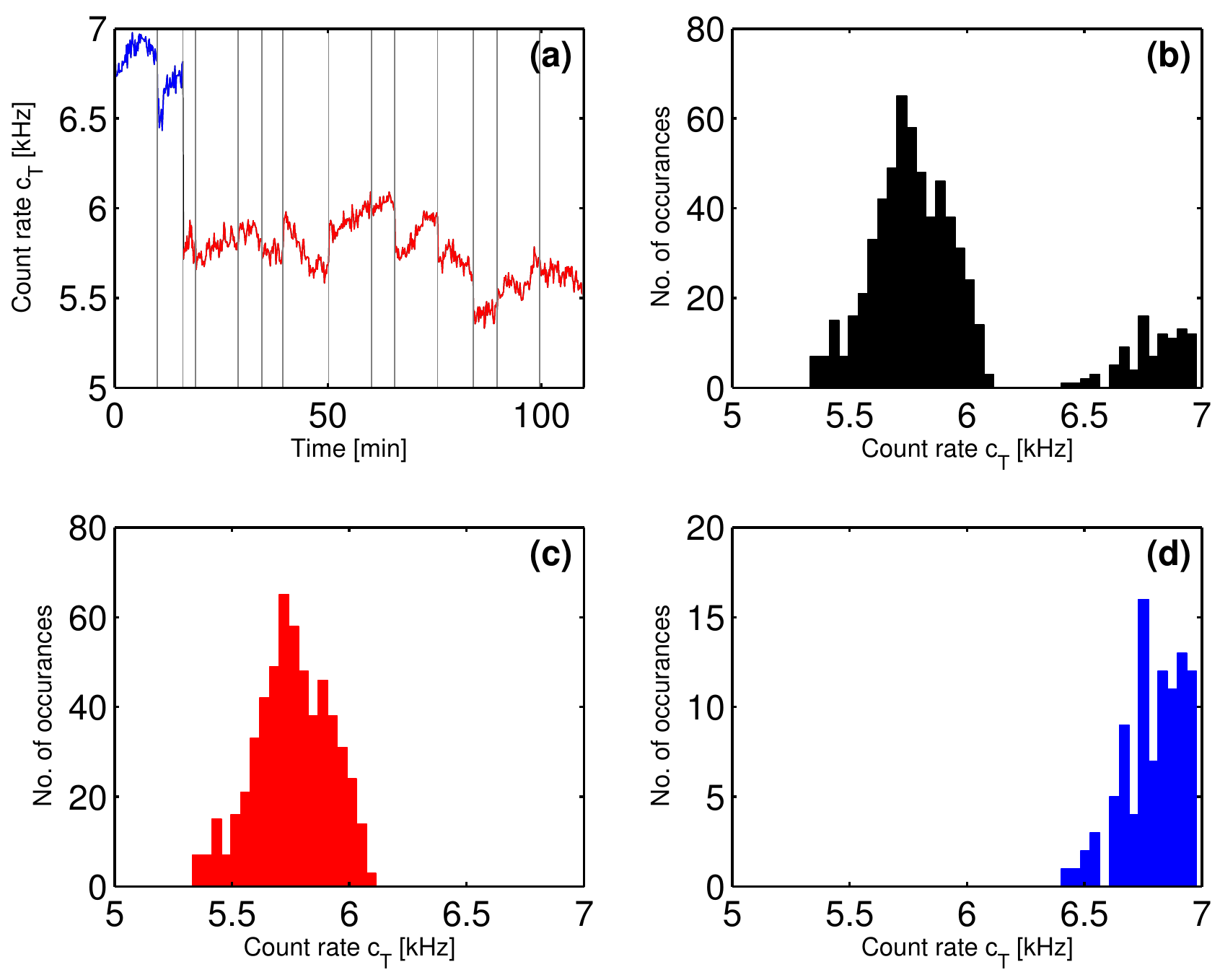}
\caption{Histogram of number of experiments $c_T$, derived from idler detection, for the measurement of setting \textit{sd} with \hsp\, input. \textbf{(a)}: the count rates $c_T$ as observed during the total measurement time $\Delta t_\text{meas} \approx 110 \, \min $, where the vertical lines separate the data belonging to different measurement runs. Regimes of the same systematic conditions are marked by colour (\textit{red} and \textit{blue}). \textbf{(b)}: Histogram of the total count rate trace shown in \textbf{(a)}. \textbf{(c)}: Histogram of the count rate regime 1 (\textit{red} in \textbf{(a)}). \textbf{(d)}: Histogram of the count rate regime 2 (\textit{blue} in \textbf{(a)}). }
\label{app_ch6_coinc_rates}
\end{figure}

\subsection{Coincidence rates observed during {\gtwo} measurements \label{app6_coinc_rates}}

The following tables list the coincidence count rates observed during the measurements of the {\gtwo} data, presented in section \ref{ch6_subsec_g2results}. 
Listed count rates ${c}^{{t}}_{i,k}$ represent the average over all count rate data points $c^{{t}}_{j,i,k}(t_m)$; as usual $t$ is the time bin, $k \in \{(H,T), (V,T), (H,V,T) \}$ denotes the coincidence type and $i$ indicates the measurement setting 
$i\in \left\{\textit{scd}, \textit{sd}, \textit{cd}\right\}$. 
As introduced in section \ref{sec_ch4_photon_detection} of the main text, the $c^{{t}}_{j,i,k}(t_m)$ are the counts registered by the FPGA in one $10\, \sec$ integration time bin $t_m$ for each measurement run $j$. 
The errors on the ${c}^{{t}}_{i,k}$ are given by the standard error of the $c^{{t}}_{i,k}(t_m)$. 

The standard error can be used as an error metric thanks to the applicability of the central limit theorem, which results in a normal distribution for the sum $S_i = \underset{j,m}{\sum} c^{{t}}_{j,i,k}(t_m)$, while the individual count rates $c^{{t}}_{j,i,k}(t_m)$ are Poisson distributed. 
Thus 
${c}^{{t}}_{k} = \frac{ \overset{N_r}{ \underset{j=1}{\sum} } \underset{m}{\sum} c^{{t}}_{j,i,k}(t_m)}{N_r}$ 
is also normally distributed with a sample standard deviation 
$\sigma^{t}_{k} = \sqrt{\frac{ \overset{N_r}{ \underset{j=1}{\sum} } \underset{m}{\sum} \left(c^t_{j,i,k}(t_m) -{c}^t_{i,k} \right)^2}{N_r - 1}}$, 
which leads to a standard error of $\Delta {c}^t_{i,k} = \frac{\sigma^{t}_{k,i} }{\sqrt{N_r}}$. 
Here, $N_r$ is the number of measurement runs $j$ per input photon number {\Nin}. 

Besides the coincidence rates, the tables below also contain the experimental repetition rates $f_\text{rep} = c_T$. For \coh\, inputs, these correspond to the {\tisa} clock signal, divided-down by the Pockels cell (see section \ref{ch6_sec:setup}). 
Consequently, the signal is stable with a negligible error $\left( \Delta c_T \right)$ on the order of mHz, which is therefore not quoted. 
For \hsp\,, $c_T$ is the number of detected idler photons\footnote{
	With a few counts lost due to the delay settings of DDG 1, used to temporally filter false triggering on dark counts 
	(see sections \ref{ch5_sec_setup} \& \ref{ch6_setup_electronics}.)
}. This number is subject to systematic errors. 
These are mainly oscillations in the UV coupling efficiency into the ppKTP waveguide and, more importantly, changes in the SPDC generation efficiency, originating from a drift in the optimal phase matching temperature of the waveguide (see appendix \ref{app_ch5_insufficiencies}). The latter effect causes count rate offsets on a daily time scale. 
As fig. \ref{app_ch6_coinc_rates} shows, this leads to step-changes in $c_T$ and in turn to a much greater variation in count rate than expected from a statistical error. 
The error $\Delta c_T$ can thus not be computed by statistical means only. Instead it is the sum of the statistical count rate fluctuation and the size of the systematic variation. The latter is determined by separating the data into regimes with equal systematic conditions and then taking the difference between the mean count rates of the regime with highest and the one with lowest rates.
Fig. \ref{app_ch6_coinc_rates} illustrates the procedure by the example of the data for setting \textit{sd}. 
The count rates $c_T(t_m)$ (\textbf{a}) are not normally distributed, as observable from their histogram (\textbf{b}). Rather there are two regimes, separated by the size of the count rate step change. Each regime shows approximately normally distributed rates\footnote{
	As one would expect for a Poissonian distribution with a large mean.
} $c_T$ (\textbf{c} and \textbf{d}) with standard deviation $\sigma_{1,2}$ and mean $\mu_{1,2}$.
The statistical error $\Delta c_T^\text{stat} = \frac{\sqrt{\sigma_1^2 + \sigma_2^2}}{\sqrt{N_r}}$ and the systematic error $\Delta c_t^\text{sys} = \frac{|\mu_1 - \mu_2|}{2}$ give the total error $\Delta c_T^\text{tot}=\Delta c_T^\text{stat}+\Delta c_T^\text{sys}$, quoted in the following tables. \newline
The tables furthermore state the total measurement time $\Delta t_\text{meas}$ for recording all runs $N_r$, contributing to the respective set of numbers ${c}^t_{i,k}$. 
The rates are split according to the settings $i$ and combine the rates for all recorded input photon numbers \Nin\, for the three measured signal types \hsp\,, \coh\, and noise. \newline
\begin{itemize}
\item \textbf{Active Raman memory:} Experimental configuration with optical pumping, i.e. {\cs} state preparation in $6^2S_\frac{1}{2}, F=4$. \newline
\begin{center}
\footnotesize
\begin{tabular}{@{} c | c | c | c | c | c | c  @{}} 
\toprule
\multicolumn{7}{c}{\textbf{Memory interaction setting \textit{scd}}} \\
\toprule
						& \hsp\, 				& \multicolumn{5}{c}{Coherent states} \\\hline
Coinc. 					& $N_{\text{in}} = 0.22$  & $N_{\text{in}} = 0.23$ 	& $N_{\text{in}} = 0.49$ 	& $N_{\text{in}} = 0.91$ 	& $N_{\text{in}} = 1.66$ 	& 
$N_{\text{in}} = 2.16$ \\
\midrule
$c_{T}$ [kHz]				& $5.9 \pm 0.5$ 		& $5.722$				& $5.722$ 			& $5.722$ 			& $5.722$ 	& $5.722$	\\
$c^{\text{in}}_{{H,T}}$ [Hz] 	& $28.37 \pm 0.08$		& $26.49 \pm 0.05$		& $51.91 \pm 0.09$		& $80.8 \pm 0.15$		& $127.7 \pm 0.35$		& $169.4 \pm 0.4$ \\
$c^{\text{in}}_{{V,T}}$ [Hz] 	& $28.6 \pm 0.08$		& $25.8 \pm 0.05$		& $51.4 \pm 0.1$		& $80.6 \pm 0.17$		& $130 \pm 0.4$		& 
$170 \pm 0.4$ \\
$c^{\text{out}}_{{H,T}}$ [Hz] 	& $29.1 \pm 0.1$		& $32.49 \pm 0.07$		& $34.4 \pm 0.1$ 		& $55.9 \pm 0.2$		& $61.7 \pm 0.2$		& $80 \pm 0.3$ \\
$c^{\text{out}}_{{V,T}}$ [Hz] 	& $29.4 \pm 0.1$		& $32.31 \pm 0.07$ 		& $34.1 \pm 0.1$		& $55.5 \pm 0.2$		& $62.6 \pm 0.2$		& $80.4 \pm 0.2$ \\
$c^{\text{in}}_{{H,V,T}}$ [Hz] 	& $0.126 \pm 0.003$		& $0.166 \pm 0.003$ 	& $0.567 \pm 0.008$		& $1.29 \pm 0.01$		& $3.11 \pm 0.03$		& $5.38 \pm 0.03$ \\
$c^{\text{out}}_{{H,V,T}}$ [Hz] & $0.229 \pm 0.004$	& $0.309 \pm 0.004$ 	& $0.325 \pm 0.006$		& $0.84 \pm 0.01$		& $0.95 \pm 0.01$		& $1.54 \pm 0.02$ \\ 
\bottomrule
\end{tabular}
\end{center}

\begin{center}
\footnotesize
\begin{tabular}{@{} c | c | c | c | c | c | c  @{}} 
\toprule
\multicolumn{7}{c}{\textbf{Input signal setting \textit{sd}}} \\
\toprule
						& \hsp\, 				& \multicolumn{5}{c}{Coherent states} \\\hline
Coinc. 					& $N_{\text{in}} = 0.22$  & $N_{\text{in}} = 0.23$ 	& $N_{\text{in}} = 0.49$ 	& $N_{\text{in}} = 0.91$ 	& $N_{\text{in}} = 1.66$ 	& 
$N_{\text{in}} = 2.16$ \\
\midrule
$c_{T}$ [kHz]				& $5.91 \pm 0.53$ 		& $5.722$ 	& $5.722$  	& $5.722$ 	& $5.722$ & $5.722$ \\
$c^{\text{in}}_{{H,T}}$ [Hz] 	& $30.51 \pm 0.13$		& $33.35 \pm 0.1$	& $74.17 \pm 0.14$		& $128.87 \pm 0.28$		& $209.97 \pm 0.56$		& $290.32 \pm 0.48$ \\
$c^{\text{in}}_{{V,T}}$ [Hz] 	& $30.56 \pm 0.14$		& $33.22 \pm 0.1$	& $73.45 \pm 0.16$		& $128.35 \pm 0.28$		& $214.2 \pm 0.74$		& $292.11 \pm 0.54$ \\
$c^{\text{out}}_{{H,T}}$ [Hz] 	& $0.23 \pm 0.01$		& $0.34 \pm 0.01$	& $0.65 \pm 0.01$ 		& $1.09 \pm 0.02$		& $1.82 \pm 0.02$		& $3.05 \pm 0.03$ \\
$c^{\text{out}}_{{V,T}}$ [Hz] 	& $0.25 \pm 0.01$		& $0.35 \pm 0.01$ 	& $0.65 \pm 0.01$		& $1.09 \pm 0.02$		& $1.85 \pm 0.02$		& $3 \pm 0.03$ \\
$c^{\text{in}}_{{H,V,T}}$ [Hz] 	& $(26 \pm 7) \cdot 10^{-4}$		& $0.2 \pm 0.005$ 	& $0.97 \pm 0.01$		& $2.87 \pm 0.02$		& $7.84 \pm 0.06$		& $14.91 \pm 0.07$ \\
$c^{\text{out}}_{{H,V,T}}$ [Hz] & $0 \pm 0$		& $0 \pm 0$ 		& $0 \pm 0$			& $0 \pm 0$			& $(5 \pm 4)\cdot 10^{-4}$			& $(11 \pm 6)\cdot 10^{-4}$ \\ 
\bottomrule
\end{tabular}
\end{center}
\begin{center}
\footnotesize
\begin{tabular}{@{} c | c | c | c  @{}} 
\toprule
\multicolumn{4}{c}{\textbf{Noise setting \textit{cd}}} \\
\toprule
Coinc. 					& \hsp\, 						& Coh. states 				& \hsp\, \& \coh\, \\ 
\midrule
$c_{T}$ [kHz]				& $4.22 \pm 1.88$				& $5.72 \pm (6 \cdot 10^{-4})$		& $5.14 \pm 1.88$ \\
$c^{\text{in}}_{{H,T}}$ [Hz] 	& $6.33 \pm 0.03$				& $8.35 \pm 0.02$				& $7.57 \pm 0.02$ \\	
$c^{\text{in}}_{{V,T}}$ [Hz] 	& $6.16 \pm 0.04$				& $7.94 \pm 0.017$				& $7.25 \pm 0.02$ \\	
$c^{\text{out}}_{{H,T}}$ [Hz] 	& $15.23 \pm 0.09$				& $21.51 \pm 0.04$				& $19.07 \pm 0.05$ \\	
$c^{\text{out}}_{{V,T}}$ [Hz] 	& $15.15 \pm 0.09$				& $21.27 \pm 0.04$ 				& $18.89 \pm 0.05$ \\	
$c^{\text{in}}_{{H,V,T}}$ [Hz] 	& $(14.7 \pm 0.6) \cdot 10^{-3}$	& $(18.9 \pm 0.6)\cdot 10^{-3}$ 	& $(17.3 \pm 0.4)\cdot 10^{-3}$	\\
$c^{\text{out}}_{{H,V,T}}$ [Hz] & $(93.9 \pm 1.6)\cdot 10^{-3}$		& $(135.7 \pm 1.5)\cdot 10^{-3}$ 	& $(119.5 \pm 1.1) \cdot 10^{-3}$ \\
\bottomrule
\end{tabular}
\end{center}

The above table for the noise setting shows the detected count rates first split up according to the input signal type (\hsp\, or \coh\,), and secondly for the combination of both datasets. The latter corresponds to the average over the two previous rates.
The separation is done, because the experimental repetition rate $f_\text{rep} = c_{T}$ is different between both signal input types, with a fluctuating rate for \hsp\, inputs and a constant rate for \coh\,.

\item \textbf{Raman memory off:} 
Experimental configuration without optical pumping, i.e. the {\cs} population is approximately evenly distributed between the $6^2S_\frac{1}{2}, F=3$ and the $6^2S_\frac{1}{2}, F=4$ ground states. 

\begin{center}
\footnotesize
\begin{tabular}{@{} c | c | c | c | c | c | c  @{}} 
\toprule
\multicolumn{7}{c}{\textbf{Signal and control setting \textit{sc}}} \\
\toprule
						& \hsp\, 				& \multicolumn{5}{c}{Coherent states} \\\hline
Coinc. 					& $N_{\text{in}} = 0.22$  & $N_{\text{in}} = 0.23$ 	& $N_{\text{in}} = 0.49$ 	& $N_{\text{in}} = 0.91$ 	& $N_{\text{in}} = 1.66$ 	& 
$N_{\text{in}} = 2.16$ \\
\midrule
$c_{T}$ [kHz]				& $11.97 \pm 0.1$		& $5.722$ 		& $5.722$ 		& $5.722$ 			& $5.722$ 		& $5.722$	\\
$c^{\text{in}}_{{H,T}}$ [Hz] 	& $144.58 \pm 0.21$		& $72.65 \pm 0.13$	& $106.56 \pm 0.16$		& $171.7 \pm 0.2$		& $230.67 \pm 0.36$		& $310.24 \pm 0.41$ \\
$c^{\text{in}}_{{V,T}}$ [Hz] 	& $144.5 \pm 0.21$		& $72.5 \pm 0.13$	& $107.89 \pm 0.15$		& $170.84 \pm 0.18$		& $237.21 \pm 0.41$		& $312.02 \pm 0.41$ \\
$c^{\text{out}}_{{H,T}}$ [Hz] 	& $86.12 \pm 0.16$		& $40.52 \pm 0.08$		& $39.42 \pm 0.11$ 		& $51.59 \pm 0.14$		& $41.02 \pm 0.09$		& $46.94 \pm 0.15$ \\
$c^{\text{out}}_{{V,T}}$ [Hz] 	& $85.16 \pm 0.16$		& $40.42 \pm 0.08$ 		& $39.9 \pm 0.12$		& $50.87 \pm 0.13$		& $42.1 \pm 0.09$		& $47.14 \pm 0.15$ \\
$c^{\text{in}}_{{H,V,T}}$ [Hz] 	& $2.65 \pm 0.02$		& $1.58 \pm 0.02$ 		& $3.07 \pm 0.02$		& $7.33 \pm 0.03$		& $12.15 \pm 0.05$		& $20.9 \pm 0.07$ \\
$c^{\text{out}}_{{H,V,T}}$ [Hz] & $1.0 \pm 0.01$		& $0.512 \pm 0.008$ 	& $0.49 \pm 0.01$		& $0.82 \pm 001$		& $0.54 \pm 0.01$		& $0.69 \pm 0.01$ \\ 
\bottomrule
\end{tabular}
\end{center}

\begin{center}
\footnotesize
\begin{tabular}{@{} c | c | c | c| c | c| c  @{}} 
\toprule
\multicolumn{7}{c}{\textbf{Input signal setting \textit{s}}} \\
\toprule
						& \hsp\, 				& \multicolumn{5}{c}{Coherent states} \\\hline
Coinc. 					& $N_{\text{in}} = 0.22$  & $N_{\text{in}} = 0.23$ 	& $N_{\text{in}} = 0.49$ 	& $N_{\text{in}} = 0.91$ 	& $N_{\text{in}} = 1.66$ 	& 
$N_{\text{in}} = 2.16$ \\
\midrule
$c_{T}$ [kHz]				& $11.85 \pm 0.06$		& $5.722$			& $5.722$ 		& $5.722$ 		& $5.722$ 			& $5.722$	\\
$c^{\text{in}}_{{H,T}}$ [Hz] 	& $59.34 \pm 0.2$		& $32.66 \pm 0.11$	& $68.32 \pm 0.18$		& $122.82 \pm 0.28$		& $201.37 \pm 0.44$		& $270.65 \pm 0.56$ \\
$c^{\text{in}}_{{V,T}}$ [Hz] 	& $58.8 \pm 0.18$		& $32.72 \pm 0.11$	& $69.5 \pm 0.19$		& $122.56 \pm 0.26$		& $207.05 \pm 0.52$		& $270.51 \pm 0.54$ \\
$c^{\text{out}}_{{H,T}}$ [Hz] 	& $0.45 \pm 0.01$		& $0.3 \pm 0.01$	& $0.62 \pm 0.02$ 		& $1.07 \pm 0.02$	 	& $1.74 \pm 0.02$		& $2.75 \pm 0.04$ \\
$c^{\text{out}}_{{V,T}}$ [Hz] 	& $0.52 \pm 0.02$		& $ 0.3 \pm 0.01$ 	& $ 0.64 \pm  0.01$		& $1.09 \pm 0.02$		& $1.81 \pm 0.03$		& $2.78 \pm 0.03$ \\
$c^{\text{in}}_{{H,V,T}}$ [Hz] 	& $0.008 \pm 0.002$		& $ 0.19 \pm 0.01$ 	& $ 0.84 \pm  0.02$		& $2.59 \pm 0.03$		& $7.28 \pm 0.06$		& $12.94 \pm 0.08$ \\
$c^{\text{out}}_{{H,V,T}}$ [Hz] & $0 \pm 0$	& $0 \pm 0$ 		& $(0.3 \pm 0.3)\cdot 10^{-3}$	& $0 \pm 0$	& $(0.3 \pm 0.3)\cdot 10^{-3}$	& $(1.6 \pm 0.8)\cdot 10^{-3}$ \\ 
\bottomrule
\end{tabular}
\end{center}

\begin{center}
\footnotesize
\begin{tabular}{@{} c | c | c | c  @{}} 
\toprule
\multicolumn{4}{c}{\textbf{Noise setting \textit{c}}} \\
\toprule
Coinc. 					& \hsp\, 						& Coh. states 				& \hsp\, \& \coh\, \\ 
\midrule
$c_{T}$ [kHz]				& $11.86 \pm 0.09$		& $5.722$				& $6.57 \pm 3.1$ \\
$c^{\text{in}}_{{H,T}}$ [Hz] 	& $83.16 \pm 0.14$		& $41.66 \pm  0.09$		& $47.36 \pm 0.21$	\\
$c^{\text{in}}_{{V,T}}$ [Hz] 	& $82.59 \pm 0.13$		& $41.36 \pm 0.09$		& $47.02 \pm 0.21$	\\
$c^{\text{out}}_{{H,T}}$ [Hz] 	& $85.51 \pm 0.14$		& $40.31 \pm 0.08$		& $46.52 \pm 0..22$ \\
$c^{\text{out}}_{{V,T}}$ [Hz] 	& $84.89 \pm 0.15$		& $40.26 \pm 0.08$ 		& $46.39 \pm 0.22$  \\
$c^{\text{in}}_{{H,V,T}}$ [Hz] 	& $1.11 \pm  0.01$		& $0.579 \pm 0.004$ 	& $0.652 \pm 0.005$	 \\
$c^{\text{out}}_{{H,V,T}}$ [Hz] & $1.12 \pm 0.01$		& $0.504 \pm 0.004$ 	& $0.588 \pm 0.005$	 \\ 
\bottomrule
\end{tabular}
\end{center}
\end{itemize}

\subsection{{\gtwo} statistic tests\label{app6_stat_tests}}

The three statistical tests used in section \ref{ch6_photon_statistics} and appendix \ref{app_ch6_g2_count_aggregation_details} for aggregating the {\gtwo} data are hypothesis tests. These operate on the principle of testing against a Null Hypothesis $H_0$, which may or may not be excluded given the statistical properties of the data. Here $H_0$ can either be a qualitative statement, e.g. the distribution of the data (Shapiro-Wilk test), or a quantitative statement, e.g. the mean of the data (Student T-test). In either case, the first step is the appropriate formulation of $H_0$, which in itself is one of the main difficulties in running hypothesis tests.
Once $H_0$ is established, the data are used to calculate a test statistic $y$. 
Depending on the test statistic, particularly for quantitative tests, information from $H_0$ may also determine $y$. 
The test statistic follows a known probability density distribution (pdf) $y(x)$. The exact form of the pdf depends on the test in use, e.g. $y(x)$ for a Student T-test follows a T-pdf. 
Knowing the functional behaviour $y(x)$, the x-coordinate $\tilde{x}$, belonging to the value of 
$t := \tilde{y} = y(\tilde{x})$ calculated from the sample data, can be determined. 
Once $\tilde{x}$ is known, one can define the $p$-value, which corresponds to the probability of obtaining a value for the test statistic $t$ at least as extreme as the one actually obtained from the sample data. 

What this means is, that $p$ is the probability of getting values $\tilde{y}$ for the test statistic with $x$-coordinates at least as far away from the centre of the pdf as $\tilde{x}$. 
Since the pdf has a total area $\int_{-\infty}^{\infty} y(x) dx = 1$, the $p$-value is the area of the pdf for all $x$-values with $|x| \ge |\tilde{x}|$.
Depending on whether a double-sided or one-sided test is to be performed, one either takes into account the pdf tails in the first and second quadrant (double-sided test), with 
$p = \int_{-\infty}^{\tilde{x}} y(x) \text{d}x + \int_{\tilde{x}}^{+\infty} y(x) \text{d}x$, 
or only in the first (right-sided test), with 
$p = \int_{-\infty}^{\tilde{x}} y(x) \text{d}x$, or second (left-sided test) quadrant, with 
$p = \int_{\tilde{x}}^{+\infty} y(x) \text{d}x$, respectively. 
The $p$-value is thus the probability for events as extreme as $y$ under the condition that $H_0$ is correct. 
It is similar to the Value-at-Risk measure used in finance\cite{Ruppert:Book}. 
Notably, since information about $H_0$ is included in the test statistic $t$, ${y}(\tilde{x})$ is based on the assumptions that $H_0$ holds true. 

If the $p$-value is large, i.e. $\tilde{x}$ lies close to the origin, it is likely that $H_0$ can explain the observed data. But, if $p$ is very small, i.e. $\tilde{x}$ is far away from the pdf centre, it is very unlikely that $H_0$ is a good hypothesis to be made about the data. 
Naturally the question arises: when can a hypothesis $H_0$ be rejected? 
To this end, the statistical significance level is used. It is commonly denoted by variables $\alpha$ or $\sigma$. The latter refers to the case of a normal pdf, where it corresponds to the number of standard deviations $\tilde{x}$ is away from $x=0$, i.e. for a statistical significance of $n \cdot \sigma$, $n$ is given as $n = \frac{|\tilde{x} - x_0|}{\sigma}$ and $x_0$ is the centre of the pdf.
The significance level $n \cdot \sigma$ is commonly used to describe a probability $P(n)$, which refers to the probability of the test statistic lying within the central area of the pdf bounded by $\tilde{x}$, i.e. for the right-sided test 
$P(n) = \int_{0}^{\tilde{x}=n\sigma} y(x) dx$ and for a double-sided test 
$P(n) = \int_{\frac{n \sigma}{2}}^{\frac{n\sigma}{2}} y(x) \text{d}x$. 
The level $\alpha$ describes the complementary area of the pdf, i.e. 
$\int_{\tilde{x}=n\sigma}^{\infty} y(x) \text{d}x$ or 
$\int_{-\frac{n\sigma}{2}}^{\frac{n\sigma}{2}} y(x) \text{d}x$ for the right- and the double-sided tests. 
Thus one gets $P(n \cdot \sigma) = 1 -\alpha$. Commonly used confidence levels are:
\begin{itemize}
\item $1 \sigma$: $P(1 \cdot \sigma) = 0.6828$; $\alpha =  0.32$ 
\item $2 \sigma$: $P(2 \cdot \sigma) = 0.9545$; $\alpha = 0.046$ 
\item $3 \sigma$: $P(3 \cdot \sigma) = 0.0027$; $\alpha = 0.997$ 
\item $4 \sigma$: $P(4 \cdot \sigma) = 6.3*10^{-5}$; $\alpha = 0.9999$ 
\end{itemize}
With these definitions it is straight forward to see, that $\alpha$ and $p$ both describe the same quantity, namely the cumulative probability distribution function (cdf) outside the interval $\left[-\tilde{x}, \tilde{x}\right]$, $\left[-\tilde{x}, 0\right]$, or $\left[0, \tilde{x}\right]$, again depending on which sides the test is performed.
Thus, through specification of a significance level $\alpha$ one can exclude $H_0$ if $p < \alpha$. If $p>\alpha$, i.e. the test statistic $x$-coordinate $\tilde{x}$ lies closer to the origin than $x_\alpha$, where $x_\alpha$ satisfies $\alpha = \int_{x'} y(x) dx$  with $|x'| \ge \tilde{x}$, 
$H_0$ cannot be rejected.\newline
Importantly, the hypothesis test can only determine if $H_0$ can be rejected. 
It cannot establish, whether $H_0$ is actually true. 
For this reason one usually tries to formulate two complementary hypotheses $H_0$ and $H_1$, where the aim of the test is to exclude $H_0$, which simultaneously means that $H_1$ is true. 
However if $p>\alpha$, $H_0$ is as good as any other hypothesis that cannot be rejected given the data. \newline
Below, each of the employed hypothesis tests is described briefly and results are stated for the noise, \hsp\, and \coh\, datasets. For \coh\, inputs, numbers are quoted for $N_\text{in}=0.23 \, \ppp$, whose {\gtwo} is to be contrasted with the ones obtained from the \hsp\, and noise data. The other \coh\, input photon numbers do not require the same amount of care in data evaluation as their differences in {\gtwo} with respect to the noise are immediately obvious from fig. \ref{fig_ch6_g2results}.

\subsubsection{Shapiro-Wilk test\cite{Ruppert:Book}}

\paragraph{Description}
The Shaipro-Wilk test is used to test if a sample of data $\{x_i \}$ is normally distributed, for which reason it is an example for a qualitative test. Its particular advantage over similar tests, such as the $\chi^{(2)}$-test\cite{Taylor:Book} or the Kolmogorow-Smirnow-test\cite{Ruppert:Book}, is good performance for small sample sizes, particularly for $n_x \le 30$ observations. It also does not suffer from binning effects, which are a problem in using the $\chi^{(2)}$-test. 
The test is by default based on the Null hypothesis $H_0$ that the data follows a normal distribution, whereby the test statistic uses a comparison between the variance obtained from the data and the expected variance the dataset would show, if it were normally distributed:
\begin{equation}
t_\text{SW} = \frac{\beta^2}{\left(n_x-1 \right) \cdot \sigma_x^2}
\label{app_ch6_ShapiroWilk}
\end{equation}
Here $n_x$ is the sample size $|\{x_i \}|$, $\sigma_x^2 = \frac{\sum_i (x_i - \mu_x)}{n -1}$, with the mean $\mu_x = \frac{\sum_i x_i}{n_x}$, is the variance of the data and $\beta^2$ is the expected variance. The latter represents the gradient of a linear fit of the data in a $QQ$-plot\cite{Ruppert:Book}. 
Thus, if the sample were normally distributed, both variances would be approximately equal. In this regard, larger values of $t_\text{SW}$ render $H_0$ more likely, such that no rejection of $H_0$ is obtained if it is above a critical, sample size dependent value $t_{SW}^{\text{crit}}$\footnote{
	The actual calculation of $\beta$ from the sample data and the subsequent 
	determination of the $p$-value from the test statistic are reasonably involved. 
	Since we utilise pre-made computer code in \textit{MatLab} to perform the test, 
	details are not provided here.
}.

\paragraph{Application}
We use the Shapiro-Wilk test as a demonstration of the applicability of the central limit theorem on the samples $\{g^{(2)}_{j,i,t} \}$ of individual {\gtwo}-values obtained for each measurement run $j$ 
(see section \ref{ch6_subsec_g2results} \& appendix \ref{app_ch6_g2_count_aggregation_details}). The central limit theorem needs to hold firstly for our error calculation on the mean of the $\{g^{(2)}_{j,i,t} \}$ and furthermore for the T-tests.
The aim is to show that a normal distribution is not rejected straight away for the $\{g^{(2)}_{j,i,t} \}$, which means the data do not exhibit fat tails. 
Notably, the central limit theorem fails particularly for fat tailed distributions, whose variances are undefined, e.g. the Cauchy/Lorentz distribution.
If the data did have such tails, this would spoil the argumentation by the central limit theorem in appendices \ref{app_ch6_g2_count_aggregation_details} and \ref{app6_data_aggregation}. 
The Shapiro-Wilk test allows to conclude their absence for the $\{g^{(2)}_{j,i,t} \}$ distributions, as these would, most likely, lead to a rejection of $H_0$. 
The associated divergence of the variance with $\sigma_x^2 \rightarrow \infty$ in the denominator of eq. \ref{app_ch6_ShapiroWilk}, would lead to $t_\text{SW} \rightarrow 0$, which means a rejection of $H_0$ as $t_\text{SW} < t_\text{SW}^\text{crit}$. 
Why is this important? Because the distribution of the $\{g^{(2)}_{j,i,t} \}$ is given by the ratio of a normal distribution over the product of two normal distributions (see appendix \ref{app6_data_aggregation}). Since the ratio of two normal distributions is Cauchy distributed, it would have fat tails and the central limit theorem would fail. Hence the argumentation in appendix \ref{app6_data_aggregation} would be flawed. Moreover also the T-test would not be applicable to the $\{g^{(2)}_{j,i,t} \}$. 
The Shaprio-Wilk test is run on all sets $\{g^{(2)}_{j,i,t} \}$, leading to the following results:

\begin{center}
\footnotesize
\begin{tabular}{@{} c | c | c | c | c | c | c  @{}} 
\toprule
Signal type & 	Setting 			& $\alpha$ 	& \multicolumn{2}{c|}{Read-in time bin} 	& \multicolumn{2}{c}{Read-out time bin}  		 \\ 

		&		& $$			& $H_0$ rejected 	& $p$-value 			& $H_0$ rejected & $p$-value 	\\
\midrule
\hsp\, 	&\textit{scd}			& $0.0455$	& \xmark	& $0.84$		& \xmark 	& $0.25$ \\
\hsp\, 	&\textit{sd}			& $0.0455$	& \xmark	& $0.09$		& - 		& -\\
\hsp\, 	&\textit{sc}			& $0.0455$	& \xmark	& $0.48$		& - 		& -\\
\hsp\, 	&\textit{s}				& $0.0455$	& \xmark	& $0.26$		& - 		& -\\
\midrule
\coh\, @ $N_\text{in} =0.23$ 	&\textit{scd}	& $0.0455$	& \xmark 		& $0.22$	&  \xmark & $0.56$ \\
\coh\, @ $N_\text{in} =0.23$ 	&\textit{sd}	& $0.0455$	& \cmark 		& $0.02$	&  - & -\\
\coh\, @ $N_\text{in} =0.23$ 	&\textit{sc}	& $0.0455$	& \xmark 		& $0.77$	&  - & -\\
\coh\, @ $N_\text{in} =0.23$	&\textit{s}		& $0.0455$	& \xmark 		& $0.71$	&  - & -\\
\midrule
noise 	&\textit{cd}	& $0.0455$	& \xmark 		& $0.29$		& \cmark & $0$ \\
noise	&\textit{c}		& $0.0455$	& \xmark 		& $0.09$		& \xmark & $0.07$\\
\bottomrule
\end{tabular}
\end{center}

\subsubsection{Student T-test\cite{Ruppert:Book}}
\paragraph{Description}
The Student T-test is used to test datasets for their mean values and, as such, is a test with a quantitative Null hypothesis $H_0$. The test can either be conducted on a single set of data samples or on two samples. 
In case of a single sample with n datapoints $x_i$, it is tested whether an assumed mean value $\tilde{\mu}$, which is provided a priori, is consistent with the mean of the sample data $\mu_x$. 
$H_0$ states that the two means are the same, i.e. $H_0: \tilde{\mu} = \mu_x$. 
For the two sample version, the mean values $\mu_x$, $\mu_y$ of the two samples $\{x_i\}$, $\{y_i\}$ are tested for disagreement. 
Analogously, $H_0$ assumes both means are equal, i.e. $H_0: \mu_x = \mu_y$. 
In both cases, the test requires either that the data in the samples are normally distributed or that they, at least, contain sufficient data points for the central limit theorem to apply. This means the sum $S_i = \Sigma_i x_i$ is normally distributed, which is true for our {\gtwo} data. \newline
While we do not want to present the calculation of the measurement statistics in detail\cite{Ruppert:Book}, a quick look at the test statistics for the one sample T-test is instructive. It explains in a simple way the normality condition and the resulting distribution of the test statistic. We consider to have sample data $\{x_i\}$ with $n$ samples, which are drawn from an arbitrary distribution. The test statistic is given by 
\begin{equation}
t = \sqrt{n} \cdot \frac{\mu_x - \tilde{\mu}}{\sigma_x},
\label{eq_app6_tstat}
\end{equation}
where $\tilde{\mu}$ is the assumed mean of the sample. 
The variable 
$\mu_x= \frac{1}{n} \overset{n}{\underset{i=1}{\sum}}x_i$ 
is the mean of the data and $\sigma_x= \sqrt{\frac{\sum_{i=1}^{n} \left(x_i - \mu_{x} \right)^2 }{n-1}}$ is the sample standard deviation. 
Thanks to the central limit theorem the variable $\mu_{x}$ is normally distributed. The variable $\sigma$ is $\chi^{(2)}$-distributed; a property which is at the heart of the $\chi^{(2)}$ statistics test for normal distributions\cite{Taylor:Book}. Since the test statistic is the quotient of both values, it is described by a T-distribution with $n-1$ degrees of freedom, which is centred around the assumed mean $\mu$. 
It is easy to see, that eq. \ref{eq_app6_tstat} simply tests how far the assumed mean $\tilde{\mu}$ is actually displaced from the proper mean of the sample data $\mu_{x}$. The same information can be obtained, e.g., from graphical investigation of the data. 
This illustrates that statistics tests do no add information, but rather provide a quantitative framework to formalise conclusions, which could also be drawn by other means (e.g. visually). 
For this reason, our analysis in appendix \ref{app_ch6_g2_count_aggregation_details} only uses these tests to check, whether the conclusion we draw from the {\gtwo} data are reasonable, and to make a quantitative statement about the {\gtwo} differences between \hsp\, and \coh\, (see section \ref{ch6_subsec_g2results}).

\paragraph{Application}
In appendix \ref{app_ch6_g2_count_aggregation_details} the double-sided, one-sample T-test is used to test whether the $g^{(2)}_{i,t}$, obtained by summing all double and triple coincidence counts over the entire measurement, can be regarded as the mean of $\left\{ g^{(2)}_{j,i,t} \right\}$, representing the {\gtwo}-values obtained for each measurement run $j$. The used $H_0$ states that the mean of the set $\{ g^{(2)}_{j,i,t} \}$ is equal to $g^{(2)}_{i,t}$, which results from summation of events over all runs $j$ (see eq. \ref{ch6_g2}). 
We run the test for the settings $\{scd, cd\}$ in both time bins $t \in \{\text{in},\text{out}\}$, and for \textit{sd} in the read-in time bin. The results are as follows:

\begin{center}
\footnotesize
\begin{tabular}{@{} c | c | c | c | c | c | c  @{}} 
\toprule
Signal type & 	Setting 			& $\alpha$ 	& \multicolumn{2}{c|}{Read-in time bin} 	& \multicolumn{2}{c}{Read-out time bin}  		 \\ 

		&		& $$			& $H_0$ rejected 	& $p$-value 			& $H_0$ rejected & $p$-value 	\\
\midrule
\hsp\, 	&\textit{scd}	& $0.0455$ 	& \xmark 	& $0.88$		& \xmark 	& $0.31$ \\
\hsp\, 	&\textit{sd}	& $0.0455$	& \xmark 	& $0.84$		& - 		& -\\
\hsp\, 	&\textit{sc}	& $0.0455$	& \xmark 	& $0.8$		& - 		& -\\
\hsp\, 	&\textit{s}		& $0.0455$	& \xmark 	& $0.81$		& - 		& -\\

\midrule
\coh\, @ $N_\text{in} =0.23$ 	&\textit{scd}	& $0.0455$ 		& \xmark	& $0.41$ 		& \xmark	& $0.88$ \\
\coh\, @ $N_\text{in} =0.23$ 	&\textit{sd}	& $0.0455$ 		& \xmark	& $0.91$ 		& -		& -\\
\coh\, @ $N_\text{in} =0.23$ 	&\textit{sc}	& $0.0455$ 		& \xmark	& $0.92$ 		& -		& -\\
\coh\, @ $N_\text{in} =0.23$	&\textit{s}		& $0.0455$ 		& \xmark	& $0.71$ 		& -		& -\\
\midrule
noise 	&\textit{cd}	& $0.0455$	&\xmark		& $0.9$ 		& \xmark	& $0.05$ \\
noise	&\textit{c}		& $0.0455$	& \xmark 		& $0.68$		& \xmark 	& $0.67$\\
\bottomrule
\end{tabular}
\end{center}

As mentioned above, the formulation of $H_0$ aims at achieving no rejection, Thus it does not allow to conclude that $g^{(2)}_{i,t}$ is actually the mean of $\{g^{(2)}_{j,i,t}\}$. 
However the data does not lead to a rejection. Based on the good visual agreement, shown in fig. \ref{fig_ch6_g2ind_g2avg}, we can regard $g^{(2)}_{i,t}$ as the mean of the $\{ g^{(2)}_{j,i,t} \}$.

\subsubsection{Welch test\cite{Ruppert:Book}}

\paragraph{Description}
To determine the statistical significance of the difference between the {\gtwo} for \hsp\, and \coh\, input signals, a Welch test is used in section \ref{ch6_subsec_g2results}.  
The Welch test is a variant of the two-sample T-test, for data samples $\{x_n \}$, $\{y_m\}$ of unequal sizes $N_n \neq N_m$, with $N_n = |\{x_n \}|$, $N_m = |\{y_m \}|$, and with different variances $\sigma_x^2 \neq \sigma_y^2$. Its test statistic is a modified version of eq. \ref{eq_app6_tstat}:  
$$
t_W = \frac{\mu_x - \mu_y - \left( \tilde{\mu}_x - \tilde{\mu}_y \right) }{\sqrt{\frac{\sigma_x^2 }{n} + \frac{\sigma_y^2 }{m}}},
$$
where $\mu_{\{ x,y\} }$ are the means of the samples $x$ and $y$, and $\tilde{\mu}_x - \tilde{\mu}_y$ is the difference between the two \textit{a priori} assumed means. The test statistic is again approximately T-distributed, however following a T-distribution with
$$
\nu =  \frac{\left( \frac{\sigma_x^2}{n} + \frac{\sigma_y^2}{m} \right)^2}{ \frac{\left(\frac{\sigma_x^2}{n} \right)^2}{n-1} +  \frac{\left( \frac{\sigma_y^2 }{m} \right)^2}{m-1}}
$$
degrees of freedom. Further details about the methodology can be found in \textit{Statistics and Finance: An Introduction} by \textit{D. Ruppert}\cite{Ruppert:Book}.

\paragraph{Application}
For our purposes, this test is chosen since, on the one hand, the {\gtwo} measurements on \hsps\,, \coh\, and noise are clearly independent of one another and have a different number of recorded runs. On the other hand, the variances of the $\{ g^{(2)}_{j,i,t} \}$ are different as well; both of these features are illustrated by the data points in fig. \ref{fig_ch6_g2ind_g2avg}. 
We are performing 3 tests, where the individual {\gtwo} values $\{ g^{(2)}_{j,i,t} \}$ of the following datasets are involved:
\begin{enumerate}
\item \hsps\, vs. \coh\, with $N_\text{in} = 0.23 \, \ppp$
\item \hsps\, vs. noise
\item \coh, vs. noise
\item \hsps\, vs. \coh\, with $N_\text{in} = 0.49 \, \ppp$
\end{enumerate}
In all cases, the Null hypothesis $H_0$ is the assumption that the means $\tilde{\mu}_x$ and $\tilde{\mu}_y$ of the two datasets involved are equal. The aim is rejection of $H_0$, which proves that there is a {\gtwo} difference with at least the chosen confidence level $\alpha$. Using $P(n \cdot \sigma) = 1 - \alpha$, the confidence levels, quoted in section \ref{ch6_subsec_g2results}, are obtained in terms of standard deviations. The test is laid out as a left-hand side test, whereby we test, whether the {\gtwo}-values of the first-named signal type in the above list has a lower mean than the second signal type.
Having observation numbers of $n=13$ for \hsp\,, $n = 14$ for \coh\, and $n=70$ for the noise measurements, the sample sizes are sufficient. The only problematic point is the greater number of noise data points, which can influence the test's reliability. Hence the results obtained when testing \hsp\, or \coh\, {\gtwo}-values against those of the noise have to be taken with a pinch of salt. Running the test yields the following results:

\begin{center}
\footnotesize
\begin{tabular}{@{} c | c | c | c | c | c | c  @{}} 
\toprule
Sample 1 	& 	Sample 2 				& $\alpha$ 	& \multicolumn{2}{c}{Read-in time bin} 	& \multicolumn{2}{c}{Read-out time bin}  		 \\ 

		&						& $$			& $H_0$ rejected 	& $p$-value 			& $H_0$ rejected & $p$-value 	\\
\midrule
\hsp\, 	& \coh\, @ $N_\text{in} =0.23$	& $0.0027$	& \cmark 			& $1.2 \cot 10^{-10}$		& \cmark & $8.7 \cdot 10^{-4}$ \\
\hsp\, 	& noise					& $0.0455$	& \cmark 			& $2.5 \cdot 10^{-20}$	& \cmark & $4.5 \cdot 10^{-3}$\\
\hsp\, 	&\coh\, @ $N_\text{in} =0.49$	& $0.3172$	& \cmark 			& $2.3 \cdot 10^{-7}$		& \xmark & $0.38$\\
\midrule
\coh\, @ $N_\text{in} =0.23$ 	&noise	& $0.3172$	& \cmark			& $1.6 \cdot 10^{-3}$		& \xmark & $0.42$ \\
\bottomrule
\end{tabular}
\end{center}

The above results justify the statements made in section \ref{ch6_subsec_g2results}. The \hsp\, {\gtwo}-values lie below those of noise and \coh\, input signals at $N_\text{in} = 0.23 \, \ppp$. 
The values for this \coh\, input photon number are not distinguishable from the noise. 
Notably, it is sufficient to choose a large $\alpha$ for demonstration, since the associated $p$-value is so large. 
The test cannot yield any rejection of $H_0$ with higher confidence levels, if it cannot do it with $1\cdot \sigma$.
Moreover, as fig. \ref{fig_ch6_g2bars} shows, the {\gtwo} for \hsps\, is approximately equal to that of \coh\, at $N_\text{in} = 0.49 \, \ppp$, so no rejection of $H_0$ is obtained for these two samples.

\subsection{Further details about {\gtwo} data aggregation\label{app6_data_aggregation}}

\paragraph{Normal distribution of coincidence detection probabilities}

Here we show, how the central limit theorem (CLT) causes the detection probabilities $p^t_{j,k,i}$ to converge to Gaussian distributions, despite the fact that the measured coincidence counts $c^t_{j,k,i}(t_m)$, contributing to these numbers, are Poissonian distributed. 
The normality arises because we are using detection probabilities $p^{t}_{j,k,i}$, which are the sum of the detection probabilities $p^{t}_{j,k,i}(t_m)$ over all $t_m$ in a run $j$, where each of these numbers corresponds to one datapoint recorded by the FPGA . The $p^{t}_{j,k,i}$ are thus each a sum of random variables.
The CLT states, that a sum $S_m =\sum_{m=1}^n x_m$ of $n$ random variables $x_m$, each with the same arbitrary distribution, converges to a normal distribution in the limit of large $n$, as long as the distribution of $\{x_m \}$ has finite mean and variance. 
Usually, the convergence can already be observed after summation of $n \sim 5$ elements\cite{Steck:LectureNotes}. In our case, each measurement run $j$ contains at least $10$ datapoints (which is the minimal number of FPGA points recorded to determine the memory efficiency). 
Yet, most runs are integrated for $\gtrsim 30\, \min$ and consequently have at least $180$ data points, i.e. $n \gtrsim 180$. 
This is more than sufficient for the CLT to hold, and the distribution of the 
$p^{t}_{j,k,i} = \Sigma_m p^{t}_{j,k,i}(t_m) = \Sigma_m c^{t}_{j,k,i}(t_m)/ c_{j,T,i}$ thus converges against a normal distribution. 

\begin{figure}[h!]
\centering
\includegraphics[width=16cm]{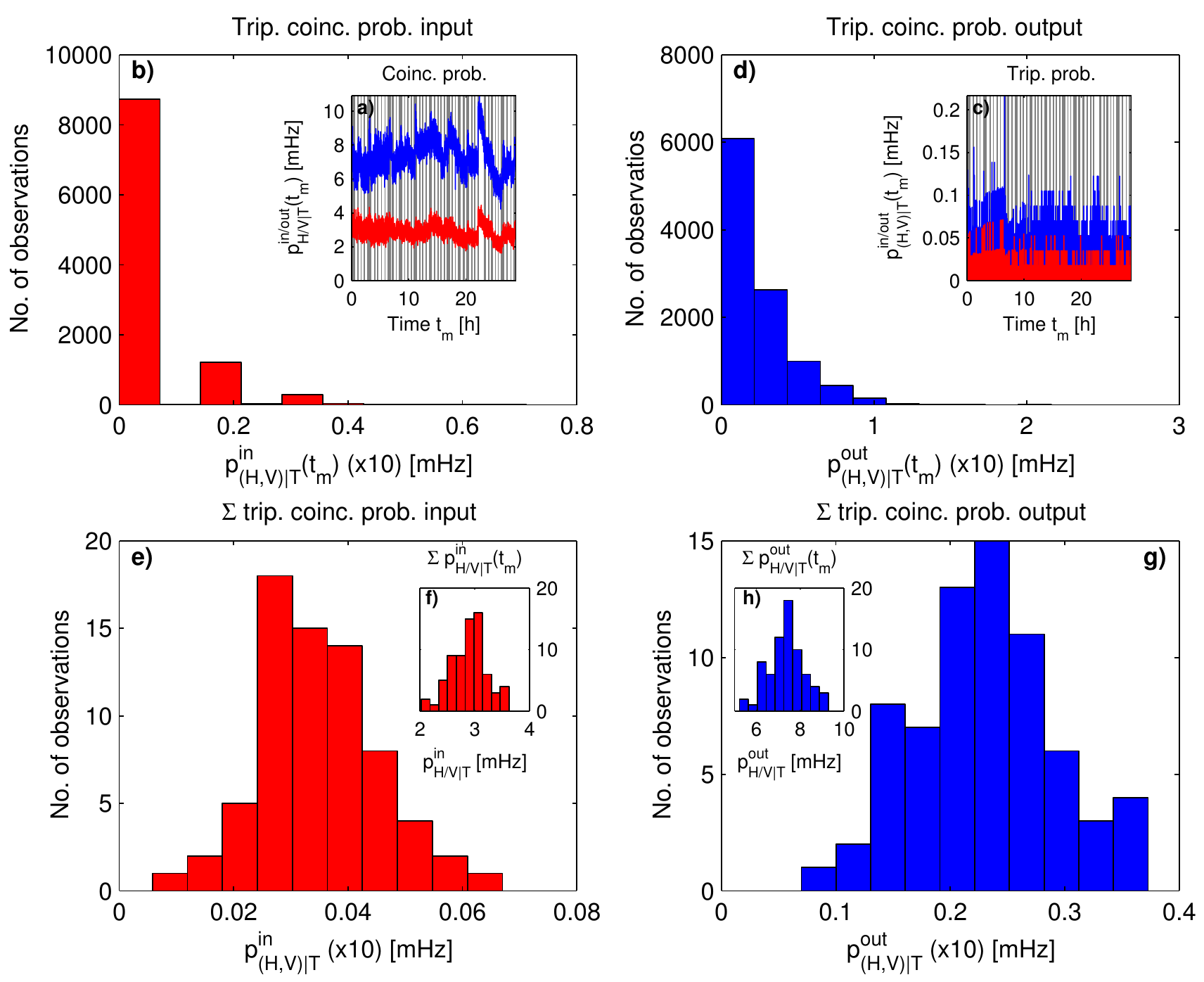}
\caption{Distributions for detection probabilities for measurements on the noise (setting \textit{cd}). 
\textbf{(a)} shows the data points for the coincidence probabilities 
$p^\text{in/out}_{(H/V)|T}(t_m) = p^\text{in/out}_{H|T}(t_m) + p^\text{in/out}_{V|T}(t_m)$ 
for the input time bin (\textit{red}) and output time bin (\textit{blue}). \textit{Vertical lines} delimit the measurement runs $j$.
\textbf{(b)} and \textbf{(d)} show the Poissonian distributions of the probabilities for triple coincidences $p^\text{in/out}_{(H,V)|T}(t_m)$ for FPGA data in the read-in and read-out bins, respectively. 
\textbf{(c)} depicts their rates as a function of measurement time $t_m$, with \textit{vertical lines} analogous to \textbf{(a)}. 
\textbf{(e)} and \textbf{(g)} illustrate the normal distributions that are obtained for the triple coincidence probabilities when summing over the $p^\text{in/out}_{(H,V)|T}(t_m)$ for all points $t_m$ per run $j$.
\textbf{(f)} and \textbf{(h)} show the analog distributions for the summed coincidence probabilities 
$p^\text{in/out}_{(H/V)|T} = \underset{m}{\sum} p^\text{in/out}_{(H/V)|T}(t_m)$.
}
\label{fig_ch6_distr_g2ind_noise}
\end{figure}

Fig. \ref{fig_ch6_distr_g2ind_noise} illustrates this argument for the noise measurements (setting \textit{cd}), which contain the largest number of runs, $N_r = 70$, and hence display the effects most clearly. These data runs have been recorded constantly alongside the measurements of all input signal types, so they are representative for the experimental conditions throughout the entire {\gtwo} experiment.
The triple coincidence probabilities $p^t_{j,((H,V)|T),cd}(t_m)$ are shown as a time trace over all measurement runs $j$ in fig. \ref{fig_ch6_distr_g2ind_noise} \textbf{c}. 
Their histograms, plotted in figs. \ref{fig_ch6_distr_g2ind_noise} \textbf{b} \& \textbf{d} for the read-in and read-out time bin, respectively, are Poissonian distributed.
However, when looking at the histograms of the probabilities $p^t_{j,((H,V)|T),cd}$, summed over all points $t_m$ per run $j$, the plots in figs. \ref{fig_ch6_distr_g2ind_noise} \textbf{e} \& \textbf{g} reveal how these converge to a normal distribution, as expected from the CLT. 
The same of course holds true for the coincidence probabilities $p^t_{j,((H/V)|T),cd}$, whose time traces $p^t_{j,((H/V)|T),cd}(t_m)$ are displayed in fig. \ref{fig_ch6_distr_g2ind_noise} \textbf{a} along with the summed probabilities per run $p^t_{j,((H/V)|T),cd}$ in figs.~\ref{fig_ch6_distr_g2ind_noise}~\textbf{f}~\&~\textbf{h}. 

\paragraph{{\gtwo}-values from averaging all counts}

Besides data aggregation based on the summation of all coincidence counts, explained in appendix \ref{app_ch6_g2_count_aggregation_details}, we have also validate the {\gtwo} results by determining the mean coincidence count detection probability over the entire dataset, i.e., over all measurement runs $j$ contributing to a specific input photon number \Nin\,. 
To this end, the variables going into eq. \ref{ch6_g2} for the {\gtwo} are $p^t_{k,i} = \frac{\Sigma_{j=1}^{N_r} \left( \Sigma_{m} p^t_{j,k,i} (t_m) \right)}{\Sigma_{j=1}^{N_r} \left( \Sigma_m 1 \right)}$, where the denominator is the total number of datapoints collected. The {\gtwo}-values are exactly the same, as the tables in \ref{app6_g2results} below show. However the errors are slightly different. While eq. \ref{ch6_g2err} is still used to obtain the {\gtwo} error, the factors $\Delta p^{t}_{k,i}$, entering the equation, are now obtained by taking the standard error $\Delta p^t_{k,i} = \frac{\sigma^t_{i,k}}{\sqrt{\Sigma_{j=1}^{N_r} \left( \Sigma_m 1 \right)}}$, with $\sigma^t_{k,i} = \sqrt{\frac{ \Sigma_{j=1}^{N_r} \left( \Sigma_{m} p^t_{j,k,i} (t_m) - p^t_{k,i} \right)^2 }{\Sigma_{j=1}^{N_r} \left( \Sigma_m 1 \right)}}$ as the sample standard deviation of each dataset.
Calculating the standard error implicitly relies on the applicability of the CLT. 
From the argument above, illustrated by fig. \ref{fig_ch6_distr_g2ind_noise}, it follows that the $p^{t}_{k,i}$ are normally distributed. 
Notably, the usage of mean count rates still depends on the absence of systematic errors throughout the dataset (see discussion in appendix \ref{app_ch6_g2_count_aggregation_details}). 
Likewise to the summation of coincidence counts, averaging fails to obtain sensible results particularly when different $N_\text{in}$ are applied, which is the case for settings \textit{sd} and \textit{s} with \coh\, inputs.

\paragraph{Averaging method of {\gtwo}-values from individual measurement runs}
In calculating the mean $\bar{g}^{(2)}_{i,t}$ over all individual $g^{(2)}_{j,k,i}$, performed in appendix \ref{app_ch6_g2_count_aggregation_details}, there is no recognition of the precision with which each $g^{(2)}_{j,k,i}$ has been determined. In other words, the measurement time $\Delta t_\text{meas}$ of each measurement run $j$, which is inversely proportional to the Poissonian error on each of these {\gtwo}-values, is not included in the standard formulae
\begin{equation}
\bar{g}^{(2)}_{i,t} = \frac{\Sigma_{j=1}^{N_r} {g}^{(2)}_{j,i,t} }{N_r}, \quad \Delta \bar{g}^{(2)}_{i,t} = \frac{1}{\sqrt{N_r}} \cdot \sqrt{\frac{\Sigma_{j=1}^{N_r} \left(g^{(2)}_{j,i,t} - \bar{g}^{(2)}_{i,t} \right)^2}{N_r -1}}
\label{app6_eq_g2_ind_unweighted}
\end{equation}
for calculating the mean and the standard error. 
As  fig. \ref{fig_ch6_g2ind_g2avg} illustrates, there is however a significant difference in the measurement times $\Delta t_\text{meas}$ between the runs $j$. 
The inclusion of the runs with $\Delta t_\text{meas} = 10\,\min$, which were primarily intended for determining the memory efficiency, into the {\gtwo} calculation is the main contributor to the varying precision of the individual datapoints. 
It is thus sensible, to extend the calculation and include a weighted average and a weighted standard deviation, given by:
\begin{align}
\tilde{g}^{(2)}_{i,t} 		&= \frac{\sum_{j=1}^{N_r} \textfrak{w}_j \cdot g^{(2)}_{j,i,t}}{\sum_{j=1}^{N_r} \textfrak{w}_j} 
\label{g2_ind_mean} \\
\Delta \tilde{g}^{(2)}_{i,t} 	&= \frac{1}{\sqrt{N_r}} \cdot \sqrt{
\underbrace{\frac{\sum_{j=1}^{N_r} \textfrak{w}_j}{ \left(\sum_{j=1}^{N_r} \textfrak{w}_j \right)^2 - \sum_{j=1}^{N_r} \textfrak{w}^2_j}}_{=: \alpha}
\cdot \sum_{j=1}^{N_r} \left(\textfrak{w}_j \cdot \left(g^{(2)}_{j,i,t} - \tilde{g}^{(2)}_{i,t} \right)^2 \right)
}
\label{g2_ind_mean_err}
\end{align}
The pre factor $\alpha$ in eq. \ref{g2_ind_mean_err} results from the usage of the sample standard deviation, i.e. using an unbiased estimator for the sample variance. 
It is the counterpart to the factor $\frac{1}{\sqrt{N-1}}$ in eq.~\ref{app6_eq_g2_ind_unweighted}.
The weighing factors $\textfrak{w}_j$ are, on the one hand, obtained from the error of the individual data points for each run as 
$\textfrak{w}^{\Delta g2}_j =\frac{ \frac{1}{\left(\Delta g^{(2)}_{j,k,i}\right)^2} }{ \overset{N_r}{\underset{j=1}{ \sum}} \frac{1}{\left(\Delta g^{(2)}_{j,k,i}\right)^2} }$, which is one standard approach for weighted averages \cite{Ruppert:Book}. 
On the other hand, we can also use weights that scale linearly with measurement time and are defined by 
 $\textfrak{w}^{\Delta t _\text{meas}}_j = \frac{\Delta t_{\text{meas},j}}{\sum_{j=1}^{N_r} \Delta t_{\text{meas},j}}$.

\paragraph{{\gtwo} measurement results\label{app6_g2results}}

Here we state the {\gtwo}-values obtained by all five calculation methods: 
\begin{enumerate}
\item Poissonian sum: count summation over all measurement runs assuming Poissonian statistics. 
\item Avg. all counts: taking the mean and standard error over the coincidence count rates for the whole dataset obtained for each setting. 
\item Avg. without weights: taking the mean and standard error for the individual $g^{(2)}_{j,k,i}$-values without weighing factors. 
\item Avg. weights $\textfrak{w}^{\Delta g2}_j$: taking the means and standard errors for the individual $g^{(2)}_{j,k,i}$-values, using the Poissonian error of individual {\gtwo}-values as weighing factors. 
\item Avg. weights $\textfrak{w}^{\Delta t_\text{meas}}_j$: taking the means and standard errors for the individual $g^{(2)}_{j,k,i}$-values, using the setting measurement times as weighing factors. 
\end{enumerate}
Note: the crossed-out values in the tables below mark results from methods 1 and 2, which are not sensible. These numbers are nevertheless stated, to demonstrate that both methods do not work when data from multiple measurements at different {\Nin} are combined.

\begin{center}
\footnotesize
\begin{tabular}{@{} c | c | c | c | c | c | c @{}} 
\toprule
\multicolumn{7}{c}{\textbf{Memory on, input time bin}} \\
\toprule
Signal 			& $N_\text{in}$ 		& Poisson. sum	 		& Avg. all counts 		& Avg., no weights 		& Avg., weights $\textfrak{w}^{\Delta g2}_j$ & Avg., weights $\textfrak{w}^{\Delta t_\text{meas}}_j$ 	\\
\midrule
\hsp\,, \textit{sd}	& $0.22 \pm 0.03$	& $0.016 \pm 0.004$		& $0.016 \pm 0.004$ 	& $0.017 \pm 0.004$ 	& $0.013 \pm 0.003$ 	& $0.013 \pm 0.003$ \\
\coh\,, \textit{sd}	& $0.23 - 2.16 $	& $\hcancel{1.565 \pm 0.005}$	& $\hcancel{1.565 \pm 0.049}$				& $1.015 \pm 0.009$		& $1.001 \pm 0.004$		& $1.017 \pm 0.004$	\\ 
noise \textit{cd}		& $- $			& $1.618 \pm 0.038$		& $1.618 \pm 0.04$		& $1.626 \pm 0.042$		& $1.552 \pm 0.033$		& $1.588 \pm 0.032$ \\
\midrule
\hsp\,, \textit{scd}	& $0.22 \pm 0.03$	& $0.917 \pm 0.02$		& $0.917 \pm 0.02$ 		& $0.912 \pm 0.031$	 	& $0.907 \pm 0.026$		& $0.913 \pm 0.026$ \\ 
\midrule
\coh\,, \textit{scd}	& $0.23 \pm 0.02$ 	& $1.392 \pm 0.024$		& $1.392 \pm 0.024$		& $ 1.364 \pm 0.032 $	& $1.38 \pm 0.026$		& $1.387 \pm 0.025$ \\
\coh\,, \textit{scd}	& $0.49 \pm 0.02$ 	& $1.217 \pm 0.017$		& $1.217 \pm 0.017$		& $1.206 \pm 0.023$		& $1.213 \pm 0.019$		& $1.215 \pm 0.019$ \\
\coh\,, \textit{scd}	& $0.91 \pm 0.04$ 	& $1.136 \pm 0.01$		& $1.136 \pm 0.011$		& $1.138 \pm 0.01$		& $1.133 \pm 0.01$		& $1.134 \pm 0.009$ \\
\coh\,, \textit{scd}	& $1.66 \pm 0.1$ 	& $1.075 \pm 0.007$		& $1.075 \pm 0.01$		& $1.086 \pm 0.015$		& $1.069 \pm 0.015$		& $1.07 \pm 0.012$ \\
\coh\,, \textit{scd}	& $2.16 \pm 0.08$ 	& $1.069 \pm 0.005$		& $1.069 \pm 0.008$		& $1.061 \pm 0.005$		& $1.064 \pm 0.004$		& $1.065 \pm 0.004$ \\
\bottomrule
\end{tabular}
\end{center}

\begin{center}
\footnotesize
\begin{tabular}{@{} c | c | c | c | c | c | c @{}} 
\toprule
\multicolumn{7}{c}{\textbf{Memory on, output time bin}} \\
\toprule
Signal 			& $N_\text{in}$ 		& Poisson. sum	 		& Avg. all counts 		& Avg., no weights 		& Avg., weights $\textfrak{w}^{\Delta g2}_j$ & Avg., weights $\textfrak{w}^{\Delta t_\text{meas}}_j$ 	\\
\midrule
noise \textit{cd}		& $-$			& $1.705 \pm 0.015$		& $1.705 \pm 0.018$		& $1.672 \pm 0.022$		& $1.673 \pm 0.015$		& $1.679 \pm 0.016$ \\
\midrule
\hsp\,, \textit{scd}	& $0.22 \pm 0.03$	& $1.586 \pm 0.025$		& $1.586 \pm 0.028$ 	& $1.553 \pm 0.032$		& $1.566 \pm 0.028$ 	& $1.569 \pm 0.028$ \\
\midrule
\coh\,, \textit{scd}	& $0.23 \pm 0.02$ 	& $1.685 \pm 0.021$		& $1.685 \pm 0.023$		& $1.683 \pm 0.016$		& $1.672 \pm 0.014$		& $1.67 \pm 0.014$ \\
\coh\,, \textit{scd}	& $0.49 \pm 0.02$ 	& $1.582 \pm 0.029$		& $1.582 \pm 0.03$		& $1.567 \pm 0.031$		& $1.572 \pm 0.027$		& $1.574 \pm 0.026$ \\
\coh\,, \textit{scd}	& $0.91 \pm 0.04$ 	& $1.546 \pm 0.017$		& $1.546 \pm 0.021$		& $1.533 \pm 0.017$		& $1.524 \pm 0.017$		& $1.524 \pm 0.016$ \\
\coh\,, \textit{scd}	& $1.66 \pm 0.1$ 	& $1.414 \pm 0.018$		& $1.414 \pm 0.02$		& $1.418 \pm 0.018$		& $1.406 \pm 0.018$		& $1.409 \pm 0.017$ \\
\coh\,, \textit{scd}	& $2.16 \pm 0.08$ 	& $1.374 \pm 0.012$		& $1.374 \pm 0.015$		& $1.362 \pm 0.02$		& $1.363 \pm 0.022$		& $1.367 \pm 0.023$ \\
\bottomrule
\end{tabular}
\end{center}

\begin{center}
\footnotesize
\begin{tabular}{@{} c | c | c | c | c | c | c @{}} 
\toprule
\multicolumn{7}{c}{\textbf{Memory off, input time bin}} \\
\toprule
Signal 			& $N_\text{in}$ 		& Poisson. sum	 		& Avg. all counts 		& Avg., no weights 		& Avg., weights $\textfrak{w}^{\Delta g2}_j$ & Avg., weights $\textfrak{w}^{\Delta t_\text{meas}}_j$ 	\\
\midrule
\hsp\,, \textit{s}		& $0.22 \pm 0.03$	& $0.026 \pm 0.006$		& $0.026 \pm 0.006$ 	& $0.028 \pm 0.007$		& $0.022 \pm 0.009$		& $0.026 \pm 0.007$ \\
\coh\,, \textit{s}		& $0.23 - 2.16 $	& $\hcancel{1.495 \pm 0.006}$		& $\hcancel{1.495 \pm 0.057}$			& $0.997 \pm 0.009$		& $1.003 \pm 0.004$	 	& $0.998 \pm 0.003$ \\
noise \textit{c}		& $- $			& $1.922 \pm 0.01$		& $1.922 \pm 0.02$		& $1.882 \pm 0.013$		& $1.892 \pm 0.011$		& $1.887 \pm 0.011$ \\
\midrule
\hsp\,, \textit{sc}	& $0.22 \pm 0.03$	& $ 1.520  \pm 0.011$	& $1.520 \pm 0.011$  	& $1.518 \pm 0.007$ 	& $1.519 \pm 0.007$ 	& $1.518 \pm 0.007$ \\
\midrule
\coh\,, \textit{sc}	& $0.23 \pm 0.02$ 	& $1.717 \pm 0.016$		& $1.717 \pm 0.017$		& $1.714 \pm 0.027$		& $1.714 \pm 0.027$		& $1.715 \pm 0.027$ \\
\coh\,, \textit{sc}	& $0.49 \pm 0.02$ 	& $1.529 \pm 0.011$		& $1.529 \pm 0.012$		& $1.537 \pm 0.017$		& $1.528 \pm 0.014$		& $1.528 \pm 0.014$ \\
\coh\,, \textit{sc}	& $0.91 \pm 0.04$ 	& $1.43 \pm 0.006$		& $1.43 \pm 0.007$		& $1.43 \pm 0.017$		& $1.429 \pm 0.017$		& $1.429 \pm 0.017$ \\
\coh\,, \textit{sc}	& $1.66 \pm 0.1$ 	& $1.27 \pm 0.005$		& $1.27 \pm 0.006$		& $1.27 \pm 0.012$		& $1.269 \pm 0.012$		& $1.27 \pm 0.012$ \\
\coh\,, \textit{sc}	& $2.16 \pm 0.08$ 	& $1.236 \pm 0.003$		& $1.236 \pm 0.005$		& $1.234 \pm 0.009$		& $1.235 \pm 0.009$		& $1.234 \pm 0.009$ \\
\bottomrule
\end{tabular}
\end{center}

\begin{center}
\footnotesize
\begin{tabular}{@{} c | c | c | c | c | c | c @{}} 
\toprule
\multicolumn{7}{c}{\textbf{Memory off, output time bin}} \\
\toprule
Signal 			& $N_\text{in}$ 		& Poisson. sum	 		& Avg. all counts 		& Avg., no weights 		& Avg., weights $\textfrak{w}^{\Delta g2}_j$ & Avg., weights $\textfrak{w}^{\Delta t_\text{meas}}_j$ 	\\
\midrule
noise \textit{c}		& $- $			& $1.789 \pm 0.01$		& $1.789 \pm 0.02$		& $1.735 \pm 0.018$		& $1.762 \pm 0.015$		& $1.743 \pm 0.016$				\\
\bottomrule
\end{tabular}
\end{center}

\chapter{Appendix: Noise characterisation and memory performance\label{app_ch7}}

\section{Fluorescence noise contribution\label{app_ch7_flour_noise}}

Table \ref{tab_app_ch7_fluor} below lists the fraction $R^t_{i,\text{FL},k}$ of the fluorescence noise contribution to the total noise level $N^t_{i,k}$ in time bin $t$ for the S and AS channels ($i \in \left\{ \text{S}, \text{AS} \right\})$. Numbers are calculated using the dataset shown in fig. \ref{fig_ch7_flourescence} and are stated for the spin-polarised {\cs} ensemble ($k = cd$) as well as the thermally distributed ensemble ($k=c$) for both storage times $\tau_\text{S} = 12.5\ns$ and $\tau_\text{S} = 312\ns$.

\begin{table}[h!]
\centering
\begin{tabular}{c|c|c|c|c|c}
\toprule
$\tau_\text{S}$ [ns] & Control pulse ($n$) & $R^{n}_{\text{S},\text{FL},cd}$ [$\%$] &  $R^{n}_{\text{S},\text{FL},c}$ [$\%$] &  $R^{n}_{\text{AS},\text{FL},cd}$ [$\%$] &  $R^{n}_{\text{AS},\text{FL},c}$ [$\%$] \\
\midrule
12.5	&  1 	&    $16 \pm 0.4$ 	& $3.69 \pm 0.09$ 	& $0.62 \pm 0.04$   	& $1.2 \pm   0.08$ \\
12.5	&  2 	&    $14 \pm 0.2$   	& $8.9 \pm 0.1$  	& $2.39 \pm  0.08 $ 	& $2.2  \pm  0.1$ \\
12.5	&  3 	&    $14.4 \pm 0.2$  	& $12.6 \pm 0.2$  	& $2.31 \pm 0.07$ 	& $2.9  \pm  0.1$ \\
12.5	&  4 	&    $14.6  \pm 0.2$  & $15.1 \pm 0.2$ 	& $2.35 \pm 0.07$ 	& $3  \pm  0.1$ \\
12.5	&  5 	&    $15.5 \pm 0.2$ 	& $17 \pm 0.2$ 	& $2.41 \pm  0.07$   & $3.3  \pm  0.1$ \\
12.5	&  6 	&    $15.9 \pm 0.2$  	& $17.9 \pm 0.2$  	& $2.75 \pm 0.08$  	& $3.6 \pm   0.1$ \\
12.5	&  7 	&    $16.3 \pm 0.2$ 	& $18.9 \pm 0.2$  	& $2.5 \pm 0.08$  	& $3.6  \pm 0.2$ \\
12.5	&  8 	&    $17  \pm 0.2$  	& $19.6 \pm 0.2$ 	& $2.7 \pm 0.08$  	& $3.6  \pm  0.1$ \\
12.5	&  9 	&    $17.9 \pm 0.2$ 	&  $20.6 \pm 0.3$ 	& $2.9 \pm 0.08$ 	& $4 \pm  0.1$ \\
\hline
312 	& 1 	& $17.5 \pm 0.43$ 	& $3.28 \pm 0.08$ 	& $0.85 \pm 0.07$   	& $1.7 \pm 0.1$ \\
312 	& 2 	& $6.8 \pm 0.2$ 	& $3.39 \pm 0.09$ 	& $0.75 \pm 0.06$  	& $2 \pm 0.2$ \\
312 	& 3 	& $11.3 \pm 0.2$ 	& $8.6 \pm 0.1$ 	& $2.5 \pm 0.1$	& $2.9 \pm 0.2$ \\
\bottomrule
\end{tabular}
\caption{Fluorescence noise fraction $R^t_{j,\text{FL},k}$ of total noise level $N^t_{j,k}$.}
\label{tab_app_ch7_fluor}
\end{table}

\section{Theory noise level prediction versus experiment\label{app_ch7_det_eff_fit}}

In section \ref{ch7_subsec_theory_prediction} of the main text, we compare the prediction for memory noise $N^t_{i,k}$ from our theory model (appendix \ref{app6_coh_model}) against the measured noise levels over a train of $9$ consecutive control pulses ($t$). 
The noise floor is compared for both, spin-polarised and thermally distributed ensembles ($k\in \left\{cd,c \right\}$) as well as for S and AS channels ($i \in \left\{\text{S}, \text{AS} \right\}$) .
Since this is a comparison of an absolute prediction with an absolute number from a measurement, all efficiency factors, relevant for the experiment, need to be known with certainty. 
For 

\begin{equation}
N^t_{i,k} = \frac{\tilde{a}^t_{i,k}}{f_\text{rep}\cdot \Delta t_\text{meas} \cdot T_\text{sig} \cdot \eta_\text{det}}
\label{eq_app7_noise_level}
\end{equation} 

\begin{figure}[h!]
\centering
\includegraphics[width=\textwidth]{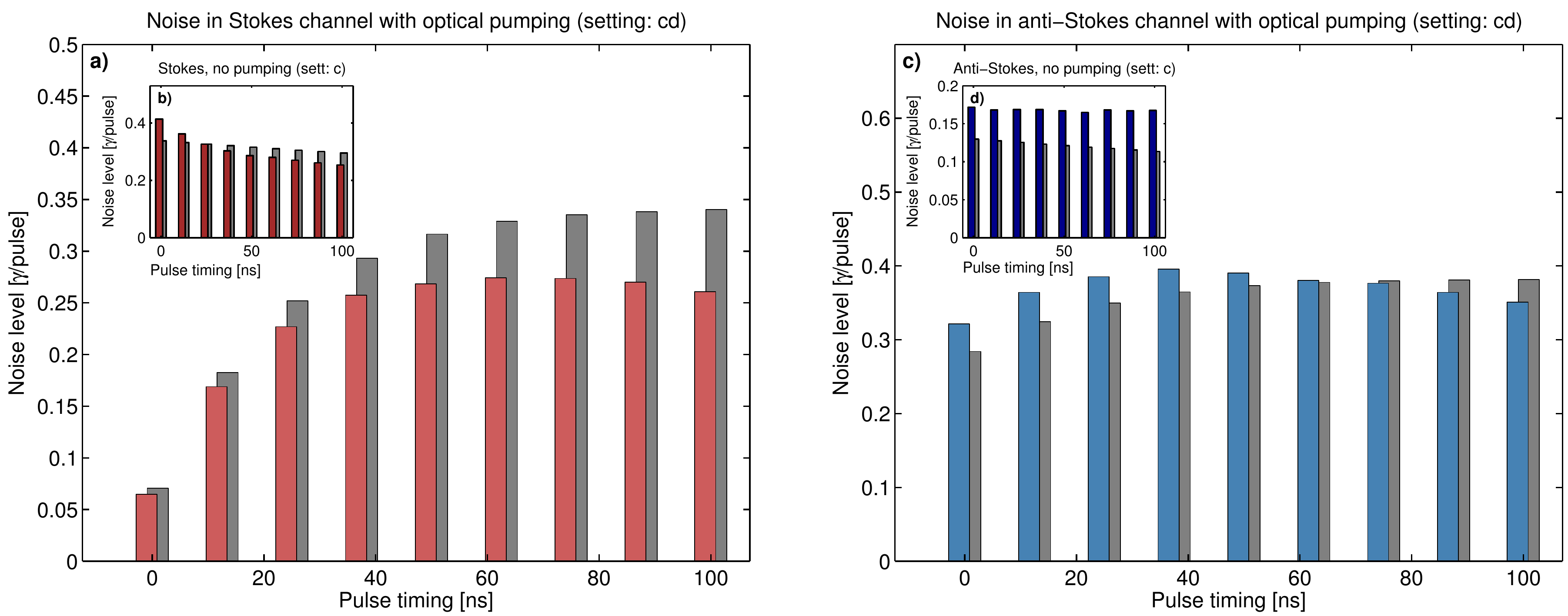}
\caption{Absolute noise levels obtained by prediction of the theory model for the {\cs} memory (\textit{grey bars}) and from the experiment (\textit{grey bars}). The measured data assumes an APD detection efficiency of $\tilde{\eta}_\text{det}  = 37.6\,\%$, which yields the best fit between observation and theory prediction. Panel content and colour codings are the same as in fig. \ref{fig_ch7_theory_prediction} of the main text.}
\label{fig_app_ch7_theory_prediction_deteff_fit}
\end{figure}

(see section \ref{ch7_subsec_theory_prediction}), which is analogue the definition of the input photon number (eq. \ref{ch6_eq_Nin}) 
$$
N_\text{in}= \frac{\bar{c}^\text{in}_{sd}}{\bar{c}_{sd,T} \cdot T_\text{sig} \cdot \eta_\text{APD, H/V}},
$$ 
this is not the case for the detector efficiency $\eta_\text{det} = \eta_\text{APD, H/V}$. 
For the work presented in this thesis, the detector efficiency of $50\,\%$ for both APDs \spcmdh\, and \spcmdv\, (fig. \ref{fig_6_setup}) has only been assumed. 
However, we have observed experimentally, that \spcmdh\, is in fact slightly less efficiency than \spcmdv\,, whereby 
all measurements on the memory noise floor in chapter \ref{ch7} were conducted using \spcmdh\,. 
It is thus reasonable, to fit the experimentally observed noise levels $\{ N^t_{i,k} \}$ onto their theory counterparts, using $\eta_\text{det}$ as a free fitting parameter. 
To this end, we use eq. \ref{eq_app7_noise_level} with the experimental data for the integrated, fluorescence noise subtracted count rates $\tilde{a}^t_{i,k}$, as obtained from fig. \ref{ch7_Fluorescence}, alongside the experimental parameters $f_\text{rep} = 4\kHz$, $\Delta t_\text{meas} = 10\min$, $T_\text{sig,S} = 9\,\%$ and $T_\text{sig,AS} = 8.97\,\%$. 
The fit is simultaneously optimised over all time bins $t$, both ensemble configurations $i$ and noise channels $k$. 
It yields a value of $\tilde{\eta}_\text{det} = 37.58\,\%$. 
Fig. \ref{fig_app_ch7_theory_prediction_deteff_fit} shows the noise levels predicted by our theory model in comparison with the updated experimental noise levels $N^t_{i,k}(\tilde{\eta}_\text{det})$, obtained under utilisation of this fitted efficiency. 
The resemblance between theory and experiment is even better for almost all configurations with the exception of the AS channel in the thermally distributed ensemble.

The detector efficiency $\tilde{\eta}_\text{det}$ is a reasonable result and might indeed be the actual detection efficiency of \spcmdh\,. To maintain consistency within this thesis and with respect to published work\cite{Michelberger:2014}, which all utilise the assumption $\eta_\text{det} \approx 50 \,\%$, the noise data has not been re-evaluated using this estimated detector efficiency.

\section{Anti-Stokes seeding by the input signal \label{app_ch7_subsec_ASseeding}}

Here we estimate an upper bound on the amount of noise gain in the S channel, which can falsely be attributed to the memory efficiency {\etamem} (eqs. \ref{eq6_err_memeff_in} - \ref{eq6_err_memeff_ret}). 
Since the dynamics of our system corresponds to a beam-splitter interaction for the Raman memory, and a two-mode squeezer for the FWM noise, we can obtain the upper bound in neglecting the memory component in our coherent interaction model (see appendix \ref{app6_coh_model}).
In this case, the {\cs} vapour system reduces to an optical parametric amplifier for FWM noise, which is a well studied system in atomic gases\cite{Philips:2009,Boyer:2008,Boyer:2008aa,Camacho:2009ao,Boyer:2013}.

Using the notation of appendix \ref{app6_coh_model} for Stokes ($\vec{S}$) and anti-Stokes ($\vec{A}$) annihilation operators, the solutions for the optical parametric amplifier\cite{Loudon:2004gd,Philips:2009} yields photon number expectation values of\cite{Carman:1966}:
\begin{align*}
N_{S,\text{out}} = \langle \vec{S}_\text{out}^\dagger \vec{S}_\text{out} \rangle &\sim 
	\langle \vec{S}_\text{in}^\dagger \vec{S}_\text{in} \rangle \left( \cosh{(\mathcal{P})} \right)^2 +  
	(\langle \vec{A}_\text{in} \vec{A}^\dagger_\text{in} \rangle ) \left( \sinh{(\mathcal{P})}\right)^2  \\
			& = \langle \vec{S}_\text{in}^\dagger \vec{S}_\text{in} \rangle \left( \cosh{(\mathcal{P})} \right)^2 +  
	(1+ \langle \vec{A}_\text{in}^\dagger \vec{A}_\text{in} \rangle ) \left( \sinh{(\mathcal{P})} \right)^2  \\
	& = N_{S,\text{in}} \left( \cosh{(\mathcal{P})} \right)^2 + (1+N_{A,\text{in}}) \left( \sinh(\mathcal{P}) \right)^2 \\ 
N_{A,\text{out}} = \langle \vec{A}_\text{out}^\dagger \vec{A}_\text{out} \rangle &\sim  
N_{A,\text{in}} \left( \cosh{(\mathcal{P})} \right)^2 + (1+N_{S,\text{in}}) \left( \sinh(\mathcal{P}) \right)^2,
\end{align*}
where the coupling constant $\mathcal{P}$ contains the details of the atomic response function and the memory control pulse parameters\cite{Carman:1966,Lukin:1998,Lauk:2013}.
This is similar to the solution for the Maxwell-Bloch equations in our coherent model (eqs. \ref{eq_app_ch6_Aout1} - \ref{eq_app_ch6_Bout1}, appendix \ref{app6_coh_model}), 
where the Greens functions have been simplified to 
$\mathbb{G}_{A,S} = \mathbb{G}_{S,A} \sim \sinh{(\mathcal{P})}$, $\mathbb{G}_{S,S} = \mathbb{G}_{A,A} \sim \cosh{(\mathcal{P})}$ and $\mathbb{G}_{B,\left\{ S,A\right\}} = 0$.

Looking at the last term for the anti-Stokes noise level $N_{A,\text{out}}$ first, we can see that,  
apart from any anti-Stokes input (term $\sim \left( \cosh{(\mathcal{P})} \right)^2 \cdot N_{A,\text{in}}$) and a fixed noise background (term $\sim \left( \sinh{(\mathcal{P})} \right)^2$), the noise number is also linearly proportional to the amount of memory input signal at the Stokes frequency (term $\sim \sinh{(\mathcal{P})} N_{S,\text{in}}$). 
This is the relationship we use for the seeding experiments described in section \ref{ch7_subsec_ASseeding} of the main text. It allows us to monitor the dependence of the anti-Stokes noise on the memory input signal $N_{S,\text{in}}$.

To obtain an upper limit for the amount of noise gain in the S channel, we use the equation for $N_{S,\text{out}}$.
Since no anti-Stokes signal is inserted, $N_{A,\text{in}}=0$, we obtain a Stokes noise photon number for setting \scd\, where $N_{S,\text{in}}\neq 0$, of 
$$N_{scd} = N_{S,\text{out}} = \left( \cosh{(\mathcal{P})} \right)^2 N_{S,\text{in}} + \left( \sinh{(\mathcal{P})} \right)^2.$$
For setting \cd\,, where $N_{S,\text{in}} = 0$, we obtain
$$N_{cd} =  N_{S,\text{out}}|_{N_{S,\text{in} =0}} = \left( \sinh{(\mathcal{P})}\right)^2.$$
The additional contribution to the signal in the Stokes mode, produced by FWM gain, is thus given by:
$$G_S= \frac{N_{scd} - N_{cd}}{N_{S,\text{in}}} = \left(\cosh{(\mathcal{P})} \right)^2 = 1+ \left( \sinh{(\mathcal{P})} \right)^2 = 1+N_{cd} = 1+N_\text{noise}^\text{out}.$$
For $G_S = 1$, the signal at the {\cs} cell output is equal to the input signal, in which case the memory efficiency $\eta_\text{mem}$ would be exactly determined by eq. \ref{eq6_memeff}. 
With our finite, unseeded noise floor of $N_\text{noise}^\text{out} = 0.15 \ppp$ (eq. \ref{ch6_eq_NoiseFloor}), this increases to $G = 1.15$. 
So at most $15\,\%$ of signal entering the memory efficiency calculation, i.e. the signal obtained after subtracting the  \cd\, background noise from the counts for \scd\,, can come from FWM Stokes gain. 
The most conservative bound on {\etamem} is thus given by $\eta^\text{obs}_\text{mem} = G \cdot \eta^\text{real}_\text{mem}$, whereby
$\eta^\text{obs}_\text{mem}$ and $\eta^\text{real}_\text{mem}$ are the experimentally observable (eq. \ref{eq6_memeff}) and the real memory efficiency. 
This means, the actual memory efficiency for \hsp\, storage would be $\eta_\text{mem}^\text{real} \approx 18\, \%$ instead of $\eta_\text{mem}^\text{obs} \approx 21\,\%$ (section \ref{ch6_subsec:MemEffCalc}). 
For \coh\, storage at similar $N_\text{in}$, we would have $\eta_\text{mem}^\text{real} \approx 24\,\%$, rather than $\eta_\text{mem}^\text{obs} \approx 28\,\%$. 

It is important to note that this is a maximum bound. After all, the system is a memory and the interaction is not pure two-mode squeezing. In fact the Raman memory process dominates. 
The gain $G$ will consequently be much better approximated by the experimental values in section \ref{ch7_subsec_ASseeding}, where the contribution to the memory signal is small. 
Ultimately, this expectation is also backed-up when comparing the memory efficiencies, obtained at the single photon level (fig. \ref{fig_6_avg_memeff}), with those for bright coherent states, measured on the Menlo PD. For the latter measurements, FWM noise is irrelevant. The values tend to coincide for both regimes. Particularly, the single photon level data does not consistently exceed the bright light numbers, as would be expected for significant noise gain.

\section{Expected memory efficiency increase from signal seeding\label{app_ch7_subsec_seedingmemeff}}

To obtain a quantitative estimate about how much of the measured memory efficiency is due to noise, the normalised memory efficiency and noise curves in fig. \ref{fig_ch7_ASseeding} \textbf{d} are used as follows: 

From the definition of the memory efficiency in chapter \ref{ch6}, we obtain the real memory efficiency in terms of the observed coincidence rates for settings \scd\, and \cd\, to (eq. \ref{eq6_memeff}; for simplicity we ignore diode laser and signal leakage, and assume a constant experimental repetition rate): 
$\eta^\text{real}_\text{mem} = \frac{c^\text{out}_{scd} - c^\text{out}_{cd} }{c^\text{in}_{sd}}.$
Noise seeding by the signal will lead to an artificially increased number of $c^\text{out}_{scd}$ counts, which contains the additional counts $c^t_\text{seeded} = c^t_{scd,\text{S}}(N_\text{in}) - c^t_{cd,\text{S}}(N_\text{in} = 0)$ due to seeding.
With the assumption of section \ref{ch7_subsec_ASseeding} that these equal the increase in AS counts, we obtain 
$c^t_\text{seeded}=  c^t_{scd,\text{AS}}(N_\text{in}) - c^t_{cd,\text{AS}}$. Division by $c^t_{cd,\text{AS}}$ yields
$$
\frac{c^t_\text{seeded}}{c^t_{cd,\text{AS}}} =  \frac{c^t_{scd,\text{AS}}(N_\text{in})}{c^t_{cd,\text{AS}} } - 1= N^t_{\text{AS,norm}}(N_\text{in}) - N^t_{\text{AS,norm}}(0),
$$ 
which are the blue lines in fig. \ref{fig_ch7_ASseeding} \textbf{c}.
The overestimated memory efficiency reads 
$\eta^\text{obs}_\text{mem} = \frac{c^\text{out}_{scd} + c^\text{out}_\text{seeded} - c^\text{out}_{cd} }{c^\text{in}_{sd}}$.
From both expressions, the relative amount of overestimation is 
$\Delta \eta_\text{mem} = \eta_\text{mem}^\text{obs} - \eta^\text{real}_\text{mem} = \frac{c^\text{out}_\text{seeded}}{c^\text{in}_{sd}}$, which in turn yields the relative amount of efficiency overestimation of
$$
\delta \eta_\text{mem} =\frac{\Delta \eta_\text{mem}}{\eta^\text{obs}_\text{mem}} = \frac{c^\text{out}_\text{seeded}}{\eta^\text{obs}_\text{mem} \cdot c^\text{in}_{sd}} = 
\frac{c^\text{out}_\text{seeded}}{c^\text{out}_{scd} + c^\text{out}_\text{seeded} - c^\text{out}_{cd}} = 
\frac{ \frac{c^\text{out}_\text{seeded} + c^\text{out}_{cd}}{ c^\text{out}_{cd}} -1 }{
\frac{c^\text{out}_{scd} + c^\text{out}_\text{seeded}}{ c^\text{out}_{cd}} - 1} = 
\frac{N^\text{out}_\text{AS,norm}(N_\text{in}) -1}{N^\text{out}_\text{S,norm}(N_\text{in}) -1}.
$$ 
In the last step, the definitions for the normalised AS noise level and the normalised \scd\, signal level in the S channel of
$N_{i,\text{norm}}(N_\text{in}) = \frac{c_i^t(N_\text{in}) }{c_{i}^t(0)} =  \frac{c_i^t(N_\text{in}) }{c_{cd}^t}$ have been used, whereby $i\in \left\{\text{S}, \text{AS} \right\}$ and $N_{i,\text{norm}}(0) =1$.
The values for $N^\text{out}_\text{S,norm}(N_\text{in})$ lie on the read lines in fig. \ref{fig_ch7_ASseeding}. 
Notably, as stated in eq. \ref{ch6_eq_Nin}, photon numbers {\Nin} correspond to the detected coincidences, scaled by signal filter transmission and detector efficiency. Hence both quantities are used interchangeably.
	
Observing the $1^\text{st}$ read-out time bin for the largest \coh\, input signal photon number, involved in the \gtwo\, measurements of chapter \ref{ch6}, and for the measured seeding value closest to {\Nin} for \hsp\, input, the following numbers are obtained: 
\begin{itemize}
\item $N_\text{in} \approx 2.23 \ppp$: \newline 
${c}_{scd,\text{S}}^\text{out1} (N_\text{in}) = 4.88$, ${c}_{scd,\text{AS}}^\text{out1} (N_\text{in}) = 1.39$
$\Rightarrow$ $c^\text{out}_\text{seeded} = 0.22$, $\eta_\text{mem}^\text{obs} = 3.88$, and $\delta \eta_\text{mem} = 9.93 \,\%$.  
\item $N_\text{in} \approx 0.38 \ppp$: \newline
${c}_{scd,\text{S}}^\text{out1} (N_\text{in}) = 1.66$, ${c}_{scd,\text{AS}}^\text{out1} (N_\text{in}) = 1.047$
$\Rightarrow$ $c^\text{out}_\text{seeded} = 0.047$, $\eta_\text{mem}^\text{obs} = 0.657$, and 
$\delta \eta_\text{mem} = 7.1 \,\%$.  
\end{itemize}

\section{Population flow model for SRS in Stokes channel\label{app_ch7_subsec_flow_model_srs_stokes}}

When discussing the noise emission from the memory without optical pumping in section \ref{ch7_subsec_c_noise}, we have seen that the Stokes emission for setting \textit{c} decreases over a consecutive train of control pulses. 
In a naive model, one could assume that this is due to population reshuffling between the F$=3$ and F$=4$ hyperfine ground states, caused by the larger Raman coupling constant $C_\text{S} = \frac{\alpha}{\Delta_\text{S}}$ for S scattering than for AS scattering with $C'_\text{AS} = \frac{\alpha}{\Delta_\text{AS}}$. 
If we consider a classical model and neglect any other efficiency factors, which should influence both FWM channels equally, the population transfer ratios into and out of the F$=3$ storage state ($N_{\ket{s}}$) can be expressed as: 
 \begin{itemize}
 \item Population transfer from F$=4$ to F$=3$ by SRS into the AS channel: $\frac{d}{dt} N_{\ket{s}} = (C'_\text{AS})^2 N_{\ket{i}}$
 \item Population transfer from F$=3$ to F$=4$ by SRS into the S channel: $\frac{d}{dt} N_{\ket{s}} = - (C_\text{S})^2 N_{\ket{s}}$,
 \end{itemize}

whereby $N_{\ket{i}}$ is the population of F$=4$.
Initially, half of the total population $N_\text{tot}$ is located in $N_{\ket{s}}$. This will be changed by the imbalance between the above transfer rates and eventually converge against a steady-state population determined by:
$$
\frac{d}{dt} N_{\ket{s}}^{ss} = (C'_\text{AS})^2 N^{ss}_{\ket{i}} - (C_\text{S})^2 N^{ss}_{\ket{s}} = 
(C'_\text{AS})^2 \left( N_\text{tot}- N^{ss}_{\ket{s}} \right) - (C_\text{S})^2 N_{\ket{s}} =  0 
\Rightarrow N^{ss}_{\ket{s}} = \frac{N_\text{tot}}{\frac{(C_\text{S})^2}{(C'_\text{AS})^2} + 1}
$$

Over the course of the control pulse sequence, the population decrease has a ration of 
$ R^{p9/p1}_{c,\text{S}} = \frac{N_{\ket{s}}^{p9}}{N_{\ket{s}}^{p1}} $. 
Here $N_{\ket{s}}^{p1} = \frac{N_\text{tot}}{2}$ is the initial population in F$=3$, that gives rise to the first S noise pulse in fig. \ref{fig_ch7_flourescence} \textbf{c}, whereby $N_{\ket{s}}^{p9}$ is the population giving rise to the $9^\text{th}$ pulse at which the S noise level has reached its steady state. Using $N^{ss}_{\ket{s}}$, we obtain

$$
R^{p9/p1}_{c,\text{S}} = \frac{N_{\ket{s}}^{p9}}{N_{\ket{s}}^{p1}} = \frac{N_{\ket{s}}^{ss}}{\frac{N_{\text{tot}}}{2}} = 
\frac{2}{\frac{(C_\text{S})^2}{(C'_\text{AS})^2} + 1} = \frac{2}{\frac{(\Delta_\text{AS})^2}{(\Delta_\text{S})^2} + 1} = 0.5591.
$$

Since the SRS intensity $I_\text{SRS}$ is linearly proportional to the population $N$ of the state it couples to\cite{Carlsten:1977}, the ratio $R^{p9/p1}_{c,\text{S}}$ can also be expressed in terms of the count rates observed in pulses 1 and 9 of the S noise emission:  
$R^{p9/p1}_{c,\text{S}} = \frac{I_\text{SRS}^{p9}}{I_\text{SRS}^{p1}} = \frac{c^{p9}_{c,\text{S}}}{c^{p1}_{c,\text{S}}}$.
 The data has a ratio of $R^{p9/p1}_{c,\text{S}} \approx 0.62$, which does not agree with the theoretical expectation. 
Balance of population transfer therefore cannot explain the decrease in S scattering for setting \textit{c}. 
One can expect a steady state pump flow model not to describe the real situation closely, since SRS in both channels is a weak process, which should not result in any significant reshuffling of population. The exemplification of this fact are the low count rates for S and AS emission.
Moreover, in convergence against the steady state, population would be moved from the storage state $\ket{\text{s}}$ into the initial state $\ket{\text{i}}$, which, in turn, should increase the amount of SRS into the AS channel. Yet, as fig. \ref{fig_ch7_flourescence} shows, this rate is constant.

\section{Parameter dependences of memory efficiency and noise\label{appendix_ch7_noise_parameter_dependence}}

When we varied some of the experimental parameters during the characterisation of the noise sources in chapter \ref{ch7} of the main text, we did not see a substantial improvement in the SNR over the values quoted in chapter \ref{ch6}. 
One can thus ask, whether any of the main setup parameters (see chapter \ref{ch2}), including storage time, control field energy, detuning, cell temperature and spatial mode size, can result in a better performance. 
For our specific set-up, the short answer to this is negative; otherwise such changes would have been implemented prior to the \gtwo\, measurements of chapter \ref{ch6}. 
Nevertheless, interesting information has been collected when we tested these additional parameters to see if there was any improvement. These findings are particularly relevant for the experimental set-up of the bulk cell {\cs} memory and might be of use to the reader, who seeks to reproduce some of our results. 
For this reason, we will discuss these measurements in the following:

\subsection{Lifetime\label{ch7_lifetime}}
As the first one of these parameters we analyse the lifetime properties. Here, we evaluate the memory efficiency and noise lifetimes at the single photon level.

\paragraph{Decoherence and memory lifetime}
The functional dependence of both observables on the storage time $\tau_\text{S}$ firstly enables to further distinguish between FWM and SRS processes. 
Secondly, it can be used to estimate the decoherence mechanism, that currently limits memory performance. 
In general, memory efficiency and noise level reduction for longer storage times $\tau_\text{S}$ predominantly 
originate either from spin-wave dephasing or atomic diffusion. 
The former is analogue to the magnetic dephasing (section \ref{ch7_subsec_mag_dephasing}), induced by stray magnetic fields.
As we have seen in fig. \ref{fig_ch7_mag_dephasing}, magnetic dephasing will lead to a Gaussian-shaped decoherence function\cite{Reim2011_supp}. 

Diffusion reduces the number of {\cs} atoms that are excited in the spin-wave superposition as these just move outside the active volume defined by the overlap region of the signal and control pulses. 
Resulting atomic loss lowers the atomic number density entering the optical depth of the Raman memory\cite{Nunn:DPhil}, leading to lower retrieval efficiencies for memory signals and FWM noise from the {\cs} spin-wave. 
Through its influence on the optical depth, diffusion decoherence leads to exponential lifetime decay, as \textit{Chrapkiewicz et. al.} have shown\cite{Chrapkiewicz:2014fk}. 
Another decoherence channel is spin-changing {\cs}-{\cs} collisions, which we expect to be unlikely (see section \ref{ch7_subsec_fluorescence}).

Notably, diffusive decoherence can be expected to yield different lifetimes between the memory efficiency and noise level. This is due to the different mode sizes of both beams in the current set-up\footnote{
	This set-up was used for the work presented in chapters \ref{ch6} and \ref{ch7}. 
	The mode sizes were different for the work in chapter \ref{ch4}, as these experiments had been conducted
	on a previous incarnation of the memory experiment. 
}. 
With a FWHM mode size of FWHM$_\text{sig} \approx 95 \mum$ the signal is focussed tighter than the control, whose mode size equals FWHM$_\text{ctrl} \approx 238 \mum$ (see fig. \ref{fig_ch7_control_focussing}). 
Despite a larger diffraction angle, the signal mode is smaller than the control mode over the entire length of the {\cs} cell. 
For the memory efficiency, the mode overlap between control and signal is critical, which in this case is limited by the signal mode. Moreover, the signal collection into SMF behind the memory (see fig. \ref{fig_6_setup}) is optimised on the signal mode. 
Memory retrieval from any atoms that have travelled outside the overlap region between signal and control will couple into the collecting SMF with reduced efficiency. 
One can see this in the SMF coupling efficiencies for signal and control pulses transmitted through the memory: while the signal mode has $\eta^\text{signal}_\text{SMF} \approx 85\,\%$, only 
$\eta^\text{control}_\text{SMF} \approx 45\,\%$ of the control couple into the SMF\footnote{
	For determining the control SMF coupling efficiency, the deflection of signal 
	and control are reversed on the PBD behind the memory (see fig. \ref{fig_6_setup}).
}. 
The loss from atomic diffusion thus reduces the amount of observable retrieved signal and therewith the memory efficiency. Importantly, this loss cannot be balanced by the inflow of atoms into the signal and control overlap region, since these atoms are not part of the spin-wave coherence.

In contrast, the relevant spatial mode for the memory noise is solely set by the control itself. 
Observed noise will still be limited to the mode selected by SMF coupling, which mainly equals the mode of the signal. 
FWM spin-wave coherence is however established amongst the atoms in the volume covered by the control. 
As a result, the net outflow of atoms from the volume covered by the signal mode can be balanced by the inflow of {\cs} atoms from the surrounding region, which is still covered by the control. 
Thus, the net amount of atoms  in the signal mode volume, which participate in the FWM spin-wave, decreases slower for larger times $\tau_\text{S}$ than the memory signal.

\paragraph{Measurement}
Determining the memory lifetime at the single photon level is similar to the measurement for bright {\cs} input signals, presented in section \ref{ch4_subsec_fidelity}. 
During the measurement, the $1^\text{st}$ pulse picking window of the \pockels\, is closed to select only a single pulse to define the read-in bin. 
The fully opened $2^\text{nd}$ \pockels\, window is delayed to sequentially increasing times $\tau_\text{S}$, setting the memory storage time. 
Note, $\tau_\text{S}$ is defined as the time between the first pulses in both \pockels\, pulse picking windows.
The exception to this procedure is $\tau_\text{S} = 12.5\ns$. Here the $1^\text{st}$ pulse picking window is opened fully, with the $1^\text{st}$ and $2^\text{nd}$ pulse therein defining the read-in and read-out bin, respectively.
For each value of $\tau_\text{S}$ the memory settings $\left\{ scd,sd,cd,d\right\}$ are measured, whereby, in contrast to experiments with bright \coh\, inputs, setting \cd\, is now required to subtract the noise background in {\etamem} (eq. \ref{eq6_memeff}). 

The more complicated task is the measurement of the noise lifetime. 
We expect decoherence to occur only for the fraction of noise that arises from coupling to a previously excited spin-wave. 
This equals the amount of noise exceeding the level generated by the $1^\text{st}$ control pulse in \pockels\, window 1 (see fig. \ref{fig_ch7_flourescence}).  
However the unseeded portions of S and AS noise build up over a sequence of $\sim 5$ consecutive pulses. 
For the above described memory control pulse sequence, 
the noise increase caused by the $2^\text{nd}$ control pulse would thus counteract decoherence decay leading to a larger noise level upon retrieval, even if the read-out control pulse is delayed by a time $\tau_\text{S} \gg 12.5\ns$. 
For this reason, we instead apply a control sequence with the two \pockels\, windows fully opened to select $9$ consecutive pulses in each window. 
Therewith, the noise level is built-up to saturation in the first window, as shown in fig. \ref{fig_ch7_flourescence} \textbf{a}. 
The first pulse of the $2^\text{nd}$ window, applied at $\tilde{\tau}_\text{S} > 12.5 \ns$ after the last pulse in the first window, now retrieves the noise spin-wave that is still present. 
Requiring the noise build-up, the effectively probed storage times $\tilde{\tau}_\text{S} = \tau_\text{S} - 112.5\ns$ are shorter than the memory storage times $\tau_\text{S}$ for the same \pockels\, window timings. 
With this control sequence the relative amount of noise dephasing 
$N^\text{rel}_{i,k} = \frac{N^\text{p10}_{i,k}}{N^\text{p9}_{i,k}}$ 
probes the change in the noise due to spin-wave coupling ($i \in \left[\text{S}, \text{AS} \right]$). 
It is determined by the noise levels $N^\text{p10}_{\text{noise}}$ of the $1^\text{st}$ pulse in picking window 2 (overall this is the $10^\text{th}$ pulse) and $N^\text{p9}_{\text{noise}}$ for the $9^\text{th}$ 
pulse of \pockels\, window 1. 
Again $\tilde{\tau}_\text{S} = 12.5\ns$ is an exception: here, the $8^\text{th}$ and $9^\text{th}$ pulse in \pockels\, window 1 take the roles of pulses 9 and 10 to determine the reference level and the retrieved noise levels.
Note that without spin-wave coupling, as for SRS, or without any decoherence, the $10^\text{th}$ pulse would just lead to the same noise level as the $9^\text{th}$, yielding $N^\text{rel}_{i,k} = 1$. 

The retrieval pulse p10 still generates noise within the pulse itself, so the limit of 
$N^\text{rel}_{i,k}$ for $\tilde{\tau}_\text{S} \rightarrow \infty$ will be $\frac{N^\text{p1}_{i,k}}{N^\text{p9}_{i,k}}$, where $N^\text{p1}_{ \{ \text{AS/S} \},cd}$ is the noise for the $1^\text{st}$ control pulse in \pockels\, window 1. 
To access solely the decoherence of the spin-wave related fraction, this instantaneous noise needs to be subtracted. This gives 
$N^\text{norm}_{i,k} = \frac{N^\text{p10}_{i,k} - N^\text{p1}_{i,k}}{N^\text{p9}_{i,k} - N^\text{p1}_{i,k}}$, 
which can directly be compared to the memory efficiency lifetime.  

Since the major part of AS noise originates from SRS, the better method to probe decoherence of spin-wave related AS noise lies in investigating seeded AS noise. To this end, the procedure of section  \ref{ch7_subsec_ASseeding} is implemented on a control sequence with only $1$ pulse in \pockels\, window 1, which is also the time bin for the S signal input with $N_\text{in} \approx 6.3 \ppp$. 
The relative noise level is defined by $N^\text{rel}_{\text{AS},scd} = \frac{N^\text{p2}_{\text{AS},scd}}{N^\text{p2}_{\text{AS},cd}}$, whereby p2 is the retrieval pulse after $\tau_\text{S}$.
The contribution to $N^\text{p2}_{\text{AS},scd}$ purely due to seeding is accessed by subtracting the amount of noise obtained without S signal input ($N^\text{p2}_{\text{AS},cd}$). By normalising the difference to the level observed for 
$\tau_{\text{S},0} = 12.5\ns$, a normalised decoherence curve 
$N^\text{norm}_{\text{AS,seed}} (\tau_\text{S}) = \frac{N^\text{p2}_{\text{AS},scd}(\tau_\text{S})-N^\text{p2}_{\text{AS},cd}(\tau_\text{S})}{N^\text{p2}_{\text{AS},scd}(\tau_{\text{S},0})-N^\text{p2}_{\text{AS},cd}(\tau_{\text{S},0})}$ is obtained, which can be contrasted directly with the normalised memory efficiency 
$\eta_\text{mem}^\text{norm}(\tau_\text{S}) = \frac{\eta_\text{mem}(\tau_\text{S})}{\eta_\text{mem}(\tau_{\text{S},0})}$ 
and the unseeded noise $N^\text{norm}_{\text{S},cd} (\tau_\text{S})$.

\begin{figure}
\includegraphics[width=\textwidth]{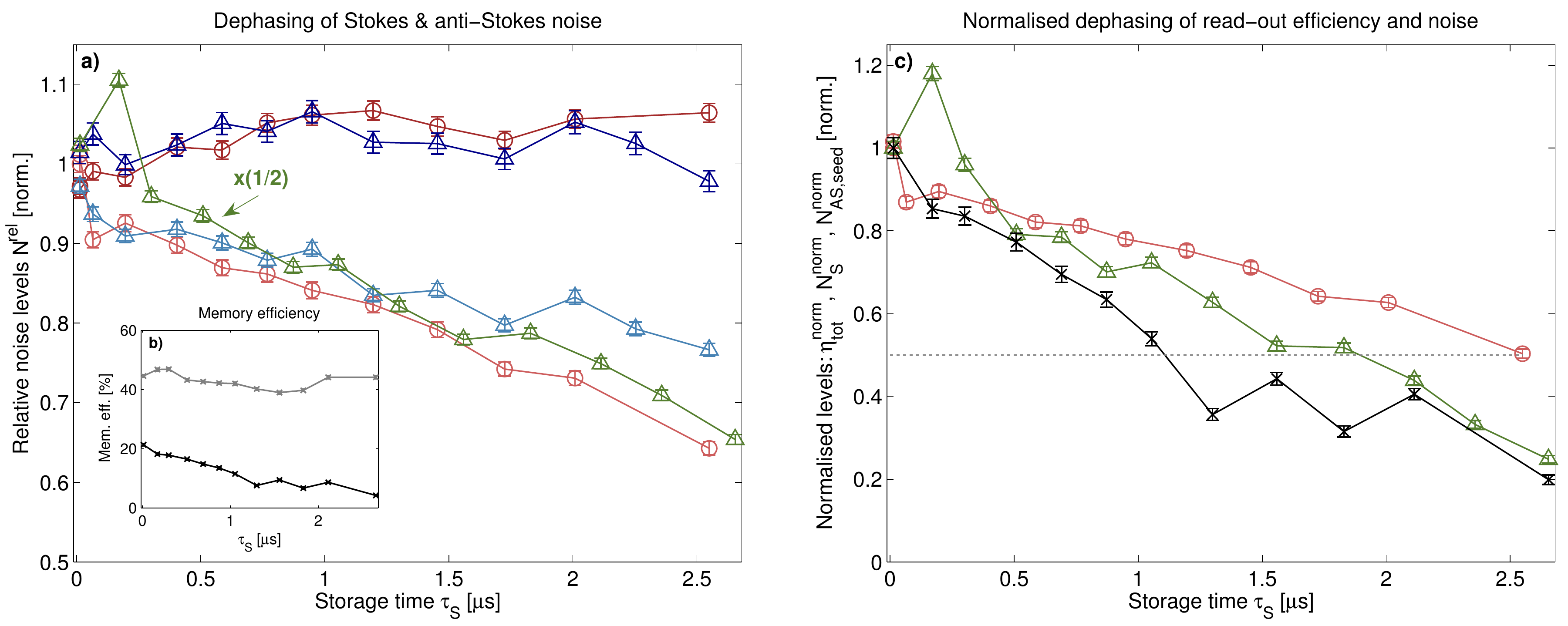}
\caption{Lifetime measurements. \textbf{(a)}: Noise decoherence expressed in terms of the relative noise levels $N^\text{rel}_{i,k}(\tau_\text{S})$. Noise emitted by the thermally distributed ensemble (setting \textit{c}) is shown in \textit{dark red} for the S and \textit{dark blue} for the AS channel. \textit{Light red} and \textit{light blue} denote the noise for the spin-polarised ensemble when probed with the control (setting \textit{cd}), while \textit{green} marks increased AS noise from S seeding, which is scaled down by a factor of 2. 
\textbf{(b)}: Memory read-in efficiency $\eta_\text{in}$ (\textit{grey}) and read-out efficiency $\eta_\text{mem}$ (\textit{black}).
\textbf{(c)}: Decoherence for the normalised memory efficiency $\eta^\text{norm}_\text{mem}$ (\textit{black}), the background subtracted normalised S noise $N^\text{norm}_{\text{S},cd}$ (\textit{red}) and the fraction of seeded AS noise $N^\text{norm}_{\text{AS,seed}}$ (\textit{green}).}
\label{fig_ch7_lifetime}
\end{figure}

\paragraph{Results}
Fig. \ref{fig_ch7_lifetime} \textbf{a} shows the results for the noise in the first read-out bin.
Plotting the relative noise levels $N^\text{rel}_{i,k}(\tau_\text{S})$,  
allows for direct comparison between the noise in both channels, S and AS, for both optical pumping configurations. 

If the {\cs} ensemble is thermally distributed (setting \textit{c}), neither S (\textit{dark red}) nor AS (\textit{dark blue}) noise show any dependence on $\tau_\text{S}$. 
This ties in with our expectation for SRS, where the control addresses incoherent {\cs} population in each ground state. 
It is therefore not sensitive to any loss of atoms that were previously excited into a spin-wave coherence. 
Decoherence by dephasing has no influence, as the absence of magnetic dephasing in section \ref{ch7_subsec_mag_dephasing} has also shown.
Diffusion is irrelevant as well because, without optical state preparation, all {\cs} atoms are, by default, thermally distributed. Diffusion inflow from outside the active volume balances atom loss, such that the population numbers of the ensemble, probed by the control, remain constant.

Contrary, noise emitted by the spin-polarised ensemble decreases for larger storage times $\tau_\text{S}$. 
All relative levels, $N^\text{rel}_{\text{S},cd}$ for the S channel (\textit{light red}), $N^\text{rel}_{\text{AS},cd}$ for the unseeded AS (\textit{light blue}) and $N^\text{rel}_{\text{AS},scd}$ for seeded AS (\textit{green}), have a similar $\tau_\text{S}$ dependence. 
Since $N^\text{rel}_{\text{AS},scd} \ge 1$ it has been rescaled for comparison with the background noise.

Displayed in fig. \ref{fig_ch7_lifetime} \textbf{b}, the memory read-out efficiency {\etamem} equally decays for longer $\tau_\text{S}$, while the read-in efficiency {\etain} remains unaffected. 
The functional dependences on $\tau_\text{S}$ do not display a Gaussian shape for either process, which implies that magnetic dephasing is unlikely to limit the current performance\footnote{
	Particularly having observed Gaussian shaped decoherence in fig. \ref{ch7_subsec_mag_dephasing}, when 
	applying a $B$-field, backs this. 
}. 
We have already seen similar results for the bright \coh\, memory in fig.~\ref{fig_ch4_fidelity_lifetime}. These measurements displayed exponential decoherence, while the curves in fig.~\ref{fig_ch7_lifetime} appear more linear. 
However this could also relate to the inability to observe the limit $\tau_\text{S} \rightarrow \infty$ at the single photon level. The maximum storage time $\tau_\text{S} \approx 2.5 \, \mu\text{s}$ in fig. \ref{fig_ch7_lifetime} is set by the dynamic range of the TAC, for which reason the observed $\tau_\text{S}$ range could still fall into the approximately linear initial region of an exponential decrease\footnote{
	In this case the 1/e lifetime in the current memory configuration would however have to be longer 
	than in the previous version used for the measurements in chapter \ref{ch4}. 
	This is unlikely for pure atomic diffusion, since the 
	signal beam waist is smaller in the current experiment. However, another possibility could be 
	the presence of additional drift currents between cold and hot spots in the {\cs} cell. Improved thermal insulation
	would cut these down, reducing atom loss rates and increasing the memory lifetime.
}. 
It is therefore more reasonable to assume that atomic diffusion rather than magnetic dephasing limits the storage time. 

In this regard, a comparison of the memory decoherence to the decoherence of the purely spin-wave dependent noise in the S and AS channel is interesting. 
To this end, the normalised values $\eta_\text{mem}^\text{norm}$, $N^\text{norm}_{\text{S},cd}$ and $N^\text{norm}_{\text{AS,seed}}$ are displayed in fig. \ref{fig_ch7_lifetime} \text{c}.

Memory efficiency and seeded AS noise have an approximate equal half-lifetime\footnote{
	These times are longer than the equivalent time of $\tau^\text{HL} \approx 0.9 \, \mu \text{s}$ 
	for the bright \coh\, memory in chapter \ref{ch4}, which supports the notion regarding 
	atomics drift currents.
}, while the S noise decays a bit slower, with lifetime numbers as follows:
\begin{center}
\begin{framed}
\begin{tabular}{ll}
\centering
Memory efficiency: 	& $\tau^\text{HL}_\text{mem} \approx 1.3\, \mu \text{s}$ 	\\
Anti-Stokes noise: 	& $\tau^\text{HL}_\text{AS} \approx 1.5\, \mu \text{s}$ 	\\
Stokes noise: 		& $\tau^\text{HL}_\text{S} \approx 2.5 \, \mus$			
\end{tabular}
\end{framed}
\end{center}
As discussed above, for decoherence by diffusion, one would expect $\tau^\text{HL}_\text{mem} < \tau^\text{HL}_{\text{S},cd}$, given the smaller signal mode size. 
Importantly, the volumes of the {\cs} cell containing the spin-wave of the seeded AS noise is also defined by the signal mode, because only atoms in the overlap region of signal and control can experience seeding. So one expects the observed similarity $\tau^\text{HL}_\text{AS} \approx \tau^\text{HL}_\text{mem}$ for a diffusion limited lifetime.

In conclusion, these findings confirm our previous identification of the relevant noise sources and indicate, that atomic diffusion\cite{Reim:2011ys} most likely limits the lifetimes of the memory efficiency and the FWM noise.

\subsection{Control pulse energy dependence \label{ch7_ctrl_power}}

The next parameter to investigate is the energy content in the control pulses. 
Here we test the scaling of the noise with the power in the control field and contrast it with the behaviour of the Raman memory efficiency. 
To obtain a rough idea of the behaviour one should expect, we conceive the different processes in terms of a sequence of Raman transitions, emitting S and AS photons. 
The transition probability for each Raman transition is proportional to the Raman coupling constant $|C|^2$ (see appendix \ref{app6_coh_model}). 
In turn, $C$ is proportional to the control field Rabi-frequency $\Omega_c =  \frac{\vec{d}_{g,e} \cdot \vec{E}_{c}}{\hbar}$ of the transitions the control field $E_{c}$ couples to\cite{Nunn:2007wj}. 
For simplicity\footnote{
	Since our experiments are not Zeeman state selective and far detuned, 
	in reality we have a sum running over all dipole allowed transitions with 
	appropriate Clebsh-Gordan coefficients\cite{Steck:2008qf, Penzkofer:1979}.
}, these have been denoted by a single transition dipole moment $\vec{d}_{g,e}$ between ground state $g$ and excited state $e$.
Therewith the photon production rates $\frac{\text{d} \langle n_i \rangle}{\text{dt}} = c_i  \sim I_{c}$ for emitted S or AS photons ($i \in \left\{ \text{S}, \text{AS} \right\}$), which are proportional to the experimentally observed photon count rates $c_i$, are linearly dependent on the control intensity $I_{c} = \frac{1}{2} \epsilon_0 c |\vec{E}_{c}|^2$.
Another way\cite{Penzkofer:1979, Raymer:1981aa, Raymer:1977} to express this uses the Raman scattering cross section $\frac{\text{d} \sigma}{\text{d} \Omega} \sim \Omega_c^2$, which determines the Raman intensity $I_i \sim \frac{\text{d} \sigma}{\text{d} \Omega} \sim I_c^2$ of emitted light in channel $i$. 
Therewith we can estimate the scaling of the memory efficiencies and the noise:
\begin{itemize}
\item Raman memory storage corresponds to a single Raman transition, stimulated by a weak signal field. The read-in efficiency should thus scale as $\eta_\text{in} \sim I_c$.
\item Memory retrieval happens via a $2^\text{nd}$ Raman transition, whose signal input (S channel) is in the vacuum state, such that $\eta_\text{ret} \sim I_c$. The total memory efficiency {\etamem} is the product of {\etain} and {\etaret}, so it scales as $\eta_\text{mem} \sim I_c^2$.
\item SRS, as expected for the thermally distributed {\cs} ensemble (setting \textit{c}), is also a Raman transition with the vacuum state as the signal input\cite{Raymer:1981aa}. Again we expect $N_{i,c} \sim I_c$ for both channels $i\in \left\{ \text{S} ,\text{AS} \right\}$.
\item The same holds for AS emission in the $1^\text{st}$ control pulse time bin (read-in) with spin-polarised {\cs} (setting \cd\,). 
As long  no previously excited spin-wave exists, the first leg of the FWM interaction is SRS into the AS channel. 
AS emission should hence also show the proportionality $N^\text{p1}_{\text{AS},cd} \sim I_c$. 
However, due to the possibility of FWM spin-wave read-out within the same control pulse, 
AS can be stimulated by seeding from the S noise emission, following the mechanism discussed in section \ref{ch7_subsec_ASseeding}.
In this case the exponent of $I_c$ would be expected to exceed $1$. 
\item FWM S noise emission resembles memory retrieval, since it can only occur after generation of a spin-wave. 
Like the memory read-out efficiency {\etamem}, S noise is a product of two Raman transitions and should hence follow $N_{\text{S},cd} \sim I_c^2$. 
Similar to the AS channel, multiple transitions within the same or subsequent control pulses can lead to variations in the scaling (i.e. the exponent). The approximation should hold best for the first pulse in a control sequence (read-in bin).
\end{itemize}
Notably, these arguments for the scaling behaviour with $I_c$ are very simple. In fact, they are too simple to describe the experimental observations accurately. For instance, the dynamic Stark shift\cite{Nunn:2007wj, Reim2010, Moiseev:2011}, introduced by the large electric field strengths of the control pulses, is completely ignored. 
It will reduce the exponents in the scaling with $I_c$, because it
shifts the atomic levels out of two-photon resonance with signal and control\footnote{
	The same argument applies to SRS, where the Stark shift also changes the 
	effective detuning.
}.   
Another possible reduction can result from saturation effects, which are, for example, expected for {\etamem} at large control pulse energies\cite{Reim2010}.
In reality, the scaling will thus be sub-linear or sub-quadratic for processes involving single or double Raman transitions, respectively. 
Importantly, the ratio between the exponents for both processes should nevertheless be conserved. 
So we expect count rates from processes involving two Raman transitions to scale with an exponent twice as large as those originating from a single Raman transition.

\paragraph{Measurement}
Experimental determination of the scaling with control intensity is straight forward. 
We use the same control pulse sequence for $\tau_\text{S} = 312\ns$ storage time as shown in figs. \ref{fig_ch7_flourescence} (insets) and \ref{fig_ch7_SNRintwin}. Therewith, the memory efficiency and the noise levels in both channels, S and AS,  are probed. 
Besides these noise measurements on the spin-polarised ensemble (setting \cd\,), the AS noise characteristics are tested on the thermally distributed {\cs} atoms (setting \textit{c}). 
Stimulated AS emission (setting \scd\,), seeded by sending input signal into the S channel in the read-in bin, is also investigated.
Likewise to the lifetime measurements in section \ref{ch7_lifetime}, we expect a different behaviour for the seeded fraction of the noise (setting \scd\,) than for the unseeded AS noise background, generated solely by the control (setting \cd\,). 
To obtain the seeded fraction $N^t_\text{AS,seed} = N^t_{\text{AS},scd} - N^t_{\text{AS},cd}$, the background noise level $N^t_{\text{AS},cd}$ is subtracted from the observed level $N^t_{\text{AS},scd}$ when noise is seeded. 

To avoid having to define a specific measurement location along the {\cs} cell at which $I_c$ is calculated, 
we instead refer to the control pulse energy $E^\text{p}_c = \frac{\bar{P}_c}{N_\text{p} \cdot f_\text{rep}}$, which is defined by the average power of the control pulse train $\bar{P}_c$, the control sequence repetition rate $f_\text{rep} = 4\kHz$, and the number $N_\text{p}$ of control pulses in each sequence\footnote{
	The average intensity $\bar{I}_c(z) = \frac{\bar{P}_c}{\pi (w_0(z))^2}$ changes over the length of the {\cs}, 
	as it is dependent on the control's beam waist $w_0(z)$. The instantaneous intensity 
	additionally depends on 
	the pulse shape, whereby, for ideal sech-shaped pulses, a peak intensity of 
	$I^\text{peak}_c(z) \approx 0.88 \cdot \frac{1}{\tau_\text{p}} \cdot \frac{1}{\pi  (w_0 (z))^2}  \cdot 
	\frac{\bar{P}_c}{N_\text{p} \cdot f_\text{rep}}$
	is reached.
}. 
$\bar{P}_c$ is measured on a power meter at the input window of the {\cs} cell. It ignores any linear absorption of the control during propagation through the {\cs}. 
To furthermore minimise the contribution of leakage pulses that are not selected by the \pockels\,, power measurements are conducted with both \pockels\, pulse picking windows open fully to select $N_\text{p} = 18$ pulses per sequence.

\begin{figure}[h!]
\includegraphics[width=\textwidth]{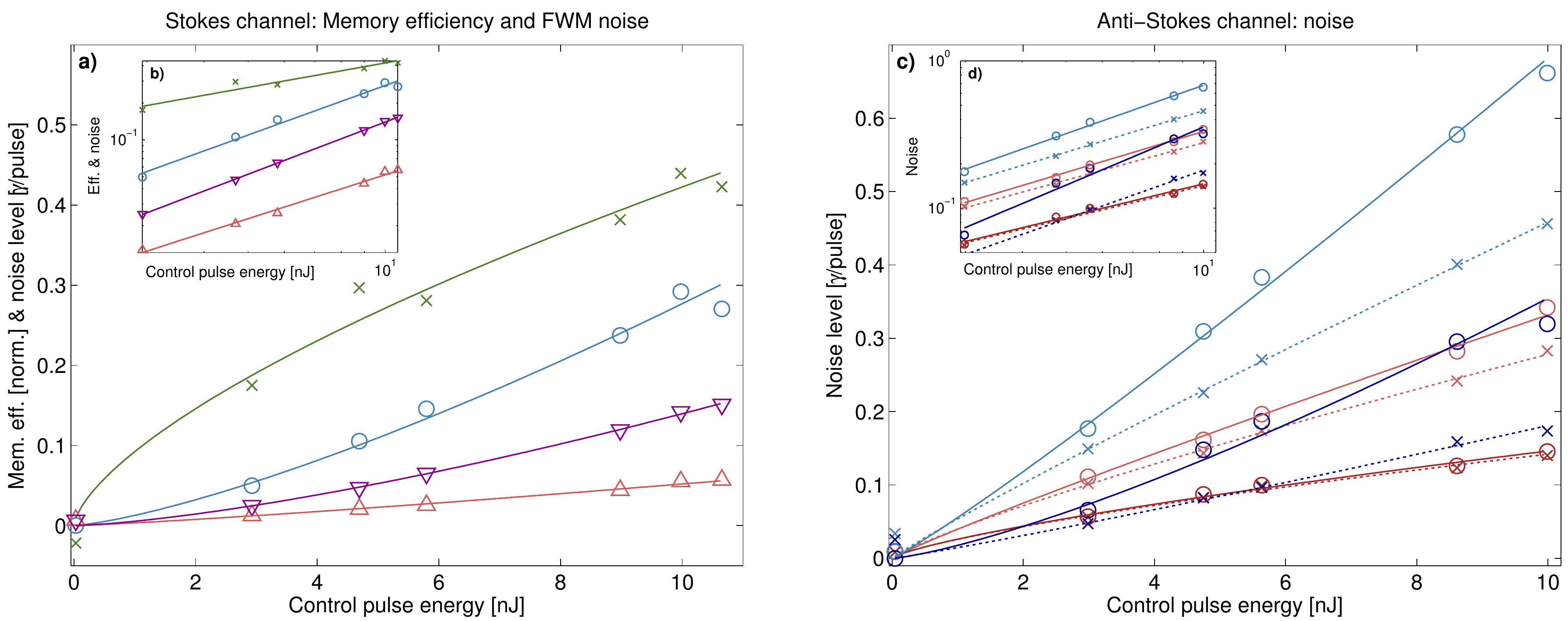}
\caption{Memory efficiency and noise dependence on control pulse energy. 
\textbf{(a)} \& \textbf{(b)}: S channel measurements, showing the memory read-in ({\etain}, \textit{green}) and the read-out ({\etamem}, \textit{blue}) efficiencies, as well as FWM noise, emitted by the spin-polarised ensemble, in the input ($N_{\text{S},cd}^\text{in}$, \textit{orange}) and the $\tau_\text{S} = 312\ns$ output ($N_{\text{S},cd}^\text{out}$, \textit{magenta}) time bin.
\textbf{(c)} \& \textbf{(d)}: AS channel, showing 
FWM noise emitted by the spin-polarised ensemble, with the background noise $N_{\text{AS},cd}^t$ in \textit{orange}, the total noise $N_{\text{AS},scd}^t$, observed when seeding, in \textit{light blue}, and the purely seeded fraction $N_{\text{AS},\text{seed}}^t$ in \textit{dark blue}. Also shown is SRS $N_{\text{AS},c}^t$ from the thermally distributed ensemble in \textit{red}. \textit{Dashed lines} and $\times$ mark the input, \textit{solid lines} and $\circ$ mark the $\tau_\text{S} = 312\ns$ output bin. 
For both channels, the main panels show data with linearly scaled axes, insets have logarithmic scaling to illustrate the pulse energy dependence. \textit{Lines} in the main panels correspond to the linear fits on the logarithmically scaled data in the insets. 
}
\label{fig_ch7_cntrl_power}
\end{figure}

\paragraph{Results}

Memory efficiency and noise level in the S channel are shown in fig. \ref{fig_ch7_cntrl_power} \textbf{a}. 
The read-in efficiency data
{\etain} (\textit{green}) increases sub-linear and with a lower gradient than the read-out efficiency {\etamem} (\textit{blue}), which is super-linear as a function of $E_c^\text{p}$.  
Similarly,  S noise $N^t_{\text{S},cd}$ for the pumped ensemble also increases super-linear in the read-in (\textit{orange}) and $312\ns$ read-out bin (\textit{magenta}). 
Since we are interested in the exact exponents $x$ of the scaling of all our observables $y\in \left\{\eta_\text{in}, \eta_\text{mem}, N^t_{i,k}  \right\}$, whereby 
$t \in \left\{ \text{in}, \text{out1} \right\}$, $i \in \left\{ \text{S}, \text{AS} \right\}$ and $k \in \left\{ scd, cd, c \right\}$, 
we can take the logarithm of the functional dependences  $y \sim (E_c^\text{p})^x$. 
Plotting the data double-logarithmically in fig. \ref{fig_ch7_cntrl_power} \textbf{b} visualises $x$ as the gradient of a simple, straight line fit $\ln (y) = \beta + x \cdot \ln(E_c^\text{p})$. 
Clearly, we can see that {\etain} has a shallower gradient than {\etamem}, whose dependence on $E_c^\text{p}$ approximately equals that of the noise $N^t_{\text{S},cd}$ in both time bins.

Performing the same analysis on the AS channel and plotting the data with linear axes scaling in fig. \ref{fig_ch7_cntrl_power} \textbf{c} first, yields a similar, linear increase with $E_c^\text{p}$ for the noise levels $N^t_{\text{AS},cd}$ of the spin-polarised ensemble (\textit{orange}) and $N^t_{\text{AS},c}$ for thermal {\cs} (\textit{red}).
$N^t_{\text{AS},scd}$ (\textit{light blue}) also grows linearly in the read-in time bin (\textit{dotted lines}), where the additional noise from seeding is small. 
However, in the read-out bin (\textit{solid lines}) the increased influence of the seeded fraction $N^t_{\text{AS},\text{seed}}$ results in super-linear growth of $N^\text{out}_{\text{AS},scd}$.
Taking $N^t_{\text{AS},\text{seed}}$ by itself (\textit{dark blue}) resembles the above-linear functional dependence of $N^t_{\text{S},cd}$, observed in fig. \ref{fig_ch7_cntrl_power} \textbf{a}.

The double-logarithmic plot in fig. \ref{fig_ch7_cntrl_power} \textbf{d} illustrates the scaling exponents. 
Straight line fits on the unseeded noise $N^t_{\text{AS},cd}$ and $N^t_{\text{AS},c}$ are collinear for both {\cs} preparation methods, while $N^\text{out}_{\text{AS},scd}$ has 
a larger gradient in both time bins. 
The part from purely seeded noise $N^t_{\text{AS},\text{seed}}$ therein clearly increases much stronger with $\ln(E_c^\text{p})$, i.e.  $x > 1$.

Phenomenologically, these observations tie in with our assumptions based on the number of Raman transitions involved to generate each signal type.
Looking at the numbers for the fitted exponents $x$, stated in table \ref{tab_ch7_fit_exp_ctrl_power}, we find all values to lie below the expectations for pure Raman transitions. 
As mentioned above, other effects, such as control field induced Stark shifts, are likely to have caused this discrepancy.  For the read-out time bin data, decoherence within the $\tau_\text{S} = 312\ns$ storage also reduces the exponent slightly. 

\begin{table}[h!]
\centering
\begin{tabular}{ l | c | c | c | c | c | c }
\toprule
$x = (\ln(y)-\beta))/\ln(E^\text{p}_c )$ 	&  	Mem. eff. 	& S noise 	&	\multicolumn{4}{|c} {AS noise} \\
\hline
Time bin  	& $\eta_\text{in}$, $\eta_\text{mem}$ & $N^t_{\text{S},cd}$ & $N^t_{\text{AS},cd}$ & $N^t_{\text{AS},c}$ & $N^t_{\text{AS},scd}$ & $N^t_{\text{AS},\text{seed}}$ \\ 
\midrule
Input bin						&  0.66 (1)	& 1.19 (2)&  0.84 (1) &  0.74 (1) & 0.93 ($\in \left[1,2\right]$) & 1.18 (2)\\
Output ($\tau_\text{S} =312\ns$) 	& 1.34  (2)& 1.41 (2)&  0.92 (1) & 0.75 (1)  & 1.09 ($\in \left[1,2\right]$) & 1.27 (2)\\
\bottomrule	
\end{tabular}
\caption{Scaling exponents ($x$) of memory efficiency and noise level dependence on control pulse energy ($\sim (E_c^\text{p})^x$) obtained by straight line fits on double-logarithmic data shown in fig. \ref{fig_ch7_cntrl_power} \textbf{b} \& \textbf{d}. Exponents expected for a decomposition of each process into a series of Raman transitions are denoted in brackets behind the fit value.}
\label{tab_ch7_fit_exp_ctrl_power}
\end{table}

Rather than looking at absolute values, we focus on the relations between the exponents of the different observables to find the following:  
\begin{itemize}
\item 
For processes relating to single Raman transitions the scaling is roughly reduced by factor $\textfrak{x} \sim  0.7$ from linearity in $E_c^\text{p}$. 
These are the observables {\etain}, $N^\text{in}_{\text{S},cd}$, $N^t_{\text{AS},cd}$ and $N^t_{\text{AS},c}$, whose fitted gradients lie in the interval ${x}_{1} \in \left[0.6, 0.8\right]$.
Their double transition counterparts {\etamem} and  $N^\text{out}_{\text{S},cd}$ have exponents ${x}_{1} \in \left[1.3,1.4\right]$, showing a reduction of $2\cdot \textfrak{x}$ with respect to the expectation of $2$. 
This is in line with the assumptions made for an increasing number of successive Raman transition.
\item
We obtain the expected similarity $x(\eta_\text{in}) \approx x(N^\text{in}_{\text{AS},c}) \approx \frac{1}{2} \cdot x(\eta_\text{mem})$. 
Since the processes for both observable {\etain} and $N^\text{in}_{\text{AS},c}$ do not involve any spin-wave coherences, they can only be based on a single Raman transition. Thus they should scale equally and with half the exponent than memory read-out.
\item S noise emission, which in FWM can only occur after prior AS emission, shows the expected exponents of a double transition process. This means, it scales similar to {\etamem}, with $x(N^\text{in}_{\text{S},cd}) \approx x(N^\text{out}_{\text{S},cd})\approx x(\eta_\text{mem})$. 
\item
Similarly, seeded AS noise must involve at least 2 Raman transitions, where the $1^\text{st}$ leg is the stimulated FWM spin-wave read-out and the $2^\text{nd}$ leg is subsequent AS scattering. 
We find the anticipated similarity for its exponents $x(N^\text{in}_{\text{AS},\text{seed}}) \approx x(N^\text{out}_{\text{AS},\text{seed}})$ with those of S noise and memory efficiency: $x(N^t_{\text{AS},\text{seed}}) \approx x(N^t_{\text{S},cd}) \approx x(\eta_\text{mem})$. 
\item
Lastly, we get $x(N^\text{in}_{\text{AS},c}) < x(N^\text{in}_{\text{AS},cd}) < x(N^\text{in}_{\text{AS},scd})$. 
We already know that $x(N^\text{in}_{\text{AS},scd})$ involves multiple FWM transitions within the control pulse, i.e., AS scattering produces a FWM spin-wave that is immediately read-out by S scattering, leading to another round of AS scattering. If this was not the case, the noise level $N^\text{in}_{\text{AS},scd}$ would not exceed $N^\text{in}_{\text{AS},cd}$ in figs. \ref{fig_ch7_flourescence}, \ref{fig_ch7_ASseeding} and \ref{fig_ch7_cntrl_power}. 
Multiple FWM cycles within the same control pulse consequently increase $x$.  
Having $x(N^\text{in}_{\text{AS},cd}) > x(N^\text{in}_{\text{AS},c})$ thus indicates, that even without seeding multiple FWM cycles occur within the same control pulse.
\end{itemize}
The relative scalings with control pulse energy between the different noise processes and the Raman memory behave as expected from our analysis.

\subsection{Detuning \label{ch7_subsec_detuning}}
The next experimental parameter is quite an easy one to modify, namely the detuning $\Delta$ of the Raman resonance from the {\cs} excited state $6^2 \text{P}_\frac{3}{2}$. 
Once more, we start by outlining the measurement methodology, followed by an interpretation of the results. 

\paragraph{Measurement}
Running the memory with different detunings requires realignment of the signal filter stage for every datapoint. 
Stage alignment is conducted with bright coherent states, sending the full {\tisa} pulse energy into the EOM (see fig. \ref{fig_6_setup}).
When turning down the signal intensity to the single photon level afterwards, the signal input photon numbers $N_\text{in}$ are not reproduced between the measurements for different detunings. 
Even for constant memory efficiency ({\etain}, {\etamem}) and noise background $N_{\text{S},cd}^t$,  
different values for $N_\text{in}$ would change the SNR.  
To benchmark the performance, we thus resort to using the ratio of memory efficiency over noise 
$R^t = \frac{\eta_\text{in/mem}}{N_{\text{S},cd}^t}$ 
instead, which is insensitive to $N_\text{in}$. 
We run this experiment with the control pulse sequence for $\tau_\text{S} = 12.5\ns$ storage time (fig. \ref{fig_ch7_flourescence}), using only the $1^\text{st}$ \pockels\, window 
to select a train of $9$ consecutive {\tisa} pulses. 

Notably, this measurement predates the addition of aluminium foil sheets to the {\cs} cell's thermal insulation, which at the time was also lacking end caps on the insulation tube (see fig. \ref{fig_ch4_int_instab_setup}). 
The memory efficiency was therefore still drifting during the measurement time, which was particularly noticeably 
when recording the $\Delta = 15\GHz$ datapoint during the first hour of operating the memory. 
This systematic insufficiency results in an artificially elevated value for {\etamem} at $\Delta = 15\GHz$.
The efficiency drift is a result of cold spots reducing the {\cs} density (see section \ref{ch7_Temp}), which also influences the noise level $N^t_{\text{S},cd}$, leaving $R^t$ approximately unaffected.

Moreover, this experiment was conducted at a time, when the 
EOM switching was still set to active high. 
In contrast to the active low switching (see section \ref{ch6_setup_electronics}), residual signal intensity leakage is present in the $2^\text{nd}$ time bin after $\tau_\text{S} = 12.5\ns$. 
Usage of the $\tau_\text{S} =12.5\ns$ control pulse sequence thus causes the presence of a small amount of input signal in the first read-out time bin\footnote{
	This is an effect of the finite rise time of the rf-switch (fig. \ref{fig_6_setup}), 
	whose falling flank suffers from exponential capacitor discharge
	behaviour. If used in the active high configuration, the residual decay voltage still allows frequency 
	modulation of some small fraction of the signal 	after $12.5\ns$, which causes signal leakage 
	in the subsequent {\tisa} pulse train time bin. When reversing the polarity, the attenuation in the
	subsequent {\tisa} pulse train time bin is substantially improved and the amount of 
	signal leakage is negligible. 
}. 
The presence of this leakage signal results in a mixture between memory read-in of this pulse and retrieval of the signal read in $12.5\ns$ beforehand, which, in turn, reduces {\etamem} in this time bin. 
We thus focus on the retrieved signal in the $2^\text{nd}$ read-out bin after $\tau_\text{S} = 25\ns$ instead.
To obtain sufficient memory read-out signal $\eta_\text{mem}^\text{out2}$, we slightly loosen the control focussing, maximising $\eta_\text{mem}^\text{out2}$.
As we will see in section \ref{ch7_control_focussing}, changing the control focussing influences {\etamem}. A looser control focus reduces the retrieval efficiency $\eta_\text{ret}^\text{out1}$ in the $1^\text{st}$ read-out time bin.
Over the control pulse train, a trade-off in the control focussing can be found that maximises {\etamem} in the $2^\text{nd}$ read-out bin\footnote{
	The trade-off is easily understood by considering the increased amount of spin-wave left in the memory upon 
	reduction of $\eta_\text{ret}^\text{out1}$, that can subsequently be retrieved by $\eta_\text{ret}^\text{out2}$.
	This spin-wave increase counter-balances the reduction in Raman coupling for $\eta_\text{ret}^\text{out2}$, 
	leading to a focus regime where $\eta_\text{mem}^\text{out2}$ is maximised.
}.

\begin{figure}[h!]
\includegraphics[width=\textwidth]{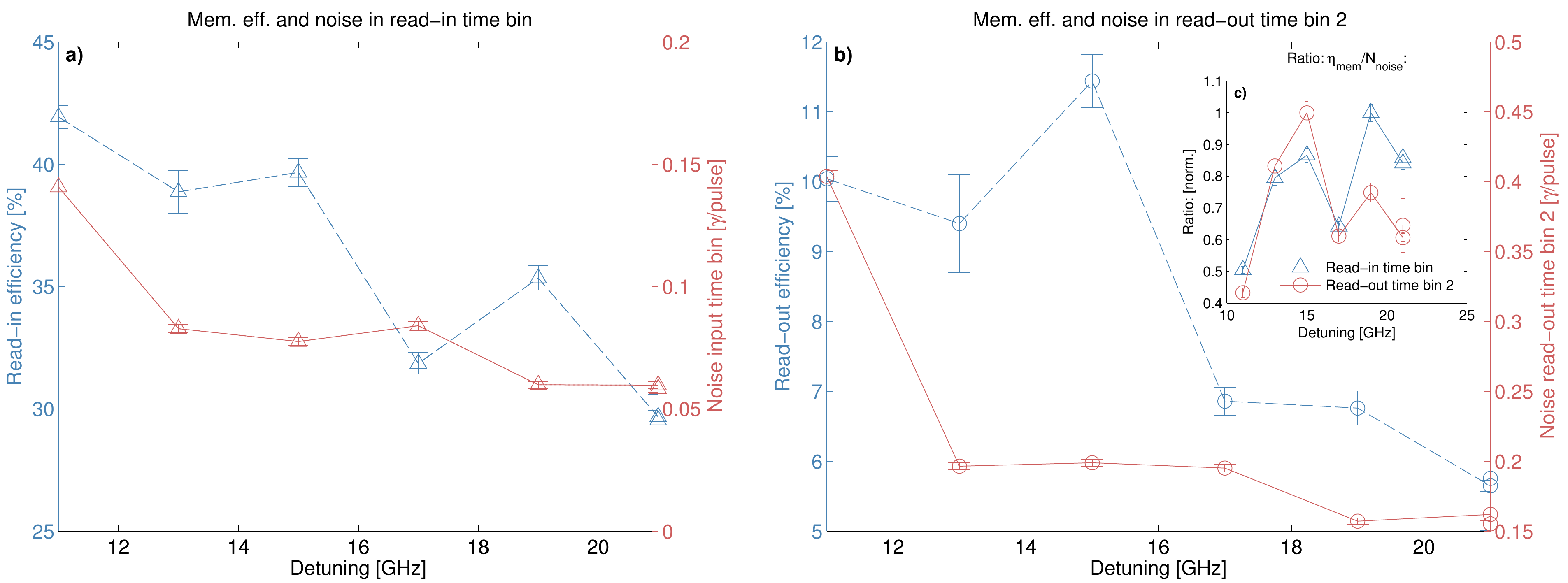}
\caption{Memory efficiency and noise level as a function of detuning: 
\textbf{(a)}: Memory read-in efficiency $\eta_\text{in}$ (\textit{blue}) and noise level in read-in time bin $N_\text{noise}^\text{in}$. 
\textbf{(b)}: Memory read-out efficiency in $2^\text{nd}$ output bin ($\tau_\text{S} = 25\ns$) $\eta_\text{mem}^\text{out2}$ (\textit{blue}) and noise level $N_\text{noise}^\text{out2}$.
\textbf{(c)}: Memory efficiencies normalised to the noise level in the respective time-bin,with 
$R^\text{in}= \frac{\eta_\text{in}}{N_\text{noise}^\text{in}}$ in \textit{blue} and 
${R^\text{out2}=\frac{\eta_\text{mem}^\text{out2}}{N_\text{noise}^\text{out2}}}$ in \textit{red}.}
\label{fig_ch7_detuning_scan}
\end{figure}

\paragraph{Results}
 Figure \ref{fig_ch7_detuning_scan} \textbf{a} and \textbf{b} show the observed values for \etain, 
 $N^\text{in}_{\text{S},cd}$ in the input time bin, and \etamem, $N^\text{out}_{\text{S},cd}$ in the $\tau_\text{S} =25\ns$ read-out time bin, respectively. 
Both, the Raman memory and the noise generation efficiency decrease as a function of detuning, which is the expected behaviour for a sequence of Raman processes\cite{Nunn:DPhil,Nunn:2007wj} (see appendix \ref{app6_coh_model}).

Here, the creation and annihilation operators for the signal mode $\hat{S}$ and the spin-wave $\hat{B}$ are coupled by the Raman coupling constant $C_\text{S} \sim \frac{1}{\Delta}$, with 
$\hat{S} \sim C_\text{S} \cdot \hat{B}$ and 
$\hat{B} \sim C_\text{S} \cdot \hat{S}$. 
In the read-in time bin, {\etain} is proportional to the number of photons stored from the signal field, which in turn is proportional to the number of spin-wave excitations $\hat{B}_\text{in}^\dagger$ generates. Hence $\eta_\text{in} \sim \langle \hat{B}_\text{in}^\dagger \hat{B}_\text{in} \rangle \sim C_\text{S}^2 \sim \frac{1}{\Delta^2}$.

The memory read-in efficiency is, in turn, determined by the number of signal photons retrieved from the spin-wave. We thus yield 
$\eta_\text{ret} \sim \langle \hat{S}_\text{out}^\dagger \hat{S}_\text{out} \rangle \sim C_\text{S}^2 \cdot \langle \hat{B}_\text{in}^\dagger \hat{B}_\text{in} \rangle \sim C_\text{S}^4 \sim \frac{1}{\Delta^4}$.
The same scaling results by considering the Raman memory interaction as a sequence of Raman transitions (see also section \ref{ch7_ctrl_power}). 
Each Raman transitions has an interaction cross section\cite{Penzkofer:1979, Raymer:1981aa} 
$\frac{\text{d} \sigma}{\text{d} \Omega} \sim \frac{1}{\Delta^2}$. 
Since {\etain} and {\etamem} effectively correspond to a sequence of a single and a double Raman transition, respectively, their success probabilities scale as $\eta_\text{in} \sim \frac{\text{d} \sigma}{\text{d} \Omega}$ and $\eta_\text{mem} \sim \left(\frac{\text{d} \sigma}{\text{d} \Omega} \right)^2$, leading to the same behaviour with $\Delta$ as obtained from our model. 

The S leg of the FWM noise should show the same proportionality.  
S noise emission is initiated by SRS in the AS channel, which itself scales as 
$N^t_{\text{AS},cd} \sim \frac{\text{d} \sigma}{\text{d} \Omega}$. 
This process is followed by retrieval of the generated spin-wave, which behaves similar to the memory signal, namely 
${\langle \hat{S}_\text{out}^\dagger \hat{S}_\text{out} \rangle \sim C_\text{S}^2 \cdot \langle \hat{B}_\text{in}^\dagger \hat{B}_\text{in} \rangle}$. 
Since the S noise generation probability is the product of the probabilities, one expects
$N^t_{\text{S},cd} \sim \frac{1}{\Delta^4}$. 

Besides FWM, the noise floor displayed in fig. \ref{fig_ch7_detuning_scan} \textbf{a} and \textbf{b} also includes the fluorescence noise fraction in the $5\ns$ pulse integration time windows. Towards smaller $\Delta$, this contribution starts to increase substantially over the levels discussed in section \ref{ch7_Fluorescence}. 
Its growth leads to the rapid increase in noise from $13\GHz$ to $11\GHz$ detuning and prevents experiments at any smaller $\Delta$ values by saturating the APD\footnote{
	Saturation is caused by the large amount of singles counts, which are detection events that do not 
	coincide with the \pockels\, trigger signal.  
	Once the singles rate exceeds the detector dead time of $\tau_\text{APD} \approx 60\ns$, 
	detection events are lost and the observable count rate saturates at 
	$c_\text{singles} = \frac{1}{\tau_\text{APD}}$. 
	The resulting blinding of the APD in turn also limits the detectable coincidences, which are
	used in this measurement. 
}. 
Experimentally, the fluorescence noise increase sets a lower bound on the detuning range. 
Towards large $\Delta$, the memory efficiency becomes too small. 
Here, the amount of retrieved signal approached the size of the Poissonian count rate fluctuations of the noise, 
making it difficult to measure and requiring long acquisition times for settings \scd\, and \cd\, to minimise uncertainty ranges. 
In practice, this sets an upper bound on the detuning. While for the $2^\text{nd}$ retrieval bin, 
the fraction of coincidence counts from the retrieved signal becomes too small at $\Delta = 21\GHz$ , switching the EOM polarity and using the $1^\text{st}$ read-out bin might allow investigation at larger $\Delta$.

Fig. \ref{fig_ch7_detuning_scan} \textbf{c} displays the ratio $R_t$. The data is normalised to the maximum value of $R^t$ for each trace $t \in \left\{ \text{in}, \text{out2}\right\}$.
Besides the fluorescence-based reduction at $\Delta = 11\GHz$, $R_\text{out2}$, which determines SNR$_\text{out}$, reduces for larger $\Delta$. So a smaller detuning appears beneficial. 
Contrary, $R_\text{in}$ does not show a trend.
$\Delta = 15\GHz$ displays good $R^t$-values in both time bins, for which reason it is chosen as the detuning for the measurements in chapter \ref{ch6} \& \ref{ch7}.

Notably the proportionality with $\Delta$ for memory efficiency and noise are only valid in the far off-resonance regime where our protocol is operating. 
For a narrowband Raman-type protocol close to resonance, e.g. GEM, the relative magnitude between the coupling strengths for Stokes ($C_\text{S}\sim \frac{1}{\Delta}$) and anti-Stokes ($C_\text{AS} \sim \frac{1}{\Delta_\text{AS}}= \frac{1}{\Delta + \delta \nu_\text{gs}}$) emissions are dominated by the ground state hyperfine splitting $\delta \nu_\text{gs}$, i.e., 
$\frac{C_\text{AS}}{C_\text{S}} = \frac{\Delta}{\Delta + \delta \nu_\text{gs}} = \frac{1}{1+\frac{\delta \nu_\text{gs}}{\Delta}} \overset{\Delta \rightarrow 0}{\longrightarrow} 0 $. 
FWM noise can be suppressed for these protocols, when tuning closer to resonance \cite{Hosseini:2011zr,Hosseini:2012}.
For our far-detuned protocol operation closer to resonance would, in principle, be possible by reducing the absorption linewidth in our cell.
The linewidth is determined by the large Neon (Ne) buffer gas pressure of $20$ Torr, whose reduction would enable to reduce $\Delta$ further. 
However, even for low buffer gas pressures, the Raman interaction still has to be operated in the adiabatic regime \cite{Nunn:DPhil}, where $\Delta \gg \Omega_\text{max}$. 
Consequently, $\Delta$ is also constrained by the peak Rabi-frequency $\Omega_\text{max}$, which is, in turn, set by the control pulse focussing parameters (see appendix \ref{ch7_control_focussing}). 
For our system, a detuning in the range of several GHz remains necessary.
It thus excludes the possibility to eliminate FWM by moving towards a smaller detuning, as it has been used in the demonstration of the GEM protocol\cite{Hosseini:2011zr,Hosseini:2012}.

\subsection{Temperature of the {\cs} vapour\label{ch7_Temp}}
Similar to the test performed in section \ref{ch7_subsec_cold_ensemble} on the AS channel for the unheated {\cs} cell, we can investigate the memory efficiency and noise level in the S channel as a function of {\cs} cell temperature. 
The cell heater is stabilised on the readings of a resistive thermal sensor, positioned at the middle of the cell and inserted between the heater belt and the outside of the glass tube (see fig. \ref{fig_ch4_int_instab_setup}). 
Because the heater belt cannot cover all parts of the cell, most notably the optical windows, parts of the cell have a lower temperature than the set-point of the heater circuit. Firstly this requires the introduction of an artificial cold-spot, mentioned in section~\ref{ch4_subsec_stability}, to avoid {\cs} condensation on the windows. Secondly, it prevents a thermal equilibrium to establish over the whole cell volume. 
Besides the possibility of introducing drift currents, the resulting temperature gradients reduce the atomic number density $n_\text{Cs}$ in the vapour. This happens because $n_\text{Cs}$ is proportional to the partial pressure $p_\text{Cs}$, which can be approximated by a Boltzman distribution\cite{Nunn:DPhil,Steck:2008qf} $p_\text{Cs} \sim \exp{\left( -1/(k_\text{B} T )\right)}$ above $25^\circ \text{C}$. 
Since the optical depth, and therewith the Raman coupling constants $C_\text{S}$ and $C_\text{AS}$, are proportional to $n_\text{Cs}$, the actual {\cs} vapour density will lie below the value expected from the heater's set-point. 
This opens two possibilities to influence the systems performance with temperature: 
On the one hand, the set-point of the heater can be modified to change the maximally achievable vapour density. In the following, we will investigate this degree of freedom first.
On the other hand, the cold spot temperature can be used to modify the average optical depth seen by signal and control. 
This will be covered in section \ref{ch7_Coldspot}.

\paragraph{Measurement}
The response of the S channel to the heater temperature was recorded alongside the detuning experiments in \ref{ch7_subsec_detuning}, i.e. the EOM was not yet operated with active low switching. 
The control sequence consisted of $9$ consecutive pulses for $\tau_\text{S} =12.5\ns$ storage time memory (see fig. \ref{fig_ch7_flourescence}) and \coh\, input signals of varying {\Nin} were used.
For these reasons, we again measure the memory efficiency in the $2^\text{nd}$ read-out bin after $\tau_\text{S} = 25\ns$ and refer to the metric $R^t = \frac{\eta_\text{in/mem}}{N_{\text{S},cd}^t}$ to test for 
memory operation with respect to noise. The experiment is run at $f_\text{rep} = 10.5 \kHz$ repetition rate and the noise background is measured for the spin-polarised (setting \cd\,) and the thermally distributed ensemble (setting 
\textit{c}).

\begin{figure}
\includegraphics[width=\textwidth]{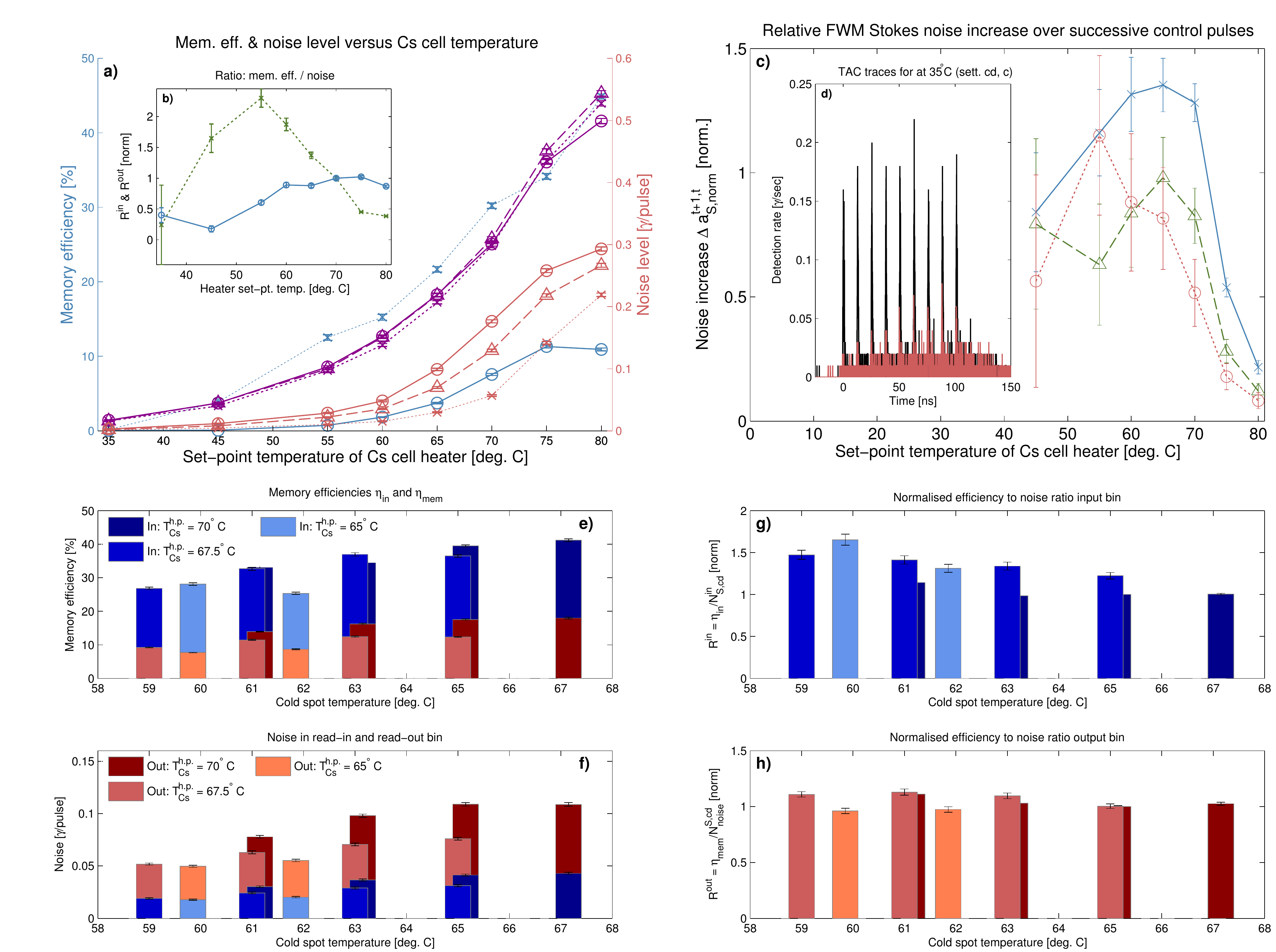}
\caption{Temperature dependence of memory efficiency and noise.
\textbf{(a)}: Memory efficiency (\textit{blue}), FWM S noise (\textit{red}), generated in the spin-polarised ensemble (setting \cd\,), and SRS into the S channel (\textit{magenta}) by thermally distributed {\cs} population (setting \textit{c}). Time bins are denoted by \textit{dotted lines} and $\times$ \textit{markers} for the read-in bin, \textit{dashed lines} and $\bigtriangleup$ \textit{markers} for the $1^\text{st}$ read-out bin, and \textit{solid lines} with $\circ$ \textit{markers} for the $2^\text{nd}$ read-out bin. 
\textbf{(b)} Memory efficiency to noise ratio for the read-in, $R^\text{in}$ (\textit{green}), and the read-out bin $R^\text{out}$ (\textit{blue}).
\textbf{(c)} Relative increases in FWM S noise over the control pulse train, showing $\Delta a^\text{p2,p1}_\text{S,norm}$ (\textit{blue}), $\Delta a^\text{p3,p2}_\text{S,norm}$ (\textit{green}) and $\Delta a^\text{p3,p4}_\text{S,norm}$ (\textit{red}), with superscript p assigning the pulse number along the train. 
\textbf{(d)} TAC count rate histograms for FWM (\textit{red}, setting \cd\,) and SRS (\textit{black}, setting \textit{c}) for a heater point temperature of $T_\text{Cs}^\text{h.p.} = 35^\circ\text{C}$. 
\textbf{(e)}: Memory read-in efficiency {\etain} (\textit{blue bars}) and read-out efficiency {\etamem} (\textit{red bars}) as a function of $T_\text{Cs}^\text{c.s.}$. 
\textbf{(f)}: FWM noise levels $N^\text{in}_{\text{S},cd}$ for the read-in (\textit{blue bars}) and $N^\text{out}_{\text{S},cd}$ for the read-out (\textit{red bars}) bin as a function of $T_\text{Cs}^\text{c.s.}$. 
\textbf{(g)} \& \textbf{(h)}: Memory efficiency to noise ratios 
$R^\text{in} = {\eta_\text{in}}/{N^\text{in}_{\text{S},cd}}$ (\textbf{g}) and 
$R^\text{out} = {\eta_\text{mem}}/{N^\text{out}_{\text{S},cd}}$ (\textbf{h}) as a function of $T_\text{Cs}^\text{h.p.}$. 
In panels \textbf{(e)} - \textbf{(f)} different heater point temperatures $T_\text{Cs}^\text{h.p.}$ are marked by the bar colour, as stated in the legends of panels \textbf{(e)} and \textbf{(f)}.
}
\label{fig_ch7_temperature}
\end{figure}

\paragraph{Results} 
As expected for decreasing vapour density, memory efficiency and noise reduce substantially and disappear when the heater set-temperature $T_\text{Cs}^\text{h.p.}$ approaches room-temperature.
Fig. \ref{fig_ch7_temperature} \textbf{a} shows the functional dependence on $T_\text{Cs}^\text{h.p.}$ for all three signal types, memory efficiency (\textit{blue lines}), $N^t_{\text{S},cd}$ (\textit{red lines}) and $N^t_{\text{S},c}$ (\textit{magenta lines}). 
All signals decay exponentially towards small values of $T_\text{Cs}^\text{h.p.}$. 
At hot temperatures they more (\etamem, $N^t_{\text{S},cd}$) or less (\etain, $N^t_{\text{S},c}$) display a saturation behaviour with the gradient tailing off. 

This ties in with the expected response, determined by the temperature dependence of the Raman coupling coefficients $C_\text{S,AS}$. 
As we have seen in appendix \ref{ch7_ctrl_power} already, our coherent model (chapter \ref{ch2}) predicts\cite{Nunn:2007wj} 
that count rates for single Raman transition processes ({\etain} and $N^\text{in}_{\text{S},c}$) are proportional to $C^2_\text{S}$. 
Memory read-out needs a $2^\text{nd}$ S channel transition, such that $\eta_\text{mem} \sim C^4_\text{S}$, while FWM involves one Raman transition in the S and the AS channel with $N^t_{\text{S},cd}\sim C^2_\text{S} \cdot C^2_\text{AS}$. 
The Raman coupling's dependence on the optical depth\footnote{
	The optical depth is defined\cite{Nunn:DPhil} as 
	$d = \frac{|\vec{d}_{i,e}^{*} \cdot \vec{e_\text{s}}|^2 \omega_\text{s} n_\text{Cs} L_\text{Cs}}
	{2\gamma \epsilon_0 \hbar c}$, where $\vec{d}_{i,e}$ is the dipole moment of the signal transition, 
	$\vec{e}_\text{s}$ is the polarisation vector of the signal electric field, $\omega_\text{s}$ is the signal
	frequency, $n_\text{Cs}$ is the atomic vapour density of {\cs}, $L_\text{Cs}$ is the vapour cell length, and
	$\gamma$ is the spontaneous emission rate of the excited state. 
} 
$C\sim \sqrt{d}$ introduces the sensitivity to temperature via the atomic vapour density $n_\text{Cs}$, which in turn is determined by the {\cs} vapour pressure $p_\text{Cs} \sim \exp{(-1/T_\text{Cs})}$. The ideal gas law yields 
$n_\text{Cs} = \frac{p_\text{Cs}}{k_\text{B} \cdot T_\text{Cs}} \sim \frac{\exp{\left(-\frac{1}{T_\text{Cs}}\right)}}{T_\text{Cs}}$, and, in turn, 
$C_\text{S,AS}^n \sim \frac{\exp{\left(-\frac{n}{2 T_\text{Cs}}\right)}}{ \left( T_\text{Cs} \right)^{ \frac{n}{2} }}$, 
which matches the $T_\text{Cs}^\text{h.p.}$ dependence of the data.

The equal scaling with $T_\text{Cs}^\text{h.p.}$ of signal and noise also leads to a roughly flat ratio\footnote{
	In the temperature range with significant vapour density, 
	$T^\text{h.p.}_\text{Cs} \in \left[ 60^\circ \text{C}, 80^\circ \text{C} \right]$, 
	$R^\text{in}$ decreases, which can be a result of fluctuations in {\etain} and 
	$N^\text{in}_{\text{S},cd}$ between measurements. 
	$R^\text{out}$ is however approximately independent of $T^\text{h.p.}_\text{Cs}$.
} 
$R^\text{out}$, illustrated in fig. \ref{fig_ch7_temperature} \textbf{b}. 
The reduction 
$\eta_\text{mem}(T_\text{Cs}^\text{h.p.}=80^\circ\text{C})<\eta_\text{mem}(T_\text{Cs}^\text{h.p.}=70^\circ \text{C})$, that can be seen in fig.~\ref{fig_ch7_temperature}~\textbf{a}, can also be a result of memory efficiency drift, caused by the cell insulation, which did not yet contain aluminium foil sheets when this data was recorded. 
In conclusion, experimentally no SNR improvement can be obtained with cell temperature changes. 

In fig. \ref{fig_ch7_flourescence} \textbf{a} we have seen the saturation of FWM noise over a train of successive control pulses. 
Since the same control pulse sequence is employed here, we can calculate the relative increase $\Delta a_\text{S}^{t+1,t}$ between the integrated areas $\tilde{a}_i^t$ of consecutive noise pulses, in the same manner as in section \ref{ch7_sec_fluor_results}. 
Normalising them to the area of the first pulse $\tilde{a}_\text{S}^\text{p1}$, $\Delta a_\text{S,norm}^{t+1,t} = \frac{\Delta a_\text{S}^{t+1,t}}{\tilde{a}_\text{S}^t}$, enables comparability between the different temperatures $T_\text{Cs}$. 
Therewith, the saturation speed, i.e. the convergence rate against the saturated FWM S noise level, can be tested. 
To this end $\Delta a_\text{S,norm}^{t+1,t}$ is plotted\footnote{
	These are the relative increases between the $1^\text{st}$ and the $2^\text{nd}$, 
	the $2^\text{nd}$ and the $3^\text{rd}$, as well as the $3^\text{rd}$ and the $4^\text{th}$ pulse 
	in the FWM noise pulse train of fig. \ref{fig_ch7_flourescence} \textbf{a}.
} 
for $t=1$ (\textit{blue}), $t=2$ (\textit{green}) and $t=3$ (\textit{red}) in fig.~\ref{fig_ch7_temperature}~\textbf{c}. 
All three traces show the largest increase between their respective pulses for the temperature range of $T_\text{Cs}^\text{h.p.} \in \left[ 60^\circ \text{C}, 70^\circ \text{C} \right]$ after which the increase reduces towards $0$. This means, the noise increase over successive control pulses is lower because the first pulse already contains a higher noise level. 
Noise elevation is expected since the larger Raman coupling coefficients $C_\text{S,AS}$ can give rise to multiple FWM cycles within the same control pulse. Additionally, due to its proportionality with both coupling constants 
${\left( \sim C_\text{AS}^2 \cdot C_\text{S}^2 \right)}$, S noise experiences a double boost by a larger density. 
This observation is interesting because, with proper calibration, the convergence rate of $N^t_{\text{S},cd}$, as a function of $t$, could be used as an indirect estimate for the optical depth, which is challenging to measure directly\cite{Sprague:2013}.

\paragraph{Estimation of control leakage}
Using the $T_\text{Cs}^\text{h.p.} = 35^\circ\text{C}$ value in the cold temperature region of our scan offers us another possibility to estimate the residual amount of control field leakage. 
Here, contrary to the experiments on the AS channel in section \ref{ch7_subsec_cold_ensemble}, FWM noise is the weaker process compared to SRS from the unprepared ensemble. 
The amount of FWM that still occurs can be estimated from the TAC count rate histograms, which are shown for settings \cd\, (\textit{red}) and \textit{c} (\textit{black}) in fig. \ref{fig_ch7_temperature} \textbf{d} for a total TAC integration time of $\Delta t_\text{meas} = 100\,\text{s}$. 
While SRS (setting \textit{c}) still results in S noise emission for all $9$ control pulses, the FWM S noise is only visible after saturation for control pulses $4-9$. 
We can therefore use $N^\text{p1}_{\text{S},cd}$, corresponding to the integrated area of the \cd\, noise in $1^\text{st}$ control pulse time bin, to directly establish a bound on control field leakage. 
The resulting value 
${ N_\text{leak}= N^\text{p1}_{\text{S},cd} = (1.3 \pm 0.2) \cdot 10^{(-3)}\ppp }$ 
is about half of the number estimated from the AS channel noise in section \ref{ch7_subsec_cold_ensemble}. 
Since the setup for control filtering, as well as the control beam path are the same here as they are for the \gtwo\, measurements, the amount of control field leakage, contributing to the total memory noise floor in the $12.5\,\text{ns}$ read-out bin of $N_\text{noise}^\text{out} = 0.15 \pm 0.05 \ppp$ (eq. \ref{ch6_eq_NoiseFloor}), can be estimated to $0.8 \pm 0.3\,\%$, which is negligible compared to fluorescence and FWM.

 \subsection{Cold-spot influence\label{ch7_Coldspot}}
The interest in investigating different cold spot temperatures lies in the prevention of {\cs} condensation elsewhere in the cell, particularly on the optical windows. 
While the reduction in optical depth from {\cs} condensation somewhere along the cell can, in principle, be balanced by increasing $T_\text{Cs}^\text{h.p.}$, {\cs} depositions on the optical windows ultimately terminates experiments, as it acts like a mirror for the input optical fields. 
For small temperature differences between $T_\text{Cs}^\text{h.p.}$ and cold spots in the cell, deposition rates are low, leading to a gradual decrease in memory efficiency on time-scale of several hours. 
The resulting experimental instability can be troublesome for measurements\footnote{
	Since the SNR is approximately constant as a function of temperature (see fig. \ref{ch7_Temp} \textbf{b}),
	such drifts do not adversely affect the \gtwo\, measurements, which rely on experimental
	stability due to long integration times.
}, as discussed in sections \ref{ch7_subsec_detuning} and \ref{ch7_Temp}. 
One thus needs to find the region of cold-spot temperatures, where the loss in memory efficiency from reducing $n_\text{Cs}$ is balanced by the gain in stability.

\paragraph{Measurement} 
Temperatures are determined along the {\cs} cell body by placing additional temperature sensors at both optical surfaces, the cell's entrance ($T_\text{Cs}^\text{in}$) and exit ($T_\text{Cs}^\text{out}$) windows, at the introduced cold spot\footnote{
	As shown in fig. \ref{fig_ch4_int_instab_setup} \textbf{b}, the cold spot, generated at the sealing nozzle 
	of the {\cs} cell, is positioned half-way along the cell's cylindrical axis. 
} ($T^\text{c.s.}_\text{Cs}$) and half-way between the cell centre and one of the windows ($T^\text{1/2}_\text{Cs}$). 
With these in place, the memory performance in terms of efficiency and noise level are investigated for different combinations of heater point ($T^\text{h.p.}_\text{Cs}$) and cold spot temperatures. 
$T^\text{h.p.}_\text{Cs}$ is changed by modifying the flow rate of pressured air, that blows onto the {\cs} cell sealing nozzle where the cold spot is created (see fig. \ref{fig_ch4_int_instab_setup} \textbf{b}).
Taking the readings of all temperature sensors yields the values stated in table \ref{ch7_tab_coldspot}. These are recorded after thermalisation of the system. 
Fig. \ref{fig_ch7_temperature} \textbf{e} - \textbf{f} shows all combinations by different columns for each $T^\text{c.s.}_\text{Cs}$ setting, where colour coding represents $T^\text{h.p.}_\text{Cs}$. 
The times $\Delta t_\text{meas}$ in table \ref{ch7_tab_coldspot} denote the intervals of the total measurement time, during which data has been taken for each temperature combination. 
To good approximation the temperatures $T^\text{in}_\text{Cs}$, $T^\text{out}_\text{Cs}$ and $T^\text{1/2}_\text{Cs}$ are stable. 
They obviously depend on the heater point setting $T^\text{h.p.}_\text{Cs}$, however they are independent of $T^\text{c.s.}_\text{Cs}$. 
Modifications of $T^\text{c.s.}_\text{Cs}$ can thus be expected to have a smaller influence on $n_\text{Cs}$ than changes in $T^\text{h.p.}_\text{Cs}$, enabling fine tuning of the memory efficiency to particular values. 
This might be of potential use for multiplexing applications of quantum memories\cite{Nunn:2013ab} and the generation of bespoke read-out pulse trains\cite{Reim2012}.

\begin{table}[h!]
\centering
\begin{tabular}{c|c|c|c|c|c}
\toprule
$T^\text{h.p.}_\text{Cs}$ [$^\circ\text{C}$] & $T^\text{c.s.}_\text{Cs}$ [$^\circ\text{C}$] & $T^\text{1/2}_\text{Cs}$ [$^\circ\text{C}$] & $T^\text{in}_\text{Cs}$ [$^\circ\text{C}$] & $T^\text{out}_\text{Cs}$ [$^\circ\text{C}$] & $\Delta t_\text{meas}$ [min.]\\
\midrule
70 		& 64.8 		& 69 		& 68.5	& 69 		& $0-22$ \\
70 		& 63 			& 69.2	& 68.7	& 68.4 	& $57-73$ \\
70 		& 61 			& 69.5	& 68.9	& 69.1 	& $101-119$ \\
70 		& 66.7		& 68.8	& 68		& 67.4 	& $139-156$ \\
70 		& 65.2		& 69		& 68.2	& 67.6 	& $172-179$ \\
\hline
67.5 		& 62.8		& 66.5	& 65.8	& 65.3 	& $206 - 216$ \\
67.5 		& 61			& 66.8	& 66.1	& 65.9 	& $219 - 236$ \\
67.5 		& 59			& 67		& 66.4	& 66.7 	& $249 - 272$ \\
67.5 		& 65			& 66.4	& 65.5	& 64.5 	& $287 - 304$ \\
\hline
65 		& 60			& 63		& 63.5	& 63 		& $337 - 348$ \\
65 		& 62			& 63.8	& 63		& 62.2 	& $357 - 372$ \\
\bottomrule
\end{tabular}
\caption{Temperatures measured along the {\cs} cell for different combinations between cell heater set-point $T^\text{h.p.}_\text{Cs}$ and cold spot $T^\text{c.s.}_\text{Cs}$ temperatures. $\Delta t_\text{meas}$ denotes the time period of the total measurement time used for measurements with the respective temperature pair $\left\{T^\text{h.p.}_\text{Cs},T^\text{c.s.}_\text{Cs} \right\}$.}
\label{ch7_tab_coldspot}
\end{table}

The measurements were performed after improving the cell insulation, with a storage time $\tau_\text{S} = 312\ns$, running experiments at $f_\text{rep} = 4\kHz$ repetition rate. Similar to the measurements in section  \ref{ch7_subsec_detuning} \& \ref{ch7_Temp}, the photon number of the {\cs} input signal was not kept constant during the measurements, for which reason memory performance with respect to noise is again evaluated by the efficiency-noise ratios $R^t$.   

\paragraph{Results}
From the experience gained in working with the system and taking the measurements presented in chapters \ref{ch6} \& \ref{ch7}, a cold spot temperature of $T^\text{c.s.}_\text{Cs} \approx T^\text{in/out}_\text{Cs} - 2 \, \text{K}$ has been found to result in sufficient experimental stability. 
At higher cold spot temperatures, {\cs} starts to condensate on the windows. 
This can, for instance, be observed visually; it looks like an oil-film, which makes the cell appear cloudy when looking through it. 
Thanks to the relatively short measurement time $\Delta t_\text{meas}$ (see table \ref{ch7_tab_coldspot}), required to record data for the cold spot scan, the measurements at high $T^\text{c.s.}_\text{Cs}$ values, shown in fig. \ref{fig_ch7_temperature} \textbf{e} \& \textbf{f}, are not affected by {\cs} condensation\footnote{
	Note, as mentioned earlier, {\cs} deposition on the glass walls could, in principle, also be avoided 
	using paraffin coated spectroscopy cells\cite{Klein:2006,Balabas:2010}. 
}. 
Hence the levels for both observables, memory efficiency and noise, increase when $T^\text{c.s.}_\text{Cs} \rightarrow T^\text{h.p.}_\text{Cs}$. 

In terms of the memory efficiency, fig. \ref{fig_ch7_temperature} \textbf{e} shows that both, {\etain} (\textit{blue bars}) and {\etamem} (\textit{red bars}), can be tuned on the $\%$-level by modification of $T^\text{c.s.}_\text{Cs}$ on the Kelvin-level. 
Comparing the dynamic range, achieved by modification of $T^\text{c.s.}_\text{Cs}$, with the effect size of changes in $T^\text{h.p.}_\text{Cs}$, we estimate a change of $\Delta T^\text{c.s.}_\text{Cs} \approx 6\,\text{K}$ to influence the {\cs} density $n_\text{Cs}$ by as much as a $\Delta T^\text{h.p.}_\text{Cs} \approx 2.5 \,\text{K}$ change of the heater temperature. 
This demonstrates the ability to fine tune the memory operation by the cold spot temperature $T^\text{c.s.}_\text{Cs}$.

For the temperature setting $T^\text{h.p.}_\text{Cs} = 70^\circ \text{C}$, the empirically found requirement for 
$T^\text{c.s.}_\text{Cs} \approx T^\text{in/out.}_\text{Cs} - 2\,\text{K} $ necessitates that 
${T^\text{c.s.}_\text{Cs}  \le 65^\circ \text{C}}$. 
This has very little effect on \etamem, reducing it from $\sim 18\, \%$ to $17.5\%$.
Similarly, also the noise level $N^\text{out}_{\text{S},cd}$ remains essentially constant at 
$N_{\text{S},cd}^\text{out1} \approx 10.9 \cdot 10^{-2} \ppp$. 
Fortunately, preventing {\cs} condensation does not come at any significant cost in terms of memory performance. 

On the flip-side, likewise to $T_\text{Cs}^\text{h.p.}$, $T^\text{c.s.}_\text{Cs}$ also does not suffice as a parameter to significantly change the SNR for the memory output. 
In the input bin small improvements can be obtained, as shown by the values for $R^\text{in}$ in fig. \ref{fig_ch7_temperature} \textbf{g}. 
Yet, similar to $R^\text{out}(T^\text{h.p.}_\text{Cs})$, $R^\text{out}(T^\text{c.s.}_\text{Cs})$ is approximately constant with $T^\text{c.s.}_\text{Cs}$. Fig. \ref{fig_ch7_temperature} \textbf{h} displays 
$R^\text{out}(T^\text{h.p.}_\text{Cs},T^\text{c.s.}_\text{Cs})$ for both degrees of freedom, which demonstrates that there is little change in the values (i.e. bar heights) for the different combinations of $T^\text{h.p.}_\text{Cs}$ and $T^\text{c.s.}_\text{Cs}$.

\paragraph{Conclusion}
Temperature tuning is important to achieve long-term stability in memory operation. Moreover, it is also a useful parameter to achieve specific efficiency levels. But, just like the detuning, it is not suitable for improving the performance with respect to noise. 
In the end, there is only one set of experimental parameters for which we can observe any significant change of the output SNR. 
These are the spatial modes, particularly the focussing parameters of the control field with respect to the signal field, whose influence is subject of the next section \ref{ch7_control_focussing}.

\subsection{Control focussing \label{ch7_control_focussing}}

The last experimental parameter set to look at are the transverse beam sizes for signal and control in the memory. 
In the following, we analyse their influence on the memory efficiency and the noise level.
Note, we use the full control pulse energy to do this. So, when changing the control mode size, we implicitly change the electric field amplitude of the control pulses. In terms of a fundamental parameter scan, this somehow mixes two effects, namely the control pulse energy and mode size dependence. 
However, its benefit lies in finding the most appropriate experimental layout for the memory system, when one has already decided to use the full control pulse power, which is the situation we are facing. 
As we have seen in table \ref{tab_ch7_fit_exp_ctrl_power}, the scaling for the memory efficiency and the noise with control pulse energy are approximately the same. Accordingly, we do not achieve an optimum when changing this parameter. It thus makes sense, in our case, to run the experiment at the highest control pulse energy we have available, as this ultimately reduces the measurement times and relaxes the requirements on the apparatus stability. 

\paragraph{Changing the control beam waist} 
During the experiment, only the control's spatial mode properties are changes, while those of the input signal and the diode laser are kept constant. 
As illustrated in fig. \ref{fig_6_setup}, the collimated signal and control beams are overlapped on a PBS and focussed into the {\cs} cell by the same $f=500\mm$ focal length lens. 
To change the control mode, we expand its beam diameter by collimating it with different telescopes beforehand. 
Four different configurations are investigated. 
The telescopes used in preparation of all three optical fields (signal, control and diode) are listed in table \ref{ch7_beams} alongside the corresponding beam waists and confocal parameters. 
The beam parameters for signal and diode are constant for all measurements. 
The number of control mode sizes is constrained by the space on the optical table and the availability of lenses with appropriate focal lengths. 
Amongst the control telescope configurations, the lens set with $f_1 = 175\mm$ and $f_2 = 150\mm$ is used in all other experiments (chapters \ref{ch6} and \ref{ch7}).

\begin{table}
\centering
\begin{tabular}{c | c | c | c | c | c}
\toprule
Field 	& \multicolumn{2}{|c|} {Telescope} 		& Waist $w_0$ [$\mum$] 	& FWM at $w_0$ [$\mum$] & Conf. par. [cm] \\
\midrule
Signal 	&  	& 		& 81 					& 95 					& 4.78	\\
\hline
Diode 	&	&	& 276				& 325				& 56		\\
\hline
Control 1	& $f_1 = 175\mm$ 	& $f_2 = 150\mm$ 	& 202				& 238 				& 30.1 	\\
Control 2	& $f_1 = 100\mm$ 	& $f_2 = 150\mm$ 	& 119				& 141				& 10.5 	\\
Control 3	& $f_1 = 250\mm$ 	& $f_2 = 150\mm$ 	& 245				& 288				& 44.3 	\\
Control 4	& None			& 	None 		& 140 				& 165 				& 14.4 	\\
\bottomrule
\end{tabular}
\caption{Table with beam parameters for signal, diode and control. The different control beams are produced by changing the telescope used to collimate the control before overlapping it with the signal at the memory input. Here $f_1$ and $f_2$ are the $1^\text{st}$ and $2^\text{nd}$ lens trespassed by the control, respectively.}
\label{ch7_beams}
\end{table}

\begin{figure}
\includegraphics[width=\textwidth]{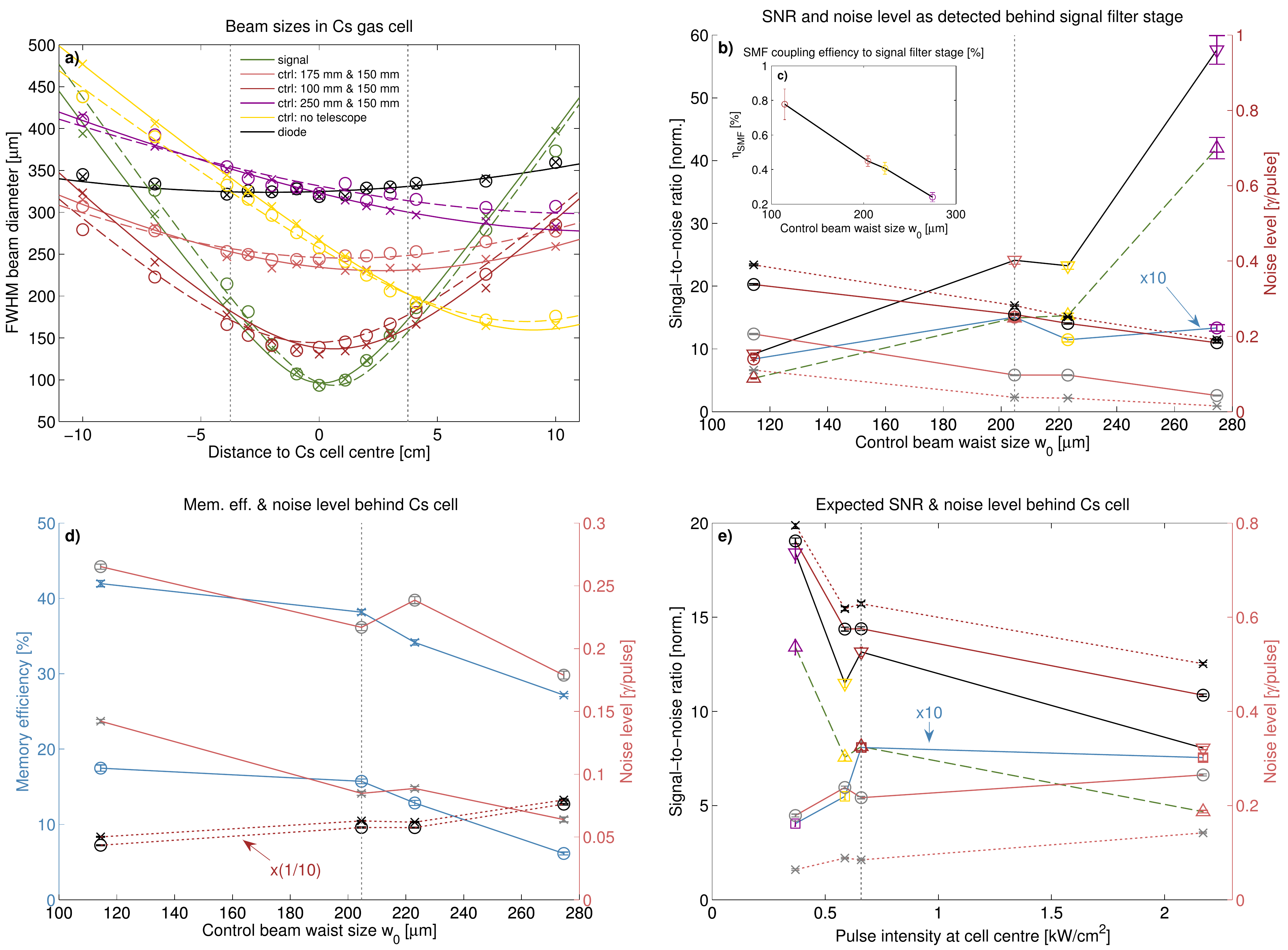}
\caption{Influence of control focussing on the memory efficiency and the noise level. 
\textbf{(a)}: Mode sizes along the {\cs} cell for the input signal (\textit{green}), the optical pumping beam (\textit{black}) and the four control telescopes defined in table \ref{ch7_beams} (telescope 1: \textit{dark red}; telescope 2: \textit{light red}; telescope 3: \textit{magenta}; no telescope: \textit{gold}). \textit{Lines} are fits for a Gaussian beam waist with $w(z) = w_0\sqrt{1+\left(\frac{\lambda z}{\pi w_0}\right)^2}$; \textit{solid lines} and $\times$ mark the horizontal, \textit{dashed lines} and $\circ$ mark the vertical direction. \textit{Vertical lines} indicate the {\cs} cell ends.
\textbf{(b)}: SNR and noise level for the 4 telescopes determined by the counts detected on APD \spcmdh\,. Colour coding: SNR$_\text{in}$ in \textit{black}, SNR$_\text{trans}$ \textit{green}, SNR$_\text{out}$ \textit{blue} with marker colours indicating the modes in panel \textbf{(a)}; $N_{\text{S},cd}^\text{in}$ and $N_{\text{S},cd}^\text{out}$ in \textit{light red} with \textit{black markers}, with \textit{dashed lines} and $\times$ for the input and \textit{solid lines} and $\circ$ \textit{markers} for the output bin; $N_{\text{S},c}^t$ in \textit{dark red} with \textit{grey markers} and otherwise the same as for $N_{\text{S},cd}^t$.
\textbf{(c)}: SMF coupling efficiency $\eta_{\text{SMF,ctrl}}$ of the control for the 4 telescopes.
\textbf{(d)}: Memory efficiencies (\textit{blue}) for {\etain} ($\times$ \textit{markers}) and {\etamem} ($\circ$ \textit{markers}) and noise levels $N_{\text{S},k}^{\text{Cs},t}$, expected by backing-out $\eta_{\text{SMF,ctrl}}$. Colour coding for $N_{\text{S},k}^{\text{Cs},t}$ as in \textbf{(b)}.
\textbf{(e)}: Noise levels $N_{\text{S},k}^{\text{Cs},t}$ (same as in panel \textbf{(d)}) and SNRs, following from these, as a function of control intensity on axis at the cell centre. Colour coding as in \textbf{(a)} \& \textbf{(b)}. 
Waist sizes $w_0$ on the x-axes of panels \textbf{(b)} - \textbf{(d)} refer to the minimum waist at the focus  of the Gaussian beams.}
\label{fig_ch7_control_focussing}
\end{figure}

Fig. \ref{fig_ch7_control_focussing} \textbf{a} illustrates the actual Gaussian modes of all three beams, including the 4 control configurations, as they propagate through the {\cs} cell.
The signal's focus is positioned at the cell's centre, whereby its confocal parameter $b = \frac{2 \pi w_0^2}{\lambda} = 4.78 \cm$ approximately stretches over half\footnote{
	Despite the mismatch between $b$ and $L_\text{Cs}$, we have not seen a significantly 
	different memory efficiency than with a configuration where $b = L_\text{Cs}$.
} of the {\cs} cell length of $L_\text{Cs}= 7.5 \cm$. 
Signal field collection behind the cell is optimised for this mode, coupling approximately $\eta_\text{SMF,sig} \approx 86\,\%$ of the signal field into the SMF leading to the filter stage.
With a $b = 55\cm$ long confocal parameter, the diode laser is essentially collimated along the entire line of optics surrounding the {\cs} cell. 
Hence, the positioning of its focal point close to the cell exit face does not matter greatly. 
Control telescopes 1 and 2 also achieve focus locations at cell centre. For the larger beam diameters, produced with telescope 3 and without any telescope, the control was focussed in front of the cell. 
However, in all cases, the control mode is consistently larger than the signal mode over the length of the {\cs} cell. 
Consequently, the {\cs} cell volume covered by the signal is also always covered by the control and all of it can, in principle, contribute to the Raman memory. 
This excludes the possibility that any of the effects we see by changing the telescopes originate from an effective reduction of the mode overlap by uncovering parts of the signal's spatial mode.

\paragraph{Expected effects from changes in the control focussing} 
Because the energy per control pulse is fixed to $E_\text{c} \approx 8.6 - 8.9 \, \text{nJ}$ during the experiment, alterations of the control mode sizes change the control intensity $I_\text{c}(z,t)$.
In turn, these also modify the Rabi-frequency going into the Raman coupling constants $C_\text{S}$ and $C_\text{AS}$, as outlined in section \ref{ch2_Raman_noise}. 
Moreover, we have seen in appendix \ref{ch7_ctrl_power}, that both, the memory efficiency {\etamem} and the S noise level $N_{\text{S},cd}^t$ scale similarly with $I_\text{c}$. 
Therefrom, changes in the control focussing should influence both observables equally, leaving the SNR in the memory read-out approximately unaffected. 
This however ignores the spatial mode selection made by SMF coupling, which differs between the control mode configurations. 
The mode matching to the SMF defines the effective volume inside the {\cs} cell, whose signal and noise emission can be observed. 
This volume approximately coincides with the spatial mode of the signal field, because, firstly, SMF coupling is optimised for the signal. 
Secondly, the Raman memory spin-wave can only be excited at locations where signal and control are present simultaneously\footnote{
	Since the control field has a larger mode than the signal, the spatial mode of the 
	Raman spin-wave will be limited by the signal's spatial mode. Note, this ignores mode
	convolution between signal and control during the Raman process as well as atomic diffusion.
}; 
a restriction that does not hold for the FWM noise.  
The S noise emitted in the forward direction is instead generated within the volume covered by the control. Since this exceeds the SMF-coupled signal volume, 
an estimate on the fraction of the total noise volume coupled into the SMF, which approximates the detected fraction of the noise emission, can be obtained using the SMF-coupling efficiency of the control. 
It is measured by maximising the control transmission into the signal output port of the PBD behind the memory (see fig. \ref{fig_6_setup}). 
The four control telescope configurations in table \ref{ch7_beams} show efficiencies of $\eta_\text{SMF,ctrl}^\text{tele1} = 45 \pm 3 \,\%$, $\eta_\text{SMF,ctrl}^\text{tele2} = 78 \pm 1 \,\%$ $\eta_\text{SMF,ctrl}^\text{tele3} = 24 \pm 3 \,\%$ and $\eta_\text{SMF,ctrl}^\text{no tele} = 41 \pm 3 \,\% $. 
These are displayed as a function of control beam waist size in fig. \ref{fig_ch7_control_focussing} \textbf{c}.

With an approximately constant collection efficiency for the signal, the different values for the observable noise fraction can result in an SNR improvement by loosening the control focussing, which, in turn, reduces $\eta_\text{SMR,ctrl}$. 
Such improvements are counteracted by a simultaneous reduction in the control intensity, leading to a lower Raman coupling $C_\text{S}(I_\text{c})$ and memory efficiency {\etamem}. 
Hence there is a trade-off, which allows for some optimisation for the SNR. 

\paragraph{Measurement}
To find the optimal trade-off between both effects, we measure the memory efficiency and the noise level behind the signal filter stage for the 4 telescope configurations. 
We can firstly use the count rates detected on the signal APD \spcmdh\, (see fig. \ref{fig_6_setup}) to determine the noise levels $N_{\text{S},k}^t$ for the spin-polarised and the thermally distributed ensemble (settings $k \in \left\{cd, c \right\}$, time bin $t$), as well as the SNR. This includes any noise reduction from SMF mode selection. 
With the above stated coupling efficiencies for $\eta_\text{SMF,ctrl}^i$, we estimate the noise levels  $N_{\text{S},k}^{\text{Cs},t} = N_{\text{S},k}^t/\eta_\text{SMF,ctrl}^i$, expected behind the {\cs} cell. 
Calculating the SNR with $N_{\text{S},k}^{\text{Cs},t}$ assumes that all noise in the control mode is collected. 
Comparison with the results after SMF-coupling indicates where SMF mode selection improves memory performance. 

Notably, in the calculation of the signal input photon number {\Nin} (eq. \ref{ch6_eq_Nin}), the signal's SMF-coupling efficiency $\eta_\text{SMF,sig}$ is also backed out to arrive at a photon number as it would be observed without spatial mode selection in the memory output. The memory efficiency, which also enters the SNR (eq. \ref{eq_ch6_SNR}), is independent of $\eta_\text{SMF,sig/ctrl}$ (eq. \ref{eq6_memeff}).

Finally, this SNR and the noise levels $N_{\text{S},k}^{\text{Cs},t}$ are tested as a function of control intensity. 
Since the peak intensity on-axis and the beam waist are inversely proportional, 
$I_\text{c} = \frac{2 \cdot E_\text{c}}{\pi f_\text{rep} w(z)^2}$, 
with the control pulse energy $E_\text{c}$ and the beam waist $w(z)$, that is diffracting along the cell, intensities are determined using the mode size at the centre of the cell, i.e. at position $0$ in fig. \ref{fig_ch7_control_focussing} \textbf{a}.

The experiment is conducted with a control pulse sequence for $\tau_\text{S} = 312\ns$ storage time at $f_\text{rep} = 5.722 \kHz$, observing only the S channel. 
The input signal is set to $N_\text{in}^\text{tele1} =1.12\ppp$, $N_\text{in}^\text{tele2} =1.15\ppp$, $N_\text{in}^\text{tele3} =1.18\ppp$ and $N_\text{in}^\text{no tele} =1.02\ppp$. 
To avoid any influences from fluorescence noise, it is subtracted from the noise count rates for both, the spin-polarised (setting \cd\,) and the thermally distributed ensemble (setting \textit{c}). 
The fluorescence contribution is measured separately, using polarisation to turn-off the two-photon transition based processes in the {\cs} (see section \ref{ch7_sec_fluor_results}).

\paragraph{Results}

Fig. \ref{fig_ch7_control_focussing} \textbf{b} shows the observed noise levels and SNRs after SMF-coupling, i.e. behind the signal filter stage. The noise levels for both optical pumping configurations (\textit{black points} for setting \textit{c}, \textit{grey points} for setting \cd\,) are decreasing with larger control waist sizes $w_0$, as expected from the double reduction by lower peak intensities $I_\text{c}$ and lower SMF-coupling efficiencies $\eta_\text{SMF,ctrl}$. 
SNR$_\text{in}$ (\textit{black line}) and SNR$_\text{trans}$ (\textit{green line}) increase, while SNR$_\text{out}$ (\textit{blue line}) remains approximately constant. 
Importantly, it shows a maximum for the $2^\text{nd}$ smallest control diameter produced with telescope 1, containing the $f_1=175\mm$ focal length lens ($w_0^\text{tele1} = 202\mum$). 
The smaller control size for telescope 2 ($f_1=100\mm$, $w_0^\text{tele2} = 119\mum$), which matches the signal 
mode more closely (fig.~\ref{ch7_control_focussing}~\textbf{a}), results in a worse SNR$_\text{out}$. 
This results from two sources:
Inspecting the memory efficiencies {\etain} ($\times$ \textit{markers}) and {\etamem} ($\circ$ \textit{markers}), displayed in fig.~\ref{ch7_control_focussing}~\textbf{d} (\textit{blue lines}), these decrease with looser control foci, which ties in with the expectation for lower Rabi-frequencies\footnote{
	For an ideal theoretical scenario, the control's confocal parameter $b$ equals the cell length.
	In this case, it can be shown\cite{Nunn:DPhil} that the spatial mode dependence drops out of the 
	Raman coupling, which reduces to $C \sim \frac{L_\text{Cs}}{\pi w(z)^2} \sim \frac{1}{\lambda}$. 
	As table \ref{ch7_beams} shows, we however investigate $b \gg L_\text{Cs}$, so this approximation
	does not hold and $C \sim \frac{1}{\pi w(z)^2}$.
}. 
Note, for the larger beam waists the reduction in memory efficiency and noise level $N_{\text{S},cd}^{\text{Cs},t}$ can be reinforced by the misplacement of the control focus to a position in front of the {\cs} cell for telescopes 3 and 4. Operating the memory with only the diverging flank of the beam reduces the control intensity further over telescopes 1 and 2, whose foci are aligned with that of the signal. 
The important point to note is the flat behaviour in {\etamem} for small waist sizes. 
Control mode reduction 
${\left( w_0^\text{tele1} = 202\mum \rightarrow w_0^\text{tele2} = 119\mum \right)}$ 
does not significantly improve the memory efficiency. 
Thus, matching the control's Rayleigh range with that of the signal is not beneficial. Instead, it increases the SMF coupling efficiency of the FWM noise (fig.~\ref{ch7_control_focussing}~\textbf{c}) and actually reduces the SNR$_\text{out}$, as we have seen in fig.~\ref{ch7_control_focussing}~\textbf{b}.

Backing out $\eta_\text{SMF,ctrl}$, the expected noise levels without SMF mode selection, $N_{\text{S},k}^{\text{Cs},t}$, are also shown in fig. \ref{ch7_control_focussing} \textbf{d}. 
Larger control beam diameters still reduce $N_{\text{S},cd}^{\text{Cs},t}$, due to reduced control intensity and Raman coupling. For this reason, the decrease of $N_{\text{S},cd}^{\text{Cs},t}$ displays a similar gradient to that of the memory efficiency. 
Conversely, SRS actually increases with a lower control intensity $I_\text{c}$ for larger control diameters. This is counter-intuitive since it should also scale with the coupling constant $C_\text{S}$, following $N_{\text{S},c}^{\text{Cs},t}\sim C_\text{S}$. The reason for this discrepancy is currently still unclear. 

Fig. \ref{ch7_control_focussing} \textbf{e} shows the noise levels as a function of intensity $I_\text{c}$ (evaluated at the control focus). 
It also displays the SNRs, expected without SMF mode selection. As we can see, for telescopes 1 and 2, where the control focus location matches the signal's location at the cell centre, SNR$_\text{out}$ is approximately independent of the control focussing. 
While this matches the initial expectation from the equal scaling with their Rabi-frequency,
the comparison with fig. \ref{ch7_control_focussing} \textbf{b} demonstrates that the different SMF coupling efficiencies for the FWM noise reduce the measured SNR for the largest control intensity point (telescope 2), compared to the values for telescope 1, which is the configuration used for all the other experiments of chapters \ref{ch6} \& \ref{ch7}. 

\paragraph{Conclusion}
The memory performance can indeed be optimised with respect to noise by using the control focussing in combination with signal mode selection by a SMF. 
Potential improvements could still be gained with better control over the focussing locations and waist sizes, producing larger beams that focus at the {\cs} cell centre. 
However the improvements in SNR$_\text{out}$ are not sufficient to facilitate single photon storage under \gtwo\, preservation.

\backmatter

\begin{spacing}{1}
\bibliography{References}

\begin{thebibliography}{100}

\bibitem{Gisin:2002aa}
N.~Gisin, G.~Ribordy, W.~Tittel, and H.~Zbinden, ``Quantum cryptography,'' {\em
  Reviews of modern physics}, vol.~74, no.~1, pp.~145--195, 2002.

\bibitem{Bennett:1998}
C.~Bennet and P.~Shor, ``Quantum information theory,'' {\em IEEE Transactions
  on Information Theory}, vol.~44, no.~6, p.~2724, 1998.

\bibitem{Bennett:2000fk}
C.~H. Bennett and D.~P. DiVincenzo, ``Quantum information and computation,''
  {\em Nature}, vol.~404, pp.~247--255, 03 2000.

\bibitem{Nielsen:2004kl}
M.~A. Nielsen and I.~L. Chuang, {\em Quantum Computation and Quantum
  Information}.
\newblock Cambridge University Press, 2004.

\bibitem{Haffner:2008vn}
H.~H{\"a}ffner, C.~F. Roos, and R.~Blatt, ``Quantum computing with trapped
  ions,'' {\em Physics Reports}, vol.~469, pp.~155--203, 12 2008.

\bibitem{Monroe:2002ys}
C.~Monroe, ``Quantum information processing with atoms and photons,'' {\em
  Nature}, vol.~416, pp.~238--246, 03 2002.

\bibitem{Wallraff:2004}
M.~H. Devoret, A.~Wallraff, and J.~M. Martinis, ``Superconducting qubits: A
  short review,'' {\em arXiv:cond-mat/041117v1}, 2004.

\bibitem{Clarke:2008zr}
J.~Clarke and F.~K. Wilhelm, ``Superconducting quantum bits,'' {\em Nature},
  vol.~453, no.~7198, pp.~1031--1042, 2008.

\bibitem{Knill:2001nx}
E.~Knill, R.~Laflamme, and G.~J. Milburn, ``A scheme for efficient quantum
  computation with linear optics,'' {\em Nature}, vol.~409, no.~6816,
  pp.~46--52, 2001.

\bibitem{Raussendorf:2001}
R.~Raussendorf and H.~Briegel, ``A one-way quantum computer,'' {\em Phys. Rev.
  Lett.}, vol.~86, pp.~5188--5191, May 2001.

\bibitem{OBrien:2003aa}
J.~L. O'Brien, G.~J. Pryde, A.~G. White, T.~C. Ralph, and D.~Branning,
  ``Demonstration of an all-optical quantum controlled-not gate,'' {\em
  Nature}, vol.~426, no.~6964, pp.~264--267, 2003.

\bibitem{Bouwmeester:1997aa}
D.~Bouwmeester, J.-W. Pan, K.~Mattle, M.~Eibl, H.~Weinfurter, and A.~Zeilinger,
  ``Experimental quantum teleportation,'' {\em Nature}, vol.~390, no.~6660,
  pp.~575--579, 1997.

\bibitem{Duan:2001vn}
L.~Duan, M.~Lukin, J.~Cirac, and P.~Zoller, ``{Long-distance quantum
  communication with atomic ensembles and linear optics},'' {\em Nature},
  vol.~414, pp.~413--418, 2001.

\bibitem{Sanguard2011}
N.~Sangouard, C.~Simon, H.~de~Riedmatten, and N.~Gisin, ``Quantum repeaters
  based on atomic ensembles and linear optics,'' {\em Rev. Mod. Phys.},
  vol.~83, pp.~33--80, Mar 2011.

\bibitem{Ursin:2007vn}
R.~Ursin, F.~Tiefenbacher, T.~Schmitt-Manderbach, H.~Weier, T.~Scheidl,
  M.~Lindenthal, B.~Blauensteiner, T.~Jennewein, J.~Perdigues, P.~Trojek,
  B.~Omer, M.~Furst, M.~Meyenburg, J.~Rarity, Z.~Sodnik, C.~Barbieri,
  H.~Weinfurter, and A.~Zeilinger, ``Entanglement-based quantum communication
  over 144 km,'' {\em Nat Phys}, vol.~3, pp.~481--486, 07 2007.

\bibitem{Chen:2008fk}
Y.-A. Chen, S.~Chen, Z.-S. Yuan, B.~Zhao, C.-S. Chuu, J.~Schmiedmayer, and
  J.-W. Pan, ``Memory-built-in quantum teleportation with photonic and atomic
  qubits,'' {\em Nat Phys}, vol.~4, pp.~103--107, 02 2008.

\bibitem{Pan:1998}
J.-W. Pan, D.~Bouwmeester, H.~Weinfurter, and A.~Zeilinger, ``Experimental
  entanglement swapping: Entangling photons that never interacted,'' {\em Phys.
  Rev. Lett.}, vol.~80, pp.~3891--3894, May 1998.

\bibitem{Lvovsky:2009ve}
A.~Lvovsky, B.~Sanders, and W.~Tittel, ``{Optical quantum memory},'' {\em
  Nature Photonics}, vol.~3, no.~12, pp.~706--714, 2009.

\bibitem{Fleischhauer:2000vn}
M.~Fleischhauer and M.~D. Lukin, ``Dark-state polaritons in electromagnetically
  induced transparency,'' {\em Phys. Rev. Lett.}, vol.~84, pp.~5094--5097, May
  2000.

\bibitem{Fleischhauer:2002}
M.~Fleischhauer and M.~D. Lukin, ``Quantum memory for photons: Dark-state
  polaritons,'' {\em Phys. Rev. A}, vol.~65, p.~022314, Jan 2002.

\bibitem{Phillips:2001}
D.~F. Phillips, A.~Fleischhauer, A.~Mair, R.~L. Walsworth, and M.~D. Lukin,
  ``Storage of light in atomic vapor,'' {\em Phys. Rev. Lett.}, vol.~86,
  pp.~783--786, Jan 2001.

\bibitem{Specht:2011}
H.~P. Specht, C.~Nolleke, A.~Reiserer, M.~Uphoff, E.~Figueroa, S.~Ritter, and
  G.~Rempe, ``A single-atom quantum memory,'' {\em Nature}, vol.~473,
  pp.~190--193, 05 2011.

\bibitem{Sete:2015}
E.~A. Sete and H.~Eleuch, ``High-efficiency quantum state transfer and quantum
  memory using a mechanical oscillator,'' {\em ArXiv:1503.00211v2}, 2015.

\bibitem{Sirois:2015}
A.~J. Sirois, M.~A. Castellanos-Beltran, M.~P. DeFeo, L.~Ranzani, F.~Q. Lecocq,
  R.~W. Simmonds, J.~D. Teufel, and J.~Aumentado, ``Coherent-state storage and
  retrieval between superconducting cavities using parametric frequency
  conversion,'' {\em ArXiv:1503.00257v1}, 2015.

\bibitem{kuzmich}
D.~N. Matsukevich and A.~Kuzmich, ``Quantum state transfer between matter and
  light,'' {\em Science}, vol.~306, pp.~663--666, 2004.

\bibitem{Choi:2008aa}
K.~S. Choi, H.~Deng, J.~Laurat, and H.~Kimble, ``Mapping photonic entanglement
  into and out of a quantum memory,'' {\em Nature}, vol.~452, no.~7183,
  pp.~67--71, 2008.

\bibitem{Lettner:2011}
M.~Lettner, M.~M\"ucke, S.~Riedl, C.~Vo, C.~Hahn, S.~Baur, J.~Bochmann,
  S.~Ritter, S.~D\"urr, and G.~Rempe, ``Remote entanglement between a single
  atom and a bose-einstein condensate,'' {\em Phys. Rev. Lett.}, vol.~106,
  p.~210503, May 2011.

\bibitem{Tittel:2010}
W.~Tittel, M.~Afzelius, T.~Chaneli{\'e}re, R.~Cone, S.~Kr{\"o}ll, S.~Moiseev,
  and M.~Sellars, ``Photon-echo quantum memory in solid state systems,'' {\em
  Laser and Photonics Reviews}, vol.~4, no.~2, pp.~244--267, 2010.

\bibitem{Wrachtrup:2006cr}
J.~Wrachtrup and F.~Jelezko, ``{Processing quantum information in diamond},''
  {\em Journal of Physics-Condensed Matter}, vol.~18, no.~21, p.~807, 2006.

\bibitem{Sabooni2013}
M.~Sabooni, S.~T. Kometa, A.~Thuresson, S.~Kr{\"o}ll, and L.~Rippe,
  ``Cavity-enhanced storage---preparing for high-efficiency quantum memories,''
  {\em New Journal of Physics}, vol.~15, no.~3, p.~035025, 2013.

\bibitem{Longdell:2005aa}
J.~Longdell, E.~Fraval, M.~Sellars, and N.~Manson, ``Stopped light with storage
  times greater than one second using electromagnetically induced transparency
  in a solid,'' {\em Physical review letters}, vol.~95, no.~6, p.~063601, 2005.

\bibitem{Dudin:2013aa}
Y.~Dudin, L.~Li, and A.~Kuzmich, ``Light storage on the time scale of a
  minute,'' {\em Physical Review A}, vol.~87, no.~3, p.~031801, 2013.

\bibitem{Heinze:2013aa}
G.~Heinze, C.~Hubrich, and T.~Halfmann, ``Stopped light and image storage by
  electromagnetically induced transparency up to the regime of one minute,''
  {\em Physical review letters}, vol.~111, no.~3, p.~033601, 2013.

\bibitem{Reim:2011ys}
K.~Reim, P.~Michelberger, K.~Lee, J.~Nunn, N.~Langford, and I.~Walmsley,
  ``Single-photon-level quantum memory at room temperature,'' {\em Physical
  Review Letters}, vol.~107, no.~5, p.~53603, 2011.

\bibitem{Bao:2012aa}
X.-H. Bao, A.~Reingruber, P.~Dietrich, J.~Rui, A.~D{\"u}ck, T.~Strassel, L.~Li,
  N.-L. Liu, B.~Zhao, and J.-W. Pan, ``Efficient and long-lived quantum memory
  with cold atoms inside a ring cavity,'' {\em Nature Physics}, vol.~8, no.~7,
  pp.~517--521, 2012.

\bibitem{Nunn:2013ab}
J.~Nunn, N.~Langford, W.~Kolthammer, T.~Champion, M.~Sprague, P.~Michelberger,
  X.-M. Jin, D.~England, and I.~Walmsley, ``Enhancing multiphoton rates with
  quantum memories,'' {\em Physical review letters}, vol.~110, no.~13,
  p.~133601, 2013.

\bibitem{Nunn:DPhil}
J.~Nunn, ``Quantum memory in atomic ensembles,'' {\em DPhil thesis, University
  of Oxford}, 2008.

\bibitem{Gorshkov:2007}
A.~V. Gorshkov, A.~Andr\'e, M.~D. Lukin, and A.~S. S\o{}rensen, ``Photon
  storage in $\lambda$-type optically dense atomic media. i. cavity model,''
  {\em Phys. Rev. A}, vol.~76, p.~033804, Sep 2007.

\bibitem{Gorshkov:2007aa}
A.~V. Gorshkov, A.~Andr{\'e}, M.~D. Lukin, and A.~S. S{\o}rensen, ``Photon
  storage in $\lambda$-type optically dense atomic media. ii. free-space
  model,'' {\em Physical Review A}, vol.~76, no.~3, p.~033805, 2007.

\bibitem{Gorshkov:2007rw}
A.~V. Gorshkov, A.~Andre, M.~D. Lukin, and A.~S. Sorensen, ``Photon storage in
  $\lambda$-type optically dense atomic media. iii. effects of inhomogeneous
  broadening,'' {\em Physical Review A}, vol.~76, p.~033806, 2007.

\bibitem{Gorshkov:2008rz}
A.~V. Gorshkov, T.~Calarco, M.~D. Lukin, and A.~S. Sorensen, ``Photon storage
  in $\lambda$-type optically dense atomic media. iv. optimal control using
  gradient ascent,'' {\em Physical Review A (Atomic, Molecular, and Optical
  Physics)}, vol.~77, no.~4, p.~043806, 2008.

\bibitem{Novikova:2012}
I.~Novikova, R.~Walsworth, and Y.~Xiao, ``Electromagnetically induced
  transparency-based slow and stored light in warm atoms,'' {\em Laser and
  Photonics Reviews}, vol.~6, no.~3, pp.~333--353, 2012.

\bibitem{Nilsson:2005kl}
M.~Nilsson and S.~Kroell, ``Solid state quantum memory using complete
  absorption and re-emission of photons by tailored and externally controlled
  inhomogeneous absorption profiles,'' {\em Optics Communications}, vol.~247,
  no.~4-6, pp.~393--403, 2005.

\bibitem{Alexander:2006tg}
A.~Alexander, J.~Longdell, M.~Sellars, and N.~Manson, ``{Photon Echoes Produced
  by Switching Electric Fields},'' {\em Physical Review Letters}, vol.~96,
  no.~4, p.~43602, 2006.

\bibitem{Staudt:2007hc}
M.~U. Staudt, S.~R. Hastings-Simon, M.~Nilsson, M.~Afzelius, V.~Scarani,
  R.~Ricken, H.~Suche, W.~Sohler, W.~Tittel, and N.~Gisin, ``Fidelity of an
  optical memory based on stimulated photon echoes,'' {\em Physical Review
  Letters}, vol.~98, no.~11, p.~113601, 2007.

\bibitem{Hetet:2008dp}
G.~H{\'e}tet, J.~Longdell, A.~Alexander, P.~Lam, and M.~Sellars,
  ``{Electro-Optic Quantum Memory for Light Using Two-Level Atoms},'' {\em
  Physical Review Letters}, vol.~100, no.~2, p.~23601, 2008.

\bibitem{Hetet:2008kx}
G.~Hetet, J.~J. Longdell, M.~J. Sellars, P.~K. Lam, and B.~C. Buchler,
  ``Multimodal properties and dynamics of gradient echo quantum memory,'' {\em
  Physical Review Letters}, vol.~101, no.~20, p.~203601, 2008.

\bibitem{Hetet:2008jt}
G.~H\'{e}tet, M.~Hosseini, B.~M. Sparkes, D.~Oblak, P.~K. Lam, and B.~C.
  Buchler, ``Photon echoes generated by reversing magnetic field gradients in a
  rubidium vapor,'' {\em Opt. Lett.}, vol.~33, no.~20, pp.~2323--2325, 2008.

\bibitem{Hetet:2008kl}
G.~H{\'e}tet, J.~Longdell, M.~Sellars, P.~Lam, and B.~Buchler, ``{Bandwidth and
  Dynamics of the Gradient Echo Memory},'' {\em Arxiv preprint
  arXiv:0801.3860}, 2008.

\bibitem{Hosseini:2011aa}
M.~Hosseini, B.~Sparkes, G.~Campbell, P.~Lam, and B.~Buchler, ``High efficiency
  coherent optical memory with warm rubidium vapour,'' {\em Nature
  communications}, vol.~2, p.~174, 2011.

\bibitem{Nunn:2007wj}
J.~Nunn, I.~A. Walmsley, M.~G. Raymer, K.~Surmacz, F.~C. Waldermann, Z.~Wang,
  and D.~Jaksch, ``Mapping broadband single-photon wave packets into an atomic
  memory,'' {\em Physical Review A (Atomic, Molecular, and Optical Physics)},
  vol.~75, no.~1, p.~011401, 2007.

\bibitem{Reim2010}
K.~F. Reim, J.~Nunn, V.~O. Lorenz, B.~J. Sussman, K.~C. Lee, N.~K. Langford,
  D.~Jaksch, and I.~A. Walmsley, ``{Towards high-speed optical quantum
  memories},'' {\em Nature Photonics}, vol.~4, pp.~218--221, Mar. 2010.

\bibitem{England:2014}
D.~G. England, G.~Fisher, Kent~A.\, J.-P.~W. MacLean, P.~J. Bustard,
  R.~Lausten, K.~J. Resch, and B.~J. Sussman, ``Storage and retrieval of
  $\text{THz}$-bandwidth single photons using a room-temperature diamond
  quantum memory,'' {\em Phys. Rev. Lett.}, vol.~114, p.~053602, Feb 2015.

\bibitem{Weedbrook:2012}
C.~Weedbrook, S.~Pirandola, R.~Garcia-Patron, N.~J. Cerf, T.~C. Ralph, J.~H.
  Shapiro, and S.~Lloyd, ``Gaussian quantum information,'' {\em Rev. Mod.
  Phys.}, vol.~84, pp.~621--669, May 2012.

\bibitem{Kozhekin:2000bs}
A.~E. Kozhekin, K.~M{\o}lmer, and E.~Polzik, ``Quantum memory for light,'' {\em
  Phys. Rev. A}, vol.~62, p.~033809, 2000.

\bibitem{Hammerer:2010vn}
K.~Hammerer, A.~S{\o}rensen, and E.~Polzik, ``{Quantum interface between light
  and atomic ensembles},'' {\em Reviews of Modern Physics}, vol.~82, no.~2,
  p.~1041, 2010.

\bibitem{Julsgaard:2004lr}
B.~Julsgaard, J.~Sherson, J.~I. Cirac, J.~Fiurasek, and E.~S. Polzik,
  ``Experimental demonstration of quantum memory for light,'' {\em Nature},
  vol.~432, no.~7016, pp.~482--486, 2004.

\bibitem{Damon:2011}
V.~Damon, M.~Bonarota, A.~Louchet-Chauvet, T.~Chaneli{\`e}re, and J.-L.~L.
  Gou{\"e}t, ``Revival of silenced echo and quantum memory for light,'' {\em
  New Journal of Physics}, vol.~13, no.~9, p.~093031, 2011.

\bibitem{Afzelius:2009qf}
M.~Afzelius, C.~Simon, H.~De~Riedmatten, and N.~Gisin, ``{Multimode quantum
  memory based on atomic frequency combs},'' {\em Physical Review A}, vol.~79,
  no.~5, p.~52329, 2009.

\bibitem{Clausen:2011kx}
C.~Clausen, I.~Usmani, F.~Bussieres, N.~Sangouard, M.~Afzelius,
  H.~de~Riedmatten, and N.~Gisin, ``Quantum storage of photonic entanglement in
  a crystal,'' {\em Nature}, vol.~469, pp.~508--511, 01 2011.

\bibitem{Rielander:2014aa}
D.~Riel{\"a}nder, K.~Kutluer, P.~M. Ledingham, M.~G{\"u}ndo{\u{g}}an,
  J.~Fekete, M.~Mazzera, and H.~de~Riedmatten, ``Quantum storage of heralded
  single photons in a praseodymium-doped crystal,'' {\em Physical Review
  Letters}, vol.~112, no.~4, p.~040504, 2014.

\bibitem{Bussieres:2014aa}
F.~Bussieres, C.~Clausen, A.~Tiranov, B.~Korzh, V.~B. Verma, S.~W. Nam,
  F.~Marsili, A.~Ferrier, P.~Goldner, H.~Herrmann, {\em et~al.}, ``Quantum
  teleportation from a telecom-wavelength photon to a solid-state quantum
  memory,'' {\em arXiv preprint arXiv:1401.6958}, 2014.

\bibitem{Sinclair:2014}
N.~Sinclair, E.~Saglamyurek, H.~Mallahzadeh, J.~A. Slater, M.~George,
  R.~Ricken, M.~P. Hedges, D.~Oblak, C.~Simon, W.~Sohler, and W.~Tittel,
  ``Spectral multiplexing for scalable quantum photonics using an atomic
  frequency comb quantum memory and feed-forward control,'' {\em Phys. Rev.
  Lett.}, vol.~113, p.~053603, Jul 2014.

\bibitem{Saglamyurek:2011fk}
E.~Saglamyurek, N.~Sinclair, J.~Jin, J.~Slater, D.~Oblak, F.~Bussi{\`e}res,
  M.~George, R.~Ricken, W.~Sohler, and W.~Tittel, ``Broadband waveguide quantum
  memory for entangled photons,'' {\em Nature}, vol.~469, no.~7331,
  pp.~512--515, 2011.

\bibitem{Afzelius:2010fk}
M.~Afzelius, I.~Usmani, A.~Amari, B.~Lauritzen, A.~Walther, C.~Simon,
  N.~Sangouard, J.~Min{\'a}{\v{r}}, H.~De~Riedmatten, N.~Gisin, {\em et~al.},
  ``{Demonstration of atomic frequency comb memory for light with spin-wave
  storage},'' {\em Physical review letters}, vol.~104, no.~4, p.~40503, 2010.

\bibitem{Timoney:2013}
N.~Timoney, I.~Usmani, P.~Jobez, M.~Afzelius, and N.~Gisin,
  ``Single-photon-level optical storage in a solid-state spin-wave memory,''
  {\em Phys. Rev. A}, vol.~88, p.~022324, Aug 2013.

\bibitem{Guendogan:2013}
M.~G{\"u}ndo{\u g}an, M.~Mazzera, P.~M. Ledingham, M.~Cristiani, and
  H.~de~Riedmatten, ``Coherent storage of temporally multimode light using a
  spin-wave atomic frequency comb memory,'' {\em New Journal of Physics},
  vol.~15, no.~4, p.~045012, 2013.

\bibitem{Jobez:2014}
P.~Jobez, I.~Usmani, N.~Timoney, C.~Laplane, N.~Gisin, and M.~Afzelius,
  ``Cavity-enhanced storage in an optical spin-wave memory,'' {\em New Journal
  of Physics}, vol.~16, no.~8, p.~083005, 2014.

\bibitem{Kim:2010}
Y.-W. Cho and Y.-H. Kim, ``Atomic vapor quantum memory for a photonic
  polarization qubit,'' {\em Opt. Express}, vol.~18, pp.~25786--25793, Dec
  2010.

\bibitem{Gruendogan:2012}
M.~G\"undo\ifmmode~\breve{g}\else \u{g}\fi{}an, P.~M. Ledingham, A.~Almasi,
  M.~Cristiani, and H.~de~Riedmatten, ``Quantum storage of a photonic
  polarization qubit in a solid,'' {\em Phys. Rev. Lett.}, vol.~108, p.~190504,
  May 2012.

\bibitem{Clausen:2012}
C.~Clausen, F.~Bussi\`eres, M.~Afzelius, and N.~Gisin, ``Quantum storage of
  heralded polarization qubits in birefringent and anisotropically absorbing
  materials,'' {\em Phys. Rev. Lett.}, vol.~108, p.~190503, May 2012.

\bibitem{Zhang:2011aa}
H.~Zhang, X.-M. Jin, J.~Yang, H.-N. Dai, S.-J. Yang, T.-M. Zhao, J.~Rui, Y.~He,
  X.~Jiang, F.~Yang, {\em et~al.}, ``Preparation and storage of
  frequency-uncorrelated entangled photons from cavity-enhanced spontaneous
  parametric downconversion,'' {\em Nature Photonics}, vol.~5, no.~10,
  pp.~628--632, 2011.

\bibitem{Migdall:book}
A.~Migdall, S.~V. Polyakov, J.~Fan, and J.~C. Bienfang, {\em Single-Photon
  Generation and Detection: Physics and Applications}.
\newblock Academic Press; 1 edition, 2013.

\bibitem{Steck:LectureNotes}
D.~A. Steck, ``Quantum and atom optics lecture notes,'' 2007.

\bibitem{Penzkofer:1979}
A.~Penzkofer, A.~Laubereau, and W.~Kaiser, ``High intensity raman
  interactions,'' {\em Progress in Quantum Electronics}, vol.~6, no.~2, pp.~55
  -- 140, 1979.

\bibitem{Raymer:1985bv}
M.~G. Raymer and I.~A. Walmsley, ``Quantum theory of spatial and temporal
  coherence properties of stimulated {R}aman scattering,'' {\em Phys. Rev. A},
  vol.~32(1), pp.~332--344, 1985.

\bibitem{ShoreBook}
B.~W. Shore, {\em Manipulating Quantum States Using Laser Pulses}.
\newblock Cambridge University Press, 2011.

\bibitem{England:2013}
D.~G. England, P.~J. Bustard, J.~Nunn, R.~Lausten, and B.~J. Sussman, ``From
  photons to phonons and back: A $\text{THz}$ optical memory in diamond,'' {\em
  Phys. Rev. Lett.}, vol.~111, p.~243601, Dec 2013.

\bibitem{Hosseini:2012}
M.~Hosseini, B.~M. Sparkes, G.~T. Campbell, P.~K. Lam, and B.~C. Buchler,
  ``Storage and manipulation of light using a raman gradient-echo process,''
  {\em Journal of Physics B: Atomic, Molecular and Optical Physics}, vol.~45,
  no.~12, p.~124004, 2012.

\bibitem{Zeuthen:2011}
E.~Zeuthen, A.~Grodecka-Grad, and A.~S. S\o{}rensen, ``Three-dimensional theory
  of quantum memories based on $\ensuremath{\Lambda}$-type atomic ensembles,''
  {\em Phys. Rev. A}, vol.~84, p.~043838, Oct 2011.

\bibitem{Loudon:2004gd}
R.~Loudon, {\em The Quantum Theory of Light}.
\newblock Oxford University Press, 2004.

\bibitem{Steck:2008qf}
D.~A. Steck, ``Cesium {D} line data,'' {\em Theoretical Division (T-8), Los
  Alamos National Laboratory}, 2008.

\bibitem{Lukin:2003vn}
M.~Lukin, ``{Colloquium: Trapping and manipulating photon states in atomic
  ensembles},'' {\em Reviews of Modern Physics}, vol.~75, no.~2, pp.~457--472,
  2003.

\bibitem{Dicke:1954fk}
R.~Dicke, ``Coherence in spontaneous radiation processes,'' {\em Physical
  Review}, vol.~93, no.~1, p.~99, 1954.

\bibitem{Wieczorek:2009kx}
W.~Wieczorek, R.~Krischek, N.~Kiesel, P.~Michelberger, G.~T{\'o}th, and
  H.~Weinfurter, ``Experimental entanglement of a six-photon symmetric dicke
  state,'' {\em Physical review letters}, vol.~103, no.~2, p.~20504, 2009.

\bibitem{Raymer:1977}
M.~G. Raymer and J.~L. Carlsten, ``Simultaneous observations of stimulated
  raman scattering and stimulated collision-induced fluorescence,'' {\em Phys.
  Rev. Lett.}, vol.~39, pp.~1326--1329, Nov 1977.

\bibitem{Wu:2010}
C.~Wu, M.~G. Raymer, Y.~Y. Wang, and F.~Benabid, ``Quantum theory of phase
  correlations in optical frequency combs generated by stimulated raman
  scattering,'' {\em Phys. Rev. A}, vol.~82, p.~053834, Nov 2010.

\bibitem{Saunders:2015}
D.~J. Saunders, J.~H.~D. Munns, T.~F.~M. Champion, C.~Qiu, K.~T. Kaczmarek,
  E.~Poem, P.~M. Ledingham, I.~A. Walmsley, and N.~J., ``A cavity-enhanced
  room-temperature broadband raman memory,'' {\em arXiv:1510.04625}, 2015.

\bibitem{Nunn:2008xr}
K.~Surmacz, J.~Nunn, K.~Reim, K.~C. Lee, V.~O. Lorenz, B.~Sussman, I.~A.
  Walmsley, and D.~Jaksch, ``Efficient spatially resolved multimode quantum
  memory,'' {\em Physical Review A (Atomic, Molecular, and Optical Physics)},
  vol.~78, no.~3, p.~033806, 2008.

\bibitem{Carlsten:1977}
J.~L. Carlsten, A.~Sz\"oke, and M.~G. Raymer, ``Collisional redistribution and
  saturation of near-resonance scattered light,'' {\em Phys. Rev. A}, vol.~15,
  pp.~1029--1045, Mar 1977.

\bibitem{Raymer:1981aa}
M.~Raymer and J.~Mostowski, ``Stimulated raman scattering: unified treatment of
  spontaneous initiation and spatial propagation,'' {\em Physical Review A},
  vol.~24, no.~4, p.~1980, 1981.

\bibitem{Raymer:1979}
M.~G. Raymer, J.~L. Carlsten, and G.~Pichler, ``Comparison of collisional
  redistribution and emission line shapes,'' {\em Journal of Physics B: Atomic
  and Molecular Physics}, vol.~12, no.~4, p.~L119, 1979.

\bibitem{Carman:1966}
R.~L. Carman, R.~Y. Chiao, and P.~L. Kelley, ``Observation of degenerate
  stimulated four-photon interaction and four-wave parametric amplification,''
  {\em Phys. Rev. Lett.}, vol.~17, pp.~1281--1283, Dec 1966.

\bibitem{Philips:2009}
N.~B. Phillips, A.~V. Gorshkov, and I.~Novikova, ``Slow light propagation and
  amplification via electromagnetically induced transparency and four-wave
  mixing in an optically dense atomic vapor,'' {\em Journal of Modern Optics},
  vol.~56, no.~18-19, pp.~1916--1925, 2009.

\bibitem{Phillips:2011}
N.~B. Phillips, A.~V. Gorshkov, and I.~Novikova, ``Light storage in an
  optically thick atomic ensemble under conditions of electromagnetically
  induced transparency and four-wave mixing,'' {\em Physical Review A},
  vol.~83, no.~6, p.~063823, 2011.

\bibitem{Camacho:2009ao}
R.~Camacho, P.~Vudyasetu, and J.~Howell, ``{Four-wave-mixing stopped light in
  hot atomic rubidium vapour},'' {\em Nature Photonics}, vol.~3, no.~2,
  pp.~103--106, 2009.

\bibitem{Kumar1994}
P.~Kumar and M.~I. Kolobov, ``Degenerate four-wave mixing as a source for
  spatially-broadband squeezed light,'' {\em Optics Communications}, vol.~104,
  no.~4--6, pp.~374 -- 378, 1994.

\bibitem{Grice:1997}
W.~P. Grice and I.~A. Walmsley, ``Spectral information and distinguishability
  in type-ii down-conversion with a broadband pump,'' {\em Phys. Rev. A},
  vol.~56, pp.~1627--1634, Aug 1997.

\bibitem{Reim2012}
K.~F. Reim, J.~Nunn, X.-M. Jin, P.~S. Michelberger, T.~F.~M. Champion, D.~G.
  England, K.~C. Lee, W.~S. Kolthammer, N.~K. Langford, and I.~A. Walmsley,
  ``Multipulse addressing of a raman quantum memory: Configurable beam
  splitting and efficient readout,'' {\em Phys. Rev. Lett.}, vol.~108,
  p.~263602, Jun 2012.

\bibitem{Michelberger:2014}
P.~S. Michelberger, T.~F.~M. Champion, M.~R. Sprague, K.~T. Kaczmarek,
  M.~Barbieri, X.~M. Jin, D.~G. England, W.~S. Kolthammer, D.~J. Saunders,
  J.~Nunn, and I.~A. Walmsley, ``Interfacing $\text{GHz}$-bandwidth heralded
  single photons with a warm vapour raman memory,'' {\em New Journal of
  Physics}, vol.~17, no.~4, p.~043006, 2015.

\bibitem{Demtroeder:ExPhys3}
W.~Demtr{\"o}der, {\em Experimentalphysik 3 - Atome, Molek{\"u}le und
  Festk{\"o}rper}.
\newblock Springer-Verlag, 2010.

\bibitem{Lewis:1980fk}
E.~L. Lewis, ``Collisional relaxation of atomic excited states, line broadening
  and interatomic interactions,'' {\em Physics Reports}, vol.~58, pp.~1--71, 2
  1980.

\bibitem{Liran:1980}
N.~D. Bhaskar, J.~Pietras, J.~Camparo, W.~Happer, and J.~Liran, ``Spin
  destruction in collisions between cesium atoms,'' {\em Phys. Rev. Lett.},
  vol.~44, pp.~930--933, Apr 1980.

\bibitem{Sushkov:2008}
A.~O. Sushkov and D.~Budker, ``Production of long-lived atomic vapor inside
  high-density buffer gas,'' {\em Phys. Rev. A}, vol.~77, p.~042707, Apr 2008.

\bibitem{Chrapkiewicz:2014fk}
R.~Chrapkiewicz, W.~Wasilewski, and C.~Radzewicz, ``How to measure diffusional
  decoherence in multimode rubidium vapor memories?,'' {\em Optics
  Communications}, vol.~317, pp.~1--6, 4 2014.

\bibitem{Omont:1972}
A.~Omont, W.~Smith, E, and J.~Cooper, ``Redistribution of resonance rediation:
  1. the effect of collisions,'' {\em The Astrophysical Journal}, vol.~175,
  pp.~185--199, 1972.

\bibitem{Krause:1966}
L.~Krause, ``Collisional excitation transfer between the $\text{P}^{2}_{1/2}$
  and $\text{P}^{2}_{3/2}$ levels in alkali atoms,'' {\em Appl. Opt.}, vol.~5,
  pp.~1375--1382, Sep 1966.

\bibitem{Rousseau:1975}
D.~L. Rousseau, G.~D. Patterson, and P.~F. Williams, ``Resonance raman
  scattering and collision-induced redistribution scattering in
  $\text{I}_{2}$,'' {\em Phys. Rev. Lett.}, vol.~34, pp.~1306--1309, May 1975.

\bibitem{Raymer:1983}
P.~D. Kleiber, J.~Cooper, K.~Burnett, C.~V. Kunasz, and M.~G. Raymer, ``Theory
  of time-dependent intense-field collisional resonance fluorescence,'' {\em
  Phys. Rev. A}, vol.~27, pp.~291--301, Jan 1983.

\bibitem{Manz:2007}
S.~Manz, T.~Fernholz, J.~Schmiedmayer, and J.-W. Pan, ``Collisional decoherence
  during writing and reading quantum states,'' {\em Phys. Rev. A}, vol.~75,
  p.~040101, Apr 2007.

\bibitem{Klein:2006}
M.~Klein, I.~Novikova, D.~F. Phillips, and R.~L. Walsworth, ``Slow light in
  paraffin-coated rb vapour cells,'' {\em Journal of Modern Optics}, vol.~53,
  no.~16-17, pp.~2583--2591, 2006.

\bibitem{Balabas:2010}
M.~V. Balabas, T.~Karaulanov, M.~P. Ledbetter, and D.~Budker, ``Polarized
  alkali-metal vapor with minute-long transverse spin-relaxation time,'' {\em
  Phys. Rev. Lett.}, vol.~105, p.~070801, Aug 2010.

\bibitem{Jiang:2009}
S.~Jiang, X.-M. Luo, L.-Q. Chen, B.~Ning, S.~Chen, J.-Y. Wang, Z.-P. Zhong, and
  J.-W. Pan, ``Observation of prolonged coherence time of the collective spin
  wave of an atomic ensemble in a paraffin-coated $^{87}\text{R}\text{b}$ vapor
  cell,'' {\em Phys. Rev. A}, vol.~80, p.~062303, Dec 2009.

\bibitem{Happer:1972}
W.~Happer, ``Optical pumping,'' {\em Rev. Mod. Phys.}, vol.~44, pp.~169--249,
  Apr 1972.

\bibitem{Bergman:2001}
N.~V. Vitanov, T.~Halfmann, B.~W. Shore, and K.~Bergmann, ``Laser-induced
  population transfer by adiabatic passage techniques,'' {\em Annual Review of
  Physical Chemistry}, vol.~52, no.~1, pp.~763--809, 2001.
\newblock PMID: 11326080.

\bibitem{Vurgaftman:2013aa}
I.~Vurgaftman and M.~Bashkansky, ``Suppressing four-wave mixing in
  warm-atomic-vapor quantum memory,'' {\em Physical Review A}, vol.~87, no.~6,
  p.~063836, 2013.

\bibitem{Walther:2005uq}
P.~Walther, K.~J. Resch, T.~Rudolph, E.~Schenck, H.~Weinfurter, V.~Vedral,
  M.~Aspelmeyer, and A.~Zeilinger, ``Experimental one-way quantum computing,''
  {\em Nature}, vol.~434, pp.~169--176, 03 2005.

\bibitem{OBrien:2007zr}
J.~L. O'Brien, ``Optical quantum computing,'' {\em Science}, vol.~318,
  pp.~1567--1570, 12 2007.

\bibitem{Kok:2007}
P.~Kok, W.~Munro, K.~Nemoto, T.~Ralph, J.~Dowling, and G.~Milburn, ``Linear
  optical quantum computing with photonic qubits,'' {\em Rev. Mod. Phys.},
  vol.~79, pp.~135--174, Jan 2007.

\bibitem{OBrien:2009fk}
J.~L. O'Brien, A.~Furusawa, and J.~Vuckovic, ``Photonic quantum technologies,''
  {\em Nat Photon}, vol.~3, pp.~687--695, 12 2009.

\bibitem{Bouwmeester:1999}
D.~Bouwmeester, J.-W. Pan, M.~Daniell, H.~Weinfurter, and A.~Zeilinger,
  ``Observation of three-photon greenberger-horne-zeilinger entanglement,''
  {\em Phys. Rev. Lett.}, vol.~82, pp.~1345--1349, Feb 1999.

\bibitem{Jennewein:2000}
T.~Jennewein, C.~Simon, G.~Weihs, H.~Weinfurter, and A.~Zeilinger, ``Quantum
  cryptography with entangled photons,'' {\em Phys. Rev. Lett.}, vol.~84,
  pp.~4729--4732, May 2000.

\bibitem{Lim:2005}
Y.~Lim, A.~Beige, and L.~Kwek, ``Repeat-until-success linear optics distributed
  quantum computing,'' {\em Phys. Rev. Lett.}, vol.~95, p.~030505, Jul 2005.

\bibitem{Lim:2006}
Y.~Lim, S.~Barrett, A.~Beige, P.~Kok, and L.~Kwek, ``Repeat-until-success
  quantum computing using stationary and flying qubits,'' {\em Phys. Rev. A},
  vol.~73, p.~012304, Jan 2006.

\bibitem{Schumacher:1995}
B.~Schumacher, ``Quantum coding,'' {\em Phys. Rev. A}, vol.~51, pp.~2738--2747,
  Apr 1995.

\bibitem{Zhou:2012}
Z.-Q. Zhou, W.-B. Lin, M.~Yang, C.-F. Li, and G.-C. Guo, ``Realization of
  reliable solid-state quantum memory for photonic polarization qubit,'' {\em
  Phys. Rev. Lett.}, vol.~108, p.~190505, May 2012.

\bibitem{Ding:2014}
D.-S. Ding, Z.~Wei, Z.-Y. Zhou, S.~Shi, S.~Bao-Sen, and G.~Guang-Can, ``Raman
  quantum memory of photonic polarised entanglement,'' {\em arXiv:1410.7101
  [quant-ph]}, 2014.

\bibitem{England2012}
D.~G. England, P.~S. Michelberger, T.~F.~M. Champion, K.~F. Reim, K.~C. Lee,
  M.~R. Sprague, X.-M. Jin, N.~K. Langford, W.~S. Kolthammer, J.~Nunn, and
  I.~A. Walmsley, ``High-fidelity polarization storage in a gigahertz bandwidth
  quantum memory,'' {\em Journal of Physics B: Atomic, Molecular and Optical
  Physics}, vol.~45, p.~124008, June 2012.

\bibitem{Higginbottom:2012}
D.~B. Higginbottom, B.~M. Sparkes, M.~Rancic, O.~Pinel, M.~Hosseini, P.~K. Lam,
  and B.~C. Buchler, ``Spatial-mode storage in a gradient-echo memory,'' {\em
  Phys. Rev. A}, vol.~86, p.~023801, Aug 2012.

\bibitem{Nunn:2008oq}
J.~Nunn, K.~Reim, K.~C. Lee, V.~O. Lorenz, B.~J. Sussman, I.~A. Walmsley, and
  D.~Jaksch, ``Multimode memories in atomic ensembles,'' {\em Physical Review
  Letters}, vol.~101, no.~26, p.~260502, 2008.

\bibitem{OBrien:2003fk}
J.~L. O'Brien, G.~J. Pryde, A.~G. White, T.~C. Ralph, and D.~Branning,
  ``Demonstration of an all-optical quantum controlled-not gate,'' {\em
  Nature}, vol.~426, pp.~264--267, 11 2003.

\bibitem{Bhandari:1988}
R.~Bhandari and J.~Samuel, ``Observation of topological phase by use of a laser
  interferometer,'' {\em Phys. Rev. Lett.}, vol.~60, pp.~1211--1213, Mar 1988.

\bibitem{Fiorentino:2008}
M.~Fiorentino and R.~G. Beausoleil, ``Compact sources of polarization-entangled
  photons,'' {\em Opt. Express}, vol.~16, pp.~20149--20156, Nov 2008.

\bibitem{Broom:2010}
M.~Broome, A.~Fedrizzi, B.~Lanyon, I.~Kassal, A.~Aspuru-Guzik, and A.~White,
  ``Discrete single-photon quantum walks with tunable decoherence,'' {\em Phys.
  Rev. Lett.}, vol.~104, p.~153602, Apr 2010.

\bibitem{Langford:PhD}
N.~K. Langford, ``Encoding, manipulating and measuring quantum information in
  optics,'' {\em PhD thesis, University of Queensland}, 2007.

\bibitem{Novikova:2007kc}
I.~Novikova, A.~V. Gorshkov, D.~F. Phillips, A.~S. Sorensen, M.~D. Lukin, and
  R.~L. Walsworth, ``Optimal control of light pulse storage and retrieval,''
  {\em Physical Review Letters}, vol.~98, no.~24, p.~243602, 2007.

\bibitem{Jenkins:2006}
S.~D. Jenkins, D.~N. Matsukevich, T.~Chaneli\`ere, A.~Kuzmich, and T.~A.~B.
  Kennedy, ``Theory of dark-state polariton collapses and revivals,'' {\em
  Phys. Rev. A}, vol.~73, p.~021803, Feb 2006.

\bibitem{Mohapatra:2008cl}
A.~K. Mohapatra, M.~G. Bason, B.~Butscher, K.~J. Weatherill, and C.~S. Adams,
  ``A giant electro-optic effect using polarizable dark states,'' {\em Nat
  Phys}, vol.~4, pp.~890--894, 11 2008.

\bibitem{Curty:2005}
M.~Curty and N.~L\"utkenhaus, ``Intercept-resend attacks in the
  bennett-brassard 1984 quantum-key-distribution protocol with weak coherent
  pulses,'' {\em Phys. Rev. A}, vol.~71, p.~062301, Jun 2005.

\bibitem{Felix:2001}
S.~F{\'e}lix, N.~Gisin, A.~Stefanov, and H.~Zbinden, ``Faint laser quantum key
  distribution: Eavesdropping exploiting multiphoton pulses,'' {\em Journal of
  Modern Optics}, vol.~48, no.~13, pp.~2009--2021, 2001.

\bibitem{Massar:1995}
S.~Massar and S.~Popescu, ``Optimal extraction of information from finite
  quantum ensembles,'' {\em Phys. Rev. Lett.}, vol.~74, pp.~1259--1263, Feb
  1995.

\bibitem{Dusek:2000}
M.~Du\ifmmode~\check{s}\else \v{s}\fi{}ek, M.~Jahma, and N.~L\"utkenhaus,
  ``Unambiguous state discrimination in quantum cryptography with weak coherent
  states,'' {\em Phys. Rev. A}, vol.~62, p.~022306, Jul 2000.

\bibitem{Taylor:1909}
G.~Taylor, {\em Interference fringes with feeble light}, vol.~15.
\newblock Proc. Camb. Phil. Soc., 1909.

\bibitem{Joensson:1961}
C.~J{\"o}nsson, ``Elektroneninterferenzen an mehreren k{\"u}nstlich
  hergestellten feinspalten,'' {\em Zeitschrift f{\"u}r Physik}, vol.~161,
  no.~4, pp.~454--474, 1961.

\bibitem{Nielsen:2002}
M.~A. Nielsen, ``A simple formula for the average gate fidelity of a quantum
  dynamical operation,'' {\em Physics Letters A}, vol.~303, no.~4, pp.~249 --
  252, 2002.

\bibitem{Kupchak:PhD}
C.~Kupchak, {\em Complete Characterisation of Quantum Optical Processes with a
  Focus on Quantum Memory}.
\newblock PhD thesis, University of Calgary, 2013.

\bibitem{Bowdrey:2002}
M.~D. Bowdrey, D.~K. Oi, A.~J. Short, K.~Banaszek, and J.~A. Jones, ``Fidelity
  of single qubit maps,'' {\em Physics Letters A}, vol.~294, no.~5--6, pp.~258
  -- 260, 2002.

\bibitem{Lounis:2005}
B.~Lounis and M.~Orrit, ``Single-photon sources,'' {\em Reports on Progress in
  Physics}, vol.~68, no.~5, p.~1129, 2005.

\bibitem{Eisaman2011}
M.~D. Eisaman, J.~Fan, A.~Migdall, and S.~V. Polyakov, ``Invited review
  article: Single-photon sources and detectors,'' {\em Review of Scientific
  Instruments}, vol.~82, no.~7, pp.~--, 2011.

\bibitem{Takeuchi:2014}
S.~Takeuchi, ``Recent progress in single-photon and entangled-photon generation
  and applications,'' {\em Japanese Journal of Applied Physics}, vol.~53,
  no.~3, p.~030101, 2014.

\bibitem{Brecht:2011}
B.~Brecht, A.~Eckstein, A.~Christ, H.~Suche, and C.~Silberhorn, ``From quantum
  pulse gate to quantum pulse shaper---engineered frequency conversion in
  nonlinear optical waveguides,'' {\em New Journal of Physics}, vol.~13, no.~6,
  p.~065029, 2011.

\bibitem{Moseley:2008}
P.~J. Mosley, J.~S. Lundeen, B.~J. Smith, and I.~A. Walmsley, ``Conditional
  preparation of single photons using parametric downconversion: a recipe for
  purity,'' {\em New Journal of Physics}, vol.~10, no.~9, p.~093011, 2008.

\bibitem{Scholz:2009}
M.~Scholz, L.~Koch, and O.~Benson, ``Statistics of narrow-band single photons
  for quantum memories generated by ultrabright cavity-enhanced parametric
  down-conversion,'' {\em Phys. Rev. Lett.}, vol.~102, p.~063603, Feb 2009.

\bibitem{Hong:1986}
C.~K. Hong and L.~Mandel, ``Experimental realization of a localized one-photon
  state,'' {\em Phys. Rev. Lett.}, vol.~56, pp.~58--60, Jan 1986.

\bibitem{Eckstein:2011}
A.~Eckstein, A.~Christ, P.~J. Mosley, and C.~Silberhorn, ``Highly efficient
  single-pass source of pulsed single-mode twin beams of light,'' {\em Phys.
  Rev. Lett.}, vol.~106, p.~013603, Jan 2011.

\bibitem{Fernandez-Gonzalvo:2013}
X.~Fernandez-Gonzalvo, G.~Corrielli, B.~Albrecht, M.~Grimau, M.~Cristiani, and
  H.~de~Riedmatten, ``Quantum frequency conversion of quantum memory compatible
  photons to telecommunication wavelengths,'' {\em Opt. Express}, vol.~21,
  pp.~19473--19487, Aug 2013.

\bibitem{Fekete:2013}
J.~Fekete, D.~Riel\"ander, M.~Cristiani, and H.~de~Riedmatten,
  ``Ultranarrow-band photon-pair source compatible with solid state quantum
  memories and telecommunication networks,'' {\em Phys. Rev. Lett.}, vol.~110,
  p.~220502, May 2013.

\bibitem{Eisaman:2005}
M.~Eisaman, A.~Andr{\'e}, F.~Massou, M.~Fleischhauer, A.~Zibrov, and M.~Lukin,
  ``Electromagnetically induced transparency with tunable single-photon
  pulses,'' {\em Nature}, vol.~438, no.~7069, pp.~837--841, 2005.

\bibitem{Muecke:2013}
M.~M\"ucke, J.~Bochmann, C.~Hahn, A.~Neuzner, C.~N\"olleke, A.~Reiserer,
  G.~Rempe, and S.~Ritter, ``Generation of single photons from an atom-cavity
  system,'' {\em Phys. Rev. A}, vol.~87, p.~063805, Jun 2013.

\bibitem{Chaneliere:2005jh}
T.~Chaneliere, D.~N. Matsukevich, S.~D. Jenkins, S.~Y. Lan, T.~A.~B. Kennedy,
  and A.~Kuzmich, ``Storage and retrieval of single photons transmitted between
  remote quantum memories,'' {\em Nature}, vol.~438, pp.~833--836, 12 2005.

\bibitem{Sangouard:2008xr}
N.~Sangouard, C.~Simon, B.~Zhao, Y.-A. Chen, H.~de~Riedmatten, J.-W. Pan, and
  N.~Gisin, ``Robust and efficient quantum repeaters with atomic ensembles and
  linear optics,'' {\em Physical Review A (Atomic, Molecular, and Optical
  Physics)}, vol.~77, no.~6, p.~062301, 2008.

\bibitem{Lee:2012}
K.~C. Lee, B.~J. Sussman, M.~R. Sprague, P.~Michelberger, K.~F. Reim, J.~Nunn,
  N.~K. Langford, P.~J. Bustard, D.~Jaksch, and I.~A. Walmsley, ``Macroscopic
  non-classical states and terahertz quantum processing in room-temperature
  diamond,'' {\em Nat Photon}, vol.~6, pp.~41--44, 01 2012.

\bibitem{Saglamyurek:2014aa}
E.~{Saglamyurek}, N.~{Sinclair}, J.~A. {Slater}, D.~{Oblak}, and W.~{Tittel},
  ``{An integrated processor for photonic quantum states using a broadband
  light-matter interface},'' {\em ArXiv e-prints}, Feb. 2014.

\bibitem{Timoney:2012}
N.~Timoney, B.~Lauritzen, I.~Usmani, M.~Afzelius, and N.~Gisin, ``Atomic
  frequency comb memory with spin-wave storage in 153 eu 3 + :y 2 sio 5,'' {\em
  Journal of Physics B: Atomic, Molecular and Optical Physics}, vol.~45,
  no.~12, p.~124001, 2012.

\bibitem{Jobez:2015}
P.~Jobez, C.~Laplane, N.~Timoney, N.~Gisin, A.~Ferrier, P.~Goldner, and A.~M.,
  ``Coherent spin control at the quantum level in an ensemble-based optical
  memory,'' {\em ArXiv:1501.0398v1}, 2015.

\bibitem{Guendogan:2015}
M.~G{\"u}ndogan, P.~M. Ledingham, K.~Kutluer, M.~Mazzera, and H.~de~Riedmatten,
  ``A solid state spin-wave quantum memory for time-bin qubits,'' {\em
  arXiv:1501.03980v1}, 2015.

\bibitem{Mandel}
L.~Mandel and E.~Wolf, {\em Optical Coherence and Quantum Optics}.
\newblock 1995.

\bibitem{Yariv:hc}
A.~Yariv, {\em Quantum Electronics}.
\newblock Wiley, 1975.

\bibitem{Trebino:book}
R.~Trebino, {\em Frequency-resolved optical gating: The measurement of
  ultrashort laser pulses}.
\newblock Kluwer Academic Publishers, 2000.

\bibitem{Weinberg:1970}
D.~C. Burnham and D.~L. Weinberg, ``Observation of simultaneity in parametric
  production of optical photon pairs,'' {\em Phys. Rev. Lett.}, vol.~25,
  pp.~84--87, Jul 1970.

\bibitem{Rarity:1987}
J.~Rarity, P.~Tapster, and E.~Jakeman, ``Observation of sub-poissonian light in
  parametric downconversion,'' {\em Optics Communications}, vol.~62, no.~3,
  pp.~201 -- 206, 1987.

\bibitem{Loudon:1993}
J.~R. Jeffers, N.~Imoto, and R.~Loudon, ``Quantum optics of traveling-wave
  attenuators and amplifiers,'' {\em Phys. Rev. A}, vol.~47, pp.~3346--3359,
  Apr 1993.

\bibitem{Bocquillon:2009}
E.~Bocquillon, C.~Couteau, M.~Razavi, R.~Laflamme, and G.~Weihs, ``Coherence
  measures for heralded single-photon sources,'' {\em Phys. Rev. A}, vol.~79,
  p.~035801, Mar 2009.

\bibitem{Eisenberg:2004}
H.~S. Eisenberg, G.~Khoury, G.~A. Durkin, C.~Simon, and D.~Bouwmeester,
  ``Quantum entanglement of a large number of photons,'' {\em Phys. Rev.
  Lett.}, vol.~93, p.~193901, Nov 2004.

\bibitem{Yurke:1987}
B.~Yurke and M.~Potasek, ``Obtainment of thermal noise from a pure quantum
  state,'' {\em Phys. Rev. A}, vol.~36, pp.~3464--3466, Oct 1987.

\bibitem{Kurtsiefer:2001}
C.~Kurtsiefer, M.~Oberparleiter, and H.~Weinfurter, ``Generation of correlated
  photon pairs in type-ii parametric down conversion---revisited,'' {\em
  Journal of Modern Optics}, vol.~48, no.~13, pp.~1997--2007, 2001.

\bibitem{Kurtsiefer:2008}
A.~Ling, A.~Lamas-Linares, and C.~Kurtsiefer, ``Absolute emission rates of
  spontaneous parametric down-conversion into single transverse gaussian
  modes,'' {\em Phys. Rev. A}, vol.~77, p.~043834, Apr 2008.

\bibitem{Branczyk:2010}
A.~M. Branczyk, T.~C. Ralph, W.~Helwig, and C.~Silberhorn, ``Optimized
  generation of heralded fock states using parametric down-conversion,'' {\em
  New Journal of Physics}, vol.~12, no.~063001, 2010.

\bibitem{Boyd:2003nx}
R.~Boyd, {\em {Nonlinear Optics}}.
\newblock Academic Press, 2003.

\bibitem{HandbookNonlinearCrystals}
V.~G. Dmitriev, G.~G. Gurzadyan, and D.~N. Nikogosyan, {\em Handbook of
  nonlinear crystals}.
\newblock Springer-Verlag, 1999.

\bibitem{Moseley:2009}
P.~J. Mosley, A.~Christ, A.~Eckstein, and C.~Silberhorn, ``Direct measurement
  of the spatial-spectral structure of waveguided parametric down-conversion,''
  {\em Phys. Rev. Lett.}, vol.~103, p.~233901, Dec 2009.

\bibitem{Karpinski:2012}
M.~Karpi\'{n}ski, C.~Radzewicz, and K.~Banaszek, ``Dispersion-based control of
  modal characteristics for parametric down-conversion in a multimode
  waveguide,'' {\em Opt. Lett.}, vol.~37, pp.~878--880, Mar 2012.

\bibitem{Fiorentino:2007}
M.~Fiorentino, S.~M. Spillane, R.~G. Beausoleil, T.~D. Roberts, P.~Battle, and
  M.~W. Munro, ``Spontaneous parametric down-conversion in periodically poled
  $\text{KTP}$ waveguides and bulk crystals,'' {\em Opt. Express}, vol.~15,
  pp.~7479--7488, Jun 2007.

\bibitem{Ramelow:2013}
S.~Ramelow, A.~Mech, M.~Giustina, S.~Gr\"{o}blacher, W.~Wieczorek, J.~Beyer,
  A.~Lita, B.~Calkins, T.~Gerrits, S.~W. Nam, A.~Zeilinger, and R.~Ursin,
  ``Highly efficient heralding of entangled single photons,'' {\em Opt.
  Express}, vol.~21, pp.~6707--6717, Mar 2013.

\bibitem{Fedrizzi:2007}
A.~Fedrizzi, T.~Herbst, A.~Poppe, T.~Jennewein, and A.~Zeilinger, ``A
  wavelength-tunable fiber-coupled source of narrowband entangled photons,''
  {\em Opt. Express}, vol.~15, pp.~15377--15386, Nov 2007.

\bibitem{Bierlein:1989}
J.~D. Bierlein and H.~Vanherzeele, ``Potassium titanyl phosphate: properties
  and new applications,'' {\em J. Opt. Soc. Am. B}, vol.~6, pp.~622--633, Apr
  1989.

\bibitem{Vanherzeele:1992}
H.~Vanherzeele and J.~D. Bierlein, ``Magnitude of the nonlinear-optical
  coefficients of $\text{KTiOPO}_{4}$,'' {\em Opt. Lett.}, vol.~17,
  pp.~982--984, Jul 1992.

\bibitem{Kwiat:1995}
P.~G. Kwiat, K.~Mattle, H.~Weinfurter, A.~Zeilinger, A.~V. Sergienko, and
  Y.~Shih, ``New high-intensity source of polarization-entangled photon
  pairs,'' {\em Phys. Rev. Lett.}, vol.~75, pp.~4337--4341, Dec 1995.

\bibitem{Satyanarayan:1999}
M.~N. Satyanarayan, A.~Deepthy, and H.~L. Bhat, ``Potassium titanyl phosphate
  and its isomorphs: Growth, properties, and applications,'' {\em Critical
  Reviews in Solid State and Materials Sciences}, vol.~24, no.~2, pp.~103--191,
  1999.

\bibitem{Boulanger:1994}
B.~Boulanger, M.~M. Fejer, R.~Blachman, and P.~F. Bordui, ``Study of
  $\text{KTiOPO}_{4}$ gray‐tracking at 1064, 532, and 355 nm,'' {\em Applied
  Physics Letters}, vol.~65, no.~19, pp.~2401--2403, 1994.

\bibitem{Bierlein:1987}
J.~D. Bierlein, A.~Ferretti, L.~H. Brixner, and W.~Y. Hsu, ``Fabrication and
  characterization of optical waveguides in $\text{KTiOPO}_{4}$,'' {\em Applied
  Physics Letters}, vol.~50, no.~18, pp.~1216--1218, 1987.

\bibitem{Roelofs:1994}
M.~G. Roelofs, A.~Suna, W.~Bindloss, and J.~D. Bierlein, ``Characterization of
  optical waveguides in $\text{KTiOPO}_{4}$ by second harmonic spectroscopy,''
  {\em Journal of Applied Physics}, vol.~76, no.~9, pp.~4999--5006, 1994.

\bibitem{Eger:1997}
D.~Eger, M.~Oron, M.~Katz, A.~Reizman, G.~Rosenman, and A.~Skilar,
  ``Quasi-phase-matched waveguides in electric field poled, flux grown ktp,''
  {\em Electronics Letters}, vol.~33, no.~18, pp.~1548--1550, 1997.

\bibitem{Christ:2009}
A.~Christ, K.~Laiho, A.~Eckstein, T.~Lauckner, P.~J. Mosley, and C.~Silberhorn,
  ``Spatial modes in waveguided parametric down-conversion,'' {\em Phys. Rev.
  A}, vol.~80, p.~033829, Sep 2009.

\bibitem{Karpinski:2009}
M.~Karpi{\'n}ski, C.~Radzewicz, and K.~Banaszek, ``Experimental
  characterization of three-wave mixing in a multimode nonlinear ktiopo4
  waveguide,'' {\em Applied Physics Letters}, vol.~94, no.~18, pp.~--, 2009.

\bibitem{Fallahkhair:2008}
A.~Fallahkhair, K.~S. Li, and T.~E. Murphy, ``Vector finite difference
  modesolver for anisotropic dielectric waveguides,'' {\em J. Lightwave
  Technol.}, vol.~26, pp.~1423--1431, Jun 2008.

\bibitem{Kato:2002}
K.~Kato and E.~Takaoka, ``Sellmeier and thermo-optic dispersion formulas for
  $\text{KTP}$,'' {\em Appl. Opt.}, vol.~41, pp.~5040--5044, Aug 2002.

\bibitem{Christ:PhD}
A.~Christ, {\em Theory of ultrafast waveguided parametric down-conversion: From
  fundamentals to applications}.
\newblock PhD thesis, Universit{\"a}t Paderborn, 2013.

\bibitem{URen:PhD}
A.~U'Ren, {\em Multi-photon state engineering for quantum information
  processing applications}.
\newblock PhD thesis, University of Oxford, 2004.

\bibitem{Moseley:PhD}
P.~J. Mosley, {\em Generation of Heralded Single Photons in Pure Quantum
  States}.
\newblock PhD thesis, University of Oxford, 2007.

\bibitem{Krischek:2010ys}
R.~Krischek, W.~Wieczorek, A.~Ozawa, N.~Kiesel, P.~Michelberger, T.~Udem, and
  H.~Weinfurter, ``Ultraviolet enhancement cavity for ultrafast nonlinear
  optics and high-rate multiphoton entanglement experiments,'' {\em Nat
  Photon}, vol.~4, pp.~170--173, 03 2010.

\bibitem{URen:2006}
A.~B. U'Ren, C.~Silberhorn, R.~Erdmann, K.~Banaszek, W.~P. Grice, I.~A.
  Walmsley, and M.~G. Raymer, ``Generation of pure-state single-photon
  wavepackets by conditional preparation based on spontaneous parametric
  downconversion,'' {\em arXiv:quant-ph/0611019}, 2006.

\bibitem{Mosley:2008hs}
P.~J. Mosley, J.~S. Lundeen, B.~J. Smith, P.~Wasylczyk, A.~B. U'Ren,
  C.~Silberhorn, and I.~A. Walmsley, ``Heralded generation of ultrafast single
  photons in pure quantum states,'' {\em Physical Review Letters}, vol.~100,
  no.~13, p.~133601, 2008.

\bibitem{Law:2000}
C.~K. Law, I.~A. Walmsley, and J.~H. Eberly, ``Continuous frequency
  entanglement: Effective finite hilbert space and entropy control,'' {\em
  Phys. Rev. Lett.}, vol.~84, pp.~5304--5307, Jun 2000.

\bibitem{Eberly:2005oq}
J.~H. {Eberly}, ``{Schmidt Analysis of Pure-State Entanglement},'' {\em ArXiv
  Quantum Physics e-prints}, Aug. 2005.

\bibitem{Brecht:2014}
B.~Brecht, A.~Eckstein, R.~Ricken, V.~Quiring, H.~Suche, L.~Sansoni, and
  C.~Silberhorn, ``Demonstration of coherent time-frequency schmidt mode
  selection using dispersion-engineered frequency conversion,'' {\em Phys. Rev.
  A}, vol.~90, p.~030302, Sep 2014.

\bibitem{Lvovsky:2012}
P.~Palittapongarnpim, A.~MacRae, and A.~I. Lvovsky, ``Note: A monolithic filter
  cavity for experiments in quantum optics,'' {\em Review of Scientific
  Instruments}, vol.~83, no.~6, 2012.

\bibitem{Witlef:PhD}
W.~Wieczorek, {\em Multi-Photon Entanglement: Experimental Observation,
  Characterization, and Application of up to Six-Photon Entangles States}.
\newblock PhD thesis, Ludwig-Maximilians Universit{\"a}t M{\"u}nchen, 2009.

\bibitem{Palik:1996}
E.~D. Palik, H.~Boukari, and R.~W. Gammon, ``Experimental study of the effect
  of surface defects on the finesse and contrast of a fabry--perot
  interferometer,'' {\em Appl. Opt.}, vol.~35, pp.~38--50, Jan 1996.

\bibitem{McKay:1999}
J.~A. McKay, ``Single and tandem fabry--perot etalons as solar background
  filters for lidar,'' {\em Appl. Opt.}, vol.~38, pp.~5851--5858, Sep 1999.

\bibitem{Diplomarbeit}
P.~S. Michelberger, {\em Femtosecond pulsed enhancement cavity for mutli-photon
  entanglement Femtosecond pulsed enhancement cavity for multi-photon
  entanglement studies}.
\newblock PhD thesis, Technische Universit{\"a}t M{\"u}nchen, 2009.

\bibitem{Kiesel:PhD}
N.~Kiesel, ``Experiments on multiphoton entanglement,'' {\em PhD thesis, Ludwig
  Maximilians Universit{\"a}t, M{\"u}nchen}, 2007.

\bibitem{Reim:PhD}
K.~Reim, {\em Broadband optical quantum memory}.
\newblock PhD thesis, University of Oxford, 2011.

\bibitem{Spring:PhD}
J.~B. Spring, {\em Single photon generation and quantum computing with
  integrated photonics}.
\newblock PhD thesis, University of Oxford, 2014.

\bibitem{HBT:1956}
R.~Q.~T. R.~Hanbury~Brown, ``Correlation between photons in two coherent beams
  of light,'' {\em Nature}, vol.~177, p.~3, January 1956.

\bibitem{Lee:2011}
K.~C. Lee, M.~R. Sprague, B.~J. Sussman, J.~Nunn, N.~K. Langford, X.-M. Jin,
  T.~Champion, P.~Michelberger, K.~F. Reim, D.~England, D.~Jaksch, and I.~A.
  Walmsley, ``Entangling macroscopic diamonds at room temperature,'' {\em
  Science}, vol.~334, no.~6060, pp.~1253--1256, 2011.

\bibitem{Christ:2011}
A.~Christ, K.~Laiho, A.~Eckstein, K.~N. Cassemiro, and C.~Silberhorn, ``Probing
  multimode squeezing with correlation functions,'' {\em New Journal of
  Physics}, vol.~13, no.~3, p.~033027, 2011.

\bibitem{Spring:2013ab}
J.~Spring, P.~Salter, B.~Metcalf, P.~Humphreys, M.~Moore, N.~Thomas-Peter,
  M.~Barbieri, X.~Jin, N.~Langford, W.~Kolthammer, {\em et~al.}, ``On-chip low
  loss heralded source of pure single photons.,'' {\em Optics express},
  vol.~21, no.~11, pp.~13522--13532, 2013.

\bibitem{Christ:2013}
A.~Christ, B.~Brecht, W.~Mauerer, and C.~Silberhorn, ``Theory of quantum
  frequency conversion and type-ii parametric down-conversion in the high-gain
  regime,'' {\em New Journal of Physics}, vol.~15, no.~5, p.~053038, 2013.

\bibitem{Grangier:1986}
P.~Grangier, G.~Roger, and A.~Aspect, ``Experimental evidence for a photon
  anticorrelation effect on a beam splitter: A new light on single-photon
  interferences,'' {\em EPL (Europhysics Letters)}, vol.~1, no.~4, p.~173,
  1986.

\bibitem{Appel:2008aa}
J.~Appel, E.~Figueroa, D.~Korystov, M.~Lobino, and A.~Lvovsky, ``Quantum memory
  for squeezed light,'' {\em Physical review letters}, vol.~100, no.~9,
  p.~093602, 2008.

\bibitem{Jensen:2010aa}
K.~Jensen, W.~Wasilewski, H.~Krauter, T.~Fernholz, B.~M. Nielsen, M.~Owari,
  M.~Plenio, A.~Serafini, M.~Wolf, and E.~Polzik, ``Quantum memory for
  entangled continuous-variable states,'' {\em Nature Physics}, vol.~7, no.~1,
  pp.~13--16, 2010.

\bibitem{Bussieres:2014ab}
F.~Bussieres, C.~Clausen, A.~Tiranov, B.~Korzh, V.~B. Verma, S.~W. Nam,
  F.~Marsili, A.~Ferrier, P.~Goldner, H.~Herrmann, {\em et~al.}, ``Quantum
  teleportation from a telecom-wavelength photon to a solid-state quantum
  memory,'' {\em arXiv preprint arXiv:1401.6958}, 2014.

\bibitem{Bimbard:2014aa}
E.~Bimbard, R.~Boddeda, N.~Vitrant, A.~Grankin, V.~Parigi, J.~Stanojevic,
  A.~Ourjoumtsev, and P.~Grangier, ``Homodyne tomography of a single photon
  retrieved on demand from a cavity-enhanced cold atom memory,'' {\em Physical
  Review Letters}, vol.~112, no.~3, p.~033601, 2014.

\bibitem{Spring2013}
J.~B. Spring, P.~S. Salter, B.~J. Metcalf, P.~C. Humphreys, M.~Moore,
  N.~Thomas-Peter, M.~Barbieri, X.-M. Jin, N.~K. Langford, W.~S. Kolthammer,
  M.~J. Booth, and I.~A. Walmsley, ``On-chip low loss heralded source of pure
  single photons,'' {\em Opt. Express}, vol.~21, pp.~13522--13532, Jun 2013.

\bibitem{goldschmidt2013mode}
E.~A. Goldschmidt, F.~Piacentini, I.~R. Berchera, S.~V. Polyakov, S.~Peters,
  S.~K{\"u}ck, G.~Brida, I.~P. Degiovanni, A.~Migdall, and M.~Genovese, ``Mode
  reconstruction of a light field by multiphoton statistics,'' {\em Physical
  Review A}, vol.~88, no.~1, p.~013822, 2013.

\bibitem{Ruppert:Book}
D.~Ruppert, {\em Statistics and Finance: An Introduction}.
\newblock Springer-Verlag, 2006.

\bibitem{Walther:2007}
P.~Walther, M.~D. Eisaman, A.~Andre, F.~Massou, M.~Fleischauer, A.~S. Zibrov,
  and M.~D. Lukin, ``Generation of narrow-band polarization-entangled photon
  pairs for atomic quantum memories,'' {\em International Journal of Quantum
  Information}, vol.~05, no.~01n02, pp.~51--62, 2007.

\bibitem{Zhang:2014}
K.~Zhang, J.~Guo, L.~Q. Chen, C.~Yuan, Z.~Y. Ou, and W.~Zhang, ``Suppression of
  the four-wave-mixing background noise in a quantum memory retrieval process
  by channel blocking,'' {\em Phys. Rev. A}, vol.~90, p.~033823, Sep 2014.

\bibitem{England:2013aa}
D.~England, P.~Bustard, J.~Nunn, R.~Lausten, and B.~Sussman, ``From photons to
  phonons and back: A $\text{THz}$ optical memory in diamond,'' {\em Physical
  Review Letters}, vol.~111, no.~24, p.~243601, 2013.

\bibitem{Sprague:2014aa}
M.~R. Sprague, P.~S. Michelberger, T.~F.~M. Champion, D.~G. England, J.~Nunn,
  X.~M. Jin, W.~S. Kolthammer, A.~Abdolvand, P.~S. Russell, and I.~A. Walmsley,
  ``Broadband quantum memory in a hollow-core photonic-cyrstal fibre,'' {\em
  Nature Photonics}, to appear.

\bibitem{Simon:2007ct}
C.~Simon, H.~de~Riedmatten, M.~Afzelius, N.~Sangouard, H.~Zbinden, and
  N.~Gisin, ``Quantum repeaters with photon pair sources and multimode
  memories,'' {\em Phys. Rev. Lett.}, vol.~98, p.~190503, 2007.

\bibitem{Migdall:2002}
A.~L. Migdall, D.~Branning, and S.~Castelletto, ``Tailoring single-photon and
  multiphoton probabilities of a single-photon on-demand source,'' {\em Phys.
  Rev. A}, vol.~66, p.~053805, Nov 2002.

\bibitem{Meany:2014}
T.~Meany, L.~A. Ngah, M.~J. Collins, A.~S. Clark, R.~J. Williams, B.~J.
  Eggleton, M.~J. Steel, M.~J. Withford, O.~Alibart, and S.~Tanzilli, ``Hybrid
  photonic circuit for multiplexed heralded single photons,'' {\em Laser and
  Photonics Reviews}, vol.~8, no.~3, pp.~L42--L46, 2014.

\bibitem{Sprague:2013}
M.~R. Sprague, D.~G. England, A.~Abdolvand, J.~Nunn, X.-M. Jin, W.~S.
  Kolthammer, M.~Barbieri, B.~Rigal, P.~S. Michelberger, T.~F.~M. Champion,
  P.~S.~J. Russell, and I.~A. Walmsley, ``Efficient optical pumping and high
  optical depth in a hollow-core photonic-crystal fibre for a broadband quantum
  memory,'' {\em New Journal of Physics}, vol.~15, no.~5, p.~055013, 2013.

\bibitem{Phillips:2008uq}
N.~Phillips, A.~Gorshkov, and I.~Novikova, ``Optimal light storage in atomic
  vapor,'' {\em Physical Review A}, vol.~78, no.~2, p.~023801, 2008.

\bibitem{Ritter:2012fk}
S.~Ritter, C.~Nolleke, C.~Hahn, A.~Reiserer, A.~Neuzner, M.~Uphoff, M.~Mucke,
  E.~Figueroa, J.~Bochmann, and G.~Rempe, ``An elementary quantum network of
  single atoms in optical cavities,'' {\em Nature}, vol.~484, pp.~195--200, 04
  2012.

\bibitem{Sparkes:2013}
B.~M. Sparkes, J.~Bernu, M.~Hosseini, J.~Geng, Q.~Glorieux, P.~A. Altin, P.~K.
  Lam, N.~P. Robins, and B.~C. Buchler, ``Gradient echo memory in an ultra-high
  optical depth cold atomic ensemble,'' {\em New Journal of Physics}, vol.~15,
  no.~8, p.~085027, 2013.

\bibitem{Hansel:2001fk}
W.~Hansel, P.~Hommelhoff, T.~W. Hansch, and J.~Reichel, ``Bose-einstein
  condensation on a microelectronic chip,'' {\em Nature}, vol.~413,
  pp.~498--501, 10 2001.

\bibitem{Brugger:2000}
K.~Brugger, T.~Calarco, D.~Cassettari, R.~Folman, A.~Haase, B.~Hessmo,
  P.~Kr{\"u}ger, T.~Maier, and J.~Schmiedmayer, ``Nanofabricated atom optics:
  Atom chips,'' {\em Journal of Modern Optics}, vol.~47, no.~14-15,
  pp.~2789--2809, 2000.

\bibitem{Kozuma:2009}
K.~Akiba, K.~Kashiwagi, M.~Arikawa, and M.~Kozuma, ``Storage and retrieval of
  nonclassical photon pairs and conditional single photons generated by the
  parametric down-conversion process,'' {\em New Journal of Physics}, vol.~11,
  no.~1, p.~013049, 2009.

\bibitem{Lauk:2013}
N.~Lauk, C.~O'Brien, and M.~Fleischhauer, ``Fidelity of photon propagation in
  electromagnetically induced transparency in the presence of four-wave
  mixing,'' {\em Physical Review A}, vol.~88, no.~013823, p.~11, 2013.

\bibitem{Geng:2014}
J.~Geng, G.~T. Campbell, J.~Bernu, D.~B. Higginbottom, B.~M. Sparkes, S.~M.
  Assad, W.~P. Zhang, N.~P. Robins, P.~K. Lam, and B.~C. Buchler,
  ``Electromagnetically induced transparency and four-wave mixing in a cold
  atomic ensemble with large optical depth,'' {\em New Journal of Physics},
  vol.~16, no.~11, p.~113053, 2014.

\bibitem{Schmidt:1994}
O.~Schmidt, K.-M. Knaak, R.~Wynands, and D.~Meschede, ``Cesium saturation
  spectroscopy revisited: How to reverse peaks and observe narrow resonances,''
  {\em Applied Physics B}, vol.~59, no.~2, pp.~167--178, 1994.

\bibitem{RousseauJCP:1976}
D.~L. Rousseau and P.~F. Williams, ``Resonance raman scattering of light from a
  diatomic molecule,'' {\em The Journal of Chemical Physics}, vol.~64, no.~9,
  pp.~3519--3537, 1976.

\bibitem{Hosseini:2009lq}
M.~Hosseini, B.~M. Sparkes, G.~Hetet, J.~J. Longdell, P.~K. Lam, and B.~C.
  Buchler, ``Coherent optical pulse sequencer for quantum applications,'' {\em
  Nature}, vol.~461, pp.~241--245, 09 2009.

\bibitem{Reim2011_supp}
K.~F. Reim, P.~Michelberger, K.~C. Lee, J.~Nunn, N.~K. Langford, and I.~A.
  Walmsley, ``{Single-Photon-Level Quantum Memory at Room Temperature:
  Supplementary Material},'' {\em Physical Review Letters}, vol.~107, no.~5,
  2011.

\bibitem{Matsukevich:2006}
D.~N. Matsukevich, T.~Chaneli\`ere, S.~D. Jenkins, S.-Y. Lan, T.~A.~B. Kennedy,
  and A.~Kuzmich, ``Observation of dark state polariton collapses and
  revivals,'' {\em Phys. Rev. Lett.}, vol.~96, p.~033601, Jan 2006.

\bibitem{Lita:08}
A.~E. Lita, A.~J. Miller, and S.~W. Nam, ``Counting near-infrared
  single-photons with 95\% efficiency,'' {\em Opt. Express}, vol.~16,
  pp.~3032--3040, Mar 2008.

\bibitem{Thomas:2017}
S.~E. Thomas, J.~H.~D. Munns, K.~T. Kaczmarek, C.~Qui, B.~Brecht, A.~Feizpour,
  P.~M. Ledingham, I.~A. Walmsley, J.~Nunn, and D.~J. Saunders, ``High
  efficiency raman memory by suppressing radiation trapping,'' {\em Arxiv
  preprint arXiv:1610.03743v2}, 2017.

\bibitem{Sprague:PhD}
M.~R. Sprague, {\em Raman Memory for Entanglement in Diamonds and Light Storage
  in Optical Fibres}.
\newblock PhD thesis, University of Oxford, 2014.

\bibitem{Michelberger:2012}
P.~Michelberger, R.~Krischek, W.~Wieczorek, A.~Ozawa, and H.~Weinfurter,
  ``Interferometric autocorrelation in the ultraviolet utilizing spontaneous
  parametric down-conversion inside an enhancement cavity,'' {\em Opt. Lett.},
  vol.~37, pp.~1223--1225, Apr 2012.

\bibitem{Grechin:1999}
V.~I. Pryalkin, V.~A. Dyakov, V.~G. Dmitriev, and S.~G. Grechin,
  ``Temperature-independent phase matching for second-harmonic generation in a
  $\text{KTP}$ crystal,'' {\em Quantum Electronics}, vol.~29, no.~1,
  pp.~77--81, 1999.

\bibitem{MacAdam:1992}
K.~B. MacAdam, A.~Steinbach, and C.~Wieman, ``A narrow‐band tunable diode
  laser system with grating feedback, and a saturated absorption spectrometer
  for $\text{Cs}$ and $\text{Rb}$,'' {\em American Journal of Physics},
  vol.~60, no.~12, pp.~1098--1111, 1992.

\bibitem{Hori:1983}
H.~Hori, Y.~Kitayama, M.~Kitano, T.~Yabuzaki, and T.~Ogawa, ``Frequency
  stabilization of gaalas laser using a doppler-free spectrum of the
  $\text{Cs}$-$\text{D}_{2}$ line,'' {\em Quantum Electronics, IEEE Journal
  of}, vol.~19, pp.~169--175, Feb 1983.

\bibitem{Boyer:2013}
M.~T. Turnbull, P.~G. Petrov, C.~S. Embrey, A.~M. Marino, and V.~Boyer, ``Role
  of the phase-matching condition in nondegenerate four-wave mixing in hot
  vapors for the generation of squeezed states of light,'' {\em Phys. Rev. A},
  vol.~88, p.~033845, Sep 2013.

\bibitem{Chuang:1997}
I.~L. Chuang and M.~A. Nielsen, ``Prescription for experimental determination
  of the dynamics of a quantum black box,'' {\em Journal of Modern Optics},
  vol.~44, no.~11-12, pp.~2455--2467, 1997.

\bibitem{James:2001}
D.~F.~V. James, P.~G. Kwiat, W.~J. Munro, and A.~G. White, ``Measurement of
  qubits,'' {\em Phys. Rev. A}, vol.~64, p.~052312, Oct 2001.

\bibitem{Paris:2004kx}
M.~Paris and J.~Rehacek, ``{Quantum State Estimation (Lecture Notes in Physics,
  vol 649)},'' 2004.

\bibitem{Hecht}
E.~Hecht, {\em Optics}.
\newblock Addison-Wesley, 1998.

\bibitem{Langford:2004}
N.~Langford, R.~Dalton, M.~Harvey, J.~O'Brien, G.~Pryde, A.~Gilchrist,
  S.~Bartlett, and A.~White, ``Measuring entangled qutrits and their use for
  quantum bit commitment,'' {\em Phys. Rev. Lett.}, vol.~93, p.~053601, Jul
  2004.

\bibitem{James:2009}
M.~S. Kaznady and D.~F.~V. James, ``Numerical strategies for quantum
  tomography: Alternatives to full optimization,'' {\em Phys. Rev. A}, vol.~79,
  p.~022109, Feb 2009.

\bibitem{Gilchrist:2005}
A.~Gilchrist, N.~K. Langford, and M.~A. Nielsen, ``Distance measures to compare
  real and ideal quantum processes,'' {\em Phys. Rev. A}, vol.~71, p.~062310,
  Jun 2005.

\bibitem{boyd:convexopt}
S.~Boyd and V.~L., {\em Convex Optimization}.
\newblock Cambridge University Press, 2004.

\bibitem{Vandenberghe:1996}
L.~Vandenberghe and S.~Boyd, ``Semidefinite programming,'' {\em SIAM Review},
  vol.~38, no.~1, pp.~49--95, 1996.

\bibitem{Barreiro:2005}
J.~Barreiro, N.~Langford, N.~Peters, and P.~Kwiat, ``Generation of
  hyperentangled photon pairs,'' {\em Phys. Rev. Lett.}, vol.~95, p.~260501,
  Dec 2005.

\bibitem{Langford:2005}
N.~Langford, T.~Weinhold, R.~Prevedel, K.~Resch, A.~Gilchrist, J.~O'Brien,
  G.~Pryde, and A.~White, ``Demonstration of a simple entangling optical gate
  and its use in bell-state analysis,'' {\em Phys. Rev. Lett.}, vol.~95,
  p.~210504, Nov 2005.

\bibitem{nielsen2000qca}
M.~Nielsen and I.~Chuang, {\em {Quantum Computation and Quantum Information}}.
\newblock Cambridge University Press, 2000.

\bibitem{Jozsa:1994}
R.~Jozsa, ``Fidelity for mixed quantum states,'' {\em Journal of Modern
  Optics}, vol.~41, no.~12, pp.~2315--2323, 1994.

\bibitem{OBrien:2004}
J.~L. O'Brien, G.~J. Pryde, A.~Gilchrist, D.~F.~V. James, N.~K. Langford, T.~C.
  Ralph, and A.~G. White, ``Quantum process tomography of a controlled-not
  gate,'' {\em Phys. Rev. Lett.}, vol.~93, p.~080502, Aug 2004.

\bibitem{Kraus:1983}
K.~Kraus, A.~B{\"o}hm, J.~D. Dollard, and W.~Wootters, {\em States, effects and
  operations: fundamental notions of quantum theory}.
\newblock No.~190 in Lecture Notes in Physics, Springer, 1983.

\bibitem{Specht:PhD}
H.~P. Specht, {\em Einzelatom-Quantenspeicher f{\"u}r Polarisations-Qubits}.
\newblock PhD thesis, Technische Universit{\"a}t M{\"u}nchen,
  Max-Planck-Institut f{\"u}r Quantenoptik, 2010.

\bibitem{Lobino:2008}
M.~Lobino, D.~Korystov, C.~Kupchak, E.~Figueroa, B.~C. Sanders, and A.~I.
  Lvovsky, ``Complete characterization of quantum-optical processes,'' {\em
  Science}, vol.~322, no.~5901, pp.~563--566, 2008.

\bibitem{Lobino:2009fk}
M.~Lobino, C.~Kupchak, E.~Figueroa, and A.~Lvovsky, ``Memory for light as a
  quantum process,'' {\em Physical review letters}, vol.~102, no.~20,
  p.~203601, 2009.

\bibitem{Lvovsky:2011}
S.~Rahimi-Keshari, A.~Scherer, A.~Mann, A.~T. Rezakhani, A.~I. Lvovsky, and
  B.~C. Sanders, ``Quantum process tomography with coherent states,'' {\em New
  Journal of Physics}, vol.~13, no.~1, p.~013006, 2011.

\bibitem{Kupchak:2014}
C.~Kupchak, T.~Mittiga, B.~Jordaan, M.~Namazi, C.~N{\"o}lleke, and E.~Figueroa,
  ``Room-temperature quantum memory for polarization states,'' {\em
  arXiv:1405.6117 [quant-ph]}, 2014.

\bibitem{Hosseini:2011zr}
M.~Hosseini, G.~Campbell, B.~Sparkes, P.~Lam, and B.~Buchler, ``Unconditional
  room-temperature quantum memory,'' {\em Nature Physics}, vol.~7, no.~10,
  pp.~795--799, 2011.

\bibitem{Rhode:2007}
P.~P. Rohde, W.~Mauerer, and C.~Silberhorn, ``Spectral structure and
  decompositions of optical states, and their applications,'' {\em New Journal
  of Physics}, vol.~9, no.~4, p.~91, 2007.

\bibitem{Eberly:2006}
J.~Eberly, ``Schmidt analysis of pure-state entanglement,'' {\em Laser
  Physics}, vol.~16, no.~6, pp.~921--926, 2006.

\bibitem{Chen:2009}
J.~Chen, A.~J. Pearlman, A.~Ling, J.~Fan, and A.~L. Migdall, ``A versatile
  waveguide source of photon pairs for chip-scale quantum information
  processing,'' {\em Opt. Express}, vol.~17, pp.~6727--6740, Apr 2009.

\bibitem{Siegman}
A.~E. Siegman, {\em Lasers}.
\newblock University Sceince Books, 1990.

\bibitem{Stipcevic:10}
M.~Stip\v{c}evi\'{c}, H.~Skenderovi\'{c}, and D.~Gracin, ``Characterization of
  a novel avalanche photodiode for single photon detection in vis-nir range,''
  {\em Opt. Express}, vol.~18, pp.~17448--17459, Aug 2010.

\bibitem{Taylor:Book}
J.~R. Taylor, {\em An Introduction to Error Analysis: The Study of
  Uncertainties in Physical Measurements}.
\newblock University Science Books, 1996.

\bibitem{Boyer:2008}
C.~F. McCormick, A.~M. Marino, V.~Boyer, and P.~D. Lett, ``Strong low-frequency
  quantum correlations from a four-wave-mixing amplifier,'' {\em Phys. Rev. A},
  vol.~78, p.~043816, Oct 2008.

\bibitem{Boyer:2008aa}
V.~Boyer, A.~M. Marino, R.~C. Pooser, and P.~D. Lett, ``Entangled images from
  four-wave mixing,'' {\em Science}, vol.~321, no.~5888, pp.~544--547, 2008.

\bibitem{Lukin:1998}
M.~D. Lukin, P.~R. Hemmer, M.~L\"offler, and M.~O. Scully, ``Resonant
  enhancement of parametric processes via radiative interference and induced
  coherence,'' {\em Phys. Rev. Lett.}, vol.~81, pp.~2675--2678, Sep 1998.

\bibitem{Moiseev:2011}
S.~A. Moiseev, ``Photon-echo quantum memory with complete use of natural
  inhomogeneous broadening,'' {\em Phys. Rev. A}, vol.~83, p.~012307, Jan 2011.

\end{thebibliography}
\end{spacing}

\end{document}